MŰEGYETEM 1782

Budapest University of Technology and Economics
Department of Telecommunications
Mobile Communications and Quantum Technologies Laboratory

# Information Geometric Superactivation of Asymptotic Quantum Capacity and Classical Zero-Error Capacity of Zero-Capacity Quantum Channels

Ph.D. Thesis
of

Laszlo Gyongyosi

Thesis Supervisor:
Prof. Dr. Sandor Imre

Submitted to the Department of Telecommunications
in fulfillment of the requirements for the degree of

Doctor of Philosophy

at the

BUDAPEST UNIVERSITY OF TECHNOLOGY AND ECONOMICS

2013

# Abstract

The superactivation of zero-capacity quantum channels makes it possible to use two zero-capacity quantum channels with a positive joint capacity at the output. Currently, we have no theoretical background for describing all possible combinations of superactive zero-capacity channels; hence, there may be many other possible combinations. In this PhD Thesis I provide an algorithmic solution to the problem of superactivation and prove that superactivation effect is rooted in information geometric issues.

I show a fundamentally new method of finding the conditions for the superactivation of asymptotic quantum capacity of zero-capacity quantum channels. To discover these superactive zero-capacity channel-pairs in practice, we have to analyze an extremely large set of possible quantum states, channel models and channel probabilities. My proposed method can be a valuable tool for improving the results of fault-tolerant quantum computation and possible communication techniques over very noisy quantum channels. I introduce a method of finding the superactive quantum channel combinations for which the classical zero-error capacity can be superactivated. The zero-error capacity of the quantum channel describes the amount of information that can be transmitted perfectly through a noisy quantum channel. My proposed algorithmical framework is the first such solution to discover the still unknown superactive quantum channel combinations. The results make it possible to determine efficiently the conditions of superactivation of the zero-error capacity of quantum channels, without the extremely high computational costs exploiting the possibilities in information geometry.

The success of future long-distance quantum communications and global quantum key distribution systems largely depends on the development of efficient quantum repeaters. Using the proposed results on the superactivation of zero-error capacity of quantum channels, the efficiency of the quantum repeater can be increased greatly, which opens new possibilities for future long-distance quantum communications.

# Acknowledgements

I would like to express my thank and gratitude to my supervisor, *Prof. Dr. Sandor Imre*, for continuous and inestimable encouragement, support and advices.

I would like to thank the support of Sandor Csibi Ph.D. Researcher Scholarship at the Budapest University of Technology and Economics, the grant of TAMOP-4.2.2.B-10/1--2010-0009, COST Action MP1006, and the grant of the Hungarian Academy of Sciences.

I would like to thank my family for the support and assistance.

# Theses

**Thesisgroup 1.** *I proved that the superactivation of arbitrary dimensional quantum channels can be determined by means of an appropriate information geometric object. I discovered that the superactivation effect is rooted in information geometric issues.*

> **Thesis 1.1.** *I showed that the superactivation of arbitrary dimensional quantum channels can be determined by an abstract geometrical object called the quantum informational superball* [P1, P2, P3, B1, C1-C15], [Chapter 5, Appendix D, Appendix E].

> **Thesis 1.2.** *I proved that the radius of the quantum superball measures the superactivated capacities of the joint channel structure, where the elements of the joint structure are arbitrary dimensional quantum channels* [P1, P2, P3, B1, C1-C15], [Chapter 5, Appendix D, Appendix E].

> **Thesis 1.3.** *I proved that the superactivation of the joint structure of arbitrary quantum channels is determined by the properties of the quantum relative entropy function* [P1, P2, P3, P9, B1, C1-C15], [Chapter 5, Appendix D, Appendix E].

**Thesisgroup 2.** *I constructed an algorithm to determine the conditions of superactivation of the asymptotic quantum capacity of arbitrary dimensional quantum channels.*

> **Thesis 2.1.** *I showed that the superactivated single-use and asymptotic quantum capacity of the joint structure of arbitrary quantum channels can be determined by the proposed information geometric object* [P1, P2, P7, B1, C11-C15], [Chapter 6, Appendix D, Appendix E, Appendix F, Appendix G].

**Thesis 2.2.** *I proved that the proposed geometric properties can be exploited to construct an information geometric algorithm for the algorithmic superactivation of arbitrary dimensional quantum channels* [P1, P2, P7, P9, B1, C11-C15], [Chapter 6, Appendix D, Appendix E, Appendix F, Appendix G].

**Thesis 2.3.** *I constructed an efficient information geometric algorithm to study the superactivation of the quantum capacity of arbitrary quantum channels* [P1, P2, P7, B1, C11-C15], [Chapter 6, Appendix D, Appendix E, Appendix F, Appendix G].

**Thesisgroup 3.** *I proposed an algorithm to determine the conditions of superactivation of the classical zero-error capacity of arbitrary dimensional quantum channels. My proposed polynomial approximation method avoids the problem of NP-completeness.*

**Thesis 3.1.** *I showed that the superactive channel combinations and the input conditions of the superactivation of classical zero-error capacity of quantum channels can be discovered by the proposed information geometric approach* [P3, B1, C1-C10], [Chapter 7, Appendix D, Appendix E, Appendix F, Appendix G].

**Thesis 3.2.** *I proved that the superactivation of classical zero-error capacity of quantum channels can be analyzed by the proposed algorithm with minimized error by using the smaller subset of input density matrices* [P3, B1, C1-C10], [Chapter 7, Appendix D, Appendix E, Appendix F, Appendix G].

**Thesis 3.3.** *I proved that by using $\mu$-similar quantum informational distances and the weak core-set of quantum states, the superactivation of zero-error capacity of quantum channels can be determined by a polynomial approximation algorithm without the problem of NP-completeness* [P3, B1, C1-C10], [Chapter 7, Appendix D, Appendix E, Appendix F, Appendix G].

# Contents

“Marvelous, what ideas the young people have these days.

But I don't believe a word of it.”

*Albert Einstein (1927)*

# Chapter 1

# Introduction

"Nothing exists except atoms and empty space; everything else is opinion."

*Democritus of Abdera (ca. 400 BC)*

According to Moore's Law, the physical limitations of classical semiconductor-based technologies could be reached by 2020. We will then step into the *Quantum Age*. When first quantum computers become available on the shelf, today's encrypted information will not remain secure. Classical computational complexity will no longer guard this information. Quantum communication systems exploit the quantum nature of information offering new possibilities and limitations for engineers when designing protocols. In the first decade of the 21st century, many revolutionary properties of *quantum channel*s have been discovered. These phenomena were previously completely unimaginable. However, the picture has been changed. In the near future, advanced quantum communication and networking technologies driven by Quantum Information Processing will revolutionize the traditional methods. Quantum information will help to resolve still open scientific and technical prob-lems, as well as expand the boundaries of classical computation and communication systems.

The capacity of a communication channel describes the capability of the channel for delivering information from the sender to the receiver, in a faithful and recoverable way. The different capacities of quantum channels have been discovered just in the '90s, and there are still many open questions about the different capacity measures. Thanks to Shannon we can calculate the capacity of classical channels within the frames of classical Information Theory[1] [Shannon48]. However, for quantum channels, many new capacity definitions exist in comparison to a classical communication channel. In the case of a classical channel, we can send only classical information. Quantum channels extend the possibilities, and besides the classical information we can send entanglement-assisted classical information, private classical information, and of course, quantum information [Imre12]. On the other hand, the elements of classical Information Theory cannot be applied in general for quantum information—in other words, they can be used only in some special cases. There is no general formula to describe the capacity of every quantum channel model, but one of the main results of the recent researches was the "very simplified" picture, in which the various capacities of a quantum channel (i.e., the classical, private, quantum) are all non-additive. Contrary to classical channels, quantum channels can be used to construct more advanced communication primitives. Entanglement or the superposed states carry quantum information, which cannot be described classically. Moreover, in the quantum world there exist quantum transformations which can create entanglement or can control the properties of entanglement. Quantum channels can be implemented in practice very easily e.g. via optical fiber networks or by wireless optical channels, and make it possible to send various types of information. The errors are a natural interference from the noisy environment, and the can be much diverse due to the extended set of quantum channel models. The advanced properties of quantum channels were discovered mainly in the end of the

[1] *Quantum Shannon Theory (QST) has deep relevance concerning the information transmission and storage in quantum systems. It can be regarded as a natural generalization of classical Information Theory ore more precisely classical Information Theory represents a special, orthogonality-restricted case of Quantum Information Theory.*

2000s. These results of Quantum Information Theory (QIT) were completely unimaginable before, and the researchers were shocked, rather than just surprised.

Recently, the one of the most surprising discoveries in QIT was the possibility of the *superactivation* of quantum communication channels. The superactivation makes it possible to use zero-capacity quantum channels to transmit information. The effect of superactivation was discovered in 2008 [Smith08], and later, in 2009 it has been shown that both the classical zero-error [Duan09], [Cubitt09] and the quantum zero-error [Cubitt09a] capacities of quantum channels can be superactivated in certain cases. The complete theoretical background of the superactivation is currently unsolved, however, it is already known that it is based on the non-additivity (i.e., on the extreme violation of additivity) of the various quantum channel capacities.

This Ph.D Thesis is organized as follows. In the remaining parts of Chapter 1 we overview the content of the Ph.D Thesis. In Chapter 2 we introduce the reader to the representation of information stored in quantum states according to Quantum Information Theory. In Chapter 3, we study the classical capacities and the quantum capacity of a noisy quantum channel. We also discuss here the encoding and decoding of quantum information and the properties of the quantum capacity. In Chapter 4, we study the additivity problem of quantum channel capacities and the superactivation effect. Chapter 5 discusses the information geometric interpretation of the superactivation property. In Chapter 6 the proposed algorithmical solution to the superactivation of asymptotic quantum capacity is presented. Chapter 7 introduces the information geometric method of the superactivation of classical zero-error capacity. Finally, we conclude the results in Chapter 8. Supplementary material is included in the Appendix.

The complete 'historical' background with the description of the most relevant works can be found in the Related Work subsections in the Appendix of the Ph.D Thesis.

## 1.1 Emerging Quantum Influences

Quantum Computing has demonstrated its usefulness in the last decade with many new scientific discoveries. The quantum algorithms were under intensive research during the end of the twentieth century. But after Shor published the prime factorization method in 1994 [Shor94], and Grover introduced the quantum searching algorithm in 1996 [Grover96], results in the field of quantum algorithms tapered off somewhat. In the middle of the 90s, there was a silence in the field of quantum algorithms and this did not change until the beginning of the present century. This silence has been broken by the solution of some old number theoretic problems, which makes it possible to break some—but not just those which are RSA (Rivest-Shamir-Adleman [Rivest78]) based—very strong cryptosystems. Notably, these hard mathematical problems can now be solved by polynomial-time quantum algorithms. Later, these results have been extended to other number theoretic problems, and the revival of quantum computing is more intensive than ever [Imre12]. Public key classical cryptography relies heavily on the complexity of factoring integers (or similar problems such as discrete logarithm). Quantum Computers can use the Shor algorithm to break efficiently today's cryptosystems. We will need a new kind of cryptography in the future. Because classical cryptographic methods in wired and wireless systems are suffering vulnerabilities, new methods based on quantum mechanical principles have been developed. To break classical cryptosystems, several new different quantum algorithms (besides Shor's algorithm) can be developed and used in the future. After quantum computers are built, today's encrypted information will not stay secure anymore, because although the computational complexity of these classical schemes makes it hard for classical computers to solve them, they are not hard for quantum computers. Using classical computers, the efficiency of code breaking is restricted to polynomial time, however, with a quantum computer this tasks can be terminated exponentially faster.

On the other hand, these very important quantum algorithms cannot be used if there is no a stable framework of physical implementations which stands behind these theoretical results. There are many new results which have been published in the last decade on the development of such physical implementations, and many new paradigms have been provided. These physical implementations make it possible to use the theoretical results of quantum computation, such as quantum algorithms, and these developments give the theoretical background to the processing of quantum information. At the end of the twentieth century many new practical developments were realized, and many novel results introduced into the field of quantum computation and Quantum Information Processing (QIP). Another important research field regarding the properties of the physical implementations of quantum information is related to the decoherence and the preciseness of the measurement outcomes. Many researchers started to analyze the question, whether entanglement can help to increase the precision of quantum computation and the probabilities of the right measurement outcomes. The limitations of these quantum algorithms are a different question. This problem has brought about the need for the evolution of a new field in quantum computation: quantum complexity theory. The main task of this field is to clarify the computational limitations of quantum computation, and to analyze the relationship between classical problem classes and quantum problem classes. As the quantum computer becomes a reality, the classical problem classes have to be regrouped, and new subclasses have to be defined. The most important question is the description of the effects of quantum computational power on NP-complete problems. According to our current knowledge, quantum computers cannot solve NP-complete problems, hence if a problem is NP-complete in the terms of classical complexity theory, then it will remain NP-complete in terms of quantum complexity theory. On the other hand, as has been shown by Mosca and Stebila [Mosca06], there are still many open questions, and it is conceivable that new results will be born in the near future regarding this problem field. The last decade has introduced some new physical approaches to realize quantum circuits in practice. The design of quantum circuits involves the physical manipulation techniques of quantum states, the development of quantum states and the various techniques of measurement of the output. In the beginning of the evolution of this field in quantum computation, quantum states were identified with spins or other special degrees of freedom, with the ability to realize a two-level quantum system. In the last decade this concept has been changed, and it has been shown that quantum systems can be realized by collective system manipulation. Many new techniques have been developed in the last decade to implement a quantum computer in practice, using linear optics, adiabatic systems and entangled physical particles. At the end of the twentieth century many new practical developments were realized, and many novel results introduced into the field of quantum computation and Quantum Information Processing.

## 1.2 Quantum Information Theory

Chapter 2 summarizes the elements of Quantum Information Theory as they are known today, and we show their usage in practical communication. The theoretical background of communication over quantum channels is based on the fundamental results of QIT. We also analyze the still open questions in this field. The actual state of Quantum Information Theory reflects our current knowledge of the quantum world, and it also determines the success of quantum communication protocols and techniques. The primary employment of Quantum Information Theory is to describe quantum channel capacities, to measure entanglement, and to analyze the information-theoretic security of quantum cryptographic primitives. An important question in Quantum Information Theory is the description of the capacities of noisy quantum channels. In the case of Quantum Information Theory, we have to distinguish between classical and quantum information, either of which could be sent through the channel. If we would like to handle the errors of a quantum channel, and would like to construct efficient error-correcting schemes, or would like to describe the benefits of entanglement, then we have to know the fundamental theoretical background which allows us to realize these advanced results in practice. Quantum Information Theory is the corner-stone of quantum communication and Quantum Information Processing. The current state of Quantum Information Theory draws a picture from the currently available limits and possibilities in Quantum Information Processing, such as from the applications of these results in practice. The security of quantum cryptographic protocols and other private quantum communication schemes are also limited by the actual state of Quantum Information Theory. The phenomena of the quantum world cannot be described by the fundamental results of classical Information Theory. Quantum Information Theory is the natural extension of the results of classical Information Theory. But Quantum Information Theory brings something new into the global picture and helps to complete the missing, classically indescribable and even unimaginable parts. Quantum Information Theory lays down the theoretical background of Quantum Information Processing and synthesizes it with other aspects of quantum mechanics, such as quantum communication, secure and private quantum channels, or quantum error correction [Imre12].

The primary employment of Quantum Information Theory is to describe quantum channel capacities, to measure entanglement, and to analyze the information-theoretic security of quantum cryptographic primitives. In Fig. 1.1, we highlighted some important parts of Quantum Information Theory.

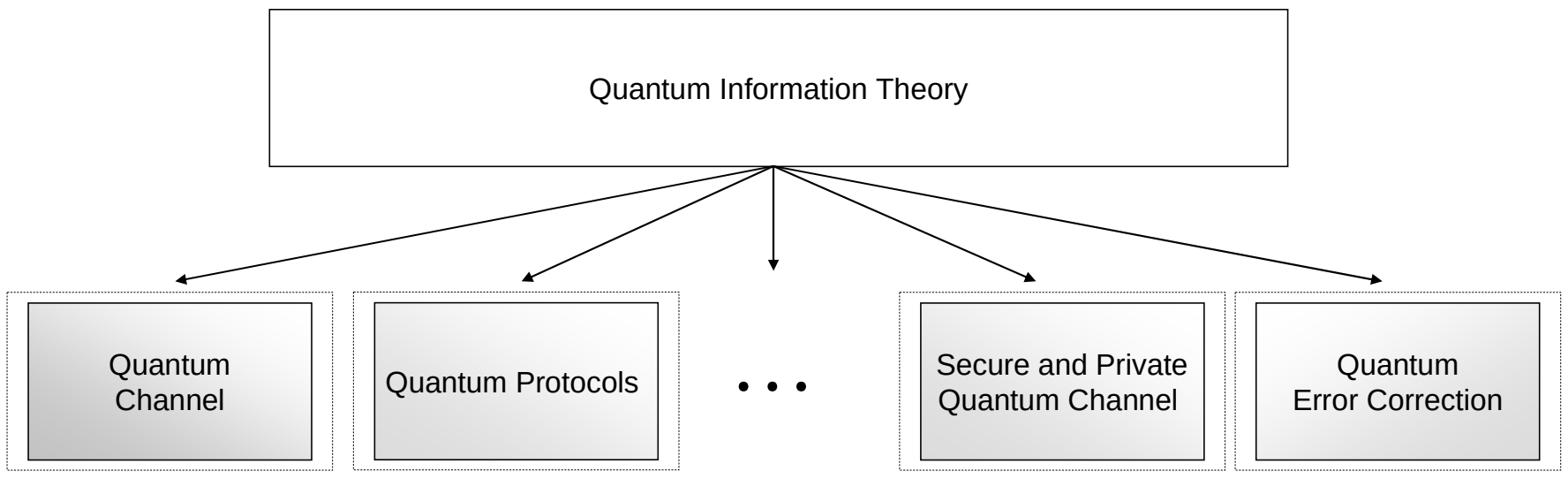

**Fig. 1.1.** Quantum Information Theory provides the theoretical background for various subjects in Quantum Information Processing.

With the help of Quantum Information Theory the information transmission through the quantum channel can be discussed for both classical and quantum information. The transmission of classical information through a quantum channel can be defined by a formula very similar to the classical Shannon channel coding theorem. On the other hand, the transmission of quantum information through a quantum channel has opened new dimensions in the transmission of information. As we will see, there are still many open questions in Quantum Information Theory. The various channel capacities of the quantum channels have been proven to be non-additive, however there are many special cases for which strict additivity holds. These fundamental questions will be discussed in detail in the Appendix of Ph.D Thesis. As follows from the connection defined between classical and Quantum Information Theory, every classical and quantum protocol can be described by using the elements of Quantum Information Theory. The definitions and main results of Quantum Information Theory, such as the density matrix, entanglement, measurement operators, von Neumann entropy, quantum relative entropy, Holevo bound, fidelity, and quantum informational distance, will be discussed in Chapter 2 and in Appendix B. The Related Work subsection is also included in Appendix B.

## 1.3 Quantum Channel Capacities

The concept of a quantum channel models communication on an abstract level, thus it does not require the deep analysis of the various physical systems, and instead it is enough to distill their essence from information transmission point of view. The *capacity* of a quantum channel gives us the rate at which classical or quantum information increase with each use of the quantum channel. We can define the *single-use (or single-letter)* and the *asymptotic capacity* of the quantum channel: the first quantifies the information which can be sent through a single use of the channel, the latter quantifies the information which can be transmitted if arbitrarily many uses of the quantum channel are allowed (*Note*: In this Ph.D dissertation the term *single-use* will be used). Many capacities can be defined for a quantum channels: it has a *classical capacity*, a *quantum capacity*, a *private capacity*, an *entanglement assisted capacity* and a *zero-error capacity* (classical and quantum). Some of these have also been defined in classical information theory, but many of these are completely new. The classical capacity of a quantum channel was first investigated by Holevo, who showed that from a two-level quantum state, or qubit, at most one bit classical information can be extracted. This theory is not contradictory to the fact that the description of a quantum state requires an infinite number of classical bits. As we will see, this one classical bit bound holds just for two-level quantum states (the *qubits*) since, in the case of a $d$-level quantum state (called *qudits*) this bound can be exceeded. The classical capacity of a quantum channel can be measured in different settings, depending on whether the input contains tensor product or entangled quantum states, and the output is measured by *single* or by *joint measurement settings*. These input and output combinations allow us to construct different channel settings, and the capacities in each case will be different. This is a completely new phenomenon in comparison with classical communication systems, where this kind of differentiation is not possible. The *additivity* of a quantum channel depends on the encoding scheme and on the measurement apparatus which is used for measuring the quantum states. If we use product input states, hence there is no entanglement among them, and if we do not apply joint measurement on the output then the classical capacity of a quantum channel will be additive, which means that the capacity can be achieved by a single use. If we use joint measurement on the outputs then such additivity is not guaranteed, which also suggests that in general the classical capacity is not additive. We note that many questions are still not solved in this field, as we will see later Chapter 3 where the properties of classical capacity of quantum channels will be discussed in details. The *classical capacity* of a quantum channel was formulated by Holevo, Schumacher and Westmoreland [Holevo98], [Schumacher97] and it is known in QIT as the *HSW channel capacity.* While the classical capacity measures classical information transmission over a noisy quantum channel, the *quantum capacity* of a quantum channel describes the amount of quantum information which can be transmitted through a noisy quantum channel. The formula of quantum capacity was introduced by Lloyd, Shor and Devetak in [Lloyd97], [Shor02], [Devetak03], and after the inventors it is called the *LSD channel capacity.* Both the HSW and the LSD channel capacities provide lower bounds on the ultimate limit for a noisy quantum channel to transmit classical or quantum information. One of the most important applications of quantum capacity calculations is the transmission of entanglement. The quantum error-correction techniques are developed for the optimization of quantum capacity in a noisy environment. As in the case of classical channel capacity where we will use the *Holevo information* as measure, for quantum capacity we will introduce a completely different correlation measure, the concept of *quantum coherent information.* We note that the generalized quantum channel capacity cannot be measured by the *single-use* version of quantum coherent information (or at least, it works just for some special channels), hence we have to compute the *asymptotic* form. This fact also implies that the additivity of the quantum capacities will be violated, too. These questions will be described in Chapter 4. Further supplementary information with the Related Work subsections are included in Appendices C and D. In Chapter 5 we show, that capacity of the quantum channels can be described in geometrical interpretation, and it can be used in the superactivation of the quantum channel capacities.

## 1.4 Motivation

Quantum channel additivity and the superactivation of the quantum channels are currently active areas of research in Quantum Information Theory. To this day, strict additivity for quantum channel capacity has been conjectured, but not proven. The additivity property of quantum channels is still an exciting subject of current research. The equality of channel capacities is known for some special cases, but the generalized rule is still *unknown.* Chapter 4 analyzes the advanced properties of the quantum channels, – such as the additivity property of quantum channel capacity and superactivation of zero-capacity quantum channels. In the proposed work we also investigate whether entanglement across input states could help to enhance the transmission of information on quantum channels – as entanglement can help in other problems in quantum computation. In practical applications, unentangled quantum states can be constructed easier than entangled states, hence transmission through optical or wireless quantum channels is mostly based on unentangled input states. If entanglement cannot help to improve the capacity of quantum channels, then unentangled input states could be used to achieve maximum channel capacity, which statement has deep relevance in practice.

The other, rather challenging nowadays problem of Quantum Information Theory – and the main subject of this Ph.D Thesis – is called *superactivation*. The problem of superactivation of quantum channel capacities will be introduced in Chapter 4. As shown in Fig. 1.2, the problem of superactivation can be discussed as part of a larger problem – the problem of quantum channel additivity. The superactivation property can be discussed from the viewpoint of the superactivation of quantum capacity or the classical and quantum zero-error capacities of the quantum channel.

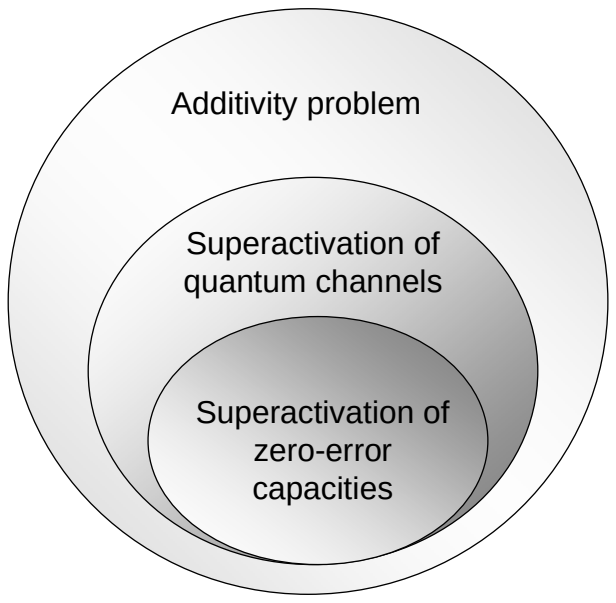

**Fig. 1.2.** The problem of superactivation of zero-capacity quantum channels as a sub domain of a larger problem set.

Supplementary information is included in Appendix D. The theses on the proposed algorithmic superactivation of quantum channel capacities will be presented in Chapter 5, Chapter 6 and Chapter 7. Further supplementary information and the Related Works subsections are included in Appendices E, F and G.

# 1.5 Research Objectives

Exploiting the fusion of the elements of Quantum Information Theory and computational geometry, many still open questions regarding on quantum channel capacities can be an-swered in a rather different way in comparison of the well-known methods. A plausible geometrical picture can be assigned to each channel model and instead of numerical calculations on their capacities; one can utilize the much straightforward geometric representation. The reader will be introduced to this interesting and rather surprising field in Chapter 5. Chapter 6 and Chapter 7 discuss the proposed information geometric approach for the superactivation of asymptotic quantum capacity and the classical zero-error capacity of quantum channels.

Initially, the superactivation property was proven for just one combination of two zero-capacity quantum channels, which can be used for the transmission of quantum information. In this combination, each quantum channel has zero quantum capacity individually, however their joint quantum capacity is strictly greater than zero. Later, these results have been extended. The superactivation has also opened a very large gap between the single-use quantum capacity and the asymptotic quantum capacity, which will be demonstrated in Chapter 4. With the help of superactivation the difference between the single-use and the asymptotic quantum capacity of a channel can be made arbitrarily large. (Since maximized quantum coherent information describes only the single-use quantum capacity of a quantum channel, in general it cannot be used to describe the asymptotic quantum capacity of a quantum channel.) The superactivation has opened the door which could clear up the question of the ability to transmit classical and quantum information through a noisy quantum channel. In 2009 it was discovered that the classical zero-error [Duan09], [Cubitt09] and the quantum zero-error [Cubitt09a] capacities of the quantum channel can also be superactivated. These could have many revolutionary practical consequences in the quantum communication networks of the future. With the help of superactivation, temporarily useless quantum channels (i.e., channels with individually zero quantum capacity or zero zero-error capacities) can be used together to avoid communication problems, and the capacities of the quantum channels can be increased. In the initial discovery of phenomenon of superactivation, only two classes of superactive zero-capacity quantum channels were known [Smith08]. Later the superactivation was extended to classes of generic channels which can be used for superactivation [Brandao11].

### 1.5.1 Research Methodology

The number of efficient approximation algorithms for quantum informational distances is very small, because of the special properties of quantum informational generator functions and of asymmetric quantum informational distances. If we wish to analyze the properties of quantum channels using today's classical computer architectures, an extremely efficient algorithm is needed. The numerical computation of the Holevo capacity and the classical zero-error capacity of quantum channels is an extremely difficult and hard computational problem (NP-complete), as has been stated by Beigi and Shor in 2007 [Beigi07]. The proposed solution avoids this problem. I show a fundamentally new method of finding the conditions for the superactivation of asymptotic quantum capacity and classical zero-error capacity of zero-capacity quantum channels, based on efficient information geometrical algorithms. To discover the *superactive* zero-capacity quantum channel-pairs, we have to analyze an extremely large set of possible quantum states, channel models and channel probabilities. Smith and Yard have found only one possible combination for superactivation of quantum capacity [Smith08]. Here, I show that this result can be confirmed with the proposed method, and this framework can be extended to discover other possible channels [Gyongyosi11b]. Currently, we have no theoretical results for describing all possible combinations of superactive zero-capacity channels; hence there should be many other possible combinations. With the help of my efficient computational geometric approach, the superactivation of zero-capacity quantum channels can be analyzed very efficiently.

I would like to analyze the properties of the quantum channel using classical computer architectures and algorithms since, currently, we have no quantum computers. To this day, the most efficient classical algorithms for this purpose are computational geomet-ric methods. I use these classical computational geometric tools to discover the still unknown *superactive* zero-capacity quantum channels, as depicted in Fig. 1.3.

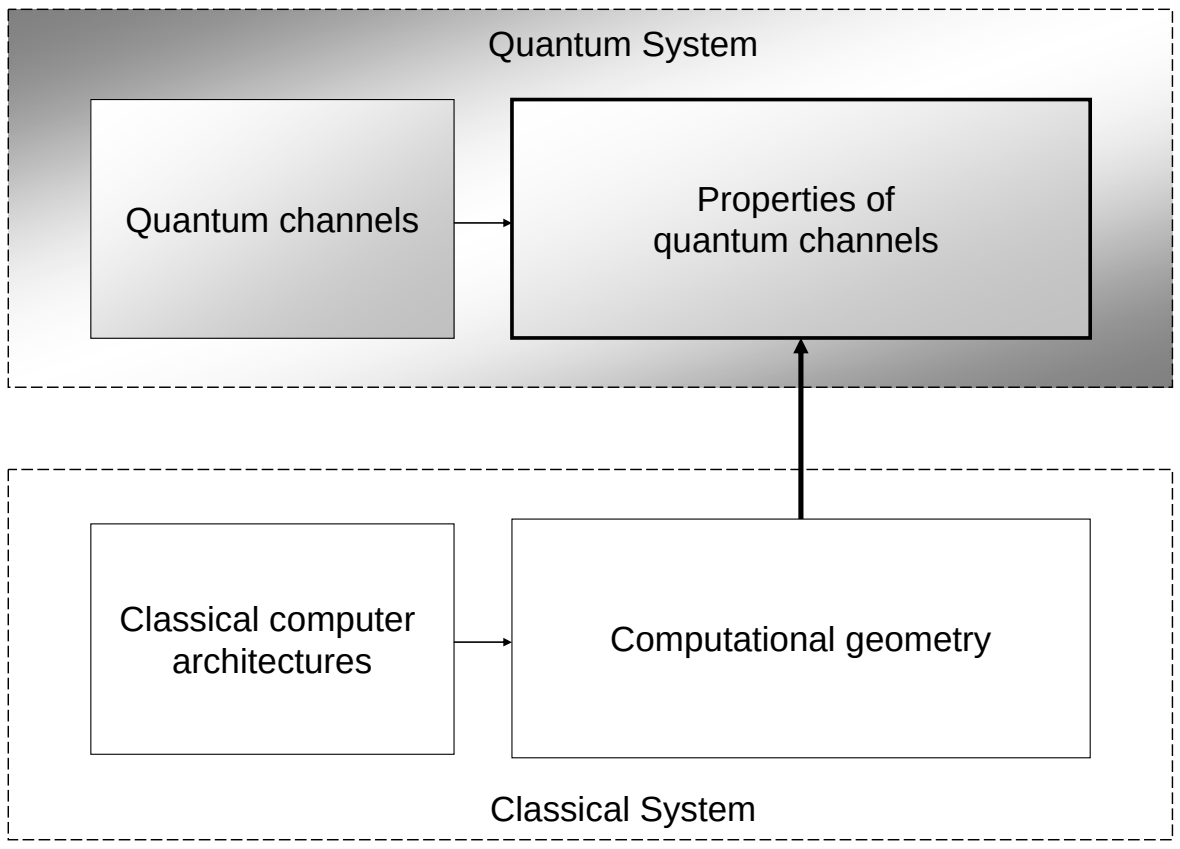

**Fig. 1.3.** The logical structure of the proposed analysis. I use current classical architectures to analyze the properties of quantum channels.

Computational Geometry was originally focused on the construction of efficient algorithms and provides a very valuable and efficient tool for computing hard tasks. In many cases, traditional linear programming methods are not very efficient.

In Chapter 6 and Chapter 7, I present that advanced geometric methods play a fundamental role in the analysis of the superactivation of zero-capacity quantum channels. To analyze a quantum channel for a large number of input quantum states with classical computer architectures, very fast and efficient algorithms are required. Here, I use these classical computational geometric tools to discover the still unknown superactive zero-capacity quantum channels. Unlike ordinary geometric distances, the quantum informational distance is not a metric and is not symmetric, hence this pseudo-distance features as a measure of informational distance. In my Ph.D Thesis I combine the results of Quantum Information Geometry (for example the quantum relative entropy-based distance functions) and the fast methods of Classical Computational Geometry (exploiting the elements and tools of information geometry) as illustrated in Fig. 1.4.

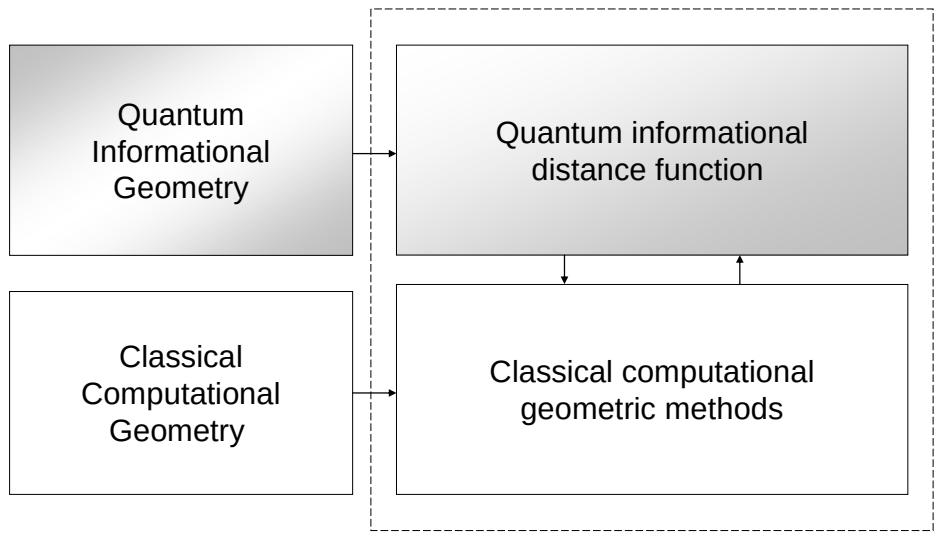

**Fig. 1.4.** Quantum information as distance measure in classical computational geometric methods.

At present, computational geometry algorithms are an active, widely used and integrated research field. Many difficult problems can be extended to computational geometric methods, however these geometric problems require well-designed and efficient algorithms [Gyongyosi11b]. In my work, I will use quantum informational function as distance measure instead of classical Euclidean distance function. The distances between the density matrices are calculated by the quantum relative entropy function. I combined the elements of computational geometry with the elements of the Hilbert space and the mathematical framework of quantum mechanics. I also used analytical and numerical analysis.

The results of the Ph.D Thesis demonstrate that computational geometric methods can support the analysis of superactivation of zero-capacity quantum channels very efficiently.

# Chapter 2

# Quantum Information Theory

"You don't understand quantum mechanics, you just get used to it."

*von Neumann*

Communication through a quantum channel cannot be described by the results of classical information theory; it requires the generalization of classical information theory by quantum perception of the world. In the general model of communication over a quantum channel $\mathcal{N}$, the encoder encodes the message in some coded form, and the receiver decodes it, however in this case, the whole communication is realized through a quantum system.

Chapter 2 is organized as follows. In the first part, we summarize the basic definitions and formulas of Quantum Information Theory. Next, we describe the encoding of quantum states and the meaning of Holevo information, the quantum mutual information and quantum conditional entropy. The description of a noisy quantum channel, purification, isometric extension, Kraus representation and the related works with the list of references can be found in Appendix B.

# 2.1 Introduction

The information sent through quantum channels is carried by quantum states, hence the encoding is fundamentally different from any classical encoder scheme. The encoding here means the preparation of a quantum system, according to the probability distribution of the classical message being encoded. Similarly, the decoding process is also different: here it means the measurement of the received quantum state. The preparation of quantum states, the measurement of the received states and the properties of quantum communication channel, and the fundamental differences between the classical and quantum communication channel cannot be described without the elements of Quantum Information Theory.

The model of the quantum channel represents the physically allowed transformations which can occur on the sent qubit. The result of the channel transformation is another density matrix. The physically allowed channel transformations could be very different; nevertheless they are always *Completely Positive Trace Preserving* (CPTP) transformations. The trace preserving property means that the corresponding density matrices at the input and output of the channel have the same trace. The input of a quantum channel is a quantum state, which encodes information into a physical property. The quantum state is sent through a quantum communication channel, which in practice can be implemented e.g. by an optical-fiber channel, or by a wireless quantum communication channel. To extract any information from the quantum state, it has to be measured at the receiver's side. The outcome of the measurement of the quantum state (which might be perturbed) depends on the transformation of the quantum channel, since it can be either totally probabilistic or deterministic. In contrast to classical channels, a quantum channel transforms the information coded into quantum states, which can be e.g. the spin state of the particle, the ground and excited state of an atom, or several other physical approaches.

# 2.2 The Quantum Channel

Besides the fact that the Bloch sphere provides a very useful geometrical approach to describe the density matrices, it also can be used to analyze the capacities of the various quantum channel models. From algebraic point of view, quantum channels are linear CPTP maps, while from a geometrical viewpoint, the quantum channel is an affine transformation. While, from the algebraic view the transformations are defined on density matrices, in the geometrical approach, the transformations are interpreted as Bloch vectors. Since, density matrices can be expressed in terms of Bloch vectors, hence the map of a quantum channel also can be analyzed in the geometrical picture.

The image of the quantum channel's linear transform is an *ellipsoid* on the Bloch sphere (see Fig. 2.1). To preserve the condition for a density matrix $\rho$, the noise on the quantum channel $\mathcal{N}$ must be trace-preserving (TP), i.e.,

$$Tr\left(\mathcal{N}\left(\rho\right)\right) = Tr\left(\rho\right), \tag{2.1}$$

and it must be Completely Positive (CP), i.e., for any identity map *I*, the map $I \otimes \mathcal{N}$ maps a semi-positive Hermitian matrix to a semi-positive Hermitian matrix.

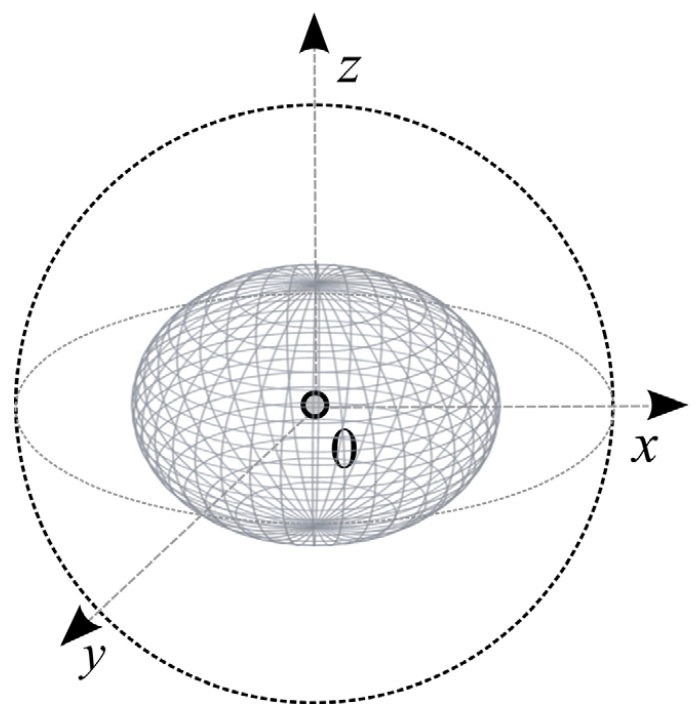

**Fig. 2.1.** Geometrically the image of the quantum channel is an ellipsoid.

The distinction of terms *unital* and *non-unital* quantum channels means the following thing: for a unital quantum channel $\mathcal{N}$, the channel map transforms the *I* identity matrix to the *I* identity matrix, while this condition does not hold for a non-unital channel. To express it, for a unital quantum channel, we have

$$\mathcal{N}\left(I\right) = I\,, \tag{2.2}$$

while for a non-unital quantum channel,

$$\mathcal{N}\left(I\right) \neq I\,. \tag{2.3}$$

As we will see, this difference can be rephrased in a geometrical interpretation, and the properties of the maps of the quantum channels can be analyzed using informational geometry.

For a unital quantum channel, the center of the geometrical interpretation of the channel ellipsoid is equal to the center of the Bloch sphere. This means that a unital quantum channel preserves the average of the system states. On the other hand, for a non-unital quantum channel, the center of the channel ellipsoid will differ from the center of the Bloch sphere. The main difference between unital and non-unital channels is that the non-unital channels do not preserve the average state in the center of the Bloch sphere. It follows from this that the numerical and algebraic analysis of non-unital quantum channels is more complicated than in the case of unital ones. While unital channels shrink the Bloch sphere in different directions with the center preserved, non-unital quantum channels shrink both the original Bloch sphere and move the center of the ball from the origin of the Bloch sphere. This fact makes our analysis more complex, however, in many cases, the physical systems cannot be described with unital quantum channel maps.

Unital channel maps can be expressed as convex combinations of the four unitary Pauli operators ($X$, $Y$, $Z$ and $I$), hence unital quantum maps are also called *Pauli channels*. Since the unital channel maps can be expressed as the convex combination of the basic unitary transformations, the unital channel maps can be represented in the Bloch sphere as different rotations with shrinking parameters [Nielsen2000], [Imre05]. On the other hand, for a non-unital quantum map, the map cannot be decomposed into a convex combination of unitary rotations and the transformation not just shrinks the ball, but also moves its center from the origin of the Bloch sphere [Imre12].

The geometrical interpretation of a unital and a non-unital quantum channels are illustrated in Fig. 2.2.

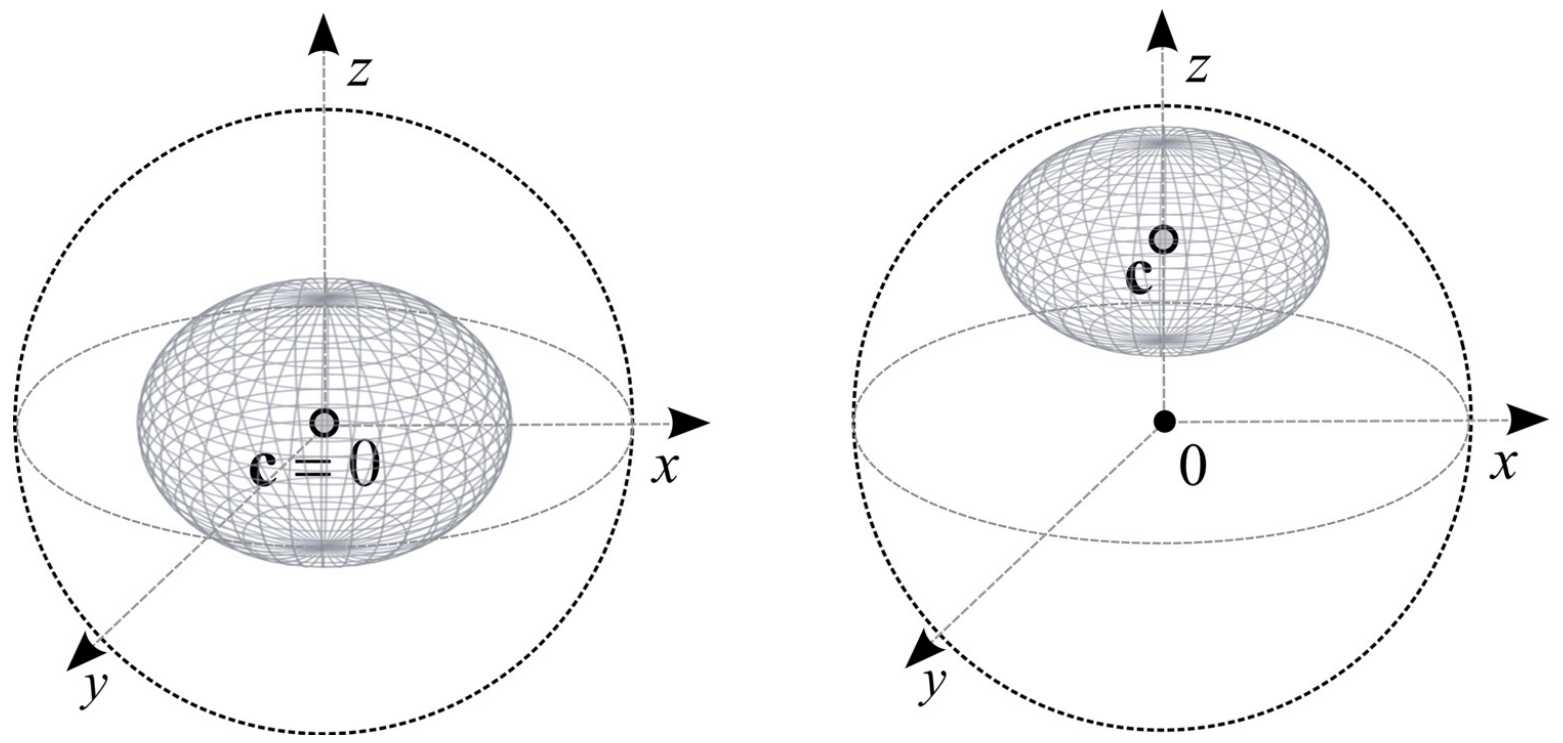

**Fig. 2.2.** The geometrical interpretation of a unital and a non-unital quantum channels.

The unital channel maps can be expressed as convex combinations of the basic unitary transformations, while non-unital quantum maps cannot be decomposed into a convex combination of unitary rotations, because of the geometrical differences between the two kinds of maps. The geometrical approaches can help to reduce the complexity of the analysis of the different quantum channel models, and as we will show, many algebraic results can be converted into geometrical problems. The connection between the channel maps and their geometrical interpretation on the Bloch sphere makes it possible to give a simpler and more elegant solution for several hard, and still unsolved problems.

For further supplementary information see Appendix B.

## 2.3 Quantum Channel Capacity

The capacity of a communication channel describes the capability of the channel for sending information from the sender to the receiver, in a faithful and recoverable way. The perfect idealistic communication channel realizes an identity map. In the case of a quantum communication channel, it means that the channel can transmit the quantum states perfectly. The capacity of the quantum channel measures the closeness to the idealistic identity transformation.

Quantum Information Processing exploits the quantum nature of information. It offers fundamentally new solutions in the field of computer science and extends the possibilities to a level that cannot be imagined in classical communication systems. For quantum communication channels, many new capacity definitions were developed in comparison to classical counterparts. A quantum channel can be used to realize classical information transmission or to deliver quantum information, such as quantum entanglement. To describe the information transmission capability of the quantum channel, we have to make a distinction between the various capacities of a quantum channel. The encoded quantum states can carry classical messages or quantum messages. In the case of classical messages, the quantum states encode the output from a classical information source, while in the case of quantum messages the source is a quantum information source. On one hand in the case of a classical communication, only one type of capacity measure can be defined, on the other hand for a quantum communication channel a number of different types of quantum channel capacities can be applied, with different characteristics. There are plenty of open questions regarding these various capacities. The *single-use* capacity of a quantum channel is not equal to the *asymptotic* capacity of the quantum channel, in general (We note, it also depends on the type of the quantum channel).

The encoding and the decoding mathematically can be described by the operators $\mathcal{E}$ and $\mathcal{D}$, realized on the blocks of quantum states. The model of communication through noisy quantum channel with encoding, delivery and decoding phases is illustrated in Fig. 2.3.

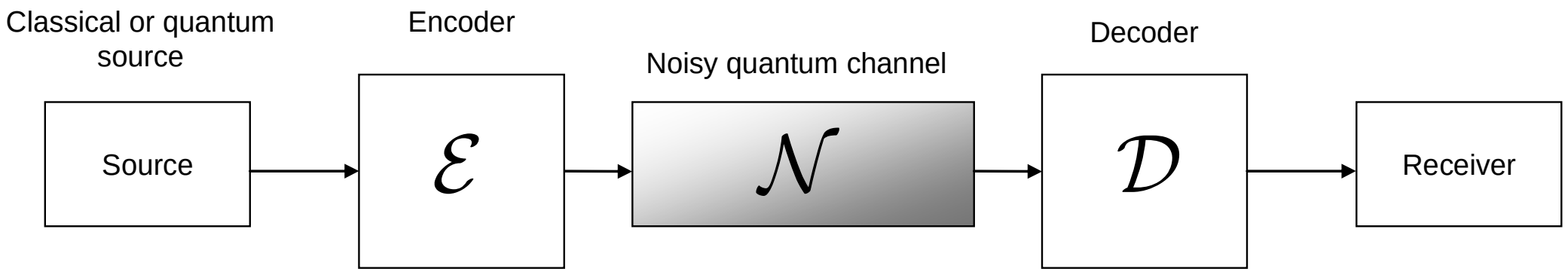

**Fig. 2.3.** Communication over noisy quantum channel. According to the noise of the quantum channel, the pure input state becomes a mixed state.

In the Ph.D Thesis the terms *classical quantity* and *quantum quantity* with relation to the quantum channel $\mathcal{N}$ are used as follows:

- *classical quantity*: it is a measure of the *classical transmission capabilities* of a quantum channel. (For example the Holevo information, quantum mutual information, etc.)
- *quantum quantity*: it is a measure of the *quantum transmission capabilities* of a quantum channel. (For example the quantum coherent information, see Chapter 3.)

If we mention classical quantity we will do this with relation to the quantum channel $\mathcal{N}$, i.e., for example the Holevo information is also not a "typical" classical quantity since it is describes a quantum system not a classical one, but with relation to the quantum channel we can use the *classical* mark.

## 2.4 Basic Definitions

Quantum Information Theory also has relevance to the discussion of the capacity of quantum channels and to information transmission and storage in quantum systems. While the transmission of product states can be described similar to classical information, on the other hand, the properties of quantum entanglement cannot be handled by the elements of classical Information Theory. Of course, the elements of classical Information Theory can be viewed as a subset of the larger and more complex Quantum Information Theory. Before starting the discussion on various capacities of quantum channels and the related consequences we summarize the basic definitions and formulas of Quantum Information Theory intended to represent the information stored in quantum states.

### 2.4.1 The von Neumann Entropy

Quantum Information Processing exploits the quantum nature of information. It offers fundamentally new solutions in the field of computer science and extends the possibilities to a level that cannot be imagined in classical communication systems. On the other hand, it requires the generalization of classical information theory through a quantum perception of the world. In order to measure the maximum amount of classical information that can be sent over quantum channels, we have to redefine the well-known formulas of classical information theory. In case of a quantum communication system, the information sent through quantum channels is carried by quantum states, hence the encoding is fundamentally different from any classical encoder scheme. Encoding here means the preparation of a quantum system, according to the probability distribution of the classical message being encoded. Similarly, the decoding process is also different: here it means the appropriate measurement of the received quantum state. These fundamental differences between the classical and quantum systems cannot be described without the elements of Quantum Information Theory.

As Shannon entropy plays fundamental role in classical information theory, the von Neumann entropy does the same for quantum information. The von Neumann entropy $\mathrm{S}(\rho)$ of quantum state $\rho$ can be viewed as an extension of classical entropy for quantum systems [Nielsen2000], [Wilde11]. It measures the information of the quantum states in the form of the uncertainty of a quantum state. The classical Shannon entropy $H(X)$ of a variable $X$ with probability distribution $p(x)$ can be defined as

$$H(X) = -\sum_{x \in X} p(x) \log(p(x)), \tag{2.4}$$

with $1 \leq H(X) \leq \log(|X|)$, where $|X|$ is the cardinality of the set $X$.

The von Neumann entropy

$$\mathrm{S}(\rho) = -Tr(\rho \log(\rho)) \tag{2.5}$$

measures the information contained in the quantum system $\rho$. Furthermore $\mathrm{S}(\rho)$ can be expressed by means of the Shannon entropy for the eigenvalue distribution

$$\mathrm{S}(\rho) = H(\lambda) = -\sum_{i=1}^{d} \lambda_i \log(\lambda_i), \tag{2.6}$$

where $d$ is the level of the quantum system and $\lambda_i$ are the eigenvalues of density matrix $\rho$.

### 2.4.2 The Holevo Quantity

The *Holevo bound* determines the amount of information that can be extracted from a single qubit state. If Alice sends a quantum state $\rho_i$ with probability $p_i$ over an idealistic quantum channel, then at Bob's receiver a mixed state

$$\rho_B = \rho_A = \sum_i p_i \rho_i \tag{2.7}$$

appears. Bob constructs a measurement $\{M_i\}$ to extract the information encoded in the quantum states. If he applies the measurement to $\rho_A$, the probability distribution of Bob's classical symbol $B$ will be $\Pr[b|\rho_A] = Tr\left(M_b^\dagger M_b \rho_A\right)$.

As had been shown by Holevo [Holevo73], the bound for the maximal classical mutual information between Alice and Bob is

$$I(A:B) \leq \mathrm{S}(\rho_A) - \sum_i p_i \mathrm{S}(\rho_i) \equiv \chi, \tag{2.8}$$

where $\chi$ is called the *Holevo quantity.* The Holevo quantity can be taken over all ensembles $\{p_i, \rho_i\}$ of input quantum states as

$$\chi = \mathrm{S}\left(\sum_i p_i \rho_i\right) - \sum_i p_i \mathrm{S}(\rho_i). \tag{2.9}$$

In classical information theory and classical communication systems, the mutual information $I(A:B)$ is bounded only by the classical entropy of $H(A)$, hence $I(A:B) \leq H(A)$.

The mutual information $I(A:B)$ is bounded by the classical entropy of $H(A)$, hence $I(A:B) \leq H(A)$. On the other hand, for mixed states and pure non-orthogonal states the Holevo quantity $\chi$ can be greater than the mutual information $I(A:B)$, however, it is still bounded by $H(A)$, which is the bound for the pure orthogonal states

$$I(A:B) \leq \chi \leq H(A). \tag{2.10}$$

The *Holevo bound* highlights the important fact that one qubit can contain at most one classical bit i.e., cbit of information.

### 2.4.3 The Quantum Relative Entropy

The relative entropy function measures the “distance” between two probability distributions. In information theory, the relative entropy function is also known as the Kullback-Leibler divergence [Kullback51]. An other interpretation of the relative entropy function was introduced by Bregman, known as the class of Bregman divergences [Bregman67]. Since the relative entropy function is not symmetric, the distance between $A$ and $B$ is not necessarily the same as the “informational distance” from $B$ to $A$. The quantum relative entropy measures the *informational distance* between quantum states, and introduces a deeper characterization of the quantum states than the von Neumann entropy. The quantum relative entropy function was originally introduced by Umegaki, and later modified versions have been defined by Ohya, Petz and Watanbe [Ohya97]. Some possible applications of quantum relative entropy in QIT were introduced by Schumacher and Westmoreland [Schumacher2000] and Vedral [Vedral2000]. For the complete list of related works see Appendix B.

The relative entropy in classical systems is a measure that quantifies how close a probability distribution $p$ is to a model or candidate probability distribution $q$. For probability distributions $p$ and $q$, the classical relative entropy is given by

$$D(p\|q) = \sum_i p_i \log\left(\frac{p_i}{q_i}\right), \tag{2.11}$$

while the quantum relative entropy between quantum states $\rho$ and $\sigma$ is

$$D(\rho\|\sigma) = Tr(\rho\log(\rho)) - Tr(\rho\log(\sigma)) = Tr[\rho(\log(\rho) - \log(\sigma))]. \tag{2.12}$$

In the definition above, the term $Tr(\rho\log(\sigma))$ is finite only if $\rho\log(\sigma) \geq 0$ for all diagonal matrix elements. If this condition is not satisfied, then $D(\rho\|\sigma)$ could be infinite, since the trace of the second term could go to infinity. The *quantum informational distance* (i.e., quantum relative entropy) has some distance-like properties (for example, the quantum relative entropy function between a maximally mixed state and an arbitrary quantum state is symmetric, hence in this case it is not just a pseudo distance), however it is *not commutative*, thus $D(\rho\|\sigma) \neq D(\sigma\|\rho)$, and $D(\rho\|\sigma) \geq 0$ iff $\rho \neq \sigma$, and $D(\rho\|\sigma) = 0$ iff $\rho = \sigma$. Note, if $\sigma$ has zero eigenvalues, $D(\rho\|\sigma)$ may diverge, otherwise it is a finite and continuous function. Furthermore, the quantum relative entropy function has another interesting property, since if we have two density matrices $\rho$ and $\sigma$, then the following property holds for the traces used in the expression of $D(\rho\|\sigma)$

$$Tr(\rho\log(\rho)) \geq Tr(\rho\log(\sigma)). \tag{2.13}$$

The symmetric Kullback-Leibler distance is widely used in classical systems, for example in computer vision and sound processing. For further information on the basic functions and definitions of Quantum Information Theory see Appendix B.

### 2.4.4 Brief Summary

The character of classical information and quantum information is significantly different. There are many phenomena in quantum systems which cannot be described classically, such as entanglement, which makes it possible to store quantum information in the correlation of quantum states. Similarly, a quantum channel can be used with pure orthogonal states to realize classical information transmission, or it can be used to transmit non-orthogonal states or even quantum entanglement. Information transmission also can be approached using the question, whether the input consists of unentangled or entangled quantum states. This leads us to say that for quantum channels many new capacity definitions exist in comparison to a classical communication channel. In possession of the general communication model and the quantities which are able to represent information content of quantum states we can begin to investigate the possibilities and limitations of information transmission through quantum channels.

The Related Work subsection of Appendix B introduces the reader to the referred and cited works of the above discussed results.

# Chapter 3

# Quantum Channel Capacities

"It's always fun to learn something new about quantum mechanics."

*Benjamin Schumacher*

Communication over quantum channels is bounded by the corresponding capacities. This chapter lays down the fundamental theoretic results on the classical capacities and the quantum capacity of quantum channels. These results are all required to analyze the advanced and more promising properties of quantum communication channels.

Chapter 3 is organized as follows. In the first part of this chapter, we introduce the reader to formal description of a noisy quantum channel. Then we start to discuss the classical capacity of a quantum channel. Next, we show the various encoder and decoder settings for transmission of classical information. We define the exact formula for the measure of maximal transmittable classical information. In the second part, first we discuss the transmission of quantum information over a nosy quantum channel. Next, we define the quantum coherent information and overview its main properties. Finally the formula for the measure of maximal transmittable quantum information over a quantum channel will be introduced. Supplementary information and the description of the most relevant works can be found in the Related Work subsection of Appendix C.

# 3.1 Introduction

We introduce a general model which allows considering the non-idealistic effects of the quantum channel. These effects modify the quantum states traveling through the channel and thus restrict the information at the receiver side. The discussed model is general enough to analyze the limitations for information transfer over quantum channels. Each quantum channel can be represented as a CPTP map (*Completely Positive Trace Preserving Map*), hence the process of information transmission through a quantum communication channel can be described as a quantum operation.

The general model of a quantum channel describes the transmission of an input quantum bit, and its interaction with the environment (see Fig. 3.1.). Assuming Alice sends quantum system $\rho_A$ into the channel this state becomes entangled with the environment $\rho_E$, which is initially in a pure state $\rho_E = |0\rangle\langle 0|$. For a mixed input state a so called *purification state* $P$ can be defined, from which the original mixed state can be restored by a partial trace (see Appendix C) of the pure system $\rho_A P$. The unitary operation $U_{AE}$ of a quantum channel $\mathcal{N}$ entangles $\rho_A P$ with the environment $\rho_E$, and outputs Bob's mixed state as $\rho_B$, and the purification state as $P$. The purification state is a reference system, it cannot be accessed, it remains the same after the transmission [Nielsen2000].

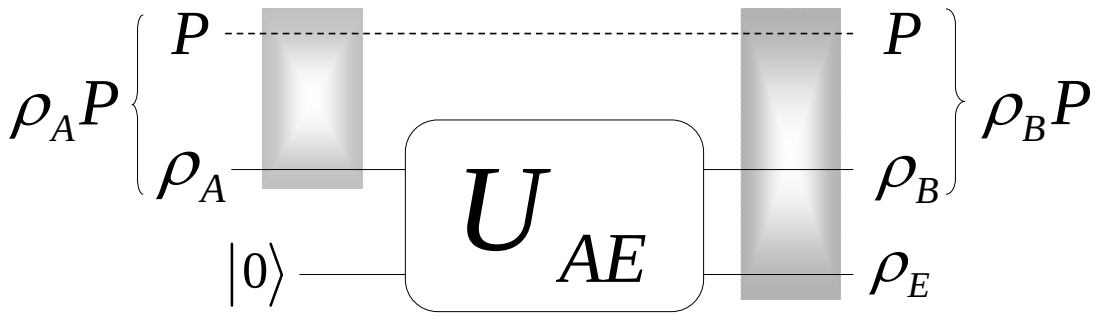

**Fig. 3.1.** The formal model of a noisy quantum communication channel. The pure purification state and Alice's input state are entangled. The output of the channel is a mixed state.

The output of the noisy quantum channel is denoted by $\rho_B$, the post state of the environment by $\rho_E$, while the post-purification state after the output realized on the channel output is depicted by $P$.

## 3.2 Transmission of Classical Information over Noisy Quantum Channels

As the next step during our journey towards the quantum information transfer through quantum channels (which is the most general case) we are leaving the well-known classical (macro) world and just entering into the border zone. Similar to the ancient Romans - who deployed a sophisticated wide border defense system (called *the limes* which consisted of walls, towers, rivers, etc.), instead of drawing simply a red line between themselves and the barbarians – we remain classical in terms of inputs and outputs but allow the channel operating in a quantum manner. Quantum channels can be used in many different ways to transmit information from Alice to Bob. Alice can send classical bits to Bob, but she also has the capability of transmitting quantum bits. In the first case, we talk about the classical capacity of the quantum channel, while in the latter case, we have a different measure - the quantum capacity.

Compared to classical channels – which have only one definition for capacity – the transmittable classical information and thus the corresponding capacity definition can be different when one considers quantum channels. This fact splits the classical capacity of quantum channels into three categories, namely the *classical capacity* $C(\mathcal{N})$ (also known as the *product-state* classical capacity, or the HSW (Holevo-Schumacher-Westmoreland), *private classical capacity* $P(\mathcal{N})$ and *entanglement-assisted classical capacity* $C_E(\mathcal{N})$ (see Fig. 3.2.). The product-state classical capacity $C(\mathcal{N})$ is a natural extension of the capacity definition from classical channels to the quantum world. For the sake of simplicity the term classical capacity will refer to the unentangled version in the forthcoming pages of this Ph.D Thesis. (The entangled version will be referred as the entanglement-assisted classical capacity.) As we will see, the HSW classical capacity $C(\mathcal{N})$ is defined for product state inputs; however it is possible to define its asymptotic version for entangled input states. The private classical capacity $P(\mathcal{N})$ has deep relevance in secret quantum communications and quantum cryptography. It describes the rate at which the channel is able to send classical information through the channel in secure manner. Security here means that an eavesdropper will not be able to access the encoded information without revealing her/himself. The entanglement-assisted classical capacity $C_E(\mathcal{N})$ measures the classical information which can be transmitted through the channel, if Alice and Bob have already shared entanglement before the transmission. A well-known example of such protocols is superdense coding [Imre05].

Next, we discuss the above listed various classical capacities of quantum channels in detail.

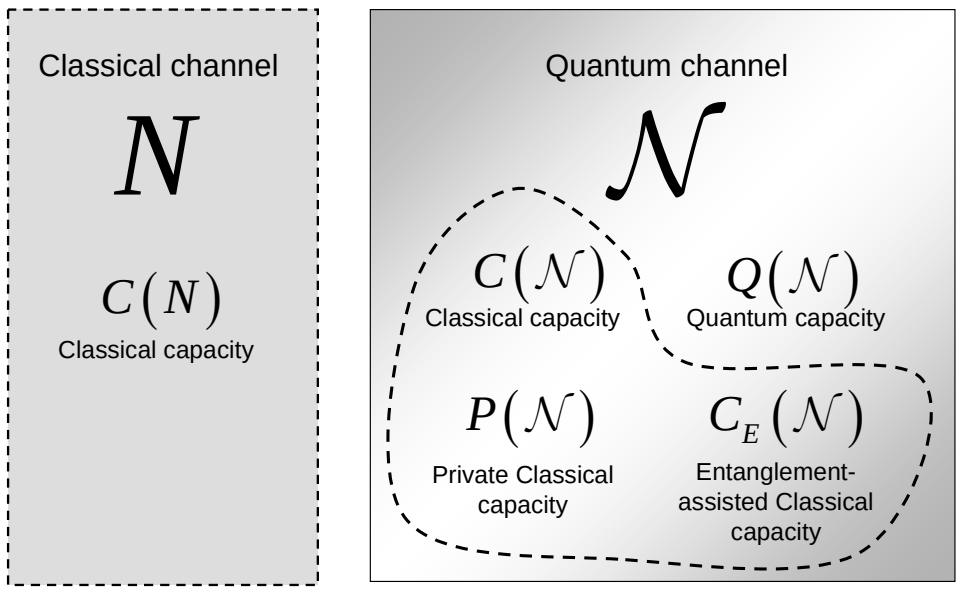

**Fig. 3.2.** The "zoo" (taxonomy) of different capacities of classical and the quantum communication channels. The classical capacities of the quantum channel are separated by a dashed line from the quantum capacity.

As the first obvious generalization of classical channel capacity definition is if we maximize the quantum mutual information over all possible input ensembles

$$C(\mathcal{N}) = \max_{all\ p_i,\rho_i} I(A:B). \tag{3.1}$$

Next, we start to discuss the classical information transmission capability of a noisy quantum channel.

### 3.2.1 The Holevo-Schumacher-Westmoreland Capacity

The HSW-theorem (Holevo-Schumacher-Westmoreland) defines the maximum of classical information which can be transmitted through a noisy quantum channel $\mathcal{N}$ if the input contains product states (i.e., entanglement is not allowed, also known as the *product-state classical capacity*) and the output is measured by joint measurement setting (see the *second* measurement setting in Appendix C). In this setting, for the quantum noisy communication channel $\mathcal{N}$, the classical capacity can be expressed as follows

$$\begin{aligned} C(\mathcal{N}) &= \max_{all\ p_i,\rho_i} \chi = \max_{all\ p_i,\rho_i}\left[\mathrm{S}(\sigma_{out}) - \sum_i p_i \mathrm{S}(\sigma_i)\right] \\ &= \max_{all\ p_i,\rho_i}\left[\mathrm{S}\left(\mathcal{N}\left(\sum_i p_i\rho_i\right)\right) - \sum_i p_i \mathrm{S}(\mathcal{N}(\rho_i))\right], \end{aligned} \tag{3.2}$$

where the maximum is taken over all ensembles $\{p_i, \rho_i\}$ of input quantum states, while for $\sigma_{out}$ see Appendix C. This capacity reaches its maximum for a perfect noiseless quantum channel. Since we know the Holevo quantity $\chi$ (see Appendix C), it can be stated, that the HSW channel capacity is just the maximization of $\chi$, hence $C(\mathcal{N}) = \max_{all\ p_i,\rho_i} \chi$.

We might ask, what is the fundamental difference between the Holevo bound which we have introduced previously (see Chapter 2), and the HSW capacity, defined just now. This difference can be seen if in the decoding process Bob uses POVM (Positive Operator Valued Measure) [Imre05] measurement on the transmitted *codewords* (joint measurement), instead of applying it on every qubit one-by-one (single measurement). If Alice chooses among a set of quantum codewords, then is it possible to transmit these codewords through the noisy quantum channel $\mathcal{N}$ to Bob with arbitrary small error, if

$$R < C\left(\mathcal{N}\right) = \max_{all\ p_i,\rho_i}\left[\mathrm{S}\left(\mathcal{N}\left(\sum_i p_i\rho_i\right)\right) - \sum_i p_i\mathrm{S}\left(\mathcal{N}\left(\rho_i\right)\right)\right]; \tag{3.3}$$

if Alice adjusts $R$ to be under $\max_{all\ p_i,\rho_i}\chi$, then she can transmit her codewords with arbitrarily small error. On the other hand, the channel capacity for a classical (i.e., not a quantum) channel $N$ can be expressed as $C\left(N\right) = \max_{all\ p_i,\rho_i} I\left(A:B\right)$, hence, for a classical communication channel, the channel capacity can be defined as the maximum of the classical mutual information. Similar to the quantum case, there exists a classical code rate $R < C\left(N\right)$ for channel $N$, which allows Alice to transmit information through a classical channel with arbitrarily low error. If Alice chooses $R > C\left(\mathcal{N}\right)$, then she cannot select a quantum code of arbitrary size, which was needed for her to realize an error-free communication. The HSW channel capacity guarantees an error-free quantum communication only if $R < C\left(\mathcal{N}\right) = \max_{all\ p_i,\rho_i}\chi$ is satisfied for her code rate $R$.

For the different measurement settings and the asymptotic HSW capacity see Appendix C and the book of Imre and Gyongyosi [Imre12]. Next we discuss the private classical capacity of quantum channels.

### 3.2.2 The Private Classical Capacity

The private classical capacity $P\left(\mathcal{N}\right)$ of a quantum channel $\mathcal{N}$ describes the maximum rate at which the channel is able to send *classical information* through the channel reliably and *privately* (i.e., without any information leaked about the original message to an eavesdropper). Privately here means that an eavesdropper will not be able to access the encoded information without revealing her/himself i.e., the private classical capacity describes the maximal secure information that can be obtained by Bob on an eavesdropped quantum communication channel. The generalized model of the private communication over quantum channels is illustrated in Fig. 3.3. The first output of the channel is denoted by $\sigma_B = \mathcal{N}\left(\rho_A\right)$, the second "receiver" is the eavesdropper $E$, with state $\sigma_E$. The single-use private classical capacity from these quantities can be expressed as the maximum of the difference between two mutual information quantities. The eavesdropper, Eve, attacks the quantum channel, and she steals $I\left(A:E\right)$ from the mutual information $I\left(A:B\right)$ between Alice to Bob, therefore the *single-use* private classical capacity (*private information*) can be determined as

$$P^{(1)}\left(\mathcal{N}\right) = \max_{all\ p_i,\rho_i}\left(I\left(A:B\right) - I\left(A:E\right)\right). \tag{3.4}$$

while the *asymptotic* private classical capacity is

$$P\left(\mathcal{N}\right) = \lim_{n\to\infty}\frac{1}{n}P^{(1)}\left(\mathcal{N}^{\otimes n}\right) = \lim_{n\to\infty}\frac{1}{n}\max_{all\ p_i,\rho_i}\left(I\left(A:B\right) - I\left(A:E\right)\right), \tag{3.5}$$

where $\otimes n$ represents the $n$ uses of the quantum channel.

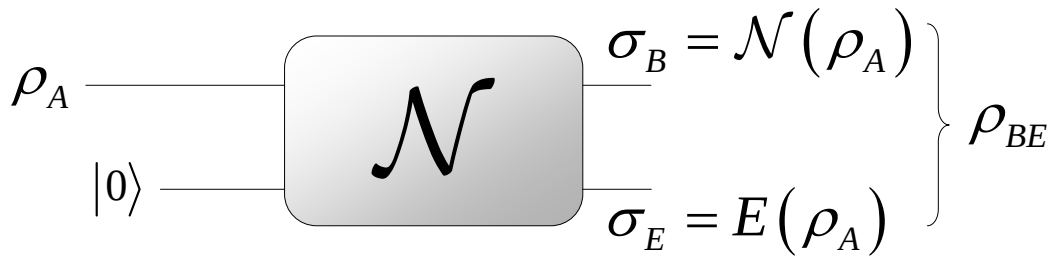

**Fig. 3.3.** The private classical capacity of a quantum channel. The environment is not depicted.

The asymptotic and the single-use private classical capacity can be expressed as the difference of two quantum mutual information functions (see Appendix B), see (3.4) and (3.5). Here, we give an equivalent definition for private classical capacity and show, that it also can be rewritten using the Holevo quantity, as follows:

$$P\left(\mathcal{N}\right) = \lim_{n\to\infty}\frac{1}{n}\max_{all\ p_i,\rho_i}\left(\mathcal{X}_{AB} - \mathcal{X}_{AE}\right), \tag{3.6}$$

where

$$\mathcal{X}_{AB} = \mathrm{S}\left(\mathcal{N}_{AB}\left(\rho_{AB}\right)\right) - \sum_i p_i \mathrm{S}\left(\mathcal{N}_{AB}\left(\rho_i\right)\right) \tag{3.7}$$

and

$$\mathcal{X}_{AE} = \mathrm{S}\left(\mathcal{N}_{AE}\left(\rho_{AE}\right)\right) - \sum_i p_i \mathrm{S}\left(\mathcal{N}_{AE}\left(\rho_i\right)\right) \tag{3.8}$$

measure the Holevo quantities between Alice and Bob, and Alice and the eavesdropper Eve, respectively, where $\rho_{AB} = \sum_i p_i \rho_i$ and $\rho_{AE} = \sum_i p_i \rho_i$ .

An important corollary from (3.5), while the quantum mutual information itself is additive (see the properties of quantum mutual information function in Appendix B), the difference of two quantum mutual information functions is not (i.e., we need the asymptotic version to compute the true private classical capacity of a quantum channel.)

### 3.2.3 The Entanglement-assisted Classical Capacity

The last capacity regarding classical communication over quantum channels is called entanglement-assisted classical capacity $C_E\left(\mathcal{N}\right)$, which measures the classical information which can be transmitted through the channel, if Alice and Bob have shared entanglement before the transmission i.e., entanglement is applied not between the input states like in case of the HSW-theorem (i.e., the product-state capacity). This capacity measures classical information, and it can be expressed with the help of the *quantum mutual information function* (see Appendix B) as

$$C_E\left(\mathcal{N}\right) = \max_{all\ p_i, \rho_i} I\left(A : B\right). \tag{3.9}$$

The main difference between the classical capacity $C\left(\mathcal{N}\right)$ and the entanglement-assisted classical capacity $C_E\left(\mathcal{N}\right)$, is that in the latter case the maximum of the transmittable classical information is equal to the quantum mutual information, - hence the entanglement-assisted classical capacity can be derived from the *single-use* version. From (3.9) the reader can conclude, there is no need for the asymptotic version to express the entangle-ment-assisted classical capacity. It also can be concluded, that shared entanglement does not change the additivity of quantum mutual information - or with other words, it remains true if the parties use shared entanglement for the transmission of classical information.

We note an important property of shared entanglement: while it does not provide any benefits in the improving of the asymptotic classical capacity of the quantum channel, (see (3.9)), it can be used to increase the single-use classical capacity. It was shown, that with the help of shared entanglement the transmission of a single quantum bit can be realized with higher success probability, - this strategy is known as the CHSH (*Clauser-Horne-Shimony-Holt*) game, for details see [Imre05].

## 3.3 The Classical Zero-Error Capacity

Shannon's results on capacity [Shannon48] guarantees transmission rate only in average when using multiple times of the channel. The zero-error capacity of the quantum channel describes the amount of (classical or quantum) information which can be transmitted *perfectly* (*zero probability of error*) through a noisy quantum channel. The zero-error capacity of the quantum channel could have an overriding importance in future quantum communication networks. The zero-error capacity stands a very strong requirement in comparison to the standard capacity where the information transmission can be realized with asymptotically small *but non-vanishing* error probability, since in the case of zero-error communication the *error probability of the communication has to be zero*, hence the transmission of information has to be perfect and no errors are allowed. While in the case of classical non zero-error capacity for an $n$-length code the error probabilities after the decoding process are $\Pr\left[error\right] \to 0$ as $n \to \infty$, in case of an $n$-length zero-error code, $\Pr\left[error\right] = 0$. In this subsection we give the exact definitions which required for the characterization of a quantum zero-error communication system. We will define the classical and quantum zero-error capacities and in Appendix C the connection between zero-error quantum codes and the elements of graph theory is also presented. For further information see the book of Imre and Gyongyosi [Imre12].

### 3.3.1 Classical Zero-Error Capacities of Quantum Channels

Let us assume that Alice has information source $\{X_i\}$ encoded into quantum states $\{\rho_i\}$ which will be transmitted through a quantum channel $\mathcal{N}$ (see Fig. 3.4.). The quantum state will be measured by a set of POVM operators [Imre05], [Medeiros05] $\mathcal{P}=\{\mathcal{M}_1,\ldots,\mathcal{M}_k\}$. The zero-error transmission of quantum states requires perfect distinguishability of the quantum codewords, i.e., they have to be pairwise non-adjacent (orthogonal). *Non-adjacent codewords can be distinguished perfectly.* Two inputs are called adjacent if they can result in the same output. The number of possible non-adjacent codewords determines the rate of maximal transmittable classical information through quantum channels. In the $d$ dimensional Hilbert space (e.g. $d$=2 for qubits) at most $d$ pairwise distinguishable quantum states exist, thus for a quantum system which consist of $n$ pieces of $d$ dimensional quantum states at most $d^n$ pairwise distinguishable $n$-length quantum codewords are available. Obviously if two quantum codewords are not orthogonal, then they cannot be distinguished perfectly. We note, if we would like to distinguish between $K$ *pairwise orthogonal* quantum codewords (the length of each codewords is $n$) in the $d^n$ dimensional Hilbert space, then we have to define the POVM set

$$\mathcal{P}=\left\{\mathcal{M}^{(1)},\ldots,\mathcal{M}^{(K)}\right\}, \tag{3.10}$$

where $\mathcal{M}^{(i)}$ are set of $d$ *dimensional* projectors on the individual quantum systems (e.g. qubits) which distinguish the $n$-length codewords

$$\mathcal{M}^{(i)}=\{\mathcal{M}_1,\ldots,\mathcal{M}_m\}, \tag{3.11}$$

where $m = d^n$. The probability that Bob gives measurement outcome $j$ from quantum state $\rho_i$ is

$$\Pr\left[\ j \middle| \rho_i \right] = Tr\left( \mathcal{M}_j \mathcal{N}\left( \rho_i \right) \right). \tag{3.12}$$

where $\rho_i = \left| \psi_{X_i} \right\rangle \left\langle \psi_{X_i} \right|$. Quantum *codeword* $\left| \psi_{X_i} \right\rangle$ encodes the $n$-length classical codeword $X_i = \left\{ x_{i,1}, x_{i,2}, \ldots, x_{i,n} \right\}$ consisting of $n$ product input quantum states:

$$\left| \psi_{X_i} \right\rangle = \left[ \left| \psi_{i,1} \right\rangle \otimes \left| \psi_{i,2} \right\rangle \otimes \left| \psi_{i,3} \right\rangle \cdots \otimes \left| \psi_{i,n} \right\rangle \right],\ i = 1..K\ . \tag{3.13}$$

The quantum block code consist of $K$ codewords (each quantum codeword has length $n$),

$$\begin{aligned} \left| \psi_{X_1} \right\rangle &= \left[ \left| \psi_{1,1} \right\rangle \otimes \left| \psi_{1,2} \right\rangle \otimes \left| \psi_{1,3} \right\rangle \cdots \otimes \left| \psi_{1,n} \right\rangle \right] \\ &\vdots \qquad\qquad\qquad\qquad \vdots \\ \left| \psi_{X_K} \right\rangle &= \left[ \left| \psi_{K,1} \right\rangle \otimes \left| \psi_{K,2} \right\rangle \otimes \left| \psi_{K,3} \right\rangle \cdots \otimes \left| \psi_{K,n} \right\rangle \right], \end{aligned} \tag{3.14}$$

where $K$ is the number of classical ($n$ length) messages. The decoder will obtain the output codeword $X_i' = \left\{ x_{i,1}', x_{i,2}', \ldots, x_{i,n}' \right\}$ generated by the POVM measurement operators, where the POVM $\mathcal{M}^{(i)}$ can distinguish $m$ messages $\left\{ X_1', X_2', \ldots X_m' \right\}$ ($n$-length) at the output. Bob would like to determine each output word with unit probability; that is, message $i \in [1, K]$ has to be identified with unit probability.

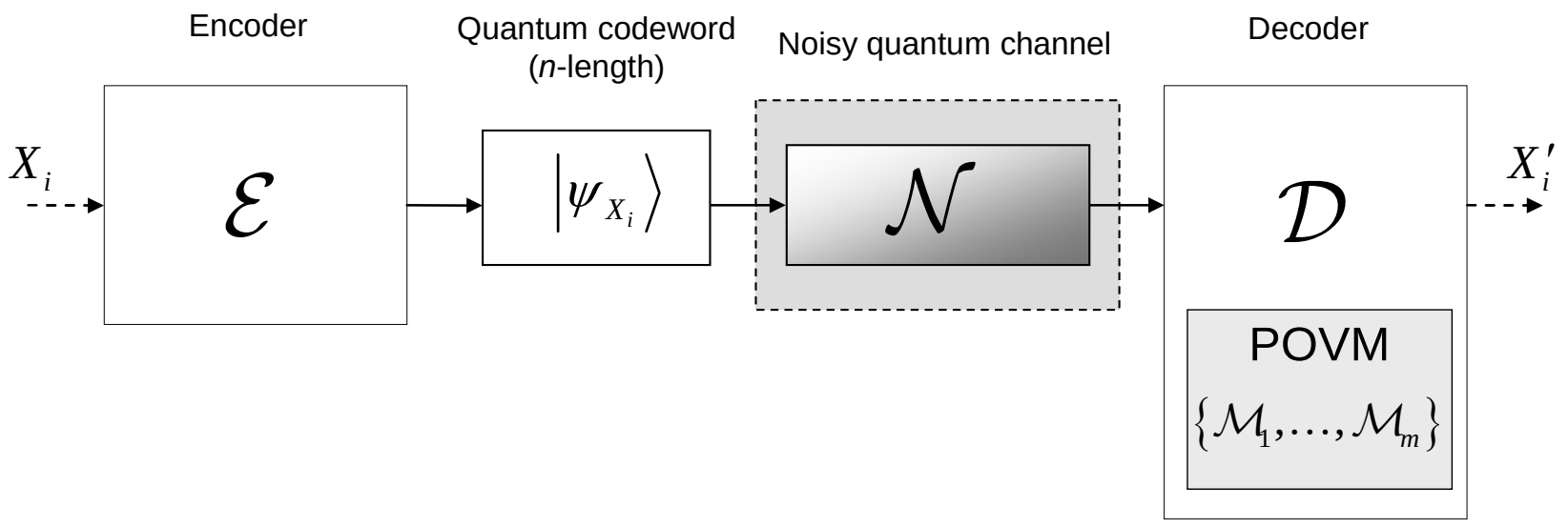

**Fig. 3.4.** A quantum zero-error communication system.

The zero probability of error means that for the input code $\left|\psi_{X_i}\right\rangle$ the decoder has to identify the classical output codeword $X_i'$ with classical input codeword $X_i$ perfectly for each possible $i$, otherwise the quantum channel has no zero-error capacity; that is, for the zero-error quantum communication system

$$\Pr\left[X_i' \middle| X_i\right] = 1. \quad (3.15)$$

For the formal definitions of quantum zero-error communication and the properties of the classical zero-error capacity and supplementary material see Appendix C.

## 3.4 The Quantum Capacity of a Quantum Channel

Having discussed the general model of quantum channels and introduced various classical capacities in this section we focus on the *quantum information* transfer over quantum channels. Two new quantities will be explained. By means of *fidelity* one can describe the differences between two quantum states e.g. between the input and output states of a quantum channel. On the other hand *quantum coherent information* represents the quantum information loss to the environment during quantum communication similarly as mutual information did for classical information. Exploiting this latter quantity we can define the maximal quantum information transmission rate through quantum channels analogously to Shannon's noisy channel theorem. The classical capacity of a quantum channel is described by the maximum of quantum mutual information and the Holevo information (see Appendix C). The quantum capacity of the quantum channels is described by the maximum of *quantum coherent information*. The concept of quantum coherent information plays a fundamental role in the computation of the *LSD (Lloyd-Shor-Devetak)* channel capacity [Lloyd97], [Shor02], [Devetak03] which measures the asymptotic quantum capacity of the quantum capacity in general.

## 3.4.1 Quantum Capacity of Quantum Channels

In the case of quantum communication, the source is a quantum information source and the *quantum information* is encoded into quantum states. When transmitting quantum information, the information is encoded into non-orthogonal superposed or entangled quantum states chosen from the ensemble $\{\rho_k\}$ according to a given probability $\{p_k\}$. If the states $\{\rho_k\}$ are pure and mutually orthogonal, we talk about classical information; that is, in this case the quantum information reduces to classical. Bob decodes the received the quantum state which is typically modified according to the noise of the quantum channel. As depicted in Fig. 3.5, the encoding and the decoding mathematically can be described by the operators $\mathcal{E}$ and $\mathcal{D}$ realized on the blocks of quantum states [Bennett98], [Nielsen2000], [Imre12]. The input of the encoder consists of $m$ pure quantum states, and the encoder maps the $m$ quantum states into the joint state of $n$ intermediate systems. Each of them is sent through an independent instance of the quantum channel $\mathcal{N}$ and decoded by the decoder $\mathcal{D}$, which results in $m$ quantum states again. The output of the decoder $\mathcal{D}$ is typically mixed, according to the noise of the quantum channel. The rate of the code is equal to $m/n$.

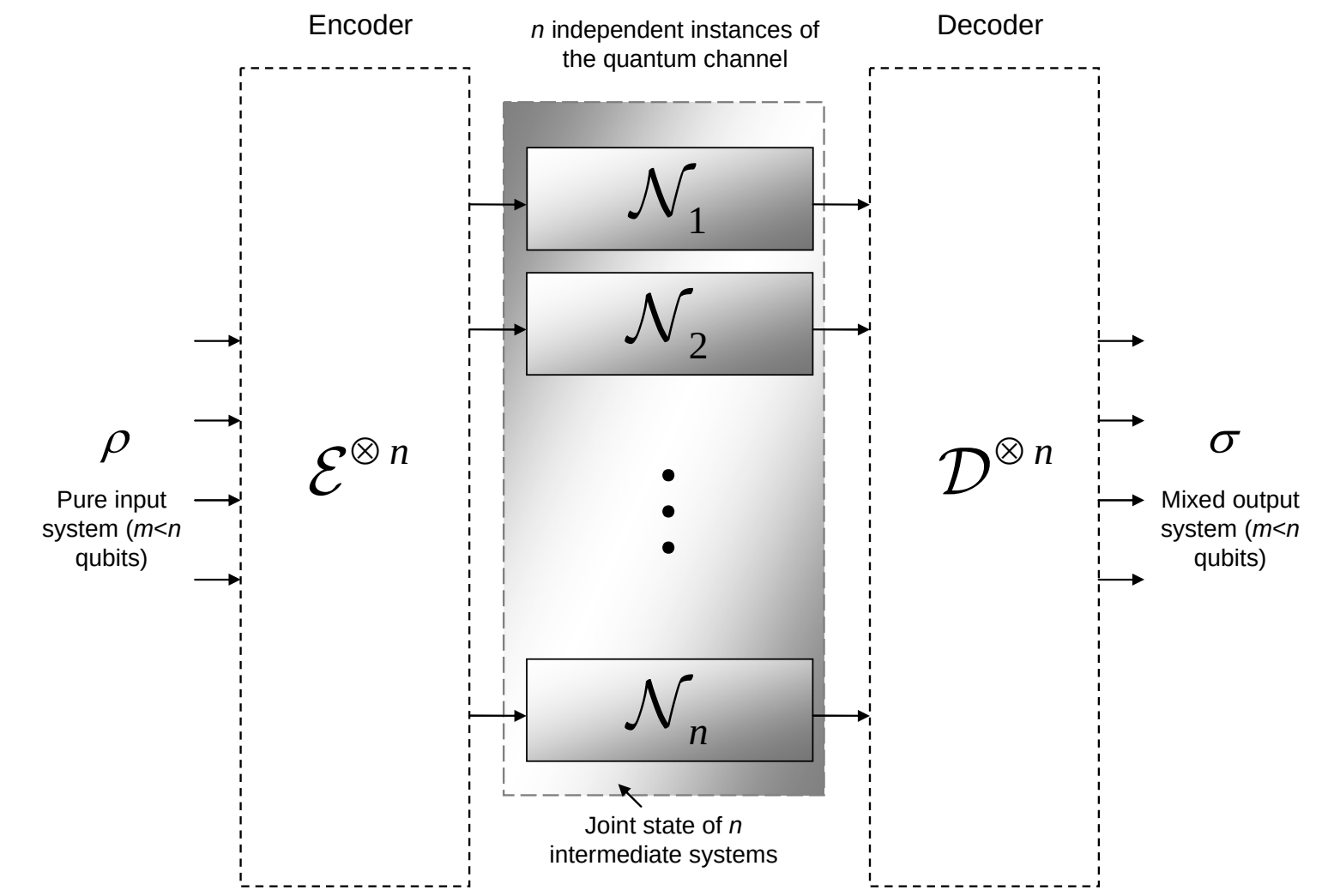

**Fig. 3.5.** Transmission of codewords through the quantum channel. The pure input quantum state consists of $m$ qubits, and the encoder produces a joint state of $n$ intermediate systems. The encoded qubits are passed through the independent instances of the quantum channel.

Theoretically quantum states have to preserve their original superposition during the whole transmission, without the disturbance of their actual properties. Practically, quantum channels are entangled with the environment which results in mixed states at the output. Mixed states are classical probability weighted sum of pure states where these probabilities appear due to the interaction with the environment.

## 3.4.2 Quantum Coherent Information

In case of the classical capacity $C(\mathcal{N})$, the correlation between the input and the output is measured by the Holevo information and the quantum mutual information function. In case of the quantum capacity $Q(\mathcal{N})$, we have a completely different correlation measure with completely different behaviors: it is called the *quantum coherent information*. There is a *very important difference* between the maximized quantum mutual information: the maximized quantum mutual information of a quantum channel is always additive (see Appendix B), but not the maximized quantum coherent information. The $\mathrm{S}_E$ *entropy exchange* between the initial system $PA$ and the output system $PB$ is defined as follows. The entropy that is acquired by $PA$ when input system $A$ is transmitted through the quantum channel $\mathcal{N}$ can be expressed with the help of the von Neumann entropy function as follows

$$\mathrm{S}_E = \mathrm{S}_E\left(\rho_A : \mathcal{N}\left(\rho_A\right)\right) = \mathrm{S}\left(\rho_{PB}\right), \tag{3.16}$$

or in other words the von Neumann entropy of the output system $\rho_{PB}$. As can be concluded, the value of entropy exchange depends on $\rho_A$ and $\mathcal{N}$ and is independent from the purification system $P$. Now, we introduce the environment state $E$, and we will describe the map of the quantum channel as a unitary transformation. The environment is initially in a pure state $\left|0\right\rangle$. After the unitary transformation $U_{A\rightarrow BE}$ has been applied to the initial system $A\left|0\right\rangle$, it becomes

$$U_{A\to BE}\left(A\left|0\right\rangle\right) = BE\,. \tag{3.17}$$

The map of the quantum channel as a unitary transformation on the input system and the environment is shown in Fig. 3.6.

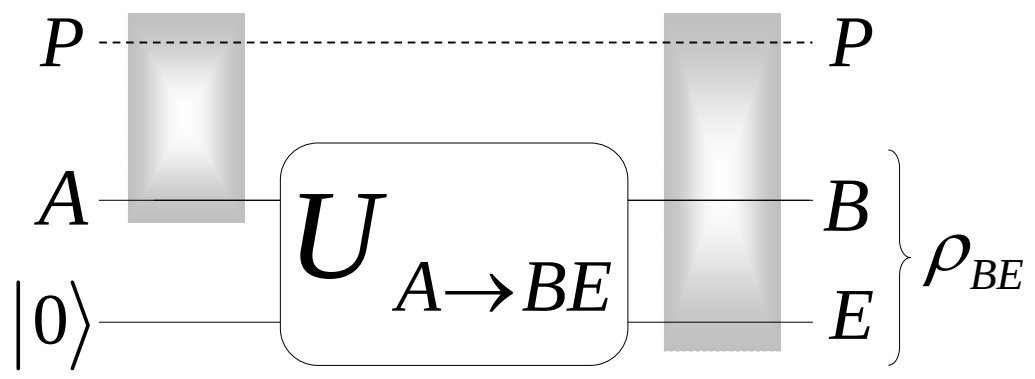

**Fig. 3.6.** The map of the quantum channel as a unitary transformation on the input system and the environment. The unitary transformation entangles $AP$ with the environment $E$, which is initially in a pure state.

Now, from the entropy of the *final state* of the environment $\rho_E$, the *entropy exchange* $\mathrm{S}_E$ can be expressed, since

$$\mathrm{S}\left(\rho_{PB}\right) = \mathrm{S}\left(\rho_E\right) = \mathrm{S}_E\,. \tag{3.18}$$

$\mathrm{S}_E$ measures the increase of entropy of the environment $E$, or with other words the entanglement between $PA$ and $E$, after the unitary transformation $U$ had been applied to the system. This entropy exchange $\mathrm{S}_E$ is analogous to the classical conditional entropy; however in this case we talk about quantum instead of classical information. The $I_{coh}\left(\rho_A : \mathcal{N}\left(\rho_A\right)\right)$ quantum coherent information can be expressed as

$$\begin{aligned} I_{coh}\left(\rho_A : \mathcal{N}\left(\rho_A\right)\right) &= \mathrm{S}\left(\mathcal{N}\left(\rho_A\right)\right) - \mathrm{S}_E\left(\rho_A : \mathcal{N}\left(\rho_A\right)\right) \\ &= \mathrm{S}\left(\rho_B\right) - \mathrm{S}\left(\rho_{PB}\right) \\ &= \mathrm{S}\left(\rho_B\right) - \mathrm{S}\left(\rho_E\right), \end{aligned} \tag{3.19}$$

where $\mathrm{S}_E\left(\rho_A : \mathcal{N}\left(\rho_A\right)\right)$ is the entropy exchange as defined in (3.16). Using the definition of quantum coherent information (3.19), it can be verified that quantum coherent informa-tion takes its maximum if systems $A$ and $P$ are *maximally entangled* and the quantum channel $\mathcal{N}$ is *completely noiseless.* This can be presented easily

$$\mathrm{S}\left(\rho_B\right)=\mathrm{S}\left(\rho_A\right), \tag{3.20}$$

since the input state $\rho_A$ is maximally mixed, and

$$\mathrm{S}\left(\rho_{PB}\right)=0, \tag{3.21}$$

because the input system $\left|\psi^{PA}\right\rangle\left\langle\psi^{PA}\right|$ will be pure after the state had been transmitted through the idealistic quantum channel. If the input system $\left|\psi^{PA}\right\rangle\left\langle\psi^{PA}\right|$ is not a maximally entangled state, or the quantum channel is not idealistic, then the value of quantum coherent information will decrease. Considering another expressive picture, quantum coherent information measures the quantum capacity as the difference between the von Neumann entropies of two channel output states. The first state is received by Bob, while the second one is received by a "second receiver" - called the environment. If we express the transformation of a quantum channel as the partial trace of the overall system, then

$$\mathcal{N}\left(\rho_A\right)=Tr_E\left(U\rho_A U^\dagger\right), \tag{3.22}$$

and similarly, due to the "effect" of the environment $E$, we will get

$$E\left(\rho_A\right)=\rho_E=Tr_B\left(U\rho_A U^\dagger\right). \tag{3.23}$$

It can be concluded that the quantum coherent information measures the capability of transmission of entanglement over a quantum channel. For the exact measure of quantum coherent information of some important quantum channels see Appendix C.

### 3.4.3 Connection between Classical and Quantum Information

As it has been shown by Schumacher and Westmoreland [Schumacher2000], the quantum coherent information also can be expressed with the help of Holevo information, as follows

$$I_{coh}\left(\rho_A : \mathcal{N}\left(\rho_A\right)\right) = \left(\mathcal{X}_{AB} - \mathcal{X}_{AE}\right), \tag{3.24}$$

where

$$\mathcal{X}_{AB} = \mathrm{S}\left(\mathcal{N}_{AB}\left(\rho_{AB}\right)\right) - \sum_i p_i \mathrm{S}\left(\mathcal{N}_{AB}\left(\rho_i\right)\right) \tag{3.25}$$

and

$$\mathcal{X}_{AE} = \mathrm{S}\left(\mathcal{N}_{AE}\left(\rho_{AE}\right)\right) - \sum_i p_i \mathrm{S}\left(\mathcal{N}_{AE}\left(\rho_i\right)\right) \tag{3.26}$$

are the Holevo quantities between Alice and Bob, and between Alice and environment $E$, where $\rho_{AB} = \sum_i p_i \rho_i$ and $\rho_{AE} = \sum_i p_i \rho_i$ are the average states. The definition of (3.24) also draws a very important connection between the transmission of quantum information and classical information - since the amount of transmittable quantum information can be derived by the Holevo information, which measures classical information. The *single-use* quantum capacity $Q^{(1)}\left(\mathcal{N}\right)$ can be expressed as

$$\begin{aligned} Q^{(1)}\left(\mathcal{N}\right) &= \max_{all\ p_i,\rho_i}\left(\mathcal{X}_{AB} - \mathcal{X}_{AE}\right) = \\ &= \max_{all\ p_i,\rho_i}\left(\mathrm{S}\left(\mathcal{N}_{AB}\left(\sum_{i=1}^{n} p_i\left(\rho_i\right)\right)\right) - \sum_{i=1}^{n} p_i \mathrm{S}\left(\mathcal{N}_{AB}\left(\rho_i\right)\right)\right. \\ &\qquad \left. - \mathrm{S}\left(\mathcal{N}_{AE}\left(\sum_{i=1}^{n} p_i\left(\rho_i\right)\right)\right) + \sum_{i=1}^{n} p_i \mathrm{S}\left(\mathcal{N}_{AE}\left(\rho_i\right)\right)\right), \end{aligned} \tag{3.27}$$

where $\mathcal{N}\left(\rho_i\right)$ represents the $i$-th output density matrix obtained from the quantum channel input density matrix $\rho_i$.

The *asymptotic* quantum capacity $Q(\mathcal{N})$ can be expressed by

$$\begin{aligned} Q(\mathcal{N}) &= \lim_{n\to\infty} \frac{1}{n} Q^{(1)}\left(\mathcal{N}^{\otimes n}\right) \\ &= \lim_{n\to\infty} \frac{1}{n} \max_{all\ p_i,\rho_i} I_{coh}\left(\rho_A : \mathcal{N}^{\otimes n}\left(\rho_A\right)\right) \\ &= \lim_{n\to\infty} \frac{1}{n} \max_{all\ p_i,\rho_i} \left(\mathcal{X}_{AB} - \mathcal{X}_{AE}\right). \end{aligned} \tag{3.28}$$

As follows, the quantum coherent information can be computed as the difference between the Holevo information of Alice and Bob, and the Holevo information, which is passed from Alice to the environment. It is summarized in Fig. 3.7.

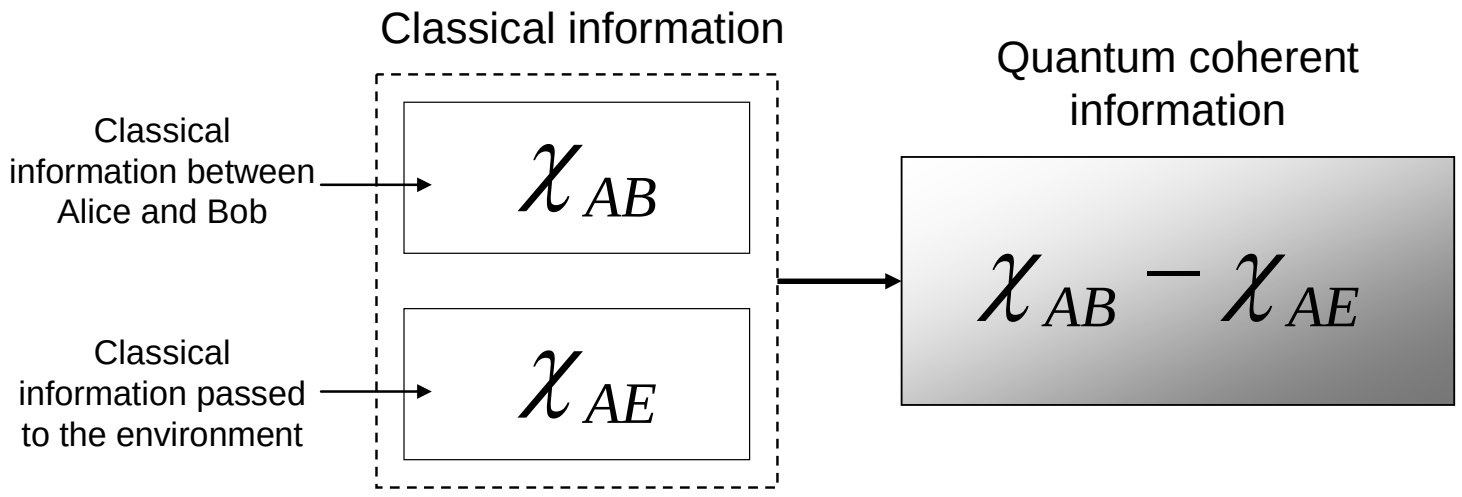

**Fig. 3.7.** Computation of the quantum coherent information from classical Holevo quantity. The first quantity measures the transmitted classical information from Alice to Bob, the second quantity measures the classical information which passed from Alice to the environment during the transmission.

As summarize, the quantum capacity $Q(\mathcal{N})$ of a quantum channel $\mathcal{N}$ can be defined by $\mathcal{X}_{AB}$, the *Holevo quantity* of Bob's output and by $\mathcal{X}_{AE}$, the information leaked to the environment during the transmission.

The most important works related on the quantum capacity of quantum channels and the complete historical background can be found in the Related Work subsection of Appendix C.

# Chapter 4

# Superactivation of Quantum Channels

"Whatever the answers, it is clear is that the structure of Quantum Information Theory is much richer than most of us ever anticipated."
*Jonathan Oppenheim,* University of Cambridge

In the first decade of the 21st century, many revolutionary properties of quantum channels were discovered. These phenomena are purely quantum mechanical and completely unimaginable in classical systems. Recently, one of the most important discoveries in Quantum Information Theory was the possibility of transmitting quantum information over zero-capacity quantum channels.

Chapter 4 is organized as follows. In the first part we introduce the problem of additivity of quantum channel capacities. In the second part we overview the superactivation of quantum capacity of zero-capacity quantum channels. Next we show, that there is a huge difference between the superactivation of single-use and asymptotic quantum capacity. Further information regarding the background of superactivation of quantum channels can be found in Appendix D. The cited works are summarized in the Related Work subsection of Appendix D.

## 4.1 Introduction

The superactivation of zero-capacity quantum channels makes it possible to use two zero-capacity quantum channels with a positive joint capacity for their output. The phenomenon called superactivation is rooted in the extreme violation of additivity of the channel capacities of quantum channels. Currently, we have no theoretical background to describe all possible combinations of superactive zero-capacity channels. In practice, to discover such superactive zero-capacity channel-pairs, we must analyze an extremely large set of possible quantum states, channel models, and channel probabilities. An efficient algorithmical method of finding such combinations will be presented in Chapter 6 and Chapter 7. Additivity of quantum channels in terms of their capacity is one of the very important questions in Quantum Information Theory. The problem statement is whether entangled input states and/or joint measurements could improve the joint capacity of the quantum channels? In the case of classical channels, the correlation between the inputs of the channel does not improve the channel capacity, hence strict additivity holds [Imre12]. Furthermore, additivity is strongly related to the single-use and asymptotic capacities. Obviously these capacities for a certain channel are equal if additivity holds. The equality of the various channel capacities (i.e., classical, private and quantum) is known for some special cases, but the generalized rule is still *unknown*. At present, the main questions connected the quantum channel additivity have not solved yet, some of them are only confirmed for some classes of quantum channels. Recently, the question of the additivity property of a quantum channel has been studied exhaustively, however the most basic question on the classical capacity of a quantum channel - namely, the additivity of classi-cal channel capacity in the asymptotic setting for *different channel maps* - still remain open.

## 4.2 Preliminaries

In 2009, a counterexample to strict additivity was shown by Matt Hastings [Hastings09]. Hastings has analyzed the additivity of Holevo information. He has constructed a channel pair, using a random construction scheme, which can produce an output which is a counterexample to the additivity of the minimum output entropy—a theorem which was introduced in the middle of the 1990s. The counterexample of Hastings's [Hastings09] implies that the entangled states are more resistant to noise, and the output entropy of the channel output states will be lower.

## 4.3 Superactivation of Quantum Capacity

In 2008, Smith and Yard [Smith08] have found only one possible combination for superactivation of quantum capacity. Since the properties of superactivation quantum channel capacities were first reported on, many further quantum informational results have been achieved. Recently, Duan [Duan09] and Cubitt *et al.* [Cubitt09] found a possible combination for the superactivation of the classical zero-error capacity of quantum channels, which has opened up a debate regarding the existence of other possible channel combinations. Later, these results were extended to the superactivation of quantum zero-error capacity by Cubitt and Smith [Cubitt09a]. In the next subsection the superactivation of quantum-capacity will be studied, the superactivation of zero-error capacities will be discussed in Chapter 7.

### 4.3.1 Superactivation of Single-use and Asymptotic Quantum Capacity

As was shown by Smith and Yard [Smith08] for the combination of any quantum channel $\mathcal{N}_1$ that has some private classical capacity $P(\mathcal{N}_1) > 0$ and a 50% erasure symmetric channel $\mathcal{N}_2$, the following connection holds between the $Q^{(1)}(\mathcal{N}_1 \otimes \mathcal{N}_2)$ single-use, the $Q(\mathcal{N}_1 \otimes \mathcal{N}_2)$ asymptotic quantum capacity of the joint structure $\mathcal{N}_1 \otimes \mathcal{N}_2$ and the private classical capacity $P(\mathcal{N}_1)$ of the first channel:

$$Q^{(1)}(\mathcal{N}_1 \otimes \mathcal{N}_2) = \frac{1}{2} P(\mathcal{N}_1), \tag{4.1}$$

and

$$Q(\mathcal{N}_1 \otimes \mathcal{N}_2) \geq \frac{1}{2} P(\mathcal{N}_1). \tag{4.2}$$

The channel combination for the superactivation of the asymptotic quantum capacity of zero-capacity quantum channels is shown in Fig. 4.8.

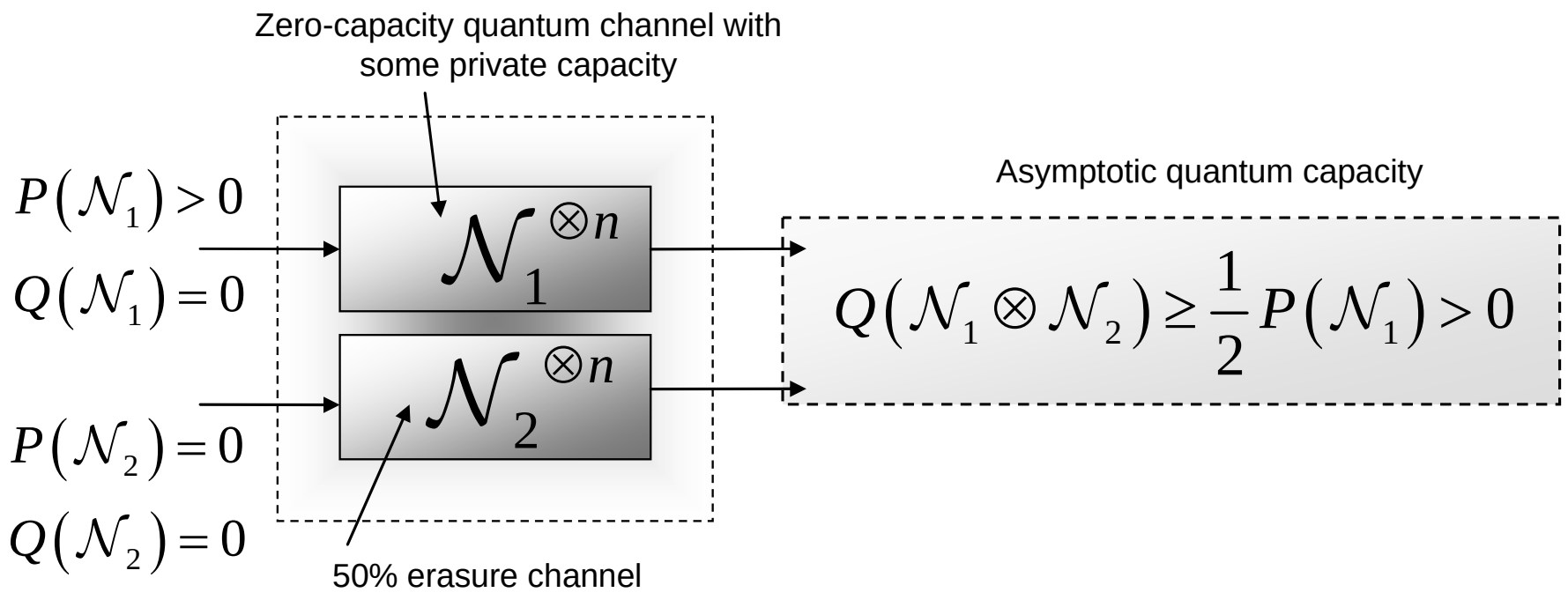

**Fig. 4.8.** The first channel has some positive private classical capacity, and the second quantum channel is a 50% erasure channel with zero quantum capacity.

It is possible to find other combinations of d dimensional quantum channels $\mathcal{N}_1 \in \mathcal{H}^d$ and $\mathcal{N}_2 \in \mathcal{H}^d$, which has individually "zero-capacity" in the sense that

$$Q\left(\mathcal{N}_1\right) = Q\left(\mathcal{N}_2\right) = 0, \tag{4.3}$$

and still satisfy

$$Q\left(\mathcal{N}_1 \otimes \mathcal{N}_2\right) > 0\,. \tag{4.4}$$

This rather strange phenomena is called *superactivation*. For the channel combination $\mathcal{N}_1 \otimes \mathcal{N}_2 \in \mathcal{H}^{d^2}$ the positive single-use quantum capacity $Q^{(1)}\left(\mathcal{N}_1 \otimes \mathcal{N}_2\right)$ was proven using simple algebra [Smith08]. In the channel construction of Smith and Yard's, the superactivation of the quantum capacity of the two quantum channels requires two EPR states (In their proof [Smith08], the first channel is the four-dimensional Horodecki channel $\mathcal{N}_H$ with $P\left(\mathcal{N}_H\right) > 0$, the second is the four-dimensional 50% erasure channel $\mathcal{A}_e$.).

### 4.3.2 Large Gap between Single-use and Asymptotic Quantum Capacity

The difference between the superactivated single-use quantum capacity $Q^{(1)}\left(\mathcal{N}_1 \otimes \mathcal{N}_2\right)$ and superactivated asymptotic quantum capacity $Q\left(\mathcal{N}_1 \otimes \mathcal{N}_2\right)$ can be made arbitrarily high, if we use a different channel combination [Smith08], [Smith09b]. The results on the superactivation of quantum capacity also implied the fact that the quantum capacity is not convex, hence for the combination of two quantum channels $\mathcal{N}_1$ and $\mathcal{N}_2$, the following property holds between their joint quantum capacity $Q^{(1)}\left(\mathcal{N}_1 \otimes \mathcal{N}_2\right)$ and their individual capacities $Q^{(1)}\left(\mathcal{N}_1\right)$ and $Q^{(1)}\left(\mathcal{N}_2\right)$

$$Q^{(1)}\left(\left(1-p\right)\mathcal{N}_1 + p\mathcal{N}_2\right) > \left(1-p\right)Q^{(1)}\left(\mathcal{N}_1\right) + pQ^{(1)}\left(\mathcal{N}_2\right), \tag{4.5}$$

with probability $0 \leq p \leq 1$, which means the following: the single-use joint quantum capacity of the channel combination could be greater than the sum of their individual quan-tum capacities. Finally, we show a channel combination example for which there is a large difference between the *single-use* and the *asymptotic* quantum capacity. This can be achieved by the combination of a $d$ dimensional random phase coupling channel $\mathcal{R}_d$ (defined by random unitary maps) [Smith09b], and a 50% erasure channel, denoted by $\mathcal{N}_2$. The random phase coupling channel and is defined as follows:

$$\mathcal{N}_1 = \mathcal{R}_d = \mathcal{R}^{U_1,U_2}{}_d \otimes \left|U_1 U_2\right\rangle\left\langle U_1 U_2\right|, \tag{4.6}$$

where $\mathcal{R}^{U_1,U_2}{}_d$ denotes the *d dimensional* random phase coupling channel. The random phase coupling channel consists of two unitary transformations $U_1$ and $U_2$, where both unitary transformations are unknown to Alice, while Bob knows both $U_1$ and $U_2$. The asymptotic quantum capacity of this structure will be denoted by

$$Q\left(\mathcal{N}_1 \otimes \mathcal{N}_2\right) = Q\left(\mathcal{R}_d \otimes \mathcal{N}_2\right), \tag{4.7}$$

where $\mathcal{R}_d$ is the random phase coupling channel and $\mathcal{N}_2$ is the 50% erasure channel. In the case of the single-use quantum capacity, this channel realizes the following map as depicted in Fig. 4.9:

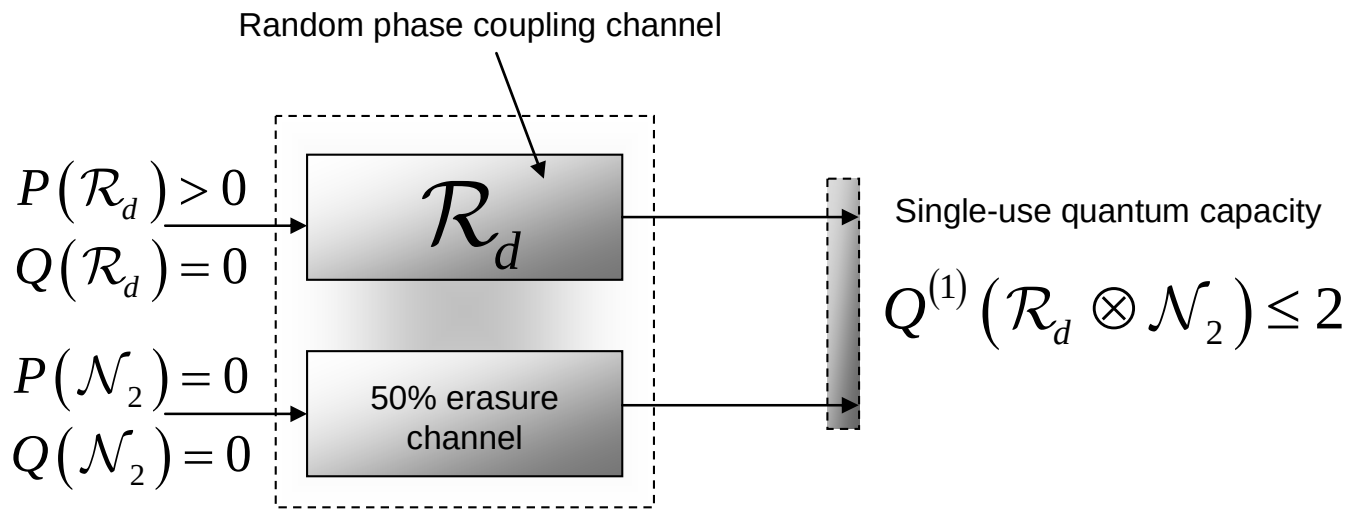

**Fig. 4.9.** The single-use quantum capacity of the channel construction, which consist of the random phase coupling channel and the 50% erasure quantum channel.

In this case, the *single-use* quantum capacity of the joint channel is measured by the maximized quantum coherent information, as

$$Q^{(1)}\left(\mathcal{R}_d \otimes \mathcal{N}_2\right) = \max_{all\ p_i,\rho_i} I_{coh}\left(\rho_A : \mathcal{R}_d \otimes \mathcal{N}_2\right) \leq 2\,, \tag{4.8}$$

since $Q^{(1)}\left(\mathcal{R}_d\right) \leq 2$ or $Q^{(1)}\left(\mathcal{N}_2\right) = 0$. On the other hand, if we measure the *asymptotic* quantum capacity for the same channel construction $\left(\mathcal{R}_d \otimes \mathcal{N}_2\right)$, then we will find that

$$Q\left(\mathcal{R}_d \otimes \mathcal{N}_2\right) \geq \frac{1}{8}\log\left(d\right) \gg Q^{(1)}\left(\mathcal{R}_d \otimes \mathcal{N}_2\right), \tag{4.9}$$

where $d$ is the input dimension. This asymptotic version of the previously seen construction is shown in Fig. 4.10.

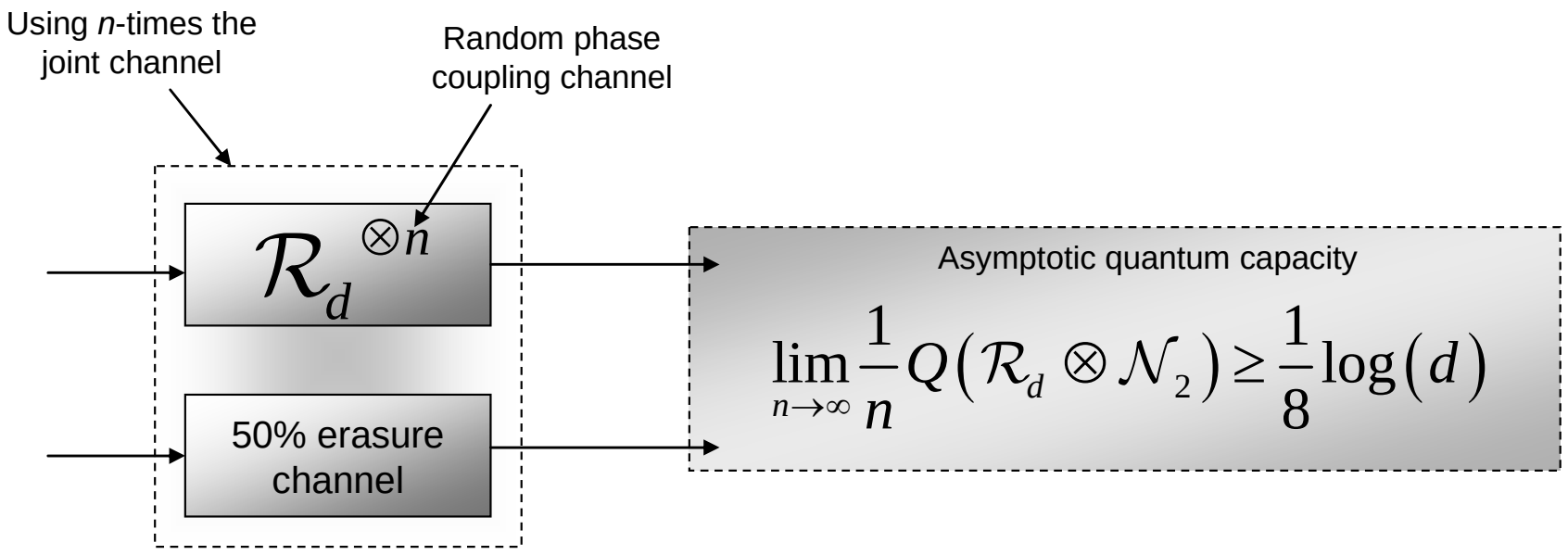

**Fig. 4.10.** The asymptotic capacity of the channel combination. There is a big gap between the quantum maximized coherent information and the asymptotic quantum capacity of the analyzed channel construction.

Summarize, if we have a joint channel combination which contains a random phase coupling channel and a 50% erasure channel, then the convexity of quantum capacity (see (4.5)) will be also satisfied, since for this combination the joint quantum capacity greater than the sum of individual capacities. Moreover, for this channel combination while the single-use quantum capacity of the structure is bounded by 2, the asymptotic quantum capacity can be significantly increased.

Further supplementary information can be found in Appendix D. In the Related Work subsection of Appendix D the most important works regarding on the superactivation of quantum channel capacities are summarized.

# Chapter 5

# Geometric Interpretation of Superactivation

> "The book of nature is written in the characters of geometry."
>
> *Galileo Galilei*

In this chapter I define an informational geometric object to analyze the superactivation of quantum channel capacities. The theoretical background of the construction of the proposed informational geometric approach is also shown. The algorithmical framework for the computation of the quantum superball to the superactivation analysis will be presented in Chapter 6 and Chapter 7. The theses of Chapter 5 lay down the theoretical background of the further analysis on the superactivation of the asymptotic quantum capacity and classical zero-error capacity of quantum channels.

Chapter 5 is organized as follows. In the first section I present the theoretical background of the geometric interpretation of quantum channel capacities. Then I introduce the theses on the proposed quantum informational *superball* object for the analysis of the superacti-vation effect of quantum channels. Then I illustrate the fitting steps of the quantum superball. Finally I conclude the results. Supplementary information and the Related Work subsection are included in Appendix E.

# 5.1 Introduction

The problem of superactivation roots in the problem of additivity of quantum channel capacities. As we have seen in the previous chapters, various channel capacities can be defined for a quantum channel. Most of these capacities are known to be non-additive (see Appendix D) - however the additivity of the general HSW capacity is still unknown. Here, we apply the geometric interpretation to solve the problem of superactivation of quantum channel capacities and define a new informational geometric object based on the fact that the capacity of quantum channels can be described by geometrical tools. As it will be shown in Chapter 6 and Chapter 7, using the proposed information geometric approach the NP-completeness of computation of Holevo information - which was shown by Beigi and Shor [Beigi07] can be avoided and efficient algorithms can be constructed. The geometrical interpretation of the capacities of the quantum channels was studied by Hayashi *et al.* [Hayashi03-05], Cortese [Cortese02-03], Ruskai et al. [Ruskai01-03] and King *et al.* [King99-09]. While in case of the problem of the quantum channel additivity the main problem can be summarized as that whether entanglement between states can help to send information on quantum channel $\mathcal{N}_1 \otimes \mathcal{N}_2$, in case of the superactivation problem we have a different problem [Imre12]. The quantum channels $\mathcal{N}_1$ and $\mathcal{N}_2$ are zero-capacity channels from the viewpoint of the capacity being superactivated. For example, in the channel combination of Smith and Yard [Smith08] the quantum channels $\mathcal{N}_1$ and $\mathcal{N}_2$ have zero quantum capacities $Q\left(\mathcal{N}_1\right) = Q\left(\mathcal{N}_2\right)$, but it is not true for the private classical capacity of the first channel, since $P\left(\mathcal{N}_1\right) > 0$ in the proposed channel construction. The fact that $P(\mathcal{N}_1) > 0$, makes possible to achieve $Q(\mathcal{N}_1 \otimes \mathcal{N}_2) > 0$ positive joint quantum capacity for the joint channel $\mathcal{N}_1 \otimes \mathcal{N}_2$.

In comparison to the additivity problem, in case of superactivation of quantum channel capacities the situation is more complex, since the possibility of superactivation depends on the properties of the quantum channels employed in the joint channel construction, on the correlation among the channels, the dimensions of the channels and on the probability of the channel maps. As we will show in this chapter the superactivation problem can be described by the elements of information geometry.

## 5.2 Geometric Interpretation of Quantum Channel Capacity

In this section we summarize the theoretical background of the proposed information geometric analysis of superactivation of quantum channel capacities. As it was mentioned in Section 2., the maps of physical quantum channels are Completely Positive (CP), trace-preserving maps (i.e., CPTP). The noise of the quantum channel performs a linear transformation, and maps the original Bloch sphere to a distorted Bloch sphere, as an affine map. The noise could cause rotation, and other unital and non-unital distortions, which alterations can be described by the distortion vector. Generally, the noise transforms the Bloch sphere into an ellipsoid and the center of the transformed Bloch sphere can be shifted from the origin of the original Bloch sphere. As it was discussed in Section 2, for a unital quantum channel, the center of the transformed Bloch sphere is the same as for the original Bloch sphere, while for non-unital channels it differs. To describe the geometric representation of a general quantum channel model, we need *twelve* dimensions. The *shape* of the transformed Bloch sphere requires three parameters, the *center* of the distorted Bloch sphere requires another three parameters, the *orientation* of the Bloch sphere requires three more parameters, while the last three parameters are required to *rotate* the original Bloch sphere into a standard position relative to the transformed Bloch sphere. The physically allowed transformations of the quantum channel define a convex polytope—a tetrahedron—which can be used to analyze these transformations [Hayashi06], [Petz08], see Appendix E. All allowed maps on the tetrahedron are Completely Positive maps, inside the tetrahedron we can find the Positive maps. For a certain Positive map, the center of the transformed distorted Bloch sphere will be inside the original Bloch sphere, however, not every transformed ball inside the original Bloch sphere is Completely Positive (see Appendix E).

This section purposes to lay down the fundamental theoretical background of the geometric interpretation of quantum channels' capacity. In Chapter 6, we will extend these results to analyze the superactivation of the asymptotic quantum capacity and the asymptotic classical zero-error capacity of quantum channels in Chapter 7. For the geometric interpretation of the quantum informational distance and the mathematical background see Appendix E.

### 5.2.1 Quantum Delaunay Triangulation

One important tool in the computation of the quantum informational ball is the Delaunay structure. Generally speaking it maximizes the minimum angle of all the angles of the triangles [Aurenhammer2000], [Boissonnat07], [Nielsen07], [Goodman04], [Rajan94], [Imre12] and it can be constructed not only on Euclidean metrics. In the three dimensional representation, the dual of the Delaunay diagram is a Voronoi diagram, where a Voronoi vertex coincidences with the circumcenter of the Delaunay cell, and the dual of the Delaunay facet is a Voronoi edge. Similarly, the dual of the Delaunay edge is a Voronoi facet, and the dual of a Delaunay vertex is a Voronoi cell [Goodman04], [Rajan94]. For further information and for the list of the cited works see Appendix E.

### 5.2.1.1 Delaunay Triangulation in the Quantum Space

In this section, we introduce the definition of Delaunay triangulation, and we extend it to the quantum space. The quantum Delaunay triangulation of set of quantum states $\mathcal{S}$ denoted by $Del(S)$ is the geometric dual of quantum Voronoi diagrams $vo(\mathcal{S})$ [Goodman04]. The Delaunay triangulations can be constructed from Laguerre diagrams (diagram of balls) [Aurenhammer2000], [Boissonnat07], [Nielsen07]. If the set of balls are Euclidean balls $\mathbb{R}^d$ , then this triangulation is called regular triangulation of the balls, and the vertices of this triangulation are the centers of the balls, whose cell is non empty [Goodman04]. The circumcenter of the given quantum states is the center of the *circle* that passes through the density matrices $\rho_1$ and $\rho_2$ of the edge $\rho_1\rho_2$ and endpoints $\rho_1$, $\rho_2$ and $\rho_3$ of the triangle $\rho_1\rho_2\rho_3$ , see Fig. 5.1 [Gyongyosi11b], [Imre12].

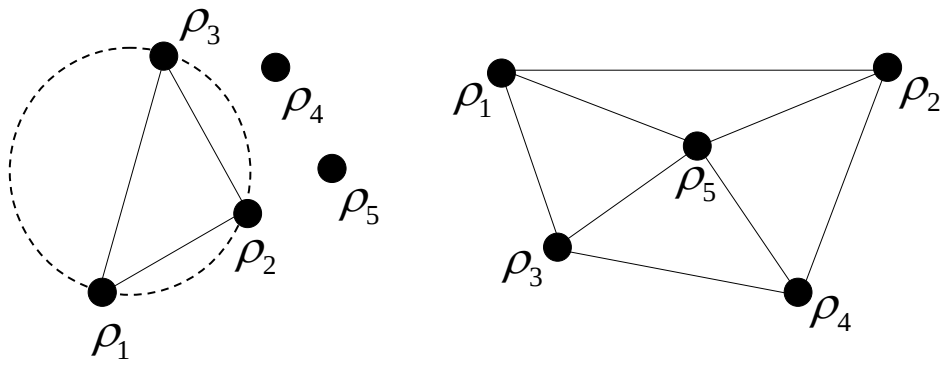

**Fig. 5.1.** The Delaunay triangulation of a certain set of density matrices.

The triangle $t$ is said to be *Delaunay*, when its circumcenter is *empty*. In Fig. 5.2(a) the circle centered at a vertex $\mathbf{c}$, gives an *empty* circumcenter for quantum state set $\mathcal{S} = \{\rho_1, \rho_2, \ldots \rho_n\}$. The Delaunay triangulation of set $\mathcal{S}$, denoted by $Del(\mathcal{S})$, is unique, if at most three quantum states $\rho \in \mathcal{S}$ are co-circular [Gyongyosi11b], [Imre12]. The *Delaunay triangulation* $Del(\mathcal{S})$ of $\mathcal{S}$ *maximizes* the *minimum* angle among all triangulation of $\mathcal{S}$, Fig. 5.2(b).

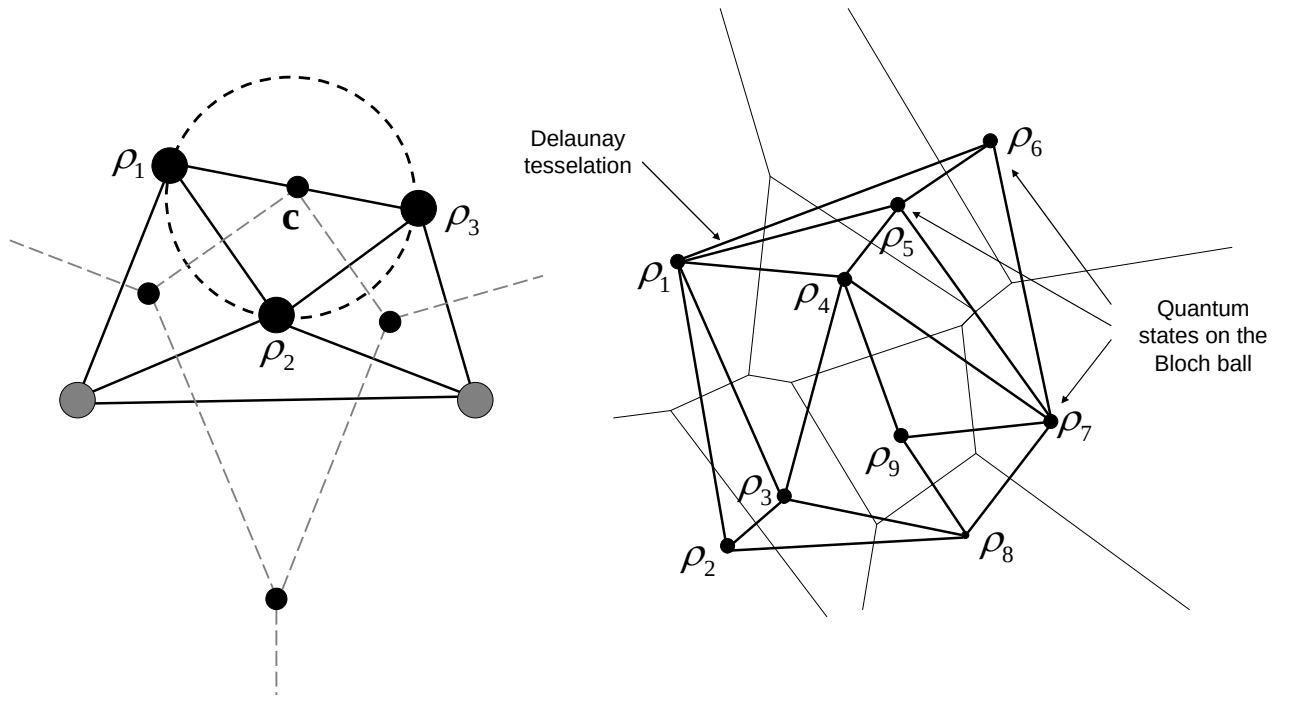

**Fig. 5.2.** a: The triangle of quantum states corresponds to the vertex $\mathbf{c}$, which is the center of its circumcenter. b: Delaunay tessellation between the density matrices.

If we choose a subset $\gamma$ of at most $d+1$ states in $\mathcal{S} = \left\{\rho_1, \ldots, \rho_n\right\}$, then the convex hull of the associated quantum states $\rho_i, i \in \gamma$, is a simplex of the quantum triangulation of $\mathcal{S}$, if and only if there exists an *empty* quantum informational ball $B$ passing through the $\rho_i, i \in \gamma$. This is called the *empty ball* property, see Fig. 5.3. The quantum Delaunay triangulation of the density matrices has a distorted structure, since the distance calculations are based on the quantum relative entropy, instead of Euclidean functions [Gyongyosi11b].

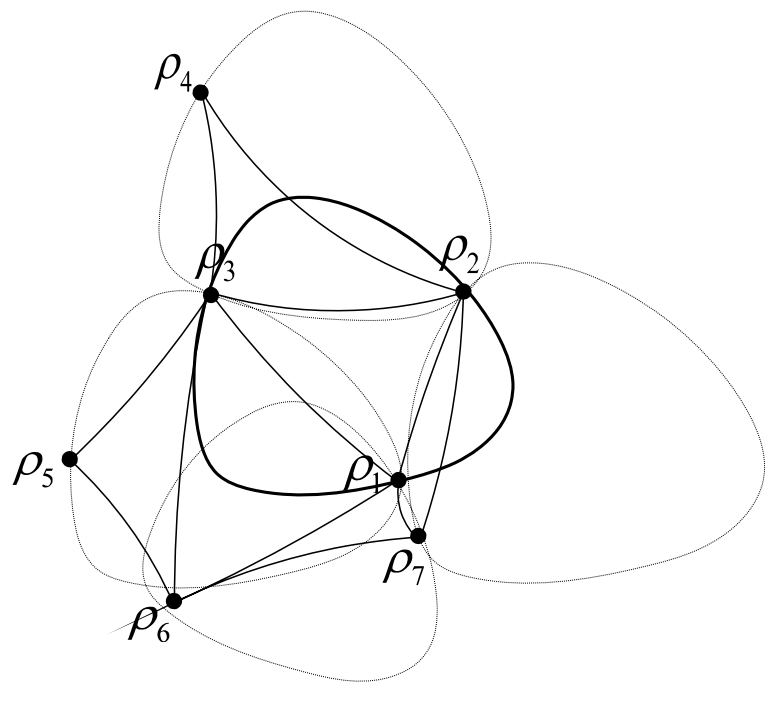

**Fig. 5.3.** The empty ball property of quantum Delaunay triangulation.

Using the results of Rajan from classical information geometry [Rajan94] for a given triangulation $T$ and set of quantum states $\mathcal{S}$, the radius of the smallest quantum informational ball, (see Appendix E) can be computed by quantum Delaunay triangulation $Del\left(\mathcal{S}\right)$ from triangulation $T\left(\mathcal{S}\right)$ of the quantum states as

$$Del\left(\mathcal{S}\right)=\min_{T\in T\left(\mathcal{S}\right)}\max_{\zeta\in T}r\left(\zeta\right), \tag{5.1}$$

where $\zeta$ is the $d$-simplex, and $r\left(\zeta\right)$ is the radius of the smallest quantum informational ball containing $d$-simplex $\zeta$.

In Fig. 5.4(a) we show the Delaunay triangulation on the Bloch sphere with respect to quantum relative entropy function $D\left(\cdot\|\cdot\right)$. The *mixed* quantum states are denoted by $\rho_1$, $\rho_2$ and $\rho_3$, the quantum Delaunay triangle is denoted by $Del\left(\rho_1,\rho_2,\rho_3\right)$. The bisector points between the quantum states with respect to quantum relative entropy denoted by points $v_1$, $v_2$ and $v_3$. The bisectors intersect the center of the smallest quantum informational ball, denoted by $\mathbf{c}^*$. In Fig. 5.4(b) we show the radius $r^*$ of the smallest enclosing quantum informational ball, centered at point $\mathbf{c}^*$. The distorted structure of the smallest enclosing quantum relative entropy ball is can be seen, the quantum ball is derived from the quantum Delaunay triangulation.

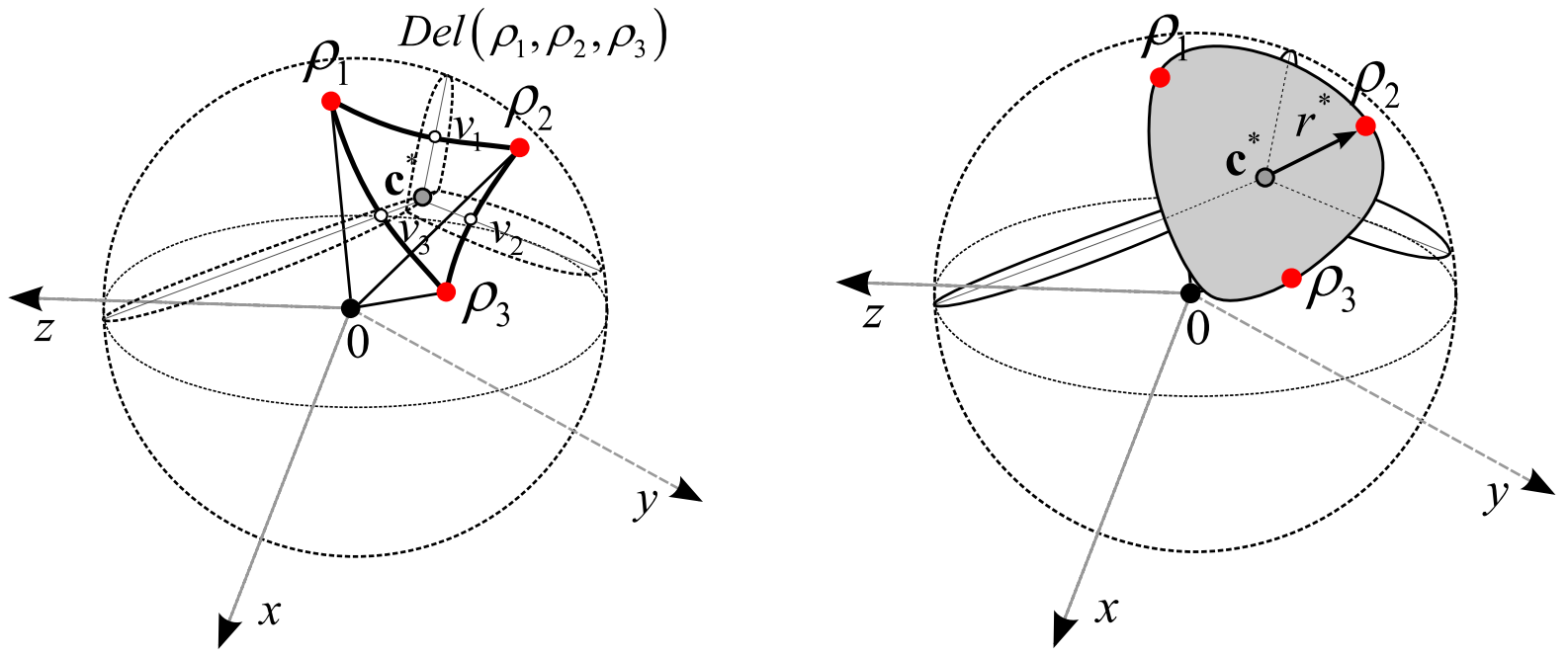

**Fig. 5.4.** a: Delaunay triangle with respect to quantum informational distance. (b): Smallest enclosing quantum informational ball and its radius.

Based on the results of Section 5.2, in Section 5.3 we define a new informational geometric object for the analysis of the superactivation of quantum channels. The algorithm for the determination of the proposed information geometric object will be introduced in Chapter 6 and Chapter 7. Further supplementary material is included in Appendix E. We also suggest the book of Imre and Gyongyosi [Imre12]. The Related Work subsection is included in Appendix E.

# 5.3 The Quantum Informational Superball

In this section, the theses are presented regarding the information geometric approach for the superactivation of quantum channels. The fundamental result of the information geometric analysis for the superactivation effect is summarized in Thesisgroup 1.

**Thesisgroup 1.** *I proved that the superactivation of arbitrary dimensional quantum channels can be determined by means of an appropriate information geometric object. I discovered that the superactivation effect is rooted in information geometric issues.*

**Definition 1** (On the quantum informational superball). The superball is an abstract geometrical object which measures the superactivated capacity of the joint channel structure $\mathcal{N}_1 \otimes \mathcal{N}_2$ in the Hilbert space $\mathcal{H}^{d_1 \cdot d_2} = \mathcal{H}^{d_1} \otimes \mathcal{H}^{d_2}$, formulated by channels $\mathcal{N}_1 \in \mathcal{H}^{d_1}$ and $\mathcal{N}_2 \in \mathcal{H}^{d_2}$. The quantum informational superball is defined over the space $\mathbb{C}^{d_1 \cdot d_2} = \mathbb{C}^{d_1} \otimes \mathbb{C}^{d_2}$. The $\mathbb{C}^{d_1 \cdot d_2}$ superball structure uses function $D(\cdot\|\cdot)$ to measure the superactivation of the joint channel $\mathcal{N}_1 \otimes \mathcal{N}_2$ of arbitrary dimensional quantum channels with Hilbert space $\mathcal{H}^{d_1 \cdot d_2} = \mathcal{H}^{d_1} \otimes \mathcal{H}^{d_2}$.

**Remark 1** (On the structure of the geometrical object). *The iteration steps are realized in the space $\mathcal{H}^{d_1} \otimes \mathcal{H}^{d_2}$ of the $\mathcal{N}_1 \otimes \mathcal{N}_2$ joint structure with $d_1$ and $d_2$ dimensional channels,* $\mathcal{N}_1 \in \mathcal{H}^{d_1}$ and $\mathcal{N}_2 \in \mathcal{H}^{d_2}$. *The informational geometric algorithms will use the space of $\mathbb{C}^{d_1 \cdot d_2} = \mathbb{C}^{d_1} \otimes \mathbb{C}^{d_2}$ for the iterations. The result of the informational theoretic distances in $\mathbb{C}^{d_1 \cdot d_2}$ will be measured by the quantum relative entropy function in $\mathbb{C}^{d_1 \cdot d_2}$. The distance calculations in the space of $\mathcal{H}^{d_1 \cdot d_2}$ of $\mathcal{N}_1 \otimes \mathcal{N}_2$ are based on the use of the $D(\cdot\|\cdot)$ quantum relative entropy function. The superactivated capacities of $\mathcal{N}_1 \otimes \mathcal{N}_2$ in $\mathcal{H}^{d_1 \cdot d_2} = \mathcal{H}^{d_1} \otimes \mathcal{H}^{\cdot d_2}$ will be measured as the quantum informational distance in* $\mathbb{C}^{d_1 \cdot d_2} = \mathbb{C}^{d_1} \otimes \mathbb{C}^{d_2}$, *between the* $\sigma_{12}$ *average channel output density matrix and the* $\rho_{12}$ *optimal channel output matrix of the joint channel* $\mathcal{N}_1 \otimes \mathcal{N}_2 \in \mathcal{H}^{d_1 \cdot d_2} = \mathcal{H}^{d_1} \otimes \mathcal{H}^{d_2}$.

**Corollary 1** (On the informational distance-measuring property of the superball structure). *The radius of the* $\mathbb{C}^{d_1 \cdot d_2}$ *superball construction measures the informational distances* $D(\sigma_{12} \| \rho_{12})$ *in the space* $\mathbb{C}^{d_1 \cdot d_2}$, *where* $\sigma_{12}$ *and* $\rho_{12}$ *are the* $\mathbb{C}^{d_1 \cdot d_2}$ *dimensional average and optimal density matrices of* $\mathcal{N}_1 \otimes \mathcal{N}_2$, *where* $\mathcal{N}_1 \in \mathcal{H}^{d_1}$ and $\mathcal{N}_2 \in \mathcal{H}^{d_2}$.

*Note:* In the superactivation of quantum capacity, the two channels in $\mathcal{N}_1 \otimes \mathcal{N}_2$ are both assumed to be $d$ dimensional, i.e., $d = d_1 = d_2$ and $\mathcal{N}_1 \in \mathcal{H}^{d}$, $\mathcal{N}_2 \in \mathcal{H}^{d}$ *and* $\mathcal{N}_1 \otimes \mathcal{N}_2 \in \mathcal{H}^{d^2} = \mathcal{H}^{d_1} \otimes \mathcal{H}^{d_2}$.

The results on the geometric interpretation of the superactivation of quantum channels are illustrated with the $Q^{(1)}(\mathcal{N}_1 \otimes \mathcal{N}_2)$ single-use quantum capacity of the joint structure $\mathcal{N}_1 \otimes \mathcal{N}_2$. The proposed theses hold for all channel capacities of the joint channel $\mathcal{N}_1 \otimes \mathcal{N}_2$ for which the superactivation is possible. These capacities are $Q^{(1)}(\mathcal{N}_1 \otimes \mathcal{N}_2)$, $Q(\mathcal{N}_1 \otimes \mathcal{N}_2)$, $C_0^{(1)}(\mathcal{N}_1 \otimes \mathcal{N}_2)$, $C_0(\mathcal{N}_1 \otimes \mathcal{N}_2)$ and $Q_0^{(1)}(\mathcal{N}_1 \otimes \mathcal{N}_2)$, $Q_0(\mathcal{N}_1 \otimes \mathcal{N}_2)$. For simplicity in this chapter's figures and theses, we assumed the use of the $Q^{(1)}(\mathcal{N}_1 \otimes \mathcal{N}_2)$ single-use quantum capacity of the joint structure $\mathcal{N}_1 \otimes \mathcal{N}_2$. The results will be extended to the asymptotic quantum capacity in Chapter 6 and to the single-use and asymptotic classical zero-error capacities in Chapter 7. Further supplementary information on the mathematical background of the proposed information geometric approach is included in Appendix E.

## 5.3.1 The Geometrical Structure of the Superball

The geometric interpretation of the superactivated joint capacity of the joint channel structure $\mathcal{N}_1 \otimes \mathcal{N}_2 \in \mathcal{H}^{d_1 \cdot d_2} = \mathcal{H}^{d_1} \otimes \mathcal{H}^{d_2}$ will be given using an abstract quantum informational object over $\mathbb{C}^{d_1 \cdot d_2}$. The result on the information geometric interpretation of the superactivated channel capacities is summarized in Thesis 1.1.

> **Thesis 1.1.** *I showed that the superactivation of arbitrary dimensional quantum channels can be determined by an abstract geometrical object called the quantum informational superball.*

The superactivation property of the joint channel $\mathcal{N}_{12} = \mathcal{N}_1 \otimes \mathcal{N}_2 \in \mathcal{H}^{d_1 \cdot d_2} = \mathcal{H}^{d_1} \otimes \mathcal{H}^{d_2}$ will be analyzed by means of the superball structure over $\mathbb{C}^{d_1 \cdot d_2}$. The superball has radius $r_{12} \in \mathbb{C}^{d_1 \cdot d_2}$ defined over $\mathbb{C}^{d_1 \cdot d_2}$ of $\mathcal{N}_1 \otimes \mathcal{N}_2 \in \mathcal{H}^{d_1} \otimes \mathcal{H}^{d_2}$, formed by balls with $r_1 \in \mathbb{C}^{d_1}$, $r_2 \in \mathbb{C}^{d_2}$ of channels $\mathcal{N}_1 \in \mathcal{H}^{d_1}$ and $\mathcal{N}_2 \in \mathcal{H}^{d_2}$ with quantum informational theoretic (i.e., defined by the quantum relative entropy function) radii lengths $r_1^* = 0 \in \mathbb{R}$, $r_2^* = 0 \in \mathbb{R}$. The $r_{12}^* \in \mathbb{R}$ informational theoretic length of radius $r_{12} \in \mathbb{C}^{d_1 \cdot d_2}$ is

$$r_{12}^* > r_1^* + r_2^* \in \mathbb{R}. \tag{5.2}$$

**Remark 2** *The radius $r_{12} \in \mathbb{C}^{d_1 \cdot d_2}$ of the superball defined over the space of $\mathbb{C}^{d_1 \cdot d_2}$ of the joint channel structure $\mathcal{N}_{12} = \mathcal{N}_1 \otimes \mathcal{N}_2 \in \mathcal{H}^{d_1 \cdot d_2} = \mathcal{H}^{d_1} \otimes \mathcal{H}^{d_2}$. The radius $r_{12} \in \mathbb{C}^{d_1 \cdot d_2}$ has magnitude $|r_{12}| = m_{12} \in \mathbb{R}$. The informational theoretic length $r_{12}^* \in \mathbb{R}$ will be referred as the informational theoretic radius.*

As will be shown in Chapter 6 and Chapter 7, the superball representation can be extended to analyze the $Q(\mathcal{N}_1 \otimes \mathcal{N}_2)$ quantum capacity and the $C_0(\mathcal{N}_1 \otimes \mathcal{N}_2)$ zero-error capacity of quantum channels. The steps of construction for the quantum superball are shown for the superactivation of the $Q^{(1)}\left(\mathcal{N}_1 \otimes \mathcal{N}_2\right)$ single-use quantum capacity (see the LSD-theorem in Chapter 3) of the joint structure. In Chapters 6 and 7, the results will be extended for the $Q\left(\mathcal{N}_1 \otimes \mathcal{N}_2\right)$ superactivated asymptotic quantum capacity and the superactivated $C_0\left(\mathcal{N}_1 \otimes \mathcal{N}_2\right)$ classical zero-error capacity of $\mathcal{N}_1 \otimes \mathcal{N}_2 \in \mathcal{H}^{d_1} \otimes \mathcal{H}^{d_2}$.

In the construction of the quantum superball, we rely on the fact that superactivation of quantum channels is an extreme violation of the additivity of quantum channels. As was shown by Shor [Shor04a], the additivity problem of quantum channel capacities results in the classical capacity of the tensor product of two quantum channels. $\mathcal{N}_1$ and $\mathcal{N}_2$ is additive if and only if the minimum output entropy (see Appendix D) of the joint channel $\mathcal{N}_1 \otimes \mathcal{N}_2 \in \mathcal{H}^{d_1 \cdot d_2} = \mathcal{H}^{d_1} \otimes \mathcal{H}^{d_2}$ is the strict sum of the minimum output entropy of the analyzed channels $\mathcal{N}_1 \in \mathcal{H}^{d_1}$ and $\mathcal{N}_2 \in \mathcal{H}^{d_2}$, taken separately. This result can be extended to studying the geometric background of the joint channel structure's superactivation of $\mathcal{N}_{12} = \mathcal{N}_1 \otimes \mathcal{N}_2 \in \mathcal{H}^{d_1 \cdot d_2} = \mathcal{H}^{d_1} \otimes \mathcal{H}^{d_2}$ by our abstract geometric object. The distance calculations inside the $\mathbb{C}^{d_1 \cdot d_2}$ object are defined by the quantum relative entropy function $D\left(\cdot\middle\|\cdot\right)$ over $\mathbb{C}^{d_1 \cdot d_2}$. The theoretical background of this new geometric representation can be found in Section 5.2. Further information is included in Appendix E.

If we assume that the joint structure $\mathcal{N}_{12} = \mathcal{N}_1 \otimes \mathcal{N}_2 \in \mathcal{H}^{d_1 \cdot d_2} = \mathcal{H}^{d_1} \otimes \mathcal{H}^{d_2}$ consists of two zero-capacity quantum channels $\mathcal{N}_1 \in \mathcal{H}^{d_1}$ and $\mathcal{N}_2 \in \mathcal{H}^{d_2}$, the single-use joint quantum capacity $Q^{(1)}\left(\mathcal{N}_1 \otimes \mathcal{N}_2\right)$ will be computed by the radius $r_{12}^{(1)^*} \in \mathbb{R}$ of the $\mathbb{C}^{d_1 \cdot d_2}$ smallest enclosing quantum informational *superball* over the space $\mathbb{C}^{d_1 \cdot d_2}$ as follows:

$$\begin{aligned} r_{12}^{(1)^*} &= Q^{(1)}\left(\mathcal{N}_1 \otimes \mathcal{N}_2\right) \\ &= \min_{\sigma} \max_{\rho} D\left(\rho_{12}^{AB} \middle\| \sigma_{12}^{AB}\right) - \min_{\sigma} \max_{\rho} D\left(\rho_{12}^{AE} \middle\| \sigma_{12}^{AE}\right) \\ &= \min_{\sigma} \max_{\rho} D\left(\rho_{12}^{AB-AE} \middle\| \sigma_{12}^{AB-AE}\right), \end{aligned} \tag{5.3}$$

$\rho_{12}^{AB} \in \mathbb{C}^{d_1} \otimes \mathbb{C}^{d_2}$ is the optimal output state of joint channel $\mathcal{N}_{12} \in \mathcal{H}^{d_1 \cdot d_2}$, and $\sigma_{12}^{AB} \in \mathbb{C}^{d_1} \otimes \mathbb{C}^{d_2}$ is the average state of joint channel $\mathcal{N}_{12} \in \mathcal{H}^{d_1 \cdot d_2}$ between Alice and Bob, also referred by $\mathcal{N}_{AB}$ (for exact definitions and formulas, see Chapter 6). The term $E$ denotes the environment, and $AE$ denotes the channel $\mathcal{N}_{AE}$ between Alice and the environment with the optimal state $\rho_{12}^{AE} \in \mathbb{C}^{d_1} \otimes \mathbb{C}^{d_2}$ (referred as the environment's optimal state), and average state $\sigma_{12}^{AE} \in \mathbb{C}^{d_1} \otimes \mathbb{C}^{d_2}$ (environment's average state). The final optimal output channel state is depicted by $\rho_{12}^{AB-AE} \in \mathbb{C}^{d_1} \otimes \mathbb{C}^{d_2}$, while $\sigma_{12}^{AB-AE} \in \mathbb{C}^{d_1} \otimes \mathbb{C}^{d_2}$ is the final output average state of the channel between Alice and the environment. From (5.3) follows that $r_{12}^{(1)^*} > 0$ only if the $\mathcal{N}_{12} = \mathcal{N}_1 \otimes \mathcal{N}_2 \in \mathcal{H}^{d_1 \cdot d_2}$ joint structure is superactive, i.e., $Q^{(1)}(\mathcal{N}_{12}) > 0$ otherwise $r_{12}^{(1)^*} = Q^{(1)}(\mathcal{N}_1) = Q^{(1)}(\mathcal{N}_2) = Q^{(1)}(\mathcal{N}_{12}) = 0$. In the space of $\mathbb{C}^{d_1 \cdot d_2}$, the average state $\sigma_{12}^{AB-AE} \in \mathbb{C}^{d_1} \otimes \mathbb{C}^{d_2}$ of the joint channel $\mathcal{N}_{12} \in \mathcal{H}^{d_1 \cdot d_2}$ is the *center* of the $\mathbb{C}^{d_1 \cdot d_2}$ quantum superball, generated from the optimal averages $\sigma_1^{AB-AE} \in \mathbb{C}^{d_1} \otimes \mathbb{C}^{d_2}$ and $\sigma_2^{AB-AE} \in \mathbb{C}^{d_1} \otimes \mathbb{C}^{d_2}$ of $\mathcal{N}_1 \in \mathcal{H}^{d_1}$ and $\mathcal{N}_2 \in \mathcal{H}^{d_2}$ as follows:

$$\sigma_{12}^{AB-AE} = \arg\max_{\sigma_i} \left\{ \mathrm{S}\left(\sigma_1^{AB-AE}\right), \mathrm{S}\left(\sigma_2^{AB-AE}\right) \right\}, \tag{5.4}$$

where $\mathrm{S}\left(\sigma_1^{AB-AE}\right)$ and $\mathrm{S}\left(\sigma_2^{AB-AE}\right)$ denote the von Neumann entropy of optimal average states $\sigma_1^{AB-AE} \in \mathbb{C}^{d_1} \otimes \mathbb{C}^{d_2}$ and $\sigma_2^{AB-AE} \in \mathbb{C}^{d_1} \otimes \mathbb{C}^{d_2}$ of $\mathcal{N}_1$ and $\mathcal{N}_2$ in the joint channel $\mathcal{N}_1 \otimes \mathcal{N}_2 \in \mathcal{H}^{d_1 \cdot d_2} = \mathcal{H}^{d_1} \otimes \mathcal{H}^{d_2}$. In other words, we have to find the minimal length radius $r^*_{\sigma_{12}^{AB-AE}} \in \mathbb{R}$ of optimal average states with magnitudes $m_{\sigma_1^{AB-AE}} \in \mathbb{R}$ and $m_{\sigma_2^{AB-AE}} \in \mathbb{R}$ as follows:

$$\left| r_{\sigma_{12}^{AB-AE}} \right| = \min\left\{ r^*_{\sigma_1^{AB-AE}}, r^*_{\sigma_2^{AB-AE}} \right\} = \min\left\{ m_{\sigma_1^{AB-AE}}, m_{\sigma_2^{AB-AE}} \right\} \in \mathbb{R}. \tag{5.5}$$

The joint optimal state $\rho_{12}^{AB-AE} \in \mathbb{C}^{d_1} \otimes \mathbb{C}^{d_2}$ is on the boundary of the $\mathbb{C}^{d_1 \cdot d_2}$ quantum superball, defined as follows:

$$\rho_{12}^{AB-AE} = \arg\min_{\rho_i} \left\{ \mathrm{S}\left(\rho_1^{AB-AE}\right), \mathrm{S}\left(\rho_2^{AB-AE}\right) \right\}, \tag{5.6}$$

where $\rho_i^{AB-AE} \in \mathbb{C}^{d_1} \otimes \mathbb{C}^{d_2}$ are the optimal states of $\mathcal{N}_1$ and $\mathcal{N}_2$ in the joint channel $\mathcal{N}_1 \otimes \mathcal{N}_2 \in \mathcal{H}^{d_1 \cdot d_2} = \mathcal{H}^{d_1} \otimes \mathcal{H}^{d_2}$. The optimal joint state $\rho_{12}^{AB-AE} \in \mathbb{C}^{d_1} \otimes \mathbb{C}^{d_2}$ has minimal von Neumann entropy $\mathrm{S}_{\min}\left(\rho_{12}^{AB-AE}\right)$ with a magnitude of $m_{\rho_{12}^{AB-AE}} \in \mathbb{R}$. The joint optimal output state is defined as radius $r^*_{\rho_{12}^{AB-AE}} \in \mathbb{R}$ with magnitude

$$\left| r_{\rho_{12}^{AB-AE}} \right| = m_{\rho_{12}^{AB-AE}} \in \mathbb{R}. \tag{5.7}$$

If the joint state is a *product* state, then $\rho_{12}^{AB-AE} \in \mathbb{C}^{d_1} \otimes \mathbb{C}^{d_2}$ can be decomposed as follows:

$$\rho_{12}^{AB-AE} = \rho_{12,(1)}^{AB-AE} \otimes \rho_{12,(2)}^{AB-AE} \in \mathbb{C}^{d_1} \otimes \mathbb{C}^{d_2}, \tag{5.8}$$

with magnitudes $\left| r_{\rho_{12,(1)}^{AB-AE}} \right| = m_{\rho_{12,(2)}^{AB-AE}} \in \mathbb{R}$ and $\left| r_{\rho_{12,(2)}^{AB-AE}} \right| = m_{\rho_{12,(2)}^{AB-AE}} \in \mathbb{R}$, and

$$\left| r_{\rho_{12}^{AB-AE}} \right| = m_{\rho_{12}^{AB-AE}} = \left| r_{\rho_{12,(1)}^{AB-AE}} \right| + \left| r_{\rho_{12,(2)}^{AB-AE}} \right| = m_{\rho_{12,(1)}^{AB-AE}} + m_{\rho_{12,(2)}^{AB-AE}} \in \mathbb{R}. \tag{5.9}$$

Let $\rho_{12}^{AB-AE} = \rho_{12,(1)}^{AB-AE} \otimes \rho_{12,(2)}^{AB-AE} \in \mathbb{C}^{d_1} \otimes \mathbb{C}^{d_2}$. In this case, for the von Neumann entropy of the $\rho_{12}^{AB-AE} \in \mathbb{C}^{d_1} \otimes \mathbb{C}^{d_2}$ boundary state the following equality holds:

$$\mathrm{S}_{\min}\left(\rho_{12}^{AB-AE}\right) = \mathrm{S}_{\min}\left(\rho_{12,(1)}^{AB-AE}\right) + \mathrm{S}_{\min}\left(\rho_{12,(2)}^{AB-AE}\right). \tag{5.10}$$

It also follows that if the optimal joint state $\rho_{12}^{AB-AE} \in \mathbb{C}^{d_1} \otimes \mathbb{C}^{d_2}$ is an *entangled* state, then

$$S_{\min}\left(\rho_{12}^{AB-AE}\right) \neq S_{\min}\left(\rho_{12,(1)}^{AB-AE}\right) + S_{\min}\left(\rho_{12,(2)}^{AB-AE}\right) \tag{5.11}$$

and

$$\left|r_{\rho_{12}^{AB-AE}}\right| = m_{\rho_{12}^{AB-AE}} \neq \left|r_{\rho_{12,(1)}^{AB-AE}}\right| + \left|r_{\rho_{12,(2)}^{AB-AE}}\right| = m_{\rho_{12,(1)}^{AB-AE}} + m_{\rho_{12,(2)}^{AB-AE}} \in \mathbb{R}. \tag{5.12}$$

For an *entangled* joint optimal state $\rho_{12}^{AB-AE} \in \mathbb{C}^{d_1} \otimes \mathbb{C}^{d_2}$, it also follows from (5.6) that

$$S_{\min}\left(\rho_{12}^{AB-AE}\right) = S\left(\rho_{12}^{AB-AE}\right). \tag{5.13}$$

If the $\sigma_{12}^{AB-AE}$ joint average and the $\rho_{12}^{AB-AE}$ joint optimal states in $\mathbb{C}^{d_1} \otimes \mathbb{C}^{d_2}$ are product states, then the superactivation of the joint structure $\mathcal{N}_1 \otimes \mathcal{N}_2 \in \mathcal{H}^{d_1 \cdot d_2} = \mathcal{H}^{d_1} \otimes \mathcal{H}^{d_2}$ is not possible. *It has strict consequences on the properties of the radius* $r_{12}^{*} \in \mathbb{R}$ *over the space* $\mathbb{C}^{d_1} \otimes \mathbb{C}^{d_2}$, as we will see in the theses of this chapter. Using (5.7) and (5.4), the quantum informational radius can be expressed as we have shown in (5.2). An important conclusion from this is as follows: The superactivation of a joint channel structure with arbitrary dimensional channels $\mathcal{N}_1 \in \mathcal{H}^{d_1}$ and $\mathcal{N}_2 \in \mathcal{H}^{d_2}$ can be analyzed by a *min- and max-searching* problem, since in (5.7), we have to find those quantum states in $\mathcal{H}^{d_1} \otimes \mathcal{H}^{d_2}$ that have minimal entropy. Meanwhile, our task in (5.4) is to select the maximal entropy average state in $\mathcal{H}^{d_1} \otimes \mathcal{H}^{d_2}$. As we have shown, the problem of superactivation can be rephrased in a mathematically equivalent form using $\mathbb{C}^{d_1} \otimes \mathbb{C}^{d_2}$, and very efficient algorithmical solutions can be constructed to analyze the space of $\mathbb{C}^{d_1} \otimes \mathbb{C}^{d_2}$, without the NP-complete numerical calculations, as will be presented in Chapters 6 and 7.

The quantum superball is illustrated in Fig. 5.5. The quantum informational ball is an abstract object over $\mathbb{C}^{d_1} \otimes \mathbb{C}^{d_2}$; the structure of the ball is distorted according to the relative entropy function.

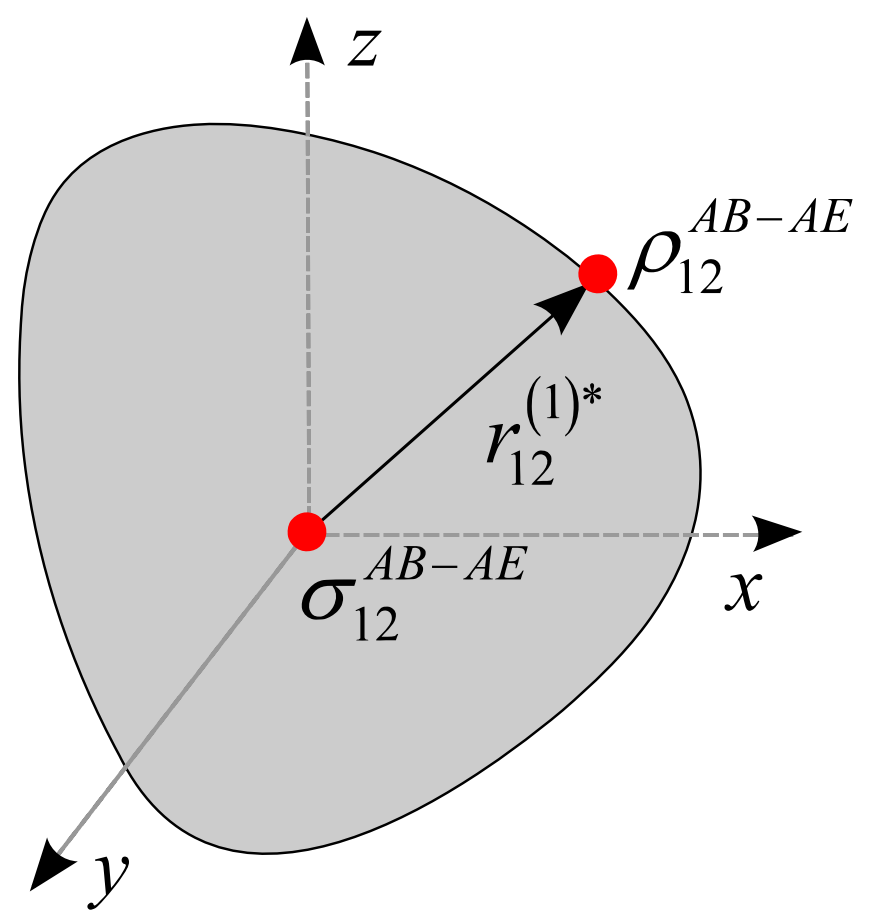

**Fig. 5.5.** Geometric interpretation of superactivation of quantum capacity in the quantum superball representation. The superball is an abstract object defined over $\mathbb{C}^{d_1} \otimes \mathbb{C}^{d_2}$ of joint channel structure using the quantum relative entropy function as distance measure.

The smallest enclosing quantum informational superball represents the convex set of the possible quantum states of channel output of joint channel construction $\mathcal{N}_{12}$. The convex hull over $\mathbb{C}^{d_1} \otimes \mathbb{C}^{d_2}$ includes the interior of the smallest quantum informational ball as well as the boundary of the $\mathbb{C}^{d_1 \cdot d_2}$ object.

#### 5.3.1.1 The Single Channel View

To illustrate the results of the information geometric approach in the space of $\mathcal{H}^{d_1 \cdot d_2} = \mathcal{H}^{d_1} \otimes \mathcal{H}^{d_2}$ of the channels $\mathcal{N}_1 \in \mathcal{H}^{d_1}$ and $\mathcal{N}_2 \in \mathcal{H}^{d_2}$, we introduce the *single channel view representation,* defined in $\mathbb{R}^3$ over the reduced subspace $\mathbb{C}^2$ of $\mathbb{C}^{d_1 \cdot d_2}$.

**Definition 2** (On the single channel view representation). The $\mathbb{R}^3$ single channel view is a reduction of space $\mathcal{H}^{d_1 \cdot d_2}$ of $\mathcal{N}_1 \otimes \mathcal{N}_2$ onto subspace $\mathcal{H}^2$. The single channel view is aimed to illustrate the results of the $\mathbb{C}^{d_1 \cdot d_2}$ dimensional iterations in $\mathbb{R}^3$ over $\mathbb{C}^2$.

**Remark 3** (On the connection between single channel view representation and channel space decomposition). *While the superball structure is defined over* $\mathbb{C}^{d_1 \cdot d_2}$, *the* $\mathbb{R}^3$ *single channel view is a reduction from* $\mathbb{C}^{d_1 \cdot d_2}$ *onto a subspace* $\mathbb{C}^2$. *It makes possible to study the superactivation of the* $\mathcal{N}_1 \otimes \mathcal{N}_2$ *joint structure of quantum channels in* $\mathcal{H}^{d_1 \cdot d_2} = \mathcal{H}^{d_1} \otimes \mathcal{H}^{d_2}$, *where* $\mathcal{N}_1 \in \mathcal{H}^{d_1}$ *and* $\mathcal{N}_2 \in \mathcal{H}^{d_2}$ *in the reduced subspace* $\mathbb{C}^2$. *The single channel representation has a connection with the decomposition of Hilbert space* $\mathcal{H}^{d_1 \cdot d_2} = \mathcal{H}^{d_1} \otimes \mathcal{H}^{d_2}$ *into qubit-spaces* $\mathcal{H}^2$, *if the channels* $\mathcal{N}_1 \in \mathcal{H}^{d_1}$ *and* $\mathcal{N}_2 \in \mathcal{H}^{d_2}$ *in* $\mathcal{N}_1 \otimes \mathcal{N}_2$ *have dimension* $d_1 = d_2 = 2^l$. *Assuming channels with dimension* $d = d_1 = d_2 = 2^l$, $\mathcal{N}_1 \in \mathcal{H}^{2^l}$, $\mathcal{N}_2 \in \mathcal{H}^{2^l}$ *and* $\mathcal{N}_1 \otimes \mathcal{N}_2$ *with* $\mathcal{H}^{2^{2l}} = \mathcal{H}^{2^l} \otimes \mathcal{H}^{2^l}$, *the space* $\mathcal{H}^{d^2} = \mathcal{H}^d \otimes \mathcal{H}^d$ *of the joint structure* $\mathcal{N}_1 \otimes \mathcal{N}_2$ *can be decomposed into the tensor product of* $2l$ *qubit-spaces as follows:* $\mathcal{H}^d \otimes \mathcal{H}^d = \mathcal{H}^{d^2} = \mathcal{H}_1^2 \otimes \mathcal{H}_2^2 \otimes \ldots \otimes \mathcal{H}_{2l}^2$.
*Assuming a joint channel* $\mathcal{N}_1 \otimes \mathcal{N}_2$ *with* $d = 2^2$ dimensional channels $\mathcal{N}_1 \in \mathcal{H}^4$ *and* $\mathcal{N}_2 \in \mathcal{H}^4$, *the space* $\mathcal{H}^4 \otimes \mathcal{H}^4 = \mathcal{H}^{16}$ *of the joint structure* $\mathcal{N}_1 \otimes \mathcal{N}_2$ *can be decomposed into* $2l = 2 \cdot 2$ *qubit-spaces as follows:* $\mathcal{H}^4 \otimes \mathcal{H}^4 = \mathcal{H}^{16} = \mathcal{H}^2 \otimes \mathcal{H}^2 \otimes \mathcal{H}^2 \otimes \mathcal{H}^2$.

*Note: If* $d \neq 2^l$ and $d_1 \neq d_2$ the reduction onto $\mathcal{H}^2$ *from* $\mathcal{H}^{d_1 \cdot d_2} = \mathcal{H}^{d_1} \otimes \mathcal{H}^{d_2}$ *also possible, however in this case the connection between the decomposition of* $\mathcal{H}^{d^2}$ *into* $2l$ *qubit-spaces will not hold.*

In Fig. 5.6(a), we depict the single-channel view of the joint structure $\mathcal{N}_{12} \in \mathcal{H}^{d_1 \cdot d_2} = \mathcal{H}^{d_1} \otimes \mathcal{H}^{d_2}$, where $\mathcal{N}_1 \in \mathcal{H}^{d_1}$ and $\mathcal{N}_2 \in \mathcal{H}^{d_2}$ are arbitrary dimensional channels. The optimal state of channel $\mathcal{N}_{AB}$ between Alice and Bob is denoted by $\rho_1^{AB} \in \mathbb{C}^2$; the average state is $\sigma_1^{AB} \in \mathbb{C}^2$. For the channel $\mathcal{N}_{AE}$ between Alice and the environment, these states are depicted by $\rho_1^{AE} \in \mathbb{C}^2$ and $\sigma_1^{AE} \in \mathbb{C}^2$, respectively. From these states, the single-use quantum capacity $Q^{(1)}(\mathcal{N}_1)$ is determined by the optimal state $\rho_1^{AB-AE} \in \mathbb{C}^2$ and the average state $\sigma_1^{AB-AE} \in \mathbb{C}^2$. For simplicity, the average states are assumed to be the center. The radii between the average states and the optimal states of channel $\mathcal{N}_{AE}$ between Alice and the environment, and $\mathcal{N}_{AB}$, between Alice and Bob are depicted by $r_1^{(AE)(1)^*} \in \mathbb{R}$ and $r_1^{(AB)(1)^*} \in \mathbb{R}$, from which the single-use quantum capacity $Q^{(1)}(\mathcal{N}_1)$ is described by the informational theoretic length $r_1^{(AB-AE)(1)^*} \in \mathbb{R}$.

In Fig. 5.6(b), the quantum superball representation is shown for the joint channel structure $\mathcal{N}_1 \otimes \mathcal{N}_2 \in \mathcal{H}^{d_1 \cdot d_2} = \mathcal{H}^{d_1} \otimes \mathcal{H}^{d_2}$. The joint optimal and average states for the channel $\mathcal{N}_{AB}$ between Alice and Bob are denoted by $\rho_{12}^{AB}$ and $\sigma_{12}^{AB}$. For the channel $\mathcal{N}_{AE}$ between Alice and the environment, these states are depicted by $\rho_{12}^{AE}$ and $\sigma_{12}^{AE}$. The $\mathbb{C}^{d_1 \cdot d_2}$ quantum superball is constructed with radius $r_{12}^{(AB-AE)(1)^*} \in \mathbb{R}$, which measures the $Q^{(1)}(\mathcal{N}_1 \otimes \mathcal{N}_2)$ single-use quantum capacity of the joint channel structure $\mathcal{N}_1 \otimes \mathcal{N}_2$. The quantum ball in $\mathbb{C}^{d_1 \cdot d_2}$ is determined by the average joint state $\sigma_{12}^{AB-AE} \in \mathbb{C}^{d_1 \cdot d_2}$ and the optimal joint state $\rho_{12}^{AB-AE} \in \mathbb{C}^{d_1 \cdot d_2}$ in the space $\mathbb{C}^{d_1 \cdot d_2}$ of the joint structure $\mathcal{N}_1 \otimes \mathcal{N}_2 \in \mathcal{H}^{d_1 \cdot d_2} = \mathcal{H}^{d_1} \otimes \mathcal{H}^{d_2}$.

In Fig. 5.6, we compare the two constructions; however, the joint channel capacity can be described only in the superball representation. In Fig. 5.6, the joint average states are assumed to be in the center. In Fig. 5.6(b) the center is the average state $\sigma_{12}^{AB-AE} = \frac{1}{d_1 \cdot d_2} I$.

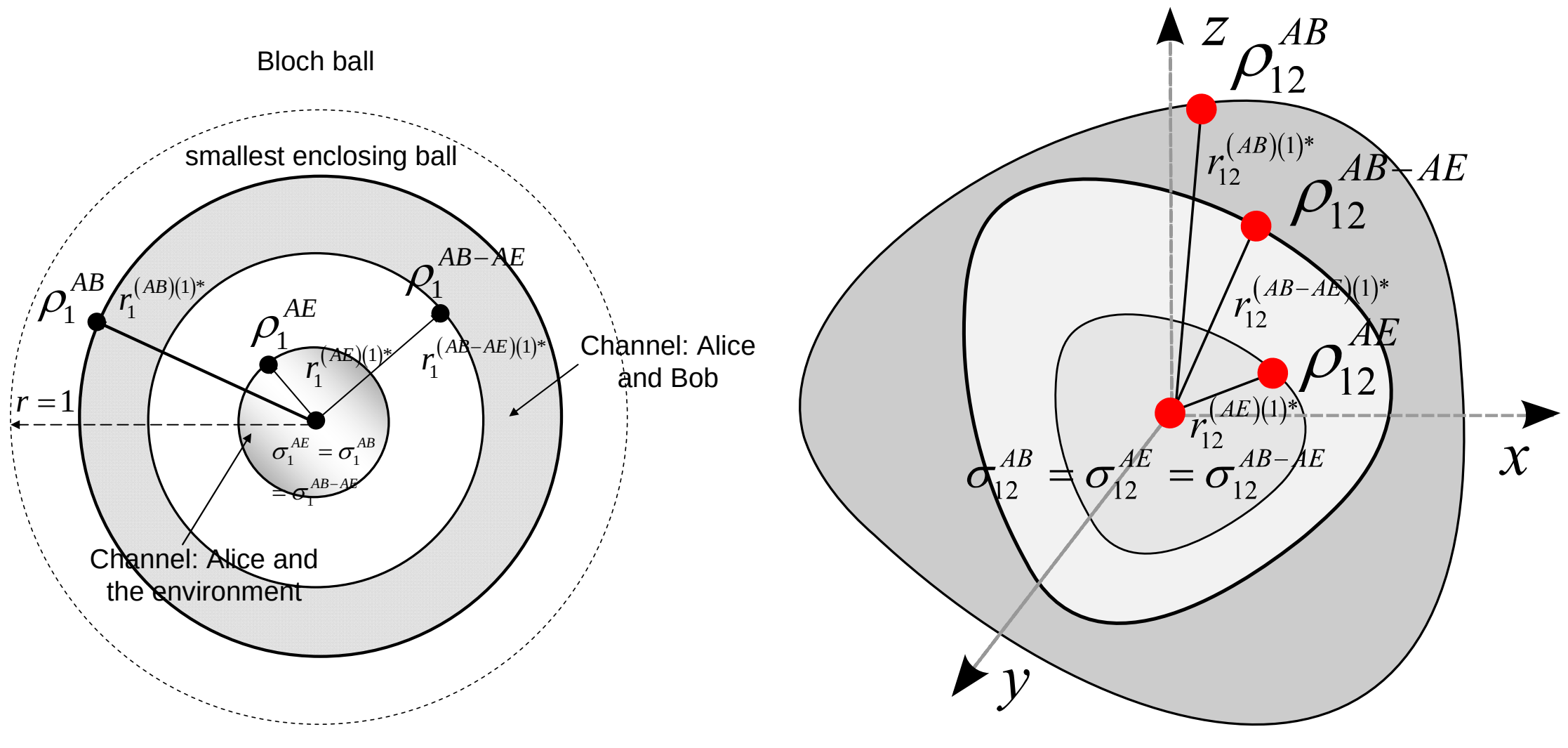

**Fig. 5.6.** Comparison of the single channel view (a) and the superball representation (b) for single-use quantum capacity of the joint channel structure. The single channel view is a reduction of $\mathbb{C}^{d_1 \cdot d_2}$ onto the subspace $\mathbb{C}^2$. The superball is defined over $\mathbb{C}^{d_1 \cdot d_2}$. The quantum ball cannot be decomposed as the radii of independent channels; it represents the superactivated capacity of $\mathcal{N}_1 \otimes \mathcal{N}_2 \in \mathcal{H}^{d_1 \cdot d_2} = \mathcal{H}^{d_1} \otimes \mathcal{H}^{d_2}$.

In the next subsection we define the properties of the defined geometrical object.

## 5.3.2 Characterization of the Quantum Superball

The optimal joint states have an important role in the computation of the quantum superball. The connection between the additivity of HSW capacity and the minimal entropy channel output states was shown in [Shor04a]. The optimal channel output states have a nice property: these states have minimal entropy among all possible channel output states. These results can be used in the construction of the quantum superball; however, both the optimal state $\rho_{12}^{AB-AE} \in \mathbb{C}^{d_1 \cdot d_2}$ and the average state $\sigma_{12}^{AB-AE} \in \mathbb{C}^{d_1 \cdot d_2}$ are joint states and can be defined only for the joint channel $\mathcal{N}_1 \otimes \mathcal{N}_2 \in \mathcal{H}^{d_1 \cdot d_2} = \mathcal{H}^{d_1} \otimes \mathcal{H}^{d_2}$ instead of the structure's individual quantum channels $\mathcal{N}_1 \in \mathcal{H}^{d_1}$ and $\mathcal{N}_2 \in \mathcal{H}^{d_2}$. These joint states can maximize the channel capacities. For the joint channel $\mathcal{N}_1 \otimes \mathcal{N}_2$, if the joint optimal output state is a product state $\rho_{12}^{AB-AE} = \rho_{12,(1)}^{AB-AE} \otimes \rho_{12,(2)}^{AB-AE} \in \mathbb{C}^{d_1 \cdot d_2}$, then the state that has minimal von Neumann entropy can be determined as follows:

$$\mathrm{S}_{\min}\left(\rho_{12}^{AB-AE}\right) = \arg \min_{\rho_{1(i)},\rho_{2(i)}} \left\{ \mathrm{S}\left(\rho_{12,(1)_{(1)}}^{AB-AE}\right), \mathrm{S}\left(\rho_{12,(2)_{(1)}}^{AB-AE}\right), \ldots, \mathrm{S}\left(\rho_{12,(1)_{(n)}}^{AB-AE}\right), \mathrm{S}\left(\rho_{12,(2)_{(n)}}^{AB-AE}\right) \right\}, \tag{5.14}$$

where $\rho_{12,(1)_{(i)}}^{AB-AE}$ and $\rho_{12,(2)_{(i)}}^{AB-AE}$ are the $i$-th optimal states of channels $\mathcal{N}_1$ and $\mathcal{N}_2$ in $\mathcal{N}_1 \otimes \mathcal{N}_2 \in \mathcal{H}^{d_1 \cdot d_2}$. If the joint optimal state $\rho_{12}^{AB-AE} \in \mathbb{C}^{d_1 \cdot d_2}$ is entangled, then the minimal entropy state can be determined as follows:

$$\mathrm{S}_{\min}\left(\rho_{12}^{AB-AE}\right) = \arg \min_{\rho_{12(i)}} \left\{ \mathrm{S}\left(\rho_{12_{(1)}}^{AB-AE}\right), \mathrm{S}\left(\rho_{12_{(2)}}^{AB-AE}\right), \ldots, \mathrm{S}\left(\rho_{12_{(n)}}^{AB-AE}\right) \right\}, \tag{5.15}$$

where $\rho_{12_{(i)}}^{AB-AE}$ are the $i$-th states of the joint channel $\mathcal{N}_1 \otimes \mathcal{N}_2 \in \mathcal{H}^{d_1 \cdot d_2}$. Geometrically, the states in (5.14) and (5.15) are the most distant from the center. There is a strict connection between the channel capacity optimization problem and the determination of the minimal entropy quantum states and it remains true, using the space of $\mathbb{C}^{d_1 \cdot d_2}$. The $\rho_{12}^{AB-AE} \in \mathbb{C}^{d_1 \cdot d_2}$ optimal joint state of the $\mathbb{C}^{d_1 \cdot d_2}$ quantum superball also have similar properties because the minimal entropy channel-output state $\mathrm{S}_{\min}\left(\rho_{12}^{AB-AE}\right)$ of joint channel $\mathcal{N}_1 \otimes \mathcal{N}_2 \in \mathcal{H}^{d_1 \cdot d_2}$ can be found on the boundary of the $\mathbb{C}^{d_1 \cdot d_2}$ quantum informational superball.

The result of the geometric determination of the superactivation of the joint structure $\mathcal{N}_1 \otimes \mathcal{N}_2 \in \mathcal{H}^{d_1 \cdot d_2}$ is summarized in Thesis 1.2.

**Thesis 1.2.** *I proved that the radius of the quantum superball measures the superactivated capacities of the joint channel structure, where the elements of the joint structure are arbitrary dimensional quantum channels.*

The radius of the $\mathbb{C}^{d_1 \cdot d_2}$ superball can be greater than zero if and only if the channels in $\mathcal{N}_1 \otimes \mathcal{N}_2 \in \mathcal{H}^{d_1 \cdot d_2}$ can activate each other, i.e., the joint channel structure is superactive. The quantum informational radius between the $\sigma_{12}^{AB-AE} \in \mathbb{C}^{d_1 \cdot d_2}$ center of the superball and the optimal joint state $\rho_{12}^{AB-AE} \in \mathbb{C}^{d_1 \cdot d_2}$ is equal to the joint single-use quantum capacity, see Figs. 5.5 and 5.6(b). An important conclusion will be made in space $\mathbb{C}^{d_1 \cdot d_2}$ of $\sigma_{12}^{AB-AE} \in \mathbb{C}^{d_1 \cdot d_2}$ joint average output state and the $\rho_{12}^{AB-AE} \in \mathbb{C}^{d_1 \cdot d_2}$ optimal output states of the superactivated joint structure $\mathcal{N}_{12} = \mathcal{N}_1 \otimes \mathcal{N}_2$. Single-use quantum capacity of the joint channel $\mathcal{N}_1 \otimes \mathcal{N}_2$ will not be superactive if the average output joint state $\sigma_{12}^{AB-AE} \in \mathbb{C}^{d_1 \cdot d_2}$ can be given as a product state $\sigma_{12}^{AB-AE} = \sigma_{12,(1)}^{AB-AE} \otimes \sigma_{12,(2)}^{AB-AE} \in \mathbb{C}^{d_1 \cdot d_2}$. In this case, the optimal output state $\rho_{12}^{AB-AE} \in \mathbb{C}^{d_1 \cdot d_2}$ can also be given as a product state $\rho_{12}^{AB-AE} = \rho_{12,(1)}^{AB-AE} \otimes \rho_{12,(2)}^{AB-AE} \in \mathbb{C}^{d_1 \cdot d_2}$, the $Q^{(1)}\left(\mathcal{N}_1 \otimes \mathcal{N}_2\right)$. In other words, if $\sigma_{12}^{AB-AE} \in \mathbb{C}^{d_1 \cdot d_2}$ and $\rho_{12}^{AB-AE} \in \mathbb{C}^{d_1 \cdot d_2}$ can be given in a product state formula the radius in $\mathbb{C}^{d_1 \cdot d_2}$ is decomposable and the $Q^{(1)}\left(\mathcal{N}_1 \otimes \mathcal{N}_2\right)$ joint single-use quantum capacity will be zero and the joint structure $\mathcal{N}_{12} = \mathcal{N}_1 \otimes \mathcal{N}_2$ will not be superactive. If these two states cannot be given in tensor product representations, then strict additivity of $Q^{(1)}\left(\mathcal{N}_1\right)$ and $Q^{(1)}\left(\mathcal{N}_2\right)$ will fail and the channel construction $\mathcal{N}_{12} = \mathcal{N}_1 \otimes \mathcal{N}_2 \in \mathcal{H}^{d_1 \cdot d_2}$ will be superactive, which leads to $Q^{(1)}\left(\mathcal{N}_1 \otimes \mathcal{N}_2\right) > 0$ and radius in $\mathbb{C}^{d_1 \cdot d_2}$ cannot be decomposed. As follows, if density matrices $\sigma_{12}^{AB-AE} \in \mathbb{C}^{d_1 \cdot d_2}$ and $\rho_{12}^{AB-AE} \in \mathbb{C}^{d_1 \cdot d_2}$ are product states, then $\mathrm{S}_{\min}\left(\rho_{12}^{AB-AE}\right) = \mathrm{S}_{\min}\left(\rho_{12,(1)}^{AB-AE}\right) + \mathrm{S}_{\min}\left(\rho_{12,(2)}^{AB-AE}\right)$ and

$Q^{(1)}(\mathcal{N}_1) = Q^{(1)}(\mathcal{N}_2) = Q^{(1)}(\mathcal{N}_1 \otimes \mathcal{N}_2) = 0$, and the $\mathbb{C}^{d_1 \cdot d_2}$ radius will be decomposable. These results along with the previous sections conclude the thesis.

These results on the superactivation of the joint structure $\mathcal{N}_1 \otimes \mathcal{N}_2 \in \mathcal{H}^{d_1 \cdot d_2}$ are extended to the properties of the joint optimal and average states in Thesis 1.3.

> **Thesis 1.3.** *I proved that the superactivation of the joint structure of arbitrary quantum channels is determined by the properties of the quantum relative entropy function.*

Using the results derived by Cortese [Cortese02], [Cortese03], King and Ruskai [King01a] and Hayashi *et al.* [Hayashi05] on the superactivation problem of quantum channel capacities, the following statements can be made. The "product state" form expresses that the channels $\mathcal{N}_1 \in \mathcal{H}^{d_1}$ and $\mathcal{N}_2 \in \mathcal{H}^{d_2}$ of the joint structure $\mathcal{N}_{12}$ cannot activate each other, and the capacity of the joint structure $\mathcal{N}_{12} \in \mathcal{H}^{d_1 \cdot d_2}$ will be zero. We use the $\mathbb{C}^{d_1 \cdot d_2}$ superball construction and the *minimax* criterion for the joint states $\rho_{12}^{AB-AE} \in \mathbb{C}^{d_1 \cdot d_2}$ and $\sigma_{12}^{AB-AE} \in \mathbb{C}^{d_1 \cdot d_2}$ along with (5.3). If the joint average state and the joint optimal output state are entangled states, then the joint channel structure $\mathcal{N}_{12} \in \mathcal{H}^{d_1 \cdot d_2}$ is superactive and the radius in $\mathbb{C}^{d_1 \cdot d_2}$ of the superball will be greater than zero. As we have concluded, from these results, an important property follows. If the quantum channels $\mathcal{N}_1 \in \mathcal{H}^{d_1}$ and $\mathcal{N}_2 \in \mathcal{H}^{d_2}$ of the joint structure $\mathcal{N}_{12} \in \mathcal{H}^{d_1 \cdot d_2}$ can activate each other, the superball radius in $\mathbb{C}^{d_1 \cdot d_2}$ cannot be decomposed. If we prove that the $\mathbb{C}^{d_1 \cdot d_2}$ radius of the quantum superball in space of $\mathbb{C}^{d_1 \cdot d_2}$ cannot be decomposed, we also have proved Thesis 1.3.

To demonstrate, we use the result from Section 3.4.3 regarding the expression of quantum capacity; we will use the fact that quantum capacity can be expressed from the Holevo information. The channel $\mathcal{N}_{AB}$ between Alice and Bob is denoted by the term *AB*, and the channel $\mathcal{N}_{AE}$ between Alice and the environment is depicted by $AE$. The term $AB$-$AE$ denotes that information is leaked to the environment, $E$, during the transmission from Alice to Bob. As we will show next, the superactivation depends on the properties of the joint optimal and average states as well as the properties of the quantum relative entropy function (see Section 2.4.3).

If joint states $\rho_{12}^{AB-AE} \in \mathbb{C}^{d_1 \cdot d_2}$ and $\sigma_{12}^{AB-AE} \in \mathbb{C}^{d_1 \cdot d_2}$ of the joint channel $\mathcal{N}_1 \otimes \mathcal{N}_2 \in \mathcal{H}^{d_1 \cdot d_2}$ are product states, i.e., $\rho_{12}^{AB-AE} = \rho_{12,(1)}^{AB-AE} \otimes \rho_{12,(2)}^{AB-AE} \in \mathbb{C}^{d_1} \otimes \mathbb{C}^{d_2}$ and $\sigma_{12}^{AB-AE} = \sigma_{12,(1)}^{AB-AE} \otimes \sigma_{12,(2)}^{AB-AE} \in \mathbb{C}^{d_1} \otimes \mathbb{C}^{d_2}$, then the $Q^{(1)}(\mathcal{N}_1 \otimes \mathcal{N}_2)$ joint capacity will be zero, since the radius $r_{12}{}^{*} \in \mathbb{R}$ in (5.2) can be decomposed as follows [P9]:

$$
\begin{aligned}
r_{12}^{(1)*} &= Q^{(1)}\left(\mathcal{N}_1 \otimes \mathcal{N}_2\right) \\
&= \min_{\sigma_{12}} \max_{\rho_{12}} D\left(\rho_{12}^{AB} \middle\| \sigma_{12}^{AB}\right) - \min_{\sigma_{12}} \max_{\rho_{12}} D\left(\rho_{12}^{AE} \middle\| \sigma_{12}^{AE}\right) = \\
&= \min_{\sigma_{12}^{AB-AE}} \max_{\rho_{12}^{AB-AE}} D\left(\rho_{12}^{AB-AE} \middle\| \sigma_{12}^{AB-AE}\right) \\
&= \min_{\sigma_{12}^{AB-AE}} \max_{\rho_{12}^{AB-AE}} Tr_{12}\left(\left(\rho_{12}^{AB-AE}\right)\log\left(\rho_{12}^{AB-AE}\right) - \left(\rho_{12}^{AB-AE}\right)\log\left(\sigma_{12}^{AB-AE}\right)\right) \\
&= \min_{\sigma_{12,(1)}^{AB-AE}} \min_{\sigma_{12,(2)}^{AB-AE}} \max_{\rho_{12,(1)}^{AB-AE}} \max_{\rho_{12,(2)}^{AB-AE}} Tr_{12}\left(\begin{array}{l} \left(\rho_{12,(1)}^{AB-AE} \otimes \rho_{12,(2)}^{AB-AE}\right)\log\left(\left(\rho_{12,(1)}^{AB-AE}\right) \otimes \left(\rho_{12,(2)}^{AB-AE}\right)\right) \\ -\left(\left(\rho_{12,(1)}^{AB-AE}\right) \otimes \left(\rho_{12,(2)}^{AB-AE}\right)\right)\log\left(\sigma_{12,(1)}^{AB-AE} \otimes \sigma_{12,(2)}^{AB-AE}\right) \end{array}\right) \\
&= \min_{\sigma_{12,(1)}^{AB-AE}} \min_{\sigma_{12,(2)}^{AB-AE}} \max_{\rho_{12,(1)}^{AB-AE}} \max_{\rho_{12,(2)}^{AB-AE}} Tr_{12}\left(\begin{array}{l} \left(\rho_{12,(1)}^{AB-AE} \otimes \rho_{12,(2)}^{AB-AE}\right)\left(\log\left(\rho_{12,(1)}^{AB-AE}\right) \otimes I_2\right) + \\ \left(\rho_{12,(1)}^{AB-AE} \otimes \rho_{12,(2)}^{AB-AE}\right)\left(I_1 \otimes \log\left(\rho_{12,(2)}^{AB-AE}\right)\right) \end{array}\right) \\
&\quad - \min_{\sigma_{12,(1)}^{AB-AE}} \min_{\sigma_{12,(2)}^{AB-AE}} \max_{\rho_{12,(1)}^{AB-AE}} \max_{\rho_{12,(2)}^{AB-AE}} Tr_{12}\left(\begin{array}{l} \left(\rho_{12,(1)}^{AB-AE} \otimes \rho_{12,(2)}^{AB-AE}\right)\left(\log\left(\sigma_{12,(1)}^{AB-AE}\right) \otimes I_2\right) \\ +\left(\rho_{12,(1)}^{AB-AE} \otimes \rho_{12,(2)}^{AB-AE}\right)\left(I_1 \otimes \log\left(\sigma_{12,(2)}^{AB-AE}\right)\right) \end{array}\right) \\
&= \min_{\sigma_{12,(1)}^{AB-AE}} \min_{\sigma_{12,(2)}^{AB-AE}} \max_{\rho_{12,(1)}^{AB-AE}} \max_{\rho_{12,(2)}^{AB-AE}} Tr_1\left(\left(\rho_{12,(1)}^{AB-AE}\right)\log\left(\rho_{12,(1)}^{AB-AE}\right)\right) Tr_2\left(\left(\rho_{12,(2)}^{AB-AE}\right) I_2\right) \\
&\quad + \min_{\sigma_{12,(1)}^{AB-AE}} \min_{\sigma_{12,(2)}^{AB-AE}} \max_{\rho_{12,(1)}^{AB-AE}} \max_{\rho_{12,(2)}^{AB-AE}} Tr_1\left(\left(\rho_{12,(1)}^{AB-AE}\right) I_1\right) Tr_2\left(\left(\rho_{12,(2)}^{AB-AE}\right)\log\left(\rho_{12,(2)}^{AB-AE}\right)\right) \\
&\quad - \min_{\sigma_{12,(1)}^{AB-AE}} \min_{\sigma_{12,(2)}^{AB-AE}} \max_{\rho_{12,(1)}^{AB-AE}} \max_{\rho_{12,(2)}^{AB-AE}} Tr_1\left(\left(\rho_{12,(1)}^{AB-AE}\right)\log\left(\sigma_{12,(1)}^{AB-AE}\right)\right) Tr_2\left(\left(\rho_{12,(2)}^{AB-AE}\right) I_2\right) \\
&\quad - \min_{\sigma_{12,(1)}^{AB-AE}} \min_{\sigma_{12,(2)}^{AB-AE}} \max_{\rho_{12,(1)}^{AB-AE}} \max_{\rho_{12,(2)}^{AB-AE}} Tr_1\left(\left(\rho_{12,(1)}^{AB-AE}\right) I_1\right) Tr_2\left(\left(\rho_{12,(2)}^{AB-AE}\right)\log\left(\sigma_{12,(2)}^{AB-AE}\right)\right)
\end{aligned}
$$

$$
\begin{aligned}
&= \min_{\sigma_{12,(1)}^{AB-AE}} \min_{\sigma_{12,(2)}^{AB-AE}} \max_{\rho_{12,(1)}^{AB-AE}} \max_{\rho_{12,(2)}^{AB-AE}} \begin{pmatrix} Tr_1\left(\left(\rho_{12,(1)}^{AB-AE}\right)\log\left(\rho_{12,(1)}^{AB-AE}\right)\right) \\ -Tr_1\left(\left(\rho_{12,(1)}^{AB-AE}\right)\log\left(\sigma_{12,(1)}^{AB-AE}\right)\right) \end{pmatrix} \\
&\qquad + \min_{\sigma_{12,(1)}^{AB-AE}} \min_{\sigma_{12,(2)}^{AB-AE}} \max_{\rho_{12,(1)}^{AB-AE}} \max_{\rho_{12,(2)}^{AB-AE}} \begin{pmatrix} Tr_2\left(\left(\rho_{12,(2)}^{AB-AE}\right)\log\left(\rho_{12,(2)}^{AB-AE}\right)\right) \\ -Tr_2\left(\left(\rho_{12,(2)}^{AB-AE}\right)\log\left(\sigma_{12,(2)}^{AB-AE}\right)\right) \end{pmatrix} \\
&= \min_{\sigma_{12,(1)}^{AB-AE}} \min_{\sigma_{12,(2)}^{AB-AE}} \max_{\rho_{12,(1)}^{AB-AE}} \max_{\rho_{12,(2)}^{AB-AE}} \left( D\left(\rho_{12,(1)}^{AB-AE} \middle\| \sigma_{12,(1)}^{AB-AE}\right) + D\left(\rho_{12,(2)}^{AB-AE} \middle\| \sigma_{12,(2)}^{AB-AE}\right)\right) \\
&= \min_{\sigma_{12,(1)}^{AB-AE}} \min_{\sigma_{12,(2)}^{AB-AE}} \max_{\rho_{12,(1)}^{AB-AE}} \max_{\rho_{12,(2)}^{AB-AE}} D\left(\rho_{12,(1)}^{AB-AE} \middle\| \sigma_{12,(1)}^{AB-AE}\right) \\
&\qquad + \min_{\sigma_{12,(1)}^{AB-AE}} \min_{\sigma_{12,(2)}^{AB-AE}} \max_{\rho_{12,(1)}^{AB-AE}} \max_{\rho_{12,(2)}^{AB-AE}} D\left(\rho_{12,(2)}^{AB-AE} \middle\| \sigma_{12,(2)}^{AB-AE}\right) \\
&= \min_{\sigma_{12,(1)}^{AB-AE}} \max_{\rho_{12,(1)}^{AB-AE}} D\left(\rho_{12,(1)}^{AB-AE} \middle\| \sigma_{12,(1)}^{AB-AE}\right) + \min_{\sigma_{12,(2)}^{AB-AE}} \max_{\rho_{12,(2)}^{AB-AE}} D\left(\rho_{12,(2)}^{AB-AE} \middle\| \sigma_{12,(2)}^{AB-AE}\right) \\
&= r_1^{(1)^*} + r_2^{(1)^*} = r_{12}^{(1)^*} \in \mathbb{R} \\
&= Q^{(1)}\left(\mathcal{N}_1 \otimes \mathcal{N}_2\right) \\
&= Q^{(1)}\left(\mathcal{N}_1\right) + Q^{(1)}\left(\mathcal{N}_2\right) = 0,
\end{aligned}
\tag{5.16}
$$

where $I_1$ and $I_2$ are the $d_1$ and $d_2$ dimensional identity matrices ($d$=2 for the qubit case), $\rho_{12}^{AB}$ is the optimal output state of the joint channel $\mathcal{N}_{AB}$ between Alice and Bob, and $\sigma_{12}^{AB} = \sum_i p_i \rho_{12,(i)}^{AB}$ is the average state of the joint channel $\mathcal{N}_{AB}$ between Alice and Bob, referred as $\mathcal{N}_{AB}$. The term $E$ denotes the environment (for exact definitions and formulas, see Chapters 6 and 7.), $\rho_{12}^{AE}$ is the optimal state of the channel $\mathcal{N}_{AE}$ between Alice and the environment (referred to as the environment's optimal state), $\sigma_{12}^{AE} = \sum_i p_i \rho_{12,(i)}^{AE}$ is the average state of the channel $\mathcal{N}_{AE}$ between Alice and the environment (environment's average state), $\rho_{12}^{AB-AE} \in \mathbb{C}^{d_1 \cdot d_2}$ is the final optimal output channel state, while $\sigma_{12}^{AB-AE} \in \mathbb{C}^{d_1 \cdot d_2}$ is the final output average state of the joint channel $\mathcal{N}_1 \otimes \mathcal{N}_2 \in \mathcal{H}^{d_1 \cdot d_2}$. The $r_1^{(1)^*} \in \mathbb{R}$ and $r_2^{(1)^*} \in \mathbb{R}$ represent the $Q^{(1)}\left(\mathcal{N}_1\right)$ and $Q^{(1)}\left(\mathcal{N}_2\right)$ of two individual quantum channels $\mathcal{N}_1 \in \mathcal{H}^{d_1}$ and $\mathcal{N}_2 \in \mathcal{H}^{d_2}$. The decomposition of (5.3) in (5.16) implies that the single-use joint quantum capacity $Q^{(1)}\left(\mathcal{N}_1 \otimes \mathcal{N}_2\right)$ can be derived from the strict sum of independent channel quantum capacities $Q^{(1)}\left(\mathcal{N}_1\right)$ and $Q^{(1)}\left(\mathcal{N}_2\right)$, which is equal to the following:

$$r_{12}^{(1)*} = Q^{(1)}\left(\mathcal{N}_1\right) = Q^{(1)}\left(\mathcal{N}_2\right) = Q^{(1)}\left(\mathcal{N}_{12}\right) = 0\,. \tag{5.17}$$

As follows, if $r_{12}^{*} \in \mathbb{R}$ can be decomposed, then the joint states $\sigma_{12}$ and $\rho_{12}$ of the joint channel $\mathcal{N}_{12}$ cannot be entangled states in $\mathbb{C}^{d_1 \cdot d_2}$; the superactivation of the joint channel structure $\mathcal{N}_1 \otimes \mathcal{N}_2 \in \mathcal{H}^{d_1 \cdot d_2}$ is possible if and only if the joint states $\rho_{12}^{AB-AE} \in \mathbb{C}^{d_1 \cdot d_2}$ and $\sigma_{12}^{AB-AE} \in \mathbb{C}^{d_1 \cdot d_2}$ of the joint channel $\mathcal{N}_{12}$ are entangled states. The results derived in (5.16) trivially follow for the $Q\left(\mathcal{N}_1 \otimes \mathcal{N}_2\right)$ asymptotic quantum capacity, with $n \to \infty$.

From Thesis 1.3 also follows that possible set of superactive of quantum channels $\mathcal{N}_1 \otimes \mathcal{N}_2$ is also limited by the mathematical properties of the quantum relative entropy function, i.e., the superactivation effect is related to information geometric properties. The main results already are shown in (5.16); however, further statements can be derived from these decompositions. The properties of the quantum relative function also determine the superactivation of quantum channels, since the results have demonstrated the effect of superactivation also depends not only on the channel maps and the properties of the quantum channels of the joint structure as was known before, but on the basic properties of the quantum relative entropy function. Decomposing relative entropy function $D\left(\cdot\middle\|\cdot\right)$ in (5.3) does not work if the quantum channels in $\mathcal{N}_{12}$ can activate each other (i.e., if the capacity of the joint channel $\mathcal{N}_{12}$ is positive but individually, the channel capacities are equal to zero); thus, for entangled states $\rho_{12}^{AB-AE}$ and $\sigma_{12}^{AB-AE}$, the strict channel additivity will not hold for the zero-capacity channels $\mathcal{N}_1 \in \mathcal{H}^{d_1}$ and $\mathcal{N}_2 \in \mathcal{H}^{d_2}$. In that case, the joint channel $\mathcal{N}_1 \otimes \mathcal{N}_2 \in \mathcal{H}^{d_1 \cdot d_2}$ is superactive, and the joint capacity of $\mathcal{N}_1 \otimes \mathcal{N}_2$ will be positive. If the average output state $\sigma_{12}^{AB-AE}$ is a product state, and if one or more from the set of optimal output states $\rho_{12}^{AB-AE}$ is a product state, then the factorization of the quantum relative entropy function $D(\cdot\|\cdot)$ indicates that the quantum channels $\mathcal{N}_1$ and $\mathcal{N}_2$ cannot activate each other and the capacity of the joint structure $\mathcal{N}_{12}$ will be zero. From these results, it follows that the superactivated joint capacity (the $Q^{(1)}(\mathcal{N}_1 \otimes \mathcal{N}_2)$ quantum capacity or the $C_0^{(1)}(\mathcal{N}_1 \otimes \mathcal{N}_2)$ classical zero-error capacity in our case and the asymptotic versions) will be positive if and only if the decomposition of the quantum relative entropy function cannot be made.

The result on the classical zero-error capacity will be shown in Chapter 7, with the same outcome $C_0^{(1)}(\mathcal{N}_1 \otimes \mathcal{N}_2) = C_0^{(1)}(\mathcal{N}_1) + C_0^{(1)}(\mathcal{N}_2) = 0$. These results also can be extended to the asymptotic capacities, i.e., $Q(\mathcal{N}_1 \otimes \mathcal{N}_2) = Q(\mathcal{N}_1) + Q(\mathcal{N}_2) = 0$ and $C_0(\mathcal{N}_1 \otimes \mathcal{N}_2) = C_0(\mathcal{N}_1) + C_0(\mathcal{N}_2) = 0$ also holds.

### 5.3.2.1 Brief Summary

As was shown by Shor [Shor04a], the HSW channel capacity $C(\mathcal{N}_1 \otimes \mathcal{N}_2)$ of the quantum channels $\mathcal{N}_1 \in \mathcal{H}^{d_1}$ and $\mathcal{N}_2 \in \mathcal{H}^{d_2}$ is additive if and only if the minimum output entropy of the joint channel $\mathcal{N}_{12} \in \mathcal{H}^{d_1 \cdot d_2}$ is the strict sum of the minimum output entropies of the two channels, $\mathcal{N}_1$ and $\mathcal{N}_2$. These results also can be exploited in the space $\mathbb{C}^{d_1 \cdot d_2}$ for the construction of the superball over $\mathbb{C}^{d_1 \cdot d_2}$; however, the conditions on the classical and quantum capacities of quantum channels are completely different. In the superactivation problem of the joint channel structure $\mathcal{N}_{12}$, to determine the joint optimal state $\rho_{12}^{AB-AE} \in \mathbb{C}^{d_1 \cdot d_2}$ with minimal entropy $\mathrm{S}_{\min}(\rho_{12}^{AB-AE})$, we seek states on the boundary of the quantum informational superball in $\mathbb{C}^{d_1 \cdot d_2}$ (see Figs. 5.5 and 5.6(b)). If channels $\mathcal{N}_1$ and $\mathcal{N}_2$ can activate each other, then these boundary states $\rho_{12}^{AB-AE} \in \mathbb{C}^{d_1 \cdot d_2}$ are entangled states. If the channels $\mathcal{N}_1$ and $\mathcal{N}_2$ cannot activate each other, then the joint capac-ity of the joint channel structure $\mathcal{N}_{12} \in \mathcal{H}^{d_1 \cdot d_2}$ cannot be increased, and the $Q^{(1)}(\mathcal{N}_1 \otimes \mathcal{N}_2)$ joint single-use quantum capacity of $\mathcal{N}_1 \otimes \mathcal{N}_2$ will be zero, which also can be determined by the analysis of the space $\mathbb{C}^{d_1 \cdot d_2}$.

According to these currently shown results, our statements on the geometric interpretation of the superactivation of quantum channel capacities can be summarized as follows.

**Corollary 2** *For two quantum channels $\mathcal{N}_1 \in \mathcal{H}^{d_1}$ and $\mathcal{N}_2 \in \mathcal{H}^{d_2}$ from the joint structure $\mathcal{N}_{12} \in \mathcal{H}^{d_1 \cdot d_2}$ with $\mathbb{C}^{d_1 \cdot d_2}$, the radius length $r_{12}^* \in \mathbb{R}$ of the quantum superball over space $\mathbb{C}^{d_1 \cdot d_2}$ is equal to the strict sum of the radii lengths $r_1^* \in \mathbb{R}$ and $r_2^* \in \mathbb{R}$, then the arbitrary dimensional quantum channels $\mathcal{N}_1$ and $\mathcal{N}_2$ cannot activate each other, and the joint structure $\mathcal{N}_{12}$ is not superactive. If the quantum channels $\mathcal{N}_1$ and $\mathcal{N}_2$ from $\mathcal{N}_{12} \in \mathcal{H}^{d_1 \cdot d_2}$ can activate each other, then the radius $r_{12}$ over $\mathbb{C}^{d_1 \cdot d_2}$ cannot be decomposed.*

As we have seen in Thesis 1.3, if Thesis 1.2 holds then the joint average states $\sigma_{12}$ and $\rho_{12}$ are entangled states. From these theses also follows that for any channel combinations of $\mathcal{N}_1 \in \mathcal{H}^{d_1}$ and $\mathcal{N}_2 \in \mathcal{H}^{d_2}$, the superactivation of the joint structure $\mathcal{N}_{12} \in \mathcal{H}^{d_1 \cdot d_2}$ with space $\mathbb{C}^{d_1 \cdot d_2}$ cannot be achieved if the radius length of the quantum superball in $\mathbb{R}$ can be decomposed.

## 5.4 Geometric Way to Fit the Quantum Superball

Here we show the steps of geometrical iteration for the fit of the quantum superball to determine the $Q^{(1)}(\mathcal{N}_1 \otimes \mathcal{N}_2)$ superactivated single-use quantum capacity of the joint channel structure $\mathcal{N}_1 \otimes \mathcal{N}_2$. For simplicity the steps of the iteration process are illustrated by the single channel view using $\mathbb{R}^3$ over $\mathbb{C}^2$. The figures illustrate in $\mathbb{R}^3$ the space $\mathbb{C}^{d_1 \cdot d_2}$ of the joint structure $\mathcal{N}_1 \otimes \mathcal{N}_2 \in \mathcal{H}^{d_1} \otimes \mathcal{H}^{d_2}$, and the states $\rho \in \mathbb{C}^{d_1 \cdot d_2}$ and $\sigma \in \mathbb{C}^{d_1 \cdot d_2}$ represent the joint optimal and joint average states of the channel structure $\mathcal{N}_1 \otimes \mathcal{N}_2$.

In the proposed example, the $\mathcal{N}_1 \otimes \mathcal{N}_2 \in \mathcal{H}^{d_1 \cdot d_2}$ joint channel consists of an $d_1$ dimensional $\mathcal{N}_1$ depolarizing channel (unital channel) which channel has some positive private classical capacity $P^{(1)}(\mathcal{N}_1) > 0$, and $d_2$ dimensional 50% erasure channel $\mathcal{N}_2$, where $d_1 = d_2$. The quantum capacities of these channel are zero, i.e., $Q^{(1)}(\mathcal{N}_1) = Q^{(1)}(\mathcal{N}_2) = Q^{(1)}(\mathcal{N}_1 \otimes \mathcal{N}_2) = 0$ and $P^{(1)}(\mathcal{N}_2) = 0$. In Appendix E we show that these results can be extended to an arbitrary quantum channel $\mathcal{N}_1$ with $P^{(1)}(\mathcal{N}_1) > 0$ from $\mathcal{N}_1 \otimes \mathcal{N}_2$, and the $Q^{(1)}(\mathcal{N}_1 \otimes \mathcal{N}_2)$ superactivated single-use quantum capacity of $\mathcal{N}_1 \otimes \mathcal{N}_2 \in \mathcal{H}^{d_1 \cdot d_2}$ can be determined by proposed superball construction in $\mathbb{C}^{d_1 \cdot d_2}$. The asymptotic quantum capacity will be analyzed in Chapter 6, while in Chapter 7 the classical zero-error capacity will be studied. The fitting steps of the quantum superball are rooted in the basic requirements of Schumacher and Westmoreland [Schumacher99]. The location of the optimum joint state $\sigma \in \mathbb{C}^{d_1 \cdot d_2}$ of $\mathcal{N}_1 \otimes \mathcal{N}_2 \in \mathcal{H}^{d_1 \cdot d_2}$ can be determined geometrically, using the optimum quantum informational superball, which achieves the min-max criteria [Cortese02] (see Appendix E). The joint average state $\sigma$ must be expressible as a convex combination of the joint optimal states $\rho_k$ of $\mathcal{N}_1 \otimes \mathcal{N}_2$ as $\sigma = \sum_k p_k \rho_k$, which satisfies the min-max criteria and will result in the $Q^{(1)}(\mathcal{N}_1 \otimes \mathcal{N}_2)$ single-use quantum capacity of the joint structure $\mathcal{N}_1 \otimes \mathcal{N}_2$:

$$\begin{aligned} Q^{(1)}(\mathcal{N}_1 \otimes \mathcal{N}_2) &= \min_{\sigma_{12}} \max_{\rho_{12}} D\left(\rho_{12}^{AB} \middle\| \sigma_{12}^{AB}\right) - \min_{\sigma_{12}} \max_{\rho_{12}} D\left(\rho_{12}^{AE} \middle\| \sigma_{12}^{AE}\right) \\ &= \min_{\sigma_{12}^{AB-AE}} \max_{\rho_{12}^{AB-AE}} D\left(\rho_{12}^{AB-AE} \middle\| \sigma_{12}^{AB-AE}\right) \\ &= \min_{\sigma} \max_{\rho_k} D\left(\rho_k \middle\| \sigma = \sum_k p_k \rho_k\right), \end{aligned} \tag{5.18}$$

where

$$\max_{\rho_k} \rho_k = \max_{\rho_{12}^{AB-AE}} \rho_{12}^{AB-AE}, \tag{5.19}$$

and

$$\min_{\sigma} \sigma = \sum_k p_k \rho_k = \sum_k p_k \rho_{12(k)}^{AB-AE} = \min_{\sigma_{12}^{AB-AE}} \sigma_{12}^{AB-AE}. \tag{5.20}$$

In the quantum superball representation, the joint average state of $\mathcal{N}_1 \otimes \mathcal{N}_2 \in \mathcal{H}^{d_1 \cdot d_2}$ denoted by $\sigma = \sum_k p_k \rho_k$ associated with the superball vector $\mathbf{r}_\sigma$, and the joint optimal state $\rho_k$ of $\mathcal{N}_1 \otimes \mathcal{N}_2$ are associated with superball vectors $\mathbf{r}_{\rho_k}$:

$$\mathbf{r}_\sigma \equiv \sigma = \sum_k p_k \rho_k, \tag{5.21}$$

and

$$\mathbf{r}_{\rho_k} \equiv \rho_k. \tag{5.22}$$

To find the best solution, the we seeking the optimum joint average $\sigma$ and the quantum informational superball corresponding to the $Q^{(1)}(\mathcal{N}_1 \otimes \mathcal{N}_2)$ single-use quantum channel capacity of the joint structure $\mathcal{N}_1 \otimes \mathcal{N}_2$. The optimal initial state $\sigma$ can be given as a state vector $\mathbf{r}_\sigma$, in the iteration process, the algorithm modifies vector $\mathbf{r}_\sigma$ to find its optimum value. In Fig. 5.7, we illustrated an unacceptable situation, where the center $\mathbf{c}^* = \sigma = \sum_k p_k \rho_k$ of the superball does not lie inside the channel ellipsoid of the joint channel $\mathcal{N}_1 \otimes \mathcal{N}_2 \in \mathcal{H}^{d_1 \cdot d_2}$.

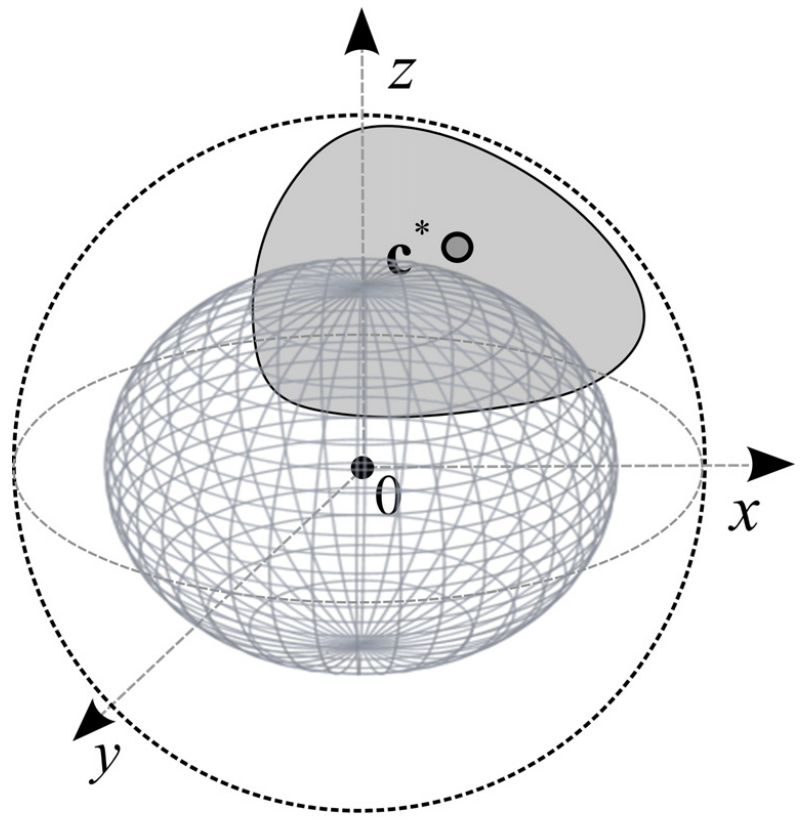

**Fig. 5.7.** An unacceptable situation, where the average state does not lie inside the channel ellipsoid (single channel view).

In Fig. 5.8, we illustrated an other unacceptable configuration, in which there are no permissible $\rho_k$ joint state, because the smallest enclosing quantum informational superball does not intersect the channel ellipsoid of the joint channel $\mathcal{N}_1 \otimes \mathcal{N}_2 \in \mathcal{H}^{d_1 \cdot d_2}$.

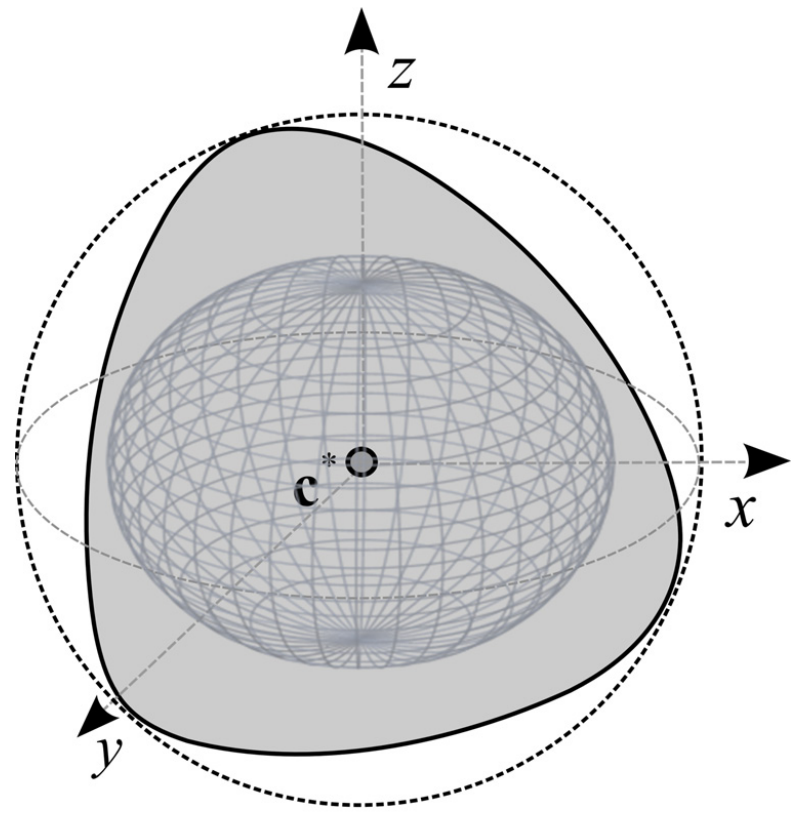

**Fig. 5.8.** Unacceptable situation, in which the quantum informational superball does not intersect the channel ellipsoid (single channel view).

In the next example, we show a situation, in which the computed length of the radius is not acceptable, because state joint average state $\sigma$ of $\mathcal{N}_1 \otimes \mathcal{N}_2 \in \mathcal{H}^{d_1 \cdot d_2}$ must be expressible as a convex combination of the joint optimal states $\rho_k$ of the joint channel $\mathcal{N}_1 \otimes \mathcal{N}_2$ which satisfy the min-max criteria. As we have illustrated in Fig. 5.9, there is only one permissible joint state denoted by $\rho_2$, and $\sigma \neq \rho_1$. It means, that only one optimal vector $\mathbf{r}_\sigma$ exists in the superball. If we find the point, where any movement of $\mathbf{r}_\sigma$ will increase $D_{\max}(\sigma) = D_{\max}(\mathbf{c}^*)$, we have found the final state of $\sigma = \mathbf{c}^*$ of $\mathcal{N}_1 \otimes \mathcal{N}_2$. In this situation, the quantum informational superball can not be used in the superactivation analysis of the joint channel structure $\mathcal{N}_1 \otimes \mathcal{N}_2 \in \mathcal{H}^{d_1 \cdot d_2}$.

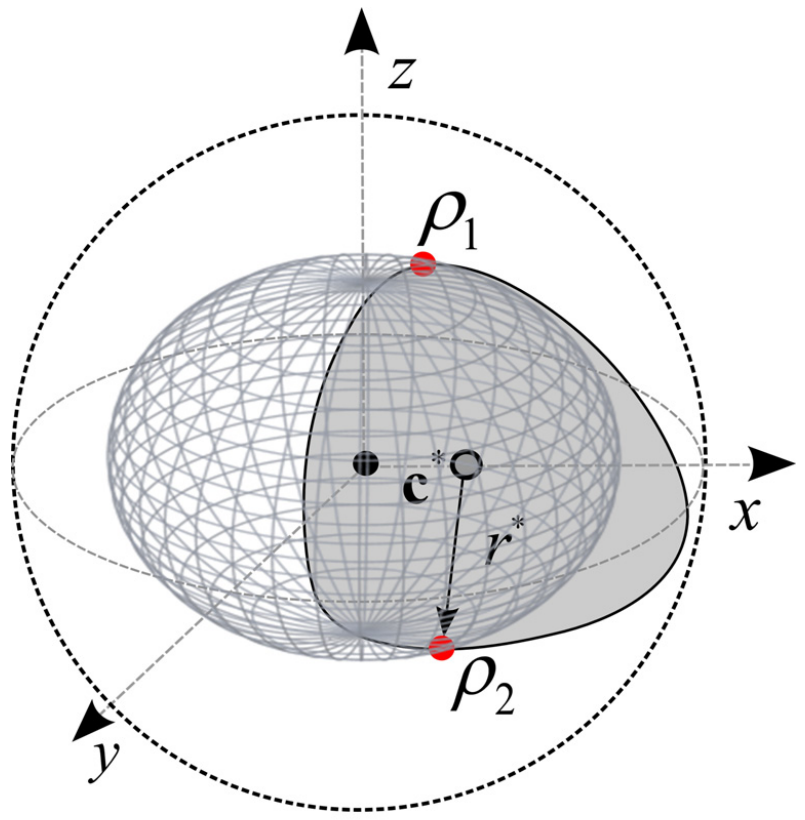

**Fig. 5.9.** The computed length of the radius is not acceptable, the average has to be expressible as a convex combination of density matrices of output states, which satisfies the min-max criteria (single channel view).

In Fig. 5.10, we illustrated the situation, if the radius of the quantum informational superball is also not acceptable, because of the algorithm did not realize the maximization criteria of [Schumacher99] in the iteration. Choosing a quantum informational superball with larger radius $r^*$ could solve the situation.

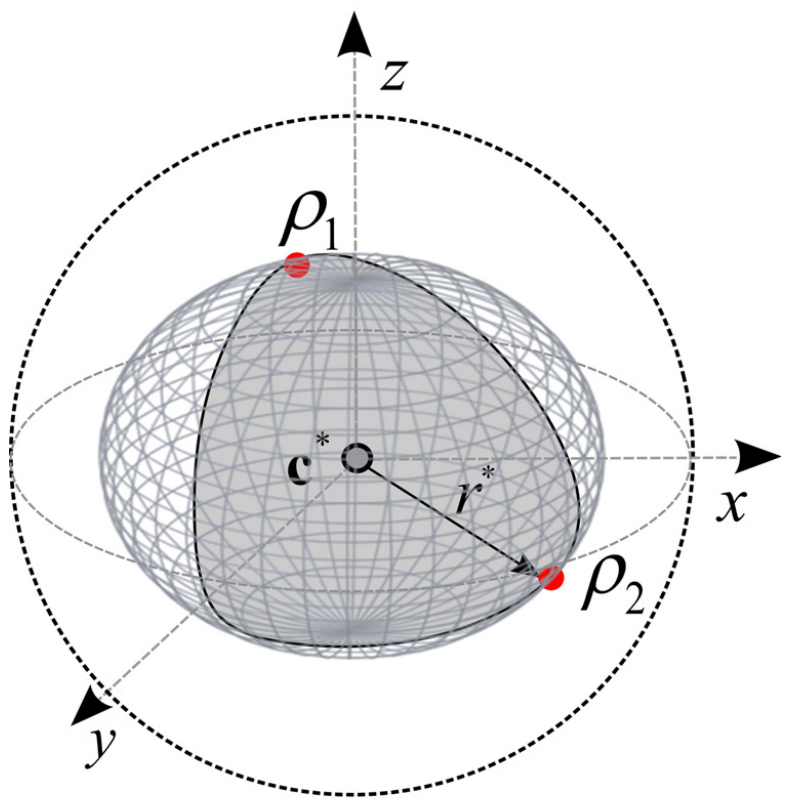

**Fig. 5.10.** The radius of the quantum superball is not acceptable, because of the algorithm did not realize the maximization criteria in the iteration (single channel view).

In Fig. 5.11, we illustrated an acceptable situation, where quantum superball intersects both of the joint states $\rho_1$ and $\rho_2$ of $\mathcal{N}_1 \otimes \mathcal{N}_2 \in \mathcal{H}^{d_1 \cdot d_2}$. The optimal joint average state of $\mathcal{N}_1 \otimes \mathcal{N}_2$ is denoted by $\sigma = \mathbf{c}^*$. The two joint states denoted by $\rho_1, \rho_2$ lie at the intersection of the quantum superball and the joint channel ellipsoid of $\mathcal{N}_1 \otimes \mathcal{N}_2 \in \mathcal{H}^{d_1 \cdot d_2}$, and these states could be used to form a convex combination $\rho_k$ that is equivalent to $\sigma = \mathbf{c}^* = \sum_k p_k \rho_k$.

In this state, the quantum superball satisfies the basic requirements of Schumacher and Westmoreland [Schumacher99], and the quantum superball can be used to analyze the superactivation of the $Q^{(1)}\left(\mathcal{N}_1 \otimes \mathcal{N}_2\right)$ single-use quantum capacity of the joint structure $\mathcal{N}_1 \otimes \mathcal{N}_2$. The radius $r^*$ represents $Q^{(1)}\left(\mathcal{N}_1 \otimes \mathcal{N}_2\right)$.

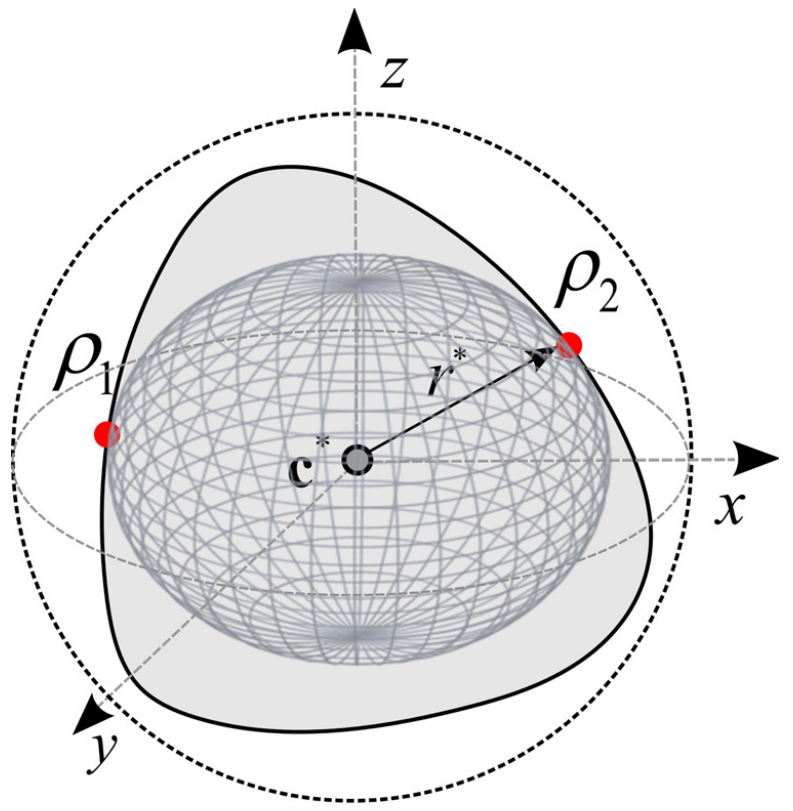

**Fig. 5.11.** An acceptable situation, where quantum informational superball intersects both of the channel endpoints (single channel view).

In this case, we get the largest smallest enclosing quantum superball with maximum radius $r^*$. The length of radius is equal to the $Q^{(1)}(\mathcal{N}_1 \otimes \mathcal{N}_2)$ superactivated single-use quantum capacity of the joint structure $\mathcal{N}_1 \otimes \mathcal{N}_2 \in \mathcal{H}^{d_1 \cdot d_2}$:

$$\begin{aligned} r^* &= Q^{(1)}(\mathcal{N}_1 \otimes \mathcal{N}_2) = \min_{\sigma} \max_{\rho} D(\rho \| \sigma) \\ &= \min_{\sigma} \max_{\rho_k} D\left(\rho_k \middle\| \sigma = \sum_k p_k \rho_k\right). \end{aligned} \tag{5.23}$$

The vector $\mathbf{r}_\sigma$ must lie in the channel ellipsoid between the two endpoints of the joint channel $\mathcal{N}_1 \otimes \mathcal{N}_2 \in \mathcal{H}^{d_1 \cdot d_2}$ denoted by the joint states $\rho_1$ and $\rho_2$. The optimum joint states found by the iteration for the joint channel $\mathcal{N}_1 \otimes \mathcal{N}_2$, are equal to the $\min_{\rho_{12}} \mathrm{S}(\rho_{12})$ minimum output von Neumann entropy joint states of $\mathcal{N}_1 \otimes \mathcal{N}_2$. Using the optimal joint state $\sigma$, we can increase the radius of the length of the quantum superball by the moving of the state. In this case we will get the unacceptable situation, as we have illustrated it in Fig. 5.12. The original position and the optimal ball is denoted by dashed line and state with joint channel output state $\rho_1$ and joint average state $\sigma = \mathbf{c}^*$ of $\mathcal{N}_1 \otimes \mathcal{N}_2$. The length of the non-optimal superball radius is denoted by $r$. The radii represent $Q^{(1)}(\mathcal{N}_1 \otimes \mathcal{N}_2)$ of the joint structure $\mathcal{N}_1 \otimes \mathcal{N}_2 \in \mathcal{H}^{d_1 \cdot d_2}$.

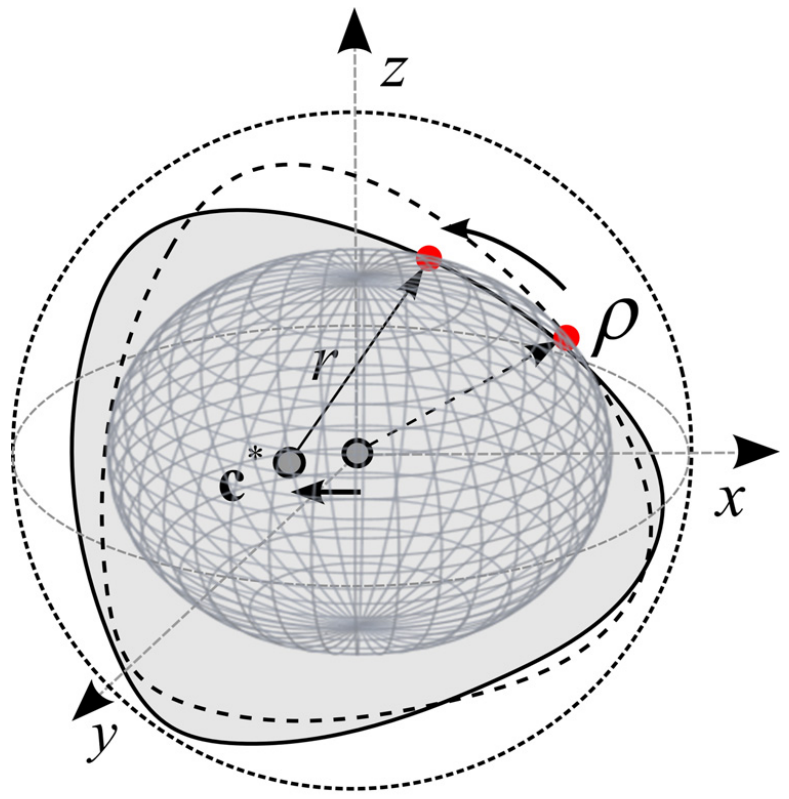

**Fig. 5.12.** By the moving of the optimal average state, the radius of the smallest enclosing quantum superball will increase, which will result an unacceptable situation (single channel view).

As the quantum superball is moved, the relative entropy function $D(\rho_1 \| \sigma)$ between joint states $\sigma = \mathbf{c}^*$ and $\rho_1$ of $\mathcal{N}_1 \otimes \mathcal{N}_2$ will increase. If we apply these iterations for the joint channel $\mathcal{N}_1 \otimes \mathcal{N}_2$ which is the combination of a depolarizing quantum channel $\mathcal{N}_1$ and an 50% erasure channel $\mathcal{N}_2$, where channel $\mathcal{N}_1$ has some positive private classical capacity $P^{(1)}(\mathcal{N}_1) > 0$, along with $Q^{(1)}(\mathcal{N}_1) = Q^{(1)}(\mathcal{N}_2) = Q^{(1)}(\mathcal{N}_1 \otimes \mathcal{N}_2) = 0$ and $P^{(1)}(\mathcal{N}_2) = 0$, then at the end of the fitting process the radius of the $\mathbb{C}^{d_1 \cdot d_2}$ quantum superball in terms of the reduced space $\mathbb{C}^2$ will be equal to

$$
\begin{aligned}
r^{*} &= Q^{(1)}\left(\mathcal{N}_1 \otimes \mathcal{N}_2\right) = \min_{\sigma}\max_{\rho} D\left(\rho \| \sigma\right) = \min_{\sigma}\max_{\rho_k} D\left(\rho_k \left\| \sigma = \sum_k p_k \rho_k \right.\right) \\
&= \min_{\sigma_{12}}\max_{\rho_{12}} D\left(\rho_{12}^{AB} \left\| \sigma_{12}^{AB}\right.\right) - \min_{\sigma_{12}}\max_{\rho_{12}} D\left(\rho_{12}^{AE} \left\| \sigma_{12}^{AE}\right.\right) \\
&= \min_{\sigma_{12}^{AB-AE}}\max_{\rho_{12}^{AB-AE}} D\left(\rho_{12}^{AB-AE} \left\| \sigma_{12}^{AB-AE}\right.\right) \\
&= \frac{1}{2}\left(\sum_{all\ p_i} p_i D\left(\rho_i \left\| \frac{1}{2} I\right.\right)\right) = \frac{1}{2}\left(1 - \sum_i p_i \mathrm{S}\left(\mathcal{N}\left(\rho_i\right)\right)\right) \\
&= \frac{1}{2}\left(1 - \mathrm{S}\left(p\frac{1}{2}I\right)\right) = \frac{1}{2}\left(1 - H\left(\frac{1}{2}p\right)\right) \\
&= \frac{1}{2} P^{(1)}\left(\mathcal{N}_1\right) > 0.
\end{aligned}
\tag{5.24}
$$

As can be concluded from (5.24), the $Q^{(1)}\left(\mathcal{N}_1 \otimes \mathcal{N}_2\right)$ superactivated single-use quantum capacity of the joint structure $\mathcal{N}_1 \otimes \mathcal{N}_2 \in \mathcal{H}^{d_1 \cdot d_2}$ derived in numerical way [Smith08] is equivalent to the geometrical formula of (5.23), which describes the $Q^{(1)}\left(\mathcal{N}_1 \otimes \mathcal{N}_2\right)$ superactivated single-use quantum capacity of $\mathcal{N}_1 \otimes \mathcal{N}_2 \in \mathcal{H}^{d_1 \cdot d_2}$ using the geometric interpretation.

These results will be extended to the $Q\left(\mathcal{N}_1 \otimes \mathcal{N}_2\right)$ asymptotic quantum capacity of $\mathcal{N}_1 \otimes \mathcal{N}_2 \in \mathcal{H}^{d_1} \otimes \mathcal{H}^{d_2}$ in Chapter 6. In Appendix E we demonstrate the results on the superactivated asymptotic joint quantum capacity of $\mathcal{N}_1 \otimes \mathcal{N}_2 \in \mathcal{H}^{d_1 \cdot d_2}$. We show that the $Q^{(1)}\left(\mathcal{N}_1 \otimes \mathcal{N}_2\right)$ superactivated single-use capacity can be determined by the $\mathbb{C}^{d_1 \cdot d_2}$ quantum superball structure if the first channel $\mathcal{N}_1$ of the joint channel structure $\mathcal{N}_1 \otimes \mathcal{N}_2 \in \mathcal{H}^{d_1 \cdot d_2}$ is an *arbitrary* quantum channel with $P^{(1)}\left(\mathcal{N}_1\right) > 0$, while the second channel is assumed to be an 50 % erasure channel.

As follows, there is an elegant alternative way to discover the superactive channel combinations and to avoid the hard numerical computations. These results will be extended to the superactivation of the classical zero-error capacity in Chapter 7.

*The proposed theses of Chapter 5 conclude Thesisgroup 1.*

For further information on the geometric interpretation of quantum channel capacities see Appendix E and the book of Imre and Gyongyosi [Imre12].

# Chapter 6

# Information Geometric Superactivation of Quantum Capacity

"If that turns out to be true, I'll quit physics."

*Max von Laue,* Nobel laureate *(1914)*

In this chapter I introduce an information geometric approach to determine the superactivation of quantum capacity of zero-capacity quantum channels. My proposed informational geometric solution is the first efficient algorithmic solution to discover the still unknown combinations to determine the superactivation of the zero-error capacity of quantum channels, without the extremely high computational costs.

This chapter is organized as follows. In the first part of this chapter, I give an introduction to the preliminaries of superactivation of quantum capacity. Next I discuss the proposed algorithm for the superactivation of single-use and asymptotic quantum capacities of quan-tum channels. The related works and supplementary information can be found in Appendix F.

## 6.1 Introduction

As it was shown in Chapter 4, the superactivation of zero-capacity quantum channels makes it possible to use two zero-capacity quantum channels with a positive joint capacity at the output. Currently, we have no theoretical background for describing all possible combinations of superactive zero-capacity channels, hence there should be many other possible combinations. This chapter shows a fundamentally new method of finding the conditions for the superactivation of asymptotic quantum capacity of zero-capacity quantum channels. In practice, to discover these superactive zero-capacity channel-pairs, we have to analyze an extremely large set of possible quantum states, channel models and channel probabilities. An extremely efficient algorithmic tool is still missing for this purpose. In this chapter, we present an algorithmic solution to reveal these still undiscovered superactive channel combinations. To analyze the superactivation of zero-capacity channels, we use our geometrical representation from Chapter 5, called the "quantum informational superball". Our method can be a very valuable tool for improving the results of fault-tolerant quantum computation and possible communication techniques over very noisy quantum channels. For the complete description of the theoretical background of superactivation see the book of Imre and Gyongyosi [Imre12].
The main result of Chapter 6 is summarized in Thesisgroup 2.

**Thesisgroup 2.** *I constructed an algorithm to determine the conditions of superactivation of the asymptotic quantum capacity of arbitrary dimensional quantum channels.*

In this chapter, first the theoretical background of the proposed information geometrical solution is presented, then we give the details of the algorithm.

### 6.1.1 Related Works

The mathematical class of quantum informational distance was introduced in [Kullback51], for convex programming, then this distance measure has been integrated into a many scientific area, such as text analysis, image and speech analysis, speech recognition, artificial intelligence, machine learning, and other fields of data analysis [Kullback87], [Onishi97], [Yoshizawa99]. For example, the analysis of informational databases can be achieved more efficiently using Kullback-Leibler distance, and it also can be applied to image comparison. The classical Kullback-Leibler divergence has been applied also in motion tracking and frame analysis. In classical systems, the Kullback-Leibler distance has been also used in clustering problems [Hiai91], [Kullback87], [Petz96], [Yoshizawa99]. Recently, the possibilities of the application of computational geometric methods in quantum space have been studied by Kato *et al.* [Kato06] and Nielsen *et al.* [Nielsen07-09], and Nock and Nielsen [Nock05]. Nielsen *et al.* [Nielsen08b] and Kato *et al.* [Kato06] have shown a method to compute the Voronoi diagrams between the quantum states. In the literature the geometric interpretation of quantum channels was also studied by King *et al.* [King99, King2000, King01], Petz *et al.* [Petz96], [Petz07-10a], Cortese [Cortese02-03], Hayashi *et al.* [Hayashi03-05] and by Ruskai *et al.* [Ruskai01]. The complexity of the numerical calculation of the Holevo capacity and the zero-error classical capacity was studied by Beigi and Shor [Beigi07]. For further information see the Related Work subsection of Appendix F and the book of Imre and Gyongyosi [Imre12].

### 6.1.2 Information Geometric Interpretation of the Quantum Capacity of Quantum Channels

The problem of superactivation of zero-capacity quantum channels can be viewed as a smaller subset of a larger problem set involving the additivity of quantum channels. When superactivating the quantum capacity of the quantum channels we can use the fact, that the quantum coherent information can be expressed in terms of the Holevo quantity. Both the single-use and the asymptotic quantum capacities can be superactivated, however the asymptotic version is the strongest version involving the single-use version, i.e., we will focus on the asymptotic quantum capacity. As we have also seen in Chapter 6, there could be a very large difference between the single-use and the superactivated asymptotic quantum capacity, i.e., from the knowledge of single-use quantum capacity we cannot study the true power of superactivation. The quantum coherent information can be computed as the difference between the Holevo information $\mathcal{X}_{AB}$, which measures the information transmitted from Alice to Bob, and the Holevo information $\mathcal{X}_{AE}$, which measures the information passed from Alice to the environment during the transmission of the quantum state, see Section 3.4.3. Like the Holevo quantity in the HSW (Holevo-Schumacher-Westmoreland) capacity of the quantum channel – the quantum coherent information determines the asymptotic LSD (Lloyd-Shor-Devetak) capacity of the quantum channels [Lloyd97], [Shor02], [Devetak03]. To define the *asymptotic* quantum capacity, we have to regularize the maximum of the quantum coherent information $I_{coh} = I\left(\rho_A : \mathcal{N}\left(\rho_A\right)\right)$, employing the parallel use of $n$ copies of channel $\mathcal{N}$ as follows

$$Q\left(\mathcal{N}\right) = \lim_{n\to\infty}\frac{1}{n}Q^{(1)}\left(\mathcal{N}^{\otimes n}\right) = \lim_{n\to\infty}\frac{1}{n}\max_{\rho_A} I\left(\rho_A : \mathcal{N}^{\otimes n}\left(\rho_A\right)\right), \tag{6.1}$$

where $Q^{(1)}\left(\mathcal{N}\right) = \max_{\rho_A} I\left(\rho_A : \mathcal{N}\left(\rho_A\right)\right)$ is the *single-use* quantum capacity of the quantum channel $\mathcal{N}$. We use the fact that the *asymptotic* quantum capacity can be expressed as

$$Q\left(\mathcal{N}\right) = \lim_{n\to\infty}\frac{1}{n}\max_{A^{\otimes n},\rho_x^A}\left(\mathcal{X}_{AB} - \mathcal{X}_{AE}\right), \tag{6.2}$$

where

$$\mathcal{X}_{AB} = \mathrm{S}\left(\mathcal{N}_{AB}\left(\rho_{AB}\right)\right) - \sum_i p_i \mathrm{S}\left(\mathcal{N}_{AB}\left(\rho_i\right)\right) \tag{6.3}$$

and

$$\mathcal{X}_{AE} = \mathrm{S}\left(\mathcal{N}_{AE}\left(\rho_{AE}\right)\right) - \sum_i p_i \mathrm{S}\left(\mathcal{N}_{AE}\left(\rho_i\right)\right) \tag{6.4}$$

measure the Holevo quantities between Alice and Bob, and between Alice and environment $E$, where $\rho_{AB} = \sum_i p_i \rho_i$ and $\rho_{AE} = \sum_i p_i \rho_i$ are the averaged states. As can be concluded, the quantum coherent information $I_{coh}$, – or with other words, the difference of $\mathcal{X}_{AB} - \mathcal{X}_{AE}$ – depends only on the noise of the quantum channel $\mathcal{N}$, and the average input state $\rho_{AB} = \sum_i p_i \rho_i$. The details of the environment and the exact choice of the pure input system have no influence on it. To see clear this connection see Section 3.4.3 and the results of Schumacher and Westmoreland [Schumacher2000].

### 6.1.3 Quantum Relative Entropy based Interpretation of Superactivated Quantum Capacity

As we have discussed in Chapter 5, both the classical and the quantum capacities of a quantum channel can be measured geometrically, using quantum relative entropy function as a distance measure. Schumacher and Westmoreland have shown in [Schumacher2000] that for a given quantum channel $\mathcal{N}$, the Holevo quantity for every optimal output state $\rho_k$ can be expressed as $\mathcal{X}\left(\mathcal{N}\right) = D\left(\rho_k \| \sigma\right)$, where $\sigma = \sum_k p_k \rho_k$ is the *optimal average output state* and the relative entropy function of two density matrices can be defined as $D\left(\rho_k \| \sigma\right) = Tr\left(\rho_k \log\left(\rho_k\right) - \rho_k \log\left(\sigma\right)\right)$. Finally, for non-optimal output states $\delta$ and optimal average output state $\sigma = \sum_k p_k \rho_k$, we have $\mathcal{X}\left(\mathcal{N}\right) = D\left(\delta \| \sigma\right) \leq D\left(\rho_k \| \sigma\right)$. We will use the geometrical interpretation of the Holevo quantity, using the quantum relative entropy function as a distance measure, to express the asymptotic quantum capacity of the quantum channel. The results on the information geometric superactivation of the $Q^{(1)}\left(\mathcal{N}_1 \otimes \mathcal{N}_2\right)$ single-use and the $Q\left(\mathcal{N}_1 \otimes \mathcal{N}_2\right)$ asymptotic quantum capacity of the joint channel structure $\mathcal{N}_1 \otimes \mathcal{N}_2 \in \mathcal{H}^{d_1 \cdot d_2}$ are summarized in Thesis 2.1.

> **Thesis 2.1.** *I showed that the superactivated single-use and asymptotic quantum capacity of the joint structure of arbitrary quantum channels can be determined by the proposed information geometric object.*

**Definition 3** I define the single-use quantum capacity $Q^{(1)}\left(\mathcal{N}\right)$ of a quantum channel $\mathcal{N} \in \mathcal{H}^d$ as the radius length $r^{(1)^*} \in \mathbb{R}$ over $\mathbb{C}^d$ of the smallest quantum informational ball as follows:

$$\begin{aligned} r^{(1)^*} &= Q^{(1)}\left(\mathcal{N}\right) = \max_{p_1,\ldots,p_n,\rho_1,\ldots,\rho_n} I_{coh}\left(\rho_A : \mathcal{N}\left(\rho_A\right)\right) = \max_{p_1,\ldots,p_n,\rho_1,\ldots,\rho_n}\left(\mathcal{X}_{AB} - \mathcal{X}_{AE}\right) \\ &= \max_{p_1,\ldots,p_n,\rho_1,\ldots,\rho_n}\left(\mathrm{S}\left(\mathcal{N}_{AB}\left(\sum_{i=1}^{n} p_i\rho_i\right)\right) - \sum_{i=1}^{n} p_i \mathrm{S}\left(\mathcal{N}_{AB}\left(\rho_i\right)\right)\right. \\ &\qquad\left. - \mathrm{S}\left(\mathcal{N}_{AE}\left(\sum_{i=1}^{n} p_i\rho_i\right)\right) + \sum_{i=1}^{n} p_i \mathrm{S}\left(\mathcal{N}_{AE}\left(\rho_i\right)\right)\right) \qquad (6.5) \\ &= \min_{\sigma}\max_{\rho} D\left(\rho_k^{AB} \middle\| \sigma^{AB}\right) - \min_{\sigma}\max_{\rho} D\left(\rho_k^{AE} \middle\| \sigma^{AE}\right) \\ &= \min_{\sigma}\max_{\rho} D\left(\rho_k^{AB-AE} \middle\| \sigma^{AB-AE}\right), \end{aligned}$$

where $\mathcal{X}_{AB}$ is the *Holevo quantity* of Bob's output, $\mathcal{X}_{AE}$ is the information leaked to the environment during the transmission, $\rho_k^{AB} \in \mathbb{C}^d$ is Bob's optimal output state, $\rho_k^{AE} \in \mathbb{C}^d$ is the environment's optimal state, $\sigma^{AB} \in \mathbb{C}^d$ is Bob's optimal output average state, $\sigma^{AE} \in \mathbb{C}^d$ is the environment's average state, while $\rho_k^{AB-AE} \in \mathbb{C}^d$ is the final optimal output channel state and $\sigma^{AB-AE} \in \mathbb{C}^d$ is the final output average state. The term *AB-AE* denotes the information which is transmitted from Alice to Bob minus the information which is leaked to the environment during the transmission. Based on (6.5) the quantum capacity can be expressed geometrically as the difference of two quantum informational balls. The first quantum ball measures the Holevo information between Alice and Bob, i.e., for this ball we define radius

$$\begin{aligned} r_{AB}{}^{*} &= \mathcal{X}_{AB} \\ &= \mathrm{S}\left(\mathcal{N}_{AB}\left(\sum_{i=1}^{n} p_i\rho_i\right)\right) - \sum_{i=1}^{n} p_i \mathrm{S}\left(\mathcal{N}_{AB}\left(\rho_i\right)\right) \\ &= D\left(\rho_k^{AB} \middle\| \sigma^{AB}\right). \end{aligned} \tag{6.6}$$

The second quantum informational ball measures the Holevo information which is leaked to the environment during the transmission as

$$\begin{aligned} r_{AE}{}^{*} &= \mathcal{X}_{AE} \\ &= \mathrm{S}\left(\mathcal{N}_{AE}\left(\sum_{i=1}^{n} p_i\rho_i\right)\right) - \sum_{i=1}^{n} p_i \mathrm{S}\left(\mathcal{N}_{AE}\left(\rho_i\right)\right) \\ &= D\left(\rho_k^{AE} \middle\| \sigma^{AE}\right). \end{aligned} \tag{6.7}$$

From these two radii, the single-use quantum capacity $Q^{(1)}\left(\mathcal{N}\right)$, i.e., the maximized quantum coherent information can be expressed as

$$Q^{(1)}\left(\mathcal{N}\right) = r^{\,(1)*} = \max r_{coh}{}^{*} = \max\left(r_{AB}{}^{*} - r_{AE}{}^{*}\right), \tag{6.8}$$

where

$$r_{coh}{}^{*} = r_{AB}{}^{*} - r_{AE}{}^{*} \tag{6.9}$$

measures the quantum coherent information between Alice and Bob. Using (6.8), the asymptotic quantum capacity $Q\left(\mathcal{N}\right)$ will be defined in (6.10).

These statements are summarized in Fig. 6.1. For simplicity the centers $\sigma^{AB} \in \mathbb{C}^2$, $\sigma^{AE} \in \mathbb{C}^2$ of the quantum balls are normalized into the origin of the Bloch sphere, i.e., in the next figure $\mathbf{c}^* = \sigma^{AB} = \sigma^{AE} = \sigma^{AB-AE} \in \mathbb{C}^2$.

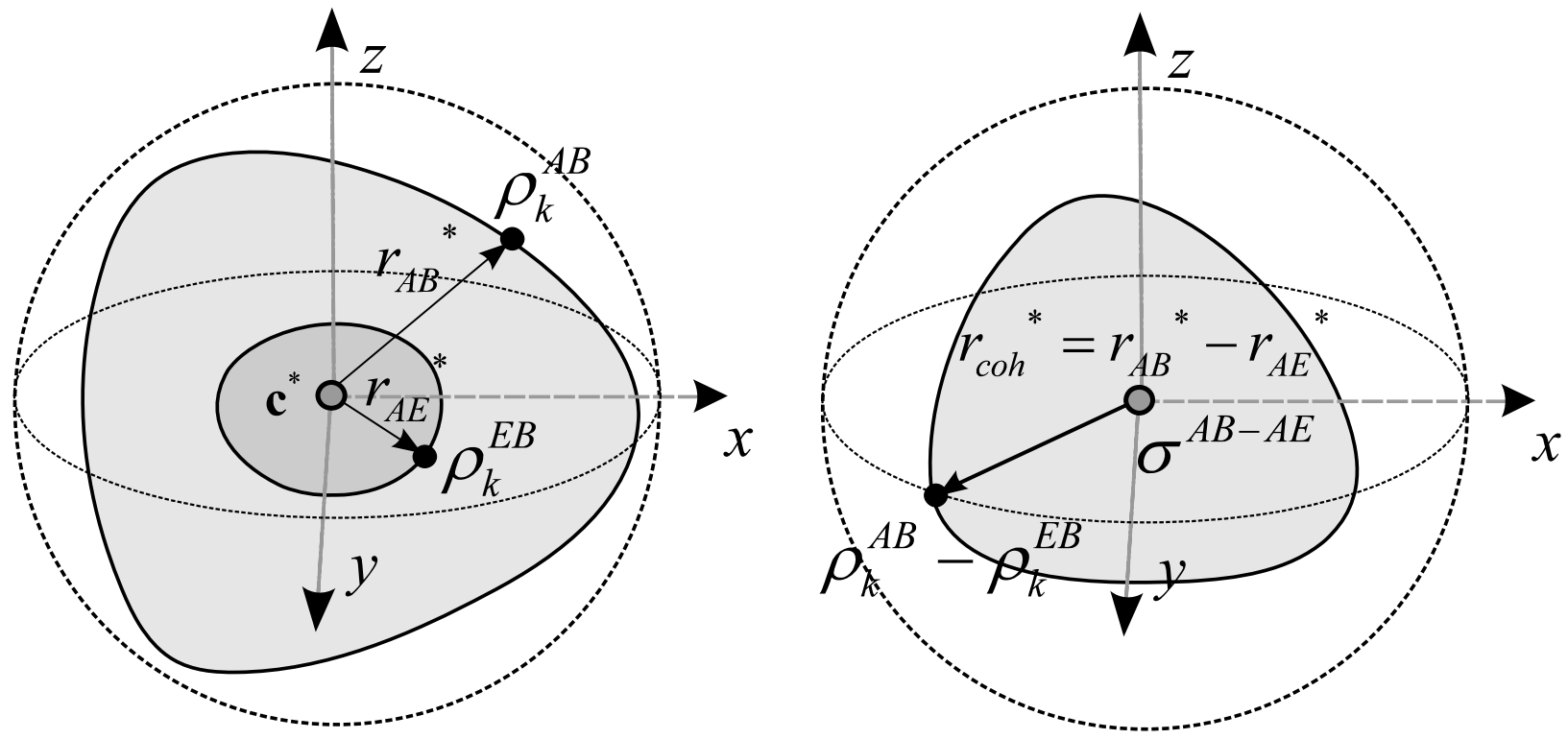

**Fig. 6.1.** (a): The two balls measure the Holevo information between Alice and Bob, and between Alice and the environment. (b): The second ball represents the quantum coherent information between Alice and Bob. The quantum capacity is expressed as the maximization of the difference of the two quantum informational balls.

The superactivation of asymptotic quantum capacity $Q(\mathcal{N}_1 \otimes \mathcal{N}_2)$ of $\mathcal{N}_1 \otimes \mathcal{N}_2 \in \mathcal{H}^{d_1 \cdot d_2}$ can also be determined by the quantum superball in the space of $\mathbb{C}^{d_1 \cdot d_2}$.

## 6.2 Information Geometric Definition of Asymptotic Quantum Capacity

Using the resulting quantum relative entropy function and the LSD-theorem the following definitions can be made.

**Definition 4** I define the asymptotic quantum capacity $Q(\mathcal{N})$ of $\mathcal{N} \in \mathcal{H}^d$ by using the $\mathbb{R}$ radii lengths of the smallest quantum informational balls over $\mathbb{C}^d$ as follows:

$$\begin{aligned} r^{*} &= Q\left(\mathcal{N}\right) = \lim_{n\to\infty}\frac{1}{n}Q^{(1)}\left(\mathcal{N}^{\otimes n}\right) = \lim_{n\to\infty}\frac{1}{n}\left(\sum_{i=1}^{n} r_i^{(1)*}\right) \\ &= \lim_{n\to\infty}\frac{1}{n}\max_{p_1,\ldots,p_n,\rho_1,\ldots,\rho_n} I_{coh}\left(\rho_A : \mathcal{N}^{\otimes n}\left(\rho_A\right)\right) = \lim_{n\to\infty}\frac{1}{n}\max_{p_1,\ldots,p_n,\rho_1,\ldots,\rho_n}\left(\mathcal{X}_{AB} - \mathcal{X}_{AE}\right) \\ &= \lim_{n\to\infty}\frac{1}{n}\max_{p_1,\ldots,p_n,\rho_1,\ldots,\rho_n}\left(\mathrm{S}\left(\mathcal{N}_{AB}^{\otimes n}\left(\sum_{i=1}^{n} p_i\rho_i\right)\right) - \sum_{i=1}^{n} p_i \mathrm{S}\left(\mathcal{N}_{AB}^{\otimes n}\left(\rho_i\right)\right)\right. \\ &\qquad \left. - \mathrm{S}\left(\mathcal{N}_{AE}^{\otimes n}\left(\sum_{i=1}^{n} p_i\rho_i\right)\right) + \sum_{i=1}^{n} p_i\,\mathrm{S}\left(\mathcal{N}_{AE}^{\otimes n}\left(\rho_i\right)\right)\right) \\ &= \lim_{n\to\infty}\frac{1}{n}\sum_{n}\left(\min_{\sigma_{1\ldots n}}\max_{\rho_{1\ldots n}} D\left(\rho_k^{AB}\middle\|\sigma^{AB}\right) - \min_{\sigma_{1\ldots n}}\max_{\rho_{1\ldots n}} D\left(\rho_k^{AE}\middle\|\sigma^{AE}\right)\right) \\ &= \lim_{n\to\infty}\frac{1}{n}\sum_{n}\left(\min_{\sigma_{1\ldots n}}\max_{\rho_{1\ldots n}} D\left(\rho_k^{AB-AE}\middle\|\sigma^{AB-AE}\right)\right), \end{aligned} \tag{6.10}$$

where $r_i^{(1)*} \in \mathbb{R}$ is the radius length of the smallest quantum informational ball, which describes the *single-use* quantum capacity of the *i*-th use of the quantum channel $\mathcal{N} \in \mathcal{H}^d$.

From these results the following conclusions can be derived regarding the quantum superball for the joint structure $\mathcal{N}_1 \otimes \mathcal{N}_2 \in \mathcal{H}^{d_1} \otimes \mathcal{H}^{d_2}$. The $r^{(1)*}_{super} \in \mathbb{R}$ of the super-ball is equal to the *single-use quantum capacity*, measured as the relative entropy function as distance measure [Gyongyosi11b].

**Definition 5** I define the superactivated single-use quantum capacity of the joint structure $\mathcal{N}_1 \otimes \mathcal{N}_2 \in \mathcal{H}^{d_1 \cdot d_2}$ by $r^{(1)*}_{super} \in \mathbb{R}$ over $\mathbb{C}^{d_1} \otimes \mathbb{C}^{d_2}$ as follows:

$$\begin{aligned} r^{(1)*}_{super}\left(\mathcal{N}_1 \otimes \mathcal{N}_2\right) &= r^{(1)*}_{i} \\ &= \max r_{coh}^{\ *} \\ &= \max\left(r_{AB}^{\ *} - r_{AE}^{\ *}\right) \\ &= \lim_{n\to\infty}\frac{1}{n}\sum_{n}\left(\min_{\sigma_{1\ldots n}}\max_{\rho_{1\ldots n}} D\left(\rho_k^{AB-AE}\middle\|\sigma^{AB-AE}\right)\right) \\ &= Q^{(1)}\left(\mathcal{N}_1 \otimes \mathcal{N}_2\right). \end{aligned} \tag{6.11}$$

As follows, the superactivation property of the quantum channel can be analyzed using the *mini-max* criterion for states $\rho_k^{AB-AE} \in \mathbb{C}^{d_1 \cdot d_2}$ and $\sigma^{AB-AE} \in \mathbb{C}^{d_1 \cdot d_2}$. The geometrical structure of the quantum information superball differs from the properties of a Euclidean ball. The quantum superball for the asymptotic quantum capacity is defined as follows. The radius $r_{super}^* \in \mathbb{R}$ of the superball is equal to the *asymptotic quantum capacity* of the superactivated joint channel structure $\mathcal{N}_1 \otimes \mathcal{N}_2 \in \mathcal{H}^{d_1} \otimes \mathcal{H}^{d_2}$, measured as the relative entropy function using $\mathbb{C}^{d_1} \otimes \mathbb{C}^{d_2}$ [Gyongyosi11b].

**Definition 6** I define the superactivated asymptotic quantum capacity of the joint structure $\mathcal{N}_1 \otimes \mathcal{N}_2 \in \mathcal{H}^{d_1 \cdot d_2}$ by $r_{super}^*\left(\mathcal{N}_1 \otimes \mathcal{N}_2\right) \in \mathbb{R}$ over $\mathbb{C}^{d_1} \otimes \mathbb{C}^{d_2}$ as follows:

$$\begin{aligned} r_{super}^*\left(\mathcal{N}_1 \otimes \mathcal{N}_2\right) &= \lim_{n\to\infty} \frac{1}{n}\left(\sum_{i=1}^{n} r_i^{(1)^*}\right) = \\ &= \lim_{n\to\infty} \frac{1}{n} \max r_{coh}{}^* \\ &= \lim_{n\to\infty} \frac{1}{n} \max\left(r_{AB}{}^* - r_{AE}{}^*\right) \\ &= \lim_{n\to\infty} \frac{1}{n} \sum_n \left(\min_{\sigma_{1\ldots n}} \max_{\rho_{1\ldots n}} D\left(\rho_k^{AB-AE} \middle\| \sigma^{AB-AE}\right)\right) \\ &= Q\left(\mathcal{N}_1 \otimes \mathcal{N}_2\right), \end{aligned} \tag{6.12}$$

where $r_i^{(1)^*} \in \mathbb{R}$ is defined over $\mathbb{C}^{d_1} \otimes \mathbb{C}^{d_2}$. The quantum informational superball of $\mathcal{N}_1 \otimes \mathcal{N}_2 \in \mathcal{H}^{d_1} \otimes \mathcal{H}^{d_2}$ with radius $r_{super}^*\left(\mathcal{N}_1 \otimes \mathcal{N}_2\right) \in \mathbb{R}$ over $\mathbb{C}^{d_1} \otimes \mathbb{C}^{d_2}$ for the case of the superactivated asymptotic quantum capacity $Q\left(\mathcal{N}_1 \otimes \mathcal{N}_2\right)$ is depicted in Fig. 6.2.

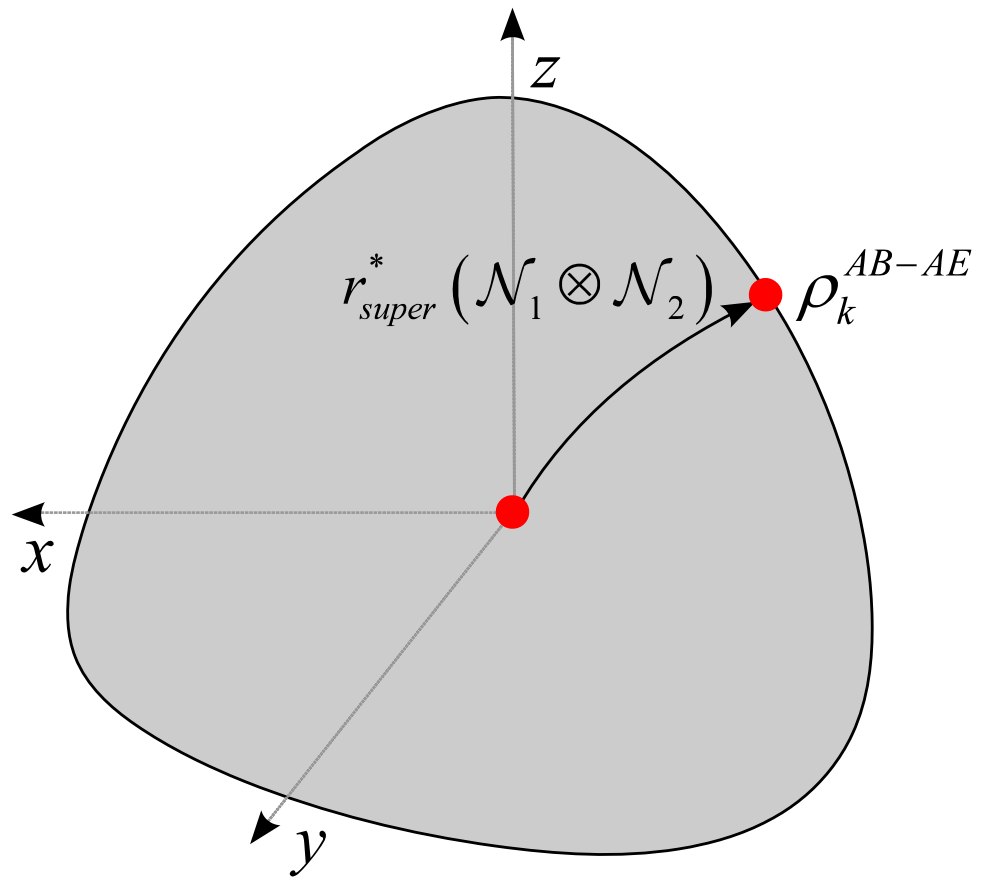

**Fig. 6.2.** The quantum superball defined for the analysis of superactivation of zero-capacity quantum channels.

From these results the following conclusions can be made. In the superactivation problem, we have to use different quantum channel models $\mathcal{N}_1$ and $\mathcal{N}_2$. For these channels, the superactivated *single-use* quantum capacity $Q^{(1)}\left(\mathcal{N}_1 \otimes \mathcal{N}_2\right)$ of $\mathcal{N}_1 \otimes \mathcal{N}_2$ can be determined by the superball radius ${r^{(1)*}}_{super}\left(\mathcal{N}_1 \otimes \mathcal{N}_2\right)$. The superactivated *asymptotic* quantum capacity $Q\left(\mathcal{N}_1 \otimes \mathcal{N}_2\right)$ of $\mathcal{N}_1 \otimes \mathcal{N}_2$ is equal to the superball radius $r^{*}_{super}\left(\mathcal{N}_1 \otimes \mathcal{N}_2\right)$. We note, using the result of Chapter 5 regarding the decomposition of the quantum relative entropy function, the following condition holds for the quantum superball of $\mathcal{N}_1 \otimes \mathcal{N}_2$. A necessary condition for the information geometric superactivation of the joint channel construction $\mathcal{N}_1 \otimes \mathcal{N}_2 \in \mathcal{H}^{d_1 \cdot d_2}$ is given in Thesis 2.2.

> **Thesis 2.2.** *I proved that the proposed geometric properties can be exploited to construct an information geometric algorithm for the algorithmic superactivation of arbitrary dimensional quantum channels.*

The result of Thesis 2.2 follows from our previously derived result on the factorization of the quantum relative entropy function. *If the joint channel structure* $\mathcal{N}_1 \otimes \mathcal{N}_2 \in \mathcal{H}^{d_1} \otimes \mathcal{H}^{d_2}$ *is superactive, then the radius of the quantum superball in* $\mathbb{C}^{d_1} \otimes \mathbb{C}^{d_2}$ *cannot be decomposed.* Using the theories of Chapter 5, if the channel combination $\mathcal{N}_1 \otimes \mathcal{N}_2$ is not superactive, (i.e., strict additivity holds) then the factorization of the quantum relative entropy function in $\mathbb{C}^{d_1} \otimes \mathbb{C}^{d_2}$ can be made, and the radius of the $\mathbb{C}^{d_1 \cdot d_2}$ superball can be expressed as the strict sum of the $\mathbb{C}^{d_1 \cdot d_2}$ radii of the independent channels $\mathcal{N}_1 \in \mathcal{H}^{d_1}$ and $\mathcal{N}_2 \in \mathcal{H}^{d_2}$. Since the superactivation effect requires the violation of additivity property, the joint channel $\mathcal{N}_1 \otimes \mathcal{N}_2 \in \mathcal{H}^{d_1} \otimes \mathcal{H}^{d_2}$ is superactive if and only if ${r^{(1)*}}_{super}\left(\mathcal{N}_1 \otimes \mathcal{N}_2\right) \in \mathbb{R}$ or $r^{*}_{super}\left(\mathcal{N}_1 \otimes \mathcal{N}_2\right) \in \mathbb{R}$ cannot be decomposed to sum of the radii of the individual channels $\mathcal{N}_1 \in \mathcal{H}^{d_1}$ and $\mathcal{N}_2 \in \mathcal{H}^{d_2}$. As follows, for the individual quantum channels $\mathcal{N}_1 \in \mathcal{H}^{d_1}$ and $\mathcal{N}_2 \in \mathcal{H}^{d_2}$ with $Q\left(\mathcal{N}_1\right) = 0$ and $Q\left(\mathcal{N}_2\right) = 0$, the relations

$$Q^{(1)}\left(\mathcal{N}_1 \otimes \mathcal{N}_2\right) > Q^{(1)}\left(\mathcal{N}_1\right) + Q^{(1)}\left(\mathcal{N}_1\right) \tag{6.13}$$

and

$$Q\left(\mathcal{N}_1 \otimes \mathcal{N}_2\right) > Q\left(\mathcal{N}_1\right) + Q\left(\mathcal{N}_1\right) \tag{6.14}$$

for the joint single-use and asymptotic quantum capacity hold if and only if ${r^{(1)*}}_{super}\left(\mathcal{N}_1 \otimes \mathcal{N}_2\right) \in \mathbb{R}$ and $r^{*}_{super}\left(\mathcal{N}_1 \otimes \mathcal{N}_2\right) \in \mathbb{R}$ defined over $\mathbb{C}^{d_1} \otimes \mathbb{C}^{d_2}$ are not decomposable.

In the case of the superactivated single-use quantum capacity, due to the impossibility of the factorization of the quantum relative entropy function (see Section 5.3.2) the superball radius length ${r^{(1)*}}_{super}\left(\mathcal{N}_1 \otimes \mathcal{N}_2\right) \in \mathbb{R}$ cannot be expressed as the sum of ${r^{(1)*}}_{super}\left(\mathcal{N}_1\right) \in \mathbb{R}$ and ${r^{(1)*}}_{super}\left(\mathcal{N}_2\right) \in \mathbb{R}$ of the smallest enclosing quantum informa-tional balls of the individual quantum channels $\mathcal{N}_1 \in \mathcal{H}^{d_1}$ and $\mathcal{N}_2 \in \mathcal{H}^{d_2}$, i.e., the following relation holds for $r^{(1)*}_{super}$:

$$\begin{aligned} r^{(1)*}_{super}\left(\mathcal{N}_1 \otimes \mathcal{N}_2\right) &= Q^{(1)}\left(\mathcal{N}_1 \otimes \mathcal{N}_2\right) \\ &\neq r^{(1)*}_{super}\left(\mathcal{N}_1\right) + r^{(1)*}_{super}\left(\mathcal{N}_2\right). \end{aligned} \tag{6.15}$$

Using the same result on the factorization of the quantum relative entropy function from Chapter 5, for the superactivated *asymptotic* quantum capacity $Q\left(\mathcal{N}_1 \otimes \mathcal{N}_2\right)$ it also follows that

$$\begin{aligned} r^{*}_{super}\left(\mathcal{N}_1 \otimes \mathcal{N}_2\right) &= Q\left(\mathcal{N}_1 \otimes \mathcal{N}_2\right) \\ &\neq r^{*}_{super}\left(\mathcal{N}_1\right) + r^{*}_{super}\left(\mathcal{N}_2\right). \end{aligned} \tag{6.16}$$

After we have laid down the theoretical background on the information geometric analysis of the superactivated quantum capacity, in the next section we present the proposed informational geometric approach and show an example for the superactivation of asymptotic quantum capacity.

## 6.3 Algorithm for the Superactivation of Quantum Capacity

Combining the theories of Chapter 5 on the information geometric description of the quantum capacity of the joint channel structure $\mathcal{N}_1 \otimes \mathcal{N}_2 \in \mathcal{H}^{d_1 \cdot d_2}$ with the theories of Chapter 6 the following statement can be made regarding the proposed information geometric algorithm as it is summarized in Thesis 2.3.

**Thesis 2.3.** *I constructed an efficient information geometric algorithm to study the superactivation of the quantum capacity of arbitrary quantum channels.*

The proposed iteration process is summarized as follows. The inputs of the geometric construction of the quantum superball are the channel output states of the two separate quantum channels $\mathcal{N}_1^{\otimes n}$ and $\mathcal{N}_2^{\otimes n}$. This process yields the superball radius as described above. The joint channel construction is denoted by $\left(\mathcal{N}_1 \otimes \mathcal{N}_2\right)^{\otimes n}$, while the output of the rounded box is the radius $r_{super}^{*}\left(\mathcal{N}_1 \otimes \mathcal{N}_2\right)$ of the quantum informational superball. The generic view of the proposed algorithm is depicted in Fig. 6.3.

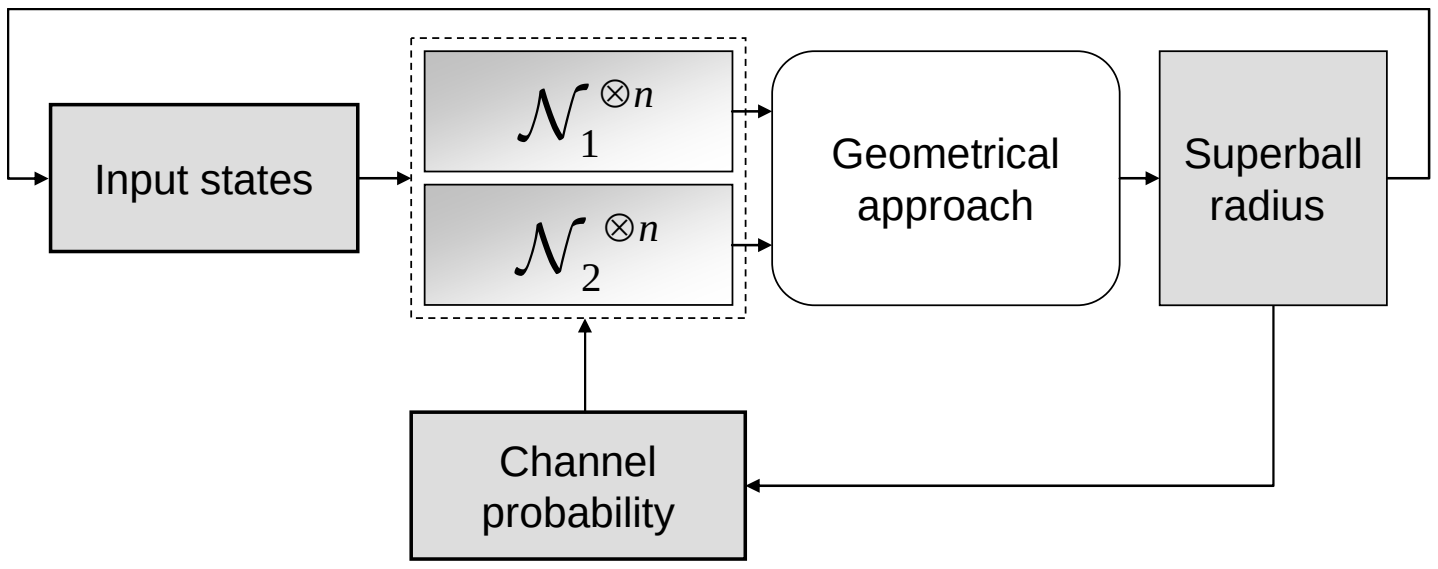

**Fig. 6.3.** The recursive algorithm iterates on the input, channel models, and error probabilities of the channels to find a combination for which superactivation holds.

The recursive iterations are made on three parameters: quantum channel models, channel probabilities (will be referred as $p_{\mathcal{C}}$), and set of input states. According to the length of the superball radius, the iteration stops if the conditions for superactivation hold. The algorithm always can determine the superactivated capacity of $\mathcal{N}_1 \otimes \mathcal{N}_2$ , the unacceptable situations can be handled as has already been interpreted in Section 5.4.

In order to present a new method to analyze geometrically the superactivation of the asymptotic quantum capacity $Q\left(\mathcal{N}_1 \otimes \mathcal{N}_2\right)$ we discuss a *Laguerre* diagrams based efficient solution to seek the *center* $\mathbf{c}$ of the set of smallest enclosing quantum information superball. The proposed algorithm consists of the following main steps:

1. Construction of quantum Delaunay triangulation (see Section 6.3.2 and Chapter 5 with the theories on the quantum informational superball, and Theses 2.1 and 2.2 from Section 6.1.3).

2. Seeking the center of the quantum superball (method in Sections 6.3.2 and 6.3.3, with theory and an illustrative example in Section 6.3.1) using an efficient core-set algorithm and the Delaunay triangulation. (The centers of the balls in the core-set approach are computed by the quantum Delaunay triangulation, see Chapter 5.)
3. Derivation of the superactivated quantum capacity $Q\left(\mathcal{N}_1 \otimes \mathcal{N}_2\right)$ of the joint channel construction $\mathcal{N}_1 \otimes \mathcal{N}_2$ from the quantum superball (based on theories of Chapter 5 and Section 6.1.3.).

*The performance analysis of the proposed algorithm is included in Appendix F.3.*

## 6.3.1 Example for the Superactivation of Quantum Capacity

Smith and Yard in 2008 [Smith08] have found only one possible combination for superactivation. At present, we have no theoretical proof to describe all possible combinations, hence there should be many other possible combinations of superactive zero-capacity quantum channels. Any quantum channel with some private capacity (for example a Horodecki channel) can be combined with an symmetric 50% erasure channel [Smith08] (for details see Appendix D), – hence two zero-capacity quantum channels can be combined together to realize information transmission. In the superactivation of the quantum capacity we can use a relationship between private and quantum channel capacities, which holds for any quantum channel $\mathcal{N}$. Considered a quantum channel for which $I\left(X:B\right) - I\left(X:E\right) \leq P^{(1)}\left(\mathcal{N}\right)$. Since the channel capacity is measured by the radius of the smallest quantum informational ball, we use the following equation to describe the $P^{(1)}\left(\mathcal{N}\right)$ *single-use* private classical capacity as

$$r_{\mathcal{N}}^{private} = \max_X I\left(X:B\right) - I\left(X:E\right) = P^{(1)}\left(\mathcal{N}\right), \tag{6.17}$$

where $r_{\mathcal{N}}^{private}$ measures the single-use private classical capacity of channel $\mathcal{N}$.

Every Horodecki channel $\mathcal{N}_H$ satisfies the relation $P^{(1)}(\mathcal{N}_H) > 0$, i.e.,

$$r_{\mathcal{N}_H}^{private} > 0. \tag{6.18}$$

There is an input for which the *superactivated single-use* joint quantum capacity $Q^{(1)}(\mathcal{N}_H \otimes \mathcal{N}_2)$ of a combination of a Horodecki channel $\mathcal{N}_H$ and a 50%-erasure channel $\mathcal{N}_2$, with private capacity $P^{(1)}(\mathcal{N}_2) = P(\mathcal{N}_2) = 0$, thus

$$r_{\mathcal{N}_2}^{private} = 0. \tag{6.19}$$

can result in the following $\mathbb{R}$ radius length over $\mathbb{C}^{d_1} \otimes \mathbb{C}^{d_2}$:

$$\begin{aligned} r^{(1)*}_{super}(\mathcal{N}_H \otimes \mathcal{N}_2) &= Q^{(1)}(\mathcal{N}_H \otimes \mathcal{N}_2) \\ &= \frac{1}{2}\left(I(X:B) - I(X:E)\right) \\ &= \frac{1}{2}P^{(1)}(\mathcal{N}_H) = \frac{1}{2}r_{\mathcal{N}_H}^{private}, \end{aligned} \tag{6.20}$$

while for the *superactivated asymptotic* joint quantum capacity

$$r^{*}_{super}(\mathcal{N}_H \otimes \mathcal{N}_2) \geq r^{(1)*}_{super}(\mathcal{N}_H \otimes \mathcal{N}_2) = \frac{1}{2}P^{(1)}(\mathcal{N}_H) = \frac{1}{2}r_{\mathcal{N}_H}^{private}. \tag{6.21}$$

The given channel combination can be verified with approximation error $\varepsilon$, using the computed radius of the superball for the $(\mathcal{N}_H \otimes \mathcal{N}_2)$ channel construction using informational geometric tools. In case of this combination the radius of the superball constriction (i.e., the superactivated asymptotic joint quantum capacity) will be $r^{*}_{super} \geq 0.01$. To describe geometrically the superactivation of zero-capacity quantum channels, we introduce the channel construction $\mathcal{C}$ which combines the two quantum channels $\mathcal{N}_H$, $\mathcal{N}_2$, and channel parameter $p_{\mathcal{C}}$ as follows

$$\mathcal{C} = p_{\mathcal{C}}\mathcal{N}_H \otimes |0\rangle\langle 0| + (1 - p_{\mathcal{C}})\mathcal{N}_2 \otimes |1\rangle\langle 1|, \tag{6.22}$$

where $0 \leq p_{\mathcal{C}} \leq 1$ gives the probability that the Horodecki channel applied in the joint construction. The defined $\mathcal{N}$ channel model is the convex combination of two zero-capacity channels $\mathcal{N}_H \otimes |0\rangle\langle 0|$ and $\mathcal{N}_2 \otimes |1\rangle\langle 1|$. Let us assume that we use these two quantum channels and their product channel representation $\mathcal{C}_1 \otimes \mathcal{C}_2$, where the channels were given in (6.22). The main goal is to find a channel probability parameter $p_{\mathcal{C}}$, for which the joint capacity of the tensor product channel $\mathcal{C}_1 \otimes \mathcal{C}_2$ is greater than zero.

The channel construction technique for superactivation of zero-capacity channels is illustrated in Fig. 6.4. To superactivate zero-capacity quantum channels, we must use the convex combination of different channel models and the *probabilistic mixtures* of these channels to realize superactivation. In this example, we assume fixed channel models and we have to iterate on the channel parameter $p_{\mathcal{C}}$.

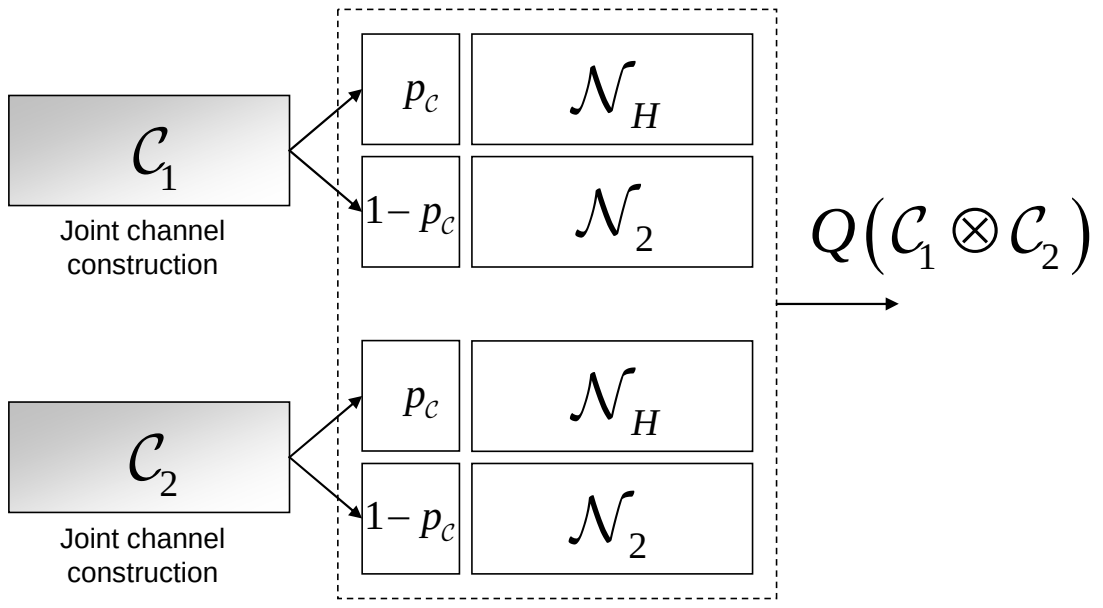

**Fig. 6.4.** Channel construction for superactivation of zero-capacity channels.

Next we will construct an algorithm to determine the superactivation and then we will apply it for the channel construction $\mathcal{C}_1 \otimes \mathcal{C}_2$ to find the superactivated quantum capacity $Q(\mathcal{C}_1 \otimes \mathcal{C}_2)$.

## 6.3.2 Step 1 - Construction of Delaunay Triangulation

We illustrate the results for the space of $\mathbb{C}^2$. The proposed algorithms are constructed for arbitrary *dimensional* channels, and the Delaunay structure between the density matrices will be generated in $\mathbb{C}^{d_1} \otimes \mathbb{C}^{d_2}$ dimensional Hilbert spaces of the joint channel structure.

To construct quantum Delaunay diagrams very efficiently in $\mathbb{C}^{d_1} \otimes \mathbb{C}^{d_2}$, we use the Laguerre diagrams [Onishi97], [Nielsen07], [Nock05], [Boissonnat07]. The *Laguerre* diagram of $b_i$ is defined as the *minimization* diagram of the corresponding $n$ distance functions $d_L^i(x) = \|\rho - x\|^2 - r^2$. The *Laguerre* bisector [Boissonnat07], [Goodman04] of the balls $b_i = b(\rho_i, r_i)$ and $b_j = b(\rho_j, r_j)$ can be expressed as

$$2\langle x, \rho_j - \rho_i \rangle + \|\rho_i\|^2 - \|\rho_j\|^2 + r_j^2 - r_i^2 = 0. \tag{6.23}$$

For pure quantum states with radii $r_j^2 = r_i^2 = 1$, and for mixed states with equal radii $r_j^2 = r_i^2$, where $r_j^2 < 1$ and $r_i^2 < 1$, the bisector is equivalent to the ordinary symmetric Euclidean bisector, hence

$$2\langle x, \rho_j - \rho_i \rangle + \langle \rho_j, \rho_j \rangle - \langle \rho_i, \rho_i \rangle = 0. \tag{6.24}$$

As it has been shown by Aurenhammer [Aurenhammer2000] and by Nielsen *et al.* [Nielsen07] and Boissonnat *et al.* [Boissonnat07], the left-sided dual quantum Delaunay tessellation of quantum states can be constructed from the Laguerre diagram of $n$ Euclidean spheres, of equations

$$\langle x - \rho_i', x - \rho_i' \rangle = \langle \rho_i', \rho_i' \rangle + 2(\mathbf{F}(\rho_i) - \langle \rho_i, \rho_i' \rangle),\ i = 1, \ldots n. \tag{6.25}$$

To underline Aurenhammer's result [Aurenhammer2000], we use quantum informational distances with $D(x\|\rho_i) \leq D(x\|\rho_j)$. In this case, we have

$$-\mathbf{F}(\rho_i) - \langle x - \rho_i, \rho_i' \rangle \leq -\mathbf{F}(\rho_j) - \langle x - \rho_j, \rho_j' \rangle. \tag{6.26}$$

Multiplying of this inequality by 2, and adding $\langle x, x \rangle$ to both sides, we get

$$\langle x, x \rangle - 2\mathbf{F}(\rho_i) - 2\langle x, \rho_i' \rangle + 2\langle \rho_i, \rho_i' \rangle \leq \langle x, x \rangle - 2\mathbf{F}(\rho_j) - 2\langle x, \rho_j' \rangle + 2\langle \rho_j, \rho_j' \rangle, \tag{6.27}$$

which can be simplified to

$$\left\langle x-\rho_i', x-\rho_i'\right\rangle - r_i^2 \le \left\langle x-\rho_j', x-\rho_j'\right\rangle - r_j^2, \tag{6.28}$$

where the squared radii can be expressed as

$$r_i^2 = \left\langle \rho_i', \rho_i'\right\rangle + 2\left(\mathbf{F}\left(\rho_i\right) - \left\langle \rho_i, \rho_i'\right\rangle\right), \tag{6.29}$$

and

$$r_j^2 = \left\langle \rho_j', \rho_j'\right\rangle + 2\left(\mathbf{F}\left(\rho_j\right) - \left\langle \rho_j, \rho_j'\right\rangle\right). \tag{6.30}$$

From (6.23), (6.29) and (6.30) we can derive the fact, that the Laguerre diagram of $x$ with respect to ordinary Euclidean ball $B\left(\rho_i', r_i\right)$ is equal to the Laguerre diagram of $x$ with respect to ordinary Euclidean ball $B\left(\rho_j', r_j\right)$ [Boissonnat07]. For the Euclidean Delaunay tessellation $Del\left(\mathcal{S}\right)$ we have function [Aurenhammer2000]

$$F\left(x\right) = \frac{1}{2}\left\|x\right\|^2. \tag{6.31}$$

The *centers* of the Euclidean spheres are denoted by $\rho_i$, and $\rho_i = \rho_i'$, thus $r_i^2 = 0$. The generator function $\mathbf{F}$ of the quantum relative entropy based diagram is the negative quantum entropy $\mathbf{F}\left(x\right) = \sum_i x_i \log x_i$, and the gradient $\nabla\mathbf{F}\left(x\right) = \left[\log\left(x_1\right)\ldots\log\left(x_d\right)\right]^T$ [Boissonnat07], [Aurenhammer2000]. The complexity of the proposed Delaunay triangulation between the density matrices is $\mathcal{O}\left(n\log n\right)$ [Goodman04], [Nielsen07], [Nock05]. As can be concluded, we reached the proposed object of Step 1 given in the introduction of Section 6.3. For further information about the algorithms of computational geometry, we suggest the works of Nielsen and Nock [Nielsen08-09], Kato *et al.* [Kato06], Onishi and Imai [Onishi97], Oto *et al.* [Oto04], Rajan [Rajan94], Sadakane *et al.* [Sadakane98], Sharir [Sharir85,94,04], Panigrahy [Panigrahy04], for detail see the Related Work subsection of Appendix E. As can be concluded, we reached the proposed object of Step 2 given in the introduction of Section 6.3. Now we step forward to Step 2.

### 6.3.3 Step 2 -The Core-Set Algorithm

To construct the quantum informational superball, we apply an approximation algorithm from classical computational geometry to determine the smallest enclosing ball of balls using *core-sets*. The core-sets have an important role in the discussed method. In the literature several approximation algorithms were presented, see the methods of Badoiu *et al.* [Badoiu02], Badoiu and Clarkson [Badoiu03], Feldman *et al.* [Feldman07], Frahling and Sohler [Frahling05], Har-Peled and Kushal [Har-Peled05], Hiai and Petz [Hiai91], Nielsen *et al.* [Nielsen08a], [Nielsen09], Nock and Nielsen [Nock05], Rajan [Rajan94]. A different approximation algorithm for qubit channels was presented by and Kato *et al.* [Kato06].
In the proposed algorithm the distance measure between quantum states is based on quantum relative entropy function $D\left(\cdot\|\cdot\right)$. The $\mathcal{E}$-core set $\mathcal{C}$ is a subset of set $\mathcal{C} \subseteq \mathcal{S}$, such that for the circumcenter $\mathbf{c}$ of the minimax ball

$$d\left(\mathbf{c},\mathcal{S}\right) \leq \left(1+\mathcal{E}\right)r, \tag{6.32}$$

where $r$ is the radius of the smallest enclosing quantum information ball for set $\mathcal{S}$.
Here, I will show, that the radius $r_{super}^{*}$ of the quantum superball of joint channel structure $\mathcal{N}_1 \otimes \mathcal{N}_2$ can be computed in an efficient way using computational geometric methods. In the single channel view, the radius will be depicted by $r^{*}$. The proposed approximation algorithm can find the *radius* $r$ of the smallest enclosing ball of balls in the quantum space in $\mathcal{O}\left(\frac{dn}{\mathcal{E}^2}\right)$ time with accuracy $\left(1+\mathcal{E}\right)$, where $n$ is the number of $d$ *dimensional* balls and the error is $0 < \mathcal{E} < 1$. In the applied approximation algorithm the *core-set* sizes are bounded by $\frac{2}{\mathcal{E}}$, independently of the dimension [Badoiu03], [Nock05]. The iterative ap-proximation algorithm, for a set of quantum states $\mathcal{S} = \{s_1, \ldots, s_n\}$ and circumcenter $\mathbf{c}$, first finds the farthest point $s_m$ of ball set $B$, and moves $\mathbf{c}$ towards $s_m$ in $\mathcal{O}(dn)$ time in every iteration step. The algorithm seeks the farthest point [Badoiu03] in the ball set $B = \{b_1 = Ball(\mathbf{c}_1, r_1), \ldots, b_n = Ball(\mathbf{c}_n, r_n)\}$ by maximizing the quantum informational distance for a current circumcenter position $\mathbf{c}$ as $\max_{i \in \{1,\ldots,n\}} D(b_i \| \mathbf{c})$. Using $\max_{x \in b_i} D(x_i \| \mathbf{c}) = D(s_i \| \mathbf{c}) + r_i$, we get $\max_{i \in \{1,\ldots,n\}} D(b_i \| \mathbf{c}) = \max_{i \in \{1,\ldots,n\}} (D(s_i \| \mathbf{c}) + r_i)$. The smallest enclosing ball of set $B = \{b_1, \ldots, b_n\}$ is the unique ball $b^* = Ball(\mathbf{c}^*, r^*)$ [Badoiu03], with minimum radius $r^*$ and center $\mathbf{c}^*$, containing all the set $B = \{b_1, \ldots, b_n\}$. The algorithm does $\left\lfloor \frac{1}{\varepsilon^2} \right\rfloor$ iterations to ensure an $(1 + \mathcal{E})$-approximation [Badoiu03], [Nielsen09], [Nock05], [Feldman07], [Frahling05], thus the overall cost of the algorithm is $\mathcal{O}\left(\frac{dn}{\varepsilon^2}\right)$. In Appendix F we present two core-set algorithms for the determination of the quantum superball, based on the theses of Chapter 5.

#### 6.3.3.1 Results

In this section we apply the previously introduced algorithm to determine $Q(\mathcal{C}_1 \otimes \mathcal{C}_2)$ of the channel construction $\mathcal{C}_1 \otimes \mathcal{C}_2$ as was shown in Fig. 6.4. The probabilistic behavior of this tensor product channel model can be described by the channel probability $p_{\mathcal{C}}$. The joint channel quantum capacity $Q(\mathcal{C}_1 \otimes \mathcal{C}_2)$ as radius $r^*_{super}(\mathcal{C}_1 \otimes \mathcal{C}_2) \in \mathbb{R}$ in function of channel probability can be described as follows:

$$\begin{aligned} r^*_{super}(\mathcal{C}_1 \otimes \mathcal{C}_2) = {} & p_{\mathcal{C}}^2 r^*_{(\mathcal{N}_H \otimes \mathcal{N}_H)} + p_{\mathcal{C}}(1 - p_{\mathcal{C}}) r^*_{(\mathcal{N}_H \otimes \mathcal{N}_2)} \\ & + (1 - p_{\mathcal{C}}) p_{\mathcal{C}} r^*_{(\mathcal{N}_2 \otimes \mathcal{N}_H)} + (1 - p_{\mathcal{C}})^2 r^*_{(\mathcal{N}_2 \otimes \mathcal{N}_2)}. \end{aligned} \tag{6.33}$$

The term $\left(1-p_{\mathcal{C}}\right)^{2} r^{*}{}_{\left(\mathcal{N}_{2} \otimes \mathcal{N}_{2}\right)}$ can be neglected, since the quantum capacity of this combination is zero, i.e., $r^{*}{}_{\left(\mathcal{N}_{2} \otimes \mathcal{N}_{2}\right)}=0$ [Smith08]. Therefore, channel model can be reduced to

$$Q\left(\mathcal{C}_{1} \otimes \mathcal{C}_{2}\right)=r_{super}^{*}\left(\mathcal{C}_{1} \otimes \mathcal{C}_{2}\right)=p_{\mathcal{C}}{}^{2} r^{*}{}_{\left(\mathcal{N}_{H} \otimes \mathcal{N}_{H}\right)}+2 p_{\mathcal{C}}\left(1-p_{\mathcal{C}}\right) r^{*}{}_{\left(\mathcal{N}_{H} \otimes \mathcal{N}_{2}\right)}, \tag{6.34}$$

and the radius of the smallest superballs can be described as

$$p_{\mathcal{C}}{}^{2} r^{*}{}_{\left(\mathcal{N}_{H} \otimes \mathcal{N}_{H}\right)}=p_{\mathcal{C}}{}^{2} Q^{(1)}\left(\mathcal{N}_{H} \otimes \mathcal{N}_{H}\right), \tag{6.35}$$

or

$$2 p_{\mathcal{C}}\left(1-p_{\mathcal{C}}\right) r^{*}{}_{\left(\mathcal{N}_{H} \otimes \mathcal{N}_{2}\right)}=2 p_{\mathcal{C}}\left(1-p_{\mathcal{C}}\right) Q^{(1)}\left(\mathcal{N}_{H} \otimes \mathcal{N}_{2}\right), \tag{6.36}$$

where $0<p_{\mathcal{C}}<1$. Here we note, the notation $\mathcal{C}_{1} \otimes \mathcal{C}_{2}$ means using the joint channel construction $\mathcal{N}_{H} \otimes \mathcal{N}_{2}$ two-times, which results in different superactivated asymptotic joint quantum capacities at the channel outputs. Hence, for this channel construction, we obtain radius length $r^{*}{}_{\left(\mathcal{N}_{H} \otimes \mathcal{N}_{H}\right)}$ with weight $p_{\mathcal{C}}{}^{2}$ and we obtain superball radius $r^{*}{}_{\left(\mathcal{N}_{H} \otimes \mathcal{N}_{2}\right)}$ with weight $2 p_{\mathcal{C}}\left(1-p_{\mathcal{C}}\right)$ in $r_{super}^{*}\left(\mathcal{C}_{1} \otimes \mathcal{C}_{2}\right)$. Using $\mathcal{N}_{H}$ and $\mathcal{N}_{2}$, the term $p_{\mathcal{C}}{}^{2} r^{*}{}_{\left(\mathcal{N}_{H} \otimes \mathcal{N}_{H}\right)}$ can never be greater than zero, because the quantum capacity of the Horodecki channel is zero [Smith08], $Q\left(\mathcal{N}_{H}\right)=r^{*}{}_{\left(\mathcal{N}_{H}\right)}=0$, i.e., radius $r^{*}{}_{\left(\mathcal{N}_{H} \otimes \mathcal{N}_{H}\right)}$ will always have zero length. Now, we consider on superball radius $r_{super}^{*}\left(\mathcal{C}_{1} \otimes \mathcal{C}_{2}\right)=r^{*}{}_{\left(\mathcal{N}_{H} \otimes \mathcal{N}_{2}\right)}$. Based on our previous results on the construction of quantum informational superball (see Chapter 5), $r_{super}^{*}\left(\mathcal{C}_{1} \otimes \mathcal{C}_{2}\right)$ can be expressed as follows

$$r_{super}^{*}\left(\mathcal{C}_{1} \otimes \mathcal{C}_{2}\right)=2 p_{\mathcal{C}}\left(1-p_{\mathcal{C}}\right) r^{*}{}_{\left(\mathcal{N}_{H} \otimes \mathcal{N}_{2}\right)} \geq 2 p_{\mathcal{C}}\left(1-p_{\mathcal{C}}\right) \frac{1}{2}\left(\left|\mathbf{r}_{1}^{*}\right|+\left|\mathbf{r}_{2}^{*}\right|\right), \tag{6.37}$$

where the radii $\mathbf{r}_1^*$ and $\mathbf{r}_2^*$ measure the single-use private classical capacities. As follows, in (6.37), the $\left|\mathbf{r}_1^*\right| \in \mathbb{R}$ and $\left|\mathbf{r}_2^*\right| \in \mathbb{R}$ represent the *private classical* capacity of the channels $\mathcal{N}_H$ and $\mathcal{N}_2$, where $P(\mathcal{N}_H) > 0$ and $P(\mathcal{N}_2) = 0$, instead of the quantum capacities $Q(\mathcal{N}_H)$ and $Q(\mathcal{N}_2)$ of $\mathcal{N}_H$ and $\mathcal{N}_2$, i.e., the decomposition of $r^*_{(\mathcal{N}_H \otimes \mathcal{N}_2)}$ can be made in that way. According to (6.22) the radius $r^*_{(\mathcal{N}_H \otimes \mathcal{N}_2)}$ is equal to zero for $p_{\mathcal{C}} \notin [0, 0.0041]$.

In Fig. 6.5, we show the smallest quantum informational balls in the range $0 < p_{\mathcal{C}} < 0.0041$. In this case, the channels have positive *superactivated* quantum capacity, i.e.,

$$0 < r^*_{(\mathcal{N}_H \otimes \mathcal{N}_2)} \geq \frac{1}{2}\left(\left|\mathbf{r}_1^*\right| + \left|\mathbf{r}_2^*\right|\right) = \frac{1}{2} P^{(1)}(\mathcal{N}_H) = \frac{1}{2} r^{private}_{\mathcal{N}_H}. \tag{6.38}$$

where $r^{private}_{\mathcal{N}_H}$ is the single-use private classical capacity of the Horodecki channel. The channels $\mathcal{N}_H$ and $\mathcal{N}_2$ with zero quantum capacities individually can be superactivated and a positive capacity can be realized on the output of the channels.

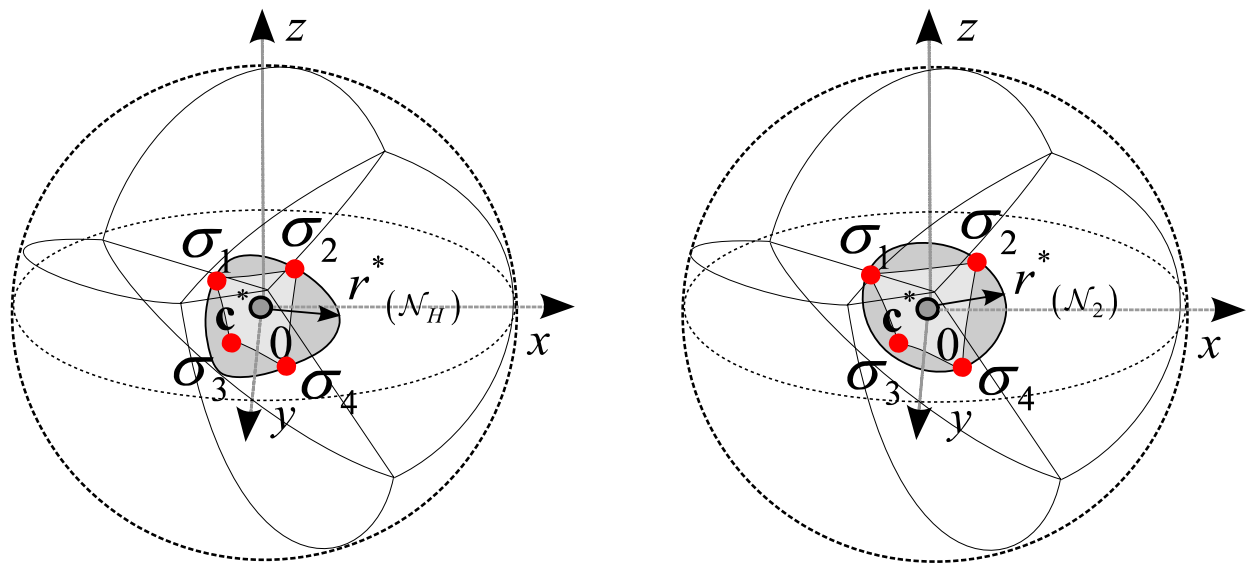

**Fig. 6.5.** The smallest enclosing balls and radii for Horodecki channel (a) and for erasure channel (b) in the single channel view. If the zero-capacity channels are superactive, the joint capacity will be positive in the given channel parameter domain $0 < p_{\mathcal{C}} < 0.0041$. (The radii represent the superactivated quantum capacity of the joint structure, using single channel view representation.)

In Fig. 6.6, we show the smallest enclosing quantum informational balls and the radii of two zero-capacity channels, $\mathcal{C}_1$ and $\mathcal{C}_2$ for channel probabilities $p_{\mathcal{C}} = 0$ and $p_{\mathcal{C}} \geq 0.0041$. The radii $r^*_{(\mathcal{N}_H)}$ and $r^*_{(\mathcal{N}_2)}$ are equal to zero for channel parameters outside the domain $0 < p_{\mathcal{C}} < 0.0041$. The radii $r^*_{(\mathcal{N}_H)}$ and $r^*_{(\mathcal{N}_2)}$ express the *quantum capacities* of the individual channels $\mathcal{N}_H$ and $\mathcal{N}_2$, $Q(\mathcal{N}_H) = Q(\mathcal{N}_2) = 0$, i.e.,:

$$r^*_{(\mathcal{N}_H \otimes \mathcal{N}_2)} = r^*_{\mathcal{N}_H} + r^*_{\mathcal{N}_2} = 0 \neq \frac{1}{2}\left(\left|\mathbf{r}_1^*\right| + \left|\mathbf{r}_2^*\right|\right). \tag{6.39}$$

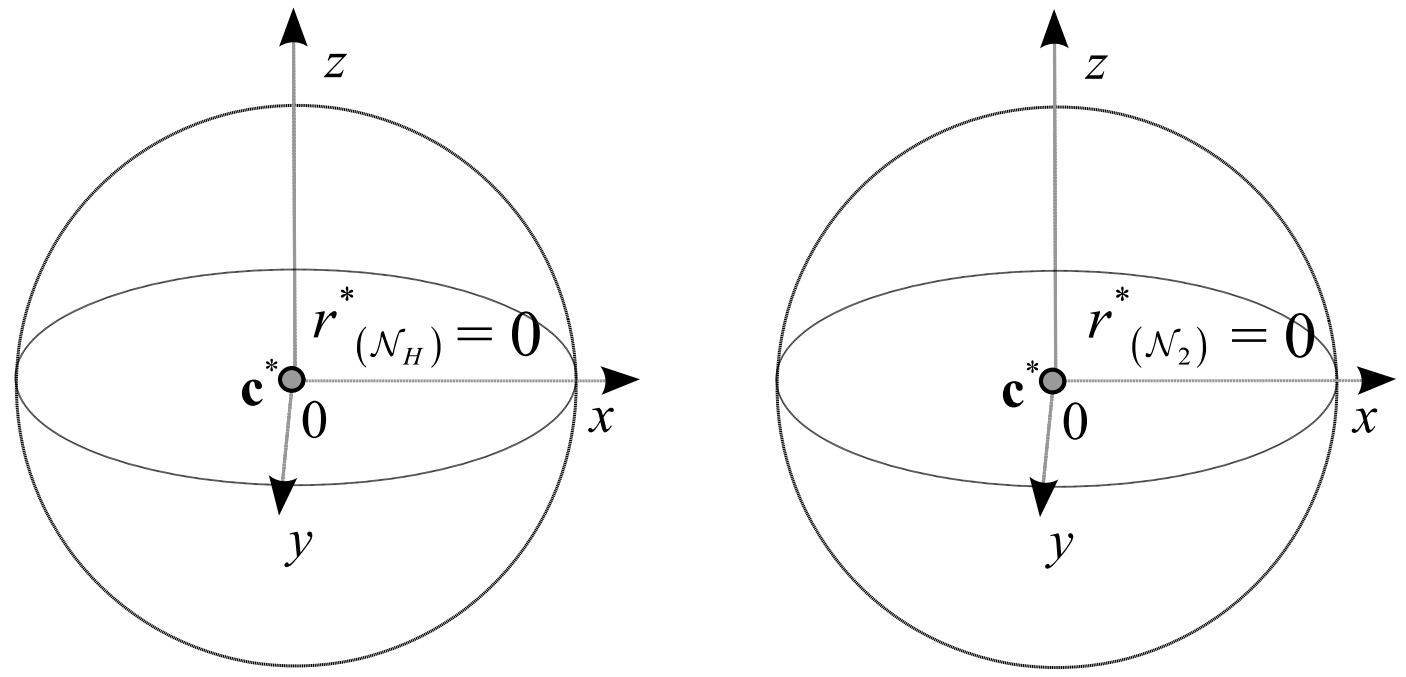

**Fig. 6.6.** Outputs of two zero-capacity quantum channels (single channel view). The radii of the smallest quantum informational balls are equal to zero. (The radii represent the quantum capacity of the joint structure, using single channel view representation.)

The results of the superactivation as a function of different $p_{\mathcal{C}}$ probabilities, where $r^*_{(\mathcal{N}_H \otimes \mathcal{N}_2)}$ is the radius of the superball, which describes the joint capacity of the joint structure $\mathcal{N}_H \otimes \mathcal{N}_2$ are shown in Fig. 6.7.

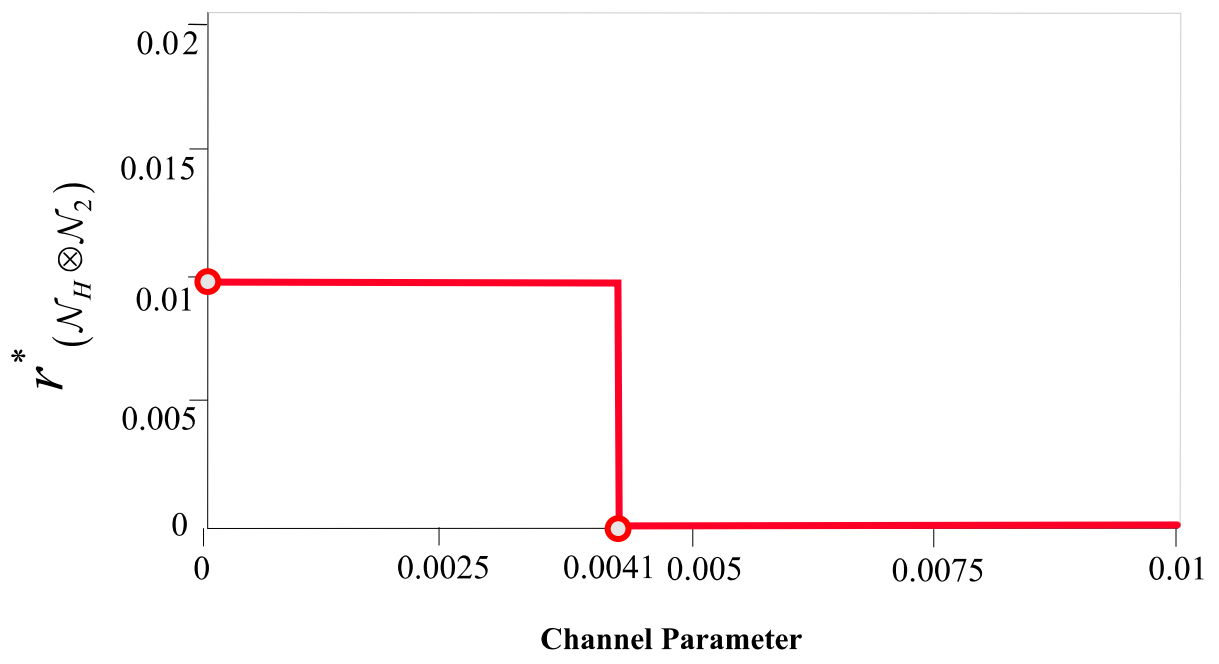

**Fig. 6.7.** The output of the optimization algorithm describes the radius of the superball, which will be positive only for a given domain of the channel parameter.

As can be observed, the length of the radius of the quantum informational superball is $r^{*}_{(\mathcal{N}_H \otimes \mathcal{N}_2)} = 0$ for channel parameters in the domain $p_{\mathcal{C}} \notin (0, 0.0041)$.

In Fig. 6.8, we show the superball radius $r^{*}_{super}(\mathcal{C}_1 \otimes \mathcal{C}_2)$, which describes $Q(\mathcal{C}_1 \otimes \mathcal{C}_2)$, see (6.34). As we have convex combinations of channels, the superball radius $r^{*}_{super}(\mathcal{C}_1 \otimes \mathcal{C}_2)$ will be measured as $r^{*}_{(\mathcal{N}_H \otimes \mathcal{N}_H)} = 0$ with zero weight, and $r^{*}_{(\mathcal{N}_H \otimes \mathcal{N}_2)} = 0.01$ with weight $2p_{\mathcal{C}}(1 - p_{\mathcal{C}})$ which leads to $r^{*}_{super}(\mathcal{C}_1 \otimes \mathcal{C}_2) = 2p_{\mathcal{C}}(1 - p_{\mathcal{C}}) r^{*}_{(\mathcal{N}_H \otimes \mathcal{N}_2)} = 2p_{\mathcal{C}}(1 - p_{\mathcal{C}}) \cdot (0.01)$.

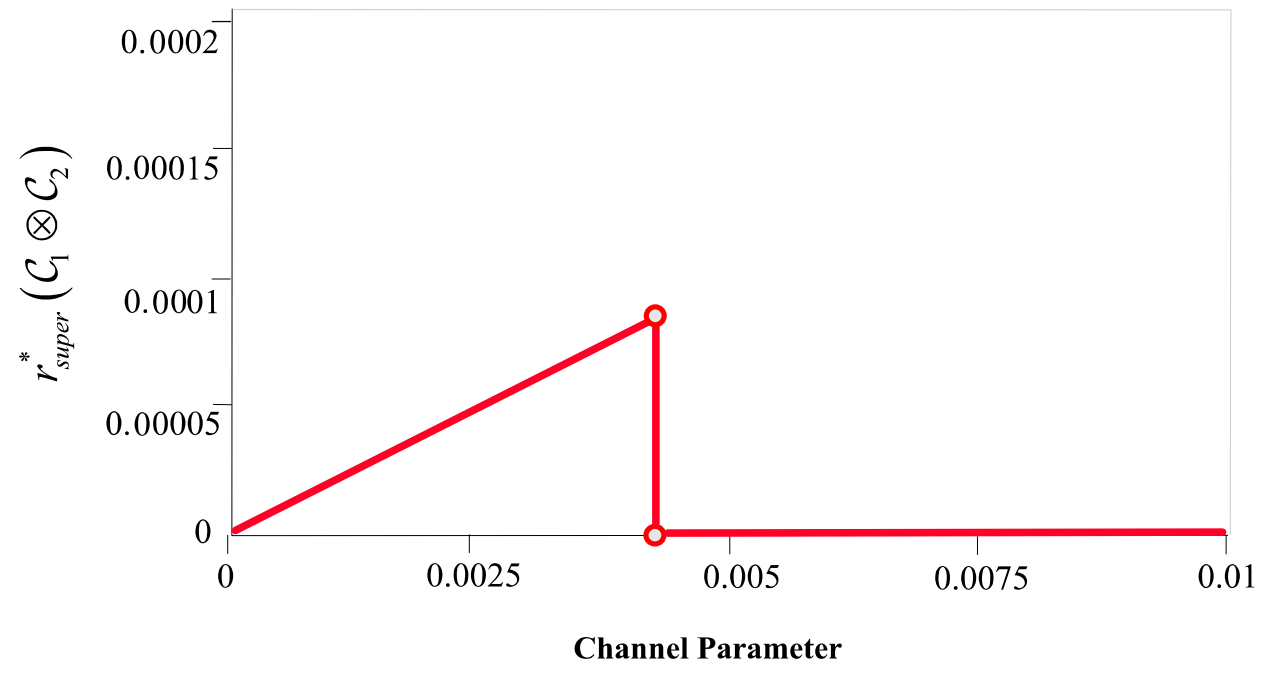

**Fig. 6.8.** The length of superball radius $r^{*}_{super}(\mathcal{C}_1 \otimes \mathcal{C}_2)$, as function of the channel parameter.

It can be concluded from our results for a channel parameter in the domain $0 < p_{\mathcal{C}} < 0.1$ that the output of the algorithm will result in a channel capacity $r^{*}_{(\mathcal{N}_H \otimes \mathcal{N}_2)} = 0.01$. We have used the sub-domain $0 < p_{\mathcal{C}} < 0.1$ of parameter $p_{\mathcal{C}}$ since the critical value is $0 < p_{\mathcal{C}} < 0.0041$, as found by our algorithm. The length of the first superball radius is $r^{*}_{(\mathcal{N}_H \otimes \mathcal{N}_H)} = 0$. The second superball radius $r^{*}_{(\mathcal{N}_H \otimes \mathcal{N}_2)}$ has a length of 0.01; this output has a much higher weight, 0.0081, for fixed channel parameter $p_{\mathcal{C}} = 0.004$, which is the upper bound of the possible range $0 < p_{\mathcal{C}} < 0.0041$. The maximum weight of the $r^{*}_{(\mathcal{N}_H \otimes \mathcal{N}_H)}$ can be obtained for channel probability $p_{\mathcal{C}} = 0.004$.

**Remark 4** *If channel* $\mathcal{N}_1 \in \mathcal{H}^{d_1}$ *belongs to the degradable family, then* $P^{(1)}(\mathcal{N}_1) = P(\mathcal{N}_1)$ *will hold, since the private classical capacity of these channels is additive [Smith08a, 09a, 09b].*

*Since superactivation is analogous to a capacity-conversion phenomenon (the* $P(\mathcal{N}_1)$ *private classical capacity of a single-channel is "converted" to the* $Q^{(1)}(\mathcal{N}_1 \otimes \mathcal{N}_2)$ *quantum capacity of the joint structure* $\mathcal{N}_1 \otimes \mathcal{N}_2 \in \mathcal{H}^{d_1 \cdot d_2}$*), it also follows that for any degradable channel, we will obtain* $Q^{(1)}(\mathcal{N}_1 \otimes \mathcal{N}_2) = Q(\mathcal{N}_1 \otimes \mathcal{N}_2) = \frac{1}{2}P^{(1)}(\mathcal{N}_1) = \frac{1}{2}P(\mathcal{N}_1)$ *for the superactivated quantum capacity of* $\mathcal{N}_1 \otimes \mathcal{N}_2 \in \mathcal{H}^{d_1 \cdot d_2}$.

*If* $\mathcal{N}_1 \in \mathcal{H}^{d_1}$ *does not belong the degradable family, then we could obtain higher values for the asymptotic superactivated joint quantum capacity, i.e.,* $Q^{(1)}(\mathcal{N}_1 \otimes \mathcal{N}_2) < Q(\mathcal{N}_1 \otimes \mathcal{N}_2)$ *holds.*

*In the example of this section, the* $\mathcal{N}_H \in \mathcal{H}^4$ *Horodecki channel is degradable (PPT-Positive Partial Transpose) channel, hence the* $Q^{(1)}(\mathcal{N}_1 \otimes \mathcal{N}_2)$ *single-use superactivated capacity cannot be increased even more in the asymptotic setting.*

We can conclude from our numerical analysis that, if we have two zero-capacity channels $\mathcal{N}_1 \otimes \mathcal{N}_2$, then the convex combination of these channels can result in greater than zero capacity for a small subset of possible parameters $p_{\mathcal{C}}$. As our geometrical analysis revealed, if we have two fixed channel models and we iterate on possible values of parameter $p_{\mathcal{C}}$, then, from the radius of the smallest superball, we can determine the possible values of the "superactivation parameter". We posit that stronger combinations for superactivation can be constructed from the larger set of quantum channel models and possible parameters. If we assume a system with $2n$ quantum channels which form two disjunctive sets $\left\{\mathcal{N}_1 \otimes \ldots \otimes \mathcal{N}_n\right\}$ and $\left\{\mathcal{M}_1 \otimes \ldots \otimes \mathcal{M}_n\right\}$, then the number of possible superactive channel combinations is $n$: $\left(\mathcal{N}_1 \otimes \mathcal{M}_1\right) \otimes \ldots \otimes \left(\mathcal{N}_n \otimes \mathcal{M}_n\right)$.

The results proposed in Sections 6.3.1, 6.3.2, 6.3.3 along with Appendix F conclude Thesis 2.3.

*The sections related to Thesis 2.1, Thesis 2.2, Thesis 2.3 and Appendix F conclude Thesisgroup 2.*

# Chapter 7

# Information Geometric Superactivation of Classical Zero-Error Capacity

"Quantum mechanics is magic."

*Daniel Greenberger*

The superactivation of zero-error capacity of quantum channels may be the starting-point of a large-scale revolution in the communication of future quantum networks. The superactivation of the zero-error capacity of quantum channels makes it possible to use two quantum channels, each with zero zero-error capacity, with a positive *joint zero-error capacity.* The possible combinations which enable the superactivation of the zero-error capacity of quantum channels are still unknown. In this chapter I introduce an information geometric approach and an algorithmical framework for the superactivation of classical zero-error capacity. The details of my proposed algorithm can be found in Appendix G.

# 7.1 Introduction

The zero-error capacity cannot be calculated in an easy way such as in the case of "non zero-error" capacity. The computation of zero-error capacity, for both the quantum and the classical case, is an extremely difficult computational problem, since the task is to decide which input states can be distinguished with zero error at the channel output after these states have passed through the noisy communication channel. Moreover, in case of a quantum communication system computing the classical zero-error capacity of a quantum channel is an NP-complete problem [Beigi07]. This section studies the superactivation property of the classical zero-error capacity of quantum channels. The problem of superactivation of zero-error capacity quantum channels can be viewed as a smaller subset of a larger problem set involving the additivity of quantum channels and the superactivation of zero-capacity quantum channels, as depicted in Fig. 1.2. Recently, Duan [Duan09] and Cubitt *et al.* [Cubitt09] have found only one possible combination for superactivation of the *classical* zero-error capacity of quantum channels, and also one combination has been shown for the *quantum* zero-error capacity by Cubitt and Smith [Cubitt09a]. Their results have opened the debate on the existence of other possible channel combinations.

The superactivation of quantum channels with individual zero-error capacity is illustrated in Fig. 7.1 [Gyongyosi11a].

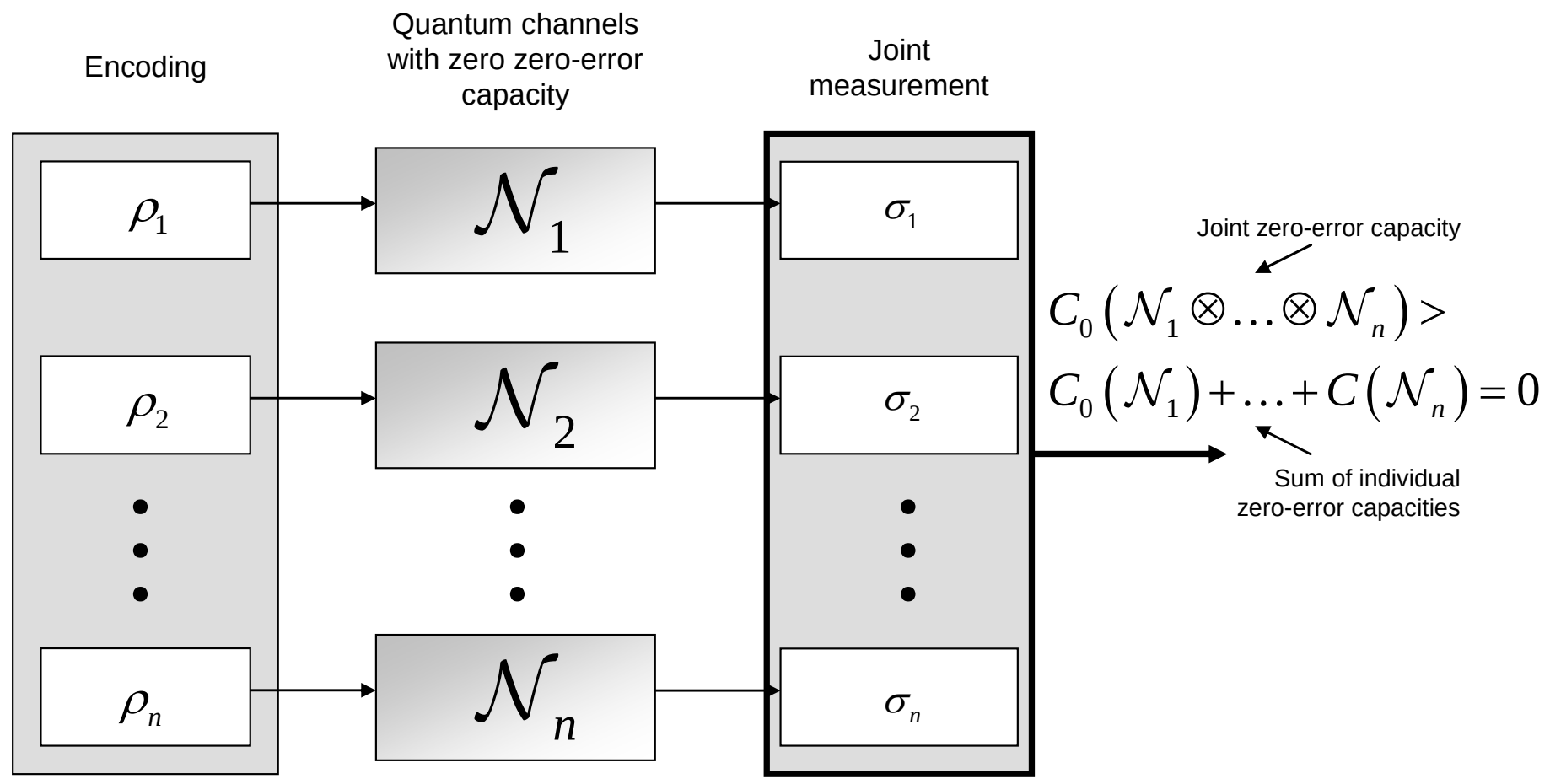

**Fig. 7.1.** Combination of quantum channels, with individual zero zero-error capacity, to realize perfect information transmission. The joint channel has positive zero-error capacity.

In Duan's work the superactive zero-error capacity quantum channels could be superactivated after two uses [Duan09]. Later, this result was improved by Cubitt *et al.* [Cubitt09], [Cubitt09a], since his combination required only a single use of the channel pair to realize the superactivation of zero-error capacities. As also stated by Duan [Duan09], superactivation can be achieved by using the entanglement, hence any classical messages can be transmitted perfectly through a zero error-capacity channel, if the classical messages are encoded by entangled states. Duan [Duan09] and Cubitt *et al.* [Cubitt09] simultaneously studied the possibility of superactivation of the classical zero-error capacity of quantum channels. Hence the superactivation of the classical zero-error capacity of quantum channels is possible, although they only found one possible solution. Later there results were extended to the quantum zero-error capacity by Cubitt *et al.* [Cubitt09a].
For further information see Appendix G or the book of Imre and Gyongyosi [Imre12].

## 7.1.1 Superactivation in Quantum Communication Networks

In future quantum communication networks, superactivation can be the ultimate weapon in situations where a quantum channel becomes totally or temporally unavailable or where quantum communication channels are extremely noisy. With the help of the superactivation of zero-error capacity of quantum channels, the perfect information transmission can be realized in a very noisy network environment. Superactivation can be applied to an optical-fiber based optical quantum communication network, or in a free-space environment, both in a dense metropolitan area or over very long distances to improve the quality of information transmission. The method of superactivation can be a very valuable tool to realize noiseless quantum communication (using both classical and quantum information) over a noisy communication environment. Using the superactivation of quantum channels, the effectiveness of the communication techniques in the future's quantum networks - in long-distances, or in a noisy metropolitan area - can be increased, and the currently used communication techniques can be further optimized [Gyongyosi10a-c], [Gyongyosi11e-f], [Gyongyosi11k-m]. To aiding metropolitan and long-distance quantum communication, the perfect information transmission through a quantum channel can have deep relevance both in military and secret government applications, or other cases where extremely high security is required.

In this chapter an information geometric algorithm for the superactivation of classical zero-error capacity of quantum channels is presented. The main result of this chapter is summarized in Thesisgroup 3.

**Thesisgroup 3.** *I proposed an algorithm to determine the conditions of superactivation of the classical zero-error capacity of arbitrary dimensional quantum channels. My proposed polynomial approximation method avoids the problem of NP-completeness.*

## 7.2 Information Geometric Superactivation of Zero-Error Capacities of a Quantum Channel

Here we extend the results of Chapter 5 and we give the definition of classical and quantum zero-error capacities in terms of the quantum informational balls. We connect the results of Chapter 3 and Chapter 5 to give an interpretation of superactivation of classical and quantum zero-error capacities.

### 7.2.1 Classical Zero-Error Capacity

In this section we define an information geometric object to express the classical zero-error capacity of the quantum channel and then we show that the $C\left(\mathcal{N}_1 \otimes \mathcal{N}_2\right)$ superactivated classical zero error capacity of the joint structure $\mathcal{N}_1 \otimes \mathcal{N}_2 \in \mathcal{H}^{d_1} \otimes \mathcal{H}^{d_2}$ can be determined by our object.

**Definition 7** I define the asymptotic classical capacity $C(\mathcal{N})$ of channel $\mathcal{N} \in \mathcal{H}^d$ by $r_{super}^{*}(\mathcal{N}) \in \mathbb{R}$ over $\mathbb{C}^d$ as follows:

$$\begin{aligned} r_{super}^{*}(\mathcal{N}) = C(\mathcal{N}) &= \lim_{n\to\infty}\frac{1}{n}C^{(1)}\left(\mathcal{N}^{\otimes n}\right) \\ &= \lim_{n\to\infty}\frac{1}{n}\left(\sum_{i=1}^{n} r_i^{(1)*}\right) \\ &= \lim_{n\to\infty}\frac{1}{n}\max_{p_1,\ldots,p_n,\rho_1,\ldots,\rho_n}\left(\mathcal{X}_{AB}\right) \\ &= \lim_{n\to\infty}\frac{1}{n}\sum_{n}\left(\min_{\sigma_{1\ldots n}}\max_{\rho_{1\ldots n}} D\left(\rho_k^{AB}\,\middle\|\,\sigma^{AB}\right)\right). \end{aligned} \tag{7.1}$$

where $r_i^{*} \in \mathbb{R}$ is the single-use capacity of the *i*-th use of quantum channel $\mathcal{N} \in \mathcal{H}^d$, $\rho_k^{AB} \in \mathbb{C}^d$ is the optimal output channel state, and $\sigma^{AB} \in \mathbb{C}^d$ is the average state. Using the result of (7.1) for the joint channel construction $\mathcal{N}_1 \otimes \mathcal{N}_2 \in \mathcal{H}^{d_1} \otimes \mathcal{H}^{d_2}$, the *asymptotic* HSW channel capacity $C(\mathcal{N}_1 \otimes \mathcal{N}_2)$ is equal to the sum of $r_{super}^{*}(\mathcal{N}_1 \otimes \mathcal{N}_2) = \lim_{n\to\infty}\frac{1}{n} r_{super}^{(1)*}(\mathcal{N}_1 \otimes \mathcal{N}_2) \in \mathbb{R}$, as it was already proven is Chapter 5.

Using the proposed results of Appendix E on the geometric interpretation of the quantum relative entropy function and the HSW channel capacity the following definition can be made.

**Definition 8** I define the $C_0(\mathcal{N})$ asymptotic classical zero-error capacity for channel $\mathcal{N} \in \mathcal{H}^d$ over $\mathbb{C}^d$ as follows:

$$\begin{aligned} r_{super(C_0)}^{*}(\mathcal{N}) = C_0(\mathcal{N}) &= \lim_{n\to\infty}\frac{1}{n}C_0^{(1)}\left(\mathcal{N}^{\otimes n}\right) \\ &= \lim_{n\to\infty}\frac{1}{n}\left(\sum_{i=1}^{n} r_{(C_0)\,i}^{(1)*}\right) \\ &= \lim_{n\to\infty}\frac{1}{n}\max_{p_1,\ldots,p_n,\rho_1,\ldots,\rho_n}\left(\mathcal{X}_{AB}^{0}\right) \\ &= \lim_{n\to\infty}\frac{1}{n}\sum_{n}\left(\min_{\sigma_{1\ldots n}}\max_{\rho_{1\ldots n}} D\left(\rho_k^{AB}\,\middle\|\,\sigma^{AB}\right)\right), \end{aligned} \tag{7.2}$$

where $r^{*}_{super(C_0)}(\mathcal{N}) \in \mathbb{R}$ denotes the classical asymptotic zero-error capacity, $r^{(1)*}_{(C_0)\,i}(\mathcal{N}) \in \mathbb{R}$ is the single-use classical zero-error capacity of the *i*-th use of quantum channel $\mathcal{N}$, $\rho_k^{AB} \in \mathbb{C}^d$ is the optimal channel output state, and $\sigma^{AB} \in \mathbb{C}^d$ is the average output state while $\mathcal{X}^0_{AB}$ measures the Holevo information which can be obtained for the zero-error input code (the input quantum zero-error code is determined by iterations made on the output).

#### 7.2.1.1 Joint Classical Zero-Error Capacity

First I define the informational geometric interpretation of the superactivated *single-use* classical zero-error capacity of $\mathcal{N}_1 \otimes \mathcal{N}_2 \in \mathcal{H}^{d_1} \otimes \mathcal{H}^{d_2}$, then the *asymptotic* version.

**Definition 9** For joint structure $\mathcal{N}_1 \otimes \mathcal{N}_2 \in \mathcal{H}^{d_1} \otimes \mathcal{H}^{d_2}$ of channels $\mathcal{N}_1 \in \mathcal{H}^{d_1}$ and $\mathcal{N}_2 \in \mathcal{H}^{d_2}$ I define the superactivated single-use classical zero-error channel capacity $C_0^{(1)}(\mathcal{N}_1 \otimes \mathcal{N}_2)$ by $r^{(1)*}_{super(C_0)}(\mathcal{N}_1 \otimes \mathcal{N}_2) \in \mathbb{R}$ over $\mathbb{C}^{d_1} \otimes \mathbb{C}^{d_2}$ as follows:

$$\begin{aligned} r^{(1)*}_{super(C_0)}(\mathcal{N}_1 \otimes \mathcal{N}_2) &= C_0^{(1)}(\mathcal{N}_1 \otimes \mathcal{N}_2) \\ &= r^{(1)*}_{(C_0)} \\ &= \max_{p_1,\ldots,p_n,\rho_1,\ldots,\rho_n} \left(\mathcal{X}^0_{AB}\right) \\ &= \min_{\sigma_{1\ldots n}} \max_{\rho_{1\ldots n}} D\left(\rho_k^{AB} \middle\| \sigma^{AB}\right), \end{aligned} \tag{7.3}$$

where $r^{(1)*}_{super(C_0)}(\mathcal{N}_1 \otimes \mathcal{N}_2) \in \mathbb{R}$ denotes the classical superactivated single-use zero-error capacity of the joint channel structure $\mathcal{N}_1 \otimes \mathcal{N}_2 \in \mathcal{H}^{d_1} \otimes \mathcal{H}^{d_2}$. It is equal to $r^{(1)*}_{(C_0)} \in \mathbb{R}$, the $C_0^{(1)}$ single-use classical zero-error capacity of the joint channel structure $\mathcal{N}_1 \otimes \mathcal{N}_2$, which trivially follows from (7.2). The $\rho_k^{AB} \in \mathbb{C}^{d_1} \otimes \mathbb{C}^{d_2}$ is the optimal channel output state of the joint structure $\mathcal{N}_1 \otimes \mathcal{N}_2$, and $\sigma^{AB} \in \mathbb{C}^{d_1} \otimes \mathbb{C}^{d_2}$ is the average output state of $\mathcal{N}_1 \otimes \mathcal{N}_2$ while $\mathcal{X}_{AB}^0$ measures the zero-error Holevo information of the superactivated structure $\mathcal{N}_1 \otimes \mathcal{N}_2$. From the result derived in detail in Appendix G and the theses of Chapter 5 also follows that the superactivation of the $C_0^{(1)}\left(\mathcal{N}_1 \otimes \mathcal{N}_2\right)$ single-use classical zero-error capacity the joint structure $\mathcal{N}_{12}$ is possible if the $\mathbb{C}^{d_1 \cdot d_2}$ radius belonging to joint states $\rho_{12}^{AB}$ and $\sigma_{12}^{AB}$ cannot in $\mathbb{C}^{d_1} \otimes \mathbb{C}^{d_2}$ be decomposed, i.e., the average joint state and the optimal joint state are both *entangled* states, otherwise the superactivation is not possible, i.e.,

$$C_0^{(1)}\left(\mathcal{N}_1\right) = C_0^{(1)}\left(\mathcal{N}_2\right) = C_0^{(1)}\left(\mathcal{N}_{12}\right) = 0. \tag{7.4}$$

The results on the decomposition of the radius of the superball (for the derivation see Appendix G) trivially follows for the $C_0\left(\mathcal{N}_1 \otimes \mathcal{N}_2\right)$ asymptotic classical zero-error capacity, i.e., for simplicity we omit the derivation (see Appendix G).

Next I discuss the information geometric representation of the superactivated *asymptotic* classical zero-error capacity of $\mathcal{N}_1 \otimes \mathcal{N}_2 \in \mathcal{H}^{d_1} \otimes \mathcal{H}^{d_2}$.

**Definition 10** For $\mathcal{N}_1 \otimes \mathcal{N}_2 \in \mathcal{H}^{d_1} \otimes \mathcal{H}^{d_2}$, I define the superactivated asymptotic classical zero-error channel capacity $C_0\left(\mathcal{N}_1 \otimes \mathcal{N}_2\right)$ by the superball radius $r_{super\left(C_0\right)}^{*}\left(\mathcal{N}_1 \otimes \mathcal{N}_2\right) \in \mathbb{R}$ over $\mathbb{C}^{d_1} \otimes \mathbb{C}^{d_2}$ as:

$$\begin{aligned} r^{*}_{super(C_0)}\left(\mathcal{N}_1 \otimes \mathcal{N}_2\right) &= C_0\left(\mathcal{N}_1 \otimes \mathcal{N}_2\right) \\ &= \lim_{n\to\infty} \frac{1}{n} C_0^{(1)}\left(\left(\mathcal{N}_1 \otimes \mathcal{N}_2\right)^{\otimes n}\right) \\ &= \lim_{n\to\infty} \frac{1}{n}\left(\sum_{i=1}^{n} r^{*}_{(C_0)\ i}\right) \\ &= \lim_{n\to\infty} \frac{1}{n} \max_{p_1,\ldots,p_n,\rho_1,\ldots,\rho_n}\left(\mathcal{X}^0_{AB}\right) \\ &= \lim_{n\to\infty} \frac{1}{n} \sum_{n}\left(\min_{\sigma_{1\ldots n}} \max_{\rho_{1\ldots n}} D\left(\rho_k^{AB} \middle\| \sigma^{AB}\right)\right), \end{aligned} \tag{7.5}$$

where $r^{*}_{super(C_0)}\left(\mathcal{N}_1 \otimes \mathcal{N}_2\right) \in \mathbb{R}$ denotes the classical superactivated asymptotic zero-error capacity of the joint channel structure $\mathcal{N}_1 \otimes \mathcal{N}_2$, $r^{(1)*}_{(C_0)\ i} \in \mathbb{R}$ is the $C_0^{(1)}$ single-use classical zero-error capacity of the joint channel structure $\mathcal{N}_1 \otimes \mathcal{N}_2$, $\rho_k^{AB} \in \mathbb{C}^{d_1} \otimes \mathbb{C}^{d_2}$ is the optimal channel output state of $\mathcal{N}_1 \otimes \mathcal{N}_2$, and $\sigma^{AB} \in \mathbb{C}^{d_1} \otimes \mathbb{C}^{d_2}$ is the average output state of $\mathcal{N}_1 \otimes \mathcal{N}_2$ while $\mathcal{X}^0_{AB}$ measures the zero-error Holevo information which can be obtained for the zero-error input code of $\mathcal{N}_1 \otimes \mathcal{N}_2$. By using the encoding scheme from Appendix G, the capacity will be normalized by $1/2$, i.e., the *asymptotic* zero-error classical capacity of the joint structure $\mathcal{N}_1 \otimes \mathcal{N}_2$ is expressed as

$$C_0\left(\mathcal{N}_1 \otimes \mathcal{N}_2\right) = \frac{1}{2} r^{*}_{super(C_0)}\left(\mathcal{N}_1 \otimes \mathcal{N}_2\right). \tag{7.6}$$

In Fig. 7.2, we illustrate the superball of the classical zero-error capacity for the analysis of superactivation in the space $\mathbb{C}^{d_1} \otimes \mathbb{C}^{d_2}$ of the joint structure $\mathcal{N}_1 \otimes \mathcal{N}_2 \in \mathcal{H}^{d_1} \otimes \mathcal{H}^{d_2}$, using the $\mathbb{R}^3$ representation. The proposed geometrical structure has similarities with the geometrical interpretation of the HSW capacity of quantum channels, which also follows from (7.2).

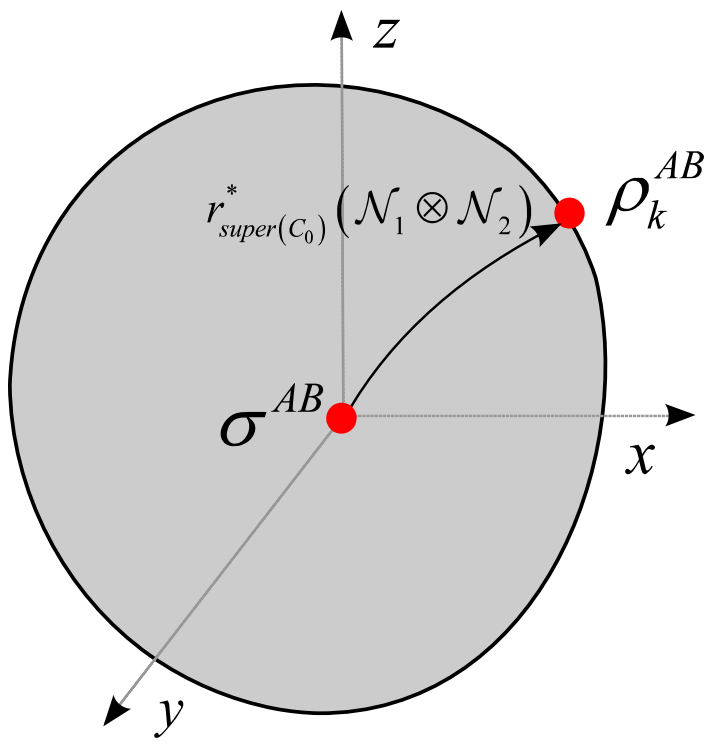

**Fig. 7.2.** The superactivation of the classical zero-error capacity is analyzed by the quantum informational superball.

The quantum zero-error capacity $Q_0(\mathcal{N})$ also can be superactivated, as it has been shown by Cubitt *et al.* [Cubitt09]. According to the correspondence between the Holevo information and the quantum coherent information of quantum channels, the proposed geometrical approach can be extended to the $Q_0(\mathcal{N}_1 \otimes \mathcal{N}_2)$ superactivated zero-error capacity of $\mathcal{N}_1 \otimes \mathcal{N}_2 \in \mathcal{H}^{d_1} \otimes \mathcal{H}^{d_2}$.

## 7.2.2 Quantum Zero-Error Capacity

**Definition 11** I define $r^*_{super(Q_0)}(\mathcal{N}_1 \otimes \mathcal{N}_2) \in \mathbb{R}$ the superactivated asymptotic quantum zero-error capacity of the joint structure $\mathcal{N}_1 \otimes \mathcal{N}_2 \in \mathcal{H}^{d_1} \otimes \mathcal{H}^{d_2}$ over $\mathbb{C}^{d_1} \otimes \mathbb{C}^{d_2}$ as:

$$
\begin{aligned}
r^{*}_{super(Q_0)}\left(\mathcal{N}_1 \otimes \mathcal{N}_2\right) &= Q_0\left(\mathcal{N}_1 \otimes \mathcal{N}_2\right) = \lim_{n\to\infty} \frac{1}{n} Q_0^{(1)}\left(\left(\mathcal{N}_1 \otimes \mathcal{N}_2\right)^{\otimes n}\right) \\
&= \lim_{n\to\infty} \frac{1}{n}\left(\sum_{i=1}^{n} r^{(1)*}_{(Q_0)i}\right) = \lim_{n\to\infty} \frac{1}{n} \max_{p_1,\ldots,p_n,\rho_1,\ldots,\rho_n} I_{coh}\left(\rho_A : \mathcal{N}^{\otimes n}\left(\rho_A\right)\right) \\
&= \lim_{n\to\infty} \frac{1}{n} \max_{p_1,\ldots,p_n,\rho_1,\ldots,\rho_n} \left(\mathcal{X}^0_{AB} - \mathcal{X}^0_{AE}\right) \\
&= \lim_{n\to\infty} \frac{1}{n} \max_{p_1,\ldots,p_n,\rho_1,\ldots,\rho_n} \Bigg(\mathrm{S}\left(\mathcal{N}^{\otimes n}_{AB}\left(\sum_{i=1}^{n} p_i\rho_i\right)\right) - \sum_{i=1}^{n} p_i \mathrm{S}\left(\mathcal{N}^{\otimes n}_{AB}\left(\rho_i\right)\right) \\
&\qquad - \mathrm{S}\left(\mathcal{N}^{\otimes n}_{AE}\left(\sum_{i=1}^{n} p_i\rho_i\right)\right) + \sum_{i=1}^{n} p_i \mathrm{S}\left(\mathcal{N}^{\otimes n}_{AE}\left(\rho_i\right)\right)\Bigg) \\
&= \lim_{n\to\infty} \frac{1}{n} \sum_{n}\left(\min_{\sigma_{1\ldots n}} \max_{\rho_{1\ldots n}} D\left(\rho_k^{AB-AE} \middle\| \sigma^{AB-AE}\right)\right),
\end{aligned}
\tag{7.7}
$$

where $r^{(1)*}_{(Q_0)i} \in \mathbb{R}$ is the single-use quantum zero-error capacity of the joint structure $\mathcal{N}_1 \otimes \mathcal{N}_2 \in \mathcal{H}^{d_1} \otimes \mathcal{H}^{d_2}$, while $\mathcal{X}^0_{AB}$ and $\mathcal{X}^0_{AE}$ measure the Holevo information which can be obtained for the zero-error quantum code (the input quantum zero-error code is determined by iterations made on the output). The $r^{*}_{super(Q_0)} \in \mathbb{R}$ of the superball is equal to the asymptotic zero-error quantum capacity.

## 7.3 Efficient Algorithm for the Superactivation of Zero-Error Capacity

In this section an efficient algorithm is constructed to discover the conditions of the superactivation of classical zero-error capacity of quantum channels, as summarized in Thesis 3.1.

**Thesis 3.1.** *I showed that the superactive channel combinations and the input conditions of the superactivation of classical zero-error capacity of quantum channels can be discovered by the proposed information geometric approach.*

The logical structure of the proposed information geometric analysis of the superactivation of the classical zero-error capacity follows the same structure as was already shown in Fig. 6.3 and in the Motivation section of Chapter 1.

As depicted in Fig. 7.3 [Gyongyosi11a], the proposed analysis combines the results of weak core-set methods [Feldman07], [Badoiu03], [Chen07], [Zhang09] the properties of efficient clustering algorithms [Ackermann08-09] and the sum of these approaches investigates a completely new framework to study the superactivation effect. To construct the weak core-set, we apply a bicreteria algorithm [Feldman07], [Gupta06] to find the required parameters of the superactivation of the classical zero-error capacity. In our method, we use similar quantum informational distance as distance measures between the density matrices. The use of standard quantum informational distance function for the construction of Voronoi diagrams have been investigated by Kato *et al.* [Kato06] and later by Nielsen and Nock [Nielsen08b]. Our contribution completely differs from these previous works since in our method, instead of these solutions the similar quantum informational function, the core-set and the weak core-set algorithms are applied which provide the most efficient algorithmical solution for the superactivation of classical zero-error capacities of quantum channels. For the details of the core-set algorithm see Appendix G. By using the core-set method, we can construct a more efficient $\left(1+\varepsilon\right)$-approximation algorithm in quantum space, using only a small subset of the original larger input set, where the input system consists of $n$ density matrices [Gyongyosi11a].

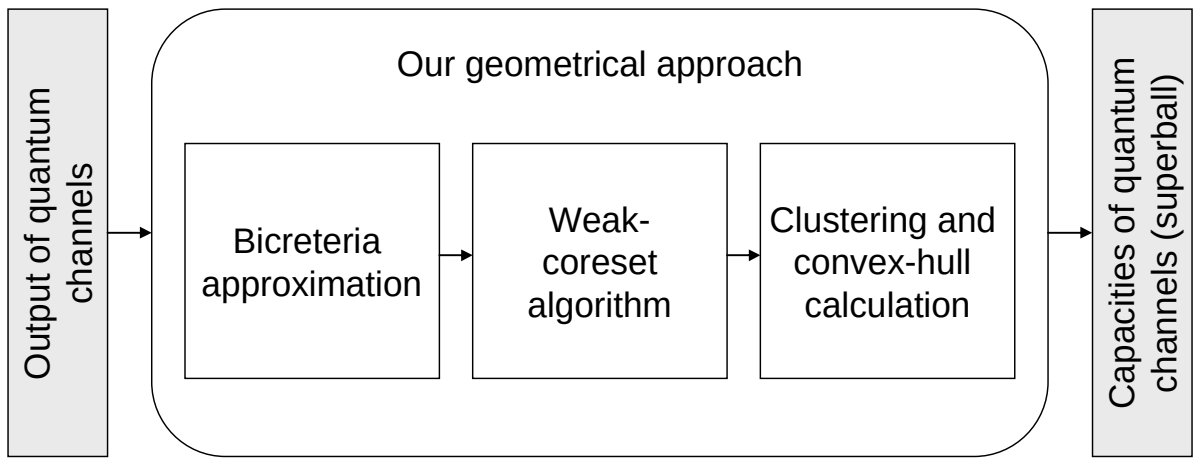

**Fig. 7.3.** Decomposition of the proposed geometrical approach. The output of the algorithm is the radius of the superball.

The core-set technique has deep relevance to classical computational geometry. A core-set of a set of output quantum states has the same behavior as the larger input set, so clustering and other approximations can be made with smaller core-sets. The core-set can be viewed as a smaller input set of channel output states, hence it can be used as the input to an approximation algorithm. The weighted sum of errors of the smaller core-set is a $\left(1 \pm \varepsilon\right)$ approximation of the larger input set. The bound on this error can be decreased only if the center points that form a finite set are used in the approximation. These core-sets are called weak core-sets [Ackermann08-09], [Chen06], [Chen06a], [Chen07], [Badoiu02], [Badoiu03], [Gupta06], [Feldman07], [Frahling05], [Har-Peled05], [Zhang09] and this method can be applied in quantum space between quantum states. Using weak core-sets, the run time of $\left(1 + \varepsilon\right)$ core-set algorithms [Ackermann08], [Feldman07], [Frahling05], [Har-Peled05] with respect to quantum informational distance can be improved. Since the proposed methods of [Ackermann08], [Chen06], [Chen06a], [Chen07], [Badoiu02], [Badoiu03], [Gupta06], [Feldman07], [Frahling05], [Har-Peled05], [Zhang09] are developed for classical systems these algorithms cannot be applied in the quantum space to calculate the distances between the density matrices. It also follows that these results cannot be used in the analysis of superactivation of quantum channels and new algorithms have to be constructed. As we will show in this chapter it is possible to construct an efficient information geometric algorithm to discover the conditions of superactivation, however it requires the definition of special functions along with an exact information geometric definition of the information geometric description of the superactivated zero-error capacity of quantum channels (see Section 7.2.1). These results will be presented in Sections 7.3.1, 7.3.2, and Appendix G. The proposed algorithmical framework is based on the theses of Chapter 5. To construct the core-set method analyzing the superactivation of zero-error capacity quantum channels, we have to introduce the definition of *similar quantum informational distance* and *weak core-sets of quantum states* [Gyongyosi11a].

## 7.3.1 Similar Quantum Informational Distance

The methods of distance calculations using the standard quantum informational distance between the density matrices have been studied by Kato *et al.* [Kato06] and by Nielsen and Nock [Nielsen08b]. However these approaches were not taken into consideration a very important problem. The quantum informational distance is asymmetric and contains singularities, since there exist density matrices $\rho$ and $\sigma$ for which $D\left(\rho,\sigma\right)=\infty$. The similar quantum informational divergence function does not contain these singularities and these distances are approximately symmetric. Our contribution takes into account this important property [Gyongyosi11a]. To use the similar quantum informational divergence function between the density matrices, first we define it as follows. The quantum informational distance function $D\left(\rho\|\sigma\right)$ between density matrices $\rho$ and $\sigma$ is $\mu$-similar for a positive real constant $\mu$, if there exists a positive definite matrix $A$ such that

$$\mu D_A\left(\rho\|\sigma\right)\leq D\left(\rho\|\sigma\right)\leq D_A\left(\rho\|\sigma\right). \tag{7.8}$$

For quantum informational distances, if the domain is given as $\kappa=\left[\lambda,\gamma\right]\subseteq R_+^d$, then $\mu=\frac{\lambda}{\gamma}$ and $\frac{1}{2\lambda}I$. If we have $0<\lambda<\gamma$, then the quantum informational distance function can be calculated by the $D\left(\rho\|\sigma\right)$ quantum relative entropy function on the domain $\kappa=\left[\lambda,\gamma\right]\subseteq R^d$. The quantum informational distance is $\mu$-similar if $\mu=\frac{\lambda}{\gamma}$ and $A=\frac{1}{2\lambda}I$. In these cases, the quantum informational distance function is $\mu$-similar, because it is restricted to a sub-domain, which *avoids the singularities*. The quantum informational distance function is strictly convex and all second-order partial derivates exist and are continuous on the domain $\kappa=\left[\lambda,\gamma\right]\subseteq R^d$ with parameters $\mu=\frac{\lambda}{\gamma}$ and $A=\frac{1}{2\lambda}I$ [Ackermann08]. The preliminary version of the applied core-set algorithm was originally presented by Chen *et al.* [Chen06], [Chen07]. We show that this method can be extended to generate a core-set based on a similar quantum informational distance function. To ob-tain an enhanced version of previously known core-set approximation algorithms, we must define weak core-sets [Feldman07] of density matrices.

## 7.3.2 Weak Core-set of Density Matrices

The weak core-sets include all the relevant information required to analyze the original extremely large input set [Feldman07]. In comparison to the original core-set approach the proposed weak core-set algorithm has significantly lower computational complexity. In our method, weak core-sets are applied to $\mu$-similar quantum informational distances since, in these subsets, the distances between quantum states are symmetric [Gyongyosi11a], hence singularities can be avoided and fast Euclidean methods can be applied [Ackermann08], [Chen06], [Chen07], [Feldman07]. We will use this subset to approximate the original input set (i.e., the density matrices of the zero-error input codewords), with approximation error $\left(1 \pm \varepsilon\right)$, hence having the results of Section 7.3.1 in our hands we can conclude Thesis 3.1 and put forward the following statement in Thesis 3.2.

> **Thesis 3.2.** *I proved that the superactivation of classical zero-error capacity of quantum channels can be analyzed by the proposed algorithm with minimized error by using the smaller subset of input density matrices.*

Using the results of Chen [Chen07], we show that by using this algorithm, the superactivation capability of quantum channels can be approximated with error $\left(1 \pm \varepsilon\right)$, using the smaller $\mu$-similar subset of input quantum states. Applying the proposed algorithm, the $\left(1 \pm \varepsilon\right)$-approximation can be obtained in run time $\mathcal{O}\left(dkn\right)$, where $k$ is the number of medians of channel output quantum states. The goal of the algorithm is to find a set of size $k$ such that the sum of errors of quantum informational distances is minimized, hence

$$error\left(\rho_i,\sigma\right)=\sum_{i=1}^{n}\min D\left(\rho_i\|\sigma\right). \tag{7.9}$$

The result in (7.9) will be referred as the $error_{\min.}\left(\rho_i,\sigma\right)$ minimized error. The algorithm solves the *k*-median problem [Ackermann08] with respect to the quantum informational distance $D$ in quantum space. The output of the algorithm is a set of *k* quantum states, for which the function $error\left(\rho_i,\sigma\right)$ is minimized. We generalize the *k*-median problem for quantum informational distances [Gyongyosi11a]. Let us assume that we have quantum states $\rho$ and $\sigma$ in domain $\mathcal{S}$, where $\sigma$ is the average state. We would like to construct an averaged subset $\sigma$ of $\mathcal{S}_{OUT}$ of *k* quantum states, for which

$$D\left(\rho,\mathcal{S}_{OUT}\right)=\min_{\sigma\in\mathcal{S}^*}D\left(\rho\|\sigma\right). \tag{7.10}$$

The *k*-median problem for quantum states can be stated as follows. We would like to use only a finite set $\mathcal{S}_{IN}$ of quantum states from the original larger space for the superactivation of the classical zero-error capacity. For a set $\mathcal{S}_{IN}$, we would like to construct a set $\mathcal{S}_{OUT}$ of *k*-quantum states, for which $error\left(\mathcal{S}_{IN},\mathcal{S}_{OUT}\right)=\sum_{\rho\in\mathcal{S}_{IN}}D\left(\rho\|\mathcal{S}_{OUT}\right)$ is minimized. As follows, using the quantum relative entropy function for the distance calculations and the finite set $\mathcal{S}_{IN}$ of density matrices, the set $\mathcal{S}_{OUT}$ of density matrices can be constructed by the proposed algorithm with the minimal error

$$error\left(\mathcal{S}_{IN},\mathcal{S}_{OUT}\right)=error_{\min.}\left(\mathcal{S}_{IN},\mathcal{S}_{OUT}\right)=\sum_{\rho\in\mathcal{S}_{IN}}\min D\left(\rho\|\mathcal{S}_{OUT}\right). \tag{7.11}$$

The error of the optimal solution for input states $\mathcal{S}_{IN}$ of the joint structure $\mathcal{N}_1\otimes\mathcal{N}_2$ is denoted by $opt_k\left(\mathcal{S}_{IN}\right)$, and the elements of the output set $\mathcal{S}_{OUT}$ are the *k* median-quantum states of set $\mathcal{S}_{IN}$. To construct a more efficient algorithm, we use only the $\mu$-similar quantum informational distances, hence the set of input quantum states $\mathcal{S}_{IN}$ is restricted to quantum states for which the singularities can be avoided [Gyongyosi11a]. We show that the superactivation properties of the classical zero-error capacity of the joint channel structure $\mathcal{N}_1 \otimes \mathcal{N}_2$ can be discovered by using $\mu$-similar quantum informational distances and the core-set construction method (for details see Appendix G). For any set $\mathcal{S}_{IN}$ of size $n$ quantum states and for any finite $\mathcal{W} \subseteq \mathcal{S}$, there exists a weak core-set of size $\mathcal{O}\left(\frac{1}{\varepsilon^2} k \log\left(n\right) \log\left(k \left|\mathcal{W}\right|^k \log n\right)\right)$. This $\mathcal{W}$-weak core-set of quantum states can be constructed in time $\mathcal{O}\left(\frac{1}{\varepsilon^2} k \log\left(n\right) \log\left(k \left|\mathcal{W}\right|^k \log n\right) + dkn\right)$, where $k$ is the number of quantum states in set $\mathcal{S}_{OUT}$, $n$ is the number of input density matrices (i.e., the number of the EPR states in the input codeword, for the encoding scheme of the superactivation of the classical zero-error capacity see Appendix G) and $d$ is the dimension. The related section 7.3.1 and 7.3.2 conclude Thesis 3.2.

We have also stated previously in Section 7.1 that the numerical calculation of the classical zero-error capacity of quantum channels is an NP-complete problem [Beigi07].

Our information geometric algorithm avoids this problem as summarized in Thesis 3.3.

> **Thesis 3.3.** *I proved that by using $\mu$-similar quantum informational distances and the weak core-set of quantum states, the superactivation of zero-error capacity of quantum channels can be determined by a polynomial approximation algorithm without the problem of NP-completeness.*

According to [Ackermann09], [Banerjee05], [Gupta06], [Zhang09] the optimal 1-median of any given input set $\mathcal{S}$ in quantum space can be uniquely defined by the centroid $c = \frac{1}{|\mathcal{S}|}\sum_{\rho \in \mathcal{S}} \rho$. From these results also follows that an optimal solution of the $k$-median clustering problem can be approached by $\left(k-1\right)$ linearly separable subsets and for any set $\mathcal{S}_{IN}$ at most $n^{dk}$ possible solutions have to considered as one of the optimal $k$-median quantum states of $\mathcal{S}_{IN}$ [Banerjee05]. Our solution avoids this problem, since we use a smaller set $\mathcal{S}$ from $\mathcal{S}_{IN}$, which is a small weighted set that has the same clustering behavior as the larger input set $\mathcal{S}_{IN}$. The core-set method used in our approach can be defined by the error of the approximation in terms of the quantum informational distance between quantum states as follows [Gyongyosi11a]:

$$error_w\left(\mathcal{S},\mathcal{S}_{OUT}\right)=\sum_{\rho\in\mathcal{S}}w\left(\rho\right)D\left(\rho\middle\|\mathcal{S}_{OUT}\right), \tag{7.12}$$

and this error is a $\left(1\pm\varepsilon\right)$-approximation of $error_{\min.}\left(\mathcal{S}_{IN},\mathcal{S}_{OUT}\right)$ for any set of quantum states $\mathcal{S}_{OUT}$ of size $\left|\mathcal{S}_{OUT}\right|=k$. For the weak core-set construction, let us assume that we have an input codeword with a set $\mathcal{S}_{IN}$ of density matrices on the joint channel $\mathcal{N}_1\otimes\mathcal{N}_2$ and a set $\mathcal{W}$. If the weight function is defined by

$$\sum_{\mathcal{S}}w\left(\rho\right)=\left|\mathcal{S}_{IN}\right|, \tag{7.13}$$

then, the weighted set $\mathcal{S}$ is a $\mathcal{W}$-weak core-set of $\mathcal{S}_{IN}$, if for all $\mathcal{S}_{OUT}\in\mathcal{W}$ of size $\left|\mathcal{S}_{OUT}\right|=k$, we have

$$\begin{aligned}&\left|error_{\min.}\left(\mathcal{S}_{IN},\mathcal{S}_{OUT}\right)-error_w\left(\mathcal{S},\mathcal{S}_{OUT}\right)\right|\\&\leq\left(\varepsilon\right)error_{\min.}\left(\mathcal{S}_{IN},\mathcal{S}_{OUT}\right).\end{aligned} \tag{7.14}$$

This $\mathcal{W}$-weak core-set is called the $\left(k,\varepsilon\right)$ weak-core-set of $\mathcal{S}_{IN}$. To get this construction with this error bound, we propose the polynomial approximation algorithm designed for the density matrices of the input codewords [Gyongyosi11a]. The details of proposed core-set method can be found in Appendix G. The overall run time of the algorithm used (see Appendix G) has been proven to be $\mathcal{O}\left(d^2 2^{\frac{k}{\varepsilon}}\log^{k+2}n+dkn\right)$ using [Feldman07], [Frahling05], [Har-Peled05] and the theoretical results of [Chen06], [Chen07], [Badoiu03] and [Ackermann08], which confirm that the algorithm can be used to solve the determination of the $C_0$ superactivated asymptotic zero-error capacity of the joint structure $\mathcal{N}_1 \otimes \mathcal{N}_2$ avoiding the NP-completeness.

The bicreteria algorithm can be computed in time $\mathcal{O}\left(dkn\right)$, hence the core-set method can be constructed in time $\mathcal{O}\left(\left|\mathcal{W}\right| + dkn\right)$. Assuming $\mathcal{E} < 1$, the algorithm always can find the capacity with polynomial complexity. Choosing $\mathcal{E} \to 0$, the running time can be optimized by decreasing the number $k$ of cluster-medians of the density matrices.

The results presented in Sections 7.3.1, 7.3.2 along with Appendix G conclude Thesis 3.3.

*The sections related to Thesis 3.1, Thesis 3.2 and Thesis 3.3 conclude Thesisgroup 3.*

# 7.4 Future Work

In this chapter I introduced a fundamentally new algorithmic solution for superactivation of the asymptotic zero-error capacity of quantum channels. With the help of my proposed informational geometric approach, the complexity of the computation of the zero-error capacity of the quantum channels can be significantly decreased. The proposed algorithmic solution can be the key to finding other possible channel models and channel parameter domains, with possible combinations being proved by theory. I have constructed an extremely fast recursive geometric algorithm to find the conditions for the computation and for the superactivation of the asymptotic classical zero-error capacity of the quantum channels.

My method is the first to solve the problem of the efficient computation of classical zero-error capacity of quantum channels. In 2011, Gyongyosi and Imre applied the algorithm presented in this chapter to find quantum channel combinations for which the classical zero-error capacity can be superactivated [Gyongyosi11h], [Gyongyosi11j]. Later, these results were extended by Gyongyosi and Imre for the superactivation of quantum zero-error capacity [Gyongyosi11g], [Gyongyosi11i] and the authors have shown that this result can be applied in the development of the quantum repeaters [Gyongyosi11o], [Gyongyosi11e] and superactivated quantum repeaters can be built for the future telecommunications [Gyongyosi11f]. The biggest problem in future quantum communications is the long-distance delivery of quantum information. Since the quantum states cannot be copied, the amplification of quantum bits is a more complex compared to classical communications [Gyongyosi11n]. The success of future long-distance quantum communications and global quantum key distribution systems strongly depends on the development of efficient *quantum repeaters*. From the viewpoint of long-distance quantum communication the development of a well-scalable quantum repeater is a cardinal question. The entanglement purification is a cardinal question from success point of view during the entanglement sharing process between the base stations of the repeaters. The fidelity of the predicated quantum states mostly depends on the noise of the quantum channel. The problem could be solved only if the fidelity can be maximized without the very expensive purification process. As my results have concluded, using very noisy quantum channels between the repeater stations, the fidelity of the states can be increased without the very inefficient and expensive purification methods. By means of the proposed solution, the efficiency of quantum repeaters can be increased even using very noisy quantum channels. Since the physical realizations of quantum communication with noisy optical fibers will be one of the most relevant questions in experimental future communications, my research work will be of interest to scientists in other fields.

As included in Appendix G the superactivation of zero-error capacity can be exploited in the quantum repeaters and *superactivated quantum repeaters* [Gyongyosi11f] can be built in the future. For the details see the articles of Gyongyosi and Imre [Gyongyosi11e], [Gyongyosi11f] and the book of Imre and Gyongyosi [Imre12].

# Chapter 8

# Conclusions

Quantum computing offers fundamentally new solutions in the field of computer science. The superactivation cannot be imagined for classical systems. In the near future, superactivation can aid long-distance quantum communications, and it can help to enhance the security of quantum communication and the performance of quantum repeaters.

In this Ph.D Thesis I presented an algorithmic solution to the problem of superactivation of quantum channels. I discussed a new field of Quantum Information Theory and investigated an algorithmic framework to the problem of superactivation of asymptotic quantum capacity and the classical zero-error capacity of quantum channels. The currently known theoretical result on superactivation is only one possible solution and the problem set of superactivation can be extended to a larger set of quantum channels. My proposed algorithmic solution can be the key to finding other possible channel models and channel parameter domains in future, with possible combinations being proved by theory. In this work we developed an efficient algorithmic solution for the study of large set of input states and channel combinations to discover superactive quantum channel combinations.

My information geometric algorithms were demonstrated for the superactivation of asymptotic quantum capacity and the classical zero-error capacity of quantum channels, without the problem of NP-completeness. The proposed methods can also be extended for the analysis of the superactivation of quantum zero-error capacity of quantum channels. In future work I would like to determine the possibility of superactivation of private classical capacity and entanglement-assisted classical capacity of quantum channels.

# Appendix

Ph.D. Thesis
of

Laszlo Gyongyosi

Thesis Supervisor:
Prof. Dr. Sandor Imre

Budapest, Hungary
2013

# Contents

# List of Publications

## Journal Papers

[P1] **Laszlo Gyongyosi**, Sandor Imre, "Algorithmic Superactivation of Asymptotic Quantum Capacity of Zero-Capacity Quantum Channels", Information Sciences, ELSEVIER, ISSN: 0020-0255; (Impact Factor: 3.28); (2011).

[P2] **Laszlo Gyongyosi**, Sandor Imre: Long-Distance Quantum Communications with Superactivated Gaussian Optical Quantum Channels, SPIE Optical Engineering, ISSN: 0091-3286, E-ISSN: 1560-2303; USA, 2011. (Impact Factor: 0.9)

[P3] **Laszlo Gyongyosi**, Sandor Imre: Information Geometric Superactivation of Classical Zero-Error Capacity of Quantum Channels, Progress in Informatics, Quantum Information Technology, Quantum Information Science Theory Group, National Institute of Informatics, Tokyo, Japan, Print ISSN : 1349-8614, Online ISSN : 1349-8606; 2011.

[P4] **Laszlo Gyongyosi**, Sandor Imre: Algorithmical Analysis of Information-Theoretic Aspects of Secure Communication over Optical-Fiber Quantum Channels, Journal of Optical and Fiber Communications Research, Springer New York, ISSN 1867-3007 (Print) 1619-8638 (Online); 2010.

[P5] Sandor Imre, **Laszlo Gyongyosi**: Quantum-assisted and Quantum-based Solutions in Wireless Systems, with Lajos Hanzo, Harald Haas, Dominic O'Brien and Markus Rupp, in: "Wireless Myths, Realities and Futures: From 3G/4G to Optical and Quantum Wireless", Proceedings of the IEEE, Issue: Special Centennial Issue (100th Year Anniversary Celebration Volume of the Proceedings of the IEEE), ISSN: 0018-9219. (Impact Factor: 5.151, IEEE Highest), 2012.

# Book

# Book Chapters

# Conference Papers

[C1] **Laszlo Gyongyosi**, Sandor Imre: Superactivated Quantum Repeaters, Quantum Information Processing 2012 (QIP2012), Dec. 2011, University of Montreal, Quebec, Canada (**PHD GRANT AWARD OF QIP2012, UNIVERSITY OF MONTREAL**).

[C2] **Laszlo Gyongyosi**, Sandor Imre: On-the-Fly Quantum Error-Correction for Space-Earth Quantum Communication Channels, First NASA Quantum Future Technologies Conference (QFT 2012), Jan. 2012, Quantum Laboratory, Center for Applied Physics, NASA Ames Research Center, Moffett Field, California, USA.

[C3] **Laszlo Gyongyosi**, Sandor Imre: Pilot Quantum Error-Correction for Noisy Quantum Channels, Second International Conference on Quantum Error Correction (QEC11), Dec. 2011, University of Southern California, Los Angeles, USA (**PHD GRANT AWARD OF QEC2011, UNIVERSITY OF SOUTHERN CALIFORNIA**).

[C4] **Laszlo Gyongyosi**, Sandor Imre: Noisy Gaussian Quantum Channels with Superactivated Zero-Error Quantum Capacity, Frontiers in Quantum Information, Computing & Communication (QICC)-2011 Meeting, 'Quantum Systems to Qubits, Optics & Semiconductors', Sept. 2011, Massachusetts Institute of Technology (MIT), Cambridge, Massachusetts, USA.

[C5] **Laszlo Gyongyosi**, Sandor Imre: Classical and Quantum Communication with Superactivated Quantum Channels, PhD Workshop, TAMOP-4.2.2/B-10/1-2010-0009, March. 2012, Doctoral School on Computer Science and Information Technologies, Budapest University of Technology and Economics.

[C6] **Laszlo Gyongyosi**, Sandor Imre: Zero-Error Transmission of Classical Information over Superactivated Optical Quantum Channels, International Conference on Quantum, Atomic, Molecular and Plasma Physics, Section on Quantum Information and Computing, Sept. 2011., Clarendon Laboratory, University of Oxford, Oxford, United Kingdom.

[C7] **Laszlo Gyongyosi**, Sandor Imre: High Performance Quantum Repeaters with Superactivated Gaussian Quantum Channels, IONS North America, IONS-NA-3, (Stanford Optical Society, Stanford Photonics Research Center), Oct. 2011, Stanford University, Stanford, California, USA (**PHD GRANT AWARD OF STANFORD UNIVERSITY, STANFORD OPTICAL SOCIETY**).

[C8] **Laszlo Gyongyosi**, Sandor Imre: Perfect Quantum Communication with Very Noisy Gaussian Optical Fiber Channels, Frontiers in Optics

(FiO) 2011, OSA's 95th Annual Meeting, Section on Quantum Computation and Communication, (American Physical Society, Optical Society of America) Oct. 2011, San Jose, California, USA.

[C9] **Laszlo Gyongyosi**, Sandor Imre: Efficient Quantum Repeaters without Entanglement Purification, International Conference on Quantum Information (ICQI) 2011, (The Optical Society of America (OSA), University of Rochester), Jun. 2011, University of Ottawa, Ottawa, Canada.

[C10] **Laszlo Gyongyosi**, Sandor Imre: Channel Capacity Restoration of Noisy Optical Quantum Channels, ICOAA '11 Conference, Section on Optical Quantum Communications, Febr. 2011, University of Cambridge, Cambridge, United Kingdom.

[C11] **Laszlo Gyongyosi**, Sandor Imre: Informational Geometric Analysis of Superactivation of Zero-Capacity Optical Quantum Channels, SPIE Photonics West OPTO 2011, Advanced Quantum and Optoelectronic Applications, "Advances in Photonics of Quantum Computing, Memory, and Communication IV", Section on Quantum Communication, Jan. 2011, The Moscone Center, San Francisco, California, USA.

[C12] **Laszlo Gyongyosi**, Sandor Imre: Algorithmic Solution to Superactivation of Zero-Capacity Optical Quantum Channels, Photonics Global Conference (PGC) 2010, Nanyang Technological University, IEEE Photonics Society, 2010, Suntec City, Singapore.

[C13] **Laszlo Gyongyosi**, Sandor Imre: Information Geometric Superactivation of Zero-Capacity Quantum Channels, The Second International Conference on Quantum Information and Technology - New Trends in Quantum Information Technology (ICQIT2010), Quantum Information Science Theory Group (QIST), National Institute of Informatics (NII), National Institute of Information and Communications Technology (NICT), 2010, QIST, NII, Tokyo, Japan.

[C14] **Laszlo Gyongyosi**, Sandor Imre: Capacity Recovery of Useless Photonic Quantum Communication Channels, ALS Conference, 2010, Lawrence Berkeley National Laboratory (Berkeley Lab), University of California, Berkeley, California, USA.

[C15] **Laszlo Gyongyosi**, Sandor Imre: Method for Discovering of Superactive Zero-Capacity Optical Quantum Channels, IONS-NA Conference, 2010, University of Arizona, Tucson (Arizona), USA. (**PHD GRANT AWARD OF UNIVERSITY OF ARIZONA, USA**).

[C16] **Laszlo Gyongyosi**, Sandor Imre: Information Geometrical Solution to Additivity of Non-Unital Quantum Channels, QCMC 2010, 10th Quantum Communication, Measurement & Computing Conference, Section on Quantum Computing and Quantum Information Theory (Centre for

# Conference Papers (accepted, in press)

# Awards

- SPIE Photonics West OPTO 2013 Award, Advances in Photonics of Quantum Computing, Memory, and Communication, San Francisco, California, USA
- Incubic/Milton Chang Award, The Optical Society of America, Rochester, New York, USA, 2012.
- BUTE PhD Researcher Fellowship 2012, Budapest University of Technology and Economics
- PhD Student Grant Award of Columbia University, New York, USA, 2012.
- PhD Student Grant Award of QCRYPT 2012, The 2nd Annual Conference on Quantum Cryptography (Centre for Quantum Technologies), National University of Singapore, Singapore, 2012.
- PhD Student Grant Award of APS DAMOP 2012 (DAMOP12), APS Division of Atomic, Molecular, and Optical Physics, American Physical Society (APS), California, USA, 2012.
- PhD Student Grant Award of Quantum Information Processing 2012 (QIP2012), University of Montreal, Canada

- PhD Student Grant Award of The Second International Conference on Quantum Error Correction (QEC2011), University of Southern California (USC), 2011, USA
- PhD Student Grant Award of Stanford University, Stanford Optical Society, 2011, USA
- PhD Candidate Scholarship 2011, Faculty of Electrical Engineering and Informatics, Budapest University of Technology and Economics
- BUTE PhD Researcher Fellowship 2011, Budapest University of Technology and Economics
- PhD Student Grant Award of University of Arizona, Optical Society of America (OSA), 2010, USA
- BUTE PhD Researcher Fellowship 2010, Budapest University of Technology and Economics
- SANDOR CSIBI PhD Researcher Scholarship 2010, Pro Progressio, Faculty of Electrical Engineering and Informatics, Budapest University of Technology and Economics
- BEST PAPER AWARD 2010 from Harvard University, Cambridge, USA
- BEST PAPER AWARD 2009, FUTURE COMPUTING, The First International Conference on Future Computational Technologies and Applications, 2009
- BEST PAPER AWARD 2009 - "POLLAK-VIRAG" - from the Scientific Association for Infocommunication (Hungary)
- Republican Scholarship, Budapest University of Technology and Economics
- Professional Scholarship, Faculty of Electrical Engineering and Informatics, BUTE (Level A, highest category)
- University Professional Scholarship, BUTE (Level A, highest category)

# Appendix A

# Preliminaries

In Appendix A, we give a brief overview of quantum mechanics, and we introduce the basic definitions of Quantum Information Processing.

## A.1 Brief Overview of Quantum Information Processing

In Quantum Information Processing, the logical values of classical bits are replaced by state vectors $|0\rangle$ and $|1\rangle$, - called the Dirac notation. Contrary to classical bits, a qubit $|\psi\rangle$ can also be in a linear combination of basis vectors $|0\rangle$ and $|1\rangle$. The state of a qubit can be expressed as

$$|\psi\rangle = \alpha|0\rangle + \beta|1\rangle, \tag{A.1}$$

where $\alpha$ and $\beta$ are complex numbers, which is also called the superposition of the basis vectors, with probability amplitudes $\alpha$ and $\beta$. A qubit $|\psi\rangle$ is a vector in a two-dimensional complex space, where the basis vectors $|0\rangle$ and $|1\rangle$ form an orthonormal basis.

The orthonormal basis $\{|0\rangle, |1\rangle\}$ forms the computational basis, in Fig. A.1 we illustrate the computational basis for the case where the probability amplitudes are real [Imre05].

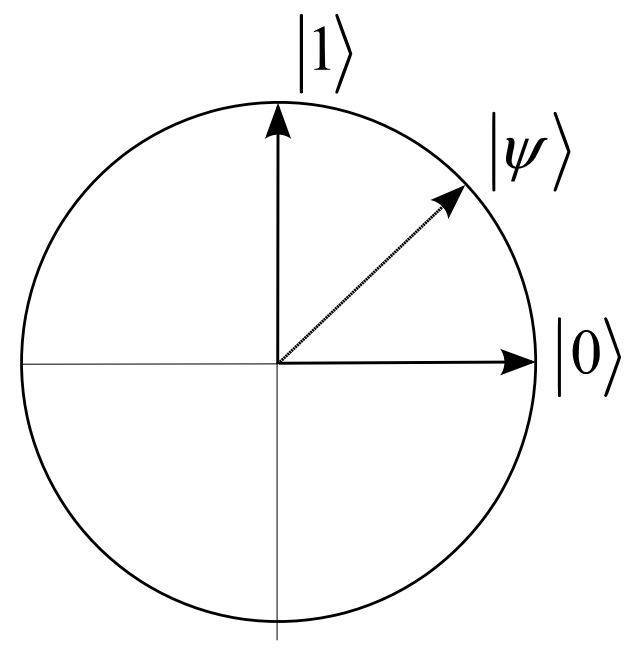

**Fig. A.1.** Computational basis and general representation of a qubit in superposition state.

The vectors or states $|0\rangle$ and $|1\rangle$ can be expressed in matrix representation by

$$|0\rangle = \begin{bmatrix} 1 \\ 0 \end{bmatrix} \text{ and } |1\rangle = \begin{bmatrix} 0 \\ 1 \end{bmatrix}. \tag{A.2}$$

If $|\alpha|^2$ and $|\beta|^2$ are the probabilities, and the number of possible outputs is only two, then for $|\psi\rangle = \alpha|0\rangle + \beta|1\rangle$ we have $|\alpha|^2 + |\beta|^2 = 1$, and the norm of $|\psi\rangle$ is

$$\||\psi\rangle\| = \sqrt{|\alpha|^2 + |\beta|^2} = 1. \tag{A.3}$$

The most general transformation of $\||\psi\rangle\|$ that respects this constraint is a linear transformation $U$ that takes unit vectors to unit vectors.

A *unitary* transformation can be defined as

$$U^\dagger U = UU^\dagger = I, \tag{A.4}$$

where $U^\dagger = \left(U^*\right)^T$, hence the adjoint is equal to the transpose of complex conjugate, and $I$ is the identity matrix.

The tensor product has an important role in quantum computation, here we quickly introduce the concept of tensor product. If we have complex vector spaces $V$ and $W$ of dimensions $m$ and $n$, then the tensor product of $V \otimes W$ is an $mn$ dimensional vector space. The tensor product is non-commutative, thus the notation preserves the ordering. The concept of a linear operator also can be defined over the vector spaces. If we have two linear operators $A$ and $B$, defined on the vector spaces $V$ and $W$, then the linear operator $A \otimes B$ on $V \otimes W$ can be defined as $\left(A \otimes B\right)\left(\left|v\right\rangle \otimes \left|w\right\rangle\right) = A\left|v\right\rangle \otimes B\left|w\right\rangle$, where $\left|v\right\rangle \in V$ and $\left|w\right\rangle \in W$. In matrix representation, $A \otimes B$ can be expressed as

$$A \otimes B = \begin{bmatrix} A_{11}B & \dots & A_{1m}B \\ \vdots & \ddots & \vdots \\ A_{m1}B & \dots & A_{mm}B \end{bmatrix}, \tag{A.5}$$

where $A$ is an $m \times m$ matrix, and $B$ is an $n \times n$ matrix, hence $A \otimes B$ has dimension $mn \times mn$. The state $\left|\psi\right\rangle$ of an $n$-qubit quantum register is a superposition of the $2^n$ states $\left|0\right\rangle, \left|1\right\rangle, \dots, \left|2^n - 1\right\rangle$, thus

$$\left|\psi\right\rangle = \sum_{i=0}^{2^n-1} \alpha_i \left|i\right\rangle, \tag{A.6}$$

with amplitudes $\alpha_i$ constrained by

$$\sum_{i=0}^{2^n-1} \left|\alpha_i\right|^2 = 1. \tag{A.7}$$

The state of an $n$-qubit length quantum register is a vector in a $2^n$-dimensional complex vector space, hence if the number of the qubits in the quantum register increases linearly, the dimension of the vector space increases exponentially.

A complex vector space $V$ is a Hilbert space $\mathcal{H}$ if there is an *inner product*

$$\langle\psi|\varphi\rangle \tag{A.8}$$

with $x,y \in \mathbb{C}$ and $|\varphi\rangle,|\psi\rangle,|u\rangle,|v\rangle \in V$ satisfying the rules $\langle\psi|\varphi\rangle = \langle\varphi|\psi\rangle^*$, $\langle\varphi|(a|v\rangle + b|v\rangle)\rangle = a\langle\varphi|u\rangle + b\langle\varphi|v\rangle$, and $\langle\varphi|\varphi\rangle > 0$ if $|\varphi\rangle \neq 0$. If we have vectors $|\varphi\rangle = a|0\rangle + b|1\rangle$ and $|\psi\rangle = c|0\rangle + d|1\rangle$, then the inner product in matrix representation can be expressed as

$$\langle\varphi|\psi\rangle = \begin{bmatrix} a^* & b^* \end{bmatrix}\begin{bmatrix} c \\ d \end{bmatrix} = a^*c + b^*d. \tag{A.9}$$

The norm of the vector $|\varphi\rangle$ can be expressed as $\||\varphi\rangle\| = \sqrt{\langle\varphi|\varphi\rangle}$, and the dual of the vector $|\varphi\rangle$ is denoted by $\langle\varphi|$. The *dual* is a linear operator from the vector space to the complex numbers, defined as $\langle\varphi|(|v\rangle) = \langle\varphi|v\rangle$. The outer product between two vectors $|\varphi\rangle$ and $|\psi\rangle$ can be defined as

$$|\psi\rangle\langle\varphi|, \tag{A.10}$$

satisfying $(|\psi\rangle\langle\varphi|)|v\rangle = |\psi\rangle\langle\varphi|v\rangle$. The matrix of the outer product $|\psi\rangle\langle\varphi|$ is obtained by usual matrix multiplication of a column matrix by a row matrix, however the matrix multiplication can be replaced by tensor product, since:

$$|\varphi\rangle\langle\psi| = |\varphi\rangle \otimes \langle\psi|. \tag{A.11}$$

If we have vectors $|\varphi\rangle = a|0\rangle + b|1\rangle$ and $|\psi\rangle = c|0\rangle + d|1\rangle$, the outer product in matrix representation can be expressed as

$$|\varphi\rangle\langle\psi| = \begin{bmatrix} a \\ b \end{bmatrix}\begin{bmatrix} c^* & d^* \end{bmatrix} = \begin{bmatrix} ac^* & ad^* \\ bc^* & bd^* \end{bmatrix}. \tag{A.12}$$

In Fig. A.2. we illustrate the general description of a unitary transformation on an $n$-length quantum state, where the input state $\left|\psi_i\right\rangle$ is either $\left|0\right\rangle$ or $\left|1\right\rangle$, generally. After the application of a unitary transformation $U$ on the input states, the state of the quantum register can be given by a state vector $\left|\psi\right\rangle$.

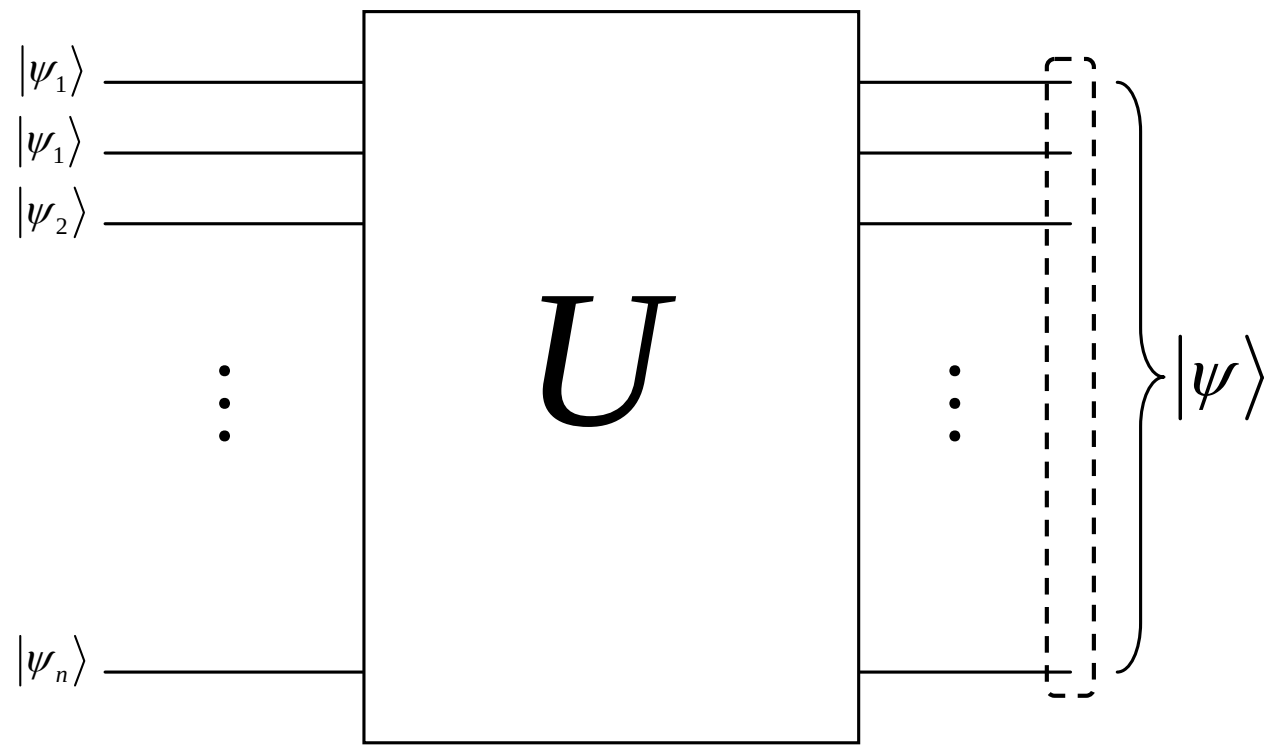

**Fig. A.2.** General sketch of a unitary transformation on an $n$-length quantum register.

The unitary operator $U$ is a $2^n \times 2^n$ matrix, with, in principle, an infinite number of possible operators. The result of the measurement of the state $\left|\psi\right\rangle$ results in zeros and ones that form the final result of the quantum computation, based on the $n$-length qubit string stores in the quantum register.

For a unitary transformation $U$, the following property holds:

$$(U^T)^* = U^{-1}, \tag{A.13}$$

where $T$ denotes transposition and * denotes complex conjugation. The inverse transformation of $U$ also can be expressed by the adjugate $U^\dagger$, which is equal to $U^{-1}$. One of the most standard quantum gates is the Controlled-NOT (CNOT) gate.

The CNOT gate is a very important gate in quantum computation, since from the one qubit quantum gates and the CNOT gates every unitary transformation can be expressed, hence these gates are universal. This gate is a two-qubit gate and it contains two qubits, called the control and the target qubit. If the control qubit is $|1\rangle$, then the gate negates the second qubit—called the target qubit.

The general CNOT gate is illustrated in Fig. A.3.

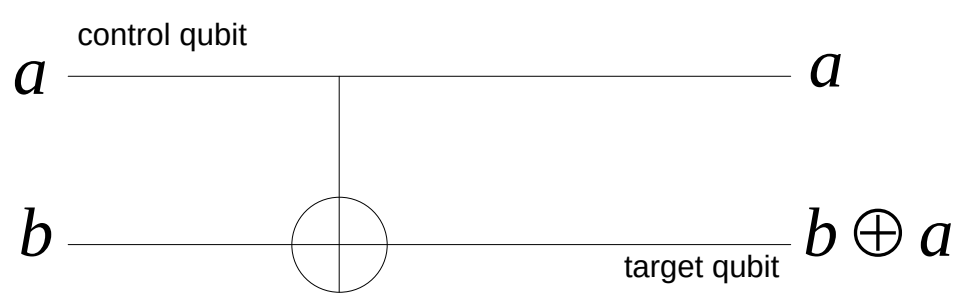

**Fig. A.3.** The Controlled-NOT (CNOT) gate.

As can be verified, the quantum CNOT gate can be regarded as the generalization of the classical XOR transformation, hence $\mathbf{CNOT}|a,\ b\rangle = |a,\ b \oplus a\rangle$, which unitary transformation can be expressed in matrix form as follows:

$$\mathbf{CNOT} = \begin{bmatrix} 1 & 0 & 0 & 0 \\ 0 & 1 & 0 & 0 \\ 0 & 0 & 0 & 1 \\ 0 & 0 & 1 & 0 \end{bmatrix}. \tag{A.14}$$

The controlled behaviour of the CNOT gate can be extended to every unitary transformation, and the generalized control quantum gate can be defined.

In Fig. A.4, we show a controlled $U$ transformation, the $U$ transformation is applied to the target qubit $b$ only if the control qubit $a$ is in high logical state. We note that the CNOT gate cannot be used to copy a quantum state, while the classical XOR gate can be applied to copy a classical bit.

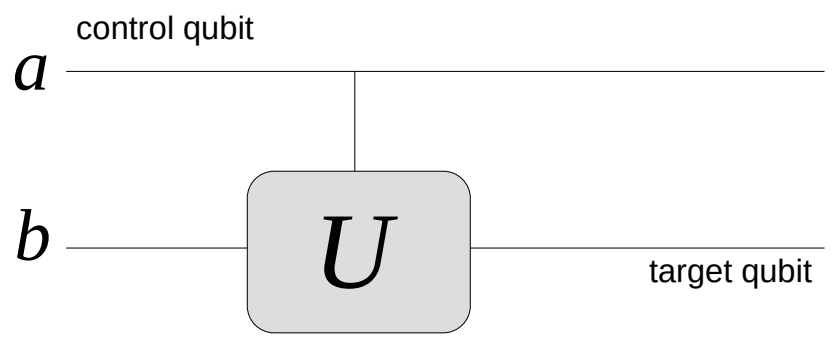

**Fig. A.4.** A controlled-$U$ gate.

As can be seen easily, for the CNOT gate, this unitary transformation $U$ is equal to the NOT-transformation, hence the CNOT gate is a controlled-$X$ gate, actually. We can also define the inverse of this transformation as the controlled-$U^\dagger$ transformation, as follows:

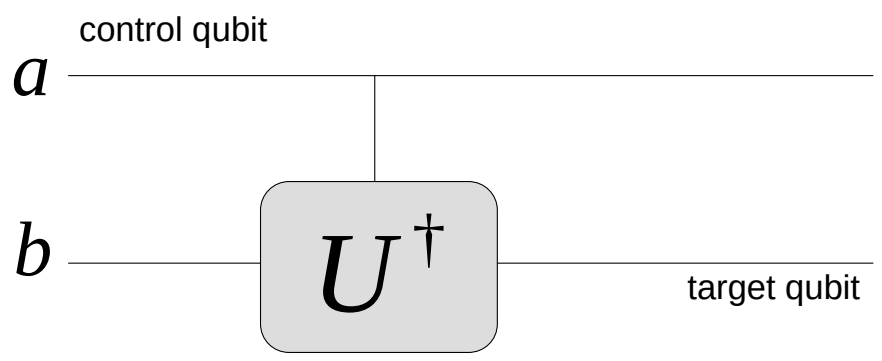

**Fig. A.5.** A controlled inverse $U$ gate.

The $M$ measurement operator converts the quantum information to classical, since after the measurement of a quantum state, the quantum information which is encoded in the quantum state becomes classical, and can be expressed as a logical 0 or 1.

The general measurement circuit is illustrated in Fig. A.6.

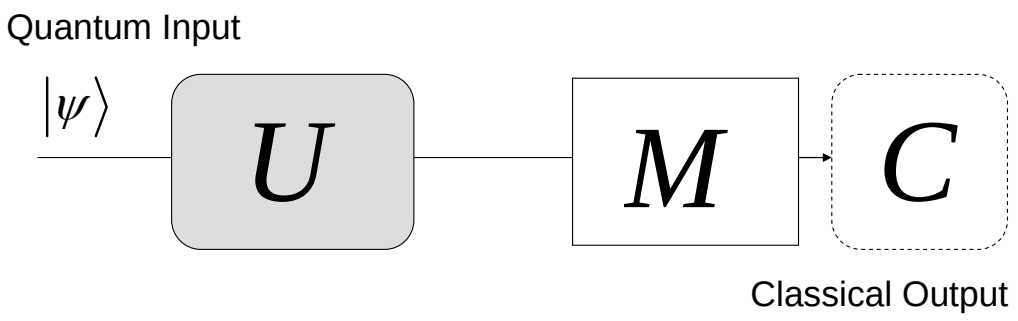

**Fig. A.6.** The measurement of quantum information. The $M$ measurement converts the quantum information to classical.

If we measure the quantum state $\left|\psi\right\rangle = \alpha\left|0\right\rangle + \beta\left|1\right\rangle$, then the output will be $M = 0$ with probability $\left|\alpha\right|^2$ or $M = 1$ with probability $\left|\beta\right|^2$.

For the general case, if we measure the $n$-length quantum register $\left|\psi\right\rangle = \sum_{i=0}^{2^n-1} \alpha_i \left|i\right\rangle$, with possible states $\left|0\right\rangle, \left|1\right\rangle, \ldots, \left|2^n - 1\right\rangle$, then the quantum measurement can be described as a set of $\left\{M_m\right\}$ of linear operators. The number of the possible

outcomes is $n$, hence the number $m$ of possible measurement operators is between $1 \leq m \leq n$.

If we measure the quantum register in state $\left|\psi\right\rangle = \sum_{i=0}^{2^n-1} \alpha_i \left|i\right\rangle$, then the outcome $i$ has a probability of

$$\Pr\left(i\right) = \left\langle\psi\middle|M_i^\dagger M_i\middle|\psi\right\rangle. \tag{A.15}$$

The sum of the probabilities of all possible outcomes is

$$\sum_{i=1}^{m} \Pr\left(i\right) = \sum_{i=1}^{m} \left\langle\psi\middle|M_i^\dagger M_i\middle|\psi\right\rangle = 1, \tag{A.16}$$

according to the completeness of the measurement operators, since

$$\sum_{i=1}^{m} M_i^\dagger M_i = I. \tag{A.17}$$

After the measurement of outcome $i$, the state of the quantum register collapses to

$$\left|\psi'\right\rangle = \frac{M_i\left|\psi\right\rangle}{\sqrt{\left\langle\psi\middle|M_i^\dagger M_i\middle|\psi\right\rangle}} = \frac{M_i\left|\psi\right\rangle}{\sqrt{\Pr\left(i\right)}}. \tag{A.18}$$

Using the previous example, if we have single quantum state $\left|\psi\right\rangle = \alpha\left|0\right\rangle + \beta\left|1\right\rangle$, then the measurement operators can be defined as

$$M_0 = \left|0\right\rangle\left\langle0\right| \text{ and } M_0 = \left|1\right\rangle\left\langle1\right|, \tag{A.19}$$

since the unknown qubit is defined in the orthonormal basis of $\left|0\right\rangle$ and $\left|1\right\rangle$.

# Appendix B

# Quantum Information Theory

## B.1 Communication over a Quantum Channel

The transmission of information through classical channels and quantum channels differs in many ways. If we would like to describe the process of information transmission through a quantum communication channel, we have to introduce the three main phases of quantum communication. In the first phase, the sender, Alice, has to encode her information to compensate the noise of the channel (i.e., for error correction), according to properties of the physical channel - this step is called *channel coding.* After the sender has encoded the information into the appropriate form, it has to be put on the quantum channel, which transforms it according to its channel map - this second phase is called the *channel evolution.* The quantum channel conveys the quantum state to the receiver; however this state is still a superposed and probably *mixed* (according to the noise of the channel) quantum state. To extract the information which is encoded in the state, the receiver has to make a measurement - this *measurement process* (with the error correction procedure) is the third phase of the communication over a quantum channel.

In Fig. B.1, we illustrate the channel coding phase. In case of transmission of classical information over a noisy quantum channel, Alice encodes her information into a physical attribute of a physical particle, such as the spin of the particles. For example, in the case of an electron or a half-spin particle, this can be an axis spin.

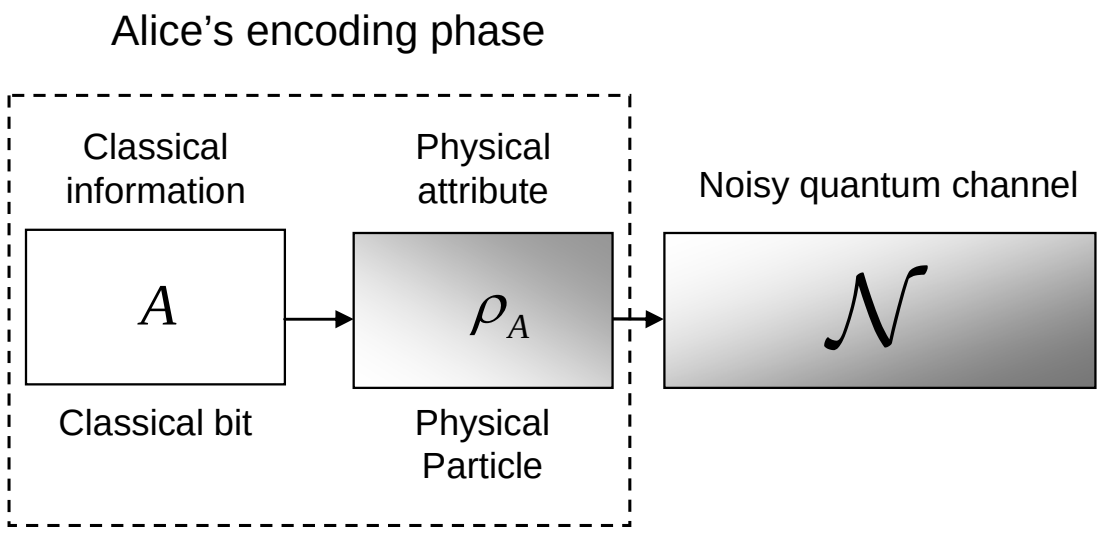

**Fig. B.1.** The channel coding phase.

The channel transformation represents the noise of the quantum channel. Physically, the quantum channel is the medium, which moves the particle from the sender to the receiver. The noise disturbs the state of the particle, in the case of a half-spin particle, it causes spin precession. The channel evolution phase is illustrated in Fig. B.2.

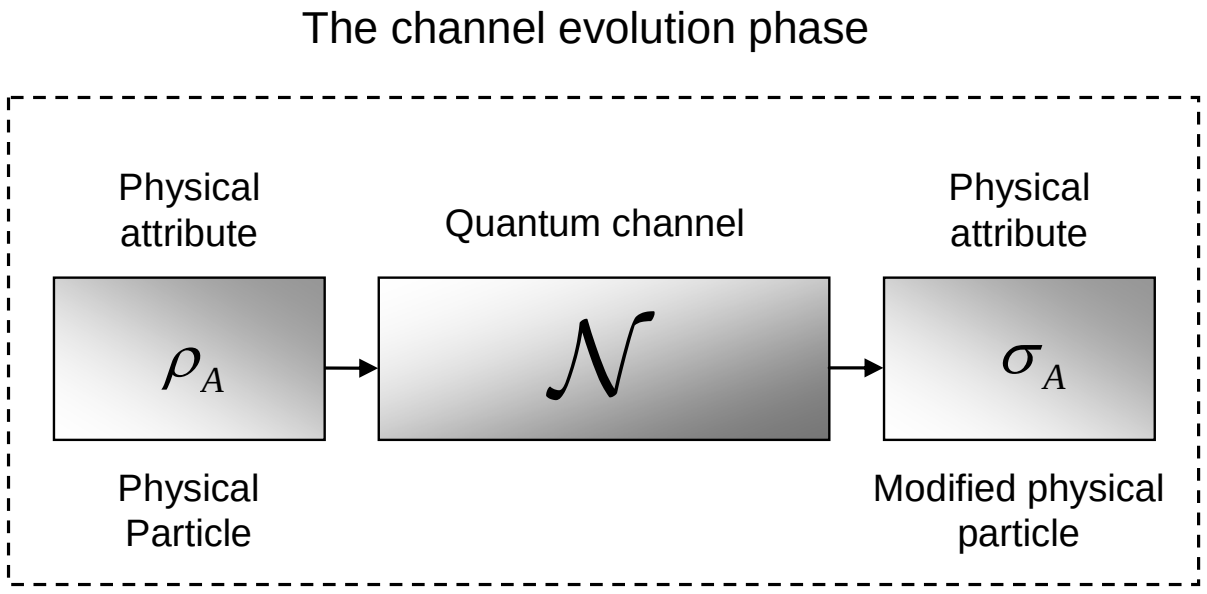

**Fig. B.2.** The channel evolution phase.

Finally, the measurement process responsible for the decoding and the extraction of the encoded information. The previous phase determines the success probability of the recovery of the original information. If the channel is completely noisy, then the receiver will get a maximally mixed quantum state. The output of the measurement of a maximally mixed state is completely undeterministic: it tells us nothing about the original information encoded by the sender. On the other hand, if the quantum channel is completely noiseless, then the information which was encoded by the sender can be recovered with probability 1: the result of the measurement will be completely deterministic and completely corre-lated with the original message. In practice, a quantum channel realizes a map which is in between these two extreme cases. A general quantum channel transforms the original pure quantum state into a mixed quantum state, - but not into a maximally mixed state - which makes it possible to recover the original message with a high - or low - probability, depending on the level of the noise of the quantum channel.

The measurement phase is illustrated in Fig. B.3.

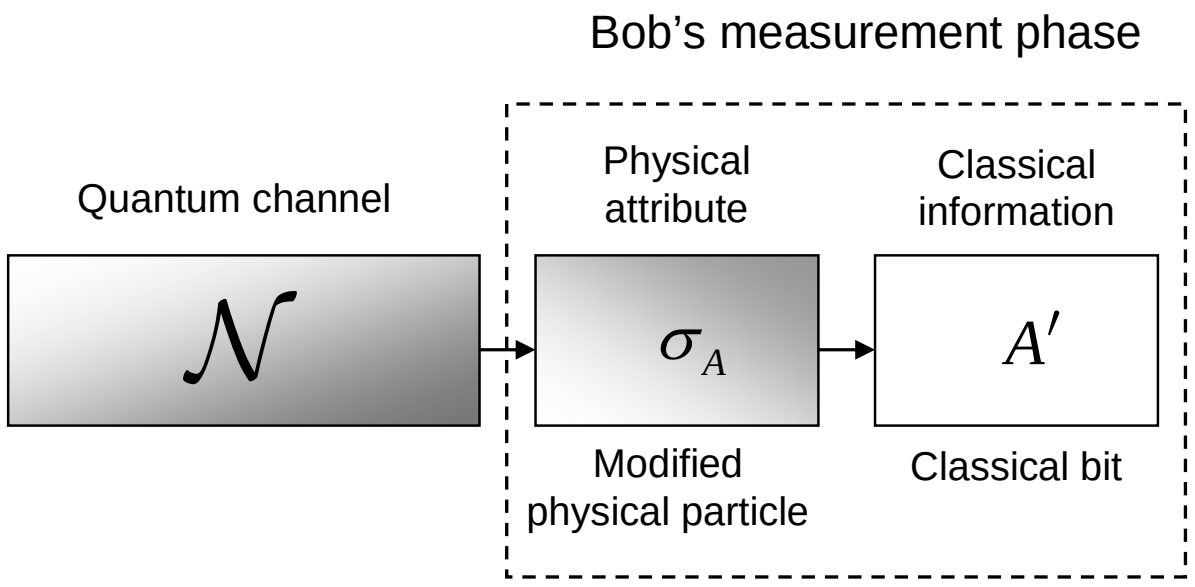

**Fig. B.3.** The measurement process.

Quantum communication channels can be divided into many different classes. As it can be found in the supplementary information of the Appendix, the different channel models modify the sent qubits in different ways.

# B.2 Interaction with the Environment

According to the noise $\mathcal{N}$ of quantum channel, Alice's sent pure quantum state $\rho_{in}$ becomes a mixed state, thus Bob will receive a mixed state denoted by $\rho_{out}$.

As shown in Fig. B.4, the information transmission through the quantum channel $\mathcal{N}$ is defined by the $\rho_{in}$ input quantum state and the initial state of the environment $\rho_E = |0\rangle\langle 0|$. In the initial phase, the environment is assumed to be in the pure state $|0\rangle$. The system state which consist of the input quantum state $\rho_{in}$ and the environment $\rho_E = |0\rangle\langle 0|$, is called the composite state $\rho_{in} \otimes \rho_E$.

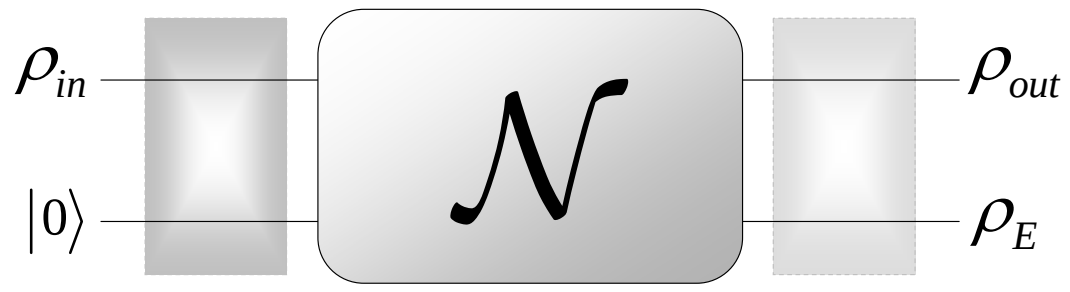

**Fig. B.4.** The general model of transmission of information over a noisy quantum channel.

If the quantum channel $\mathcal{N}$ is used for information transmission, then the state of the composite system changes unitarily, as follows:

$$U\left(\rho_{in} \otimes \rho_E\right)U^{\dagger}, \tag{B.1}$$

where $U$ is a unitary transformation and $U^{\dagger}U = I$.

After the quantum state transmitted the quantum channel $\mathcal{N}$, the $\rho_{out}$ output state can be expressed as:

$$\mathcal{N}\left(\rho_{in}\right) = \rho_{out} = Tr_E\left[U\left(\rho_{in} \otimes \rho_E\right)U^{\dagger}\right], \tag{B.2}$$

where $Tr_E$ traces out the environment $E$ from the joint state. Assuming the environment $E$ in the pure state $\left|0\right\rangle$, $\rho_E = \left|0\right\rangle\left\langle 0\right|$, the $\mathcal{N}\left(\rho_{in}\right)$ noisy evolution of the channel $\mathcal{N}$ can be expressed as:

$$\mathcal{N}\left(\rho_{in}\right) = \rho_{out} = Tr_E U\rho_{in} \otimes \left|0\right\rangle\left\langle 0\right|U^{\dagger}, \tag{B.3}$$

while the post-state $\rho_E$ of the environment after the transmission is

$$\rho_E = Tr_B U\rho_{in} \otimes \left|0\right\rangle\left\langle 0\right|U^{\dagger}, \tag{B.4}$$

where $Tr_B$ traces out the output system $B$. In general, the $i$-th input quantum state $\rho_i$ is prepared with probability $p_i$, which describes the ensemble $\left\{p_i, \rho_i\right\}$. The average of the *input* of the quantum states is expressed as

$$\sigma_{in} = \sum_i p_i\rho_i, \tag{B.5}$$

The average (or the mixture) of the *output* of the quantum channel is denoted by

$$\sigma_{out} = \mathcal{N}\left(\sigma_{in}\right) = \sum_{i} p_i \mathcal{N}\left(\rho_i\right). \tag{B.6}$$

The classical information which can be transmitted through a noisy quantum channel $\mathcal{N}$ can be expressed by the $\chi$ Holevo quantity. The Holevo quantity describes the amount of information, which can be extracted from the output about the input state. We note, this information is also referred as *accessible information* in the literature.

# B.3 Channel System Description

If we are interested in the origin of noise (randomness) in the quantum channel the model should be refined in the following way: Alice's register *X*, the purification state *P*, channel input *A*, channel output *B*, and the environment state *E*. The input system *A* is described by a quantum system $\rho_x$, which occurs on the input with probability $p_X\left(x\right)$. They together form an ensemble denoted by $\left\{p_X\left(x\right), \rho_x\right\}_{x \in X}$, where *x* is a classical variable from the register *X*. In the preparation process, Alice generates pure states $\rho_x$ according to random variable *x*, i.e., the input density operator can be expressed as $\rho_x = \left|x\right\rangle\left\langle x\right|$, where the classical states $\left\{\left|x\right\rangle\right\}_{x \in X}$ form an orthonormal basis.

According to the elements of Alice's register *X*, the input system can be characterized by the quantum system

$$\rho_A = \sum_{x \in X} p_X\left(x\right)\rho_x = \sum_{x \in X} p_X\left(x\right)\left|x\right\rangle\left\langle x\right|. \tag{B.7}$$

The system description is illustrated in Fig. B.5.

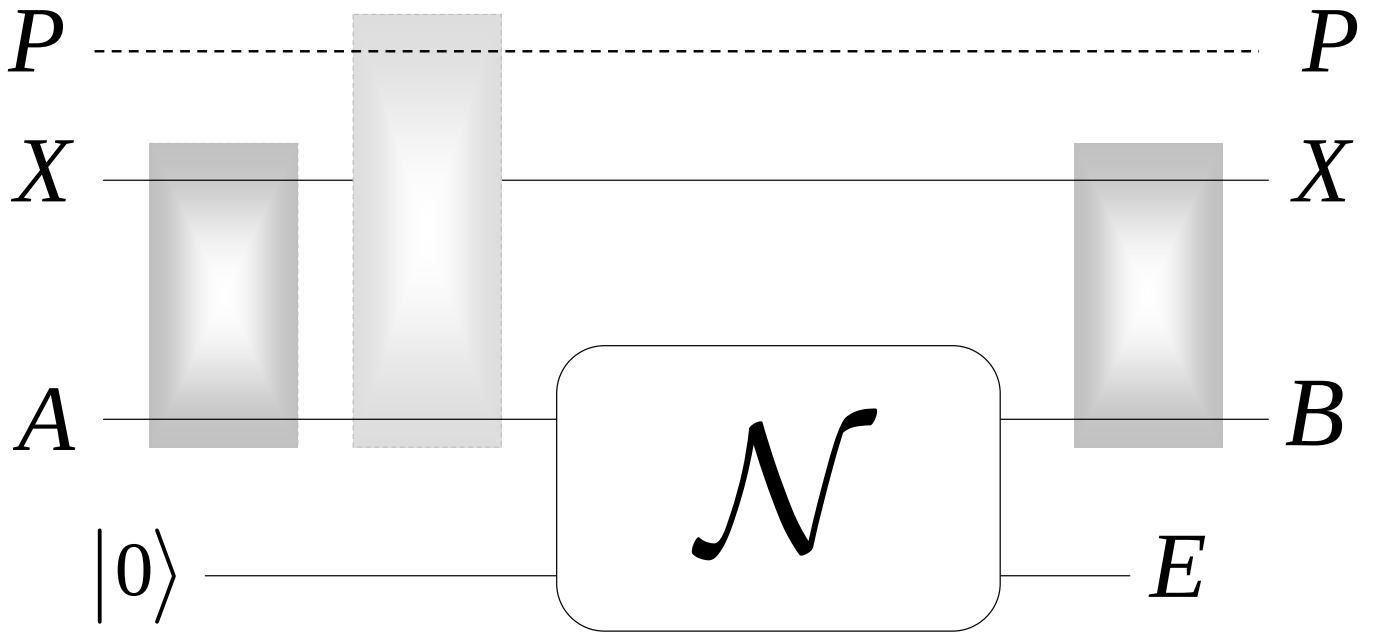

**Fig. B.5.** Detailed model of a quantum communication channel exposing the interaction with the environment. Alice's register is denoted by $X$, the input system is $A$ while $P$ is the purification state. The environment of the channel is denoted by $E$, the output of the channel is $B$. The quantum channel has positive classical capacity if and only if the channel output system $B$ will be correlated with Alice's register $X$.

The system state $\rho_x$ with the corresponding probability distribution $p_X(x)$ can be indentified by a set of measurement operators $M = \{|x\rangle\langle x|\}_{x\in X}$. If the density operators $\rho_x$ in $\rho_A$ are mixed, the probability distribution $p_X(x)$ and the classical variable $x$ from the register $X$ cannot be indentified by the measurement operators $M = \{|x\rangle\langle x|\}_{x\in X}$, since the system state $\rho_x$ is assumed to be a mixed or in a non-orthonormal state.

Alice's register $X$ and the quantum system $A$ can be viewed as a tensor product system as

$$\left\{p_X(x), |x\rangle\langle x|_X \otimes \rho_A^x\right\}_{x\in X}, \tag{B.8}$$

where the quantum state $|x\rangle$ is correlated with the quantum system $\rho_x$, using orthonormal basis $\{|x\rangle\}_{x\in X}$.

Alice's register $X$ represents a classical variable, the channel input system is generated corresponding to the register $X$ in the form of a quantum state, and it is described by the density operator $\rho_A^x$. The input system $A$ with respect to the register $X$, is described by the density operator

$$\rho_{XA} = \sum_{x \in X} p_X(x) |x\rangle\langle x|_X \otimes \rho_A^x, \tag{B.9}$$

where $\rho_A^x = |\psi_x\rangle\langle\psi_x|_A$ is the density matrix representation of Alice's input state $|\psi_x\rangle_A$.

# B.4 Purification

The *purification* gives us a new viewpoint on the noise of the quantum channel. Assuming Alice's side $A$ and Alice's register $X$, the spectral decomposition of the density operator $\rho_A$ can be expressed as

$$\rho_A = \sum_x p_X(x) |x\rangle\langle x|_A, \tag{B.10}$$

where $p_X(x)$ is the probability of variable $x$ in Alice's register $X$.

The $\{p_X(x), |x\rangle\}$ together is called an ensemble, where $|x\rangle$ is a quantum state according to classical variable $x$. Using the set of orthonormal basis vectors $\{|x\rangle_P\}_{x \in X}$ of the purification system $P$, the purification of (B.10) can be given in the following way:

$$|\varphi\rangle_{PA} = \sum_x \sqrt{p_X(x)} |x\rangle_P |x\rangle_A. \tag{B.11}$$

From the purified system state $|\varphi\rangle_{PA}$, the original system state $\rho_A$ can be expressed with the partial trace operator (see Appendix) $Tr_P(\cdot)$, which operator traces out the purification state (i.e., the environment) from the system

$$\rho_A = Tr_P(|\varphi\rangle\langle\varphi|_{PA}). \tag{B.12}$$

From joint system (B.11) and the purified state (B.12), one can introduce a new definition. The *extension* of $\rho_A$ can be given as

$$\rho_A = Tr_P(\omega_{PA}), \tag{B.13}$$

where $\omega_{PA}$ is the joint system of purification state $P$ and channel input $A$, which represents a noisy state.

# B.5 Isometric Extension

*Isometric extension* has utmost importance, because it helps us to understand what happens between the quantum channel and its environment whenever a quantum state is transmitted from Alice to Bob. Since the channel and the environment together form a closed physical system the isometric extension of the quantum channel $\mathcal{N}$ is the *unitary representation* of the channel

$$\mathcal{N} : U_{A\to BE}, \tag{B.14}$$

enabling the "one-sender and two-receiver" view: beside Alice the sender, both Bob and the environment of the channel are playing the receivers. In other words, the output of the noisy quantum channel $\mathcal{N}$ can be described only after the environment of the channel is traced out

$$\rho_B = Tr_E\left(U_{A\to BE}\left(\rho_A\right)\right) = \mathcal{N}\left(\rho_A\right). \tag{B.15}$$

# B.6 Kraus Representation

The map of the quantum channel can also be expressed by means of a special tool called the *Kraus Representation*. For a given input system $\rho_A$ and quantum channel $\mathcal{N}$, this representation can be expressed as

$$\mathcal{N}\left(\rho_A\right) = \sum_i N_i \rho_A N_i^\dagger, \tag{B.16}$$

where $N_i$ are the Kraus operators, and $\sum_i N_i^\dagger N_i = I$. The isometric extension of $\mathcal{N}$ by means of the *Kraus Representation* can be expressed as

$$\rho_B = \mathcal{N}\left(\rho_A\right) = \sum_i N_i \rho_A N_i^\dagger \to U_{A\to BE}\left(\rho_A\right) = \sum_i N_i \otimes \left|i\right\rangle_E. \tag{B.17}$$

The action of the quantum channel $\mathcal{N}$ on an operator $|k\rangle\langle l|$, where $\{|k\rangle\}$ form an orthonormal basis also can be given in operator form using the Kraus operator $N_{kl} = \mathcal{N}(|k\rangle\langle l|)$. By exploiting the property $UU^{\dagger} = P_{BE}$, for the input quantum system $\rho_A$

$$\rho_B = U_{A\to BE}(\rho_A) = U\rho_A U^{\dagger} = \left(\sum_i N_i \otimes |i\rangle_E\right)\rho_A\left(\sum_j N_j^{\dagger} \otimes \langle j|_E\right) = \sum_{i,j} N_i\rho_A N_j^{\dagger} \otimes |i\rangle\langle j|_E . \tag{B.18}$$

If we trace out the environment, we get the equivalence of the two representations

$$\rho_B = Tr_E\left(U_{A\to BE}(\rho_A)\right) = \sum_i N_i\rho_A N_i^{\dagger} . \tag{B.19}$$

# B.7 Quantum Conditional Entropy

While the classical conditional entropy function is always takes a non negative value, the *quantum conditional entropy can be negative*. The quantum conditional entropy between quantum systems $A$ and $B$ is given by

$$\mathrm{S}(A|B) = \mathrm{S}(\rho_{AB}) - \mathrm{S}(\rho_B). \tag{B.20}$$

If we have two uncorrelated subsystems $\rho_A$ and $\rho_B$, then the information of the quantum system $\rho_A$ does not contain any information about $\rho_B$, or reversely, thus

$$\mathrm{S}(\rho_{AB}) = \mathrm{S}(\rho_A) + \mathrm{S}(\rho_B), \tag{B.21}$$

hence we get $\mathrm{S}(A|B) = \mathrm{S}(\rho_A)$, and similarly $\mathrm{S}(B|A) = \mathrm{S}(\rho_B)$. The negative property of conditional entropy $\mathrm{S}(A|B)$ can be demonstrated with an *entangled* state, since in this case, the joint quantum entropy of the joint state less than the sum of the von Neumann entropies of its individual components. For a pure entangled state, $\mathrm{S}(\rho_{AB}) = 0$, while $\mathrm{S}(\rho_A) = \mathrm{S}(\rho_B) = 1$ since the two qubits are in *maximally mixed* $\frac{1}{2}I$ state, which is classically totally unimaginable. Thus, in this case

$$\mathrm{S}(A|B) = -\mathrm{S}(\rho_B) \leq 0, \tag{B.22}$$

and

$$\mathrm{S}\left(B|A\right) = -\mathrm{S}\left(\rho_A\right) \leq 0 \text{ and } \mathrm{S}\left(\rho_A\right) = \mathrm{S}\left(\rho_B\right). \tag{B.23}$$

# B.8 Quantum Mutual Information

The classical mutual information $I\left(\cdot\right)$ measures the information correlation between random variables $A$ and $B$. In analogue to classical Information Theory, $I\left(A:B\right)$ can be described by the quantum entropies of individual states and the von Neumann entropy of the joint state as follows:

$$I\left(A:B\right) = \mathrm{S}\left(\rho_A\right) + \mathrm{S}\left(\rho_B\right) - \mathrm{S}\left(\rho_{AB}\right) \geq 0\,, \tag{B.24}$$

i.e., the quantum mutual information is always a non negative function. However, there is a distinction between classical and quantum systems, since the quantum mutual information can take its value above the maximum of the classical mutual information. This statement can be confirmed, if we take into account that for an pure entangled quantum system, the quantum mutual information is

$$I\left(A:B\right) = \mathrm{S}\left(\rho_A\right) + \mathrm{S}\left(\rho_B\right) - \mathrm{S}\left(\rho_{AB}\right) = 1 + 1 - 0 = 2\,, \tag{B.25}$$

and we can rewrite this equation as

$$I\left(A:B\right) = 2\mathrm{S}\left(\rho_A\right) = 2\mathrm{S}\left(\rho_B\right). \tag{B.26}$$

This quantum function has *non-classical* properties, such as that its value for a pure joint system $\rho_{AB}$ can be

$$I\left(A:B\right) = 2\mathrm{S}\left(\rho_A\right) = 2\mathrm{S}\left(\rho_B\right) \tag{B.27}$$

while

$$\mathrm{S}\left(\rho_A\right) = \mathrm{S}\left(\rho_B\right) \text{ and } \mathrm{S}\left(\rho_{AB}\right) = 0\,. \tag{B.28}$$

As we have seen, if we use entangled states, the quantum mutual information could be *2*, while the quantum conditional entropies could be *-1*. In classical Information Theory, negative entropies can be obtained only in the case of mutual information of three or more systems. An important property of *maximized quantum mutual information*: *it is always additive for a quantum channel.*

# B.9 Partial Trace

If we have a density matrix which describes only a subset of a larger quantum space, then we talk about the reduced density matrix. The larger quantum system can be expressed as the tensor product of the reduced density matrices of the subsystems, if there is no correlation (entanglement) between the subsystems. On the other hand, if we have two subsystems with reduced density matrices $\rho_A$ and $\rho_B$, then from the overall density matrix denoted by $\rho_{AB}$ the subsystems can be expressed as

$$\rho_A = Tr_B\left(\rho_{AB}\right) \text{ and } \rho_B = Tr_A\left(\rho_{AB}\right), \tag{B.29}$$

where $Tr_B$ and $Tr_A$ refers to the partial trace operators. So, this partial trace operator can be used to generate one of the subsystems from the joint state $\rho_{AB} = \left|\psi_A\right\rangle\left\langle\psi_A\right| \otimes \left|\psi_B\right\rangle\left\langle\psi_B\right|$, then

$$\begin{aligned}\rho_A &= Tr_B\left(\rho_{AB}\right) = Tr_B\left(\left|\psi_A\right\rangle\left\langle\psi_A\right| \otimes \left|\psi_B\right\rangle\left\langle\psi_B\right|\right)\\ &= \left|\psi_A\right\rangle\left\langle\psi_A\right| Tr\left(\left|\psi_B\right\rangle\left\langle\psi_B\right|\right) = \left|\psi_A\right\rangle\left\langle\psi_A\right|\left\langle\psi_B\middle|\psi_B\right\rangle.\end{aligned} \tag{B.30}$$

Since the inner product is trivially $\left\langle\psi_B\middle|\psi_B\right\rangle = 1$, therefore

$$Tr_B\left(\rho_{AB}\right) = \left\langle\psi_B\middle|\psi_B\right\rangle\left|\psi_A\right\rangle\left\langle\psi_A\right| = \left|\psi_A\right\rangle\left\langle\psi_A\right| = \rho_A. \tag{B.31}$$

In the calculation, we used the fact that $Tr\left(\left|\psi_1\right\rangle\left\langle\psi_2\right|\right) = \left\langle\psi_2\middle|\psi_1\right\rangle$. In general, if we have to systems $A = \left|i\right\rangle\left\langle k\right|$ and $B = \left|j\right\rangle\left\langle l\right|$, then the partial trace can be calculated as

$$Tr_B\left(A \otimes B\right) = A\,Tr\left(B\right), \tag{B.32}$$

since

$$\begin{aligned}Tr_2\left(\left|i\right\rangle\left\langle k\right| \otimes \left|j\right\rangle\left\langle l\right|\right) &= \left|i\right\rangle\left\langle k\right| \otimes Tr\left(\left|j\right\rangle\left\langle l\right|\right)\\ &= \left|i\right\rangle\left\langle k\right| \otimes \left\langle l\middle|j\right\rangle\\ &= \left\langle l\middle|j\right\rangle\left|i\right\rangle\left\langle k\right|,\end{aligned} \tag{B.33}$$

where $\left|i\right\rangle\left\langle k\right| \otimes \left|j\right\rangle\left\langle l\right| = \left|i\right\rangle\left|j\right\rangle\left(\left|k\right\rangle\left|l\right\rangle\right)^T$. In this expression we have used the fact that $\left(AB^T\right) \otimes \left(CD^T\right) = \left(A \otimes C\right)\left(B^T \otimes D^T\right) = \left(A \otimes C\right)\left(B \otimes D\right)^T$.

# B.10 Quantum Entanglement

A quantum system $\rho_{AB}$ is separable if it can be written as a tensor product of the two subsystems $\rho_{AB} = \rho_A \otimes \rho_B$. Beside product states $\rho_A \otimes \rho_B$ which represent a composite system consisting of several independent states merged by means of tensor product $\otimes$ similarly to classical composite systems, quantum mechanics offers a unique new phenomenon called *entanglement*. The so-called *Bell states* (or EPR states, named after Einstein, Podolsky and Rosen) are entangled ones:

$$\begin{aligned} \left|\beta_{00}\right\rangle &= \frac{1}{\sqrt{2}}\left(\left|00\right\rangle + \left|11\right\rangle\right), \\ \left|\beta_{01}\right\rangle &= \frac{1}{\sqrt{2}}\left(\left|01\right\rangle + \left|10\right\rangle\right), \\ \left|\beta_{10}\right\rangle &= \frac{1}{\sqrt{2}}\left(\left|00\right\rangle - \left|11\right\rangle\right), \\ \left|\beta_{11}\right\rangle &= \frac{1}{\sqrt{2}}\left(\left|01\right\rangle - \left|10\right\rangle\right). \end{aligned} \tag{B.34}$$

The characterization of quantum entanglement has deep relevance in Quantum Information Theory. Quantum entanglement is the major phenomenon which distinguishes the classical from the quantum world. By means of entanglement, many classically totally unimaginable results can be achieved in Quantum Information Theory.

# B.11 Fidelity

Theoretically quantum states have to preserve their original superposition during the whole transmission, without the disturbance of their actual properties. Practically, quantum channels are entangled with the environment which results in mixed states at the output. Mixed states are classical probability weighted sum of pure states where these probabilities appear due to the interaction with the environment (i.e., noise). Therefore, we introduce a new quantity, which is able to describe the quality of the transmission of the superposed states through the quantum channel. The quantity which measures this distance is called the *fidelity*.

The fidelity for two pure quantum states is defined as

$$F\left(\left|\varphi\right\rangle,\left|\psi\right\rangle\right)=\left|\left\langle\varphi\middle|\psi\right\rangle\right|^{2}. \tag{B.35}$$

The fidelity of quantum states can describe the relation of Alice pure channel input state $\left|\psi\right\rangle$ and the received mixed quantum system $\sigma=\sum_{i=0}^{n-1}p_i\rho_i=\sum_{i=0}^{n-1}p_i\left|\psi_i\right\rangle\left\langle\psi_i\right|$ at the channel output as

$$F\left(\left|\psi\right\rangle,\sigma\right)=\left\langle\psi\middle|\sigma\middle|\psi\right\rangle=\sum_{i=0}^{n-1}p_i\left|\left\langle\psi\middle|\psi_i\right\rangle\right|^{2}. \tag{B.36}$$

Fidelity can also be defined for *mixed* states $\sigma$ and $\rho$

$$F\left(\rho,\sigma\right)=\left[Tr\left(\sqrt{\sqrt{\sigma}\rho\sqrt{\sigma}}\right)\right]^{2}=\sum_{i}p_i\left[Tr\left(\sqrt{\sqrt{\sigma_i}\rho_i\sqrt{\sigma_i}}\right)\right]^{2}. \tag{B.37}$$

Next we list the major properties of fidelity

$$0\leq F\left(\sigma,\rho\right)\leq 1, \tag{B.38}$$

$$F\left(\sigma,\rho\right)=F\left(\rho,\sigma\right), \tag{B.39}$$

$$F\left(\rho_1\otimes\rho_2,\sigma_1\otimes\sigma_2\right)=F\left(\rho_1,\sigma_1\right)F\left(\rho_2,\sigma_2\right), \tag{B.40}$$

$$F\left(U\rho U^{\dagger},U\sigma U^{\dagger}\right)=F\left(\rho,\sigma\right), \tag{B.41}$$

$$F\left(\rho,a\sigma_1+(1-a)\sigma_2\right)\geq aF\left(\rho,\sigma_1\right)+(1-a)F\left(\rho,\sigma_2\right),\ a\in\left[0,1\right]. \tag{B.42}$$

# B.12 Related Work

The field of Quantum Information Processing is a rapidly growing field of science, however there are still many challenging questions and problems. These most important results will be discussed in further chapters, but these questions cannot be exposited without a knowledge of the fundamental results of Quantum Information Theory.

### Early Years of Quantum Information Theory

Quantum Information Theory extends the possibilities of classical Information Theory, however for some questions, it gives extremely different answers Classical Information

Theory— was founded by Claude Shannon in 1948 [Shannon48]. In Shannon's paper the mathematical framework of communication was invented, and the main definitions and theorems of classical Information Theory were laid down. On the other hand, classical Information Theory is just one part of Quantum Information Theory. The other, missing part is the Quantum Theory, which was completely finalized in 1926.

The results of Quantum Information Theory are mainly based on the results of von Neumann, who constructed the mathematical background of quantum mechanics [Neumann96]. An interesting—and less well known—historical fact is that quantum entropy was discovered by Neumann before the classical information theoretic concept of entropy. Quantum entropy was discovered in the 1930s, based on the older idea of entropy in classical Statistical Mechanics, while the classical information theoretic concept was discovered by Shannon only later, in 1948. It is an interesting note, since the reader might have thought that quantum entropy is an extension of the classical one, however it is not true. Classical entropy, in the context of Information Theory, is a special case of von Neumann's quantum entropy. Moreover, the name of Shannon's formula was proposed by von Neumann [Neumann66]. More information about the connection of Information Theory and statistical mechanics can be found in work of Aspect from 1981 [Aspect81], in the book Petz [Petz08]. The elements of classical Information theory and its mathematical background were summarized in a very good book by Cover [Cover91].

The idea that the results of Quantum Information Theory can be used to solve computational problems was first claimed by Deutsch in 1985 [Deutsch85]. Later in the 90s, the answers to the most important questions of Quantum Information Theory were answered, and the main elements and the fundamentals of this field were discovered. Details about the simulation of quantum systems and the possibility of encoding quantum information in physical particles can be found in Feynman's work from 1982 [Feynman82]. Further information on quantum simulators and continuous-time automata can be found in the work of Vollbrecht and Cirac [Vollbrecht08].

### Quantum Coding and Quantum Compression

An important milestone in Quantum Information Theory is Schumacher's work from 1995 [Schumacher95a] in which he introduced the term, "qubit." In [Schumacher96a-c] the main theories of quantum source coding and the quantum compression were presented. The details of quantum data compression and quantum typical subspaces can be found in [Schumacher95a]. In this paper, Schumacher extended those results which had been presented a year before, in 1994 by Schumacher and Jozsa on a new proof of quantum noiseless coding, for details see [Schumacher94]. Schumacher in 1995 also defined the quantum coding of pure quantum states; in the same year, Lo published a paper in which he extended these result to mixed quantum states, and he also defined an encoding scheme for it [Lo95]. Schumacher's results from 1995 on the compression of quantum information [Schumacher95a] were the first main results on the encoding of quantum information—— its importance and significance in Quantum Information Theory is similar to Shannon's noiseless channel coding theorem in classical Information Theory. In this work, Schumacher also gives upper and lower bounds on the rate of quantum compression. We note, that the mathematical background of Schumacher proof is very similar to Shannon's proof, as the reader can check in [Schumacher95a] and in Shannon's proof [Shannon48].

The method of sending classical bits via quantum bits was firstly completed by Schumacher *et al.* in their famous paper form 1995 [Schumacher95]. The fundaments of noiseless quantum coding were laid down by Schumacher, one year later, in 1996 [Schumacher96]. These works cover the discussion of the relation of entropy exchange and coherent quantum information, which was completely unknown before 1996. The theory of processing of quantum information, the transmission of entanglement over a noisy quantum channel, the error-correction schemes with the achievable fidelity limits, or the classical information capacity of a quantum channel with the limits on the amount of accessible information in a quantum channel were all published in the same year, in 1996. For further information on the fidelity limits and communication capabilities of a noisy quantum channel, see the work of Barnum *et al.* also from 1996 [Barnum96]. In 1997, Schumacher and Westmoreland completed their proof on the classical capacity of a quantum channel, and they published in their famous work [Schumacher97]. These results were extended in their works from 1998 [Schumacher98a-98c]. On the experimental side of fidelity testing see the work of Radmark *et al.* [Radmark09].

About the limits for compression of quantum information carried by ensembles of mixed states, see the work of Horodecki [Horodecki98]. An interesting paper about the quantum coding of mixed quantum states was presented by Barnum *et al.* [Barnum01].

**Quantum Entanglement**

Entanglement is one of the most important differences between the classical and the quantum worlds. An interesting paper on communication via one- and two-particle operators on Einstein-Podolsky-Rosen states was published in 1992, by Bennett [Bennett92c]. About the history of entanglement see the paper of Einstein, Podolsky and Rosen from 1935 [Einstein1935]. In this work, we did not give a complete mathematical background of quantum entanglement—further details on this topic can be found in Nielsen's book [Nielsen2000] or by Hayashi [Hayashi06], or in an very good article published by the four Horodeckis in 2009 [Horodecki09]. A work on the communication cost of entanglement transformations was published by Hayden and Winter [Hayden03a]. The method of entanglement concentration was among the first quantum protocols [Bennett96b]. The method of Bennett's was improved by Nielsen in 1999, [Nielsen99]. A very important work on variable length universal entanglement concentration by local operations and its application to teleportation and dense coding was published by Hayashi and Matsumoto [Hayashi01].

**Comprehensive Surveys**

We also suggest the excellent book of Petz [Petz08] on the mathematical background of Quantum Information Theory. A very good article with the mathematical background of Quantum Mechanics was published by Bennett [Bennett95]. For a general introduction to the Quantum Information Theory and its applications see the book of Hayashi [Haya-

shi06]. We also suggest the previous part of our book from 2005, see [Imre05]. A very good introduction to Quantum Information Theory was published by Bennett and Shor [Bennett98]. We also suggest the textbook of Wilde [Wilde11].

For further details on the use of the results of quantum information theory in engineering applications and practical communications, see [Imre01], [Imre02], [Imre07], [Bacsardi10] and [Galambos10].

# Appendix C

# Quantum Channel Capacities

## C.1 Capacity of a Classical Channel

Before we start to investigate quantum channels, we survey the results of transmitting information over classical noisy channels. In order to achieve reliable (error-free) information transfer we use the so called *channel coding* which extends the payload (useful) information bits with redundancy bits so that at the receiver side Bob will be able to correct some amount of error by means of this redundancy. The channel is given an input $A$, and maps it probabilistically (it is a stochastic mapping, not a unitary or deterministic transformation) to an output $B$, and the probability of this mapping is denoted by $p\left(B|A\right)$.

The *capacity* $C\left(N\right)$ of a *classical* memoryless communication channel $N$ gives an upper bound on the number of classical bits which can be transmitted per channel use, in reliable manner, i.e., with arbitrarily small error at the receiver. The *capacity* of a *classical* memoryless communication channel $N$ gives an upper bound on the number of classical bits which can be transmitted per channel use, in reliable manner, i.e., with arbitrarily small error at the receiver. The simple memoryless classical channel model is shown in Fig. C.1.

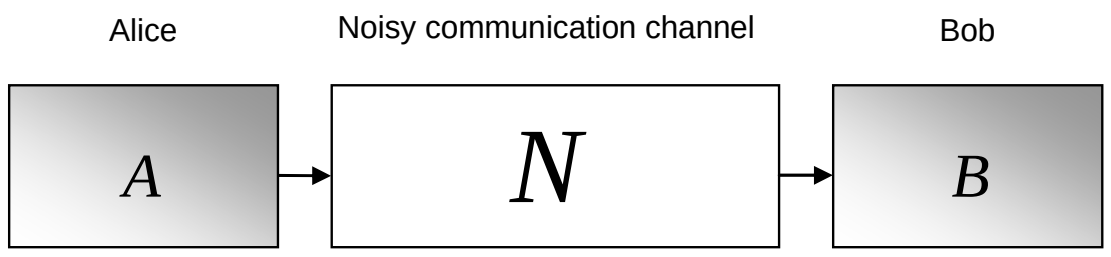

**Fig. C.1.** Simple memoryless classical channel model for Shannon's noisy channel coding theorem.

As it has been proven by Shannon the capacity $C(N)$ of a noisy classical memoryless communication channel $N$, can be expressed by means of the maximum of the mutual information $I(A:B)$ over all possible input distributions $p(x)$ of random variable $X$

$$C(N) = \max_{p(x)} I(A:B). \tag{C.1}$$

In order to make the capacity definition more plausible let us consider Fig. C.2. Here, the effect of environment $E$ is represented by the classical conditional entropies

$$H(A:E|B) > 0 \text{ and } H(B:E|A) > 0. \tag{C.2}$$

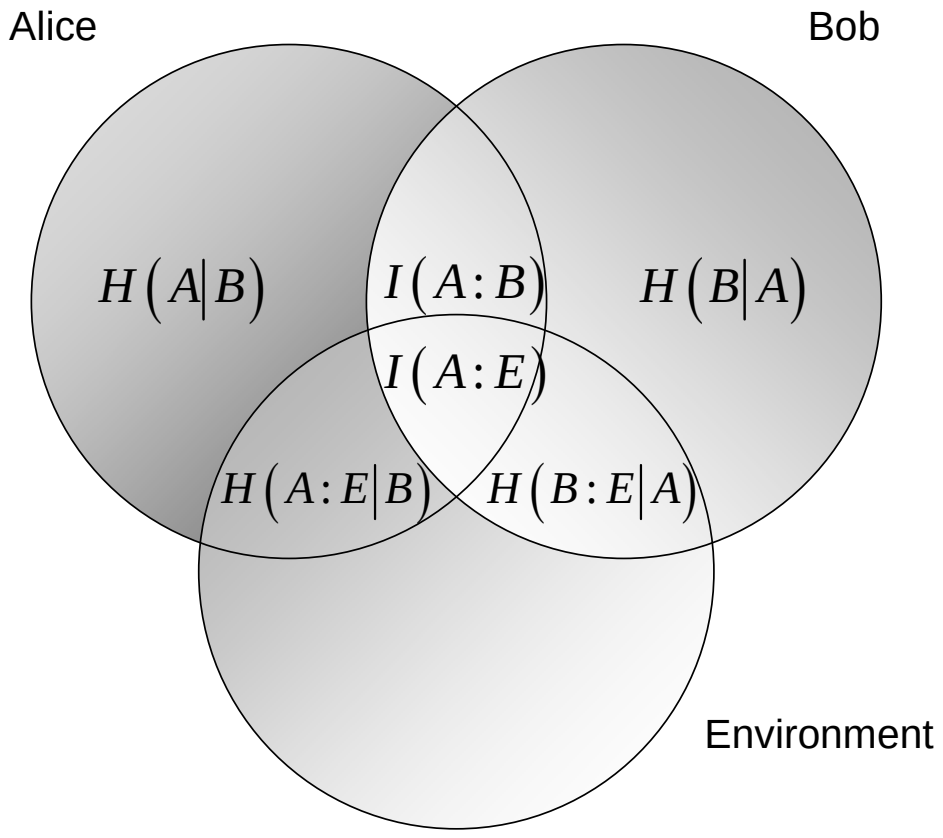

**Fig. C.2.** The effects of the environment on the transmittable information and on the receiver's uncertainty.

Now, having introduced the capacity of classical channel it is important to highlight the following distinction. The *asymptotic capacity* of any channel describes that rate, which can be achieved if the channel can be used $n$ times (denoted by $N^{\otimes n}$), where $n$ can be arbitrarily large. In case of $n = 1$ we speak about *single-use* (*single-letter*) capacity, and will be denoted by $N^{(1)}$. Multiple channel uses can be implemented in consecutive or parallel ways, however from practical reasons we will prefer the latter one. Shannon's noisy coding theorem claims that forming $K$ different codewords $m = \log K$ of length from the source bits and transmitting each of them using the channel $n$ times ($m$ to $n$ coding) the rate at which information can be transmitted through the channel is

$$R = \frac{\log\left(K\right)}{n}, \tag{C.3}$$

and exponentially small probability of error at this rate can be achieved only if $R \leq C\left(N\right)$, otherwise the probability of the successful decoding exponentially tends to zero, as the number of channel uses increases.

# C.2 Classical Capacity of a Quantum Channel

The asymptotic channel capacity is the "true measure" of the various channel capacities, instead of the single-use capacity, which characterizes the capacity only in a very special case. In the regularization step, the channel capacity is computed as a limit. In possession of this limit, we will use the following lower bounds for the single-use capacities. In Chapter 3 we have also seen, the *Holevo-Schumacher-Westmoreland* theorem gives an explicit answer to the maximal transmittable classical information over the quantum channel. Next, we show the connection between these results. As we will see in subsection C.2.1, four different measurement settings can be defined for the measurement of the *classical* capacity of the quantum channel.

Here we call the attention of the reader that Holevo bound (see Chapter 2) limits the classical information stored in a quantum bit. HSW-theorem can be regarded a similar scenario but a quantum channel deployed between Alice and Bob introduces further uncer-tainty before extracting the classical information. Obviously if we assume an idealistic channel the two scenarios become the same.

### C.2.1 Measurement Settings

Similar to classical channel encoding, the quantum states can be transmitted in codewords $n$ quit of length using the quantum channel consecutively $n$-times or equivalently we can send codewords over $n$ copies of quantum channel $\mathcal{N}$ denoted by $\mathcal{N}^{\otimes n}$. For the sake of simplicity we use $n = 2$ in the figures belonging to the following explanation. Multiple-use (asymptotic) approach offers the promise of achieving higher rates compared to the "single-use" version because in the former case optimization is performed over the joint channels instead of performing it individually. In order to make the transient smoother between the single-shot and the asymptotic approaches we depicted the scenario using *product input states* and *single* (or independent) measurement devices at the output of the channel in Fig. C.3. In that case the $C(\mathcal{N})$ classical capacity of quantum channel $\mathcal{N}$ with input $A$ and output $B$ can be expressed by the maximization of the $I(A:B)$ quantum mutual information as follows:

$$C(\mathcal{N}) = \max_{all\ p_i,\rho_i} I(A:B). \tag{C.4}$$

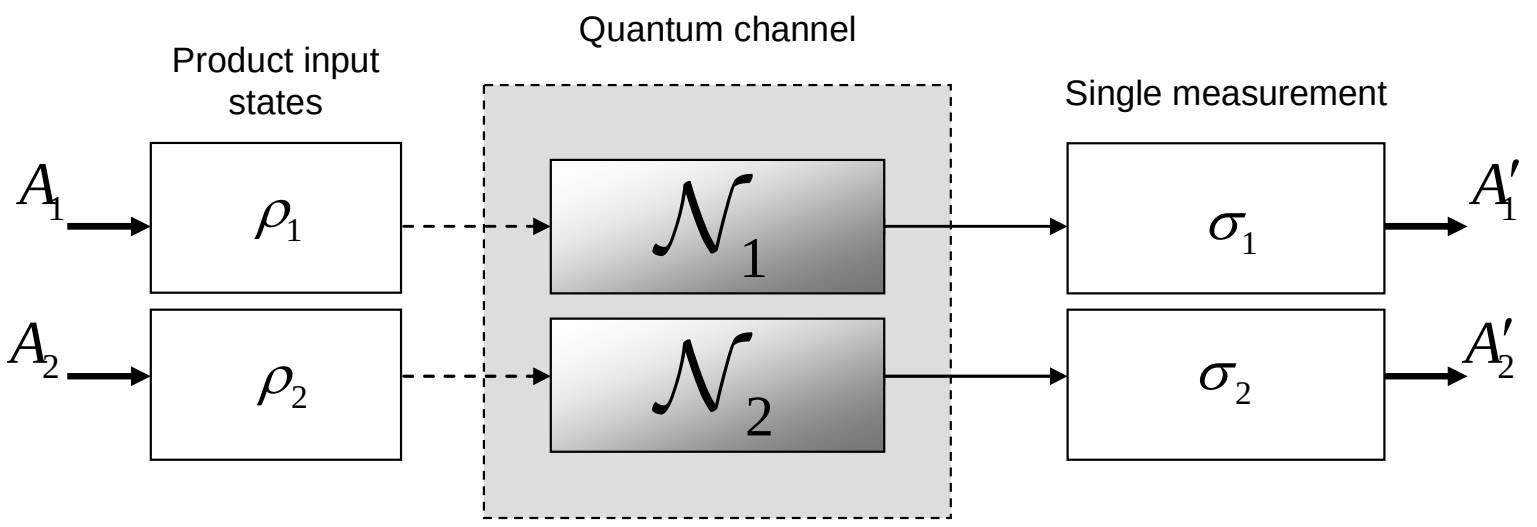

**Fig. C.3.** Transmission of classical information over quantum channel with product state inputs and single measurements. Environment is not depicted.

On the other hand, if we have *product state inputs* but we change the measurement setting from the single measurement setting to *joint measurement* setting, then the classical channel capacity cannot be given by (C.4), hence

$$C\left(\mathcal{N}\right) \neq \max_{all\ p_i,\rho_i} I\left(A:B\right). \tag{C.5}$$

If we would like to step forward, we have to accept the fact, that the quantum mutual information cannot be used to express the asymptotic version: the *maximized* quantum mutual information is *always additive* (see Section B.8) - but not the Holevo information. As follows, if we would take the regularized form of quantum mutual information to express the capacity, we will find that the asymptotic version is equal to the single-use version:

$$\lim_{n\to\infty} \frac{1}{n} \max_{all\ p_i,\rho_i} I\left(A:B\right)^{\otimes n} = \frac{1}{n} n \max_{all\ p_i,\rho_i} I\left(A:B\right) = \max_{all\ p_i,\rho_i} I\left(A:B\right). \tag{C.6}$$

From (C.6) follows, that if we have *product inputs* and *joint measurement* at the outputs, we cannot use the $\max_{all\ p_i,\rho_i} I\left(A:B\right)$ maximized quantum mutual information function to express $C\left(\mathcal{N}\right)$. If we would like to compute the classical capacity for that case, we have to leave the quantum mutual information function, and instead of it we have to use the Holevo information.

This new $C\left(\mathcal{N}\right)$ capacity (according to the *Holevo-Schumacher-Westmoreland* theorem, can be expressed by the Holevo capacity $\chi\left(\mathcal{N}\right)$, which will be equal to the maximization of Holevo information of channel $\mathcal{N}$:

$$\chi\left(\mathcal{N}\right) = \max_{all\ p_i,\rho_i} \chi = C\left(\mathcal{N}\right). \tag{C.7}$$

The Holevo capacity and the asymptotic channel capacity will be equal in this case, however, if *entangled inputs* are allowed with the *joint measurement setting* - then this equality does not hold anymore. As a conclusion, there is also a connection between the maxi-mized Holevo information and the asymptotic classical channel capacity $C(\mathcal{N})$, which has been stated by the HSW theorem:

$$\chi(\mathcal{N}) \leq C(\mathcal{N}). \tag{C.8}$$

This means that we have to redefine the asymptotic formula of $C(\mathcal{N})$ for entangled inputs and joint measurement setting, to measure the maximum transmittable classical information through a quantum channel.

The HSW-theorem gives an explicit answer for the classical capacity of the *product state input* with *joint measurement* setting, and expresses $C(\mathcal{N})$ as follows:

$$C(\mathcal{N}) = \chi(\mathcal{N}) = \max_{all\ p_i,\rho_i} \left[ \mathrm{S}\left( \mathcal{N}\left( \sum_i p_i \rho_i \right) \right) - \sum_i p_i \mathrm{S}(\mathcal{N}(\rho_i)) \right]. \tag{C.9}$$

The relation discussed above holds for the restricted channel setting illustrated in Fig. C.4, where the input consists of product states, and the output is measured by a joint measurement setting.

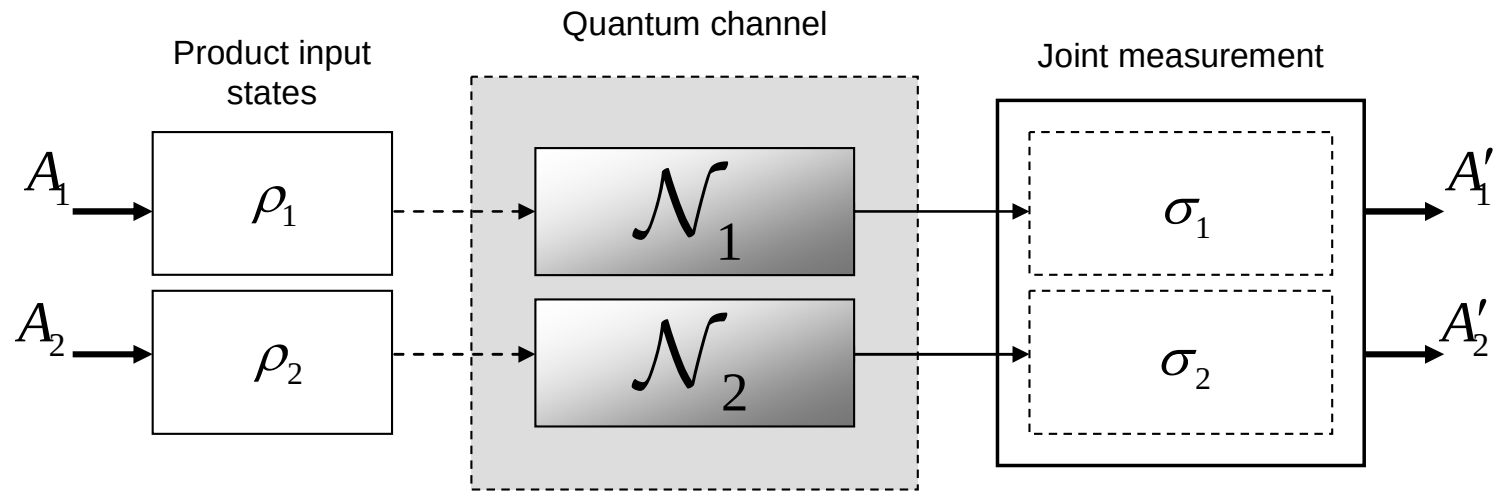

**Fig. C.4.** Transmission of classical information over quantum channel with product state inputs and joint measurements. Environment is not depicted.

In the 1990s, it was conjectured that the same formula can be applied to describe the channel capacity for entangled inputs with the *single measurement* setting; however it was an open question for a long time. Single measurement *destroys* the possible benefits arising from the entangled inputs, and joint measurement is required to achieve the benefits of entangled inputs [King2000]. In 2009 Hastings have used *entangled input states* and showed that the entangled inputs (with the *joint measurement*) can increase the amount of classical information which can be transmitted over a noisy quantum channel. In this case, $\chi(\mathcal{N}) \neq C(\mathcal{N})$ and the $C(\mathcal{N})$ can be expressed with the help of Holevo capacity as follows, using the asymptotic formula of $\chi(\mathcal{N})$:

$$C(\mathcal{N}) = \lim_{n\to\infty} \frac{1}{n} \chi\left(\mathcal{N}^{\otimes n}\right). \tag{C.10}$$

The channel construction for this relation is illustrated in Fig. C.5. The entangled input is formally denoted by $\Psi_{12}$.

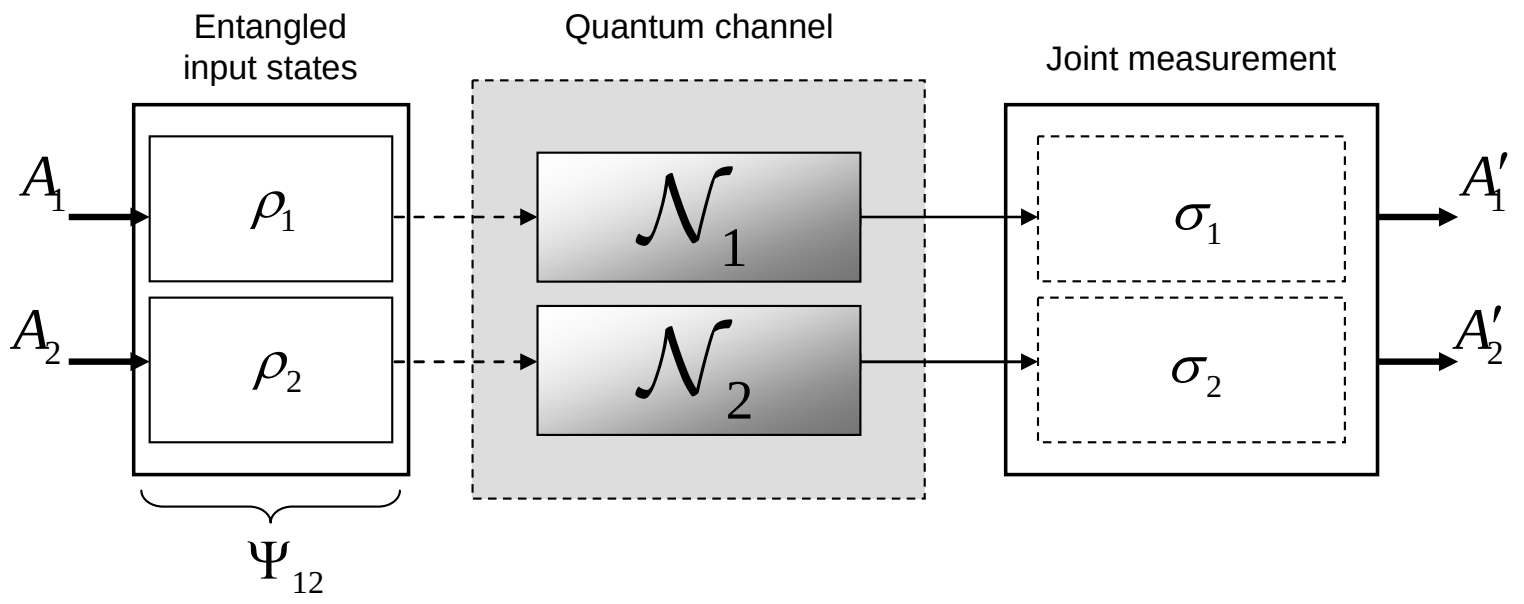

**Fig. C.5.** Transmission of classical information over quantum channel with entangled inputs $\Psi_{12}$ and joint measurements. Environment is not depicted.

We also show the channel construction of the fourth possible construction to measure the classical capacity of a quantum channel. In this case, we have entangled input states, however we use a single measurement setting instead of a joint measurement setting.

To our knowledge, currently there is no quantum channel model where the channel capacity can be increased with this setting, since in this case the benefits of entanglement vanish because the joint measurement setting has been changed into the single measurement setting. We illustrated this setting in Fig. C.6.

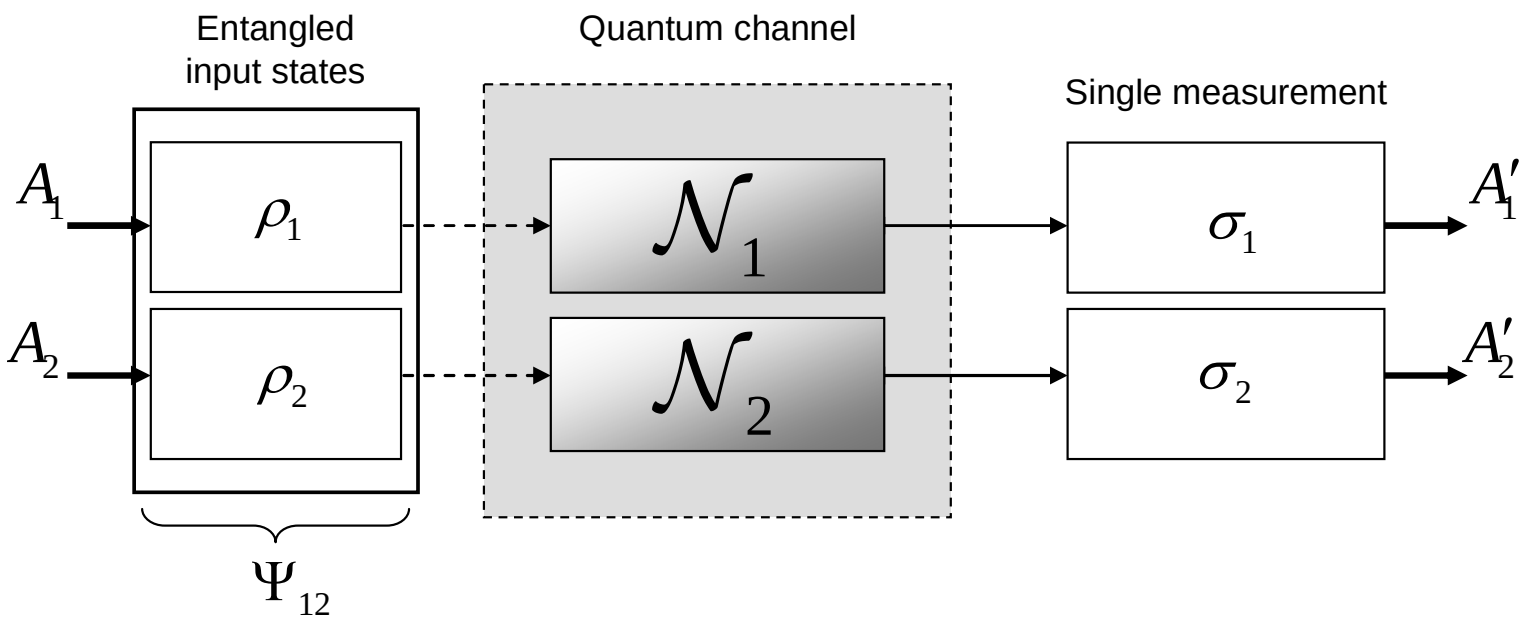

**Fig. C.6.** Transmission of classical information over quantum channel with entangled inputs and single measurements. Environment is not depicted.

To summarize, we have derived two very important results connecting the classical single-use capacity and the classical asymptotic channel formula. We have seen in (C.9), that if we have *product input states* and we change from a single to a *joint measurement* setting, then the capacity cannot be expressed by the maximized quantum mutual information function, because it is always additive (see Section B.8), hence

$$C\left(\mathcal{N}\right) \neq \lim_{n\to\infty} \frac{1}{n} \max_{all\ p_i,\rho_i} I\left(A:B\right)^{\otimes n}. \tag{C.11}$$

If we allow entangled input states and joint measurement (see (C.10)), then we have to use the asymptotic formula of the previously derived Holevo capacity, which yields

$$\chi\left(\mathcal{N}\right) \neq \lim_{n\to\infty} \frac{1}{n} \chi\left(\mathcal{N}^{\otimes n}\right). \tag{C.12}$$

The general sketch of the asymptotic $C\left(\mathcal{N}\right)$ classical capacity is illustrated in Fig. C.7. The $n$ independent uses of the quantum channel are denoted by $\mathcal{N}^{\otimes n}$.

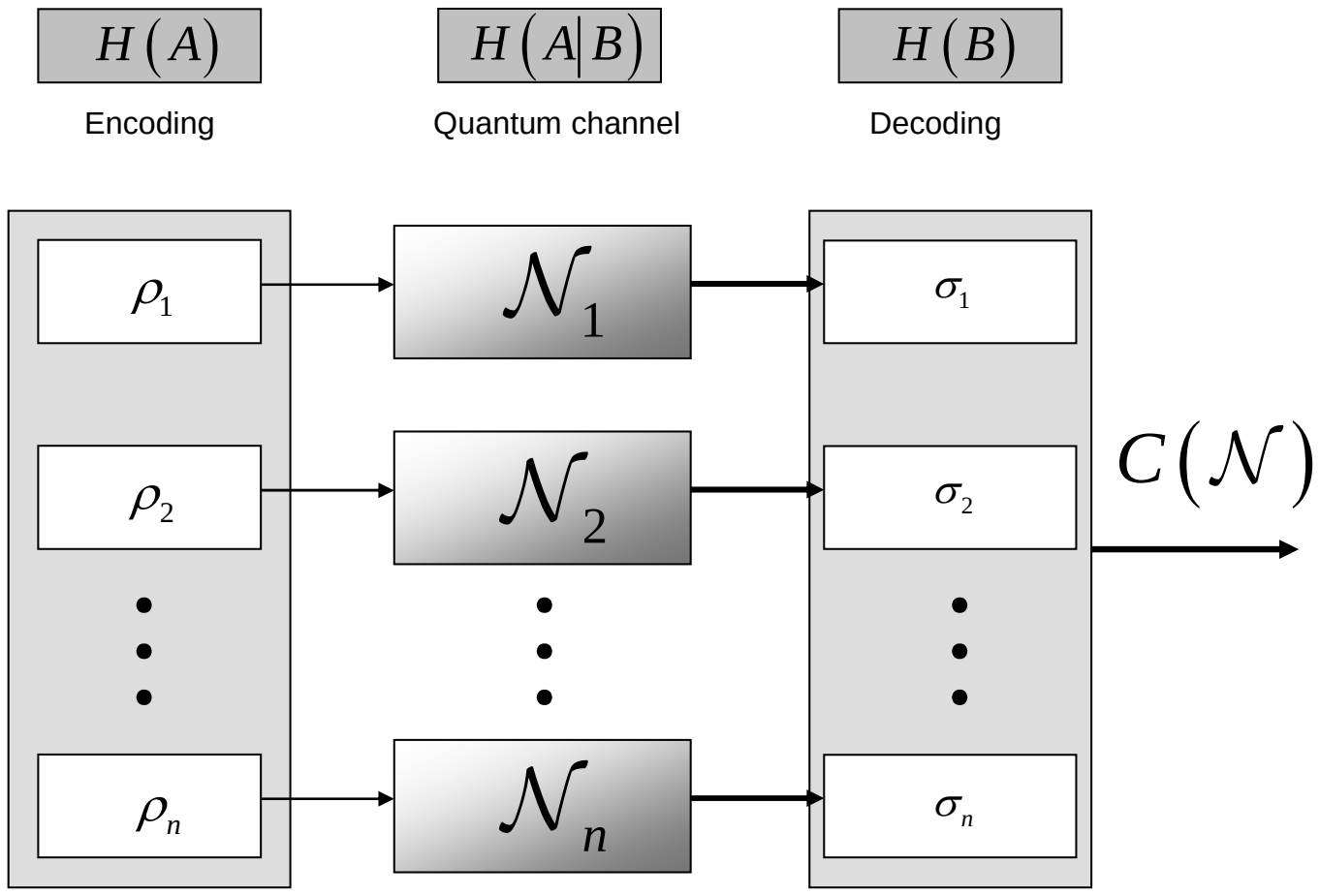

**Fig. C.7.** The asymptotic classical capacity of a quantum channel. The classical capacity measures the maximum achievable classical information by Bob using noisy quantum channel.

The strange nature of the asymptotic formula will results in some very interesting, and classically unimaginable phenomena.

## C.3 Brief Summary

The Holevo quantity measures the classical information, which remains in the encoded quantum states after they have transmitted through a noisy quantum channel. During the transmission, some information passes to the environment from the quantum state, which results in the increased entropy of the sent quantum state.

The HSW-theorem states very similar to Holevo's previous result [Holevo73]. As in the case of the Holevo quantity, the HSW capacity measures the classical capacity of a noisy quantum channel - however, the Holevo quantity also can be used to express the quantum capacity of the quantum channel, which is a not trivial fact. The HSW capacity maximizes the Holevo quantity over a set of possible input states, and expresses the classical information, which can be sent through reliably in the form of *product input states* over the noisy quantum channel, hence HSW capacity is also known as *product state channel capacity.*

As follows, the HSW-theorem defines the maximum of classical capacity of a quantum channel, which can be achieved for *product state inputs* and *joint measurement setting* (see subsection C.2.1). In this case, the input states are not entangled; hence there is no entanglement between the multiple uses of the quantum channel. As we have seen in subsection C.2.1, if the input of the channel consists of product states and we use *single measurement* setting, then the classical capacity can be expressed as the maximum of the quantum mutual information. On the other hand, if the single measurement has been changed to *joint measurement*, this statement is not true anymore; - this capacity will be equal to HSW capacity, see (C.9). Moreover, if we step forward, and we allow *entanglement* among the input states, then we cannot use anymore the HSW capacity, which was defined in Chapter 3. In this case we have to take its asymptotic formula, which was shown in (C.10). The bound of the HSW-theorem is shown in Fig. C.8.

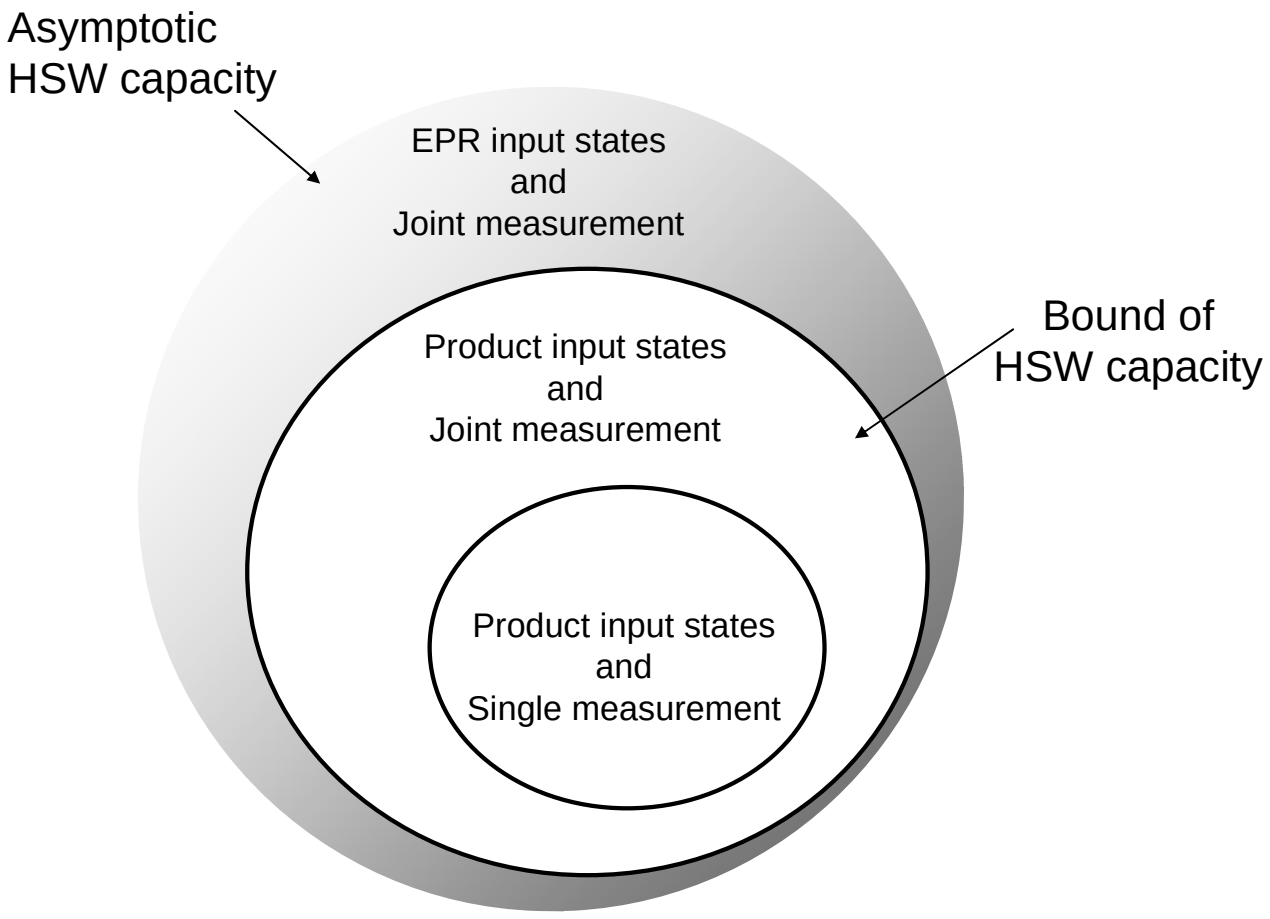

**Fig. C.8.** The bound of HSW capacity. The HSW theorem was defined for the measure of product state capacity. The capacity formula for entangled input states and the joint measurement can be described by the asymptotic version of the HSW capacity.

## C.4 Formal Definitions of Quantum Zero-Error Communication

In this subsection we review the most important definitions of quantum zero-error communication systems. The *non-adjacent* elements are important for zero-error capacity, since *only non-adjacent codewords can be distinguished perfectly.* Two inputs are called *adjacent* if they can result in the same output, while for *non-adjacent* inputs, the output of the encoder is unique. The number of possible non-adjacent codewords determines the rate of maximal transmittable classical information through quantum channels. Formally, the *non-adjacent* property of two quantum states $\rho_1$ and $\rho_2$ can be given as

$$Set_1 \cap Set_2 = \varnothing , \tag{C.13}$$

where

$$Set_i = \left\{ \Pr\left[ X_j' \middle| X_i \right] = Tr\left( \mathcal{M}_j \mathcal{N}\left( \left| \psi_{X_i} \right\rangle \left\langle \psi_{X_i} \right| \right) \right) > 0,\ j \in \left\{ 1, \ldots, m \right\} \right\},\ i = 1,2 , \tag{C.14}$$

using POVM decoder $\mathcal{P} = \left\{ \mathcal{M}_1, \ldots, \mathcal{M}_m \right\}$. In a relation of a noisy quantum channel $\mathcal{N}$, the non-adjacent property can be rephrased as follows. Two input quantum states $\rho_1$ and $\rho_2$ are non-adjacent with relation to $\mathcal{N}$, if $\mathcal{N}\left( \rho_1 \right)$ and $\mathcal{N}\left( \rho_2 \right)$ are *perfectly distinguishable*. The notation $\rho_1 \underset{\mathcal{N}}{\perp} \rho_2$ also can be used to denote the non-adjacent inputs of quantum channel $\mathcal{N}$. A quantum channel $\mathcal{N}$ has greater than zero zero-error capacity if and only if a subset of quantum states $\Omega = \left\{ \rho_i \right\}_{i=1}^{l}$ and POVM $\mathcal{P} = \left\{ \mathcal{M}_1, \ldots, \mathcal{M}_m \right\}$ exists where for at *least two states* $\rho_1$ and $\rho_2$ from subset $\Omega$, the relation (C.13) holds; that is, the non-adjacent property with relation to the POVM measurement is satisfied.

For the quantum channel $\mathcal{N}$, the two inputs $\rho_1$ and $\rho_2$ are non-adjacent if and only if the quantum channel takes the input states $\rho_1$ and $\rho_2$ into orthogonal subspaces

$$\mathcal{N}\left( \rho_1 \right) \underset{\mathcal{N}}{\perp} \mathcal{N}\left( \rho_2 \right); \tag{C.15}$$

that is, the quantum channel has positive classical zero-error capacity $C_0\left( \mathcal{N} \right)$ if and only if this property holds for the output of the channel for a given POVM $\mathcal{P} = \left\{ \mathcal{M}_1, \ldots, \mathcal{M}_m \right\}$. The previous result can be rephrased as follows. The two quantum states $\rho_1$ and $\rho_2$ are non-adjacent if and only if for the channel output states $\mathcal{N}\left( \rho_1 \right), \mathcal{N}\left( \rho_2 \right)$,

$$Tr\left(\mathcal{N}\left(\rho_1\right)\mathcal{N}\left(\rho_2\right)\right) = 0\,, \tag{C.16}$$

and if $\rho_1$ and $\rho_2$ are non-adjacent input states then

$$Tr\left(\rho_1\rho_2\right) = 0\,. \tag{C.17}$$

Non-adjacent inputs produce distinguishable outputs, as depicted in Fig. C.9.

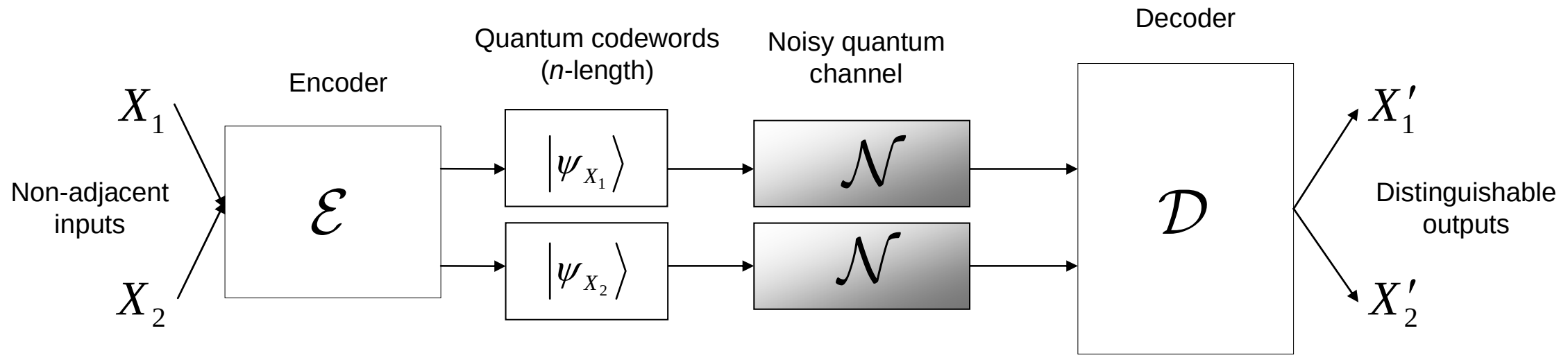

**Fig. C.9.** The non-adjacent inputs can be distinguished at the output. The quantum zero-error communication requires non-adjacent quantum codewords.

Let the two *non-adjacent* input codewords of the quantum channels be denoted by $\left|\psi_{X_1}\right\rangle$ and $\left|\psi_{X_2}\right\rangle$. These quantum codewords encode messages $X_1 = \left\{x_{1,1}, x_{1,2}, \ldots, x_{1,n}\right\}$ and $X_2 = \left\{x_{2,1}, x_{2,2}, \ldots, x_{2,n}\right\}$. For this setting, we construct the following POVM operators for the given complete set of POVM $\mathcal{P} = \left\{\mathcal{M}_1, \ldots, \mathcal{M}_m\right\}$ and the two input codewords $\left|\psi_{X_1}\right\rangle$ and $\left|\psi_{X_2}\right\rangle$ as follows (see Fig. C.10.)

$$\mathcal{M}^{(1)} = \left\{\mathcal{M}_1, \ldots, \mathcal{M}_k\right\} \tag{C.18}$$

and

$$\mathcal{M}^{(2)} = \left\{\mathcal{M}_{k+1}, \ldots, \mathcal{M}_m\right\}. \tag{C.19}$$

The groups of operators, $\mathcal{M}^{(1)}$ and $\mathcal{M}^{(2)}$, will identify and distinguish the input codewords $\left|\psi_{X_1}\right\rangle$ and $\left|\psi_{X_2}\right\rangle$. For input message $\left|\psi_{X_1}\right\rangle$ and $\left|\psi_{X_2}\right\rangle$ with the help of set $\mathcal{M}^{(1)}$ and $\mathcal{M}^{(2)}$ these probabilities are

$$\begin{aligned}\Pr\left[X_1'\middle|X_1\right] &= Tr\left(\mathcal{M}^{(1)}\mathcal{N}\left(\left|\psi_{X_1}\right\rangle\left\langle\psi_{X_1}\right|\right)\right) = 1,\\ \Pr\left[X_2'\middle|X_2\right] &= Tr\left(\mathcal{M}^{(2)}\mathcal{N}\left(\left|\psi_{X_2}\right\rangle\left\langle\psi_{X_2}\right|\right)\right) = 1,\end{aligned} \tag{C.20}$$

where $\mathcal{M}^{(1)}$ and $\mathcal{M}^{(2)}$ are orthogonal projectors, $\mathcal{M}^{(1)}$ and $\mathcal{M}^{(2)}$ are defined in (C.18) and (C.19)), and $\mathcal{M}^{(1)} + \mathcal{M}^{(2)} + \mathcal{M}^{(2+1)} = I$, to make it possible for the quantum channel to take the input states into orthogonal subspaces; that is, $\mathcal{N}\left(\left|\psi_{X_1}\right\rangle\left\langle\psi_{X_1}\right|\right) \perp \mathcal{N}\left(\left|\psi_{X_2}\right\rangle\left\langle\psi_{X_2}\right|\right)$ has to be satisfied.

The non-adjacent property also can be extended for arbitrary length of quantum codewords. For a given quantum channel $\mathcal{N}$, the two $n$-length input quantum codewords $\left|\psi_{X_1}\right\rangle$ and $\left|\psi_{X_2}\right\rangle$, which are tensor products of $n$ quantum states, then *input* codewords $\left|\psi_{X_1}\right\rangle$ and $\left|\psi_{X_2}\right\rangle$ are non-adjacent in relation with $\mathcal{N}$ if and only if *at least one* pair of quantum states $\left\{\left|\psi_{1,i}\right\rangle,\left|\psi_{2,i}\right\rangle\right\}$ from the two $n$-length sequences is perfectly distinguishable. The non-adjacency property of the codewords can be verified as follows:

$$\begin{aligned}&Tr\left(\mathcal{N}\left(\left|\psi_{X_1}\right\rangle\left\langle\psi_{X_1}\right|\right)\mathcal{N}\left(\left|\psi_{X_2}\right\rangle\left\langle\psi_{X_2}\right|\right)\right)\\ &= Tr\left(\left(\bigotimes_{i=1}^{n}\mathcal{N}\left(\left|\psi_{1,i}\right\rangle\left\langle\psi_{1,i}\right|\right)\right)\left(\bigotimes_{i=1}^{n}\mathcal{N}\left(\left|\psi_{2,i}\right\rangle\left\langle\psi_{2,i}\right|\right)\right)\right)\\ &= \prod_{i=1}^{n} Tr\left(\mathcal{N}\left(\left|\psi_{1,i}\right\rangle\left\langle\psi_{1,i}\right|\right)\mathcal{N}\left(\left|\psi_{2,i}\right\rangle\left\langle\psi_{2,i}\right|\right)\right) = 0.\end{aligned} \tag{C.21}$$

As follows from (C.21), a quantum channel $\mathcal{N}$ has non-zero zero-error capacity if and only if there exists at least two non-adjacent input quantum states $\rho_1$ and $\rho_2$. These two non-adjacent quantum states make distinguishable the two, $n$-length quantum codewords at the output of quantum channel $\mathcal{N}$ , and these input codewords will be called as *non-adjacent quantum codewords*.

The non-adjacent inputs will be distinguished by a well characterized set of POVM operators. The decoding of non-adjacent codewords to achieve zero-error communication over a quantum channel is depicted in Fig. C.10.

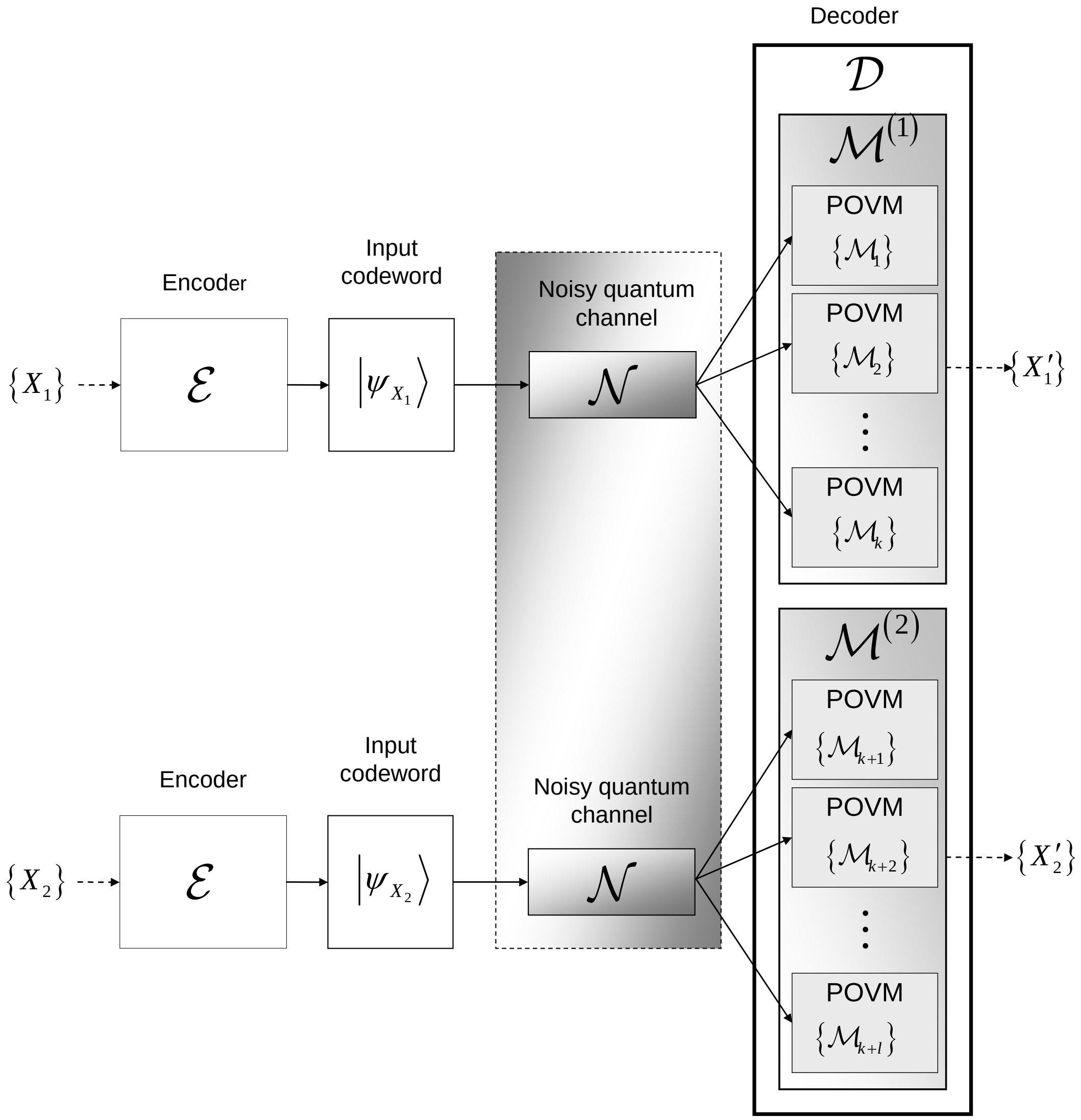

**Fig. C.10.** Each of the non-adjacent input codewords is distinguished by a set of measurement operators to achieve the zero-error quantum communication.

The joint measurement is *necessary* and *sufficient* to distinguish the input codewords with zero-error. *Necessary*, because the joint measurement is required to distinguish orthogonal general (i.e., non zero-error code) tensor product states [Bennett99a]. Sufficient, because the non-adjacent quantum states have orthogonal *supports* at the output of the noisy quantum channel, i.e., $Tr\left(\rho_i \rho_j\right) = 0$ [Medeiros05]. (The *support* of a matrix $A$ is the orthogonal complement of the kernel of the matrix. The *kernel* of $A$ is the set of all vectors $v$, for which $Av = 0$.) For the joint measurement, the $\left\{\mathcal{M}_i\right\}$, $i = 1,\ldots,m$ projectors are $d^n \times d^n$ matrices, while if we were to use a single measurement then the size of these matrices would be $d \times d$.

In Fig. C.11 we compared the difference between single and joint measurement settings for a given $n$-length quantum codeword $\left|\psi_X\right\rangle = \left[\left|\psi_1\right\rangle \otimes \left|\psi_2\right\rangle \otimes \left|\psi_3\right\rangle \cdots \otimes \left|\psi_n\right\rangle\right]$. In the case of single measurement Bob measures each of the $n$ quantum states of the $i$-th codeword states individually. In case of the joint measurement Bob waits until he receives the $n$ quantum states, then measures them together.

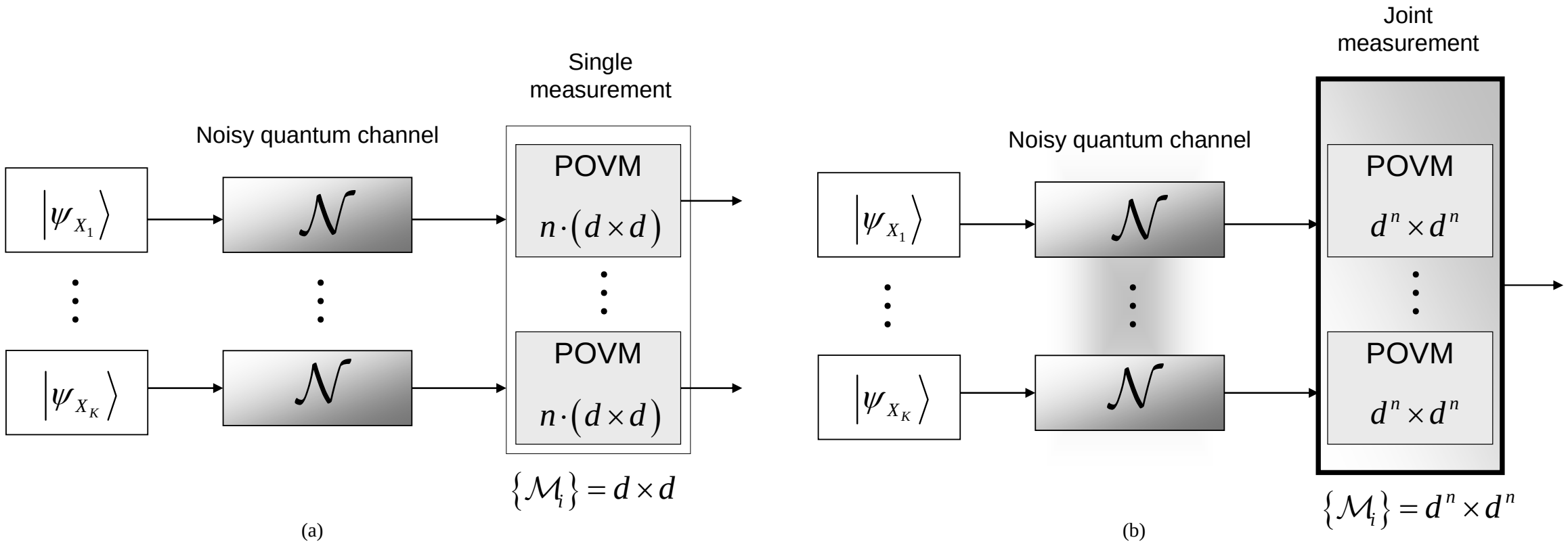

**Fig. C.11.** Comparison of single (a) and joint (b) measurement settings. The joint measurement is necessary to attain the quantum zero-error communication.

The classical zero-error quantum capacity $C_0\left(\mathcal{N}\right)$ for product input states can be reached if and only if the input states are *pure* states, similarly to the HSW capacity $C\left(\mathcal{N}\right)$.

# C.5 Achievable Zero-Error Rates in Quantum Systems

Theoretically (without making any assumptions about the physical attributes of the transmission), the *classical single-use zero-error capacity* $C_0^{(1)}(\mathcal{N})$ of the noisy quantum channel can be expressed as

$$C_0^{(1)}(\mathcal{N}) = \log(K(\mathcal{N})), \tag{C.22}$$

where $K(\mathcal{N})$ is the number of different messages which can be sent over the channel with a *single use* of $\mathcal{N}$ (or in other words the maximum size of the set of *mutually non-adjacent* inputs).

The asymptotic *zero-error capacity* of the noisy quantum channel can be expressed as

$$C_0(\mathcal{N}) = \lim_{n\to\infty} \frac{1}{n} \log\left(K\left(\mathcal{N}^{\otimes n}\right)\right), \tag{C.23}$$

where $K\left(\mathcal{N}^{\otimes n}\right)$ is the maximum number of $n$-length classical messages that the quantum channel can transmit with zero error and $\mathcal{N}^{\otimes n}$ denotes the $n$-uses of the channel (i.e., we have $n$-length classical messages). The $C_0(\mathcal{N})$ asymptotic classical zero-error capacity of a quantum channel is *upper bounded* by the HSW capacity, that is,

$$C_0^{(1)}(\mathcal{N}) \leq C_0(\mathcal{N}) \leq C(\mathcal{N}). \tag{C.24}$$

For the connection of zero-error quantum codes and graph theory see Appendix. The complete historical background can be found in the Related Work section of the Appendix.

# C.6 Connection with Graph Theory

The problem of finding *non-adjacent* codewords for the zero-error information transmission can be rephrased in terms of graph theory. The adjacent codewords are also called *confusable*, since these codewords can generate the same output with a given non-zero probability. Since we know that two input codewords $\left|\psi_{X_1}\right\rangle$ and $\left|\psi_{X_2}\right\rangle$ are *adjacent* if there is a channel output codeword $\left|\psi_{X'}\right\rangle$ which can be resulted by either of these two, that is $\Pr\left[X'\middle|X_1\right] > 0$ and $\Pr\left[X'\middle|X_2\right] > 0$.

The non-adjacent property of two quantum codewords can be analyzed by the *confusability graph* $\mathcal{G}_n$, where $n$ denotes the number of channel uses or in other words the length of the quantum codewords.

Let us take as many vertices as the number of input messages $K$, and connect two vertices if these input messages are adjacent.

For example, using the quantum version of the famous *pentagon graph* we show how the classical zero-error capacities $C_0\left(\mathcal{N}\right)$ and of the quantum channel changes if we use block codes of lengths $n$=1 and $n$=2. In the pentagon graph an input codeword from the set of non-orthogonal qubits $\left\{|0\rangle,|1\rangle,|2\rangle,|3\rangle,|4\rangle\right\}$ is connected with two other adjacent input codewords, and the number of total codewords is 5 [Lovász79].

The $\mathcal{G}_1$ *confusability* graph of the pentagon structure for block codes of length $n$=1 is shown in Fig. C.12. The vertices of the graph are the possible input messages, where $K$ = 5. The *adjacent* input messages are connected by a line. The non-adjacent inputs $|2\rangle$ and $|4\rangle$ are denoted by gray circles, and there is no connection between these two input codewords.

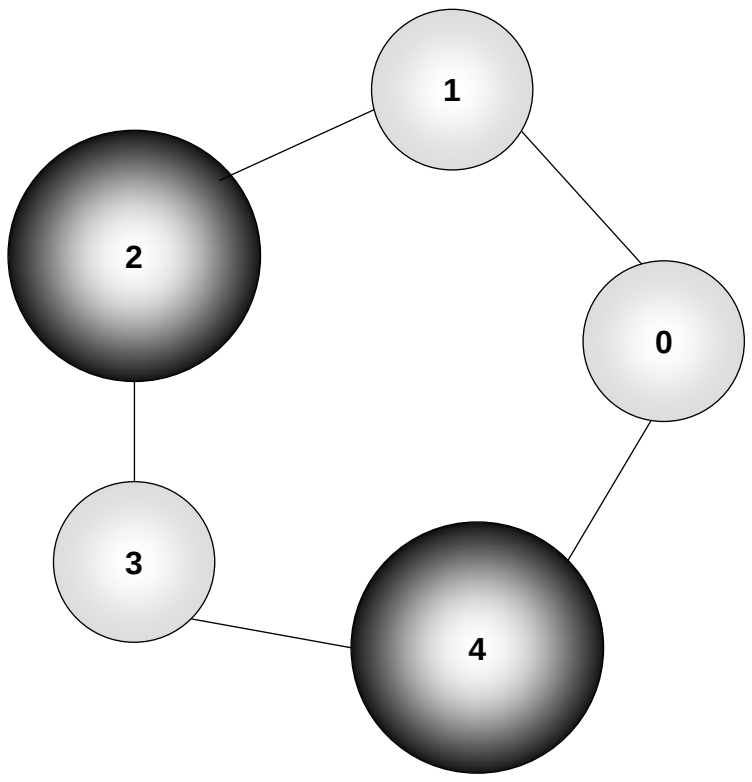

**Fig. C.12.** The confusability graph of a zero-error code for one channel use. The two possible non-adjacent codewords are denoted by the large shaded circles.

For the block codes of length one $\left(n=1\right)$, the maximal transmittable classical information with zero error for the pentagon graph is

$$C_0\left(\mathcal{N}\right)=\log\left(2\right)=1, \tag{C.25}$$

since only two non-adjacent vertices can be found in the graph. We note, other possible codeword combinations also can be used to realize the zero-error transmission, in comparison with the confusability graph in Fig. C.12, for example $\left|1\right\rangle$ and $\left|3\right\rangle$ also non-adjacent, etc. On the other hand, the maximum number of non-adjacent vertices (two, in this case) cannot be exceeded, thus $C_0\left(\mathcal{N}\right)=1$ remains in all other possible cases, too.

Let assume that we have $n=2$ length quantum codewords. First, let us see how the graph changes. The non-adjacent inputs are denoted by the large gray shaded circles. The connections between the possible codewords (which can be used as a block code) are denoted by the thick line and the dashed circle. The confusability graph $\mathcal{G}_2$ for block codes of length $n=2$ is shown in Fig. C.13. The two half-circles together on the left and right sides represent one circle and the two half circles at the top and bottom of the figure also represent one circle; thus there are five dashed circles in the figure.

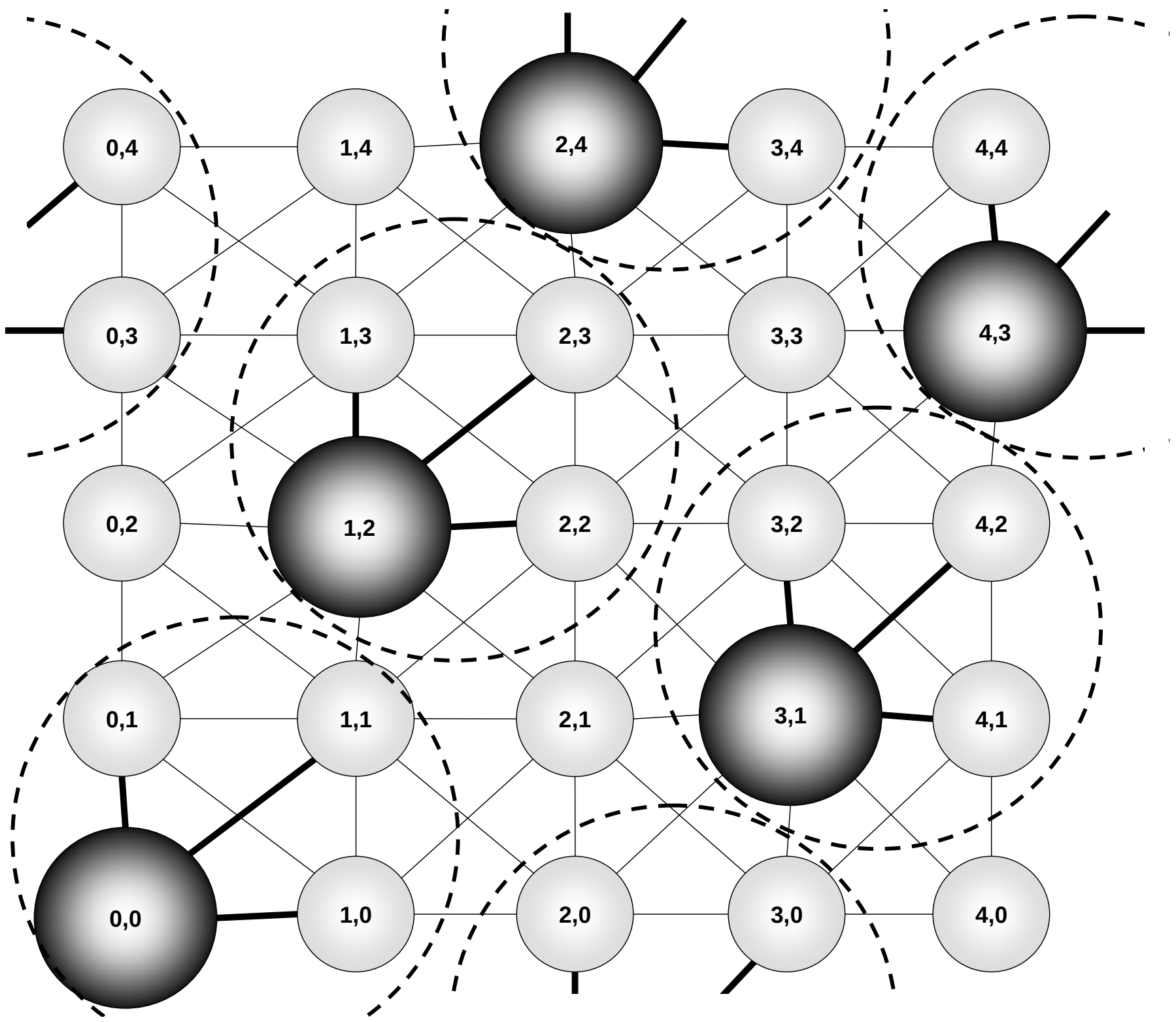

**Fig. C.13.** The graph of zero-error code for block codes of length two. The possible zero-error codewords are depicted by the thick lines and dashed circles.

It can be seen that the complexity of the structure of the graph has changed, although we have made only a small change: we increased the lengths of the block codes from $n = 1$ to $n = 2$.

The five two-length codewords and zero-error quantum block codes which can realize the zero-error transmission can be defined as follows using the computational basis $\{|0\rangle,|1\rangle,|2\rangle,|3\rangle,|4\rangle\}$. The arrows indicate those codewords which are connected in the graph with the given codeword; that is, these subsets can be used as quantum block codes as depicted in Fig. C.13.

$$\begin{aligned}
&\left|\psi_{X_1}\right\rangle = \left[\left|0\right\rangle \otimes \left|0\right\rangle\right] \to \\
&\quad \left\{\left|\psi_{X_1}\right\rangle = \left[\left|0\right\rangle \otimes \left|0\right\rangle\right], \left|\psi_{X_2}\right\rangle = \left[\left|0\right\rangle \otimes \left|1\right\rangle\right], \left|\psi_{X_3}\right\rangle = \left[\left|1\right\rangle \otimes \left|0\right\rangle\right], \left|\psi_{X_4}\right\rangle = \left[\left|1\right\rangle \otimes \left|1\right\rangle\right]\right\}, \\
&\left|\psi_{X_2}\right\rangle = \left[\left|1\right\rangle \otimes \left|2\right\rangle\right] \to \\
&\quad \left\{\left|\psi_{X_1}\right\rangle = \left[\left|1\right\rangle \otimes \left|2\right\rangle\right], \left|\psi_{X_2}\right\rangle = \left[\left|2\right\rangle \otimes \left|2\right\rangle\right], \left|\psi_{X_3}\right\rangle = \left[\left|1\right\rangle \otimes \left|3\right\rangle\right], \left|\psi_{X_4}\right\rangle = \left[\left|2\right\rangle \otimes \left|3\right\rangle\right]\right\}, \\
&\left|\psi_{X_3}\right\rangle = \left[\left|2\right\rangle \otimes \left|4\right\rangle\right] \to \\
&\quad \left\{\left|\psi_{X_1}\right\rangle = \left[\left|2\right\rangle \otimes \left|4\right\rangle\right], \left|\psi_{X_2}\right\rangle = \left[\left|3\right\rangle \otimes \left|4\right\rangle\right], \left|\psi_{X_3}\right\rangle = \left[\left|2\right\rangle \otimes \left|0\right\rangle\right], \left|\psi_{X_4}\right\rangle = \left[\left|3\right\rangle \otimes \left|0\right\rangle\right]\right\}, \\
&\left|\psi_{X_4}\right\rangle = \left[\left|3\right\rangle \otimes \left|1\right\rangle\right] \to \\
&\quad \left\{\left|\psi_{X_1}\right\rangle = \left[\left|3\right\rangle \otimes \left|1\right\rangle\right], \left|\psi_{X_2}\right\rangle = \left[\left|4\right\rangle \otimes \left|1\right\rangle\right], \left|\psi_{X_3}\right\rangle = \left[\left|3\right\rangle \otimes \left|2\right\rangle\right], \left|\psi_{X_4}\right\rangle = \left[\left|4\right\rangle \otimes \left|2\right\rangle\right]\right\}, \\
&\left|\psi_{X_5}\right\rangle = \left[\left|4\right\rangle \otimes \left|3\right\rangle\right] \to \\
&\quad \left\{\left|\psi_{X_1}\right\rangle = \left[\left|4\right\rangle \otimes \left|3\right\rangle\right], \left|\psi_{X_2}\right\rangle = \left[\left|0\right\rangle \otimes \left|3\right\rangle\right], \left|\psi_{X_3}\right\rangle = \left[\left|4\right\rangle \otimes \left|4\right\rangle\right], \left|\psi_{X_4}\right\rangle = \left[\left|0\right\rangle \otimes \left|4\right\rangle\right]\right\}.
\end{aligned} \tag{C.26}$$

The classical zero-error capacity which can be achieved by $n = 2$ length quantum codewords is

$$C_0\left(\mathcal{N}\right) = \frac{1}{2}\log\left(5\right) = 1.1609\,. \tag{C.27}$$

From an engineering point of view this result means, that for the pentagon graph, the maximum rate at which classical information can be transmitted over a noisy quantum channel with a zero error probability, can be achieved with quantum block code length of two. For the classical zero-error capacities of some typical quantum channels see Section 5 of the Ph.D Thesis.

## C.7 Entanglement-assisted Classical Zero-Error Capacity

In the previous subsection we discussed the main properties of zero-error capacity using product input states. Now, we add the entanglement to the picture. Here we discuss how the encoding and the decoding setting will change if we bring entanglement to the system and how it affects the classical zero-error capacity of a quantum channel.

If entanglement allowed between the communicating parties then the single-use and asymptotic *entanglement-assisted* classical zero-error capacities are defined as

$$C_0^{E(1)}\left(\mathcal{N}\right) = \log\left(K^E\left(\mathcal{N}\right)\right) \tag{C.28}$$

and

$$C_0^{E}\left(\mathcal{N}\right) = \lim_{n\to\infty}\frac{1}{n}\log\left(K^E\left(\mathcal{N}^{\otimes n}\right)\right). \tag{C.29}$$

where $K^E\left(\mathcal{N}^{\otimes n}\right)$ is the maximum number of $n$-length mutually non-adjacent classical messages that the quantum channel can transmit with zero error using *shared entanglement*.

Before we start to discuss the properties of the entanglement-assisted zero-error quantum communication, we introduce a new type of graph, called the *hypergraph* $\mathcal{G}_H$. The hypergraph is very similar to our previously shown *confusability* graph $\mathcal{G}_n$. The hypergraph contains a set of vertices and hyperedges. The vertices represent the *inputs* of the quantum channel $\mathcal{N}$, while the hyperedges contain all the channel inputs which could cause the same channel output with non-zero probability.

We will use some new terms from graph theory in this subsection; hence we briefly summarize these definitions:

- *maximum independent set of* $\mathcal{G}_n$: the maximum number of non-adjacent inputs ($K$),

- *clique of* $\mathcal{G}_n$: $\kappa_i$, the set of possible inputs of a given output in a confusability graph (which inputs could result in the same output with non-zero probability),

- *complete graph*: if all the vertices are connected with one another in the graph; in this case there are no non-adjacent inputs; i.e., the channel has no zero-error capacity.

Both the hypergraph and the confusability graph can be used to determine the non-adjacent inputs. However, if the number of inputs starts to increase, the number of hyperedges in the hypergraph will be significantly lower than the number of edges in the confusability graph of the same system (see Fig. C.15).

In short, the entanglement-assisted zero-error quantum communication protocol works as follows according to Fig. C.14 [Cubitt10]. Before the communication, Alice and Bob share entanglement between themselves. The $d$-dimensional shared system between Alice and Bob will be denoted by $\rho_{AB} = \left|\Phi_{AB}\right\rangle\left\langle\Phi_{AB}\right|$, where

$$\left|\Phi_{AB}\right\rangle = \frac{1}{\sqrt{d}}\sum_{i=0}^{d-1}\left|i\right\rangle_A\left|i\right\rangle_B \tag{C.30}$$

is a rank-$d$ maximally entangled qudit state (also called as *edit*). If Alice would like to send a message $q \in \left\{1,\ldots,K\right\}$, where $K$ is the number of messages, to Bob, she has to measure her half of the entangled system using a complete orthogonal basis $B_q = \left\{\left|\psi_{x'}\right\rangle\right\}$, $x' \in \kappa_q$, where $x'$ is a vertice in the hypergraph $\mathcal{G}_H$ from clique $\kappa_q$. The *orthonormal representation of a graph is a map:* the vertice $x'$ represents the unit vector $\left|\psi_{x'}\right\rangle$ such that if $x$ and $x'$ are *adjacent* then $\left\langle\psi_x\middle|\psi_{x'}\right\rangle = 0$ (*i.e., they are orthogonal in the orthonormal representation*) and $\kappa_q$ is the clique corresponding to message $q$ in the hypergraph $\mathcal{G}_H$. The hypergraph has $K$ cliques of size $d$, $\left\{\kappa_1,\ldots,\kappa_K\right\}$ (i.e., each message $q \in \left\{1,\ldots,K\right\}$ is represented by a $d$-size clique in the hypergraph $\mathcal{G}_H$.) After the measurement, Bob's state will collapse to $\left|\psi_x\right\rangle^*$. Bob will measure his state in $B_q = \left\{\left|\psi_x\right\rangle\right\}$ to get the final state $\left|\psi_{x'}\right\rangle^*$. Bob's output is denoted by $y$. Bob's possible states are determined by those vertices $x'$, for which $p\left(y\middle|x'\right) > 0$, and these *adjacent* states are *mutually orthogonal*; i.e., for any two $x_1'$ and $x_2'$, $\left\langle\psi_{x_1'}\middle|\psi_{x_2'}\right\rangle = 0$. Finally, Alice makes her measurement using $B_q = \left\{\left|\psi_{x'}\right\rangle\right\}$, then Bob measures his state $\left|\psi_x\right\rangle^*$ in $B_q = \left\{\left|\psi_{x'}\right\rangle\right\}$ to produce $\left|\psi_{x'}\right\rangle^*$.

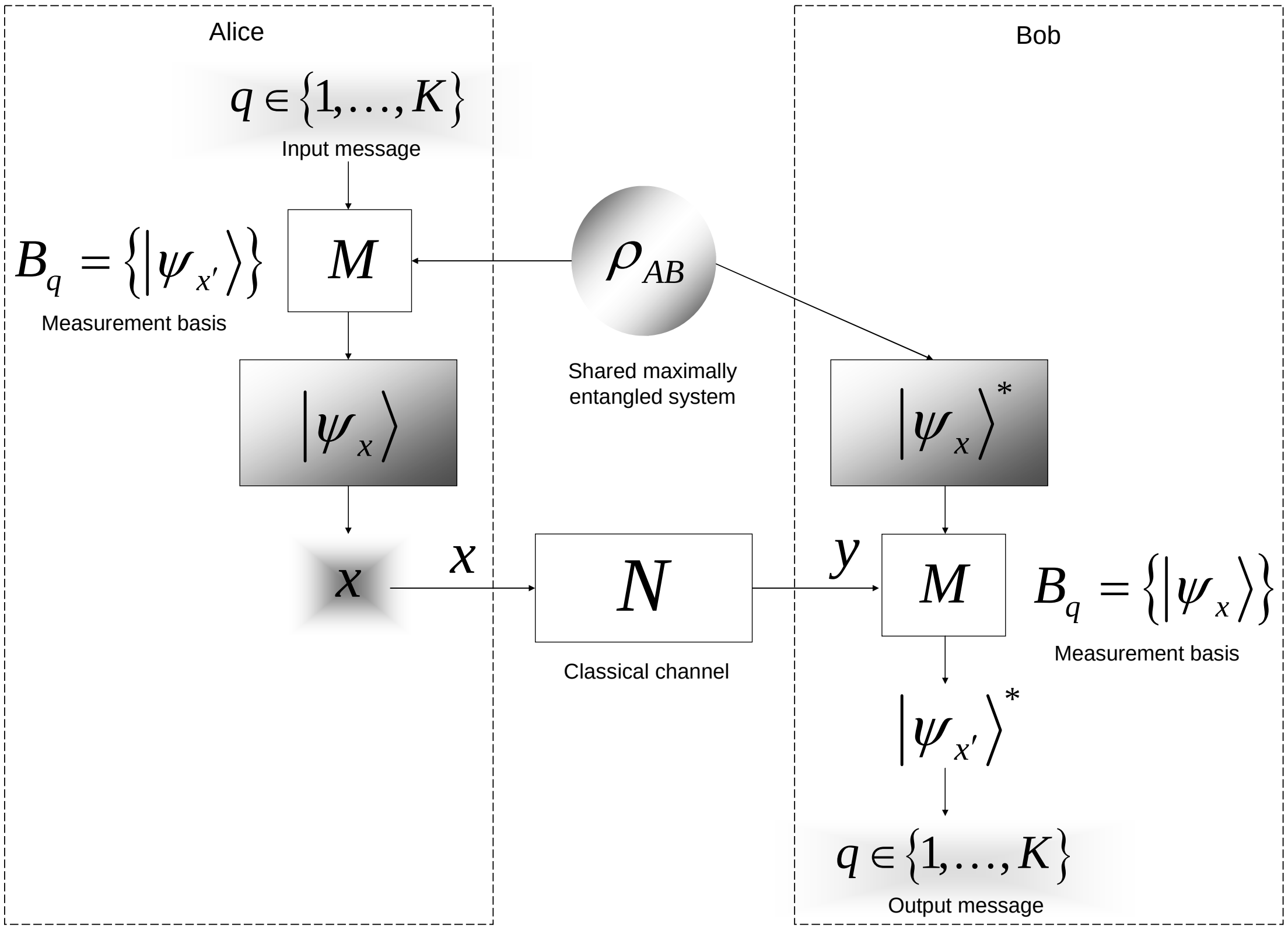

**Fig. C.14.** The steps of the entanglement-assisted zero-error quantum communication protocol.

In order to make the above explanations more plausible, let us provide an example. Supposed Alice's set contains $K = 6$ codewords and she shares a rank-four (*i.e.*, $d$=4) maximally entangled qudit state with Bob

$$\Phi_{AB} = \frac{1}{\sqrt{4}}\sum_{i=0}^{3}\left|i\right\rangle_A\left|i\right\rangle_B, \tag{C.31}$$

however, in the general case $d$ can be chosen as large as Alice and Bob would like to use. Alice measures her system from the maximally entangled state, and she chooses a basis among the $K$ possible states, according to which message $q$ she wants to send Bob. Alice's measurement outcome is depicted by $x$, which is a random value. Alice sends $q$ and $x$ to the classical channel. In the next phase, Bob performs a projective measurement to decide which $x$ value was made to the classical channel by Alice. After Bob has determined it, he can answer which one of the possible $K$ messages had been sent by Alice with the help of the maximally entangled system.

Alice makes her measurement on her side using one of the six possible bases $B_q = \left\{ \left| \psi_{x'} \right\rangle \right\}$ on her half of the state $\rho_{AB}$. Her system collapses to $\left| \psi_x \right\rangle \in B_q$, while Bob's system collapses to $\left| \psi_x \right\rangle^*$, conditioned on $x$. Alice makes $x$ to the classical channel; Bob will receive classical message $y$.

From the channel output $y = N\left(x\right)$, where $N$ is the classical channel between Alice and Bob, Bob can determine the mutually adjacent inputs (i.e., those inputs which could produce the given output). If Bob makes a measurement in basis $B_q = \left\{ \left| \psi_x \right\rangle \right\}$, then he will get $\left| \psi_{x'} \right\rangle^*$, where these states for a given set of $x'$ corresponding to possible $x$ are *orthogonal states*, so he can determine $x$ and the original message $q$. The channel output gives Bob the information that some set of mutually adjacent inputs were used on Alice's side. On his half of the entangled system, the states will be mutually orthogonal. A measurement on these mutually orthogonal states will determine Bob's state and he can tell Alice's input with certainty.

Using this protocol, the number of mutually non-adjacent input messages is

$$K^E \geq 6\,, \tag{C.32}$$

while if Alice and Bob would like to communicate with zero-error but without shared entanglement, then $K = 5$. As follows, for the single-use classical zero-error capacities we get

$$C_0^{(1)} = \log\left(5\right) \tag{C.33}$$

and

$$C_0^{E(1)} = \log\left(K^E\right) = \log\left(6\right), \tag{C.34}$$

while for the asymptotic entanglement-assisted classical zero-error capacity,

$$C_0^E \geq \log\left(K^E\right) = \log\left(6\right). \tag{C.35}$$

According to Alice's $K^E = 6$ messages, the hypergraph can be partitioned into six cliques of size $d = 4$. The adjacent vertices are denoted by a common loop. The overall system contains $6 \times 4 = 24$ basis vectors. These vectors are grouped into $K^E = 6$ orthogonal bases. Two input vectors are connected in the graph if they are adjacent vectors; i.e., they can produce the same output [Imre12].

The hypergraph $\mathcal{G}_H$ of this system is shown in Fig. C.15. The mutually non-adjacent inputs are denoted by the great shaded circles. An important property of the entanglement-assisted classical zero-error capacity is that the number of maximally transmittable messages is not equal to the number of non-adjacent inputs. While the hypergraph has five independent vertices, the maximally transmittable messages are greater than or equal to six. The confusability graph of this system for a single use of quantum channel $\mathcal{N}$ would consist of $6 \times 4 \times 9 = 216$ connections, while the hypergraph has a significantly lower number ($6 \times 6 = 36$) of hyperedges. The adjacent vertices are depicted by the loops connected by the thick lines. The six possible messages are denoted by the six, four dimensional (i.e., each contains four vertices) cliques $\left\{\kappa_1,\ldots,\kappa_K\right\}$. The cliques (dashed circles) show the set of those input messages which could result in the same output with a given probability $p > 0$. We note, the cliques are defined in the $\mathcal{G}_n$ confusability graph representation, but we also included them on the hypergraph $\mathcal{G}_H$. The adjacent vertices which share a loop represent mutually orthogonal input states. For these mutually orthogonal inputs the output will be the same.

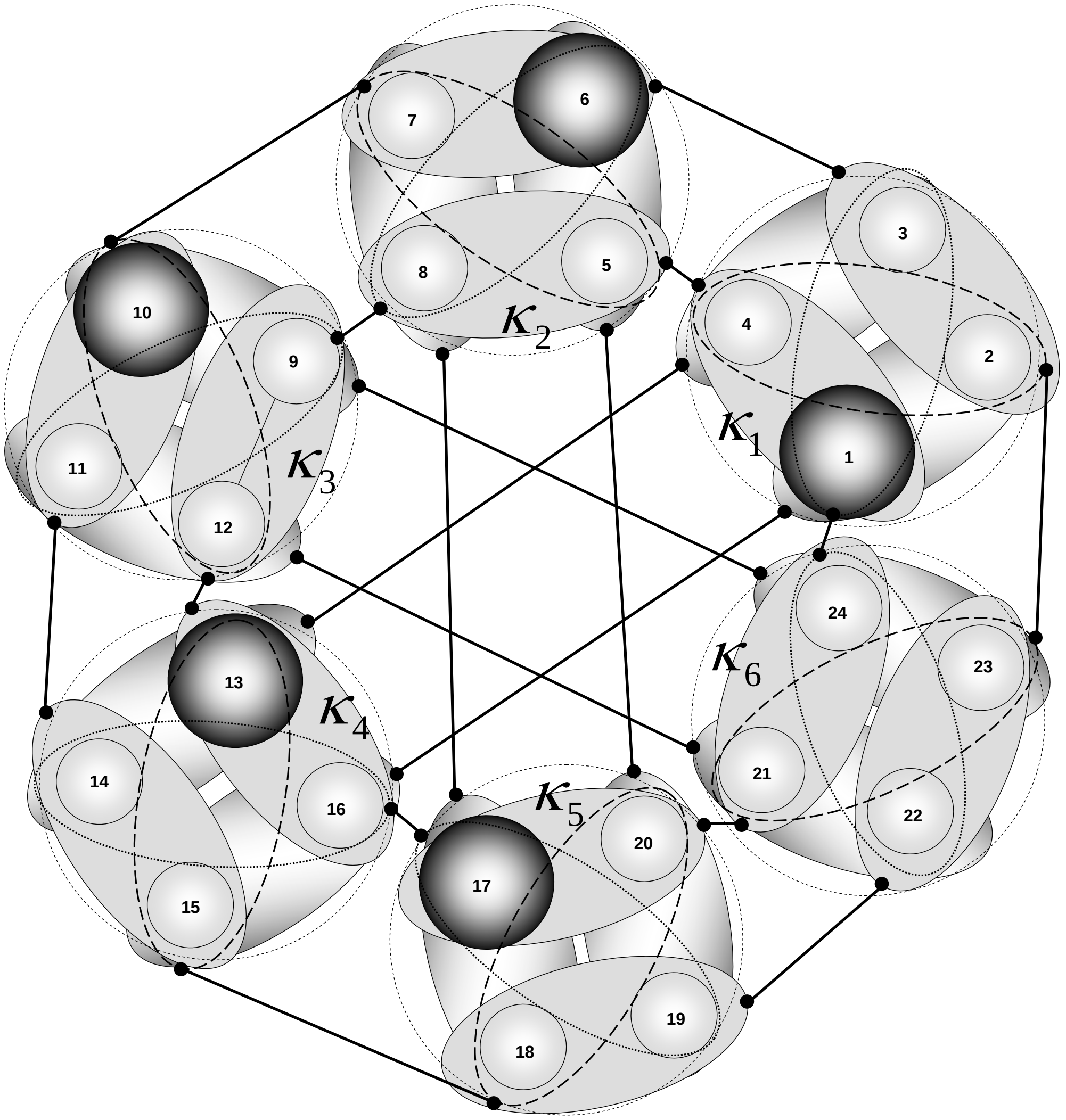

**Fig. C.15.** The hypergraph of an entanglement-assisted zero-error quantum code. The non-adjacent inputs are depicted by the great shaded circles. The adjacent vertices are depicted by loops connected by the thick lines.

The complete theoretical background of this example, i.e., the proof of the fact, that entanglement can increase the asymptotic classical zero-error capacity $C_0(\mathcal{N})$ of a quantum channel was described in [Cubitt10].

We have seen in this subsection that shared entanglement between Alice and Bob can help to increase the maximally transmittable classical messages using noisy quantum channels with zero error probability. According to the *Cubitt-Leung-Matthews-Winter* theorem (CLMW theorem) [Cubitt10] there exist entanglement-assisted quantum communication protocol which can send one of $K$ messages with *zero error*; hence for the entanglement-assisted asymptotic classical zero-error capacity

$$\log\left(K\right) \leq C_0 = \lim_{n\to\infty}\frac{1}{n}\log\left(K\left(\mathcal{N}^{\otimes n}\right)\right) < C_0^E = \lim_{n\to\infty}\frac{1}{n}\log K^E\left(\mathcal{N}^{\otimes n}\right) \geq \log\left(K^E\right). \quad \text{(C.36)}$$

Entanglement is very useful in zero-error quantum communication, since with the help of entanglement the maximum amount of perfectly transmittable information can be achieved.

As was show by Leung *et al.* [Leung10], using special input codewords (based on a special Pauli graph), entanglement can help to increase the classical zero-error capacity to the maximum achievable HSW capacity; that is, there exists a special combination for which the entanglement-assisted classical zero-error capacity $C_0^E\left(\mathcal{N}\right)$ is

$$C_0^E\left(\mathcal{N}\right) = \log\left(9\right), \quad \text{(C.37)}$$

while the classical zero-error capacity is

$$C_0\left(\mathcal{N}\right) = \log\left(7\right), \quad \text{(C.38)}$$

i.e., with the help of entanglement-assistance the number of possible input messages ($K$) can be increased.

Another important discovery is that for this special input system the entanglement-assisted classical zero-error capacity, $C_0^E\left(\mathcal{N}\right)$, is equal to the maximal transmittable classical information; that is

$$C_0^E\left(\mathcal{N}\right) = C\left(\mathcal{N}\right) = \log\left(9\right). \quad \text{(C.39)}$$

As it was shown in [Cubitt10], the maximal amount of transmittable classical information which can be sent through a noisy quantum channel increases with the number of channel uses, and with the help or EPR input states (for this special Pauli graph-based code) the classical HSW capacity can be reached, which is also the upper bound of the classical zero-error capacity. We note that the complete theoretical background on the possible impacts of entanglement on the zero-error capacities is not completely clarified, and research activities are currently in progress; on the other hand one thing is certain: *without entanglement the zero-error capacities (classical or quantum) of quantum channels cannot be superactivated* (see Section 7 of the Ph.D Thesis).

## C.8 The Quantum Capacity

As we have shown, a quantum channel can be used to transmit classical information and the amount of maximal transmittable information depends on the properties of the encoder and decoder setting, or whether the input quantum states are mixed or pure. Up to this point, we have mentioned just the transmission of classical information through the quantum channel—here we had broken this picture. The HSW-theorem was a very useful tool to describe the amount of maximal transmittable classical information over a noisy quantum channel, however we cannot use it to describe the amount of maximal transmittable *quantum information.* The fidelity for two pure quantum states is defined as

$$F\left(\left|\varphi\right\rangle,\left|\psi\right\rangle\right)=\left|\left\langle\varphi\middle|\psi\right\rangle\right|^{2}. \tag{C.40}$$

The fidelity of quantum states can describe the relation of Alice pure channel input state $\left|\psi\right\rangle$ and the received mixed quantum system $\sigma=\sum_{i=0}^{n-1}p_i\rho_i=\sum_{i=0}^{n-1}p_i\left|\psi_i\right\rangle\left\langle\psi_i\right|$ at the channel output as

$$F\left(\left|\psi\right\rangle,\sigma\right)=\left\langle\psi\middle|\sigma\middle|\psi\right\rangle=\sum_{i=0}^{n-1}p_i\left|\left\langle\psi\middle|\psi_i\right\rangle\right|^{2}. \tag{C.41}$$

Fidelity can also be defined for *mixed* states $\sigma$ and $\rho$

$$F\left(\rho,\sigma\right)=\left[Tr\left(\sqrt{\sqrt{\sigma}\rho\sqrt{\sigma}}\right)\right]^2=\sum_i p_i\left[Tr\left(\sqrt{\sqrt{\sigma_i}\rho_i\sqrt{\sigma_i}}\right)\right]^2. \tag{C.42}$$

Let us assume that we have a quantum system denoted by $A$ and a reference system $P$. Initially, the quantum system $A$ and the reference system $P$ are in a *pure entangled* state, denoted by $\left|\psi^{PA}\right\rangle$. The density matrix $\rho_A$ of system $A$ can be expressed by a partial trace over $P$, as follows

$$\rho_A=Tr_P\left(\left|\psi^{PA}\right\rangle\left\langle\psi^{PA}\right|\right). \tag{C.43}$$

The entanglement between the initial quantum system and the reference state is illustrated in Fig. C.16.

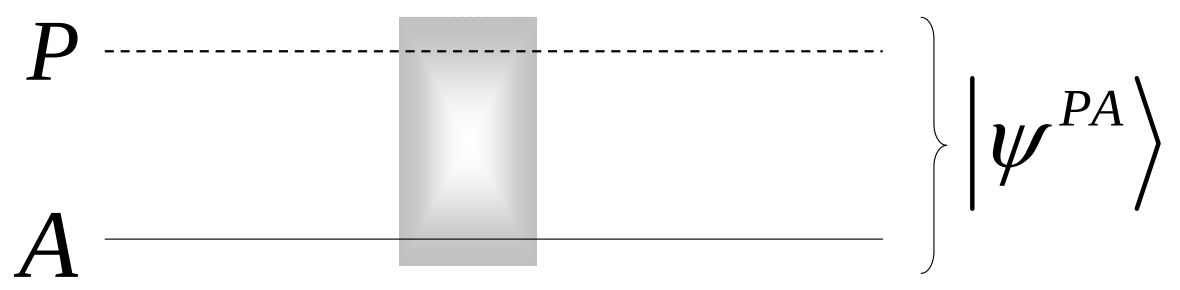

**Fig. C.16.** Initially, the quantum system and the reference system are in a pure entangled state.

In the next step, $\rho_A$ will be transmitted through the quantum channel $\mathcal{N}$, while the reference state $P$ is *isolated from the environment*, hence it has not been not modified during the transmission. After the quantum system $\rho_A$ is transmitted through the quantum channel, the final state will be

$$\rho^{PB}=\left(\mathcal{I}^P\otimes\mathcal{N}^A\right)\left(\left|\psi^{PA}\right\rangle\left\langle\psi^{PA}\right|\right), \tag{C.44}$$

where $\mathcal{I}^P$ is the identity transformation realized on the reference system $P$. After the system $A$ is sent through the quantum channel, both the quantum system $A$ and the entanglement between $A$ and $P$ are affected, as we illustrated in Fig. C.17.

The resultant output system is denoted by $B$.

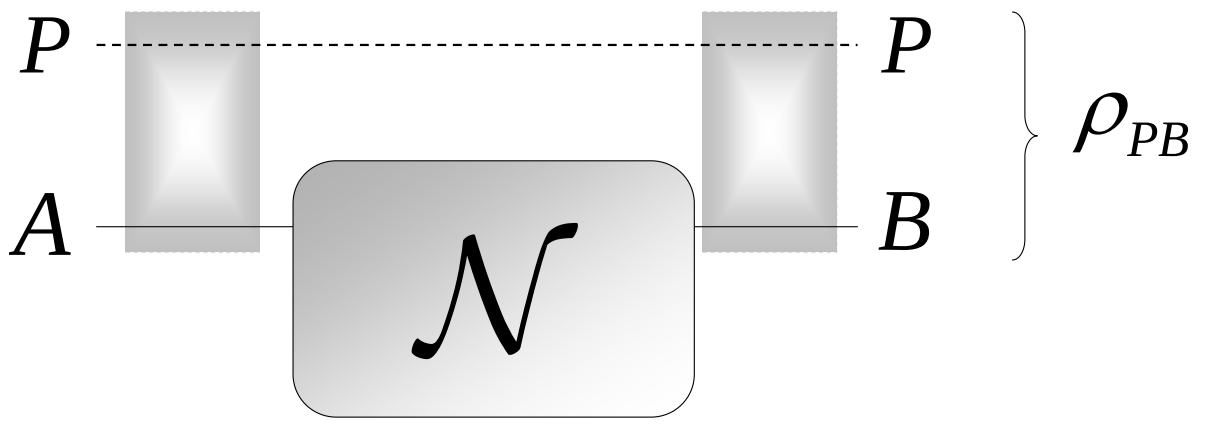

**Fig. C.17.** After the system $A$ is sent through the quantum channel, both the quantum system $A$ and the entanglement between $A$ and $P$ are affected.

Now, we can study the preserved entanglement between the two systems $A$ and $P$. Entanglement fidelity $F_E$ measures the fidelity between the initial pure system $\left|\psi^{PA}\right\rangle$ and the mixed output quantum system $\rho_{PB}$ as follows

$$F_E = F_E\left(\rho_A, \mathcal{N}\right) = F\left(\left|\psi^{PA}\right\rangle, \rho_{PB}\right) = \left\langle\psi^{PA}\right|\left(\mathcal{I}^P \otimes \mathcal{N}^A\right)\left(\left|\psi^{PA}\right\rangle\left\langle\psi^{PA}\right|\right)\left|\psi^{PA}\right\rangle. \text{ (C.45)}$$

It is important to highlight the fact that $F_E$ depends on $\left|\psi^{PA}\right\rangle$ i.e., on the reference system.

The whole process is shown in Fig. C.18. Alice can apply many independent channel uses of the same noisy quantum channel to transmit the quantum information.

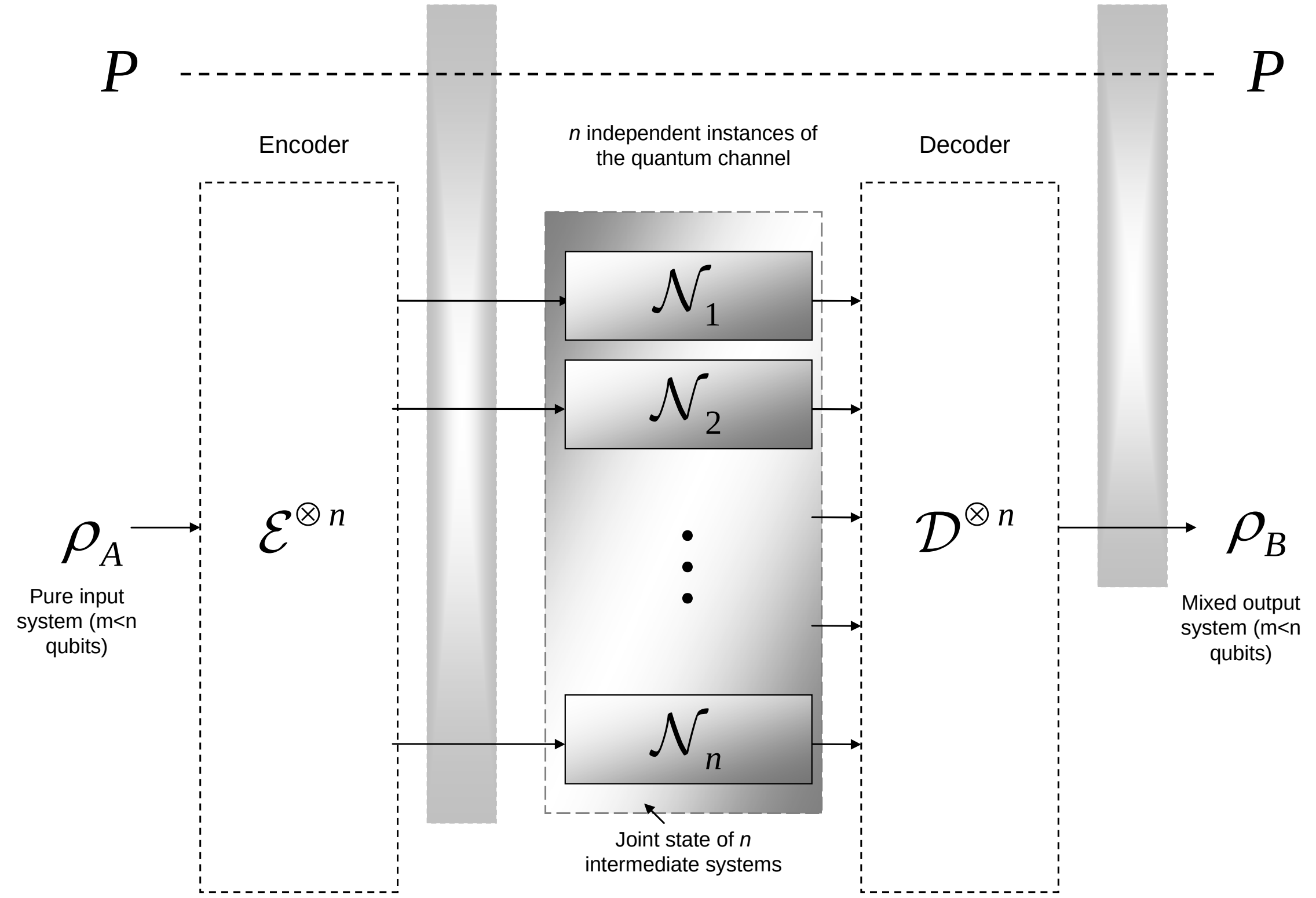

**Fig. C.18.** Transmission of quantum information with multiple uses of the quantum channel.

Similar to encoding classical information into the quantum states, the quantum messages can be transmitted over copies of a quantum channel. In this case, we have $n$ copies of a quantum channel $\mathcal{N}$ , which will be denoted as $\mathcal{N}^{\otimes n}$.

## C.8.1 Quantum Coherent Information

Finally let us make an interesting comparison between quantum coherent information and quantum mutual information. For classical information transmission, the *quantum mutual information* can be expressed according to Appendix B:

$$I(A:B) = \mathrm{S}(\rho_A) + \mathrm{S}(\rho_B) - \mathrm{S}(\rho_{AB}). \tag{C.46}$$

However, in case of *quantum coherent information,* the term $\mathrm{S}(\rho_A)$ vanishes. The channel transformation $\mathcal{N}$ modifies Alice's original state $\rho_A$, hence Alice's original density matrix cannot be used to express $\mathrm{S}(\rho_A)$, *after Alice's qubit has been sent through* the quantum channel. After the channel has modified Alice's quantum state, the initially sent qubit vanishes from the system, and we will have a different density matrix, denoted by $\rho_B = \mathcal{N}(\rho_A)$. The coherent information can expressed as $\mathrm{S}(\rho_B) - \mathrm{S}(\rho_{AB})$, where $\rho_B$ is the transformed state of Bob, and $\mathrm{S}(\rho_{AB})$ is the joint von Neumann entropy.

As follows, we will have $\mathrm{S}(\rho_B) - \mathrm{S}(\rho_{AB})$, which is equal to the *negative conditional entropy* $\mathrm{S}(A|B)$, (see Section 2 of the Ph.D Thesis) thus

$$I_{coh}(\rho_A : \mathcal{N}(\rho_A)) = \mathrm{S}(\rho_B) - \mathrm{S}(\rho_{AB}) = -\mathrm{S}(A|B). \tag{C.47}$$

This very interesting result is summarized in Fig. C.19.

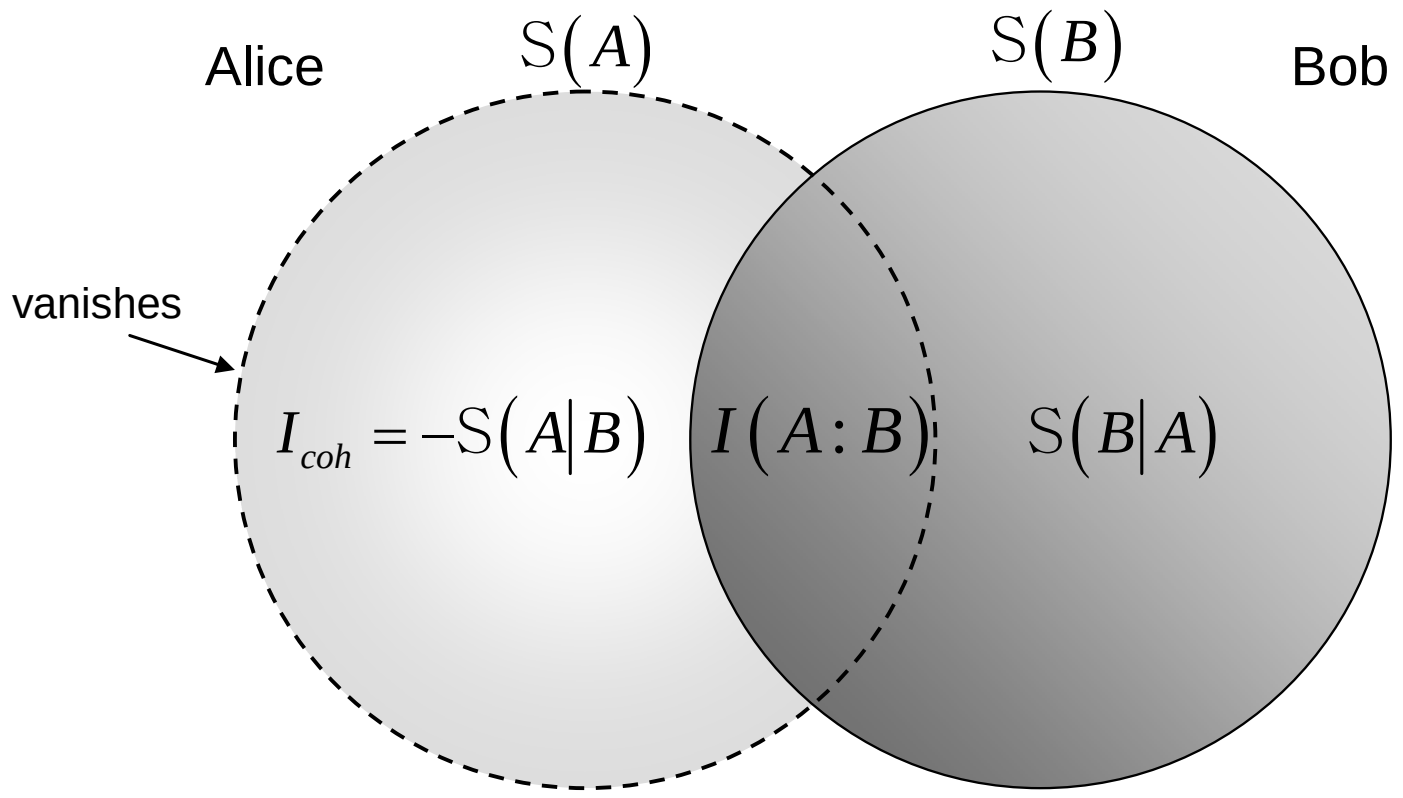

**Fig. C.19.** The expression of quantum coherent information. The source entropy of Alice's state vanishes after the state is passed to Bob.

As we have seen in this section, there is a *very important difference* between the maximized quantum *mutual information* and the maximized *quantum coherent information* of a quantum channel. While the former is always additive, it does not remain true for the latter. *The quantum coherent information* is defined as follows

$$I_{coh}(\mathcal{N}) = \mathrm{S}(\rho_B) - \mathrm{S}(\rho_E), \tag{C.48}$$

where $\rho_B$ refers to the output of the quantum channel $\mathcal{N}$, while $\rho_E$ is the state of the environment. The term $\mathrm{S}(\rho_B)$ measures how much information Bob has, while $\mathrm{S}(\rho_E)$ measures how much information environment has. As follows, the quantum coherent information $I_{coh}(\mathcal{N})$ measures that "*how much more information Bob has than the environment*" about the original input quantum state [Schumacher2000], [Nielsen2000].

## C.8.2. The Asymptotic Quantum Capacity of Quantum Channels

The concept of quantum coherent information can be used to express the *asymptotic* quantum capacity $Q(\mathcal{N})$ of quantum channel $\mathcal{N}$ called the *Lloyd-Shor-Devetak (LSD)* capacity as follows

$$\begin{aligned} Q(\mathcal{N}) &= \lim_{n\to\infty}\frac{1}{n}Q^{(1)}\left(\mathcal{N}^{\otimes n}\right) \\ &= \lim_{n\to\infty}\frac{1}{n}\max_{all\ p_i,\rho_i} I_{coh}\left(\rho_A : \mathcal{N}^{\otimes n}\left(\rho_A\right)\right) \\ &= \lim_{n\to\infty}\frac{1}{n}\max_{all\ p_i,\rho_i}\left(\mathrm{S}(\rho_B)-\mathrm{S}(\rho_E)\right), \end{aligned} \tag{C.49}$$

where $Q^{(1)}(\mathcal{N})$ represents the *single-use* quantum capacity. The asymptotic quantum capacity can also be expressed using the Holevo information, since as we have seen previously, the quantum coherent information can be derived from the Holevo information

$$Q(\mathcal{N}) = \lim_{n\to\infty}\frac{1}{n}\max_{all\ p_i,\rho_i}\left(\mathcal{X}_{AB}-\mathcal{X}_{AE}\right), \tag{C.50}$$

where $\mathcal{X}_{AB}$ denotes the classical information sent from Alice to Bob, and $\mathcal{X}_{AE}$ describes the classical information passed from Alice to the environment during the transmission.

Quantum coherent information plays a fundamental role in describing the maximal amount of transmittable quantum information through a quantum channel, and - as the Holevo quantity has deep relevance in the classical HSW capacity of a quantum channel - the quantum coherent information will play a crucial role in the LSD capacity of a quantum channel.

### C.8.3 The Assisted Quantum Capacity

There is another important quantum capacity called *assisted capacity* which measures the quantum capacity for a channel pair that contains different channel models – and it will have relevance in the *superactivation* of quantum channels (see Chapter 4). If we have a quantum channel $\mathcal{N}$, then we can find a symmetric channel $\mathcal{A}$, that results in the following assisted quantum capacity

$$Q_{\mathcal{A}}\left(\mathcal{N}\right)=Q\left(\mathcal{N}\otimes\mathcal{A}\right). \tag{C.51}$$

We note, that the symmetric channel has unbounded dimension in the strongest case, and this quantity cannot be evaluated in general. $Q_{\mathcal{A}}\left(\mathcal{N}\right)$ makes it possible to realize the superactivation of zero-capacity (in terms of LSD capacity) quantum channels. For example if we have a zero-capacity *Horodecki channel* and a zero-capacity symmetric channel, then their combination can result in positive joint capacity, as it will be shown in Chapter 4.

## C.9 The Zero-Error Quantum Capacity

Finally, let us shortly summarize the quantum counterpart of classical zero-error capacity. In the case of quantum zero-error capacities $Q_0^{(1)}\left(\mathcal{N}\right)$ and $Q_0\left(\mathcal{N}\right)$, the encoding and decoding process differs from the classical zero-error capacity: the encoding and decoding are carried out by the *coherent* encoder and *coherent* POVM decoder, whose special techniques make it possible to preserve the quantum information during the transmission [Harrow04], [Hsieh08].

The *single-use* and *asymptotic* quantum zero-error capacity is defined in a similar way

$$Q_0^{(1)}\left(\mathcal{N}\right) = \log\left(K\left(\mathcal{N}\right)\right), \tag{C.52}$$

and

$$Q_0\left(\mathcal{N}\right) = \lim_{n\to\infty}\frac{1}{n}\log\left(K\left(\mathcal{N}^{\otimes n}\right)\right), \tag{C.53}$$

where $K\left(\mathcal{N}^{\otimes n}\right)$ is the maximum number of $n$-length mutually non-adjacent quantum messages that the quantum channel can transmit with zero error. The quantum zero-error capacity is upper bounded by LSD channel capacity $Q\left(\mathcal{N}\right)$; that is, the following relation holds between the quantum zero-error capacities:

$$Q_0\left(\mathcal{N}\right) \leq Q\left(\mathcal{N}\right). \tag{C.54}$$

## C.10 Classical and Quantum Capacities of Quantum Channels

Before introducing some typical quantum channel maps let us summarize the main properties of various capacities in conjunction with a quantum channels.

First of all, the quantum capacity of a quantum channel cannot exceed the maximal classical capacity that can be measured with entangled inputs and joint measurement; at least, it is not possible in general. On the other hand, for some quantum channels, it is conjectured that the maximal *single-use* classical capacity - hence the capacity that can be reached with *product* inputs and a *single* measurement setting - is lower than the *quantum capacity* for the same quantum channel.

For all quantum channels

$$C(\mathcal{N}) \geq Q(\mathcal{N}), \quad \text{(C.55)}$$

where $C(\mathcal{N})$ is the classical capacity of the quantum channel that can be achieved with entangled input states and a joint measurement setting.

On the other hand, it is conjectured that for some quantum channels,

$$C(\mathcal{N}) < Q(\mathcal{N}) \quad \text{(C.56)}$$

holds as long as the classical capacity $C(\mathcal{N})$ of the quantum channel is measured by a classical encoder and a single measurement setting. (The classical capacities of a quantum channel can be measured in different settings, and the strongest version can be achieved with the combination of entangled inputs and joint measurement decoding.)

The relation between the various classical capacities of a quantum channel and this relation's quantum capacity are shown in Fig. C.20.

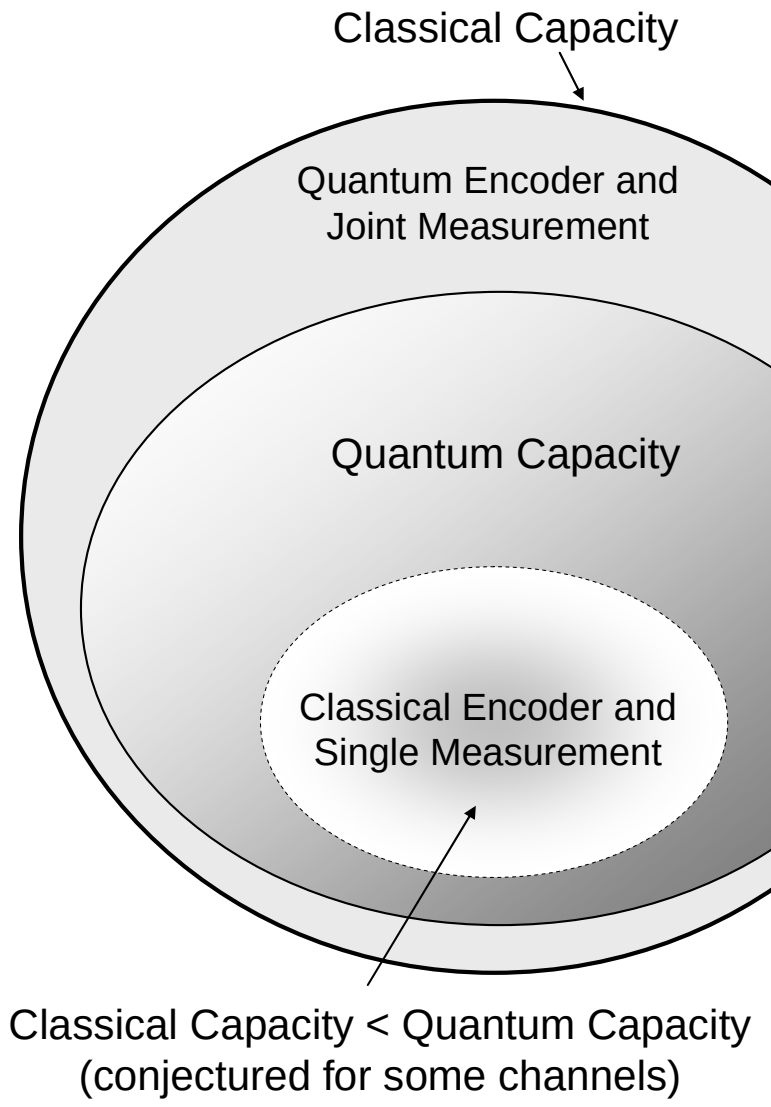

**Fig. C.20.** The relation between classical and the quantum capacity of the same quantum channel.

The description of the capacity of a quantum channel is much more complicated than it is in the case of a classical communication channel [Smith10]. According to our current knowledge, the various capacities of a quantum channel are limited to four different capacity measures that define the greatest rate at which some types of information can be sent through a quantum channel.

We summarized the fundamental differences between classical and quantum capacities in Table C.1.

| *Capacity* | *Type of information* | *Correlation measure between input and output* | *Measure of the Asymptotic channel capacity* |
|---|---|---|---|
| Classical | Classical information | Holevo information (Maximum of Quantum Mutual Information) | Holevo-Schumacher-Westmoreland formula |
| Private Classical | Private information (Classical) | Private information (Difference of Quantum Mutual Information functions) | Li-Winter-Zou-Guo, Smith-Smolin formula |
| Entanglement Assisted Classical | Classical information | Quantum mutual information | Bennett-Shor-Smolin-Thapliyal formula (Equal to single-use quantum mutual information.) |
| Quantum | Quantum information | Quantum Coherent Information | Lloyd-Shor-Devetak formula |

**Table C.1.** The measure of classical and quantum capacities of the quantum channels are different, both in the case of single-use and in the asymptotic formulas.

It can be concluded from Table C.1 that in case of a quantum communication channel we have to count with so many capacities. Each of these capacities is based on different correlation measures: for classical HSW capacity the correlation between the input and the output is measured by the Holevo information, which is a maximization of the quantum mutual information function. The private classical capacity is measured by the private information, which is the maximization of the difference between two quantum mutual information functions. For entanglement assisted capacity the correlation between input and output is also measured by the quantum mutual information, however in this case we do not have to compute the asymptotic version to get the true capacity. Finally, in the case of quantum capacity (LSD capacity) the correlation between the input and output is measured by the quantum coherent information.

# C.11 Related Work on Classical Capacity of Quantum Channels

The classical world with the classical communication channel can be viewed as a special case of a quantum channel, since classical information can be encoded into the qubits—just as into classical bits. Classical information can also be encoded in quantum states. In this section we summarize the most important works related to the classical capacity of the quantum channels.

### The Early Days

At the end of the twentieth century, the capacities of a quantum channel were still an open problem in Quantum Information Theory. Before the several, and rather different, capacities of the quantum channel were recognized, the "academic" opinion was that quantum channels could be used only for the transmission of classical information encoded in the form of quantum states [Holevo73], [Holevo73a]. As has been found later, the classical capacity of the quantum channel can be measured in several different settings. It was shown that the classical capacity depends on whether the input states are entangled or not, or whether the output is measured by single or by joint measurement setting [Bennett97], [Fuchs2000], [King09]. In a specified manner, the classical capacity has been defined for measuring the maximal asymptotic rate at which classical information can be transmitted through the quantum channel, with an arbitrarily high reliability [Barnum97a], [Schumacher97].

The first proposed capacity measure was the *classical capacity* of a quantum channel—denoted by $C(\mathcal{N})$—measures the maximum transmittable classical information—in the form of product or entangled quantum states. The idea of transmitting classical information through a quantum channel was formulated in the 1970s. The Holevo bound was introduced by Holevo in 1973, however the theorem which describes the classical capacity of the quantum channel in an explicit way appeared just about *three decades later*, in the mid 1990s.

The maximal accessible classical information from a quantum source firstly has been characterized by Levitin [Levitin69] and Holevo [Holevo73], [Holevo73a] in the early days, which were some of the first and most important results in Quantum Information Theory regarding the classical capacity of quantum channels. More information about the connection between the Holevo bound and the accessible information (which quantifies the information of the receiver after the measurement) can be found in [Holevo73], [Holevo73a]. Later this result was developed and generalized by Holevo, Schumacher, and Westmoreland, and became known in Quantum Information Theory as the *HSW channel capacity* [Schumacher97], [Holevo98]. The HSW-theorem uses the Holevo information to describe the amount of classical information which can be transmitted through a noisy quantum channel, and it makes possible to apply different measurement constructions on the sender and on the receiver's side. The proofs of the HSW-theorem, such as the direct coding theorem and the converse theorem, with the complete mathematical background can be found in the work of Holevo [Holevo98] and of Schumacher and Westmoreland [Schumacher97]. About the efficiency problems of implementation and construction of joint POVM (Positive Operator Valued Measure) measurement setting, as it was shown in the same works of the authors.

One of the most important result on the mechanism of the encoding of quantum information into physical particles was discovered by Glauber in the very early years of Quantum Information Processing [Glauber1963] and a great summarize from more than four-decades later [Glauber05]. Also from this era and field, important results on the encoding and decoding processes of quantum information were shown in the works of Gordon [Gordon1964] and Helstrom [Helstrom76]. The details of quantum coding for mixed states can be found in the work of Barnum *et al.* [Barnum01]. Later, the detection of quantum information and the process of measurement was completed in Fannes's work [Fannes73], or the work of Helstrom from 1976 [Helstrom76] or Herbert's work from 1982 [Herbert82]. Before their results, Levitin published a paper about the quantum measure of the amount of information in 1969 [Levitin69], which was a very important basis for further work.

**Classical Capacity of a Quantum Channel**

The amount of classical information which can be transmitted through a noisy quantum channel in a reliable form with product input states, using the quantum channel many times, was determined by the HSW-theorem [Holevo98], [Schumacher97]. This coding theorem is an analogue to Shannon's classical channel coding theorem, however it extends its possibilities. The inventors of the HSW-theorem—Holevo, Schumacher and Westmoreland—proved and concluded independently the same result. Holevo's result from 1998 can be found in [Holevo98], Schumacher and Westmoreland's results can be found in [Schumacher97]. They, with Hausladen *et al.* in 1995 [Hausladen95], and in 1996 [Hausladen96], have also confirmed that the maximal classical information which can be transmitted via pure quantum states is bounded by the Holevo information.

A different approach to the proof of the HSW-theorem was presented by Nielsen and Chuang in 2000 [Nielsen2000]. An interesting connection between the mathematical background of the compressibility of quantum states and the HSW-theorem was shown by Devetak in 2003 [Devetak03], who proved that a part of the mathematical background constructed for the compression of quantum information can be used to prove the HSW-theorem. Another interesting approach for proving the HSW-theorem and bounds on the error probability was presented by Hayashi and Nagaoka in 2003 [Hayashi03]. The additivity property of qubit channels which require four inputs to achieve capacity was analyzed by Hayashi *et al.* in [Hayashi05]. Very important connections regarding the transmission of classical information over noisy quantum channels was derived in the work of Schumacher and Westmoreland in 1997 [Schumacher97], and two years later, a very important work was published on the relevance of optimal signal ensembles in the classical capacity of a noisy quantum channels [Schumacher99]. (We also suggest their work on the characterizations of classical and quantum communication processes [Schumacher99a].) The classical information capacity of a class of most important practical quantum channels (Gaussian quantum channels) was shown by Wolf and Eisert [Wolf05] or the work of Lupo *et al.* [Lupo11]. The generalized minimal output entropy conjecture for Gaussian channels was studied by Giovannetti *et al.* [Giovannetti10].

About the role of feedback in quantum communication, we suggest the works of Bowen [Bowen04] and 2005 [Bowen05], the article of Bowen *et al.* [Bowen05a], and the work of Harrow [Harrow04a]. The works of Bowen provide a great introduction to the role of quantum feedback on the classical capacity of the quantum channel, it was still an open question before. As he concluded, the classical capacity of a quantum channel using quantum feedback is equal to the entanglement-assisted classical capacity, the proof was given in Bowen and Nagarajan's paper [Bowen05a].

We have also seen that the noise of a quantum channel can be viewed as a result of the entanglement between the output and the reference system called the purification state (see purification in (B.11)). Some information leaks to the environment, and to the purification state, which purification state cannot be accessed. As is implicitly woven into this section, a noisy quantum channel can be viewed as a special case of an idealistic quantum communication channel. The properties of the general quantum channel model and the quantum mutual information function can be found in the work of Adami and Cerf [Adami96]. A great analysis of completely-positive trace preserving (CPTP) maps was published by Ruskai *et al.* [Ruskai01]. Further information on the classical capacity of a quantum channel can be found in [Bennett98], [Holevo98], [King09], [Nielsen2000].

**Entanglement-assisted Classical Capacity**

In the early 1970s, it was also established that the classical capacity of a quantum channel can be higher with *shared entanglement*—this capacity is known as the *entanglement-assisted classical capacity* of a quantum channel, which was completely defined by Bennett *et al.* just in 1999 [Bennett99], and is denoted by $C_E\left(\mathcal{N}\right)$. The preliminaries of the defini-tion of this quantity were laid down by Bennett and Wiesner in 1992 [Bennett92c]. Later, in 2002 Holevo published a review paper about the entanglement-assisted classical capacity of a quantum channel [Holevo02a].

Entanglement-assisted classical communication requires a super-dense protocol-like encoding and decoding strategy [Bennett02]. About the classical capacity of a noiseless quantum channel assisted by noisy entanglement, an interesting paper was published by Horodecki *et al.* in 2001 [Horodecki01]. In the same work the authors have defined the "noisy version" of the well-known superdense coding protocol, which originally was defined by Bennett in 1992 [Bennett92c] for ideal (hence noiseless) quantum channels. As can be found in the works of Bennett *et al.* from 1999 [Bennett99] and from 2002 [Bennett02], the *entanglement-assisted classical capacity* opened the possibility to transmit more classical information using shared entanglement (in case of single-use capacity). As can be checked by the reader, the treatment of entanglement-assisted classical capacity is based on the working mechanism of the well-known superdense coding protocol—however, classical entanglement-assisted classical capacity used a noisy quantum channel instead of an idealistic one [Wilde11].

Bennett, in two papers from 1999 [Bennett99] and 2002 [Bennett02] showed that the *quantum mutual information* function (see Adami and Cerf's work [Adami96]) can be used to describe the classical entanglement-assisted capacity of the quantum channel i.e., the *maximized quantum mutual information of a quantum channel and the entanglement-assisted classical capacity are equal.* The connection between the quantum mutual information and the entanglement-assisted capacity can be found in the works of Bennett *et al.* [Bennett99] and [Bennett02]. In the latter work, the formula of the quantum-version of the well-known classical Shannon formula was generalized for the classical capacity of the quantum channel. In these two papers the authors also proved that the entanglement-assisted classical capacity is an upper bound of the HSW channel capacity.

Holevo gave an explicit upper bound on the classical information which can be transmitted through a noisy quantum channel, it is known as the Holevo-bound. The Holevo-bound states that the most classical information which can be transmitted in a qubit (i.e., two level quantum system) through a noiseless quantum channel in a reliable form, is one bit. However, as was shown later by Bennett *et al.* in 1999 [Bennett99], the picture changes, if the parties use shared entanglement (known as the *Bennett-Shor-Smolin-Thapliyal, or the BSST-* theorem). As follows, the BSST-theorem gives a closer approximation to the maximal transmittable classical information (i.e., to the "single-use" capacity) over quantum channels, hence it can be viewed as the *true "quantum version" of the well known classical Shannon capacity formula* (since it is a maximization formula), instead of the "non entanglement-assisted" classical capacity.

Moreover, the inventors of the BSST-theorem have also found a very important property of the entanglement-assisted classical capacity: *its single-use version is equal to the asymptotic version*, which implies the fact that no regularization is needed. (As we have seen in this section, we are not so lucky in the case of general classical and classical private capacities. As we will show in Section 4, we are "unlucky" in the case of quantum capacity, too.) They have also found that no classical feedback channel can increase the entanglement-assisted classical capacity of a quantum channel, and this is also true for the classical (i.e., the not entanglement-assisted one) capacity of a quantum channel. These results were also confirmed by Holevo in 2002 [Holevo02a]. It was a very important discovery in the history of the classical capacity of the quantum channel, and due to the BSST-theorem, the analogue with classical Shannon's formula *has been finally completed.* Later, it was discovered that in special cases the entanglement-assisted capacity of a quantum channel can be improved [Harrow04], [Patrón09].

The Holevo information can be attained even with pure input states, and the concavity of the Holevo information also shown. The concavity can be used to compute the classical HSW capacity of quantum channels, since the maximum of the transmittable information can be computed by a local maximum among the input states. Moreover, as was shown by Bennett *et al.* in 2002, the concavity holds for the entanglement-assisted classical capacity, too [Bennett02]— the concavity, along with the non-necessity of any computation of an asymptotic formula, and the use of classical feedback channels to improve the capacity, *makes the entanglement-assisted classical capacity the most generalized classical capacity*—and it has the same role as Shannon's formula in classical Information Theory.

The fact that the classical feedback channel does not increase the classical capacity and the entanglement-assisted classical capacity of the quantum channel, follows from the work of Bennett *et al.*, and the proof of the BSST-theorem [Bennett02]. Wang and Renner's work [Wang10] introduces the reader to the connection between the single-use classical capacity and hypothesis testing.

**The Private Classical Capacity**

The third classical capacity of the quantum channel is the *private classical capacity*, denoted by $P(\mathcal{N})$. The concept of private classical capacity was introduced by Devetak in 2003 [Devetak03], and one year later by Cai *et al.* in 2004 [Cai04]. Private classical capacity measures classical information, and it is always at least as large as the single-use quantum capacity (or the quantum coherent information) of any quantum channel. As shown in [Devetak05a], for a degradable quantum channel (see Chapter 4) the coherent information (see Chapter 3) is additive [Devetak05a],—however for a general quantum channel these statements do not hold. The additivity of private information would also imply the fact that shared entanglement cannot help to enhance the private classical capacity for degradable quantum channels. The complete proof of the private classical capacity of the quantum channel was made by Devetak [Devetak03], who also cleared up the connection between private classical capacity and the quantum capacity. As was shown by Smith *et al.* [Smith08d], the private classical capacity of a quantum channel is additive for degradable quantum channels, and closely related to the quantum capacity of a quantum channel (moreover, Smith has shown that the private classical capacity is equal to the quantum coherent information for degradable channels), since in both cases we have to "protect" the quantum states: in the case of private classical capacity the enemy is called Eve (the eavesdropper), while in the latter case the name of the enemy is “environment.” As was shown in [Devetak03], the eavesdropper in private coding acts as the environment in quantum coding of the quantum state, and vice-versa. This “gateway” or “dictionary” between the classical capacity and the quantum capacity of the quantum channel was also used by Devetak [Devetak03], by Devetak and Shor [Devetak05a] and by Smith and Smolin [Smith08d], using a different interpretation.

About the coherent communication with continuous quantum variables over the quantum channels a work was published Wilde *et al.* in [Wilde07] and [Wilde10]. On the noisy processing of private quantum states, see the work of Renes *et al.* [Renes07]. A further application of private classical information in communicating over adversarial quantum channels was shown by Leung *et al.* [Leung08]. Further information about the private classical capacity can be found in [Devetak03], [Devetak05], [Bradler09], [Li09], [Smith08d], [Smith09a], [Smith09b]. An other important work on non-additive quantum codes was shown by Smolin *et al.* [Smolin07]. A great summary on the main results of Quantum Shannon Theory was published by Wilde [Wilde11]. For further information on quantum channel capacities and advanced quantum communications see the book of Imre and Gyongyosi [Imre12] and [Gyongyosi12d].

### The Zero-Error Classical Capacity

The properties of *zero-error* communication systems are discussed in Shannon’s famous paper on the zero-error capacity of a noisy channel [Shannon56], in the work of Körner and Orlitsky on zero-error information theory [Körner98], and in the work of Bollobás on modern graph theory [Bollobas98]. We also suggest the famous proof of Lovász on the Shannon capacity of a graph [Lovász79]. The proof of the classical zero-error capacity of quantum channel can be found in Medeiros’s work [Medeiros05]. Here, he has shown, that the classical zero-error capacity of the quantum channel is also bounded above by the classical HSW capacity. The important definitions of quantum zero-error communication and the characterization of quantum states for the zero-error capacity were given by Medeiros *et al.*, in [Medeiros06]. On the complexity of computation of zero-error capacity of quantum channels see the work of Beigi and Shor [Beigi07]. The fact, that the zero-error classical capacity of the quantum channel can be increased with entanglement, was shown by Cubitt *et al.* in 2010 [Cubitt10]. The role of entanglement in the asymptotic rate of zero-error classical communication over quantum channels was shown by Leung *et al.* in 2010 [Leung10]. For further information about the theoretical background of entanglement-assisted zero-error quantum communication see [Cubitt10] and for the properties of entanglement, the proof of the Bell-Kochen-Specker theorem in [Bell1966], [Kochen1967].

## C.12 Related Work on Quantum Capacity of Quantum Channels

In this section we summarize the most important works regarding on the quantum capacity of the quantum channels. The quantum capacity is one of the most important result of Quantum Information Theory. The classical capacity of quantum channels was discovered in early years, in the beginning of the 1970s, and the researchers from this era —such as Holevo and Levitin—suggested that physical particles can encode only classical information [Levitin69], [Holevo73], [Holevo73a]. The first step in the encoding of quantum information into a physical particle was made by Feynman, in his famous work from 1982 [Feynman82]. However, the researchers did not see clearly and did not understand completely the importance of quantum capacity until the late 1990s.

#### Quantum Coherent Information

The computation of quantum capacity is based on the concept of *quantum coherent information*, which measures the ability of a quantum channel to preserve a quantum state. The definition of quantum coherent information (in an exact form) was originally introduced by Schumacher and Nielsen in 1996 [Schumacher96c]. This paper is a very important milestone in the history of the quantum capacity, since here the authors were firstly shown that the concept of quantum coherent information can be used to measure the quantum information (hence not the classical information) which can be transmitted through a quantum channel [Wilde11].

The first,—but yet not complete—definitions of the quantum capacity of the quantum channel can be found in Shor's work from 1995 [Shor95], in which Shor has introduced a scheme for reducing decoherence in quantum computer memory, and in Schumacher's articles from one year later [Schumacher96b], [Schumacher96c]. Shor's paper from 1995 mainly discusses the problem of implementation of quantum error correcting schemes - the main focus was not on the exact definition of quantum capacity. Later, Shor published an extended version with a completed proof in 2002 [Shor02].

To transmit quantum information the parties have to encode and decode coherently. An interesting engineering problem is how the receiver could decode quantum states in superposition without the destruction of the original superposition [Wilde07].

The quantum capacity of a quantum channel finally was formulated completely by the *LSD-theorem*, named after Lloyd, Shor and Devetak [Lloyd97], [Shor02], [Devetak03], and they have shown that the rate of quantum communication can be expressed by the quantum coherent information. The LSD-channel capacity states that the asymptotic quantum capacity of the quantum channel is greater than (or equal to in some special cases) the single-use capacity; hence it is not equal to the quantum coherent information [Wilde11].

The quantum capacity was first mentioned by Shor in the middle of the 1990s, who defined the quantum capacity as the highest rate at which quantum information can be transmitted through a quantum channel [Shor96]. Here, the quantum information means the fidelity of the quantum states, and the capacity describes how the channel preserves the fidelity. More information about the properties of fidelity and about the connection with other distance measures can be found in Fuch's works [Fuchs96], [Fuchs98]. An important article regarding the fidelity of mixed quantum states was published by Jozsa in 1994 [Jozsa94]. Fidelity also can be measured between entangled quantum states—a method to compute the fidelity of entanglement was published by Schumacher in 1996 [Schumacher96b]. Here, the upper bound of the quantum capacity was also mentioned, in the terms of quantum coherent information. Nielsen in 2002 [Nielsen02] defined a connection between the different fidelity measures.

**Proofs on Quantum Capacity**

The exact measure of quantum capacity was an open question for a long time. The fact that the quantum capacity cannot be increased by classical communication was formally proven by Bennett *et al.* [Bennett96a], who discussed the mixed state entanglement and quantum error correction. Barnum, in 2000 [Barnum2000], defined the connection between the fidelity and the capacity of a quantum channel, and here he also showed the same result as Bennett *et al.* did in 1996, namely that the quantum capacity cannot increased by classical communication [Bennett96a]. The works of Barnum *et al.* [Barnum2000] and Schumacher *et al.* [Schumacher98a] from the late 1990s gave very important results to the field of Quantum Information Theory, since these works helped to clarify exactly the maximum amount of transmittable quantum information over very noisy quantum channels [Wilde11].

However, a few years before Shor's proof was published, Seth Lloyd gave a different proof in 1997 on the quantum capacity of a noisy quantum channel—but his result cannot be viewed as a complete proof. The details of Lloyd's proof can be found in [Lloyd97], while Shor's results in detail can be found in [Shor02]. On the basis of Shor's results, a proof on the quantum capacity was given by Hayden *et al.* in 2008 [Hayden08b].

The next step in the history of the quantum capacity of the quantum channel was made by Devetak [Devetak03]. Devetak also gave a proof for the quantum capacity using the private classical capacity of the quantum channel, and he gave a clear connection between the quantum capacity and the private classical capacity of the quantum channel.

As in the case of the discoverers of the HSW-theorem, the discoverers gave different proofs. The quantum capacity of a quantum channel is generally lower than the classi-cal one, since in this case the quantum states encode quantum information [Wilde11]. The quantum capacity requires the transmission of arbitrary quantum states, hence not just “special” orthogonal states—which is just a subset of a more generalized case, in which the states can be arbitrary quantum states. On the several different encoder, decoder and measurement settings for quantum capacity see the work of Devetak and Winter [Devetak05], Devetak and Shor’s work [Devetak05a], and the paper of Hsieh *et al.* [Hsieh08].

In this work we have not mentioned the definition of unit resource capacity region and private unit resource capacity region, which can be found in detail in the works of Hsieh and Wilde [Hsieh10], and Wilde and Hsieh [Wilde10], [Wilde11]. In 2005, Devetak and Shor published a work which analyzes the simultaneous transmission of classical and quantum information [Devetak05a]. On the quantum capacities of bosonic channels a work was published by Wolf, Garcia and Giedke, see [Wolf06]. In 2007, Wolf and Pérez-García published a paper on the quantum capacities of channels with small environment, the details can be found in [Wolf07]. They have also determined the quantum capacity of an amplitude damping quantum channel (for the description of amplitude damping channel, see Section E.1.3), for details see the same paper from 2007 [Wolf07]. The properties of quantum coherent information and reverse coherent information were studied by Patrón in 2009 [Patrón09]. The proofs of the LSD channel capacity can be found in [Lloyd97], [Shor02], [Devetak03]. The quantum communication protocols based on the transmission of quantum information were intensively studied by Devetak [Devetak04a], and the work of the same authors on the generalized framework for quantum Shannon theory, from 2008 [Devetak08].

# Appendix D

# Superactivation of Quantum Channels

## D.1 The Additivity Problem of Quantum Channels

As was shown by Shor [Shor04a] a few years before Hastings's discovery, the additivity problem of the classical capacity of the quantum channel can be analyzed from the viewpoint of the additivity of the minimum output states. Shor has also shown that if there exists a combination for which the additivity of minimum entropy channel output states of the channels is violated, then the additivity of the Holevo information is also violated [Shor04]. Later, Hastings have found a possible combination [Hastings09], for which the additivity of the Holevo information was violated (we note, the violation in the minimum output entropies was very small, and it appears only for *sufficiently large* input dimensions), however, for the asymptotic setting the question remains opens.

To define additivity of the various capacities, we introduce a generalized notation of the various capacities of quantum channels. For this purpose, we will use the generalized "*all-in-one*" notation $C_{ALL}\left(\mathcal{N}\right)$ to describe all capacities of the quantum channel. This gener-alized notation involves the classical capacities and the quantum capacity of quantum channel. The non-additivity property consist of two categories: if *subadditivity* holds for quantum channels then the joint capacity of the channels is smaller than the sum of the individual capacities of the quantum channels

$$C_{ALL}\left(\mathcal{N}_1 \otimes \mathcal{N}_2\right) < C_{ALL}\left(\mathcal{N}_1\right) + C_{ALL}\left(\mathcal{N}_2\right). \quad \text{(D.1)}$$

On the other hand in case of *superadditivity* the joint capacity is greater than the sum of the individual capacities of the channels

$$C_{ALL}\left(\mathcal{N}_1 \otimes \mathcal{N}_2\right) > C_{ALL}\left(\mathcal{N}_1\right) + C_{ALL}\left(\mathcal{N}_2\right). \quad \text{(D.2)}$$

Finally, if strict additivity holds then

$$C_{ALL}\left(\mathcal{N}_1 \otimes \mathcal{N}_2\right) = C_{ALL}\left(\mathcal{N}_1\right) + C_{ALL}\left(\mathcal{N}_2\right). \quad \text{(D.3)}$$

To this day, there are very many conjectures on this subject. Three aspects are clear, namely additivity depends on the encoding scheme, the decoding (measurement) scheme and the properties of the map of the channels, too. For further information see the book of Imre and Gyongyosi [Imre12] and Appendix D.

## D.1.1 The Four Propositions for Additivity

For a given quantum channel $\mathcal{N}$, the minimal output entropy $\mathrm{S}$ of $\mathcal{N}$ can be defined as

$$\mathrm{S}_{min}\left(\mathcal{N}\right) = \min_{\rho \in \mathcal{S}} \mathrm{S}\left(\mathcal{N}\left(\rho\right)\right). \quad \text{(D.4)}$$

As was shown by Shor [Shor04a] for the additivity of the *minimum* entropy output of two quantum channels $\mathcal{N}_1$ and $\mathcal{N}_2$, the following property holds

$$\mathrm{S}_{min}\left(\mathcal{N}_1 \otimes \mathcal{N}_2\right) = \mathrm{S}_{min}\left(\mathcal{N}_1\right) + \mathrm{S}_{min}\left(\mathcal{N}_2\right). \quad \text{(D.5)}$$

Using quantum systems $\mathbb{S}_1$, $\mathbb{S}_2$, an entangled state is an element of the set $\mathbb{K}$ defined as

$$\mathbb{K} = \mathbb{S}_1 \otimes \mathbb{S}_2 - \left\{ \rho_1 \otimes \rho_2 \middle| \rho_1 \in \mathbb{S}_1, \rho_2 \in \mathbb{S}_2 \right\}, \tag{D.6}$$

where $\rho_1 \otimes \rho_2$ denotes the decomposable tensor product states. The entanglement of a quantum state can be defined by the entanglement of formation. For a state $\rho$ in a bipartite system $\mathbb{S}_A \otimes \mathbb{S}_B$, the entanglement of formation $E_F$ can be defined as

$$E_F\left(\rho\right) = \min_{\rho} \sum_i p_i \mathrm{S}\left(Tr_B \left|i\right\rangle\left\langle i\right|\right), \tag{D.7}$$

where the minimization is over all possible mixed $\rho$ such that $\rho = \sum_i p_i \left|i\right\rangle\left\langle i\right|$, and $\sum_i p_i = 1$. The additivity property of the entanglement formation can be defined for two states $\rho_1 \in \mathbb{S}_{A1} \otimes \mathbb{S}_{B1}$ and $\rho_2 \in \mathbb{S}_{A2} \otimes \mathbb{S}_{B2}$ in the following way

$$E_F\left(\rho_1 \otimes \rho_2\right) = E_F\left(\rho_1\right) + E_F\left(\rho_2\right), \tag{D.8}$$

where $E_F$ is calculated over bipartite $A$-$B$ partition. We can define the *strong superadditivity* of the entanglement of formation for $\rho \in \mathbb{S}_{A1} \otimes \mathbb{S}_{A2} \otimes \mathbb{S}_{B1} \otimes \mathbb{S}_{B2}$ as

$$E_F\left(\rho\right) \geq E_F\left(Tr_1(\rho)\right) + E_F\left(Tr_2(\rho)\right), \tag{D.9}$$

where $Tr_i(\cdot)$ means tracing out the space $\mathbb{S}_{Ai} \otimes \mathbb{S}_{Bi}$. The four propositions for additivity shown in this section – for the minimum output entropy, for the classical capacity, for the entanglement formation and for the strong superadditivity – *are equivalent.*

## D.1.2 Additivity of Quantum Capacity

Some very important channels – such as the erasure channel, the amplitude damping channel, or some Gaussian channels (see Appendices D and E) – are degradable quantum channels, and hence for these channels the single-use quantum capacity *will be equal* to the asymptotic quantum capacity. We note, that it might also be the case that for some non-degradable quantum channels the capacity also can be characterized by the single-use formula (i.e., quantum coherent information), however this question is still open.

### D.1.2.1 The Degradable Quantum Channel

A quantum channel $\mathcal{N}_{\mathcal{D}}$ is a *degradable* channel if the amount of information leaked to the environment is less than the amount of information which can be transmitted over it. The concept of a degradable quantum channel is illustrated in Fig. D.1.

Bob simulates the environment $E$ with $\mathcal{N}_1$ and $\mathcal{D}$. The input of the first channel is denoted by $\rho_A$ and the output of channel $\mathcal{N}_1$ is denoted by $\sigma_B$. The simulated environment is represented by $\sigma_E$. If the resultant channel $\mathcal{N}_2 = \mathcal{D}\mathcal{N}_1$ is noisier than $\mathcal{N}_1$ was, then $\mathcal{D}$ is a degrading quantum channel. The precise definition is the following: If the channel between Alice and Bob is denoted by $\mathcal{N}_1$, and the channel between Alice and the environment is $\mathcal{N}_2$, and for the noise of the two quantum channels the following relation holds

$$Noise\left(\mathcal{N}_1\right) \leq Noise\left(\mathcal{N}_2\right), \tag{D.10}$$

where $\mathcal{N}_2$ can be expressed from $\mathcal{N}_1$ as

$$\mathcal{N}_2 = \mathcal{D}\mathcal{N}_1, \tag{D.11}$$

then $\mathcal{D}$ denotes the so called *degrading channel*. Bob, having $\mathcal{N}_1$ and $\mathcal{D}$ in his hands can realize (i.e., he can "simulate") the noisier $\mathcal{N}_2$. In this case, $\mathcal{N}_1$ is called a *degradable* quantum channel.

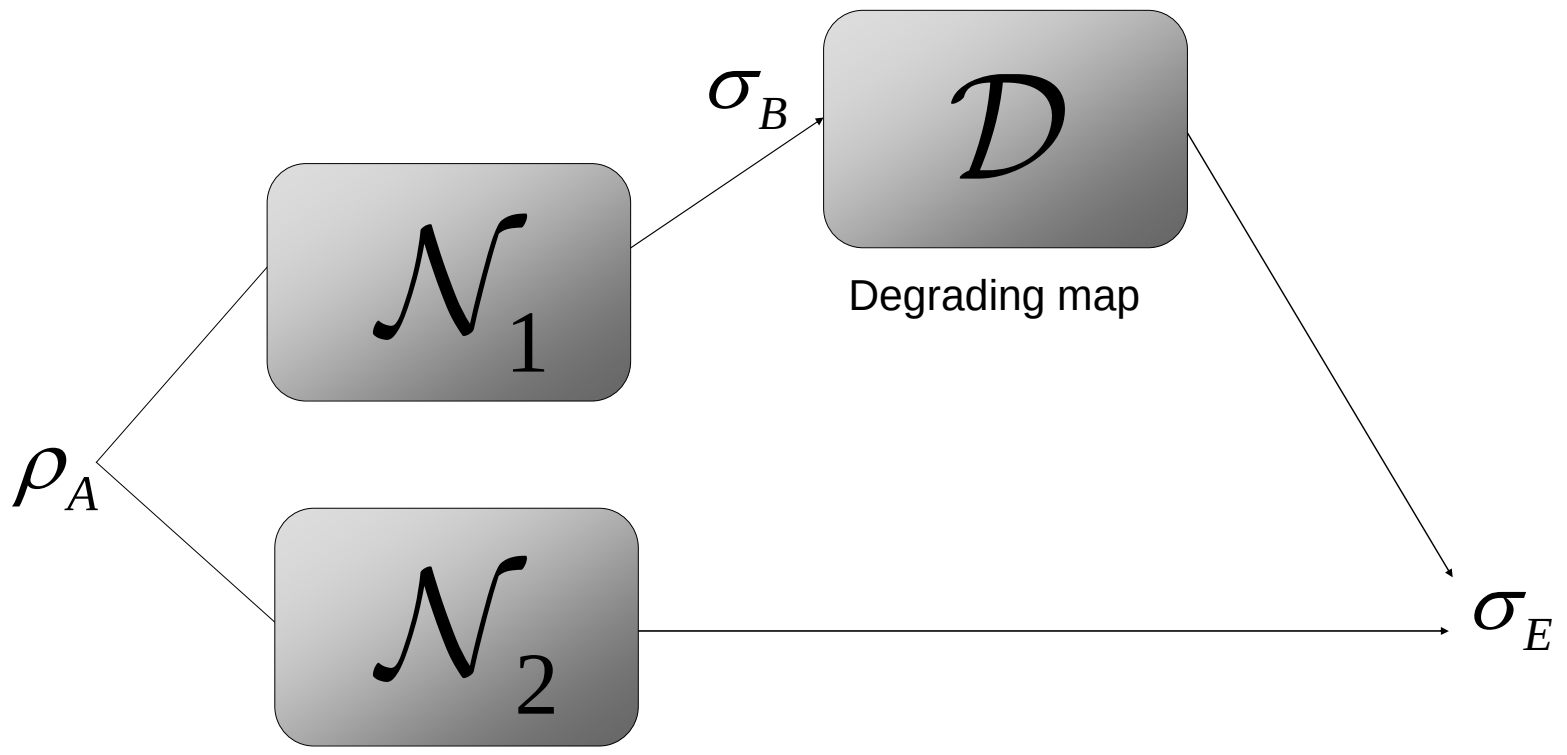

**Fig. D.1.** The concept of a degradable quantum channel. Bob can simulate the environment by means of his degrading channel. $\mathcal{N}_1$ is degradable if the simulated $\mathcal{N}_2$ is noisier than $\mathcal{N}_1$. Bob's input is the output of the $\mathcal{N}_1$; the output of $\mathcal{N}_2$ is the environment state. The environment state also can be generated by Bob with his degrading channel.

We note that the degradability of a quantum channel $\mathcal{N}_1$ can be checked easily, since if we take the inverse of $\mathcal{N}_1$ then $\mathcal{N}_1^{-1}$ is not a completely positive trace preserving map. On the other hand, if $\mathcal{N}_1$ is degradable, then there exists degrading quantum channel $\mathcal{D}$, which is equal to

$$\mathcal{D}= \mathcal{N}_2\mathcal{N}_1^{-1}, \tag{D.12}$$

since $\mathcal{N}_2 = \mathcal{D}\mathcal{N}_1$. Or, in other words, if degrading quantum channel $\mathcal{D}$ exists, then $\mathcal{N}_2\mathcal{N}_1^{-1}$ has to be a completely positive trace preserving map. In this case, $\mathcal{N}_1$ is a degradable channel. If $\mathcal{N}_2\mathcal{N}_1^{-1}$ is not a completely positive trace preserving map, then the degrading quantum channel $\mathcal{D}$ does not exist, and hence $\mathcal{N}_1$ is a not degradable. We note that the degradable quantum channel has tremendous importance in quantum communications, since the most important practical quantum channels – such as the erasure, the amplitude damping, the bosonic Gaussian channels, or the Hadamard channels – are all degradable quantum channels [Bradler09,10].

### D.1.2.2 Description of Degrading Maps

In (D.11) the degrading map $\mathcal{N}_2 = \mathcal{D}\mathcal{N}_1$ means that if the input quantum system $\rho_A$ evolutes a first-type of noise and then, a second-type of noisy transmission, the output state will be equal to the output of the "simulated" channel $\mathcal{N}_2\left(\rho_A\right) = \sigma_E$. We have concatenated two noisy quantum channels. For the input density matrix $\rho$, the output of a quantum channel $\mathcal{N}$ can be given by its Kraus representation as $\mathcal{N}\left(\rho\right) = \sum_i N_i \rho N_i^\dagger$ (see Appendix B). If we would like to describe the output of the concatenated structure of two quantum channels, then it can be done as follows:

$$\mathcal{D}\mathcal{N}_1\left(\rho\right) = \sum_i D_i \mathcal{N}_1\left(\rho\right) D_i^\dagger = \sum_{i,i'} D_i N_{i'} \rho N_{i'}^\dagger D_i^\dagger , \tag{D.13}$$

where $\left\{D_i\right\},\left\{N_i\right\}$ are the Kraus operators of the two channels $\mathcal{D}$ and $\mathcal{N}_1$. For further information see Appendices D and E.

### D.1.2.3 Additivity for Degradable Quantum Channels

For a *degradable* quantum channel $\mathcal{N}_\mathcal{D}$, the following relation holds between the single-use quantum capacity $Q^{(1)}\left(\mathcal{N}_\mathcal{D}\right)$, the asymptotic quantum capacity $Q\left(\mathcal{N}_\mathcal{D}^{\otimes n}\right)$, and the maximized quantum coherent information $\max_{all\ p_i,\rho_i} I_{coh}\left(\mathcal{N}_\mathcal{D}\right)$

$$nQ^{(1)}\left(\mathcal{N}_\mathcal{D}\right) = Q\left(\mathcal{N}_\mathcal{D}^{\otimes n}\right) = n \max_{all\ p_i,\rho_i} I_{coh}\left(\mathcal{N}_\mathcal{D}\right). \tag{D.14}$$

If we choose a non-degradable quantum channel, then we cannot say anything about the asymptotic quantum capacity from the knowledge of the single-use quantum capacity, i.e., it can be additive or non-additive; but the quantum coherent information for a non-degradable quantum channel could be superadditive. (*Note:* the same connection holds between the private classical capacity of $\mathcal{N}_\mathcal{D}$, i.e., $nP^{(1)}\left(\mathcal{N}_\mathcal{D}\right) = P\left(\mathcal{N}_\mathcal{D}^{\otimes n}\right)$ ) For example, the *depolarizing* quantum channel (see Appendix E) is a non-degradable channel, and the following relation holds for its quantum capacity

$$nQ^{(1)}\left(\mathcal{N}_{\mathcal{D}}\right) < Q\left(\mathcal{N}_{\mathcal{D}}^{\otimes n}\right), \tag{D.15}$$

which was proven for $n = 5$ [DiVincenzo98]. In the other cases, we have no superadditivity in the coherent information, or we have no knowledge about the violation of additivity of quantum coherent information. It is an interesting result, since the classical capacity of depolarizing quantum channel $\mathcal{N}$ is proven to be strictly additive.

### D.1.3 Brief Summary

In order to conclude the additivity related result we compared the additivity violation of classical and quantum channels, see Table D.1 [Imre12].

| *Capacity* | $\textit{Single - use} \stackrel{?}{=} \textit{Asymptotic}$ | *Superadditivity (Asymptotic capacity)* |
|---|---|---|
| Classical channel $C(N)$ | *Single - use = Asymptotic* | *No* |
| Classical Capacity $C(\mathcal{N})$ | *Single - use* $\neq$ *Asymptotic* | Yes, for *same* channel maps (Unknown for *different* channel maps.) |
| Quantum Capacity $Q(\mathcal{N})$ | *Single - use* $\neq$ *Asymptotic* | Yes (For same and different channels) |
| Classical Zero-Error $C_0(\mathcal{N})$ | *Single - use* $\neq$ *Asymptotic* | Yes (For same and different channels) |
| Quantum Zero-Error $Q_0(\mathcal{N})$ | *Single - use* $\neq$ *Asymptotic* | Yes (For same and different channels) |

**Table D.1.** Our current knowledge on the additivity of different capacities of quantum channels.

The strict additivity of the classical capacity of classical communication channels was proven by Shannon [Shannon48]. The classical capacity $C(\mathcal{N})$ of quantum channels (for the same channel usage) is proven to be non-additive, since Hastings has shown in 2009 [Hastings09] that there are *identical* (same channel maps) quantum channels $\mathcal{N}_1$ and $\mathcal{N}_2$, such that $C(\mathcal{N}_1 \otimes \mathcal{N}_2) > C(\mathcal{N}_1) + C(\mathcal{N}_1)$. On the other hand, if we use two different channels $\mathcal{N}$ and $\mathcal{M}$, then the additivity of $C(\mathcal{N})$ is still not clarified.

The private classical capacity $P(\mathcal{N})$ has been proven to be non-additive, for the same and for different channels, too. The question has been analyzed by Smith and Smolin [Smith09b], [Smith10], and by Li *et al.* [Li09], and they have all found that the private classical capacity is *non-additive*. The properties of the asymptotic private classical capacity have been investigated by Cai-Winter-Young [Cai04]. Interestingly, the entanglement assisted capacity $C_E(\mathcal{N})$ has been shown to be additive, for both of the channel constructions.

The main results on the additivity of entanglement assisted capacity were found by Bennett, Shor, Smolin and Thapliyal [Bennett02]. Smith and Yard showed [Smith08], that the quantum capacity is also non-additive, for two different channels $\mathcal{N}$ and $\mathcal{M}$. The superadditivity of the asymptotic zero-error classical capacity was proven by Duan [Duan09] and Cubitt *et al.* [Cubitt09]. The asymptotic quantum zero-error capacity was also found to be non-additive, as was shown by and Cubitt and Smith [Cubitt09a].

As we can conclude from these results, the classical, the quantum and the private classical capacities of the quantum channels are not additive, - at least, in general sense. This means, that counterexamples (i.e., special examples which are additive) can be found, however, this is still an open question [King09], [Wilde11]. The most important works regarding on the additivity problem of the quantum channel capacities are summarized in the Related Work subsection of Appendix D.

# D.2 Additivity of Classical Capacity

Because additivity holds for the maximized quantum mutual information $\max_{all\ p_i,\rho_i} I(A:B)$, thus, if we have tensor product input states and single measurement setting, then for the classical capacity

$$\begin{aligned} C(\mathcal{N}_{12}) &= \lim_{n\to\infty}\frac{1}{2}C(\mathcal{N}_{12}) = C(\mathcal{N}_1)+C(\mathcal{N}_2) \\ &= \max_{all\ p_i,\rho_i} I(A_1:B_1)+\max_{all\ p_i,\rho_i} I(A_2:B_2), \end{aligned} \tag{D.16}$$

where $\mathcal{N}_{12}=\mathcal{N}_1\otimes\mathcal{N}_2$ and $\{A_i,B_i\}$ denotes the input and output of the $i$-th quantum channel in the tensor product structure. On the other hand, neither the Holevo quantity $\chi$, nor the quantum coherent information $I_{coh}(\rho_A:\mathcal{N}(\rho_A))$ are additive in general, only in case of some special examples. Moreover, the picture also changes if we talk about the entanglement assisted capacity or private classical capacity of the quantum channel.

There is no general formula to describe the additivity property of every quantum channel model, but one of the main results of the recent researches was the "very simplified" picture, that

$$C_{ALL}(\mathcal{N}_{12}) \neq C_{ALL}(\mathcal{N}_1)+C_{ALL}(\mathcal{N}_2). \tag{D.17}$$

for different capacities of the most quantum channel models. From this viewpoint, the strict additivity in quantum communication can be viewed as a special case, or a counterexample. This non-additive property is deeply woven into the essence of quantum mechanical systems, as can be elucidated by the fact that, for us, and thus for an external observer, a quantum system is rather different from a classical - simply additive – system [Smith10]. The non-additive property has many advantages and disadvantages, too. We start with the most important *disadvantage*: the different capacities *cannot be described* by a simple single-use formula, like we did in the case of classical systems. Instead, we have to use an approximation, – called the asymptotic capacity. This approximation is based on the assumption, that *unlimited number* of channel copies is available. This approximation used to compute the asymptotic capacity of the quantum channel is called the *regularization* of the channel capacities, which leads us to the fact that the computation of the different quantum capacities is much more complex than for classical systems.

Now, we focus on the *advantageous side* of the regularization. The non-additive property and the regularization of the channel capacities make it possible to achieve higher channel capacities and lower error probabilities than for classical systems. Thus, these phenomena make it possible to enhance the information transmission through quantum channels or to use zero-capacity quantum communication channels by the exploitation of entangled input states, and by the combination of different channel maps.

## D.2.1 Measurement Settings

In this subsection we overview the additivity property of classical capacity of quantum channels. The classical capacities of the joint channel structure $\mathcal{N}_{12} = \mathcal{N}_1 \otimes \mathcal{N}_2$ will be referred as the *joint classical capacity*, and will be denoted by $C\left(\mathcal{N}_{12}\right) = C\left(\mathcal{N}_1 \otimes \mathcal{N}_2\right)$.

We illustrated the measurement setting for additivity analysis of tensor-product channel capacity $C\left(\mathcal{N}_1\right) + C\left(\mathcal{N}_2\right)$, with *single measurement* setting in Fig. D.2. Because of the single measurement setting – which destroys the possible benefits of entanglement – the properties of the input are irrelevant from additivity point of view. As follows, in this case strict additivity $C\left(\mathcal{N}_1\right) + C\left(\mathcal{N}_2\right)$ holds for the joint capacity $C\left(\mathcal{N}_1 \otimes \mathcal{N}_2\right)$.

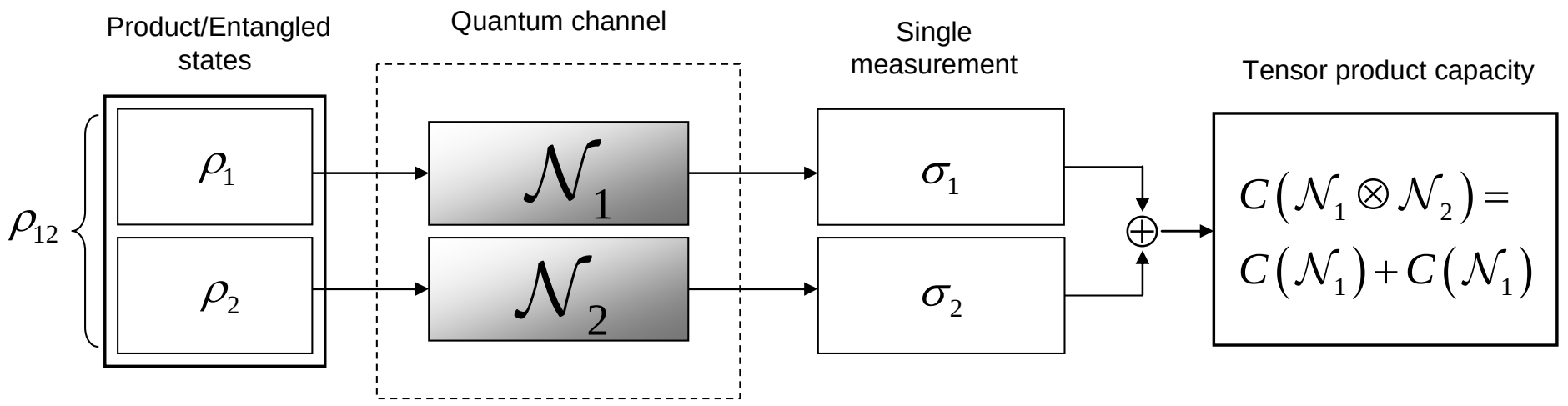

**Fig. D.2.** Setting for tensor-product channel capacity analysis, using single measurement setting. The single measurement destroys the possible benefits of entangled inputs.

Let us try to enhance the analysis of tensor-product channel capacity by *joint measurement.* $C_{PROD.}\left(\mathcal{N}_1 \otimes \mathcal{N}_2\right)$ refers to the capacity when *product input* states and *joint measurement* setting are used, as is illustrated in Fig. D.3.

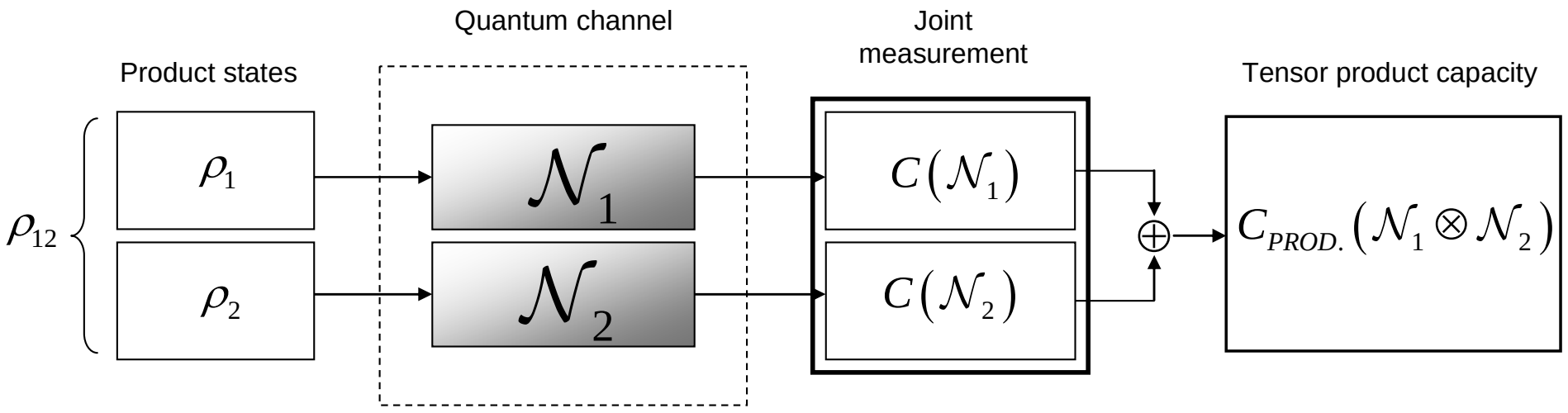

**Fig. D.3.** Setting for tensor-product channel capacity analysis, using joint measurement setting.

Finally we consider the case of EPR input states and joint measurement setting in Fig. D.4 for the classical capacity $C_{ENT.}\left(\mathcal{N}_1 \otimes \mathcal{N}_2\right)$.

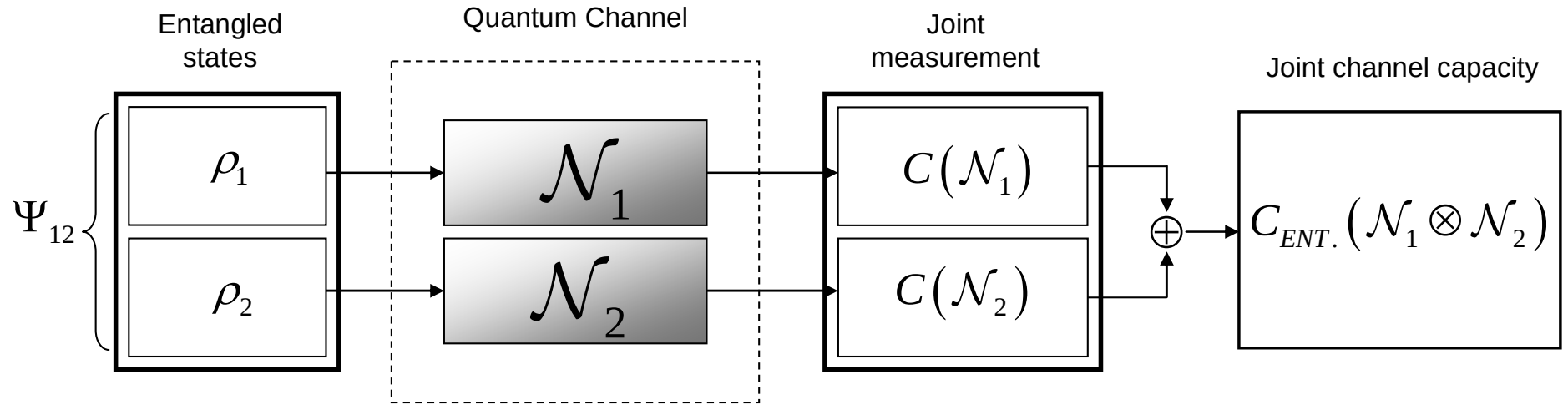

**Fig. D.4.** Setting for joint-capacity analysis using entangled input states and joint measurement setting. The joint measurement setting is required to exploit the possible benefits of entanglement.

The main question on the different classical capacities can be stated as

$$C_{ENT.}\left(\mathcal{N}_1 \otimes \mathcal{N}_2\right) \overset{?}{\geq} C_{PROD.}\left(\mathcal{N}_1 \otimes \mathcal{N}_2\right). \tag{D.18}$$

# D.3 Additivity of Private Capacity

The advanced properties of quantum channels were discovered mainly in the end of the 2000s. These results of Quantum Information Theory were completely unimaginable before, and—to put it simply—the researchers were shocked, rather than just surprised.

Recently, the most important discovery in Quantum Information Theory was the possibility of the *superactivation* of quantum communication channels.

Superactivation makes it possible to use zero-capacity quantum channels to transmit information. It was discovered in 2008 that quantum capacity of a quantum channel can be superactivated. In 2009 and 2010 it was extended to the classical zero-error capacity and the quantum zero-error capacity. (Up to 2011, the superactivation of the classical – i.e., the "non zero-error"– capacity is still open.)

The theoretical background of the superactivation of quantum capacities is currently unsolved, however, we know that it is based on the extreme violation of the additivity property, or in other words on the non-additivity of the various quantum channel capacities and the entangled input states. As we will see, currently only the classical and the quantum zero-error capacity and the quantum capacity of the quantum channel can be superactivated. On the other hand, the superactivation of the classical capacity of the quantum channels is still an open question — it can be grounded in the additivity problem of the classical capacity of the quantum channel, since Hastings' counterexample did not give an answer for the general case – so it remains also an open question and many aspects of it (such as the superactivation) are still unsolved.

Initially, the superactivation property was proven for just one combination of two zero-capacity quantum channels, which can be used for the transmission of quantum information. In this combination, each quantum channel has zero quantum capacity individually, however their joint quantum capacity is strictly greater than zero. Later, these results have been extended. The superactivation has also opened a very large gap between the single-use quantum capacity and the asymptotic quantum capacity. With the help of superactivation the difference between the single-use and the asymptotic quantum capacity of a channel can be made arbitrarily large. (Since maximized quantum coherent information describes only the single-use quantum capacity of a quantum channel, in general it cannot be used to describe the asymptotic quantum capacity of a quantum channel.) This very important result stands behind the superactivation of the quantum capacity of a quantum channel. In this section we will explain this result, and as we will see the complete theoretical background of superactivation is still missing.

The superactivation has opened the door which could clear up the question of the ability to transmit classical and quantum information through a noisy quantum channel. In 2009 it was discovered that the *classical* and *quantum zero-error capacities* of the quantum channel can also be superactivated. These could have many revolutionary practical consequences in the quantum communication networks of the future. With the help of superactivation, temporarily useless quantum channels (i.e., channels with individually zero quantum capacity or zero zero-error capacities) can be used together to avoid communication problems, and the capacities of the quantum channels can be increased.

In the initial discovery of phenomenon of superactivation, only two classes of superactive zero-capacity quantum channels were known. Later the superactivation was extended to classes of generic channels which can be used for superactivation and there could be many still unrevealed combinations. First we show the initial discovery of superactivation and then discuss the extension of the effect to the more general case. At the end of the section we will present an algorithmic solution (or at least its theoretical background) to the problem, which could help us to reveal these still undiscovered possibilities.

## D.3.1 Connection of Quantum Capacity and Private Classical Capacity

The phenomena of superactivation roots in the violation of additivity of channel capacities of quantum channels. The superactivation uses the extreme non-additive property of the quantum channels. First, we discuss briefly the non-additivity of private capacity, and then we show the connection between the superactivation of quantum capacity and the superadditivity of private classical information.

One of the most adequate measures of the security of a quantum channel is its *private capacity* (see Chapter 3). This measures the capacity of the quantum channel for se-cret quantum communication, and it's capability for quantum cryptography and private quantum communication. Private capacity gives us the maximal rate of private classical communication. We have also seen in Section 3.5.2, that the single-use formula of the private capacity is not equal to the asymptotic formula, hence the asymptotic private capacity is greater than or equal to the single-use private capacity.

These discoveries imply that the private capacity of quantum channels is not additive.

On the other hand, there exist quantum channels for which the *asymptotic* private capacity $P\left(\mathcal{N}\right)$ and the *single-use* private capacity $P^{(1)}\left(\mathcal{N}\right)$ are equal: these channels are called *degradable* channels (see Fig. D.1). (For the definition of degradable quantum channel see Chapter 4. We note, for a degradable channel this statement remains true for other capacities, too, as we will see later in this chapter.).

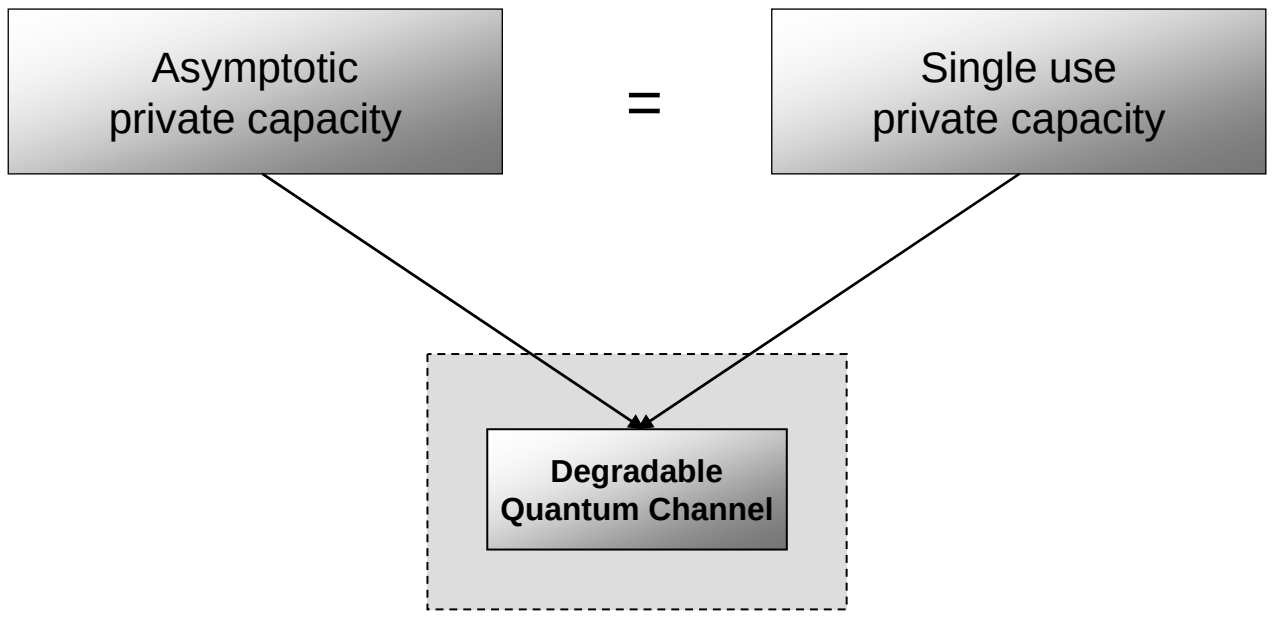

**Fig. D.1.** For a degradable quantum channel, the asymptotic private capacity is equal to the single-use private capacity. For other quantum channel models, this condition does not hold.

As in the case of the measure of classical capacity and the quantum capacity of the quantum channels, in the case of private capacity we would like to send information through a noisy quantum channel. The noise arises from the environment, or it can represent an eavesdropper on the quantum channel. While in the case of classical capacity we transmit classical information, in the case of private capacity we would like to send classical information through the channel in a form inaccessible to the environment or to an eavesdropper. The amount of maximal transmissible private information is less than or equal to the maximal classical capacity, and in general, the quantum capacity is less than or equal to the classical private capacity.

Generally, the maximum transmittable private information through a quantum channel is bounded above by the maximal amount of transmittable classical information, and bounded below by the quantum capacity of the quantum channel. This relation also shows that *the quantum information sent through the quantum channel is private information*, on the other hand *not every private information is quantum information*.

The fact that classical private information cannot exceed the "ordinary" non-private classical information is trivial. As we have discussed in Chapters 3 and 4, these capacities cannot be exceeded by the single-use capacity, except in some very special cases.

In Fig. D.2 we illustrate the relation between the classical capacity, private capacity and quantum capacity of a quantum channel.

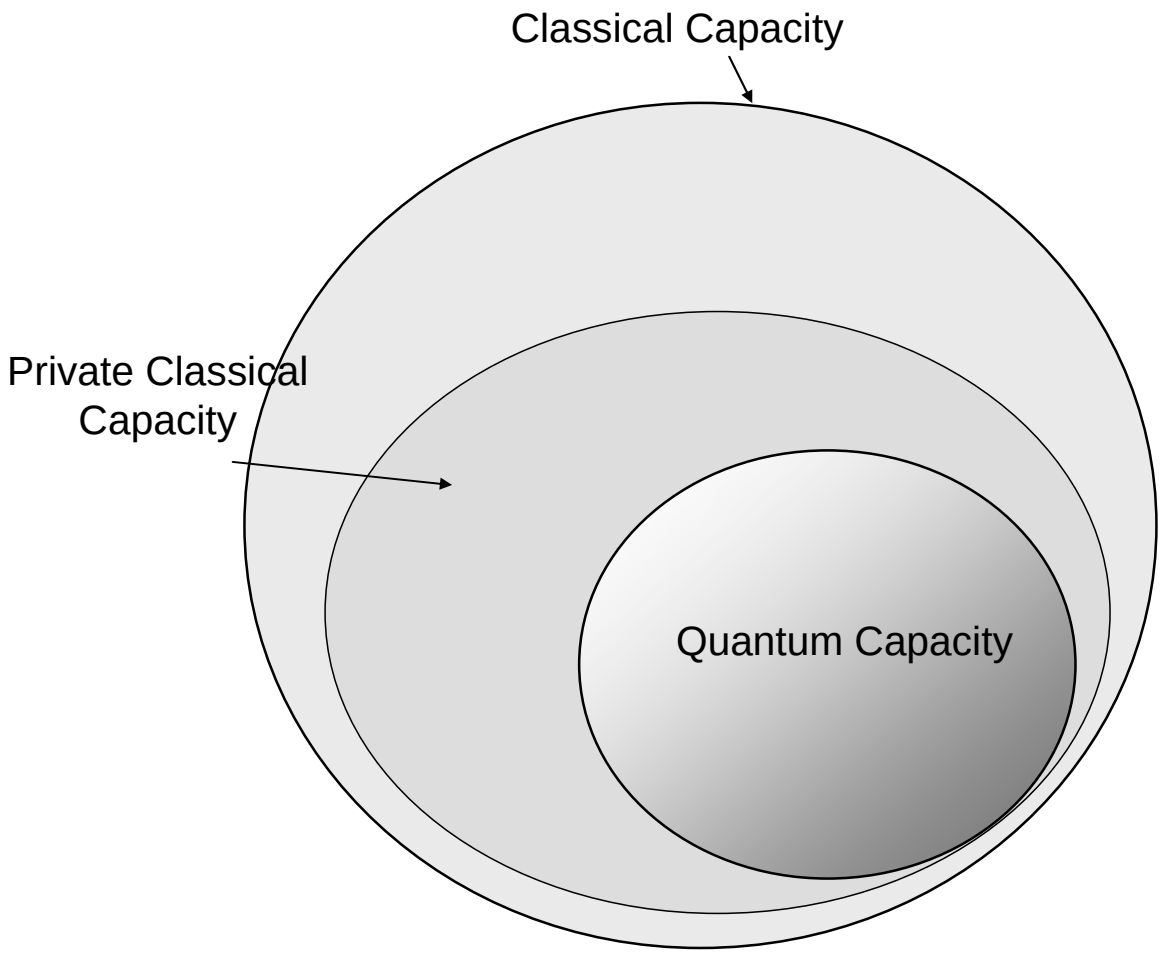

**Fig. D.2.** The generalized relation between the classical capacity, private classical capacity, and quantum capacity of a quantum channel.

The additivity of private capacity is currently an active area of research in Quantum Information Theory. We still have many open questions regarding the additivity of private capacity. Generally, it is known that additivity fails, however special cases can be found in which the single-use capacity is equal to the asymptotic capacity. This fact will have an important consequence: for those quantum channels the asymptotic private capacity will not violate the additivity property, i.e., the private capacity will be additive. This result also can be extended to the asymptotic classical capacity and the asymptotic quantum capacity. The non-additivity of the private capacity can be extended in a different way: this property can be used in the superactivation of zero-capacity quantum channels. It means that there exist quantum channel combinations for which the individual quantum capacities are equal to zero, however, the joint combination of the two channels possesses a non-zero quantum capacity. This construction uses a second channel, called the *erasure* channel, to activate the first channel.

In the next paragraph we give the definition of the *erasure* quantum channel, since it will have an important role in the superactivation.

## D.4 Erasure Quantum Channel

The erasure quantum channel will have deep relevance both in the superadditivity of classical private capacity and the superactivation of quantum capacity. Here we consider the case when the error probability is $p = 0.5$, i.e., we have an 50% erasure quantum channel, which is a *symmetric* channel.

For a symmetric channel the output will be symmetric under interchange, i.e., the output of the channel will in symmetric in $\rho_B$ (the output of the channel) and $\rho_E$ (the second output, the environment), i.e., $\rho_{BE} = \rho_{EB}$. This important property is illustrated in Fig. D.3.

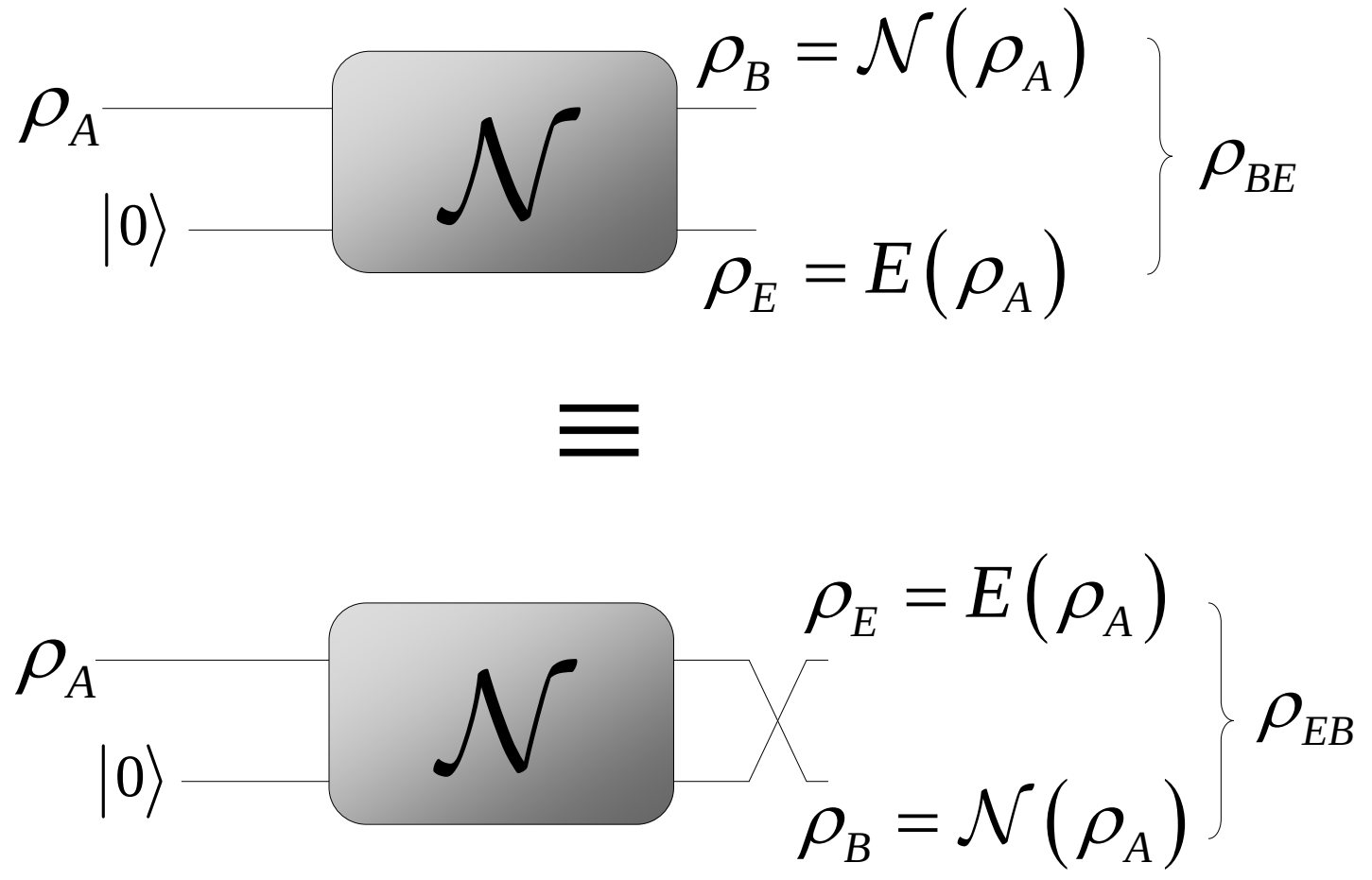

**Fig. D.3.** The output of a symmetric channel is symmetric in the joint state. The joint state is produced by the output state and the environment.

Furthermore, the symmetry makes it impossible to realize an output combination because the output state $\rho_B \rho_B$ is not possible, since any positive quantum capacity of a symmetric channel would lead to a violation of the no-cloning theorem (see Fig. D.4). This property means the following: any symmetric channels must have zero quantum capacity, i.e., $Q = 0$ for any symmetric channels. (For further details see the Further Reading.)

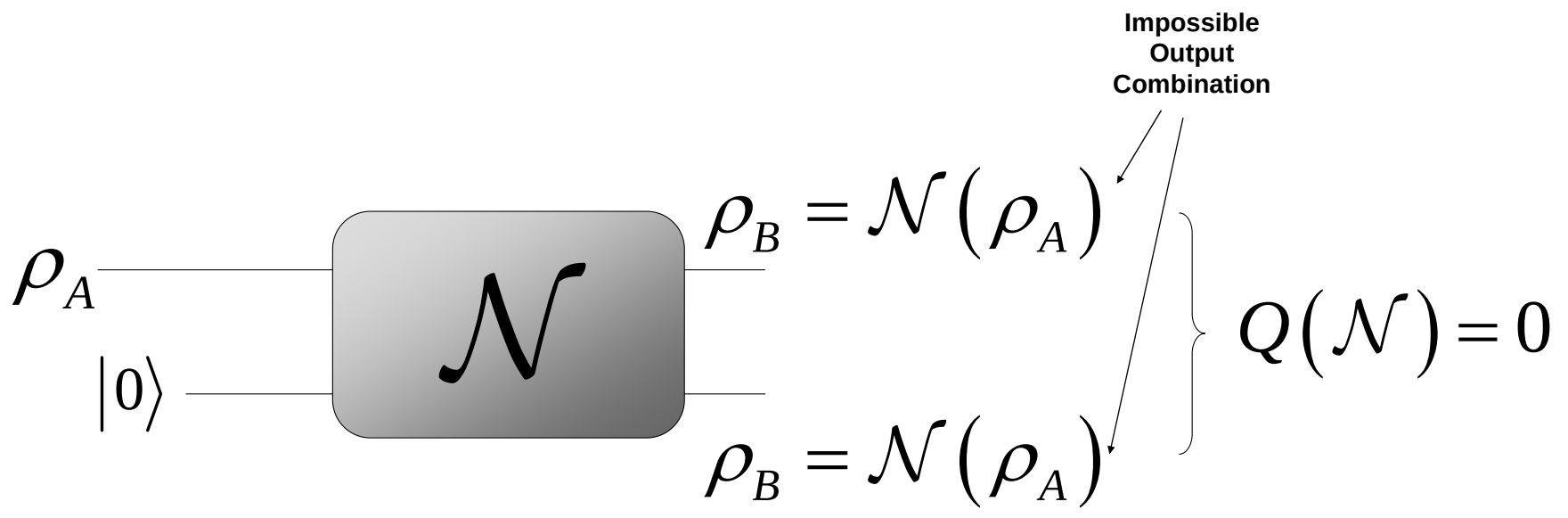

**Fig. D.4.** For a symmetric quantum channel it is not possible to produce a valid state on both of the outputs.

The *50% erasure* quantum channel is on one hand *degradable* (see Section D.1.2.1), hence the output from the environment point of view can be generated by Bob from his own output state. On the other hand the erasure quantum channel generates a valid output with 50% probability or completely erases it with 50% probability; hence it works in a symmetric way. The 50% erasure quantum channel is a symmetric quantum channel, and therefore it outputs a *valid quantum state* with 50% probability; or, it completely *gives this state to the environment.*

As it can be confirmed by the previous results, an *50% erasure channel has zero quantum capacity.* If we have a more general erasure channel with erasing probability $p$, then

$$\mathcal{N}\left(\rho\right) = \left(1-p\right)\rho + p\rho_E\,, \tag{D.19}$$

where $\rho$ is the valid output of the channel, while $\rho_E$ is the state of the environment. For the state of the environment

$$E\left(\rho\right) = p\rho + \left(1-p\right)\rho_E\,. \tag{D.20}$$

Based on (D.19) the channel map of a symmetric 50% erasure quantum channel $\mathcal{N}$ is

$$\mathcal{N}\left(\rho\right) = \frac{1}{2}\left(\rho + \left|e\right\rangle\left\langle e\right|\right), \tag{D.21}$$

where $\left|e\right\rangle$ is the erasure state. Similarly, the environment $E$ gets the following state if the state is erased:

$$E\left(\rho\right) = \frac{1}{2}\left(\rho + \left|e\right\rangle\left\langle e\right|\right). \tag{D.22}$$

If one takes an erasure channel with $p \leq \frac{1}{2}$ then we obtain a *degrading channel* $\mathcal{D}$, since in this case Bob can extract more information from the channel output than the information leaked to the environment during the transmission. It means that the environment can be simulated by Bob's output in the following way

$$E : \mathcal{N}_2 = \mathcal{D}\mathcal{N}_1\,, \tag{D.23}$$

where the environment $E$ is denoted by the second channel $\mathcal{N}_2$. (For the definition of degradable channel see Section D.1.2.1).

# D.5 Channel Combination for Superadditivity of Private Information

In this section we focus on the superadditivity property of classical private information and the theoretical background. Now, we state the following: if a special quantum channel $\mathcal{N}_1$ (currently undefined) with some classical private capacity is combined with an 50% erasure channel $\mathcal{N}_2$ (which also has zero classical private capacity), then it is possible to transmit more classical private information through the quantum channel, than the classical private capacity of the first channel originally has. The details of the first channel $\mathcal{N}_1$ can be found in [Smith08d,09b].

The second channel is an 50% erasure channel which does the following: it erases the input with probability 50%, and leaves it untouched with 50% probability, which working mechanism is theoretically equal to a zero-capacity channel, since it is able to transmit the input correctly only with 50% probability. An important conclusion is that, as in the case of quantum capacity, the classical private capacity of the joint channel construction is greater than the sum of the individual classical private capacities of the quantum channels. This means that there are special cases when the achievable joint private capacity will be greater than was initially the private capacity.

The channel combination for the realization of superadditivity is shown in Fig. D.5.

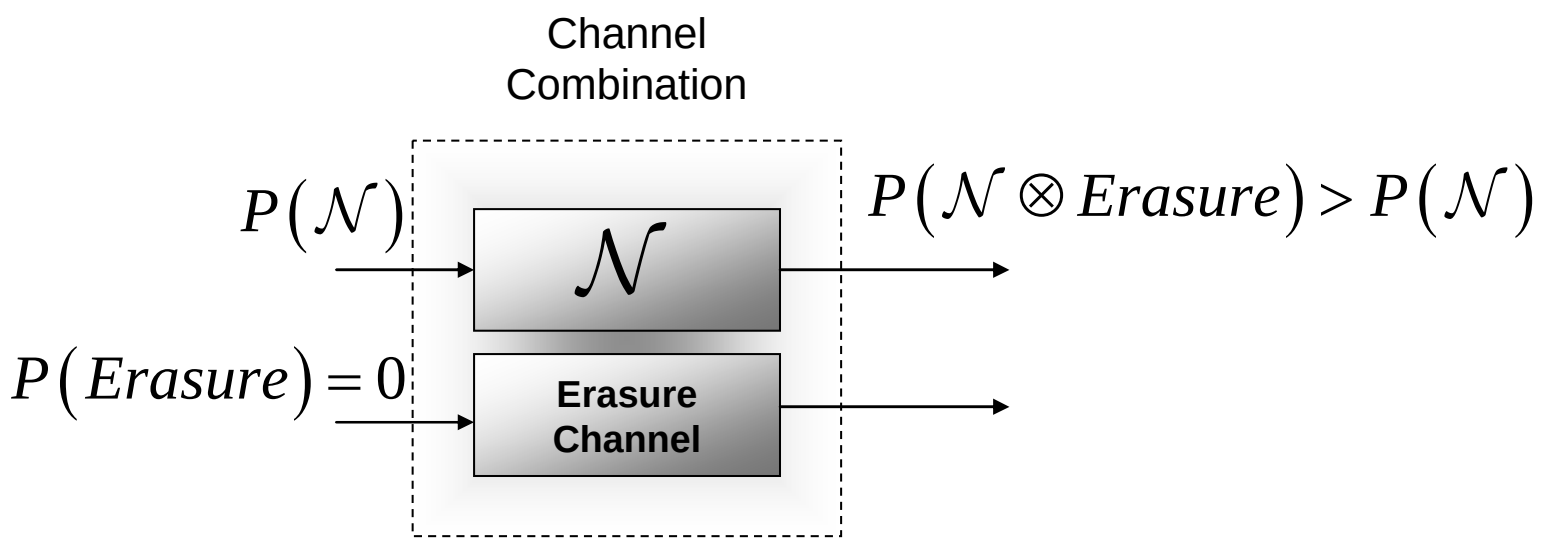

**Fig. D.5.** It is possible to combine a quantum channel with very small private capacity and a completely useless second quantum channel to realize a greater private capacity than the first channel had initially.

We note that in Fig. D.5, only the first quantum channel of the combination has some private capacity greater than zero. (The 50% erasure channel has zero private capacity, of course.) The maximum transmittable information through the quantum channel depends not just on the quantum channel itself, but also on the second channel which is used together with the original channel. By means of this construction, a counterexample to the additivity of the private capacity has also been shown. The results demonstrated that the private capacity is also *non-additive*. This fact generally makes it harder to compute the true (i.e., the asymptotic) private capacity of the quantum channel, since it cannot be given by the single-use formula and the exact determination of the asymptotic version requires high-cost computations. We note, that in the channel construction we allow to Alice and Bob to use shared entanglement. In a modified variant of the channel construction prior shared information is not allowed, hence the "randomization" of the environment is made by the quantum channel itself. This in turn implies that the receiver will not be able to distinguish between the noise of the environment and Alice's private information. Since we know that the second channel in the joint construction is the "fixed" 50% erasure quantum channel (see Fig. D.5.), next, we discuss the channel combinations which can be used as the first channel in the joint channel combination to achieve the superadditivity of private capacity. For further information see the book of Imre and Gyongyosi [Imre12]

# D.6 Behind of Superactivation - The Information Theoretic Description

In this section we give a clear information theoretic discussion of superactivation.

## D.6.1 System Model

Before we start our analysis of superactivation, we introduce a convention in the notations based on [Smith08]. According to Fig. D.6, for the first channel $\mathcal{N}_1$, the classical information which is encoded in a pure quantum system $\rho_A^x$ (occurs with probability $p^x$) will be referred to as $A$, and the map of the quantum channel will be modeled in the following way

$$\rho_{BE}^x = \mathcal{N}_1(\rho_A^x), \tag{D.24}$$

where $B$ is the output of the quantum channel, while $E$ is the environment and $\rho_{BE}^x$ describes the output as quantum system.

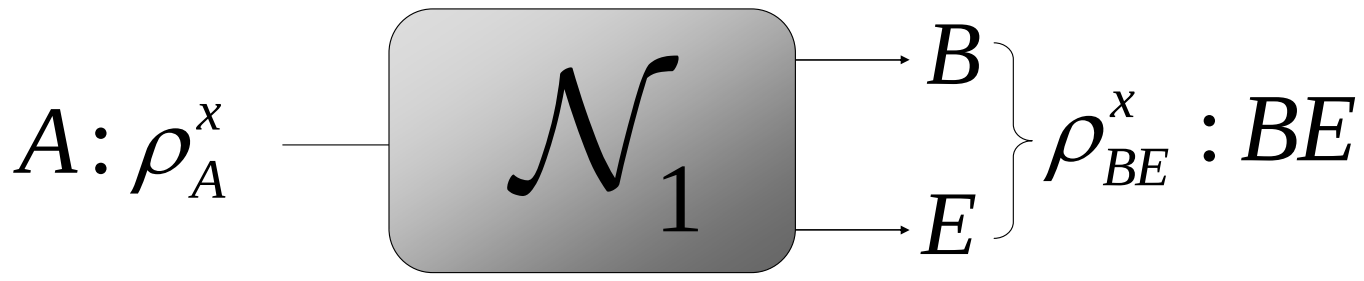

**Fig. D.6.** Description of the first quantum channel.

Similarly, for the second quantum channel $\mathcal{N}_2$, we have the following correspondence between the notations:

$$\rho_{B'F}^x = \mathcal{N}_2(\rho_{A'}^x), \tag{D.25}$$

where $B'$ is the output of the second quantum channel $\mathcal{N}_2$, while $E$ is the environment of the channel, and $\rho_{B'F}^x$ describes the output in quantum system representation (see Fig. D.7).

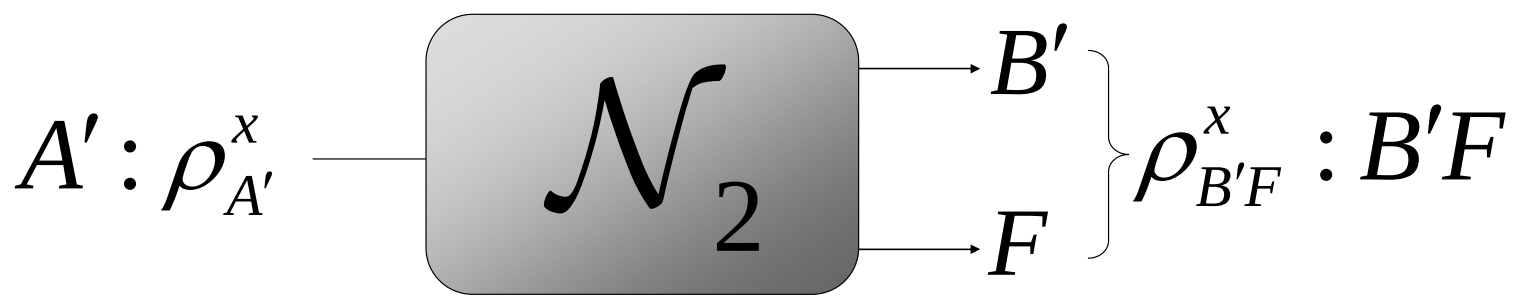

**Fig. D.7.** Description of the second quantum channel.

Now, based on (D.24) and (D.25), we will use the following parameters (in correspondence with the notations used in the figures of Chapters 2 and 3):

– the reference input with classical random variables $X$, represents Alice's classical register. Using input ensemble $\left\{p^x, \rho_A^x\right\}$ it forms a state with the input $A$, *i.e.*, $\sum_x p^x \left|x\right\rangle\left\langle x\right|^X \otimes \rho_A^x = \rho_X^x \otimes \rho_A^x$, where $\left\{\left|x\right\rangle^X\right\}$ is the orthonormal basis for $X$. After $\rho_A^x$ sent through the first channel the reference system $\rho_X^x$ will form a state with the joint state of the output and the environment state: $\sum_x p^x \left|x\right\rangle\left\langle x\right|^X \otimes \rho_{BE}^x = \rho_X^x \otimes \rho_{BE}^x$.

– *the input and output of the first channel* $\mathcal{N}_1$ *will be denoted by* $A$ *and* $B$*,*

– *the input and output of the second (the erasure) channel* $\mathcal{N}_2$ *will be referred as* $A'$ *and* $B'$*,*

– $E$ *and* $F$ *stand for the environment of the first channel* $\mathcal{N}_1$ *and the second channel (the erasure channel)* $\mathcal{N}_2$*.*

The whole system configuration is summarized in Fig. D.8.

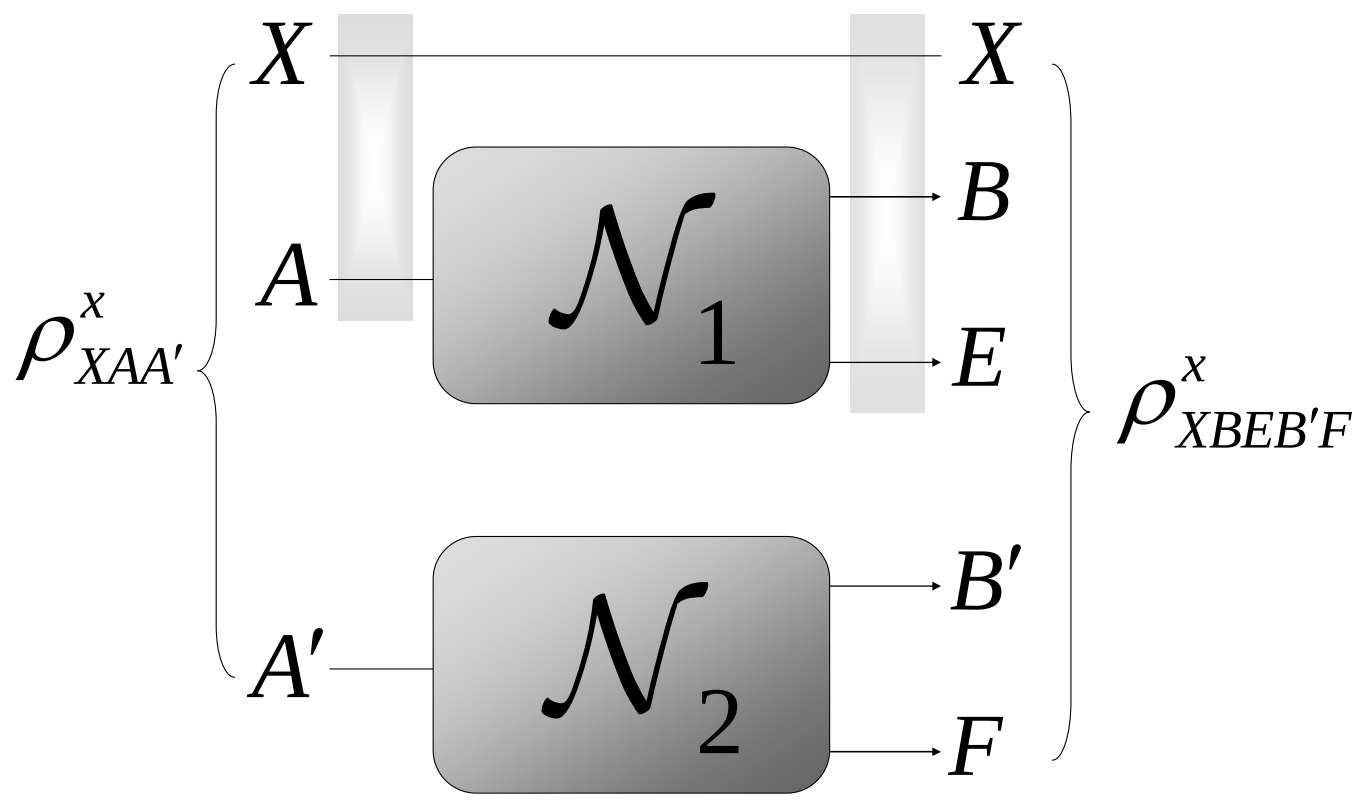

**Fig. D.8.** The complete system characterization for superactivation of quantum capacity. The first channel can be any quantum channel which has positive private classical capacity; the second quan-

tum channel is a symmetric erasure channel. (The erasure channel is finite dimensional and it can be replaced with another symmetric channel; however in this case input and output could be infinite.)

Using these notations of [Smith08], the result of can be rephrased as follows. For the channel construction $\mathcal{N}_{12} = \mathcal{N}_1 \otimes \mathcal{N}_2$ there exists an input system $\rho_{AA'}$ such that

$$I_{coh}\left(\rho_{AA'} : \mathcal{N}_{12}\left(\rho_{AA'}\right)\right) = \frac{1}{2}\left(I\left(X : B\right) - I\left(X : E\right)\right), \tag{D.26}$$

where $I\left(A : B\right)$ and $I\left(A : E\right)$ are the quantum mutual information (i.e., not the quantum coherent information, since the private information is classical) between the input and the output of the first channel and between the input and the environment.

Now, let us assume that the input states in the input ensemble $\left\{p^x, \rho_A^x\right\}$ are all *pure* states, hence for the input of the first channel $\mathcal{N}_1$:

$$\rho_A^x = \left|\varphi_A\right\rangle\left\langle\varphi_A\right|; \tag{D.27}$$

then for the quantum coherent information the following equation holds:

$$Q^{(1)}\left(\mathcal{N}_1\right) = \max_{all\ p^x, \rho_A^x} I_{coh}\left(\rho_A : \mathcal{N}_1\left(\rho_A\right)\right) = \max_{all\ p^x, \rho_A^x} I\left(X : B\right) - I\left(X : E\right), \tag{D.28}$$

where we used $\rho_A = \sum_x p_x \left|\varphi^x\right\rangle\left\langle\varphi^x\right|$. Since we restricted our attention to pure states, our case became much simpler, since due to this boundary condition the quantum coherent information $I_{coh}\left(\rho_A : \mathcal{N}_1\left(\rho_A\right)\right)$ will be equal to the maximum amount of classical private information; see (D.28).

But what happens if we have mixed inputs? This situation can be handled, too; however, we have to define the purification (see Appendix B) of the mixed quantum sys-tem. To realize this, we have to send through the second quantum channel $\mathcal{N}_2$ a purifying system which is able to purify the mixed input of the first quantum channel $\mathcal{N}_1$.

Let us assume that we have the mixed input system $\sum_x p_x \left|x\right\rangle\left\langle x\right|^X \otimes \rho_A^x$. The *purification* of this state can be defined as follows:

$$\left|\varphi_{XAA'}\right\rangle = \sum_x \sqrt{p_x} \left|x\right\rangle^X \left|\varphi_{AA'}^x\right\rangle, \tag{D.29}$$

where $\left\{\left|x\right\rangle^X\right\}$ is an orthonormal basis for *X*, and $\left|\varphi_{AA'}^x\right\rangle$ is the purification of state $\rho_A^x$. Using (D.29), $\left|\varphi_{XA}\right\rangle$, with the help of $\left|\varphi_A^x\right\rangle$, can be expressed as follows:

$$\left|\varphi_{XA}\right\rangle = \sum_x p_x \left|x\right\rangle\left\langle x\right|^X \otimes \left|\varphi_A^x\right\rangle. \tag{D.30}$$

Now, using this purified state the following question arises: How could we compute the quantum coherent information of the joint channel construction $\mathcal{N}_1 \otimes \mathcal{N}_2$ if we have input systems $A, A'$, output systems $B, B'$, and environments $E$ and $F$ (the environment of the erasure channel $\mathcal{N}_2$)?

In (D.31) we show the evolution process of positive quantum coherent information. The steps of the computation will be discussed in detail in the text right after the derivation. Using input system $\rho_{AA'}$ the quantum coherent information of the joint channel structure $\mathcal{N}_{12}$ be expressed as follows

$$\begin{aligned} &I_{coh}\left(\rho_{AA'} : \mathcal{N}_{12}\left(\rho_{AA'}\right)\right) = \\ &\quad (\textit{Step 1.}) \quad = H\left(BB'\right) - H\left(EF\right) \\ &\quad (\textit{Step 2.}) \quad = \frac{1}{2}\left(H\left(B\right) - H\left(EA'\right)\right) + \frac{1}{2}\left(H\left(BA'\right) - H\left(E\right)\right) \\ &\quad (\textit{Step 3.}) \quad = \frac{1}{2}\left(H\left(B\right) - H\left(XB\right)\right) + \frac{1}{2}\left(H\left(XE\right) - H\left(E\right)\right) \\ &\quad (\textit{Step 4.}) \quad = \frac{1}{2}\left(I\left(X : B\right) - I\left(X : E\right)\right), \end{aligned} \tag{D.31}$$

where $H$ is the Shannon entropy function. From this result the *single-use* quantum capacity can be expressed as

$$Q^{(1)}\left(\mathcal{N}_{12}\right)=\max_{all\ p,\rho_{AA'}} I_{coh}\left(\rho_{AA'}:\mathcal{N}_{12}\left(\rho_{AA'}\right)\right)=\frac{1}{2}P^{(1)}\left(\mathcal{N}_{1}\right), \tag{D.32}$$

where $P^{(1)}\left(\mathcal{N}_{1}\right)=\max_{X,\rho_A^x}\left(I\left(X:B\right)-I\left(X:E\right)\right)$.

For the *asymptotic* quantum capacity

$$Q\left(\mathcal{N}_{12}\right)=\lim_{n\to\infty}\frac{1}{n}\max_{all\ p,\rho_{AA'}} I_{coh}\left(\rho_{AA'}:\mathcal{N}_{12}\left(\rho_{AA'}\right)\right)^{\otimes n}\geq\frac{1}{2}P\left(\mathcal{N}_{1}\right), \tag{D.33}$$

where $P\left(\mathcal{N}_{1}\right)=\lim_{n\to\infty}\frac{1}{n}P^{(1)}\left(\mathcal{N}_{1}\right)=\lim_{n\to\infty}\frac{1}{n}\max_{X,\rho_A^x}\left(I\left(X:B\right)-I\left(X:E\right)\right)^{\otimes n}$. (If channel $\mathcal{N}_1$ degradable then $P\left(\mathcal{N}_{1}\right)=P^{(1)}\left(\mathcal{N}_{1}\right)$.)

Now, let us analyze what we have obtained in (D.31) as a result for $I_{coh}\left(\rho_{AA'}:\mathcal{N}_{12}\left(\rho_{AA'}\right)\right)$ using the input system as defined in (D.29).

### D.6.2 Output System Description

Here we describe in detail the steps from (D.31). The derivation will include four main steps.

*Step 1*

First, we discuss the correlation between the outputs of the first and second quantum channels and between the environments of the first and the second quantum channels

$$I_{coh}\left(\rho_{AA'}:\mathcal{N}_{12}\left(\rho_{AA'}\right)\right)=H\left(BB'\right)-H\left(EF\right). \tag{D.34}$$

After we have sent through the input system $AA'$ of $\left|\varphi_{XAA'}\right\rangle$ through the joint channel construction $\mathcal{N}_{12}$, the quantum coherent information of the output of the system can be expressed from the joint entropy of output of the two channels $BB'$ and the environments of the two channels. The detailed model is illustrated in Fig. D.9. However, in this case $\mathcal{N}_1$ can be *any* quantum channel which has some positive classical private capacity, while in the proof of Smith and Yard [Smith08b] they have used a Horodecki channel.

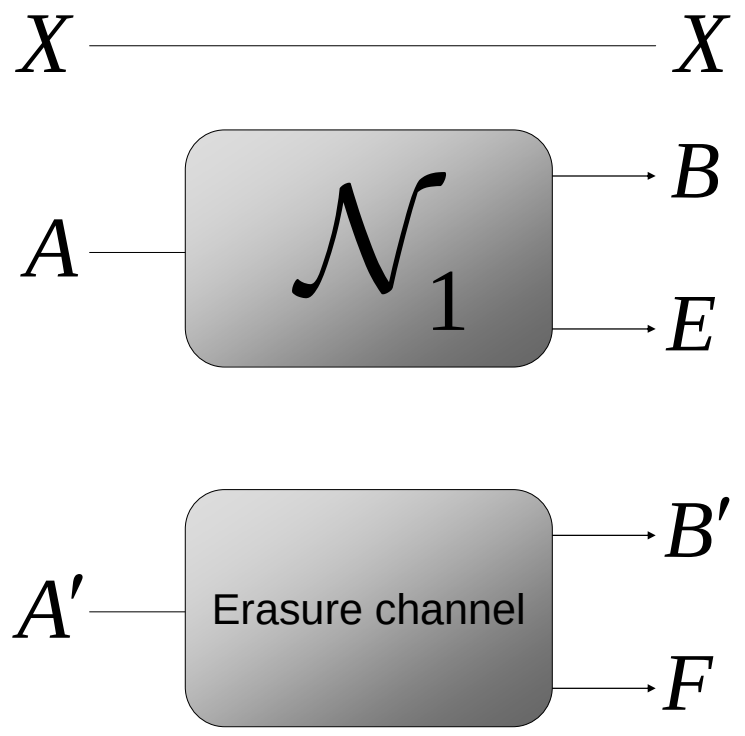

**Fig. D.9.** The whole system consists of the reference input system, the first (arbitrarily chosen) quantum channel, and the second, erasure quantum channel.

*Step 2*

Here, we use the fact that the second quantum channel $\mathcal{N}_2$ is an erasure channel, which means that either the channel can transmit input system $A'$ or it completely vanishes. This means that in half of the cases $A'$ will appear on the $B'$ output of the second channel, while in the second half of the cases $A'$ will be absorbed by environment $F$ of the second channel. *Conclusion: the whole effect is controlled by the second, erasure channel.* Based on the working mechanism of the second channel, we need to discuss two possible outcomes. As follows, the previously shown expression (D.34) will be spit into two parts:

$$I_{coh}\left(\rho_{AA'}:\mathcal{N}_{12}\left(\rho_{AA'}\right)\right)=\frac{1}{2}\left(H\left(B\right)-H\left(EA'\right)\right)+\frac{1}{2}\left(H\left(BA'\right)-H\left(E\right)\right). \quad \text{(D.35)}$$

Let us look deeper behind the *first term* of (D.35), $H\left(B\right)-H\left(EA'\right)$.

The *first system state* can be rephrased in terms of differences of entropies: in the first case (i.e., when the second channel $\mathcal{N}_2$ transmits the input system $A$) the quantum mutual information (here we discuss the first part of (D.35), and the quantum coherent information itself will be the sum of the two terms) can be expressed from the entropy of the output of the first channel $B$ and the joint entropy of the environment of the first channel $E$ and the input of the second channel $A'$. These statements are illustrated in Fig. D.10(a). Now, we continue the description with the second part of (D.35), $H(BA') - H(E)$. The *second system* state can be discussed as follows. If the second channel $\mathcal{N}_2$ erases the input system, then the quantum mutual information of the output system can be expressed from the joint entropy of the output of the first channel $B$ and the input of the second channel minus the entropy of the environment $E$ of the first quantum channel $\mathcal{N}_1$, as we illustrated in Fig. D.10(b).

(*In the next figures, the discussed parts will be denoted by dashed boxes. The correlated systems are denoted by the light gray boxes.*)

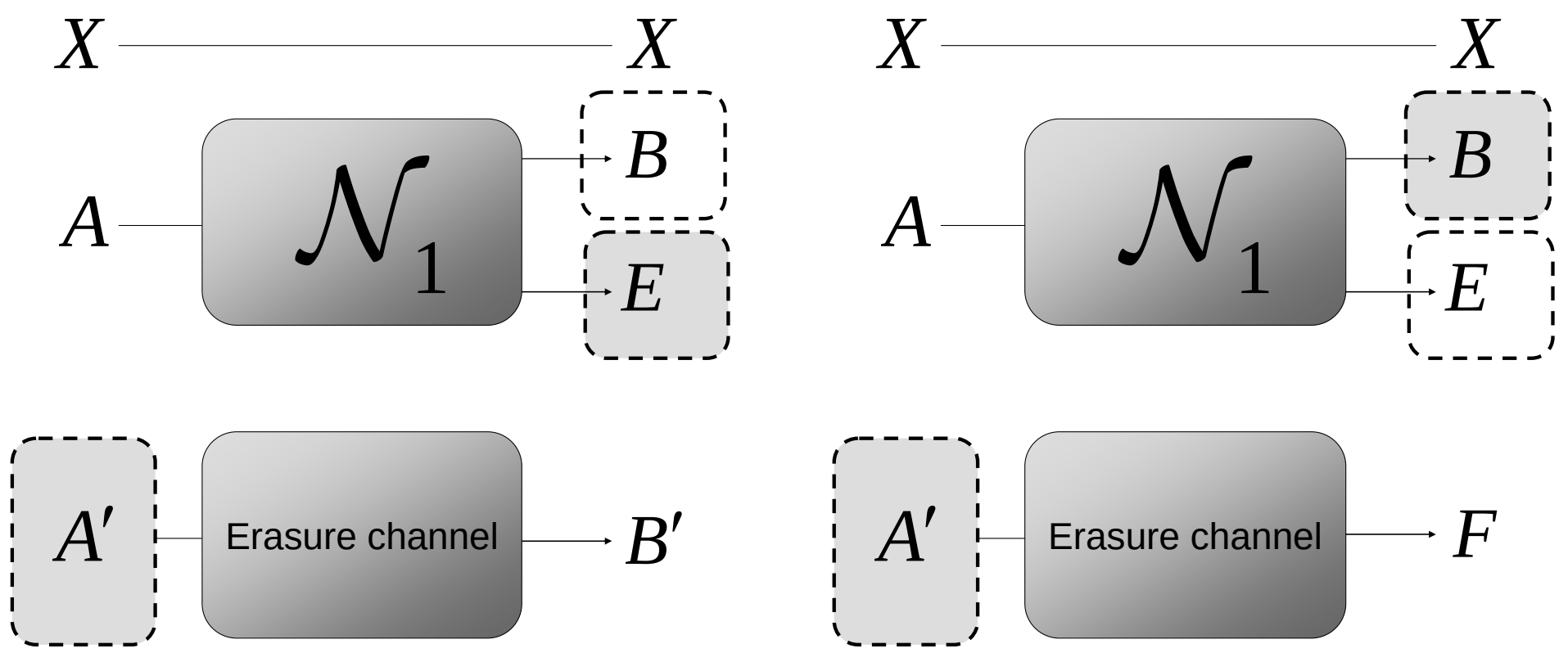

**Fig. D.10.** (a): Description in terms of the correlation between the input of the second channel and the output of the first channel and in terms of the environment of the first channel. In the first case the erasure channel transmits its input. (b): In the second case, the erasure channel erases the input and gives it to its environment $F$.

*Step 3*

In this step, we eliminate $H(EA')$ and $H(BA')$ from (D.35), and using the reference system $X$ we replace these terms with $H(XB)$ and $H(XE)$ i.e., we discuss

$$I_{coh}\left(\rho_{AA'} : \mathcal{N}_{12}\left(\rho_{AA'}\right)\right) = \frac{1}{2}\left(H(B) - H(XB)\right) + \frac{1}{2}\left(H(XE) - H(E)\right). \quad \text{(D.36)}$$

We can state the following: the term $H(B) - H(EA')$ of (D.35) can be rewritten as $H(B) - H(XB)$. Similarly, $H(BA') - H(E)$ from (D.35) can be expressed as $H(XE) - H(E)$.

It is possible, since bipartitions of any pure quantum state will have the same entropies, the entropies of the system will not change because we use the pure system $\left|\varphi_{XAA'}\right\rangle$.

Let us consider the first term of (D.36), $H(B) - H(XB)$. The joint entropy of the input system of the second, erasure channel $A'$ and the environment of the first channel $E$ is equal to the joint entropy of the reference input system $X$ and the output of the first channel $B$; see Fig. D.11(a). Now, we have arrived at the second term of (D.36), $H(XE) - H(E)$. The equivalence holds between the joint entropy of the input of the second channel $A'$ and the output of the second channel $B$ and between the joint entropy of the reference input system $X$ and the environment of the first quantum channel $E$; see Fig. D.11(b).

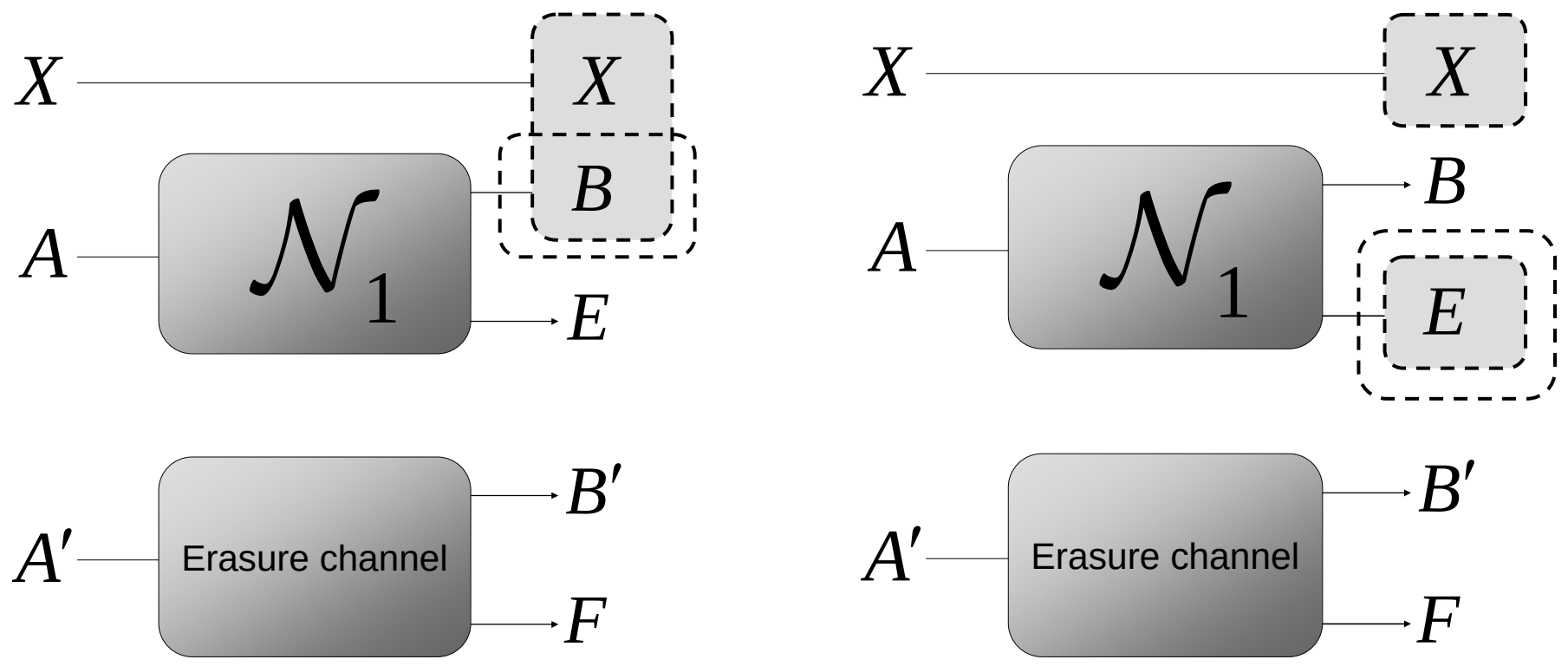

**Fig. D.11.** (a): Description in terms of the output of the first channel and in terms of the correlation between the reference input system and the output of the first quantum channel. (b): Description in terms of the correlation between the reference input and the environment of the first quantum channel and in terms of the environment of the first quantum channel.

*Step 4*

Let us see what happens if we add $H(X)$ to the first term of (D.36):

$$\begin{aligned} & H(B) - H(XB) + H(X) \\ & = H(B) - [H(X) + H(B) - I(X:B)] + H(X) \\ & = I(X:B). \end{aligned} \tag{D.37}$$

Similarly, if we subtract $H(X)$ from the second term of (D.36), we get:

$$\begin{aligned} & H(XE) - H(E) - H(X) \\ & = [H(X) + H(E) - I(X:E)] - H(E) - H(X) \\ & = -I(X:E). \end{aligned} \tag{D.38}$$

As can be checked easily, if we combine (D.37) and (D.38) we get

$$I_{coh}(\rho_{AA'} : \mathcal{N}_{12}(\rho_{AA'})) = \frac{1}{2}(I(X:B) - I(X:E)), \tag{D.39}$$

or with other words

$$I_{coh}\left(\rho_{AA'} : \mathcal{N}_{12}\left(\rho_{AA'}\right)\right) = \frac{1}{2}P\left(\mathcal{N}_1\right). \tag{D.40}$$

Based on the previous steps, the background of (D.40) can be summarized as shown in Fig. D.12.

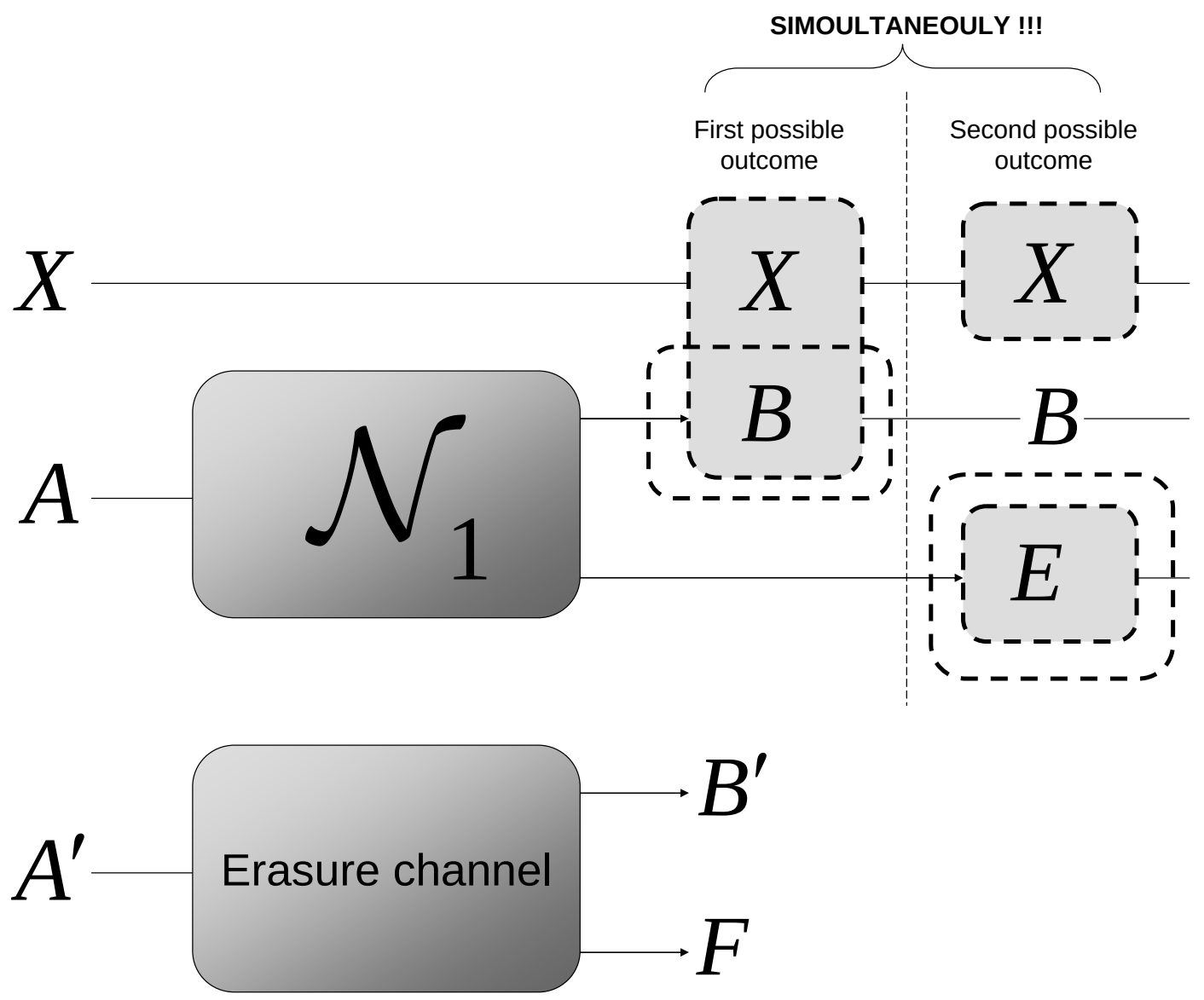

**Fig. D.12.** The final conclusion on the superactivated quantum capacity. Thanks to the erasure quantum channel, the two possible outcomes are realized simultaneously in the joint combination.

As follows, the coherent information describes the situation when the two possible outcomes of (D.36) (see Figs. D.11(a) and D.11(b)) are realized simultaneously on the output of the first quantum channel. Hence, the background of the superactivation is rooted in the *quantum parallelism*, or in more generalized description, in the working mechanism of the *quantum interferometer*. For more information about the working mechanism of the quantum interferometer, see [Imre05].

We note that in this example the first channel $\mathcal{N}_1$ is a Horodecki channel, since for this channel the $P\left(\mathcal{N}_1\right) > 0$ condition on private information is satisfied. The first quantum channel $\mathcal{N}_1$ can be any general quantum channel for which $P\left(\mathcal{N}_1\right) > 0$, and hence it can be extended for various other channel models.

The second quantum channel $\mathcal{N}_2$ was chosen to be an 50% erasure quantum channel; however, this case can also be extended to a general case where the second channel has to be a *symmetric* channel, but in this case we have to face the problem of an infinite-dimensional input and output system. Or, in other words, if we replace the 50% erasure channel (it is also symmetric channel, but finite-dimensional) with a general symmetric quantum channel with *unbounded* dimension, then we cannot make any generalized statement about the exact joint quantum capacity.

Finally, we would emphasize that the first quantum channel can be *any* quantum channel – there are no restrictions on it – which could extend the possibilities and open new perspectives in quantum communications.

## D.7 Related Work on Additivity of Quantum Channels

We summarize the most important works regarding on the additivity problem of the quantum channel capacities.

#### Additivity Problem in Quantum Information Theory

The problem of the additivity of quantum channels is rooted in the advanced properties of quantum communication channels. As we have seen in Chapter 3, in the description of the classical capacity of general quantum channels, the HSW-theorem states different capacities of the single-use and the asymptotic capacities. The HSW-theorem is a very elegant tool to describe the transmission of classical information over a quantum channel with product input states, however it does not give an answer to the additivity problem. It left open many questions, since in the general case the classical capacity of quantum channel is not additive, and the Holevo information cannot be used to describe the maximal classical capacity—for a general quantum channel, the asymptotic formula will give an explicit answer.

The role of additivity problems in Quantum Information Theory was summarized in 2000 by Amosov *et al.* [Amosov2000]. An interesting connection between the Holevo information and the minimum output entropy of a quantum channel was shown by Shor [Shor04a]. Shor proven that the additivity of Holevo information implies the additivity of the minimum output entropy, and this connection holds in the reverse direction, too. As Shor found, these various questions on additivity are the same, in particular, if one is false then all are false—for details see Shor's article from 2004 [Shor04a].

**Additivity of Holevo Information**

Shor's result on the additivity of Holevo information and minimum output entropy states made the picture so much simpler, since it made it possible to analyze the question of additivity. The strict additivity of unital quantum channels and the capacities of the most important quantum channels, such as depolarizing quantum channels (see the Appendix), was first proven, by King, in 2002 for unital quantum channels [King02] and in 2003, specially for depolarizing quantum channels [King03b] (a summary can be found in King's remarks [King09]). King also revealed the fact that the classical capacity of the depolarizing channel can be maximized without a joint measurement setting or entangled input states (hence without any "special" quantum influences, according to the channel's strict additivity). The additivity of depolarizing channel for higher dimension was studied by Amosov [Amosov07].

We also mentioned the depolarizing quantum channel regarding on the superadditivity of quantum coherent information. The depolarizing channel map has been studied exhaustively in the literature [Amosov04], [Bennett98], [Bruss2000], [Cortese02], [Datta04b], [Fujiwara02], [King03b], [Michalakis07]. The properties of erasure quantum channels can be found in the work of Bennett [Bennett97] and of Grassl from the same year [Grassl97]. The various aspects of the additivity problem in Quantum Information Theory was studied by Datta *et al.*, see their works [Datta04-04b] and [Datta05]. The additivity for covariant quantum channels was analyzed by Datta *et al.* [Datta06].

Beside the properties of encoding and decoding of quantum states, the additivity of Holevo information depends on the map of the quantum channel, and for a general quantum channel the additivity of Holevo information was an open question. King in 2003 found a quantum channel for which the Holevo information is additive, but at this time (in 2003) the general formula was still an open question.

Then *something happened* in 2009.

**Superadditivity of Holevo Information**

Until 2009, it was conjectured that the Holevo information was additive in general, however the complete theoretical background was not clear to the researchers. Then, in 2009, the picture changed, since Hastings gave a proof that the strict additivity of Holevo information does not hold in the general case (using same channel maps), and the Holevo information is superadditive. More precisely, Hastings gave a counterexample for which channel (using two identical channel maps) the classical capacity will be superadditive—hence for entangled inputs the additivity of Holevo information fails. For the details see the proof of Hastings [Hasting2009]. (We note that, besides the existence of this counterexample, it cannot be generalized—hence, the answer for the additivity of classical capacity in the general case is still open.)

We note that the preliminaries of Hastings's proof were laid down by Winter in 2007 [Winter07], and by Hayden and Winter [Hayden08]. Hastings's proof from 2009 also gave an answer to Shor's conjectures: all of the additivity conjectures—as stated in Shor's paper from 2004, see [Shor04a]—*are false*. After Hastings's proof, in 2009, Brandao and Horodecki published a paper on Hastings's counterexamples to the minimum output entropy additivity conjecture. Later, in 2010, another work was published by Aubrun *et al.*, on Hastings's additivity counterexample via Dvoretzky's theorem, see [Aubrun10]. Hastings's proof was also analyzed by Fukuda and King and Moser in 2010, see [Fukuda10], and by Fukuda and King, for details see [Fukuda10a].

While for the general case the Holevo information was found to superadditive, there are some quantum channels, for which the Holevo information remains additive. These channels are, for example, the Hadamard channels, the entanglement-breaking channel, or the identity quantum channel, which latter is also a Hadamard channel. The description of Hadamard channels and their Kraus-representation can be found in detail in the King *et al.*'s work [King07]. The capacities of these channels were also analyzed by Bradler *et al.* [Bradler10]. The details and the properties of entanglement-breaking quantum channels can be found in the work of Horodecki, Shor and Ruskai [Horodecki03]. The additivity of classical capacity for entanglement-breaking channels is proven by Shor's [Shor02a].

**Additivity of Degradable Channels**

The degradable quantum channels have many important properties. *First*, any quantum channel for which the information which can be transmitted from Alice to Bob is greater than the information leaked to the environment satisfies the requirements of a degradable quantum channel [Cubitt08b]. *Second*, the most important quantum channel models belong to this set, such as the erasure channel, the amplitude damping channel, the Hadamard quantum channel, and the bosonic quantum communication channel: these channels all have tremendous importance in practical optical communications [Wolf06]. *Third* (which is the most important for us), Devetak and Shor in 2005 [Devetak05a] have also proved that for degradable quantum channels, *quantum coherent information is additive.* This has deep relevance from the viewpoint of the computation of the capacities of degradable quantum channels, since their capacity can be derived without the computation of the asymptotic formula. (This means that degradable quantum channels are special cases, for which it is enough to know the single-use capacity. The regularization would be necessary if quantum coherent information were non-additive.)

A complementary quantum channel is also a channel, however it is an abstract channel and focuses on the information leaked to the environment, i.e., it describes the "environment's output". The description of this abstract channel model can be found in [Devetak03], [Smith08d], and [Smith09a]. The structure of degradable quantum channels was also studied by Cubitt *et al.* [Cubitt08b].

**Superadditivity of Quantum Coherent Information**

In 1997, Bennett, DiVincenzo and Smolin derived the quantum capacities of the erasure quantum channels (for details see [Bennett97]). This was an important paper, since it stated that the quantum coherent information is superadditive for the depolarizing quantum channel in a very limited domain, which was first mentioned by Shor and Smolin in 1996, in [Shor96b]. In this initial report, Shor and Smolin explicitly gave the very limited range for which the quantum coherent information will be *superadditive*. Originally, the authors studied special quantum error-correcting codes for which it is not necessary to completely reveal the error syndrome, and they finally arrived at a very important conclusion: the superadditivity of quantum coherent information. This work was a very important milestone in Quantum Information Theory, but the picture was only completed in 1998.

In this year, 1998, another important step in the history of quantum capacity was made by DiVincenzo, Shor and Smolin, who analyzed the quantum capacities of various quantum channels [DiVincenzo98]. In this paper, the authors proved that the quantum coherent information is superadditive for the depolarizing quantum channel. The authors showed an example in which they used five qubit length quantum codewords and a special encoding scheme called the repetition code concatenation; for a very limited domain, using this encoding scheme, the quantum coherent information will be additive. For details see the work of DiVincenzo, Shor and Smolin [DiVincenzo98]. This was an important result in the characterization of the quantum capacity of the quantum channel.

In the same year, Schumacher and Westmoreland derived the connection between the private information and the quantum coherent information (see [Schumacher98a]), whose connection was also used in the exact definition of the private classical capacity of the quantum channel by Devetak and Shor in 2005 [Devetak05a] and in the superactiva-tion of quantum capacity by Smith and Yard in 2008 (see [Smith08]). The connection between the Holevo information and the quantum coherent information was shown by Schumacher and Westmoreland [Schumacher2000].

The quantum capacity of another important quantum channel,—the amplitude damping channel—was proven by Giovannetti and Fazio in 2005. The quantum capacity of this channel has great relevance in practical quantum communications, since this channel describes the energy dissipation due to losing a particle. In their work, the authors studied the information-capacity description of spin-chain correlations [Giovannetti05]. On the classical and quantum capacities of Gaussian quantum channels see the works of Wolf and Eisert [Wolf05] and Wolf *et al.* [Wolf06]. About the properties of the Gaussian quantum channels see the work of Eisert and Wolf [Eisert05]. The quantum capacities of some bosonic channels were proven by Wolf and Perez-Garcia in [Wolf07]. A great paper on the classical capacity of quantum Gaussian channels was published by Lupo *et al.* in 2011 [Lupo11].

For about ten years after the superadditivity of quantum coherent information of the depolarizing quantum channel was discovered by DiVincenzo *et al.* in 1998 [DiVincenzo98], no further quantum channels were found for which the quantum coherent information is superadditive. Finally, the picture has broken in 2007, when Smith and Smolin showed new examples for the superadditivity of quantum coherent information. They introduced a new encoding scheme, called the *degenerate quantum codes*, and they proved that with the help of use of these quantum codes for some Pauli channels the quantum coherent information will be superadditive [Smith07].

# D.8 Related Work on Superactivation of Quantum Channels

In this section we summarize the most important works regarding on the superactivation of quantum channel capacities.

### Before Superactivation

The discovery of superactivation was a very important result in the characterization of the capability of the quantum channel to transmit information. The effect of superactivation roots in the extreme violation of additivity, i.e., in the superadditivity property of channel capacities of the quantum channel. Both the superadditivity of classical and quantum channel capacities were already shown before the possibility of the superactivation would had been brought to the surface. While in the case of the superactivation of quantum capacity the superadditivity of quantum coherent information, in the case of the superactivation of classical zero-error capacity, the superadditivity of Holevo information provides the theoretical background for the effect.

The superactivation is nothing more than an extreme violation of additivity property of quantum channels. As the inventors of the HSW-theorem in 1997 have conjectured, entanglement among the input states cannot help to enhance the rate of classical communication. However later, in 2009, Hastings proved that entanglement can help to increase the classical capacity, and showed that the additivity of the Holevo information can fail, i.e., the additivity works only for some very special channels and cannot be extended to the general case. For details, see Hastings's work [Hastings09].

The first important discovery—which also gave a strong background to these advanced properties—was made by Horodecki *et al.* in 2005 [Horodecki05]. As they showed, quantum information can be negative (see Section B), and they have also constructed a protocol which uses this fact. It was an important milestone from the viewpoint of the discovery of the advanced—classically unimaginable—properties of quantum channels. Further details about the meaning of the negativity of quantum information can be found in the proof of Horodecki *et al.* [Horodecki05] and 2007 [Horodecki07]. Before their results appeared, a paper about the role of negative entropy and information in Quantum Information Theory was published by Cerf [Cerf97]. In the paper of Horodecki *et al.* [Horodecki05], the authors also defined a protocol which can exploit the negativity of quantum information. The continuity of quantum conditional entropy function was proved by Alicki and Fannes in 2004 [Alicki04]. An attempt for giving a uniform framework for the currently known different quantum protocols was made by Devetak *et al.* in 2004 [Devetak04a], by Devetak and Shor in 2005 [Devetak05a], and by Devetak *et al.* in 2008 [Devetak08]. However, as they concluded, there are still many open questions. Later, in 2006, Abeyesinghe *et al.* published a paper in which they tried to give a more generalized picture of the various quantum communication protocols and their various capacities [Abeyesinghe06].

**Discovery of Superactivation**

The possibility of the superactivation of the quantum channels was discovered by Graeme Smith and Jon Yard in 2008 [Smith08]. They have shown that the quantum capacity of zero-capacity quantum channels can be superactivated, and in 2011 they demonstrated in laboratory environment that the superactivation of the quantum capacity also works in practice [Smith11]. In 2009 and 2010, Duan and Cubitt *et al.* showed that the classical zero-error capacity [Duan09], [Cubitt09], and the quantum zero-error capacity can also be superactivated [Cubitt09a]. In 2011 the effect of superactivation was extended to more general classes of quantum channels. Brandao, Oppenheim and Strelchuk have demonstrated that the superactivation of depolarizing quantum channels is also possible and can be extended for more general classes, for details see [Brandao11]. In 2010, Brandao *et al.* have also studied the connection between the public quantum communication and the effect of superactivation, and the possible impacts on quantum error correction and entanglement distillation, for details see [Brandao10]. A very good overview on the working mechanism of superactivation was published by Oppenheim in 2008, see [Oppenheim08].

In 2008 Smith and Smolin showed that the non-additivity of the private capacity can be extended in a different way [Smith08a], since this property can be used in the superactivation of zero-capacity quantum channels. Li *et al.* [Li09] constructed a channel combination for which the entanglement-assisted quantum capacity is greater than the classical capacity. The results presented by Li and Winter *et al.*, demonstrated that the private capacity is also non-additive [Li09]. The very strong non-additivity of private capacity has been shown by Smith [Smith09b], and by Li *et al.* [Li09]. Many of these discoveries were made in 2008 and 2009, and the some results were discovered just in 2010 and 2011. In both of the superadditivity property of classical private information and the superactivation of quantum capacity the erasure channel has great importance. The fact that any symmetric channel (i.e., for example a 50% erasure channel) has zero quantum capacity was shown in [Bennett97]. The proof of that any positive capacity would violate the no-cloning theorem was shown in [Bennett96a].

**Summarize**

On the superactivation of the quantum capacity of the quantum channel see [Smith11], [Smith08], [Brandao10], [Brandao11]. About the superactivation of classical zero-error capacity of a quantum channel see [Duan09], [Cubitt09]. The superactivation of quantum zero-error capacity was shown by Cubitt and Smith in [Cubitt09a]. On the algorithmical superactivation of asymptotic quantum capacity and classical zero-error capacity see the papers of Gyongyosi and Imre [Gyongyosi11a] and [Gyongyosi11b]. Further applications of relative entropy function can be found in [Kullback87], [Onishi97], [Yoshizawa99].

# Appendix E

# Geometric Interpretation of Superactivation of Quantum Channels

## E.1 Some Important Channel Maps

Here, we give a brief survey of some important quantum channel maps. We discuss the density matrix representation of these channel models and their geometric illustration on the Bloch sphere. For the corresponding definitions related to the state-vector description we advise to the reader to [Imre05]. For further information regarding the geometrical interpretation of quantum channel capacities see the book of Imre and Gyongyosi [Imre12].

### E.1.1 The Flipping Channel Models

The *bit flip* channel can be defined by means of $\sigma_X$, the Pauli $X$ transformation. The bit flip channel changes the probability amplitudes of the input qubit. The map of the bit flip channel can be expressed as

$$\mathcal{N}\left(\rho\right) = p\left(\sigma_X \rho \sigma_X\right) + \left(1-p\right)\rho \,. \tag{E.1}$$

In the geometric representation (see Fig. E.1), this channel map shrinks the original Bloch sphere along the $y$ and $z$ axes, by the factor $1-2p$.

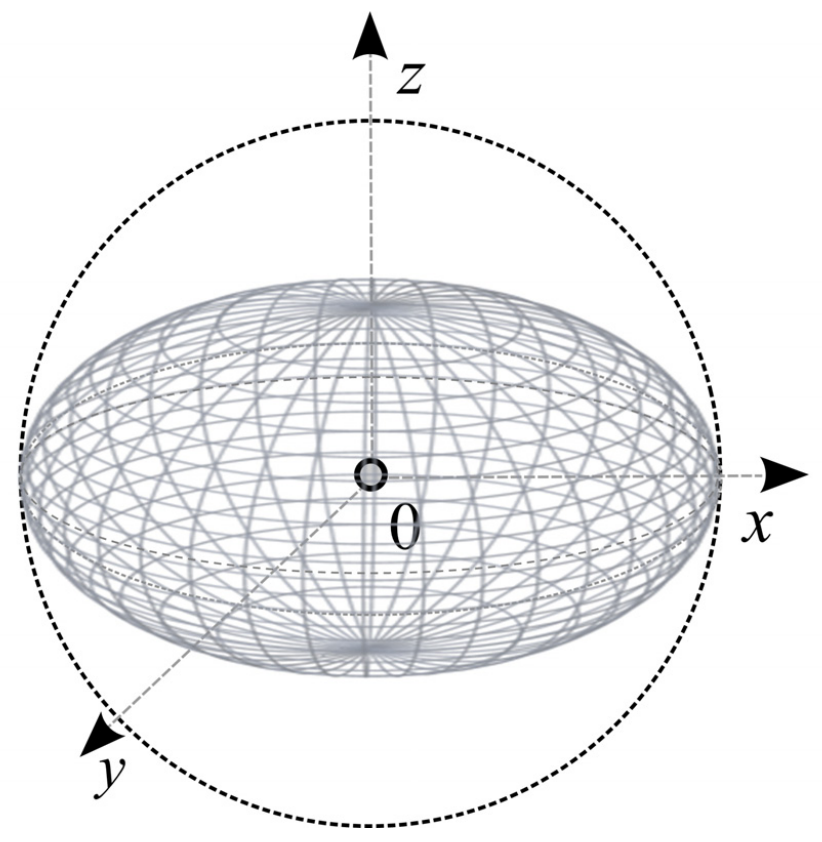

**Fig. E.1.** The bit flip channel shrinks the Bloch sphere along the $y$ and $z$ axes.

Similarly to the bit flip channel, the *phase flip* quantum channel applies the Pauli $Z$ transformation $\sigma_Z$. The phase flip channel changes the sign of the relative phase of the input qubit. The map $\mathcal{N}$ of this channel can be expressed as

$$\mathcal{N}\left(\rho\right)=p\left(\sigma_Z\rho\sigma_Z\right)+\left(1-p\right)\rho, \tag{E.2}$$

where $p$ describes the probability that the channel does a phase-flip error on the input qubit. In the Bloch sphere representation means that the width of the original Bloch sphere will be reduced by a factor of $1-2p$ in its equatorial plane. The surface of channel ellipsoid consists of the set of channel output $\rho$ vectors.

In the geometric representation (see Fig. E.2), the phase flip channel map shrinks the original Bloch sphere along the $x$ and $y$ axes, by the factor $1-2p$.

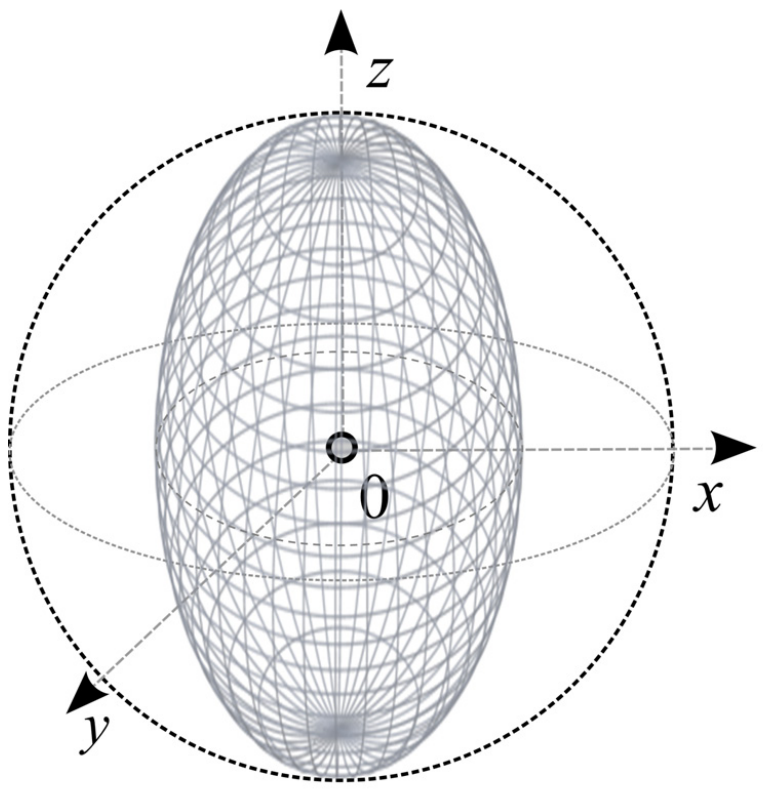

**Fig. E.2.** The geometric interpretation of the phase flip channel.

In the physical realizations, this channel map is also referred as the *phase-damping* channel model. If the channel simultaneously realizes bit flip and phase flip transformations on the input quantum state then the channel is called a *bit-phase flip* channel. The effect of the bit-phase flip channel can be described by the $\sigma_Y$, the Pauli $Y$ transformation

$$\mathcal{N}\left(\rho\right)=p\left(\sigma_Y\rho\sigma_Y\right)+\left(1-p\right)\rho\,. \tag{E.3}$$

This channel also shrinks the Bloch sphere by the factor $1-2p$ along the direction of the $x$ and $z$ axes. The geometric interpretation of the *bit-phase flip* channel is depicted in Fig. E.3.

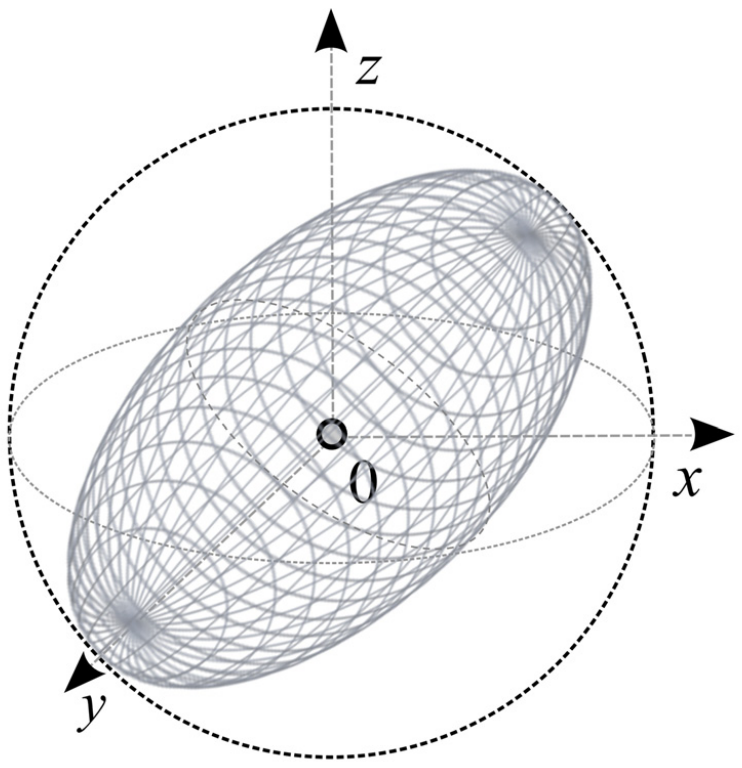

**Fig. E.3.** The image of the bit-phase flip channel.

Next, we discuss the depolarizing quantum channel model.

## E.1.2 The Depolarizing Channel Model

The last discussed unital channel model is the *depolarizing* channel which performs the following transformation

$$\mathcal{N}\left(\rho_i\right) = p\frac{I}{2} + \left(1-p\right)\rho_i, \tag{E.4}$$

where $p$ is the *depolarizing parameter* of the channel, and if Alice uses two orthogonal states $\rho_0$ and $\rho_1$ for the encoding then the mixed input state is

$$\rho = \left(\sum_i p_i\rho_i\right) = p_0\rho_0 + \left(1-p_0\right)\rho_1. \tag{E.5}$$

After the unital channel has realized the transformation $\mathcal{N}$ on state $\rho$, we will get the following result

$$\begin{aligned}\mathcal{N}(\rho) &= \mathcal{N}\left(\textstyle\sum_i p_i\rho_i\right) = \mathcal{N}\left(p_0\rho_0 + \left(1-p_0\right)\rho_1\right) \\ &= p\frac{1}{2}I + \left(1-p\right)\left(p_0\rho_0 + \left(1-p_0\right)\rho_1\right) \\ &= \begin{pmatrix} p\frac{1}{2} + \left(1-p\right)p_0 & 0 \\ 0 & p\frac{1}{2} + \left(1-p\right)\left(1-p_0\right)\end{pmatrix}.\end{aligned} \tag{E.6}$$

Geometrically, the map of the depolarizing quantum channel shrinks the original Bloch sphere in every direction by $1-p$. (The effect of the depolarizing quantum channel is shown in Fig. E.4.)

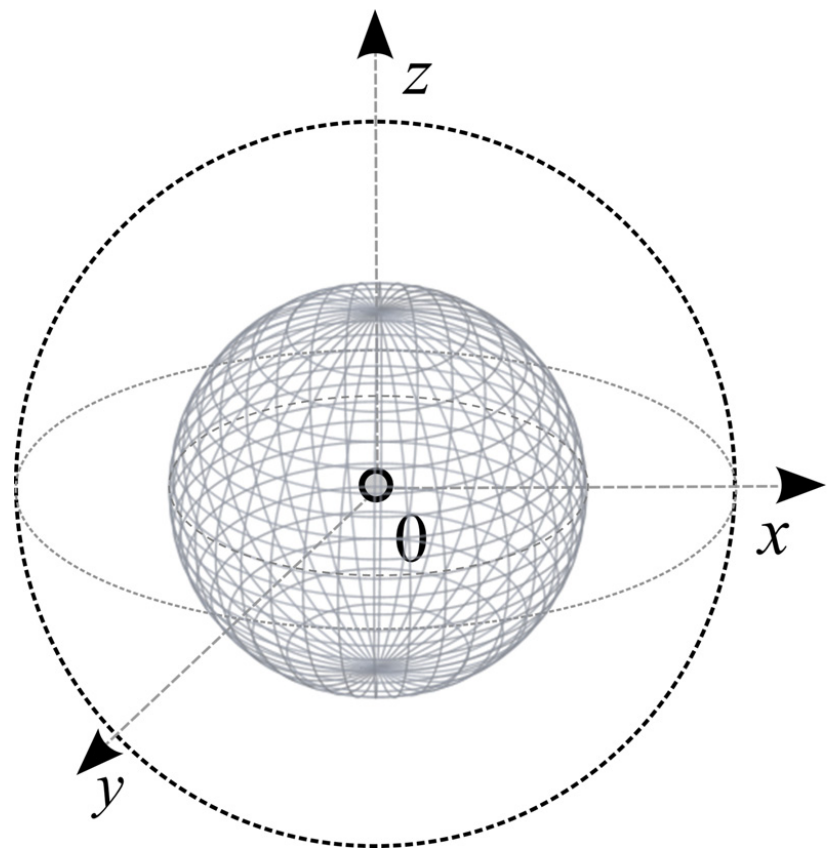

**Fig. E.4.** The depolarizing channel shrinks the original Bloch sphere in every direction.

One of the most important type of channel map to describe decoherence is called the amplitude damping channel (or decay) map. We will discuss it in the next subsection.

## E.1.3 The Amplitude Damping Channel

The result of decoherence is also a mixed quantum state, such as in the case of the previously discussed channel maps, however in this case, the density matrices of these mixed states will differ. In the case of decoherence, the non-diagonal values of the density matrix completely vanish.

The *amplitude damping* channel map shrinks the Bloch sphere in the two directions of the equatorial plane – similarly to the phase flip channel, but it also moves the center of the ellipsoid from the center of the Bloch sphere. Therefore, this channel map is not unital. The height of the scaled ellipsoid will be given by the scaling factor $1-2p$. The direction of the shift can be upward or downward. On the other hand, the width of the ellipsoid will differ from the previous cases, namely by to the factor $\sqrt{1-2p}$. The geometric interpretation of the amplitude damping quantum channel is illustrated in Fig. E.5. (We emphasize that the ball can be shrunk to the opposite direction along the $z$ axis, too.)

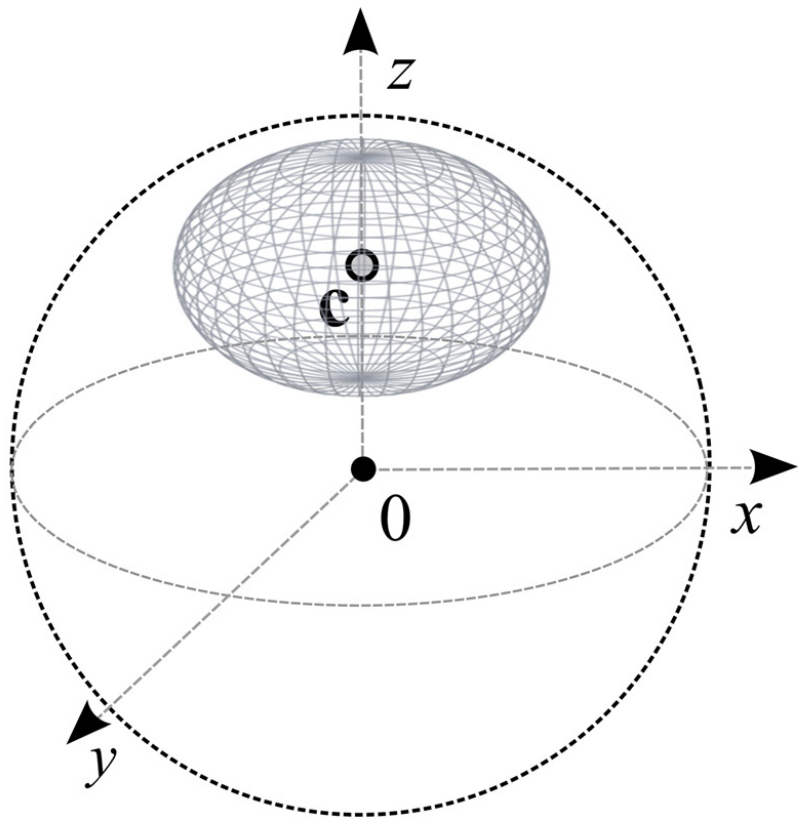

**Fig. E.5.** The picture of the amplitude damping quantum channel.

Any quantum channel $\mathcal{N}$ can be described in the Kraus-representation [King09], using a set of Kraus matrices $\mathcal{A}=\{A_i\}$ in the following form

$$\mathcal{N}(\rho) = \sum_i A_i \rho A_i^\dagger, \tag{E.7}$$

where $\sum_i A_i^\dagger A_i = I$. For an amplitude damping quantum channel

$$A_1 = \begin{bmatrix} \sqrt{p} & 0 \\ 0 & 1 \end{bmatrix}, \text{ and } A_2 = \begin{bmatrix} 0 & 0 \\ \sqrt{1-p} & 0 \end{bmatrix}, \tag{E.8}$$

where $p$ represents the probability that the channel leaves input state $|0\rangle$ unchanged. Obviously the channel flips the input state from $|0\rangle$ to $|1\rangle$ with probability $1-p$. [Cortese02], [Nielsen2000]. As can be concluded, for $p=0$, the channel output is $|1\rangle$ with probability 1. However, the channel leaves untouched the input state $|1\rangle$, hence the output of the channel will be $|1\rangle$.

For a non-unital quantum channel, the set of Kraus operators $\mathcal{A}=\{A_i\}$ can be transformed to the *King-Ruskai-Szarek-Werner* (KRSW) ellipsoid channel model with parameters $\{t_k, \lambda_k\}$, $k=1,2,3$. The effect of $\{t_k \neq 0\}$ is that the average output $\sigma = \sum_i p_i \rho_i$ of the channel moves away from the origin of the Bloch sphere, meaning that the center of the smallest enclosing quantum informational ball is not equal to the origin of the Bloch sphere.

The affine map of the amplitude damping channel can be expressed using Bloch vectors $\mathbf{r}_{in}$ and $\mathbf{r}_{out}$ in the following way

$$\mathbf{r}_{out} = \begin{pmatrix} \mathbf{r}_{out}^{(x)} \\ \mathbf{r}_{out}^{(y)} \\ \mathbf{r}_{out}^{(z)} \end{pmatrix} = \begin{pmatrix} \sqrt{1-p} & 0 & 0 \\ 0 & \sqrt{1-p} & 0 \\ 0 & 0 & 1-\frac{p}{2} \end{pmatrix} \begin{pmatrix} \mathbf{r}_{in}^{(x)} \\ \mathbf{r}_{in}^{(y)} \\ \mathbf{r}_{in}^{(z)} \end{pmatrix} + \begin{pmatrix} 0 \\ 0 \\ \frac{p}{2} \end{pmatrix}. \tag{E.9}$$

The amplitude damping channel can be visualized in the KRSW ellipsoid channel model

$$\begin{aligned} &t_x = 0,\ t_y = 0,\ t_z = 1-p, \\ &\lambda_x = \sqrt{p},\ \lambda_y = \sqrt{p},\ \lambda_z = p, \end{aligned} \tag{E.10}$$

where $p \in [0,1]$ is the channel parameter.

## E.1.4 The Dephasing Channel Model

The second type of decoherence map discussed is unitary and results in relative phase differences between the computational basis states: the channel map which realizes it is called the *dephasing* map. In contrast to the amplitude damping map, it realizes a unitary transformation. The unitary representation of the dephasing quantum channel for a given input $\rho = \sum_{i,j} \rho_{ij} |i\rangle\langle j|$ can be expressed as

$$\mathcal{N}(\rho) = \sum_i \rho_{ii} |E_i\rangle\langle E_i|, \tag{E.11}$$

where $|E_i\rangle$ are the environment states. The dephasing quantum channel acts on the density operator $\rho$ as follows

$$\mathcal{N}\left(\rho_i\right) = p\sigma_Z\rho\sigma_Z + \left(1-p\right)\rho_i, \tag{E.12}$$

where $\sigma_Z$ is the Pauli $Z$-operator. The dephasing quantum channel is also degradable, for definition of degradable quantum channel see Section D.1.2.1. The image of the dephasing channel map is similar to that of the phase flip channel map, however, the shrinkage of the original Bloch sphere is greater (see Fig. E.6). The dephasing channel transforms an arbitrary superposed quantum state $\alpha\left|0\right\rangle + \beta\left|1\right\rangle$ into a mixture

$$\mathcal{N}\left(\rho\right) \to \rho' = \begin{bmatrix} \left|\alpha\right|^2 & \alpha\beta^* e^{-\gamma(t)} \\ \alpha^*\beta e^{-\gamma(t)} & \left|\beta\right|^2 \end{bmatrix}, \tag{E.13}$$

where $\gamma\left(t\right)$ is a positive real parameter, which characterizes the coupling to the environment, using the time parameter $t$.

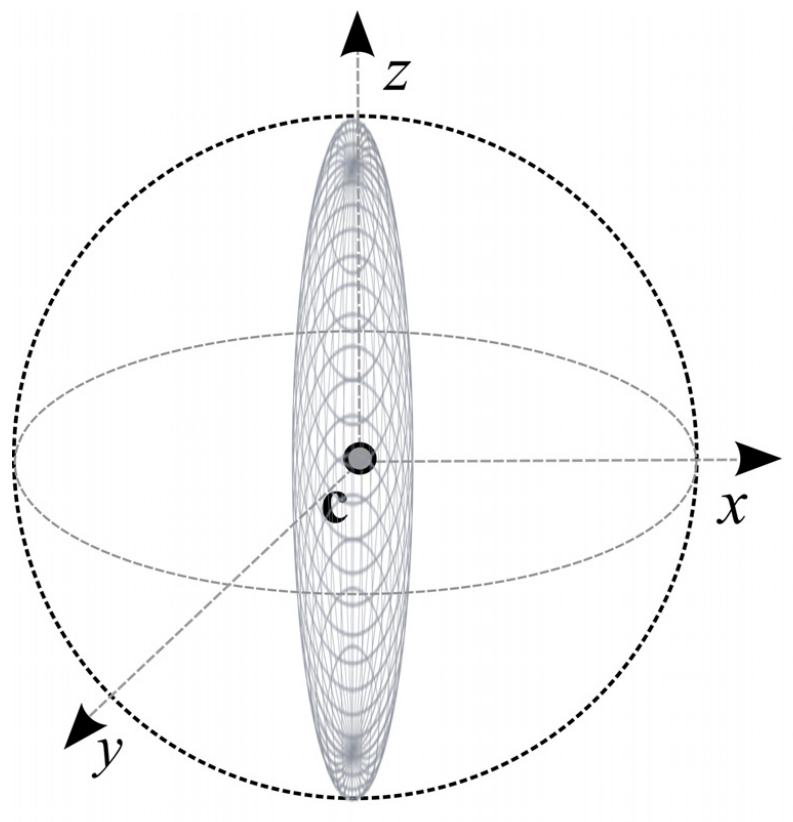

**Fig. E.6.** The image of the dephasing channel map.

Finally, next we study the pancake map, which also represents decoherence.

## E.1.5 The Pancake Map

To give an example for physically not allowed (nonphysical, non-CP) transformations, we introduce the *pancake map* in Fig. E.7. The non-CP property means, that there exists no Completely Positive Trace Preserving map, which preserves some information along the equatorial spanned by the $x$ and $y$ axes of the Bloch sphere, while it completely demolishes any information along the $z$ axis. This map is called the pancake map, and it realizes a physically not allowed (non-CP) transformation. The effect of the pancake map is similar to the bit-phase flip channel, however, this channel defines a non-CP transform: it "smears" the original Bloch sphere along the equatorial spanned by the $x$ and $y$ axes.

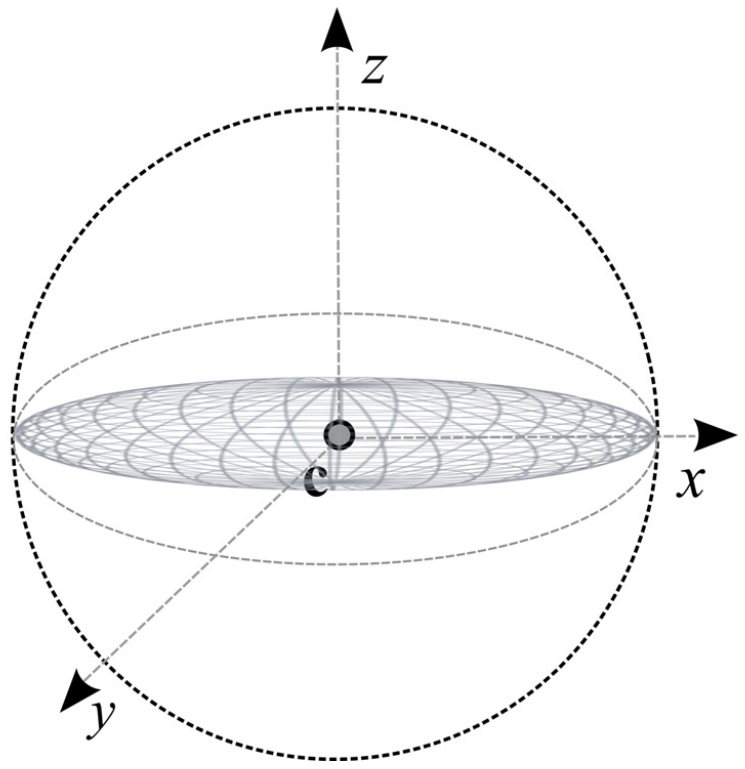

**Fig. E.7.** The pancake map is a physically not allowed map.

On the other hand, the pancake map—besides the fact that is a non-physical map—can be used theoretically to transfer some information, and some information can be transmitted through these kinds of channel maps. The reason behind decoherence is *Nature*. She cannot be perfectly eliminated from quantum systems in practice. The reduction of decoherence is also a very complex task, hence it brings us on the engineering side of the problem: the quantum systems have to be designed in such a way that the unwanted interaction between the quantum states and the environment has to be minimal [Shor95], [Shor96]. Currently - despite the efficiency of these schemes - the most important tools to reduce decoherence are quantum error-correcting codes and decoupling methods.

# E.2 Geometric Interpretation of the Quantum Channels

The map of a quantum channel compresses the Bloch sphere, by an affine map. This affine map must be Completely Positive (CP) and Trace Preserving (TP), which shrinks the Bloch sphere along the *x, y* and *z* axes. Now, we introduce a new geometric representation - called the *tetrahedron* - which also can be used to represent geometric the various channel maps. We also show that there is a connection with the Bloch sphere representation.

## E.2.1 The Tetrahedron Representation

Assuming the single-qubit case the quantum channel's output is represented by a $2\times 2$ density matrix and the operation is a trace preserving Completely Positive map. The map of the quantum channel has to be CP, thus $\mathcal{I}_n \otimes \mathcal{N}$ is a Positive map for all *n*, where $\mathcal{I}_n$ is the identity map on $n \times n$ matrices. The map of the quantum channel on a single-qubit in the Bloch sphere representation, can be given by the affine map

$$\mathcal{N}\left(\mathbf{r}\right) = \mathbf{r}_{\mathcal{N}} = A\mathbf{r} + \vec{b}, \tag{E.14}$$

where $A$ is a $3 \times 3$ diagonal matrix with entries $\vec{\eta} = \left(\eta_x, \eta_y, \eta_z\right)$ which characterizes the tetrahedron $\mathcal{T}$, $\vec{b}$ is a three-dimensional vector representing the shift of the center of the Bloch sphere, $\mathbf{r}$ is the initial Bloch vector of the sent pure quantum state, and $\mathbf{r}_{\mathcal{N}}$ is the Bloch vector of the channel output state.

The entries of $A$ specify the tetrahedron $\mathcal{T}$ in the parameter space of $\left\{\eta_x, \eta_y, \eta_z\right\}$, where $\eta_i \in \mathcal{T}$ if

$$\left|\eta_x \pm \eta_y\right| \leq \left|1 \pm \eta_z\right|. \tag{E.15}$$

The tetrahedron $\mathcal{T}$ is the *convex hull* of the points representing $\mathcal{I}$, and the three Pauli rotations, thus every transformation corresponding to a point in the tetrahedron $\mathcal{T}$ can be described as a statistical mixture of the Pauli-transformations $\mathcal{I},\sigma_x,\sigma_y$ and $\sigma_z$ where $\mathcal{I}$ is the identity transformation, and $\sigma_x,\sigma_y,\sigma_z$ are rotations by $\pi$ around the *x, y* and *z* axes. Fig. E.8 illustrates the tetrahedron $\mathcal{T}$ for of the physically allowed transformations of the quantum channel. The vertices of $\mathcal{T}$ are the Pauli-transformations $\mathcal{I},\sigma_x,\sigma_y$ and $\sigma_z$.

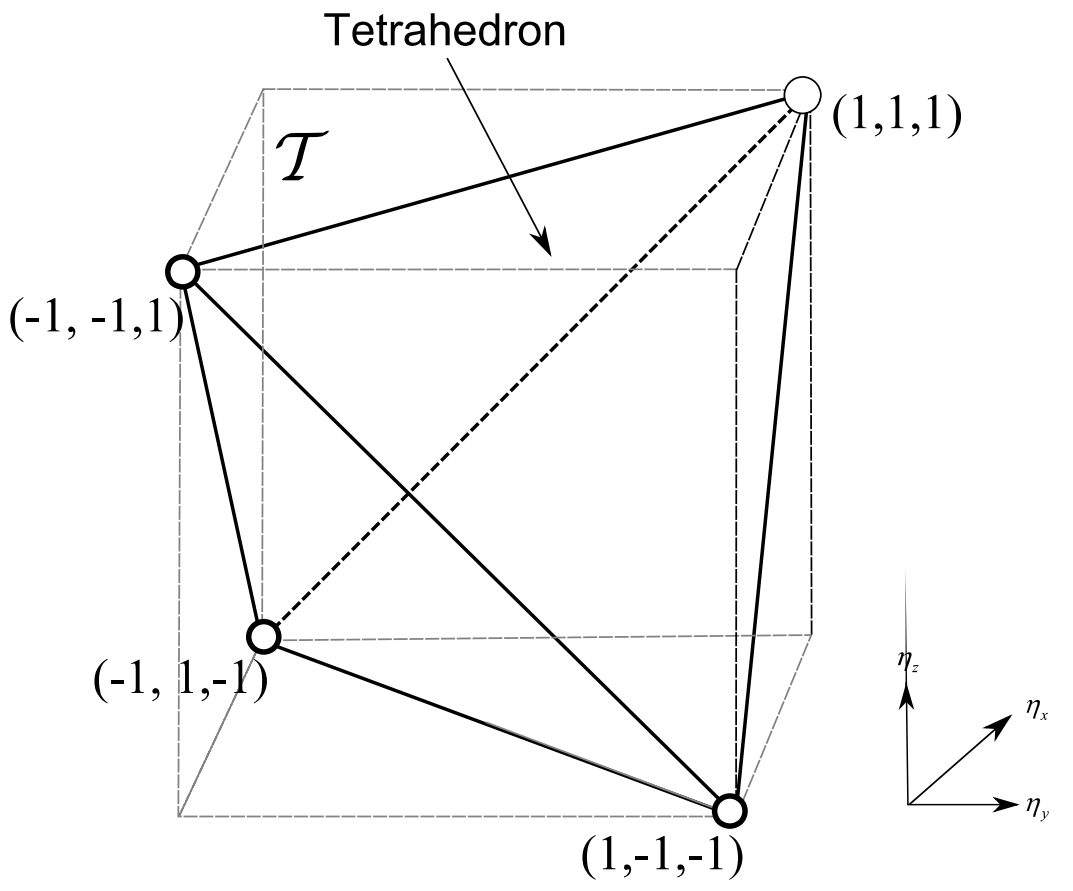

**Fig. E.8.** The Pauli transformations can be represented by the tetrahedron. The various physically allowed channel maps of the quantum channel also can be described by this representation.

The vertices of $\mathcal{T}$ correspond with the four maps which can be described as [Bengtsson06]

$$\rho \to \rho' = \sum_{j=0}^{3} \varepsilon_j \sigma_j \rho \sigma_j^{\dagger}, \tag{E.16}$$

where $\sigma_0$ is the identity matrix $\mathcal{I}$, while for $j = 1,2,3$, $\sigma_j$ denotes the *j*-th Pauli operator ($X$, $Z$, $Y$), and $\varepsilon_0 + \varepsilon_1 + \varepsilon_2 + \varepsilon_3 = 1$, where $\varepsilon_1,\varepsilon_2$ and $\varepsilon_3$ are non-negative parameters. The general transformation $\mathcal{N}$ of the quantum channel can be described as a convex sum of these maps

$$\rho' = \mathcal{N}\left(\rho\left(x,y,z\right)\right) = \varepsilon_1\sigma_1\rho\sigma_1 + \varepsilon_2\sigma_2\rho\sigma_2 + \varepsilon_3\sigma_3\rho\sigma_3 + \left(1 - \varepsilon_1 - \varepsilon_2 - \varepsilon_3\right)\rho. \tag{E.17}$$

The points forming the vertices of $\mathcal{T}$ represent unitary maps for which only one operator is required in the operator sum representation, see (E.17), while the edges of $\mathcal{T}$ depict the two-operator maps, and the faces of $\mathcal{T}$ assign the maps described by three operators. The points inside $\mathcal{T}$ require all four operators.

Since the quantum channel performs a *CP* map, the map $\mathcal{N}$ has to be *physically allowed* on all other quantum states. A unital quantum channel does not change the center of the Bloch sphere, thus $\vec{b} = 0$. In this case matrix $A$ is diagonal filled with the elements of distortion vector $\vec{\eta} = \left(\eta_x, \eta_y, \eta_z\right)$. Assuming the maximally entangled two qubit system $\left|\beta_{00}\right\rangle\left\langle\beta_{00}\right|$, where $\left|\beta_{00}\right\rangle = \sum_i \left|i\right\rangle\left|i\right\rangle$, for a CP-map $\mathcal{N}$, the condition $I \otimes \mathcal{N}\left(\left|\beta_{00}\right\rangle\left\langle\beta_{00}\right|\right) \geq 0$ has to be satisfied which leads to the channel output matrix

$$\rho' = \frac{1}{2}\begin{pmatrix} 1+\eta_z & 0 & 0 & \eta_x+\eta_y \\ 0 & 1-\eta_z & \eta_x-\eta_y & 0 \\ 0 & \eta_x-\eta_y & 1-\eta_z & 0 \\ \eta_x+\eta_y & 0 & 0 & 1+\eta_z \end{pmatrix}, \tag{E.18}$$

which is positive if and only if

$$\left(1+\eta_z\right)^2 - \left(\eta_x+\eta_y\right)^2 \geq 0, \text{ and } \left(1-\eta_z\right)^2 - \left(\eta_x-\eta_y\right)^2 \geq 0. \tag{E.19}$$

The tetrahedron representation focuses specially on the physically allowed transformations compared to the Bloch sphere. On the other hand, we have also highlighted the fact that there is a connection (see (E.14)) between Bloch sphere and tetrahedron representations.

## E.2.2 Quantum Channel Maps in Tetrahedron Representation

In this section we introduce the tetrahedron representation of the various channel maps.

### E.2.2.1 Description of Channel Maps

Now, let us investigate these quantum channel maps in the previously defined tetrahedron representation. On the edges of the tetrahedron, we can find the unital *bit flip*, *phase flip* and the *coarse graining* transformations. The bit flip and phase flip channel maps transform the original Bloch sphere into a distorted ellipsoid, which touches the original Bloch sphere at the points $\left\{\frac{1}{\sqrt{2}}\left(|0\rangle+|1\rangle\right),\frac{1}{\sqrt{2}}\left(|0\rangle-|1\rangle\right)\right\}$ and points $\left\{|0\rangle,|1\rangle\right\}$, respectively. An other important channel is the coarse graining channel; it transforms the whole Bloch sphere into a unit length vertical line segment centered at the origin of the Bloch sphere. The locations of bit flip, phase flip, bit-phase flip and the coarse graining channels are shown in Fig. E.9.

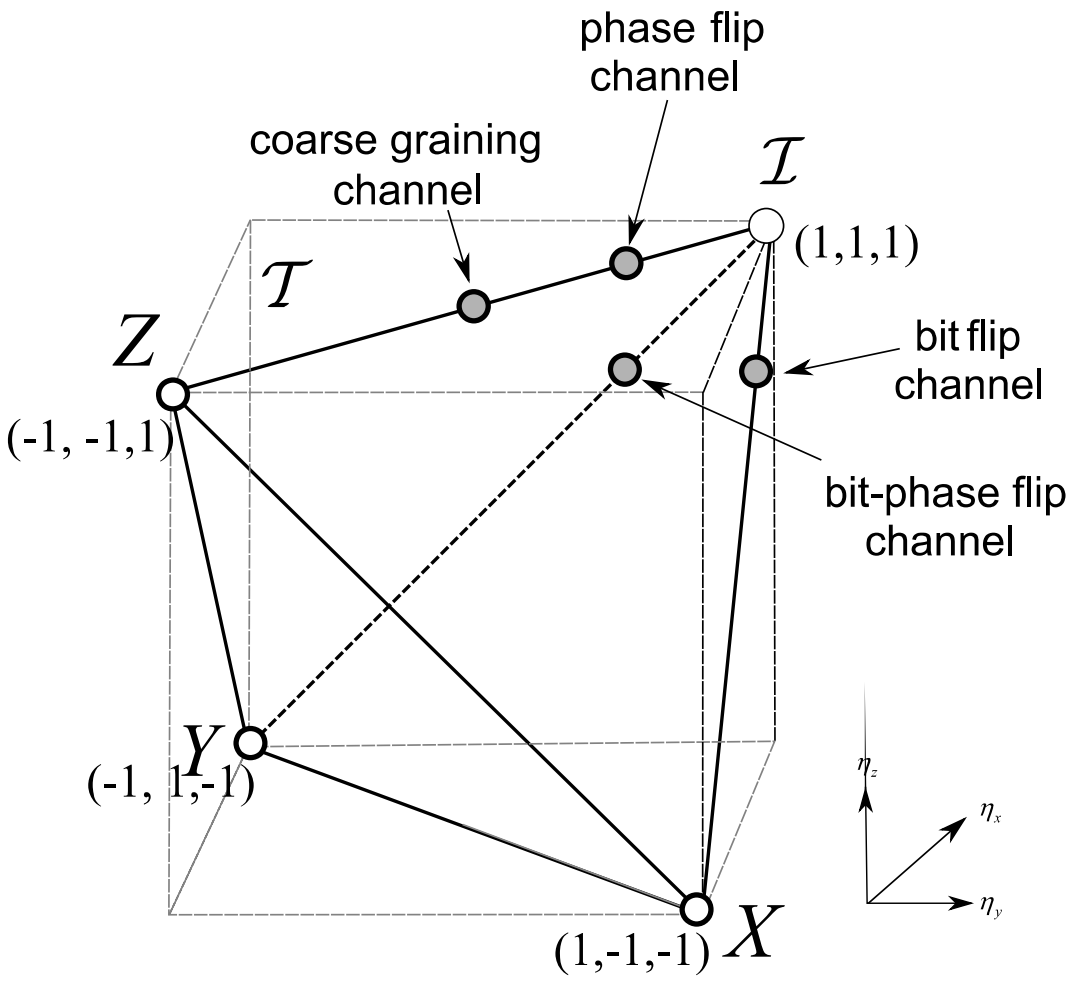

**Fig. E.9.** The bit flip, phase flip and coarse graining channels in the tetrahedron representation.

Using the distortion parameters $\left\{\eta_x,\eta_y,\eta_z\right\}$ of the tetrahedron $\mathcal{T}$, the $I$ identity transformation can be expressed as

$$\vec{\eta}_{identity}=\left(1,1,1\right). \tag{E.20}$$

The distortion vector $\vec{\eta}$ of the bit flip, phase flip, bit-phase flip and the coarse graining quantum channels in function of channel parameter $p$ can be expressed as

$$\vec{\eta}_{bit\ flip} = \left(1{,}1-2p{,}1-2p\right), \tag{E.21}$$

$$\vec{\eta}_{phase\ flip} = \left(1-2p{,}1-2p{,}1\right), \tag{E.22}$$

$$\vec{\eta}_{bit-phase\ flip} = \left(1-2p{,}1{,}1-2p\right) \tag{E.23}$$

and

$$\vec{\eta}_{coarse\ gr.} = \left(0{,}0{,}1\right). \tag{E.24}$$

In the case of bit flip, phase flip and bit-phase flip channels the "worst case scenario" occurs at $p = \frac{1}{2}$. In these cases, these channels maps are degenerated, which results in a line with unit length. Furthermore, as can be observed in Fig. E.9, the coarse graining channel can be viewed as the "worst case scenario" of a phase flip channel.

Inside the tetrahedron, we can find the *linear channel* map model and the *depolarizing* channel model, which are both unital. The linear channel transform maps the original Bloch sphere to a vertical line segment,

$$\vec{\eta}_{linear} = \left(0,0,q\right), \tag{E.25}$$

while the completely depolarizing channel maps the whole Bloch sphere to one point, namely to the center of the Bloch sphere. On the other hand, while in the case of coarse graining channel the length of the line is unit, in case of the linear quantum channel the length of the line depends on the channel parameter $q$.

The output of a completely depolarizing channel is a maximally mixed state, the channel shrinks both coordinates equally. In case of linear channels, the channel map results in a line, $2q$ of lengths, while the degenerated maps occurs at $q = 0$ and $q = 1$. If $q = 0$, the channel will be represented by a point in the center of the tetrahedron, or in other words in this case, the linear quantum channel is equivalent to a *completely depolarizing* quantum channel.

Using the tetrahedron representation, the *depolarizing* quantum channel can be expressed as

$$\vec{\eta}_{depol.} = \left[1 - x\right]\left(1,1,1\right), \tag{E.26}$$

where $x$ determines the level of the shrinking of the original Bloch sphere. In case of $x = 1$ we have a completely depolarizing channel

$$\vec{\eta}_{comp.\ depol.} = \left(0,0,0\right), \tag{E.27}$$

which map can be found in the center of the tetrahedron, as illustrated in Fig. E.10. In this figure, we also show the linear channel and the completely depolarizing channel maps in the tetrahedron representation.

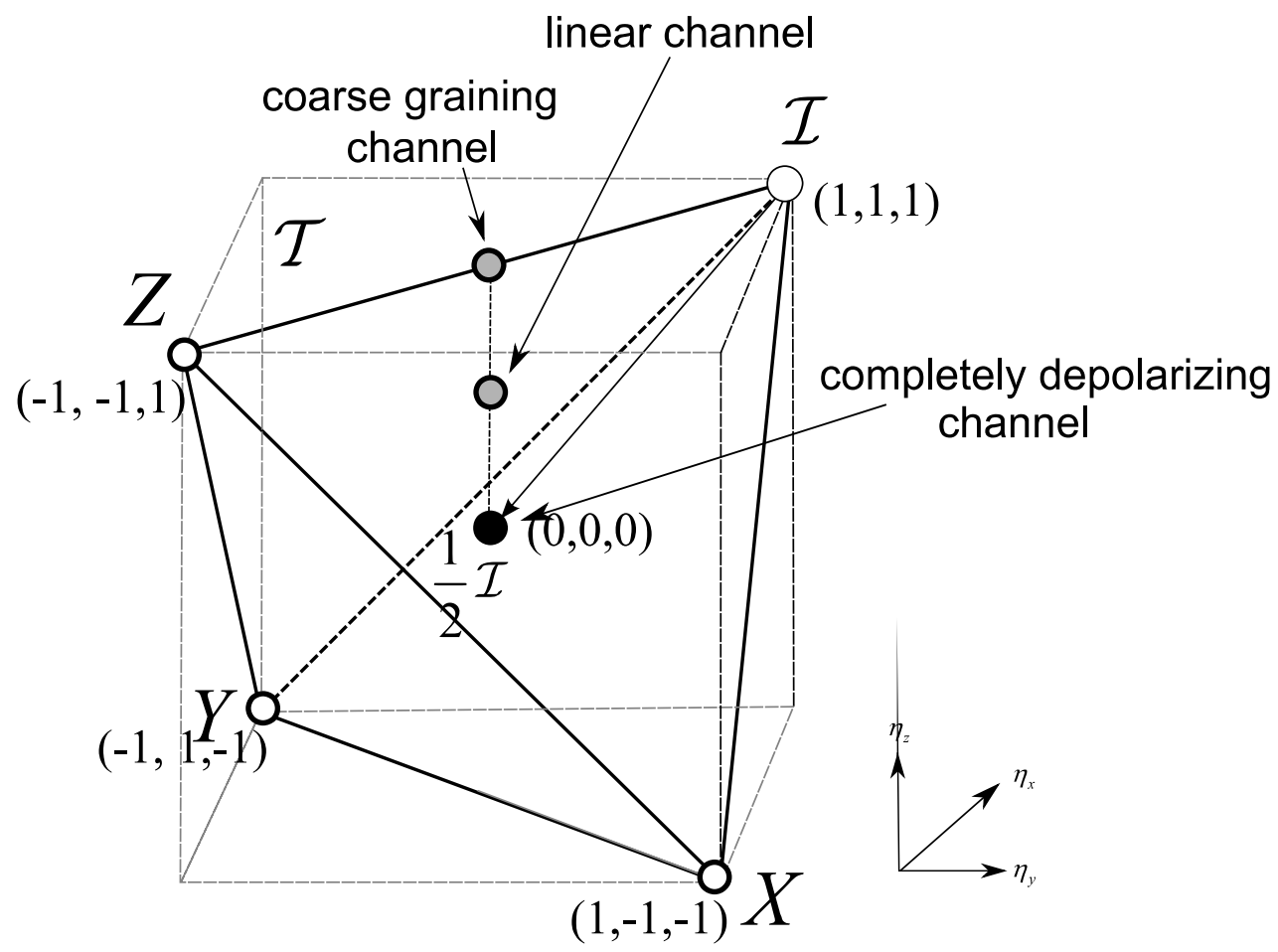

**Fig. E.10.** The linear channel and the completely depolarizing channel maps are inside the tetrahedron.

The unital quantum channels can be represented in the tetrahedron view. The most important geometric property of the unital channel maps, in comparison to the non-unital maps, is that they do not change the center of the Bloch sphere. On the other hand, the non-unital quantum channel models - such as the amplitude damping quantum channel - cannot be represented in the tetrahedron.

### E.2.2.2 Non-Unital Quantum Channel Maps

The description of the non-unital quantum channel maps requires a more complex mathematical background. But, where does the problem arise from? The problem here is that for non-unital transformations, the center of the transformed Bloch sphere will differ from the center of the original Bloch sphere. This fact might seem to be negligible at first, but it also has an important corollary: namely, in case of non-unital channels we have to define a more complex geometric structure. The reason behind this strange thing can be summarized as follows.

The cube which contains the tetrahedron $\mathcal{T}$ , defines the Positive (i.e., not Completely Positive) maps. The tetrahedron with constraints (E.19), defines the convex polytope of CP unital maps. If the map of the quantum channel is Positive, then the channel ellipsoid will lie in the original Bloch sphere. The unital quantum channels are a subset of the Positive maps, since they hold the center of the channel ellipsoid and they lie inside the original Bloch sphere. As we have stated previously, the tetrahedron defines CP maps - or with other words, the unital channel maps. For the non-unital channel maps the center differs from the origin of the Bloch sphere, which also implies that these maps cannot be represented on the tetrahedron. As we will see, for example, the very important amplitude damping channel model is a non-unital transformation, and it is a physically allowed map, but its description is more sophisticated. The amplitude damping channel model has great relevance to practical optical communications, since this channel model describes the energy dissipation due to losing a particle. Moreover, this channel can also take a mixed input state to a pure output state. Beside the fact we cannot represent the amplitude damping channel on the tetrahedron, we can give the distortion vector $\vec{\eta}$ of the amplitude damping channel as follows

$$\vec{\eta}_{ampl.\ damp.} = \left(\sqrt{1-2p},\sqrt{1-2p},1-2p\right). \tag{E.28}$$

However, it does not make complete the picture, since in this case further parameters are needed, since the center of the channel ellipsoid differs from the origin. Before discussing the most relevant channel models in detail we introduce the reader the geometric interpretation of quantum informational distance.

## E.3 The Classical Zero-Error Capacities of some Quantum Channels

Having presented some examples related to the classical and quantum capacities of quantum channels we show two illustrations for zero error capacities.

The zero-error capacity of the bit flip channel (see Section E.1.1) can be reached for the following two non-adjacent orthogonal input states (see the channel ellipsoid in Fig. E.1)

$$\left|\psi_1\right\rangle = \frac{1}{\sqrt{2}}\left(\left|0\right\rangle + \left|1\right\rangle\right) \text{ and } \left|\psi_2\right\rangle = \frac{1}{\sqrt{2}}\left(\left|0\right\rangle - \left|1\right\rangle\right) \tag{E.29}$$

Using these non-adjacent input states for the encoding, the single-use classical zero-error capacity of the bit flip channel is

$$C_0^{(1)}\left(\mathcal{N}\right) = \frac{1}{1}\log\left(2\right) = 1\,. \tag{E.30}$$

It can be verified that for these inputs the non-adjacent property holds:

$$\begin{aligned}
&Tr\left(\mathcal{N}\left(\left|\psi_1\right\rangle\left\langle\psi_1\right|\right)\mathcal{N}\left(\left|\psi_2\right\rangle\left\langle\psi_2\right|\right)\right) \\
&= Tr\left(\left(p\left(\sigma_X\left|\psi_1\right\rangle\left\langle\psi_1\right|\sigma_X\right) + \left(1-p\right)\left|\psi_1\right\rangle\left\langle\psi_1\right|\right)\left(p\left(\sigma_X\left|\psi_2\right\rangle\left\langle\psi_2\right|\sigma_X\right) + \left(1-p\right)\left|\psi_2\right\rangle\left\langle\psi_2\right|\right)\right) \\
&= 0.
\end{aligned} \tag{E.31}$$

For any other two inputs the classical zero-error capacity of the bit flip channel is trivially zero. To describe the classical zero-error capacity of the depolarizing channel, we use the channel map already shown in Section E.1.2, i.e., $\mathcal{N}\left(\rho_i\right) = p\frac{I}{2} + \left(1-p\right)\rho_i$. For a depolarizing channel any two inputs $\left\{\left|\psi_1\right\rangle\left\langle\psi_1\right|, \left|\psi_2\right\rangle\left\langle\psi_2\right|\right\}$ are *adjacent*; that is, there are no inputs for which the channel will have positive classical zero-error capacity, since

$$\begin{aligned}
& Tr\left(\mathcal{N}\left(\left|\psi_1\right\rangle\left\langle\psi_1\right|\right)\mathcal{N}\left(\left|\psi_2\right\rangle\left\langle\psi_2\right|\right)\right) \\
&= Tr\left(\left(p\left|\psi_1\right\rangle\left\langle\psi_1\right| + \left(1-p\right)\frac{1}{2}I\right)\left(\left(p\left|\psi_1\right\rangle\left\langle\psi_1\right| + \left(1-p\right)\frac{1}{2}I\right)\right)\right) \\
&= Tr\left(p^2 Tr\left(\left|\psi_1\right\rangle\left\langle\psi_1\right|\left|\psi_2\right\rangle\left\langle\psi_2\right|\right) + \frac{p\left(1-p\right)}{2}Tr\left(\left|\psi_1\right\rangle\left\langle\psi_1\right| + \left|\psi_2\right\rangle\left\langle\psi_2\right|\right) + \frac{\left(1-p\right)^2}{2}\right) \\
&> 0,
\end{aligned} \tag{E.32}$$

where $0 < p < 1$, which means that the required non-adjacent condition $Tr\left(\mathcal{N}\left(\left|\psi_1\right\rangle\left\langle\psi_1\right|\right)\mathcal{N}\left(\left|\psi_2\right\rangle\left\langle\psi_2\right|\right)\right) = 0$ is not satisfied for the output states; that is, no inputs exist for which the channel can produce maximally distinguishable outputs, and thus for the classical zero-error capacity of the depolarizing channel we have

$$C_0^{(1)}\left(\mathcal{N}\right) = C_0\left(\mathcal{N}\right) = 0. \tag{E.33}$$

## E.4 Geometric Interpretation of the Quantum Informational Distance

The aim of this section is to discuss the geometric interpretation of HSW channel capacity in general, using quantum relative entropy as a distance measure function. We will demonstrate that the HSW channel capacity can be defined by using the quantum relative entropy function as a distance measure.

Based on the results of Nielsen *et al.* [Nielsen07-07a, 08-08a] and Nock and Nielsen [Nock05] the quantum relative entropy function $D\left(\rho\middle\|\sigma\right)$ (see Chapter 2 of the Ph.D The-sis) can be described by means of a strictly convex and differentiable *generator function* $\mathbf{F}$ as

$$\mathbf{F}(\rho) = -\mathrm{S}(\rho) = Tr(\rho \log \rho), \tag{E.34}$$

where $-\mathrm{S}$ is the negative von Neumann entropy. The extended generator function $\mathbf{F}_E(\cdot)$ can be defined as

$$\mathbf{F}_E(\rho) = Tr(\rho \log \rho - \rho). \tag{E.35}$$

The *quantum relative entropy* $D(\rho \| \sigma)$ which measures the *informational* distance between quantum states with density matrices in the Bloch sphere $\rho = \rho(x, y, z)$ and $\sigma = \sigma(\tilde{x}, \tilde{y}, \tilde{z})$ can be calculated using generator function $\mathbf{F}$ in the following way

$$\begin{aligned} D(\rho \| \sigma) &= Tr(\rho(\log \rho - \log \sigma)) \\ &= \mathbf{F}(\rho) - \mathbf{F}(\sigma) - \langle \rho - \sigma, \nabla \mathbf{F}(\sigma) \rangle, \end{aligned} \tag{E.36}$$

where $\mathbf{F} : S(\mathbb{C}^d) \to \mathbb{R}$, and $S(\mathbb{C}^d)$ denotes the open convex domain, while $\langle \rho, \sigma \rangle = Tr(\rho \sigma^*) = x\tilde{x} + y\tilde{y} + z\tilde{z}$ is the *inner product* of the quantum states, and $\nabla \mathbf{F}(\cdot)$ is the gradient (i.e., the derivate of the generator function) for the quantum informational distance defined as

$$\nabla \mathbf{F}(x) = \log(x), \tag{E.37}$$

and the inverse gradient $\nabla^{-1} \mathbf{F}(\cdot)$ is

$$\nabla^{-1} \mathbf{F}(x) = e^x. \tag{E.38}$$

Similarly, the extended quantum informational function can be defined as

$$\begin{aligned} D\left(\rho\|\sigma\right) &= Tr\left(\rho\left(\log\left(\rho\right)-\log\left(\sigma\right)\right)-\rho+\sigma\right) \\ &= \mathbf{F}\left(\rho\right)-\mathbf{F}\left(\sigma\right)-\left\langle \rho-\sigma,\nabla\mathbf{F}\left(\sigma\right)\right\rangle. \end{aligned} \tag{E.39}$$

In general, function $D$ is defined by strictly convex and differentiable generator function $\mathbf{F}:S\to\mathbb{R}$ over an open convex domain $S\left(\mathbb{C}^d\right)$, however, it is not a metric, hence symmetry and triangle inequality may fail. In geometric interpretation, quantum relative entropy $D\left(\rho\|\sigma\right)$ between quantum states $\rho$ and $\sigma$ can be measured as the vertical distance between $\rho$ and the hyperplane $H_\sigma$ tangent to relative entropy function at quantum state $\sigma$ i.e., quantum relative entropy function can be expressed in geometric interpretation

$$D\left(\rho\|\sigma\right)=\mathbf{F}\left(\rho\right)-H_\sigma\left(\rho\right). \tag{E.40}$$

In Fig. E.11, we have illustrated the geometric interpretation of quantum informational distance between quantum states $\rho$ and $\sigma$. Since, we have depicted the quantum informational distance $D\left(\rho\|\sigma\right)$, as the vertical distance between the generator function $\mathbf{F}$ and $H\left(\sigma\right)$, the hyperplane tangent to $\mathbf{F}$ at $\sigma$ [Nielsen07]. The point of intersection of quantum state $\rho$ on $H\left(\sigma\right)$ is denoted by $H_\sigma\left(\rho\right)$. The tangent hyperplane to hypersurface $\mathbf{F}\left(\rho\right)$ at quantum state $\sigma$ is

$$H_\sigma\left(\rho\right)=\mathbf{F}\left(\sigma\right)+\left\langle \rho-\sigma,\nabla\mathbf{F}\left(\sigma\right)\right\rangle. \tag{E.41}$$

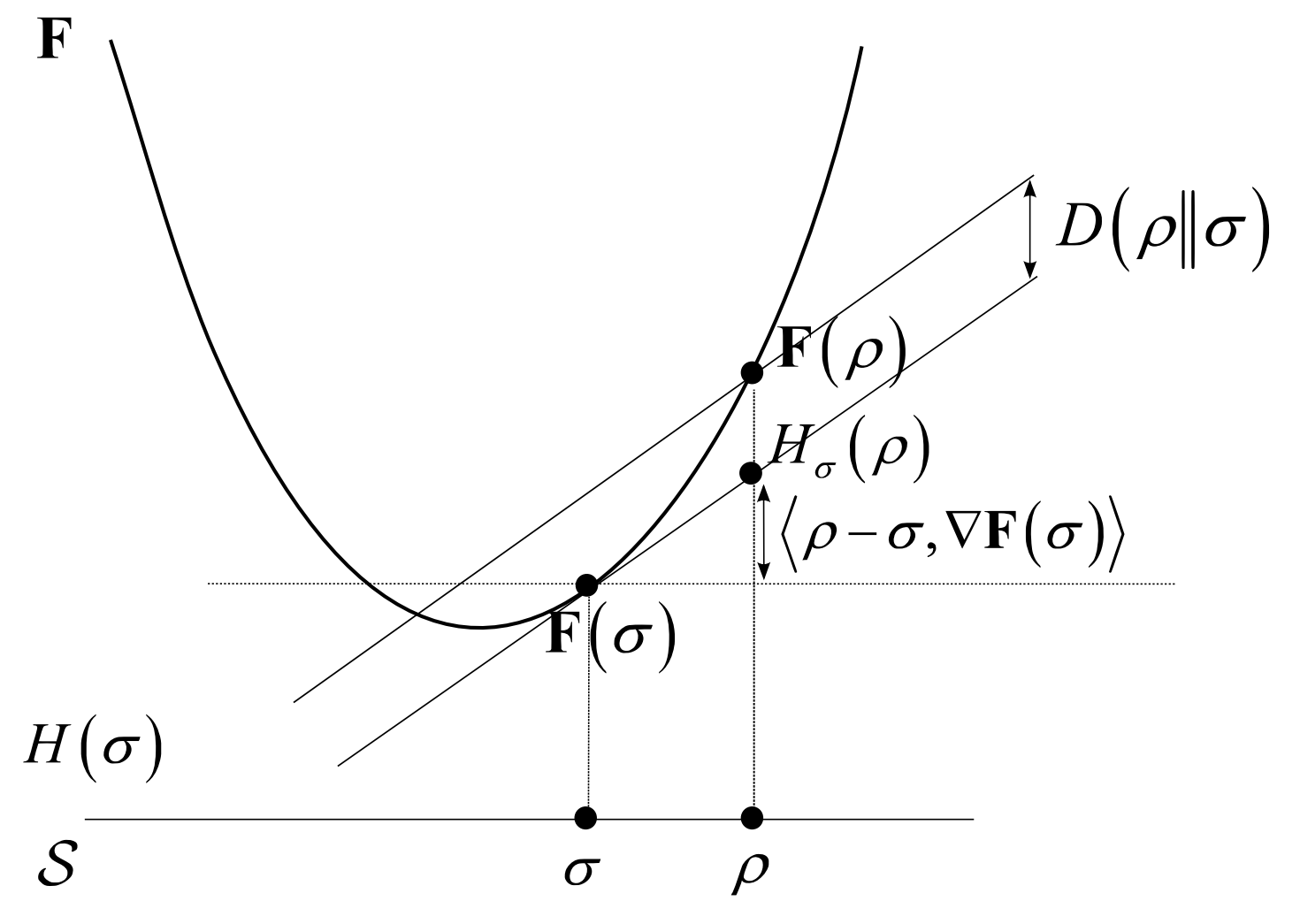

**Fig. E.11.** Geometric interpretation of quantum informational distance between quantum states.

For mixed quantum states, the associated quantum informational distance is not symmetric, i.e., $D(\rho\|\sigma) \neq D(\sigma\|\rho)$. The strict convexity of generator function $\mathbf{F}$ implies that, for any quantum state $\rho$ and $\sigma$, $D(\rho\|\sigma) \geq 0$, with $D(\rho\|\sigma) = 0$ if and only if $\rho = \sigma$. The quantum informational distance function $D(\rho\|\sigma)$ is convex in its first argument $\rho$, but not necessarily in its second argument $\sigma$. It is worth highlighting the fact that the quantum generator function has a classical analogy, because for classical probability distributions $p$, the generator function $\mathbf{F}$ is the negative Shannon entropy

$$\mathbf{F}(x) = x\log x = -x\log\frac{1}{x} = \int p(x)\log p(x)dx, \tag{E.42}$$

and

$$\nabla\mathbf{F}(x) = 1 + \log x. \tag{E.43}$$

Similarly, for classical probability distributions $p$ and $q$, the informational distance can be expressed as [Nielsen07], [Nock05]

$$D\left(p\left(x\right)\middle\|q\left(x\right)\right)=\int\left(\mathbf{F}\left(p\right)-\mathbf{F}\left(q\right)-\left\langle p-q,\nabla\mathbf{F}\left(q\right)\right\rangle\right)dx=\int p\left(x\right)\log\frac{p\left(x\right)}{q\left(x\right)}dx. \quad \text{(E.44)}$$

Finally we show the connection between the Euclidean distance and an Euclidean $\mathbf{F}$ generator function. The proof can be extended to quantum informational distances, using the quantum generator function $\mathbf{F}$. If the generator function $\mathbf{F}$ is the *squared* Euclidean distance, then the strictly convex and differentiable generator function over $\mathbb{R}^d$ can be expressed as

$$\mathbf{F}\left(x\right)=x^2=\sum_{i=0}^{d-1}x_i^2=x^Tx,\text{ with }\nabla\mathbf{F}\left(x\right)=2x\,. \quad \text{(E.45)}$$

In this case, $D\left(\rho\middle\|\sigma\right)$ can be formulated as

$$\begin{aligned}D\left(\rho\middle\|\sigma\right)&=\mathbf{F}\left(\rho\right)-\mathbf{F}\left(\sigma\right)-\left\langle\rho-\sigma,\nabla\mathbf{F}\left(\sigma\right)\right\rangle\\&=\rho^2-\sigma^2-\left\langle\rho-\sigma,2\sigma\right\rangle=\rho^2+\sigma^2-2\rho\sigma\\&=\rho^T\rho+\sigma^T\sigma-2\rho^T\sigma=\left\|\rho-\sigma\right\|^2.\end{aligned} \quad \text{(E.46)}$$

In Fig. E.12, we have illustrated the squared Euclidean distance function $D\left(\rho\middle\|\sigma\right)$, with Euclidean generator function $\mathbf{F}\left(x\right)=x^2=\sum_{i=0}^{d-1}x_i^2$.

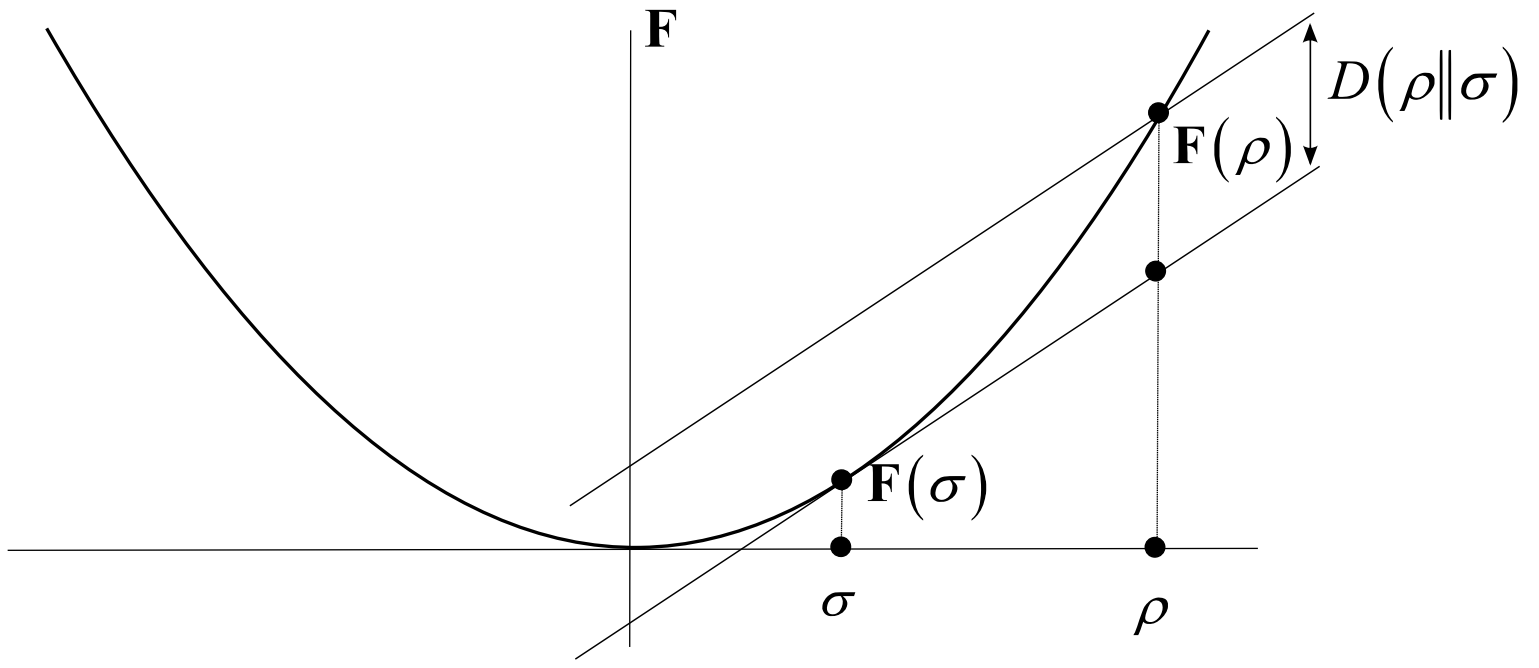

**Fig. E.12.** The squared Euclidean distance function with Euclidean generator function $\mathbf{F}$.

As we have concluded previously, the density matrices of quantum states can be represented by 3D points in the Bloch sphere. If we compute the distance between two quan-tum states in the 3D Bloch sphere representation, we compute the distance between two density matrices $\rho$ and $\sigma$. The transformation of the quantum channel $\mathcal{N}$ is modeled by an affine map, that maps quantum states to quantum states.

We have used an Euclidean generator function (E.45) in (E.46). Now, we turn our attention to the quantum informational distance function. In this case the generator function is the negative von Neumann entropy function $-\mathrm{S}$, (see (E.34)), hence the properties of the generator function will differ from (E.45). The quantum informational distance function $D(\rho\|\sigma)$ with generator function $\mathbf{F}(\rho) = -\mathrm{S}(\rho)$ is illustrated in Fig. E.13.

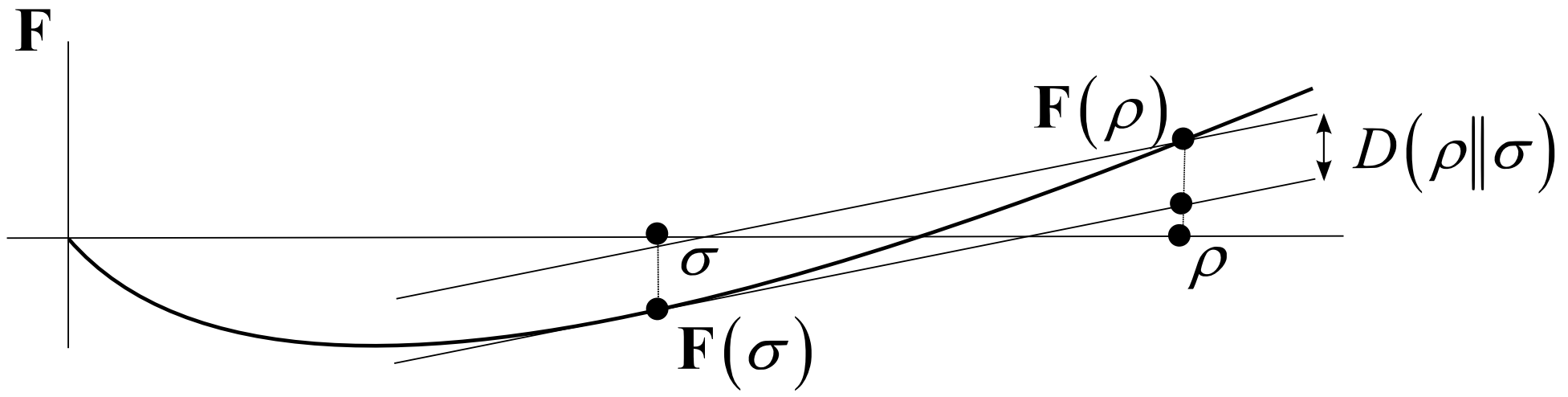

**Fig. E.13.** Negative von Neumann generator function.

The quantum informational distance function is a linear operator, thus for convex domain $\mathcal{C}d$ and convex functions $\forall \mathbf{F}_1 \in \mathcal{C}d$ and $\forall \mathbf{F}_2 \in \mathcal{C}d$, $D_{\mathbf{F}_1+\lambda\mathbf{F}_2}(\rho\|\sigma)=D_{\mathbf{F}_1}(\rho\|\sigma)+\lambda D_{\mathbf{F}_2}(\rho\|\sigma)$, for any $\lambda \geq 0$.

## E.4.1 Quantum Informational Ball

In this section we discuss the properties of quantum relative entropy based "quantum informational balls". The output states of the quantum channel can be enclosed by a ball - however between the density matrices we cannot use the Euclidean distance. If we would like to determine the capacity of the quantum channel using a geometric interpretation, then we have to seek the smallest enclosing *quantum informational ball*, which is the smallest among all possible balls. Moreover, as we have stated earlier, the computation of quan-tum channel capacity is numerically very hard, since it is an NP-Complete problem [Beigi07]. Using the smallest quantum informational ball representation the problem can be solved without the extremely high computational costs - we just have to construct an algorithm to fit the quantum ball, and we will have very good approximation of the channel capacity in our hands.

Based on [Nielsen07-07a] the geometric structure of these balls significantly differs from the geometric structure of classical Euclidean balls. The quantum informational ball $B$ with center $c$ can be defined in the Bloch sphere representation for left-sided and right-sided bisectors with respect to quantum informational distances as

$$B\left(c,r\right)=\left\{\rho\in\mathbb{C}^{d}\middle|D\left(\rho\middle\|c\right)\leq r\right\} \text{ and } B'\left(c,r\right)=\left\{\rho\in\mathbb{C}^{d}\middle|D\left(c\middle\|\rho\right)\leq r\right\}. \tag{E.47}$$

The left-sided quantum informational ball $B\left(c,r\right)$ is a convex ball, while the right-sided ball $B'\left(c,r\right)$ is not necessarily convex, see the book of Imre and Gyongyosi [Imre12]. Using inverse transformation and relation

$$D\left(\rho\middle\|\sigma\right)=\mathbf{F}\left(\rho\right)+\mathbf{F}^{*}\left(\sigma'\right)+\left\langle\rho,\sigma'\right\rangle=D^{*}\left(\sigma'\middle\|\rho'\right), \tag{E.48}$$

the connection between of left sided $B\left(c,r\right)$ and right-sided $B'\left(c,r\right)$ quantum informational balls can be expressed as

$$B'\left(c,r\right)=\nabla^{-1}\mathbf{F}\left(B\left(c',r\right)\right), \tag{E.49}$$

where $c'=\nabla\mathbf{F}\left(c\right)$. The two distances are neither *necessarily convex* nor *identical*, however, the right-sided information balls can be transformed into left-sided balls using the *inverse* transformation, we can further define a third-type quantum informational ball, by taking the symmetric distance [Nielsen07]

$$D^{symm}=\frac{D\left(\rho\middle\|\mathbf{c}\right)+D\left(\mathbf{c}\middle\|\rho\right)}{2}. \tag{E.50}$$

As we will see later, to compute the smallest enclosing quantum informational ball, we use quantum relative entropy-based Delaunay tessellation, which is *symmetric* only for pure states and *asymmetric* for mixed states. The quantum information theoretical distance is neither symmetric, nor do they satisfy the triangular inequality of metrics. The spherical Delaunay triangulation between *pure* states and between pure and mixed states with equal radii can be simply obtained as the 3D Euclidean Delaunay tessellation restricted to the Bloch sphere.

## E.4.2 Geometric Interpretation of Quantum Channel Capacity

In this section we show that the HSW [Holevo98], [Schumacher97] capacity of quantum channels can be determined in a geometrical way. Using the results of Petz [Petz96,08], Cortese [Cortese02,03], Hayashi [Hayashi05], Ruskai [Ruskai01] and King [King99-03] we refer to channel capacity as the radius $r^*$ of the smallest enclosing ball as follows

$$C(\mathcal{N}) = r^* = \max_{all\ p_i,\rho_i} \chi = \max_{all\ p_i,\rho_i} \mathrm{S}\left(\mathcal{N}\left(\sum_{i=0}^{n-1} p_i\rho_i\right)\right) - \sum_{i=0}^{n-1} p_i \mathrm{S}\left(\mathcal{N}\left(\rho_i\right)\right). \tag{E.51}$$

A quantum state can be described by its density matrix $\rho \in \mathbb{C}^{l\times l}$, which is an $l \times l$ matrix, where $d$ is the level of the given quantum system, i.e., for example for a qubit $d = 2$. For an $n$ qubit system, the level of the quantum system is $l = d^n = 2^n$. We use the fact that particle state distributions can be analyzed probabilistically by means of density matrices. A two-level quantum system can be defined by its density matrices in the following way:

$$\rho = \frac{1}{2}\begin{pmatrix} 1+z & x-iy \\ x+iy & 1-z \end{pmatrix},\ x^2+y^2+z^2 \le 1,\ x,y,z \in \mathbb{R}\,, \tag{E.52}$$

which also can be rewritten as

$$\rho = \begin{pmatrix} \frac{1+z}{2} & \frac{x-iy}{2} \\ \frac{x+iy}{2} & \frac{1-z}{2} \end{pmatrix},\ x^2+y^2+z^2 \le 1,\ x,y,z \in \mathbb{R}. \tag{E.53}$$

where $i$ denotes the complex imaginary $i^2 = -1$. The eigenvalues $\lambda_1, \lambda_2$ of $\rho(x,y,z)$ are given by

$$\lambda_1, \lambda_2 = \frac{1 \pm \sqrt{x^2+y^2+z^2}}{2}, \tag{E.54}$$

the eigenvalue decomposition $\rho$ is

$$\rho = \sum_i \lambda_i E_i, \tag{E.55}$$

where $E_i E_j$ is $E_i$ for $i = j$ and $\mathbf{0}$ for $i \neq j$. For a mixed state $\rho(x,y,z)$, $\log(\rho)$ defined by

$$\log(\rho) = \sum_i \left(\log(\lambda_i)\right) E_i. \tag{E.56}$$

The Bloch vectors $\mathbf{r}_1$ and $\mathbf{r}_2$ are real 3-dimensional vectors with length $m = 1$ for pure states, and $m < 1$ for mixed states. They can be expressed as

$$\mathbf{r} = \begin{bmatrix} r_x \\ r_y \\ r_z \end{bmatrix}. \tag{E.57}$$

Now, we define an alternate version of (E.51), using the quantum relative entropy function as distance measure between the channel output states. We will use the concept of *optimal channel output state* and *average output state (mixed state)*. The optimal channel output states are a subset of output states, i.e., they are the most distance from the origin of the Bloch sphere (The minimal von Neumann entropy channel output state also belongs to this set.). Or with other words a channel output state is called optimal, if it maximizes the Holevo quantity of the quantum channel. The *optimal average* output state is the average of the set of optimal states [Holevo98], [Schumacher97].

The Holevo quantity can be represented geometrically, using the quantum relative entropy function as a distance measure as [Petz96,08], [Schumacher99,2000], [Cortese02]

$$\chi = D\left(\rho_k \middle\| \sigma\right), \tag{E.58}$$

where $\rho_k$ denotes an *optimal* (for which the Holevo quantity will be maximal) *output state* and $\sigma = \sum_k p_k \rho_k$ is the mixture of the optimal output states [Schumacher99]. For non-optimal output states $\delta$ and optimal $\sigma = \sum_k p_k \rho_k$ we have

$$\chi = D\left(\delta \middle\| \sigma\right) \leq D\left(\rho_k \middle\| \sigma\right). \tag{E.59}$$

Using the optimal channel output density matrices $\rho_k$ and they average $\sigma$, the geometric interpretation of quantum channel capacity using the quantum relative entropy function as a distance measure can be expressed as follows [Petz96,08], [Schumacher99,2000], [Cortese02]

$$C\left(\mathcal{N}\right) = \max_{all\ p_i,\rho_i} \chi = \chi\left(\mathcal{N}\right) = r^* = \min_{\{\sigma\}} \max_{\{\rho_k\}} D\left(\rho_k \middle\| \sigma\right), \tag{E.60}$$

where the *quantum informational radius*

$$r^* = \left|\mathbf{r}^*\right| \tag{E.61}$$

is the length (with respect to quantum informational distance) of the Bloch vector $\mathbf{r}^*$. Schumacher and Westmoreland have also proven [Schumacher99], that there exists a *unique* optimum output state $\left\{p_k, \rho_k\right\}$ for every $\sigma$ that satisfies the maximization (i.e., maximizes the HSW capacity), such that $\sigma = \sum_k p_k \rho_k$. For this the Holevo information $\mathcal{X}$ can be derived in terms of the quantum relative entropy in the following way [Petz96,08], [Schumacher99,2000], [Cortese02]

$$\begin{aligned}\sum_k p_k D(\rho_k \| \sigma) &= \sum_k \left( p_k Tr(\rho_k \log(\rho_k)) - p_k Tr(\rho_k \log(\sigma)) \right) \\ &= \sum_k \left( p_k Tr(\rho_k \log(\rho_k)) \right) - Tr\left( \sum_k \left( p_k \rho_k \log(\sigma) \right) \right) \\ &= \sum_k \left( p_k Tr(\rho_k \log(\rho_k)) \right) - Tr(\sigma \log(\sigma)) \\ &= \mathrm{S}(\sigma) - \sum_k p_k \mathrm{S}(\rho_k) = \mathcal{X}.\end{aligned} \tag{E.62}$$

The result of (E.62) will have great importance later since it describes the connection between the numerical and geometric methods. The fact that the Holevo information can be described in terms of quantum relative entropy function, (see (E.58)) will provide the base of the geometric computation of quantum channel capacity.

It can therefore be concluded that the HSW channel capacity $C(\mathcal{N})$ in terms of the quantum relative entropy can be expressed as [Petz96,08], [Schumacher99,2000], [Cortese02]

$$C(\mathcal{N}) = \max_{all\ p_k, \psi_k} \sum_k p_k D\left( \mathcal{N}(\psi_k) \| \mathcal{N}(\psi) \right) = \max_{all\ p_k, \psi_k} \mathcal{X}, \tag{E.63}$$

where $\psi_k$ denotes the *pure input* quantum states of channel $\mathcal{N}$ and $\psi = \sum_k p_k \psi_k$.

### E.4.3 Quantum Relative Entropy in the Bloch Sphere Representation

As we have seen in (E.63), the HSW capacity $C(\mathcal{N})$ of quantum channel $\mathcal{N}$ can be given in a geometric representation by quantum relative entropy function $D(\cdot \| \cdot)$. The radius of the smallest quantum informational ball uses the result that the Holevo information can be measured in terms of quantum relative entropy function, (see (E.62)), and its maximized value will be equal to the capacity of the analyzed quantum channel. This is the first main element. The second: the quantum relative entropy function has a strict geometric analogy.

This quantum ball contains the channel ellipsoid, and inside the quantum ball the distances between the quantum states are measured by the quantum relative entropy function (see (E.60)), and in the geometric representation it uses the negative von Neumann entropy generator function (see (E.34)), and its extended version, the quantum relative entropy function, see (E.35). From now on, we refer the relative entropy based ball as the *smallest quantum informational ball*. The radius of this quantum ball is already defined in (E.60).

We show an example of a two-dimensional smallest enclosing quantum informational ball in Fig. E.14. This quantum relative entropy ball is a deformed ball, thus the approximation algorithm has to be tailored for quantum informational distance. The center $\mathbf{c}^*$ of the smallest enclosing quantum informational ball differs from the center of an Euclidean ball.

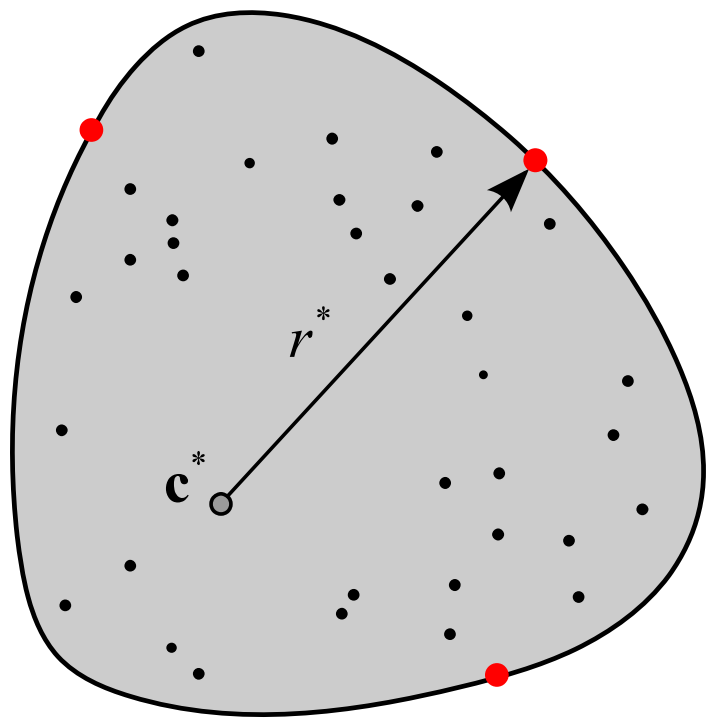

**Fig. E.14.** The smallest enclosing quantum informational ball.

In the geometric representation of $C(\mathcal{N})$, the *maximum* is taken over the *surface* of the channel ellipsoid, and the *minimum* is taken over the *interior* of the ellipsoid. The distance calculations between the quantum states are based on the quantum relative entropy function $D(\rho\|\sigma)$.

As it was shown by Cortese [Cortese02] using the work of Schumacher and Westmoreland [Schumacher99, 2000] and later by Kato *et al.* [Kato06] and Nielsen *et al.* [Nielsen07-07a,08], the quantum relative entropy function $D(\cdot\|\cdot)$ for an arbitrary quantum state $\rho=(x,y,z)$ and mixed state $\sigma=(\tilde{x},\tilde{y},\tilde{z})$, with radii $r_\rho=\sqrt{x^2+y^2+z^2}$ and $r_\sigma=\sqrt{\tilde{x}^2+\tilde{y}^2+\tilde{z}^2}$ is given by

$$\begin{aligned}D(\rho\|\sigma)&=\frac{1}{2}\log\left(\frac{1}{4}\left(1-{r_\rho}^2\right)\right)+\frac{1}{2}r_\rho\log\left(\frac{\left(1+r_\rho\right)}{\left(1-r_\rho\right)}\right)\\&\quad-\frac{1}{2}\log\left(\frac{1}{4}\left(1-{r_\sigma}^2\right)\right)-\frac{1}{2r_\sigma}\log\left(\frac{\left(1+r_\sigma\right)}{\left(1-r_\sigma\right)}\right)\langle\rho,\sigma\rangle,\end{aligned} \tag{E.64}$$

where $\langle\rho,\sigma\rangle=(x\tilde{x}+y\tilde{y}+z\tilde{z})$. For a maximally mixed state $\sigma=(\tilde{x},\tilde{y},\tilde{z})=(0,0,0)$ and $r_\sigma=0$, the quantum relative entropy can be expressed as

$$D(\rho\|\sigma)=\frac{1}{2}\log\left(\frac{1}{4}\left(1-{r_\rho}^2\right)\right)+\frac{1}{2}r_\rho\log\left(\frac{\left(1+r_\rho\right)}{\left(1-r_\rho\right)}\right)-\frac{1}{2}\log\left(\frac{1}{4}\right). \tag{E.65}$$

The quantum relative entropy between two mixed quantum states depends on the lengths of their Bloch vectors and the angle $\theta$ between them, as illustrated it in Fig. E.15.

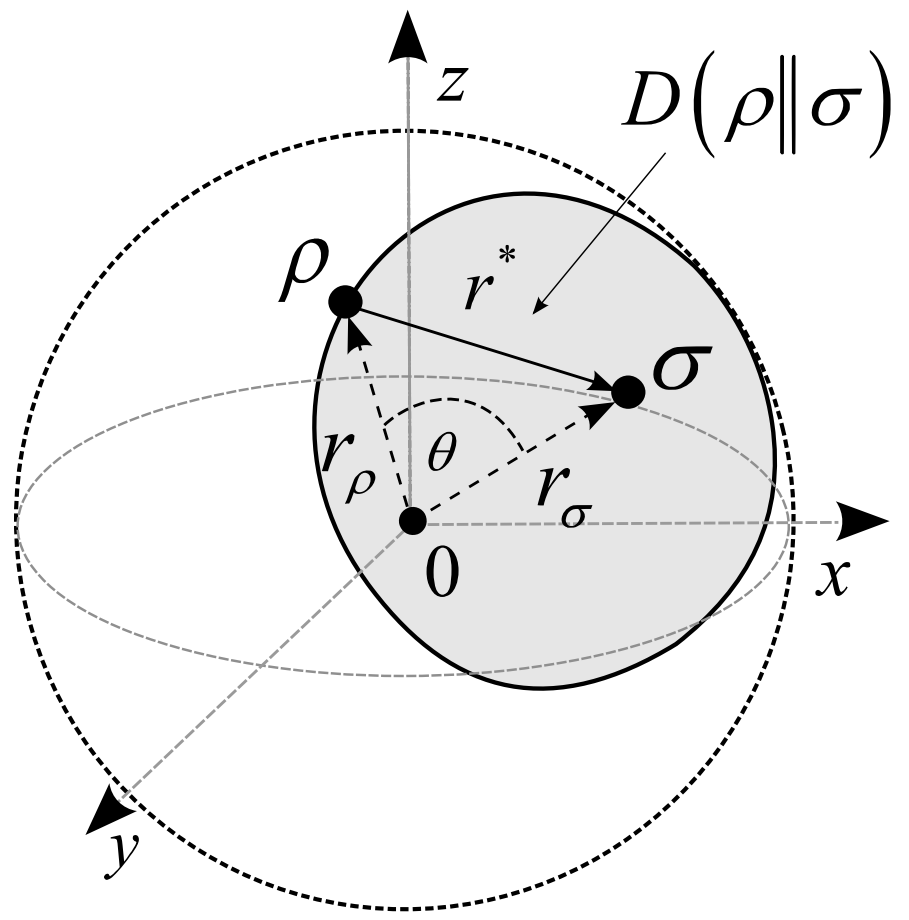

**Fig. E.15.** The quantum relative entropy between two mixed quantum states depends on the lengths of their Bloch vectors and the angle between them.

Let assume, we have a maximally mixed state $\sigma = \frac{1}{2}I$ with $r_\sigma = 0$, in this case the radius of the quantum ball (i.e., the capacity of the analyzed quantum channel) will be equal to

$$D(\rho \| \sigma) = D\left(\rho \middle\| \frac{1}{2}I\right) = \frac{1}{2}\log\left(1 - r_\rho^{\ 2}\right) + \frac{r_\rho}{2}\log\left(\frac{1 + r_\rho}{1 - r_\rho}\right) = 1 - \mathrm{S}(\rho). \tag{E.66}$$

If we have a unital quantum channel, then the average state is equal to $\sigma = \frac{1}{2}I$, hence we can use (E.66) to determine the HSW capacity of the quantum channel [Schumacher99], [Cortese02], [Kato06]. It has to be emphasized that based on (E.64) and (E.65), the relative entropy function between density matrices is equivalent to the relative entropy between the Bloch vectors $\mathbf{r}_\rho$, $\mathbf{r}_\sigma$ for the quantum states $\rho$ and $\sigma$

$$D(\rho \| \sigma) = D(\mathbf{r}_\rho \| \mathbf{r}_\sigma). \tag{E.67}$$

The results of Schumacher and Westmoreland [Schumacher2000] are based on the same fact, hence a geometric approach can be defined to measure distances on the Bloch sphere, using quantum relative entropy as distance measure function. The relative entropy between two density matrices $\rho$ and $\sigma$, can be expressed in the Bloch sphere representation. The derived formula is not symmetric in general, hence

$$D(\rho \| \sigma) \neq D(\sigma \| \rho), \tag{E.68}$$

except, if $\mathbf{r}_\rho = \mathbf{r}_\rho$. For the geometric meaning of (E.68) see Fig. E.16. The contours of $D(\rho \| \sigma)$ changes in function of $\rho$, the average state $\sigma$ is the fixed maximally mixed state in the center of the Bloch sphere. The average quantum state, $\sigma = \sum_k p_k \rho_k$, is denoted by $\mathbf{c}^*$ (and assumed to be equal to the Bloch sphere), the quantum informational radii are denoted by $r_i$, $i = 1,2,3,4$. The quantum informational distances are measured by the length of the vectors indexed from 1 to 4, with the following values $r_1 = 0.01$, $r_2 = 0.2$, $r_3 = 0.4$ and $r_4 = 0.7$.

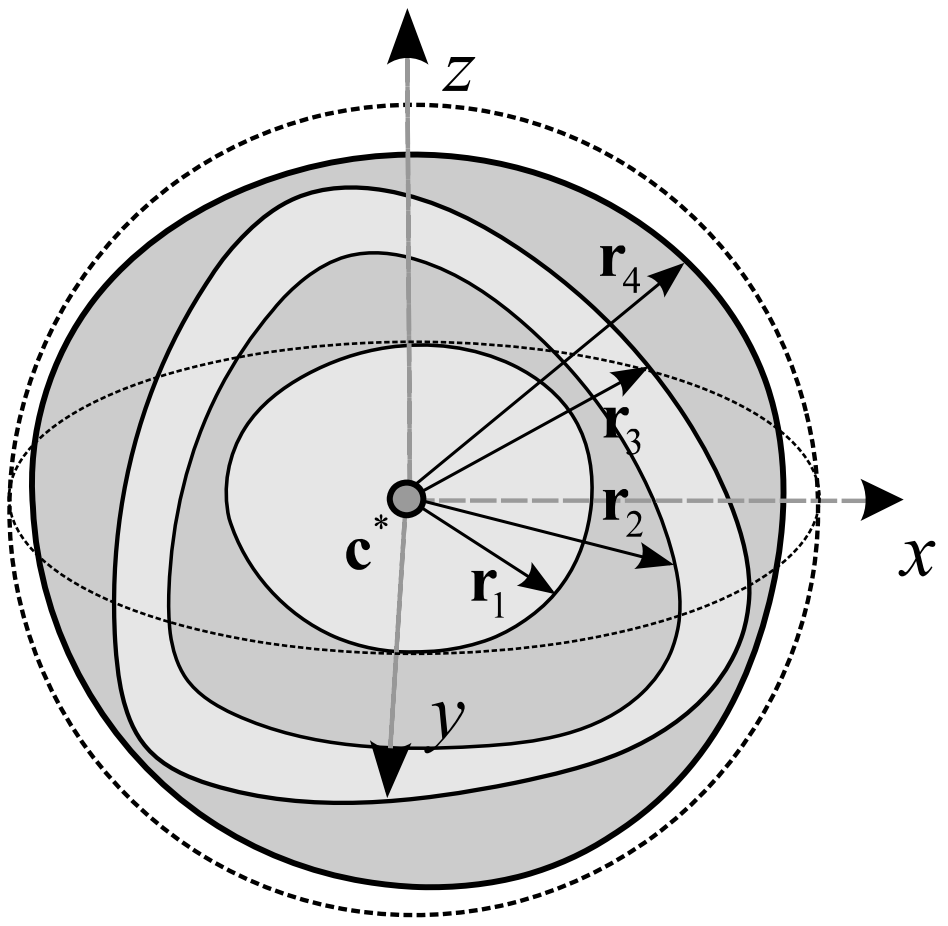

**Fig. E.16.** The smallest enclosing quantum informational ball for different radii lengths.

To see clearly the connection between the Euclidean length and the quantum informational distance of the radii, we give their comparison in Table E.1.

| *Quantum Informational length* | *Euclidean length* |
|---|---|
| 0 | 0 |
| 0.01 | 0.1 |
| 0.2 | 0.5 |
| 0.4 | 0.7 |
| 0.7 | 0.9 |
| 1 | 1 |

**Table E.1.** Comparison of Euclidean and quantum informational distances.

As it can be concluded from Fig. E.16 and Table E.1, the quantum informational distance differs from the Euclidean distance function. The length of the quantum informational radius cannot be described by the length of the Bloch vector. Moreover, the quantum ball has a distorted structure which roots in the quantum relative entropy function-based distance calculations.

### E.4.3.1 Illustration with Unital Quantum Channels

For a unital quantum channel the channel maps an identity transformation to an identity transformation, hence $\mathcal{N}\left(I\right)=I$. This property also implies some symmetries in the geometric picture of the capacity of unital quantum channels. We show the channel ellipsoid of a *unital* quantum channel and the smallest quantum informational ball (colored in grey) containing the channel ellipsoid in Fig. E.17. The information theoretic radius of this quantum informational ball describes the capacity of the analyzed quantum channel. The distorted structure of the quantum informational ball is the consequence of the distance calculations which are based on the quantum relative entropy function.

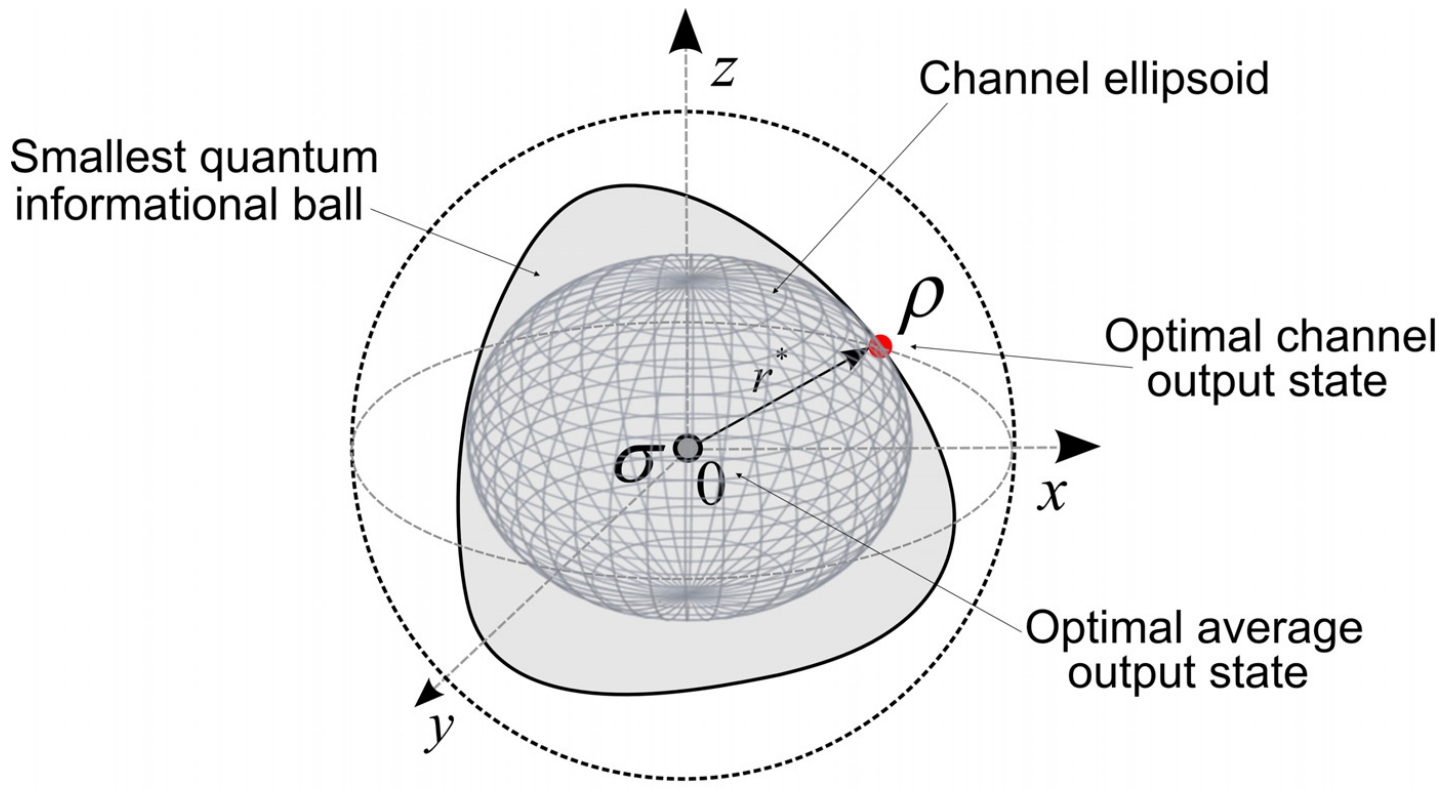

**Fig. E.17.** The channel ellipsoid of the unital quantum channel model. The center of the channel ellipsoid and the smallest quantum informational ball is equal to the center of the of the Bloch sphere.

Unital quantum channels have another important geometric property, since the average state $\sigma=\sum_k p_k\rho_k$ of the optimal *output* ensembles $\left\{p_k,\rho_k\right\}$ is equal to the center of the Bloch sphere. To make the picture clear we emphasize that here we are interested in the output of the channel i.e., our focus is on the *output* of $\mathcal{N}$, which can be analyzed by the *optimal average* output state $\sigma$, and the *optimal channel output* state $\rho$ [Petz96,08], [Schumacher99,2000], [Cortese02]. For further supplementary information see the Appendix or the book of Imre and Gyongyosi [Imre12].

### E.4.3.2 The HSW Channel Capacity and the Radius

For a given set of quantum states $\mathcal{S} = \left\{\rho_i\right\}_{i=1}^{n}$ and the distance $d\left(\cdot,\cdot\right)$ between any two quantum states of $\mathcal{S}$ is measured by the quantum relative entropy function $D\left(\cdot\|\cdot\right)$, thus a minimax optimization can be applied to the quantum relative entropy-based distances to find the center $\mathbf{c}$ of the set $\mathcal{S}$. From the center, we can compute the radius of the quantum informational ball which contains all channel output states.

We will denote the quantum relative entropy from $\mathbf{c}$ to the farthest point of $\mathcal{S}$ by

$$d\left(\mathbf{c},\mathcal{S}\right) = \max_{i} d\left(\mathbf{c},\rho_i\right) = \max_i D\left(\rho_i\|\sigma = \mathbf{c}\right). \tag{E.69}$$

Using a minimax optimization, we can minimize the maximal quantum relative entropy from $\mathbf{c}$ to the farthest point of $\mathcal{S}$ by

$$\mathbf{c}^{*} = \arg\min_{\mathbf{c}} d\left(\mathbf{c},\mathcal{S}\right). \tag{E.70}$$

The circumcenter $\mathbf{c}^{*}$ of $\mathcal{S}$ for the Euclidean distance is the optimal *minmax* center, our aim is to minimize the difference (in terms of quantum informational distance) between the approximating circumcenter $\mathbf{c}$ and the optimal circumcenter $\mathbf{c}^{*}$. The circumcenter of a set is the center of a triangle's circumcircle [Nielsen08a], [Nock05], [Boissonnat07].

As illustration the circumcenter of a random point set is shown in Fig. E.18. The circumcenter $\mathbf{c}^{*}$ minimizes the radius of the enclosing ball. The method of finding the circumcenter of a large set was shown by Welzl [Welzl91]. The triangle has a distorted structure according to the properties of quantum informational distance function.

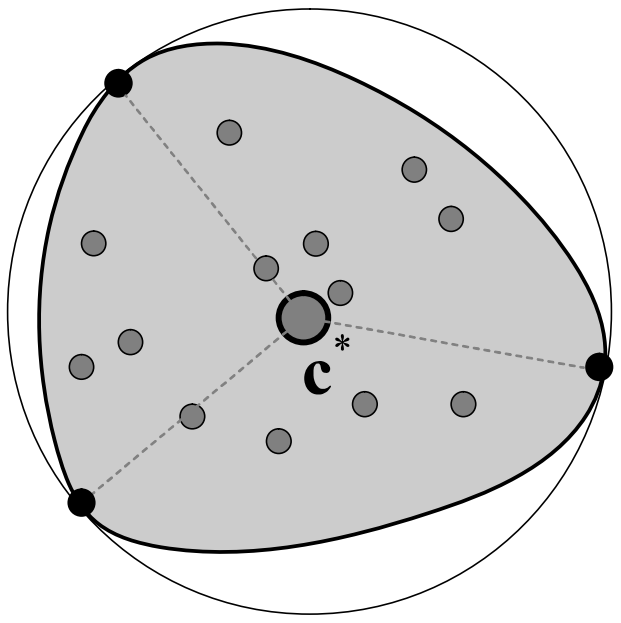

**Fig. E.18.** The circumcenter of a set of quantum states.

Using (E.69) and (E.70), the minimax optimization problem can be expressed as

$$\mathbf{c}^* = \arg\min_{\mathbf{c}} \max_i d\left(\mathbf{c}, \rho_i\right) = \arg\min_{\mathbf{c}} \max_i D\left(\rho_i \middle\| \mathbf{c}\right), \tag{E.71}$$

where function $D$ measures the quantum informational distance between channel output quantum states, see (E.62), since we use the quantum relative entropy function as distance measure. The generator function $\mathbf{F}$ is the *negative* von Neumann entropy function (E.34). In Fig. E.19 we illustrated the circumcenter $\mathbf{c}^*$ of $\mathcal{S}$ for the Euclidean distance and for *quantum informational distance* [Nock05].

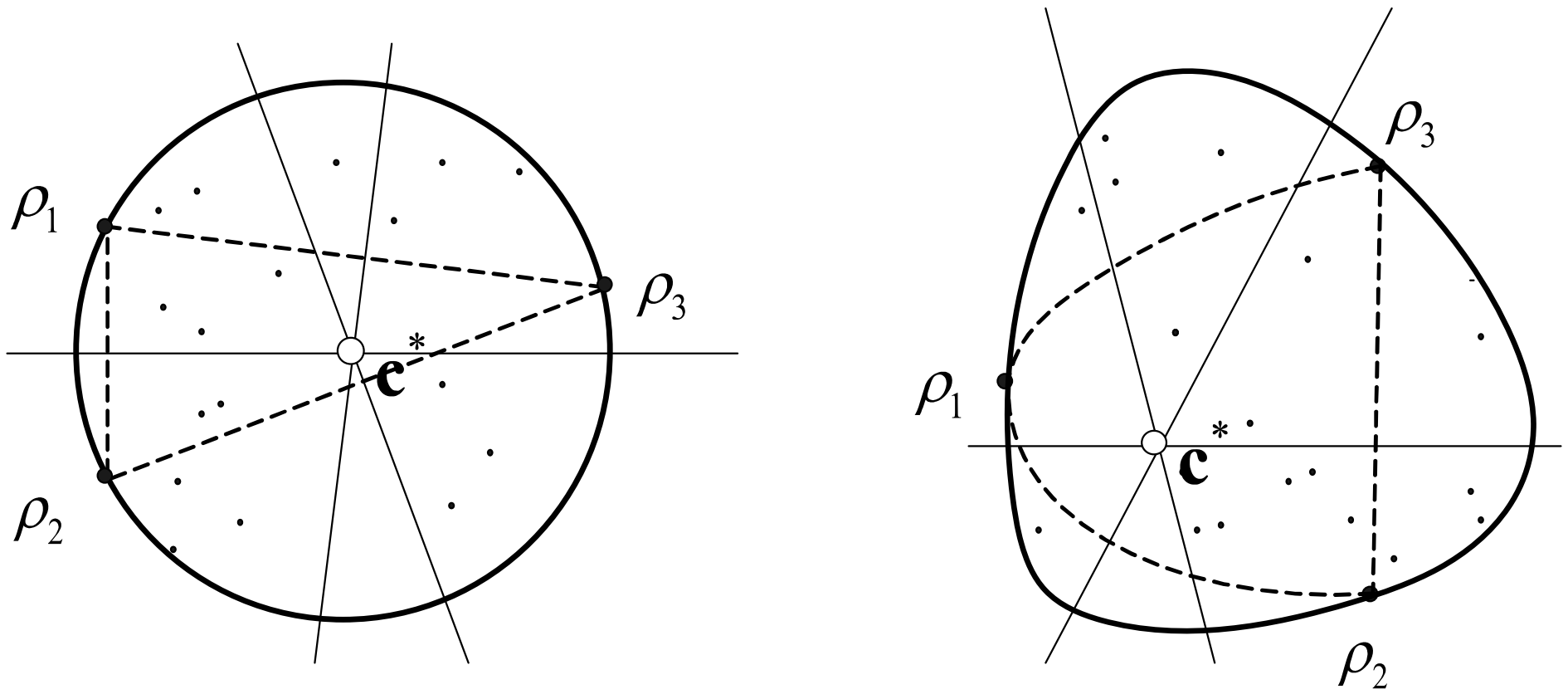

**Fig. E.19.** Euclidean distance ball and quantum relative entropy ball.

The quantum informational theoretical radius $r^*$ equals to the *maximum* quantum informational distance $D(\cdot\|\cdot)$ from the center $\mathbf{c}^*$, using input density matrices $\rho$ and $\sigma$, can be expressed as

$$r^* = \left|\mathbf{r}^*\right| = \min_{\sigma\in\mathcal{S}\left(\mathbb{C}^2\right)} \max_{\rho\in\mathcal{S}\left(\mathbb{C}^2\right)} D\left(\mathcal{N}\left(\rho\right)\middle\|\mathcal{N}\left(\sigma\right)\right), \tag{E.72}$$

where $\mathbf{r}^*$ denotes the radius vector of the smallest quantum informational ball, see (E.61). For a given set $\mathcal{S} = \left\{\rho_1, \rho_2, \ldots \rho_n\right\}$ of output states of $n$ points, we interested in computing the center $\mathbf{c}^*$ i.e., to minimize the *maximal distortion*

$$\mathbf{c}^* = \arg\min_{\mathbf{c}} \max_i d\left(\mathbf{c}, \rho_i\right) = \arg\min_{\mathbf{c}} \max_i D\left(\rho_i \middle\| \mathbf{c}\right). \tag{E.73}$$

For *squared Euclidean* distance $\left(L_2^2\right)$ the centroid is the center of minimized average distortion $MinAvg\left(L_2^2\right)$, and for Euclidean distance $L_2$, the circumcenter of $\mathcal{S}$ is the center of $MinMax\left(L_2\right)$. On the Euclidean plane $\mathbb{E}^2$, the distance measure $d\left(\rho,\sigma\right) = \left\|\rho - \sigma\right\|$ defines the $L_2$ norm metric space, and a ball $\mathbf{B} = Ball\left(\mathbf{c}, r\right)$ of center $\mathbf{c}$ and radius $r$ is defined as the set of points that are *within* distance $r$ from center c: $\mathbf{B} = \left\{\mathbf{x} \in \mathbb{E}^2 \middle| d\left(\mathbf{c},\mathbf{x}\right) \leq r\right\}$.

We denote the smallest enclosing quantum information ball of set $\mathcal{S}$ by $\mathbf{B}^*\left(\mathcal{S}\right)$ and we use $\mathbf{c}^*\left(\mathcal{S}\right)$ and $\mathbf{r}^*\left(\mathcal{S}\right)$ to refer to the *center* and *radius* of the smallest enclosing information ball $\mathbf{B}^*\left(\mathcal{S}\right)$ [Boissonnat07].

# E.5 Geometric Interpretation of the Superactivated Quantum Capacity of Arbitrary Quantum Channels

In Section 5.4 of the Ph.D Thesis, the $\mathcal{N}_1$ first channel of the joint structure was assumed to be an unital depolarizing quantum channel with $P^{(1)}(\mathcal{N}_1) > 0$. Here we extend this result, and show that the $Q^{(1)}(\mathcal{N}_1 \otimes \mathcal{N}_2)$ superactivated single-use capacity can be determined by the quantum superball structure if the first channel $\mathcal{N}_1$ of the joint channel structure $\mathcal{N}_1 \otimes \mathcal{N}_2$ is an arbitrary channel with $P^{(1)}(\mathcal{N}_1) > 0$, while the second channel is assumed to be an 50 % erasure channel.

We present in this section a mathematically equivalent, but at the same time a more practical and efficient geometric interpretation to find the $Q^{(1)}(\mathcal{N}_1 \otimes \mathcal{N}_2)$ superactivated quantum capacity of the different quantum channel models.

In the next figures, for simplicity the $Q^{(1)}(\mathcal{N}_1 \otimes \mathcal{N}_2)$ single-use quantum capacity of the joint structure $\mathcal{N}_1 \otimes \mathcal{N}_2$ will be referred as $r^*$.

## E.5.1 Extension to the Superactivated Quantum Capacity of Arbitrary Channels

The proposed geometrical approach based on the fact that the single-use classical capacity of a quantum channel can be described by the tools of information geometry, as was shown by Cortese [Cortese02], [Cortese03]. Similar results were obtained by Hayashi *et al.* [Hayashi03-05]., Ruskai *et al.* [Ruskai01] and King [King99-09]. Here we extend their results to find superactive channel combinations, assuming an arbitrary quantum channel $\mathcal{N}_1$ in the joint channel $\mathcal{N}_1 \otimes \mathcal{N}_2$. In this section we focus on the $Q^{(1)}(\mathcal{N}_1 \otimes \mathcal{N}_2)$ single-use quantum capacity of the joint channel structure $\mathcal{N}_1 \otimes \mathcal{N}_2$. In the figures for simplicity we will use the single channel view and the Bloch sphere representation; however the iterations are achieved on the abstract quantum superball.

To obtain the superactivated quantum capacity $Q^{(1)}\left(\mathcal{N}_1 \otimes \mathcal{N}_2\right)$ the *maximum* is taken over the *surface* of the superball, and the *minimum* is taken over the *interior* of the superball (see definition of quantum superball in Chapter 5). This means that vector $\mathbf{r}_\sigma$ should be adjusted to *minimize* the value initial value of $r^*$, which is $\max_{\mathbf{r}_\rho} D\left(\mathbf{r}_\rho \middle\| \mathbf{r}_\sigma\right)$ in the beginning of the iteration process. As in the previous case we will use the following notations in the figures

$$\max_{\rho_k} \rho_k = \max_{\rho_{12}^{AB-AE}} \rho_{12}^{AB-AE}, \tag{E.74}$$

and

$$\min_{\sigma} \sigma = \sum_k p_k \rho_k = \sum_k p_k \rho_{12(k)}^{AB-AE} = \min_{\sigma_{12}^{AB-AE}} \sigma_{12}^{AB-AE}. \tag{E.75}$$

In Fig. E.20(a), we illustrated the case when the radius $r^* = \max_{\mathbf{r}_\rho} D\left(\mathbf{r}_\rho \middle\| \mathbf{r}_\sigma\right)$ of quantum superball intersects the channel ellipsoid at $\mathbf{r}_\rho$. In the single channel view representation, the magnitude $m_\sigma$ describe the classical Euclidean distance from the Bloch sphere origin to center $\mathbf{c}^*$ (i.e., the length of vector $\mathbf{r}_\sigma$), while magnitude $m_\rho$ the Euclidean distance between Bloch sphere origin and quantum state $\rho$ (i.e., the length of vector $\mathbf{r}_\rho$). The vector $\mathbf{r}_\sigma$ represents the center of the quantum ball, i.e., the average state $\sigma$, while $\mathbf{r}_\rho$ is the optimal joint state $\rho$ which maximizes the quantum informational distance in the superball. To find the optimal value of vector we choose a starting point for vector $\mathbf{r}_\sigma$ in the interior of the ellipsoid (see Fig. E.20(b)). The starting point of superball vector $\mathbf{r}_\sigma$ can be

$\mathbf{r}_\sigma^T = \left[ t_x \quad t_y \quad t_z \right]$, where $\left\{ t_k \right\}$, $k = 1,2,3$ are the channel parameters in the KRSW channel representation [Cortese02-03], [Ruskai01], [Hayashi05].

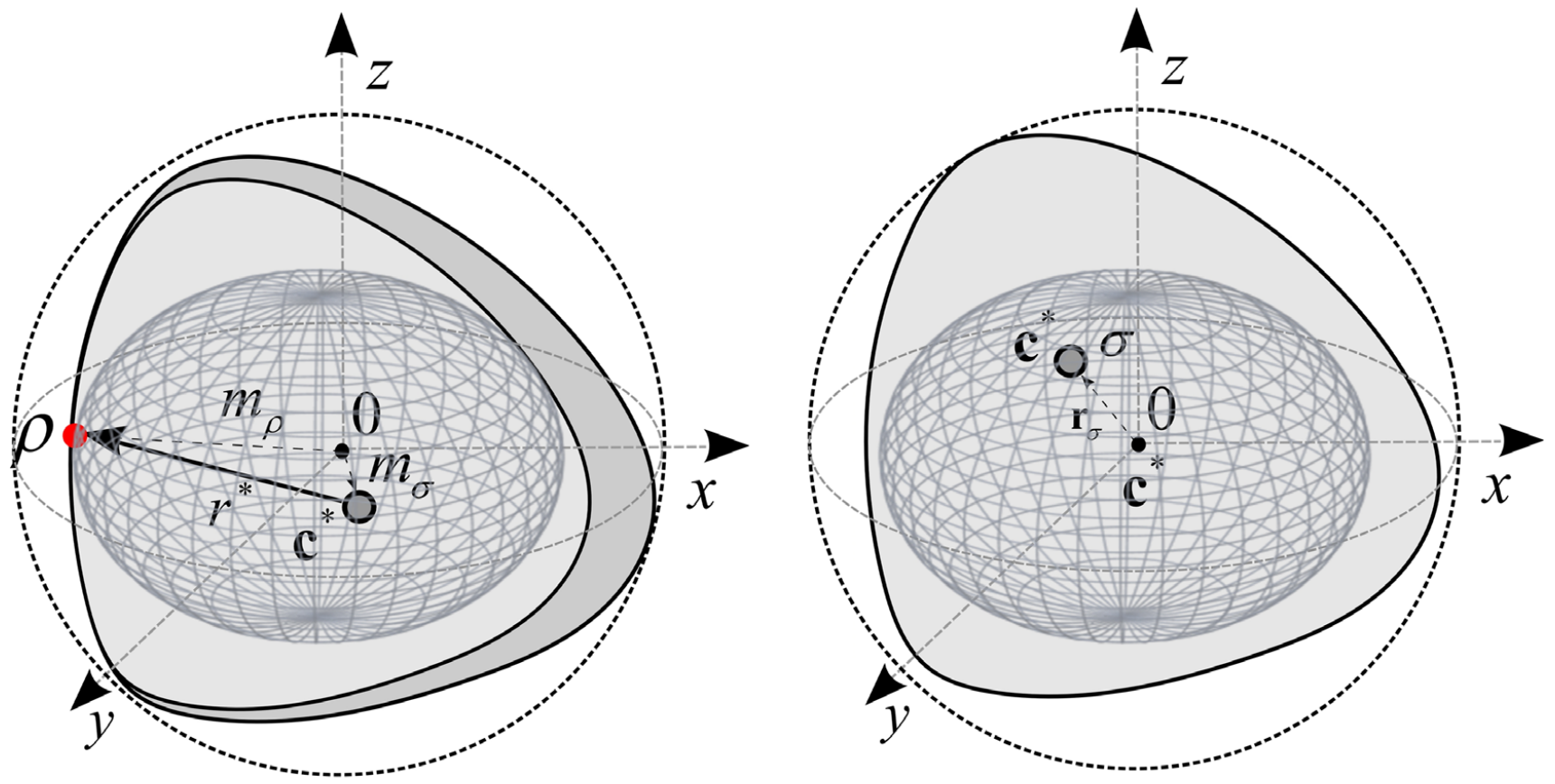

**Fig. E.20.** a: Moving of the superball vector from the optimum position will increase the value of quantum relative entropy, computed from the joint states of the joint channel structure. The optimal ball illustrated by light-grey, b: the algorithm chooses a start point in the interior of the channel ellipsoid (single channel view).

In the next step, the algorithm determines the set of joint optimal states $\left\{ \rho' \right\}$ of $\mathcal{N}_1 \otimes \mathcal{N}_2$ with corresponding $\left\{ \mathbf{r}'_\rho \right\}$ on the ellipsoid surface most distant from superball vector $\mathbf{r}_\sigma$, using the $D\left( \cdot \| \cdot \right)$ quantum relative entropy function as distance measure between the joint states of channel structure $\mathcal{N}_1 \otimes \mathcal{N}_2$. The maximum distance between the joint states $\rho$ and $\sigma$ of $\mathcal{N}_1 \otimes \mathcal{N}_2$ using the superball vectors is computed as

$$D_{\max}\left( \mathbf{r}_\sigma \right) = \max_{\mathbf{r}'_\rho} D\left( \mathbf{r}'_\rho \| \mathbf{r}_\sigma \right). \tag{E.76}$$

In Fig. E.21(a), we illustrated one of the found joint state of the joint channel $\mathcal{N}_1 \otimes \mathcal{N}_2$ by $\rho'$, the corresponding superball vector of this state is denoted by $\mathbf{r}'_\rho$. After the algorithm has selected the density matrix of the joint state, it moves the center vector $\mathbf{r}_\sigma$ towards $\rho'$ and updates the superball vector of the joint average state to $\mathbf{r}'_\sigma$ as $\mathbf{r}'_\sigma = \left( 1 - \varepsilon \right) \mathbf{r}_\sigma + \varepsilon \mathbf{r}'_\rho$. In possession of $\left\{ \rho' \right\}$, the algorithm chooses a random joint state $\rho''$ of $\mathcal{N}_1 \otimes \mathcal{N}_2$ from this set with corresponding superball vector $\mathbf{r}''_\rho$. Then, the algorithm makes a step from the center $\mathbf{r}'_\sigma$ towards the selected surface point vector $\mathbf{r}''_\rho$ in the superball according to the updating rule [Cortese02-03] $\mathbf{r}''_\sigma = \left(1-\varepsilon\right)\mathbf{r}'_\sigma + \varepsilon \mathbf{r}''_\rho$, where $\varepsilon$ denotes the size of the step (see Fig. E.21(b)).

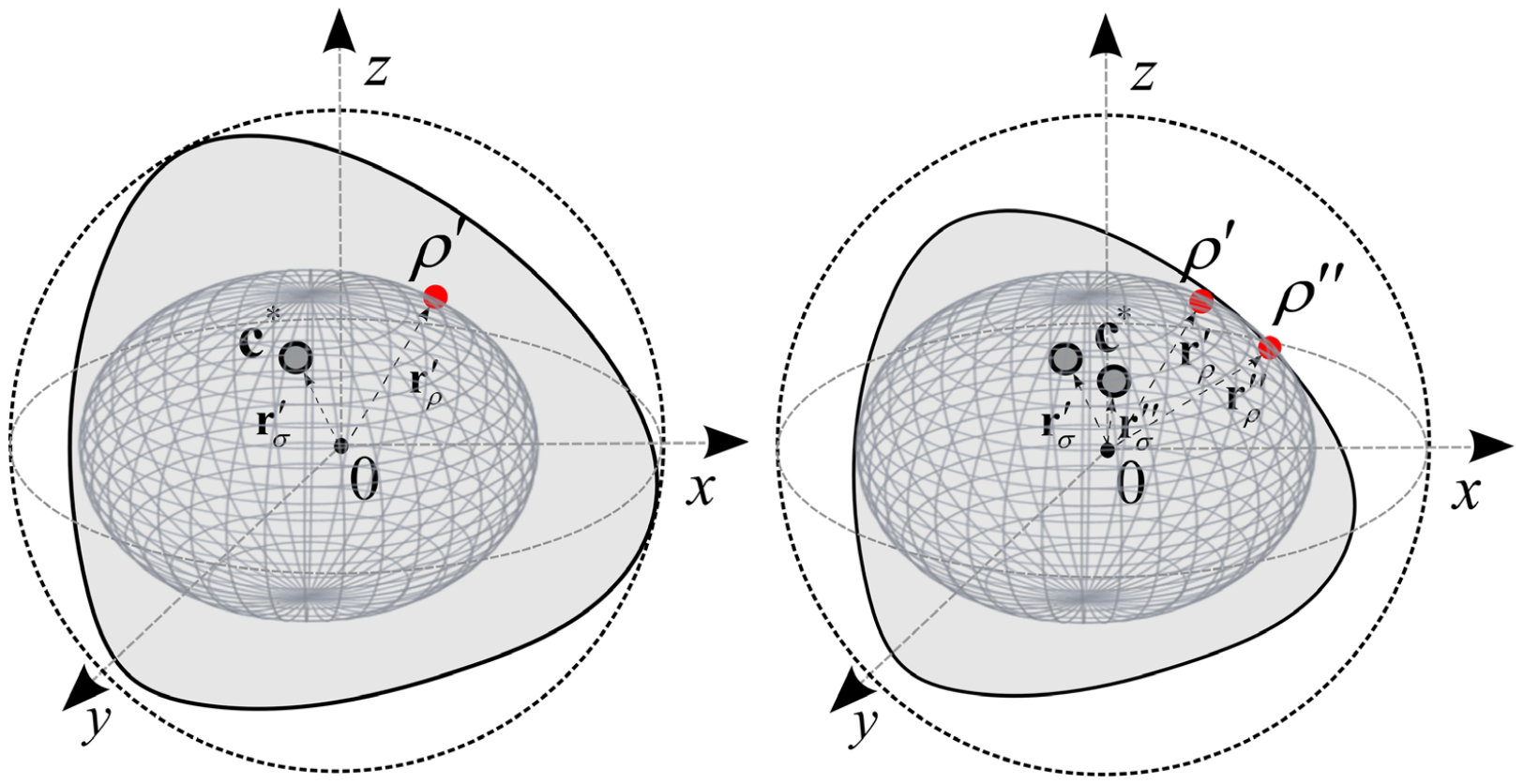

**Fig. E.21.** a: The algorithm determines the points on the ellipsoid surface most distant from the point, using the quantum relative entropy as distance measure (single channel view), b: The algorithm chooses a random vector from the maximum set of points according to previous point. The algorithm makes a step towards the found surface point vector and updates vector (single channel view).

Using the updated superball vector $\mathbf{r}''_\sigma$, the algorithm continues to loop until $D_{\max}\left(\mathbf{r}''_\sigma\right) = \max_{\mathbf{r}''_\rho} D\left(\mathbf{r}'_\rho \middle\| \mathbf{r}''_\sigma\right)$ no longer changes [Cortese02], [Coretse03], [Ruskai01]. In a general the $l+1$ iteration step of the algorithm can be summarized as

$$\mathbf{r}_\sigma[l+1] = \left(1-\varepsilon\right)\mathbf{r}_\sigma[l] + \varepsilon \mathbf{r}_\rho\left[l+1\right], \tag{E.77}$$

where $\mathbf{r}_\rho\left[l+1\right]$ denotes the vector of newly selected state from the set of possible optimal states, $\mathbf{r}_\sigma[l]$ is the vector of the previously updated average state and $\mathbf{r}_\sigma[l+1]$ is the vector of the newly updated center. The iteration converges to the optimal superball vectors $\mathbf{r}_{\rho^*}$ and $\mathbf{r}_{\sigma^*}$ of optimal joint state $\rho^*$ and optimal joint average state $\sigma^*$ of $\mathcal{N}_1 \otimes \mathcal{N}_2$. If the algorithm reaches the point, where from any movement of superball vector $\mathbf{r}_\sigma$ will increase $\max_{\mathbf{r}_\rho} D\left(\mathbf{r}_\rho \middle\| \mathbf{r}_\sigma\right)$, then the algorithm have found the optimal superball vector $\mathbf{r}_\sigma$ and the algorithm stops. The optimal vector $\mathbf{r}^*$ which measures the superactivated quantum capacity $Q^{(1)}\left(\mathcal{N}_1 \otimes \mathcal{N}_2\right)$ of the of the joint channel structure $\mathcal{N}_1 \otimes \mathcal{N}_2$ can be expressed as

$$\begin{aligned}
\mathbf{r}^* &= \min_{\mathbf{r}_{\sigma^*}} \max_{\mathbf{r}_{\rho^*}} D\left(\mathbf{r}_{\rho^*} \middle\| \mathbf{r}_{\sigma^*}\right) \\
&= \min_{\sigma_{12}^{AB-AE}} \max_{\rho_{12}^{AB-AE}} D\left(\rho_{12}^{AB-AE} \middle\| \sigma_{12}^{AB-AE}\right) \\
&= \min_{\sigma_{12}} \max_{\rho_{12}} D\left(\rho_{12}^{AB} \middle\| \sigma_{12}^{AB}\right) - \min_{\sigma_{12}} \max_{\rho_{12}} D\left(\rho_{12}^{AE} \middle\| \sigma_{12}^{AE}\right) \\
&= \frac{1}{2} P^{(1)}\left(\mathcal{N}_1\right) \\
&= Q^{(1)}\left(\mathcal{N}_1 \otimes \mathcal{N}_2\right),
\end{aligned} \tag{E.78}$$

where $\mathcal{N}_1$ can be an *arbitrary* quantum channel with $P^{(1)}\left(\mathcal{N}_1\right) > 0$, and $\mathcal{N}_2$ is an 50% erasure channel.

The final state of the algorithm with the optimal joint average state $\sigma^*$ of $\mathcal{N}_1 \otimes \mathcal{N}_2$, optimal joint channel output state $\rho^*$ of $\mathcal{N}_1 \otimes \mathcal{N}_2$, superball vectors $\mathbf{r}_{\sigma^*}$ and $\mathbf{r}_{\rho^*}$ and the optimal superball vector $\mathbf{r}^*$ in the single channel view are summarized in Fig. E.22.

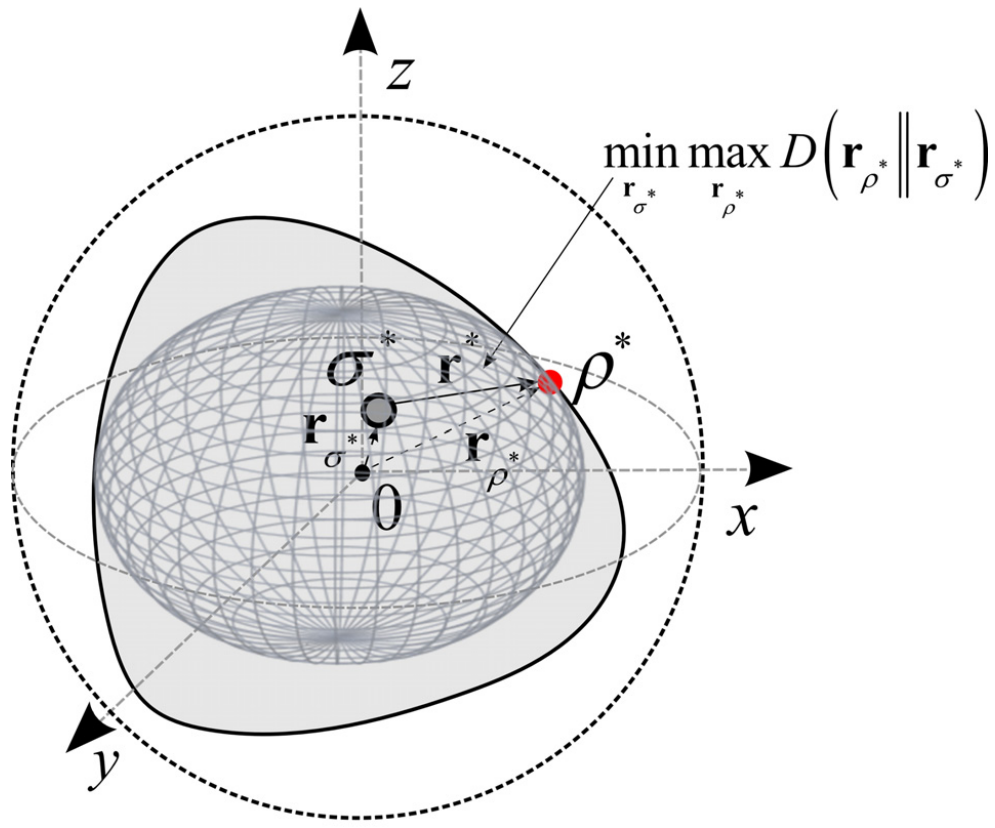

**Fig. E.22.** The final state of the iteration procedure with the optimal solution (single channel view).

Using the results of Cortese [Cortese03], the algorithm cannot become stuck in a local minimum when moving towards the optimal $\mathbf{r}_\sigma$ of the superball. To see the background of this statement, let assume we have two superball vectors denoted by $\mathbf{r}_1$ and $\mathbf{r}_2$. In this case, by means of the previously used parameter $\varepsilon$, we can define a third superball vector which defines the new joint state of $\mathcal{N}_1 \otimes \mathcal{N}_2$ as $\mathbf{r}_3 = \left(1-\varepsilon\right)\mathbf{r}_1 + \varepsilon\mathbf{r}_2$, where $0 \leq \varepsilon \leq 1$. For this superball vector, the relation $D\left(\mathbf{r}_2 \middle\| \mathbf{r}_3\right) < D\left(\mathbf{r}_2 \middle\| \mathbf{r}_1\right)$ holds for the two quantum relative entropy functions, i.e., for the joint states of the joint channel $\mathcal{N}_1 \otimes \mathcal{N}_2$. This relation makes impossible to the algorithm to stuck in a local minimum [Cortese03], and the algorithm can find the globally optimal solution, i.e., the optimal joint state of the joint channel $\mathcal{N}_1 \otimes \mathcal{N}_2$.

Based on these previous statements, geometrically the superactivated single-use quantum capacity $Q^{(1)}\left(\mathcal{N}_1 \otimes \mathcal{N}_2\right)$ can be computed from the vectors of the quantum superball as

$$\begin{aligned} r^* = \left|\mathbf{r}^*\right| = Q^{(1)}\left(\mathcal{N}_1 \otimes \mathcal{N}_2\right) &= \min_{\{\mathbf{r}_\sigma\}} \max_{\{\mathbf{r}_\rho\}} D\left(\mathbf{r}_\rho \middle\| \mathbf{r}_\sigma\right) \\ &= \min_{\sigma_{12}^{AB-AE}} \max_{\rho_{12}^{AB-AE}} D\left(\rho_{12}^{AB-AE} \middle\| \sigma_{12}^{AB-AE}\right) \end{aligned}, \quad \text{(E.79)}$$

where $\mathbf{r}_\rho$ and $\mathbf{r}_\sigma$ denote the Bloch vectors of corresponding optimal joint state $\rho$ and joint average state $\sigma$, where $\rho = \rho_{12}^{AB-AE}$ and $\sigma = \sigma_{12}^{AB-AE}$.

## E.5.2 Superactivated Quantum Capacity of the Amplitude Damping Quantum Channel

In the previous section we have concluded that the superball approach can be extended to an arbitrary quantum channel $\mathcal{N}_1$ from the joint structure $\mathcal{N}_1 \otimes \mathcal{N}_2$. Here we illustrate this result using an amplitude damping channel as the first channel $\mathcal{N}_1$ with $P^{(1)}\left(\mathcal{N}_1\right) > 0$ and the standard 50% erasure quantum channel as $\mathcal{N}_2$. We also use the result that the classical capacity of the amplitude damping channel can be determined by geometrical tools [Cortese02-03]. We do not start not describe again the whole geometric process from the beginning, since it uses the same mechanism as have already shown for the $Q^{(1)}\left(\mathcal{N}_1 \otimes \mathcal{N}_2\right)$ single-use superactivated quantum capacity of the joint structure $\mathcal{N}_1 \otimes \mathcal{N}_2$ in Chapter 5 of the Ph.D Thesis.

The geometric interpretation of the $Q^{(1)}\left(\mathcal{N}_1 \otimes \mathcal{N}_2\right)$ superactivated single-use capacity of $\mathcal{N}_1 \otimes \mathcal{N}_2$ assuming an amplitude damping channel $\mathcal{N}_1$ with $P^{(1)}\left(\mathcal{N}_1\right) > 0$ is parameterized as follows. Using the single channel view of $Q^{(1)}\left(\mathcal{N}_1 \otimes \mathcal{N}_2\right)$ of $\mathcal{N}_1 \otimes \mathcal{N}_2$ in Fig. E.23 the Euclidean distances from the origin of the Bloch sphere to center $\mathbf{c}^*$ and to point $\rho$ are denoted by $m_\sigma$ and $m_\rho$, respectively. To determine the optimal length of superball vector $\mathbf{r}_\sigma$, the algorithm moves the average state $\sigma$ of $\mathcal{N}_1 \otimes \mathcal{N}_2$, which is denoted by $\mathbf{c}^*$ in the figures. As it moves superball vector $\mathbf{r}_\sigma$ from the optimum position (Fig. E.23(a)), the larger superball corresponding to a larger value of quantum relative entropy will increase (see Fig. E.23(b)) the superactivated single-use quantum capacity $Q^{(1)}\left(\mathcal{N}_1 \otimes \mathcal{N}_2\right)$ of $\mathcal{N}_1 \otimes \mathcal{N}_2$ as $\max_{\mathbf{r}_\rho} D\left(\mathbf{r}_\rho \middle\| \mathbf{r}_\sigma\right)$. The optimal quantum superball in the single channel view is illustrated in light-grey in Fig. E.23(b).

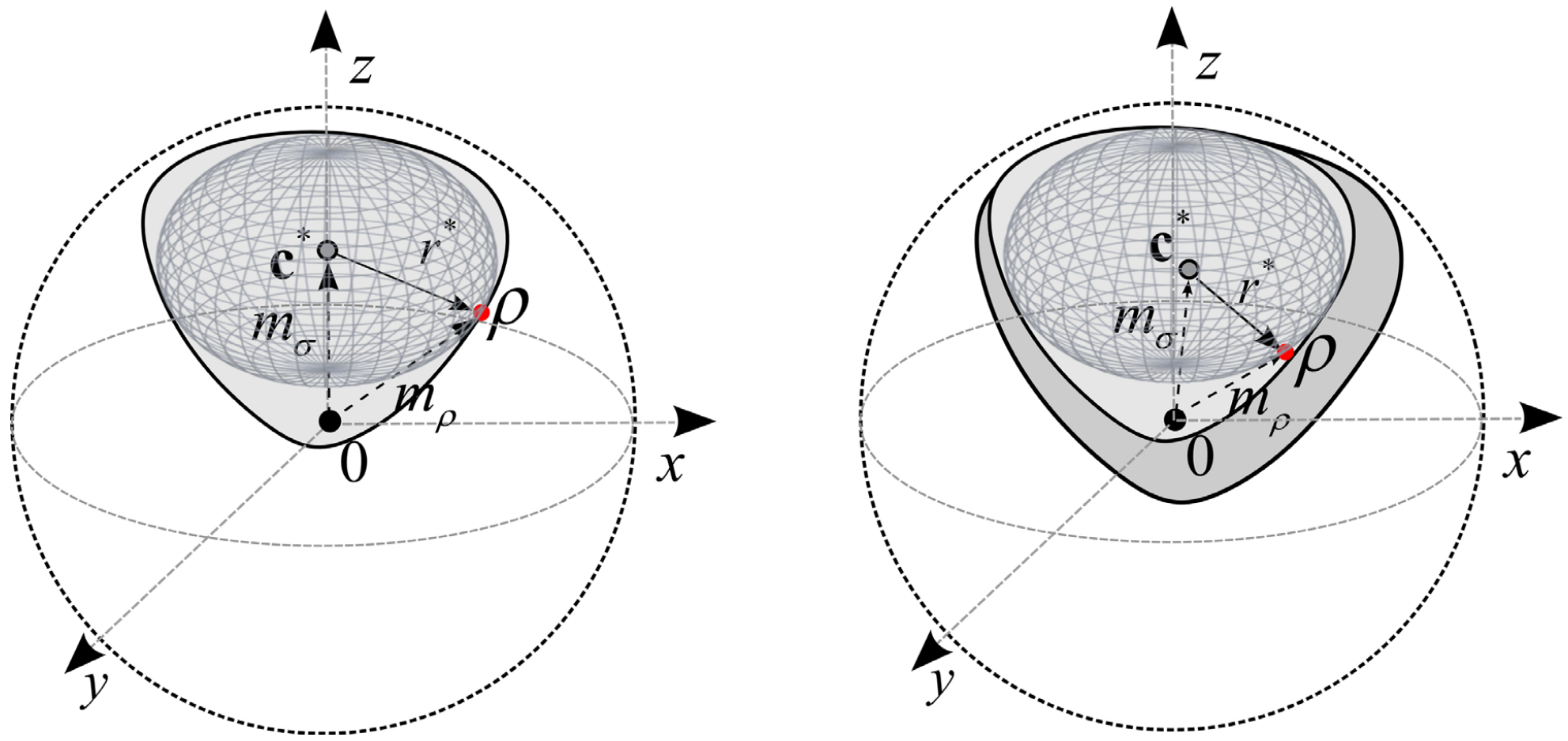

**Fig. E.23.** a: Determination of superactivated quantum capacity of the joint structure assuming an amplitude damping quantum channel as the first channel, b: the movement of the center from the optimal position increases the radius of the quantum superball (single channel view).

Using the results of Fig. E.23, in case of the $\mathcal{N}_1$ amplitude damping channel the superactivated single-use quantum capacity $Q^{(1)}\left(\mathcal{N}_1 \otimes \mathcal{N}_2\right)$ of $\mathcal{N}_1 \otimes \mathcal{N}_2$ is described by the superball approach, where $m_\sigma$ is the length of first vector $\mathbf{r}_\sigma$ which measures the Euclidean distance between the average and the center of the Bloch sphere, while the second one, $m_\rho$, represents the Euclidean distance from the center to the optimal channel output state, i.e., the length of Bloch vector $\mathbf{r}_\rho$. In case of this channel model, the average state $\sigma$ (i.e., the center $\mathbf{c}^*$ of the smallest quantum ball) of the joint structure of $\mathcal{N}_1 \otimes \mathcal{N}_2$ with the amplitude-damping $\mathcal{N}_1$ which has some private classical capacity $P^{(1)}\left(\mathcal{N}_1\right) > 0$ will differ from the center (in the single channel view and the Bloch sphere representation). Using the superball construction the superactivated single-use quantum capacity $Q^{(1)}\left(\mathcal{N}_1 \otimes \mathcal{N}_2\right)$ of $\mathcal{N}_1 \otimes \mathcal{N}_2$ can be expressed as follows. Using the angle $\theta$ between superball vectors $\mathbf{r}_\rho$ and $\mathbf{r}_\sigma$, the Euclidean length of the radii $m_\rho = \sqrt{x^2 + y^2 + z^2}$ and $m_{\sigma} = \sqrt{\tilde{x}^2 + \tilde{y}^2 + \tilde{z}^2}$, the quantum superball radius $r^*$ for an the joint channel $\mathcal{N}_1 \otimes \mathcal{N}_2$ is equal to

$$\begin{aligned} r^* &= Q^{(1)}\left(\mathcal{N}_1 \otimes \mathcal{N}_2\right) = \min_{\sigma} \max_{\rho} D\left(\rho \| \sigma\right) \\ &= \min_{\sigma} \max_{\rho_k} D\left(\rho_k \| \sigma = \sum_k p_k \rho_k\right) \\ &= \frac{1}{4}\log\left(1-\left(m_{\rho}\right)^2\right) + \frac{m_{\rho}}{4}\log\left(\frac{1+m_{\rho}}{1-m_{\rho}}\right) - \frac{1}{4}\log\left(1-\left(m_{\sigma}\right)^2\right) - \frac{\mathbf{r}_{\rho}\mathbf{r}_{\sigma}}{4r_{\sigma}}\log\left(\frac{1+m_{\sigma}}{1-m_{\sigma}}\right) \\ &= \frac{1}{4}\log\left(1-\left(m_{\rho}\right)^2\right) + \frac{m_{\rho}}{4}\log\left(\frac{1+m_{\rho}}{1-m_{\rho}}\right) - \frac{1}{4}\log\left(1-\left(m_{\sigma}\right)^2\right) - \frac{r_{\rho}\cos\left(\theta\right)}{4}\log\left(\frac{1+m_{\sigma}}{1-m_{\sigma}}\right) \\ &= \frac{1}{2}P^{(1)}\left(\mathcal{N}_1\right) > 0. \end{aligned} \tag{E.80}$$

As follows from (E.80), the superactivated single-use quantum capacity $Q^{(1)}\left(\mathcal{N}_1 \otimes \mathcal{N}_2\right)$ of $\mathcal{N}_1 \otimes \mathcal{N}_2$ can be determined by the superball approach in that case also if the first channel $\mathcal{N}_1$ is an amplitude damping channel, however in that case we cannot use a simplified geometrical formula in comparison to if the first channel $\mathcal{N}_1$ of $\mathcal{N}_1 \otimes \mathcal{N}_2$ is the depolarizing channel, see Section E.1.2. The two reasons are: First, while in the case of the depolarizing qubit channel $\mathcal{N}_1$ from $\mathcal{N}_1 \otimes \mathcal{N}_2$ the center of the superball equals to the center of the Bloch sphere and the geometrically the $Q^{(1)}\left(\mathcal{N}_1 \otimes \mathcal{N}_2\right)$ superactivated single-use quantum capacity of $\mathcal{N}_1 \otimes \mathcal{N}_2$ can be determined by a simpler formula, if $\mathcal{N}_1$ is an amplitude damping qubit channel the center of the superball will differ from the center. (For the single channel view of the superball see Fig. E.23). Second, while the depolarizing channel shrinks the Bloch sphere in every directions, the amplitude damping channel does not which also determines the geometrical interpretation of $Q^{(1)}\left(\mathcal{N}_1 \otimes \mathcal{N}_2\right)$. As follows from these results the geometrical properties of the first quantum channel $\mathcal{N}_1$ also deter-mine the geometrical interpretation of the superactivated quantum capacity $Q^{(1)}\left(\mathcal{N}_1 \otimes \mathcal{N}_2\right)$ of the joint structure $\mathcal{N}_1 \otimes \mathcal{N}_2$.

# E.6 Related Work on Information Geometry

We summarize the most important works regarding on the calculation of capacity and the geometrical interpretation of the capacities of a quantum channel.

### Fundamentals of Geometry of Quantum Channel Capacities

In the first part of this section we overviewed the most important quantum channel models, and we showed that their ability for carrying information can be measured geometrically. We analyzed just the most important channel models; further information about the various channel maps can be found in the great book of Hayashi [Hayashi06]. The description of amplitude damping channels (see Section E.1.3) with their various capacities can be found in the work of Giovanetti and Fazio [Giovannetti05].

The applications of computational geometry in the quantum space were studied in the works of Gyongyosi and Imre [Gyongyosi11a-d], in Kato's paper [Kato06] and also in the works of Nielsen *et al.* [Nielsen07], [Nielsen08], [Nielsen08a], [Nielsen08b], [Nielsen09], and Nock and Nock [Nock05]. On Voronoi diagrams by the Kullback-Leibler divergence see the works of [Onishi97] and [Nielsen07a]. The first paper on the application of computational geometry on 1-qubit quantum states was presented by Oto *et al.* in 2004 [Oto04]. The properties of quantum space with regard to different notions of "distance," and the computational cost of these constructions were studied by Onishi *et al.* [Onishi97]. The Laguerre diagrams in the quantum space were first mentioned by Nielsen *et al.* [Nielsen07]. Nielsen *et al.* have published many interesting results on the geometrical interpretation of the capacity of a quantum channel [Nielsen08a], [Nielsen08b]. An introduction to the measuring processes of continuous observables can be found in Ozawa's work [Ozawa84]. For the mathematical description of the Bloch vector for $n$-level systems see [Kimura03].

On the continuity of quantum channel capacities, a work was published by Leung and Smith in 2009 [Leung09]. On the quantum capacities of noisy quantum channels see the work of DiVincenzo, Shor and Smolin from 1998 [DiVincenzo98] and [Wilde11]. The relation of quantum capacity and superdense coding of entangled states was shown in the work of Abeyesinghe *et al.* [Abeyesinghe06]. On the quantum capacities of for quantum multiple access channels, see the work of Yard, Devetak and Hayden [Yard05b], [Yard08]. On the multiple-access bosonic communications, see the work of Yen and Shapiro [Yen05]. The connection of single-use quantum capacity and hypothesis testing was studied by Wang and Renner [Wang10].

**Geometry of Quantum Channels**

The geometrical interpretation of the Holevo-Schumacher-Westmoreland channel capacity and the proof of the relative entropy based capacity formula can be found in the works of Cortese [Cortese02] who gave some very useful results on the connection between the quantum relative entropy function and the geometric interpretation of the channel output states. The classical channel capacity and its geometric interpretation have also been analyzed by Kato [Kato06] and Nielsen *et al.* [Nielsen08b].

In the literature, many articles have investigated the question of how many input states are required to achieve the maximal channel capacity of a quantum channel [Hayashi05], [Hayashi06], [King01a], [King01c], [Ruskai01]. Ruskai *et al.* also demonstrated a very nice solution to the determination of the number of input quantum states required to achieve the optimal channel capacity [Ruskai01]. Hayashi's work is very important from the viewpoint of the geometric interpretation of HSW capacity and the maximally achievable rate for the various quantum channel models. As Hayashi *et al.* have concluded, in the cases of some quantum channel models, the optimal channel capacity can be achieved by two optimal input states, although for some other channels, the optimal capacity requires more input states [Hayashi03], [Hayashi05].

## Quantum Informational Distance

More information about the *Fundamental Information Inequality*, which states that the relative entropy function is always non-negative, and about its important mathematical corollaries, can be found in the works of Jaynes in 1957 [Jaynes57a], [Jaynes57b], and 2003 [Jaynes03]. Its "quantum version", the Fundamental Quantum Information Inequality states the monotonicity of the quantum relative entropy function. The continuity property of the quantum relative entropy function was showed by Fannes in 1973 [Fannes1973]. Further details about the properties of the quantum relative entropy function can be found in Hayashi's book [Hayashi06], or see the work of Mosonyi and Datta [Mosonyi09]. The works of Petz *et al.* [Petz96-Petz10a] also clearly presents the mathematical background of these questions. Bounds on the quantum relative entropy function and the trace distance function were determined by Schumacher and Westmoreland in 2002 [Schumacher02]. An introduction to the application of the difference distance measures can be found in Winter's work [Winter99], who showed how these measures can be used in the construction of coding theorems such as the well-known HSW-theorem and in the capacity measure of multiple access channel techniques [Czekaj08], [Yard05b], [Yard08]. The difference distance measures also can be used to quantify different parameters of the quantum communication protocols (similar to a performance measure), hence these metrics have great importance in various fields of Quantum Information Processing. Ogawa and Nagaoka in 2007 published an article [Ogawa07] in which they showed how good codes can be constructed using these various difference measures. In the computation of the classical capacities and the quantum capacity of a quantum channel, convex optimization is a very important problem. Further information about these topics can be found in the book of Boyd and Vandenberghe [Boyd04].

## Early Days

The original works of G. Voronoi from the very beginning of the twentieth century are [Voronoi1907], [Voronoi1908]. On the convex sets see [Bregman67], and the work of Bures from a different field [Bures69]. A work on the convexity of some divergence measures based on entropy functions was published by Burbea and Rao in 1982 [Burbea82]. The generalization of the complex numbers to three dimensional complex vectors originally was made by W.R. Hamilton [Goodman04]. Hamilton extended the theory of complex numbers to the three dimensional space. The formula of outer product between vectors was discovered by Grassmann. Grassmann's work was inspiration to William Kingdom Clifford, who introduced geometric algebra [Isham99]. In the nineteenth century, Clifford's motivation was to connect Grassmann's and Hamilton's work. Clifford united the inner product and outer product into a single geometric product, and he introduced the Clifford-algebra. Clifford's results have many applications in classical and quantum mechanics, and quantum physics.

**Computational Geometry**

In the literature of computational geometry, many very efficient and robust algorithms exist for computing Delaunay triangulations in two or three dimensions, whose can be applied to quantum space, with respect to quantum informational distance. In three dimensions, the combinatorial and algorithmic complexity can be computed by robust and efficient methods [Aluru05], [Amari93], [Amato01], [Arge06], [Clarkson89a], [Kitaev97], [Goodman04], [Pach02], [Rajan94]. The complexity of a Delaunay triangulation is $\mathcal{O}\left(n\log n\right)$, while the worst-case bound on the complexity of a Delaunay triangulation is $\mathcal{O}\left(n^2\right)$ [Goodman04], [Rajan94]. Computation of convex hulls was one of the first problems in computational geometry [Agarwal04], [Aluru05], [Brodal02], [Buckley88], [Chan01], [Seidel04]. The amount of literature about convex hull calculations is huge, and there are many computational geometric algorithms to solve this problem. One of the earliest algorithms was constructed by Graham and Andrew [Goodman04], then a divide-and-conquer approach was designed by Preparata and Hong [Goodman04], and later an incremental method was constructed [Aurenhammer92]. A more efficient approach was given by Over-mars and van Leeuwen [Goodman04], and similar results were introduced by Hershnerger and Suri [Goodman04], Chan [Goodman04], and Brolat and Jacob [Goodman04]. An algorithm on finding the smallest enclosing ball, and on construction of smallest enclosing disks was published by Welzl in 1991 [Welzl91].

**Complexity of Calculations**

The lower bound on the complexity of a convex hull computation was thought to $\Omega(n \log n)$, and this was not improved for a long time, until finally Jarvis introduced his "Jarvis's march" [Goodman04], [Isham99] that computes the convex hull in $\mathcal{O}(cn)$ time, where $c$ is the complexity of the convex hull [Goodman04]. The same result was obtained by Overmars and van Leeuwen, Nykat, Eddy, and finally by Green and Silverman [Goodman04], [Isham99]. The next relevant result was shown by Kirkpatrick and Seidel who further reduced the complexity of the calculation to $\mathcal{O}(\log(c)n)$, and later Chan [Goodman04] showed a simpler algorithm. As has been shown, a convex hull in three-dimensional space can be computed in $\mathcal{O}(\log(n)n)$, and for higher dimensions the complexity of the convex hull is no longer linear in the number of points [Chan01], [Chen06], [Chen07], [Clarkson89], [Clarkson89a], [Cormen01], [Cornwell97], [Feldman07].

The computation of the common intersection of half planes is dual to the computation of the convex hull of points in the plane, and it is also a well studied problem. The convex hull computation between quantum states can be derived from the problem of half-plane intersections. The problem of the common intersection of half planes was studied by Preparata and Shamos [Goodman04], and [Isham99] and they gave many solutions to the problem in $\mathcal{O}(\log(n)n)$ time. As has been shown, computing the common intersection of half-spaces is a harder problem if the dimension increases, since the common intersection can be as large as $\mathcal{O}\left(n^{\lfloor d/2 \rfloor}\right)$. Many linear programming approaches have been developed [Bregman67], [Goodman04], [Seidel04], [Shewchuck02], [Shirley05], [Wein07], [Worboys04], [Yoshizawa99], this goes beyond the scope of the current chapter. Welzl showed different methods to convex hull calculation problem of a set of points [Welzl85], [Welzl88], [Welzl91]. We also suggest the works of Sharir [Sharir85], [Sharir94], [Sharir04].

**Application of Computational Geometry**

As an introduction to basic theories and methods of computational geometry can be found in the book of Goodman and O'Rourke [Goodman04]. An interesting paper on random quantum channels, and their graphical calculus and the Bell state phenomenon was published by Collins and Nechita in 2009, for details see [Collins09]. About the core-set approaches and the properties of smallest enclosing balls, see the works of Nielsen and Nock [Nielsen07], [Nielsen08], [Nielsen08a], [Nielsen08b], [Nielsen09], [Nock05], and Kato *et al.* [Kato06]. The approximate clustering via core-sets was studied by Badoiu *et al.* see [Badoiu02]. The properties of Bregman clustering was also studied by Gupta *et al.* [Gupta06]. A work on similarity search on Bregman divergence was published by Zhang *et al.* [Zhang09]. About approximation algorithms for Bregman clustering see [Sra08].
On the optimality of Delaunay triangulation see the work of Rajan [Rajan94]. A technique on clustering based on weak coresets was published by Feldman *et al.* [Feldman07]. On the mathematical background of finding the smallest enclosing ball of balls see [Fischer04]. The problem of smaller coresets for clustering was studied by Har-Peled and Kushal [Har-Peled05]. The role of coresets in dynamic geometric data streams was studied by Frahling and Sohler in [Frahling05].

For clustering for metric and non-metric distance measures, see the works of Ackermann *et al.* [Ackermann08-09]. On range searching see the work of Agarwal *et al.* [Agarwal04] and on the properties of some important geometrical functions see [Agarwal07]. An algorithm for calculating the capacity of an arbitrary discrete memoryless channel was shown in [Arimoto72]. On the applications of Voronoi diagrams, see [Asano06], [Asano07], [Asano07a] or the works of Aurenhammer *et al.* [Aurenhammer2000], [Aurenhammer84], [Aurenhammer87], [Aurenhammer91], [Aurenhammer92], [Aurenham-

mer98], or [Boissonnat07]. An important work on clustering with Bregman divergences was published by Banerjee *et al.* [Banerjee05]. About triangulation and mesh generation see [Bern04], [Bern99].

The Delaunay triangulation and the complexity of its construction was proven by Rajan in 1994 [Rajan94]. On Delaunay triangulation and Voronoi diagram on the surface of a sphere see the work of Renka [Renka97]. A conference paper on the computation of Voronoi diagrams by divergences with additive weights was published by Sadakane *et al.* [Sadakane98]. On Delaunay refinement algorithms for triangular mesh generation see [Shewchuck02]. On streaming computation of Delaunay triangulations a work was published by Isenburg *et al.* [Isenburg06]. The minimum enclosing polytope in high dimensions was studied by Panigrahy [Panigrahy04]. A work on finding the center of large point sets in statistical space was published by Pelletier [Pelletier05].

**Comprehensive Surveys**

The mathematical background of quantum informational divergences can be found in Petz's works [Petz96], [Petz08]. A great work on modern differential geometry was published by Isham [Isham99]. A handbook on Discrete and Computational Geometry was published by Goodman and O'Rourke [Goodman04]. We also suggest the work of [Janardan04]. A summarization on convex hull computations was published by Seidel [Seidel04]. A work on dynamic planar convex hull was published by Brodal and Jacob [Brodal02]. On matrix analysis see the book of Horn and Johnson [Horn86]. A great work on modern graph theory was published by Bollobás in 1998 [Bollobas98]. The methods of information geometry were also summarized by Amari [Amari2000].

For further supplementary information see the book of Imre and Gyongyosi [Imre12].

# Appendix F

# Information Geometric Superactivation of Asymptotic Quantum Capacity

## F.1 The Basic Algorithm

As shown by Nock *et al.* [Nock05], and Nielsen *et al.* [Nielsen08a], [Nielsen09], the minimum ball of the set of balls is unique, thus the circumcenter $\mathbf{c}^*$ of the set of quantum states is

$$\mathbf{c}^* = \arg\min_{\mathbf{c}} \mathbf{F}_B\left(\mathbf{c}\right). \tag{F.1}$$

The main steps of the core-set algorithm are summarized as follows. The distance calculations between the density matrices are based on the quantum relative entropy function.

**Algorithm 1.**

1. *Select* a random center $\mathbf{c}_1$ from the set of density matrices $\mathcal{S}$

$$\mathbf{c}_1 = s_1$$

**for** $\left(i = 1,2,\ldots,\left\lceil \frac{1}{\mathcal{E}^2} \right\rceil\right)$

**do**

2. *Find* the farthest density matrix $s$ of $\mathcal{S}$ wrt. *quantum relative entropy* (F.2)

$$S \leftarrow \arg\max_{s' \in \mathcal{S}} D_F\left(\mathbf{c}_i, s'\right)$$

3. *Update* the circumcircle of quantum superball:

$$\mathbf{c}_{i+1} \leftarrow \nabla_F^{-1}\left(\frac{i}{i+1}\nabla_F\left(\mathbf{c}_i\right) + \frac{1}{i+1}\nabla_F\left(S\right)\right).$$

4. *Return* $\mathbf{c}_{i+1}$ of the quantum superball.

In Fig. F.1, we illustrated the smallest enclosing ball of balls in the single channel view using the Bloch sphere representation. We denote the set of $n$ $d$-dimensional balls by $B = \left\{b_1,\ldots,b_n\right\}$, where $b_i = Ball\left(s_i, r_i\right)$, with center $s_i$ and radius $r_i$. For $\mathcal{N}_1 \otimes \mathcal{N}_2$, the radius of the superball is $r_{super}^*$, which is referred as $r^*$ in the single channel view. The centers of the balls are computed by the quantum Delaunay triangulation as described in Chapter 6 of the Ph.D Thesis, while the complete iteration process of the fitting of the quantum superball is achieved by the core-set algorithm.

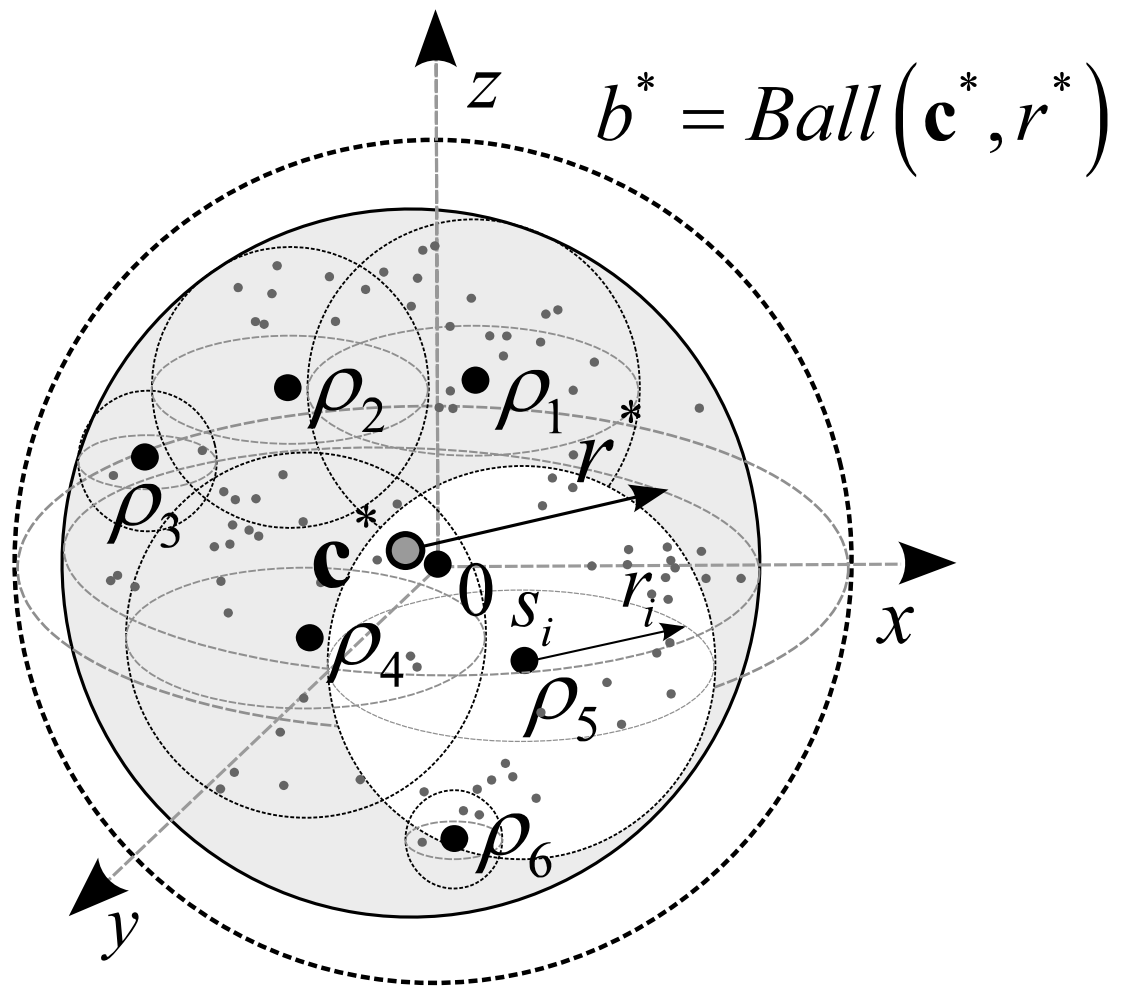

**Fig. F.1.** The smallest enclosing ball of a set of balls in the quantum space (single channel view).

At the end of the proposed scheme, the approximated value of the superactivated quantum capacity $Q\left(\mathcal{N}_1 \otimes \mathcal{N}_2\right)$ of the analyzed channel structure $\mathcal{N}_1 \otimes \mathcal{N}_2$ is obtained by the radius $r_{super}^{*}$ of the smallest quantum superball. The accuracy of the approximation is $\left(1+\mathcal{E}\right)$, where $\mathcal{E}$ can be chosen to arbitrarily low. This price negligible compared to the original complexity of the problem, since as it was shown by Beigi and Shor in 2007 [Beigi07], computing the Holevo capacity is NP-Complete (In our case, the single-use and the asymptotic quantum capacity are derived from two the Holevo quantities, see the theories of Chapter 5 of the Ph.D Thesis).

In the next figures for simplicity we will depict only the single channel view of the joint channel $\mathcal{N}_1 \otimes \mathcal{N}_2$ using the single Bloch sphere representation. The computations are performed on the joint channel structure $\mathcal{N}_1 \otimes \mathcal{N}_2$ using our quantum informational superball. In Fig. F.2 we start to describe the steps of the iterative process on approximating the smallest enclosing quantum informational ball in the $\mathcal{N}_1$ single channel view of $\mathcal{N}_1 \otimes \mathcal{N}_2$. The initial phase of the algorithm, and the sample set of quantum states are illustrated in the first figure. In the second figure the algorithm determines an initial center of the set of quantum states, and starts to fit the quantum superball.

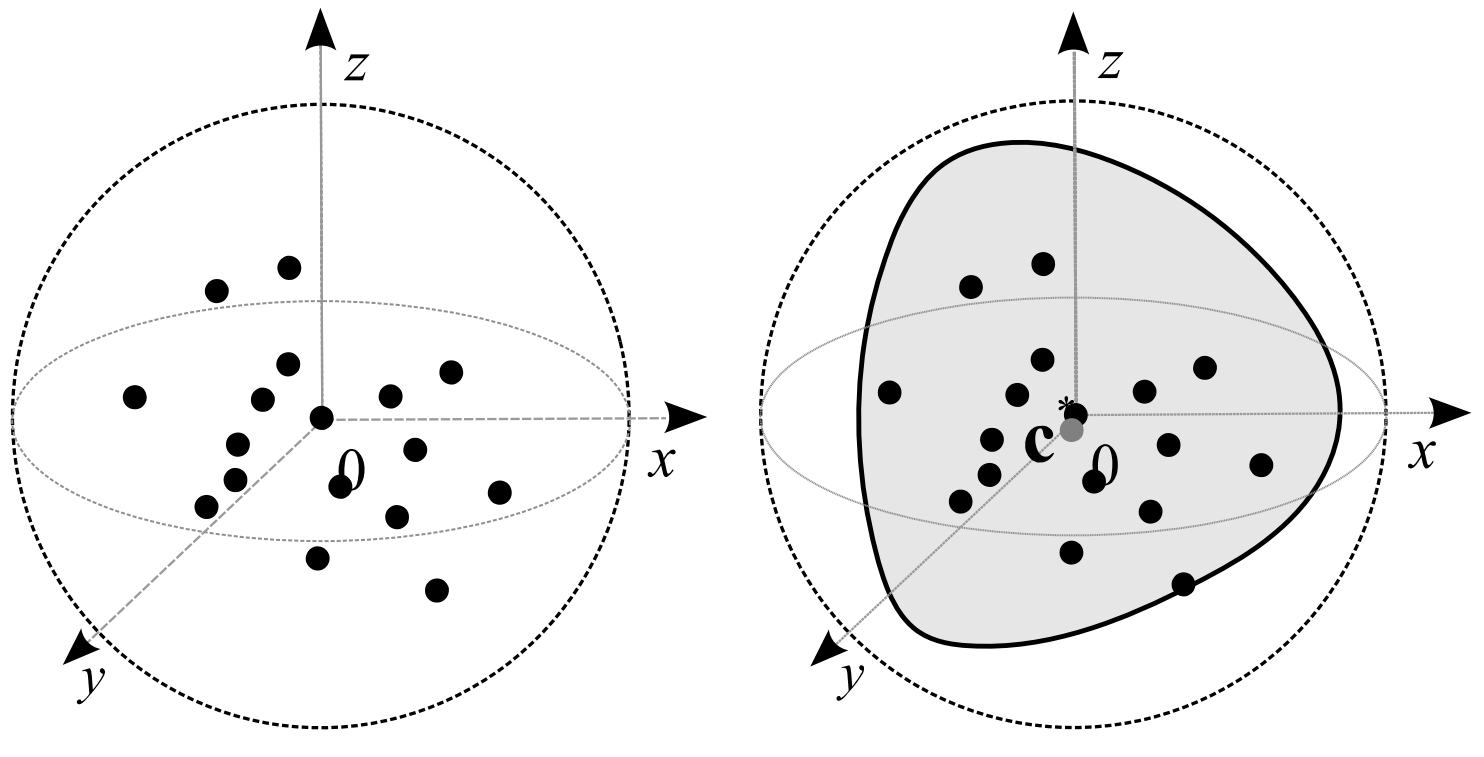

**Fig. F.2.** Fitting steps of the quantum informational ball in the Bloch sphere using random set of quantum states (single channel view).

After few iteration steps, the algorithm fits the quantum superball, which ball describes quantum channel capacity $Q\left(\mathcal{N}_1 \otimes \mathcal{N}_2\right)$ of $\mathcal{N}_1 \otimes \mathcal{N}_2$.

The fitted quantum superball will approximate the superactivated quantum capacity $Q\left(\mathcal{N}_1 \otimes \mathcal{N}_2\right)$ of the joint quantum channel structure $\mathcal{N}_1 \otimes \mathcal{N}_2$. In Fig. F.3, the quantum states with maximal distances are denoted by $\rho_1, \rho_2$ and $\rho_3$. As shown in Fig. F.3(a) the iteration moves the center of quantum ball and changes the size of the quantum superball to find the optimal solution shown in Fig. F.3(b).

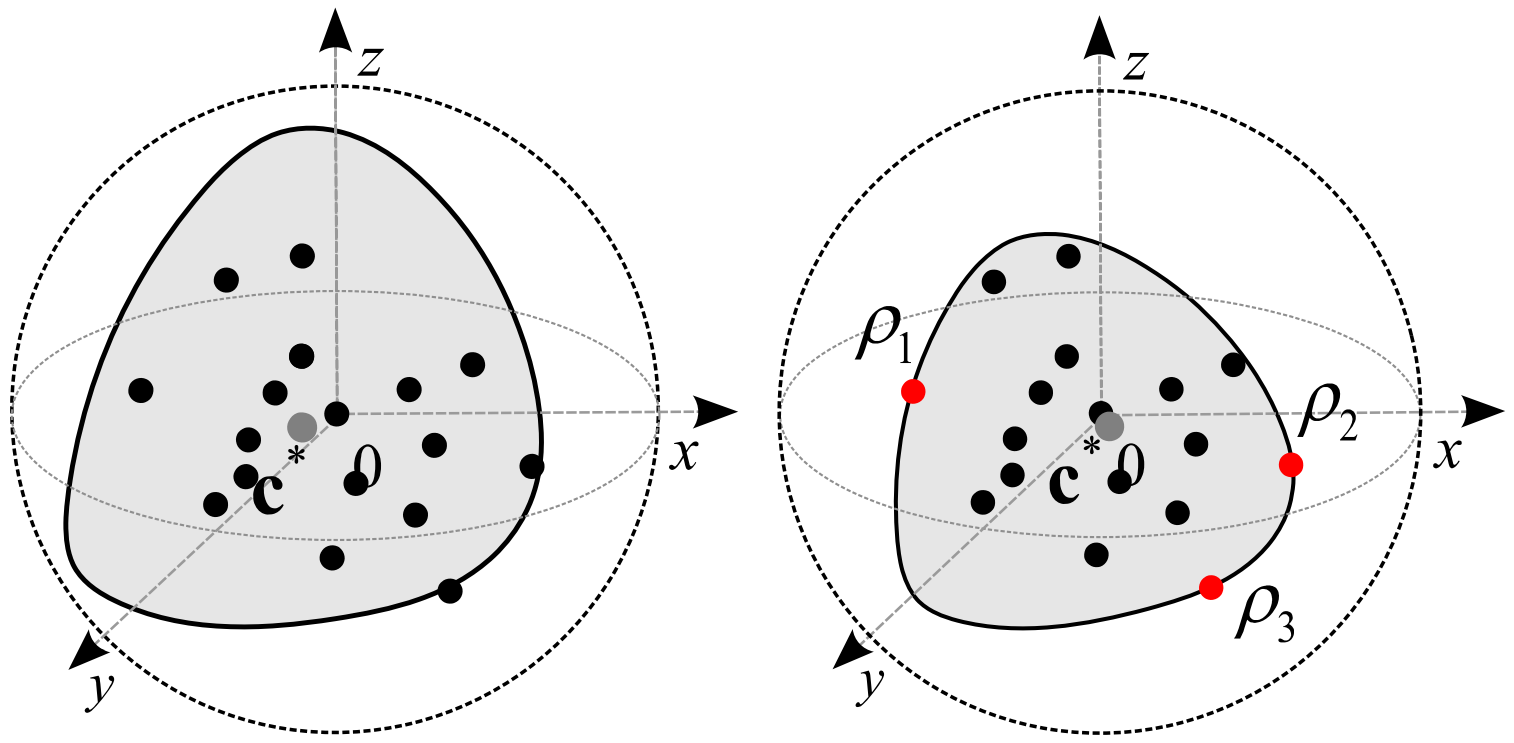

**Fig. F.3.** Fitting steps of the quantum informational ball using sample set of quantum states (single channel view).

The fitting algorithm can be extended to every quantum channel models. We compared the smallest quantum informational ball in the single channel view from $\mathcal{N}_1 \otimes \mathcal{N}_2$ and the ordinary Euclidean ball in Fig. F.4. As it can be seen, the quantum states $\rho_1, \rho_2$ and $\rho_3$ which determine the smallest enclosing ball in a Euclidean geometry differ from the states of the quantum informational ball.

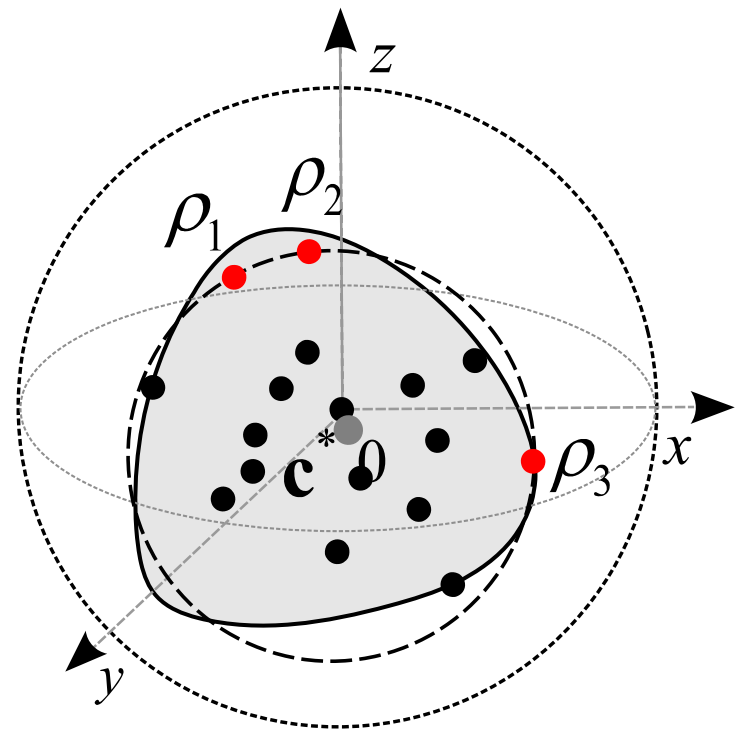

**Fig. F.4.** The maximum distance states of the smallest balls differ for the quantum informational distance and Euclidean distance (single channel view).

Using the proposed quantum superball approach to get an $(1+\mathcal{E})$-approximation of superactivated asymptotic quantum capacity $Q(\mathcal{N}_1 \otimes \mathcal{N}_2)$ of $\mathcal{N}_1 \otimes \mathcal{N}_2$, the algorithm requires to perform $\left\lfloor \frac{1}{\varepsilon^2} \right\rfloor$ iterations [Nielsen08a], [Nielsen09], [Nock05] hence the overall cost of the computation of the quantum superball of the joint channel structure $\mathcal{N}_1 \otimes \mathcal{N}_2$ is

$$\mathcal{O}\left(dn\frac{1}{\varepsilon^2}\right) = \mathcal{O}\left(\frac{dn}{\varepsilon^2}\right). \tag{F.3}$$

Now, let us see behind (F.3). It can be slipped into two parts: finding the farthest point and moving the center towards this farthest point costs $\mathcal{O}(dn)$. The second: using the quantum superball representation we need to perform $\left\lfloor \frac{1}{\varepsilon^2} \right\rfloor$ iterations to achieve the $(1+\mathcal{E})$-approximation of superactivated quantum capacity $Q(\mathcal{N}_1 \otimes \mathcal{N}_2)$ of $\mathcal{N}_1 \otimes \mathcal{N}_2$. From the two parts the overall cost is equal to (F.3). We can improve this method to get a $\mathcal{O}\left(\frac{d}{\varepsilon}\right)$ time $(1+\varepsilon)$-approximation algorithm in quantum space to compute the superactivated asymptotic quantum capacity $Q(\mathcal{N}_1 \otimes \mathcal{N}_2)$, as we will show next.

# F.2 The Improved Algorithm

We illustrate the improved core-set Algorithm in the single channel view on a set of quantum states in Fig. F.5. The iterations on the joint channel structure $\mathcal{N}_1 \otimes \mathcal{N}_2$ are performed by the quantum informational superball. For simplicity we will use the single channel view and the Bloch sphere representation.

The approximate ball has radius $r$, while the enclosing ball has radius $r+\delta$. The approximate center $\mathbf{c}$ is denoted in black, the *core-set* are colored by grey circles while $r^*$ stands for the optimal radius between the center $\mathbf{c}$ and the farthest quantum state [Nock05]. The improved algorithm increases the radius of the quantum superball from a lower bound $r$ of the optimal radius $r^*$. In this algorithm, the optimal superball radius is between $r \leq r^* \leq r+\delta$, and the process is terminates if $\delta \leq \varepsilon$, in at most $\left\lfloor \frac{1}{\varepsilon} \right\rfloor$ iterations [Nielsen08a], [Nielsen09], [Nock05].

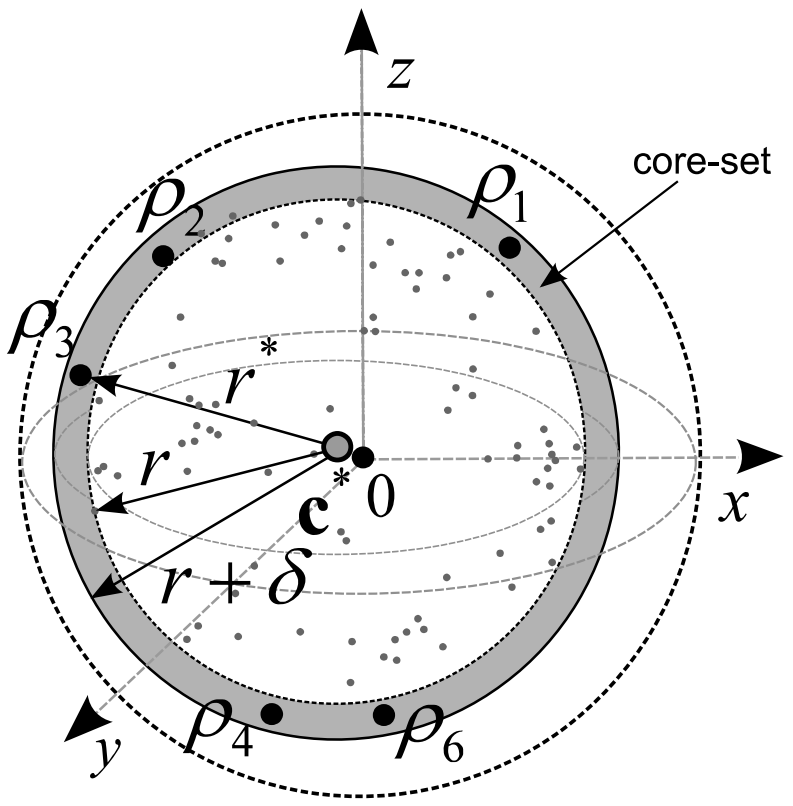

**Fig. F.5.** The approximate (light) and enclosing quantum information ball (darker) in the Bloch sphere (single channel view).

The main steps of the improved version of core-set algorithm can be summarized as follows [Nielsen08a], [Nielsen09], [Nock05]. The distance calculations between the density matrices are based on the quantum relative entropy function.

$$
\begin{aligned}
&1.\ \textit{Select} \text{ a random center } \mathbf{c}_1 \text{ from the set of density matrices } \mathcal{S} \\
&\qquad\qquad \mathbf{c}_1 \in \mathcal{S} \\
&2.\ r = \frac{1}{2}\max_i D_F\left(\mathcal{S}\middle\|\mathbf{c}_1\right); \\
&3.\ \delta = \frac{1}{2}\max_i D_F\left(\mathcal{S}\middle\|\mathbf{c}_1\right); \\
&4.\quad \textbf{for } \left(i = 1,2,\ldots,\left(\frac{1}{\delta}\right)\right) \\
&5.\qquad \textbf{do} \\
&6.\ \ S = \arg\max_i D_F\left(\mathcal{S}\middle\|c\right); \\
&7.\ \ \text{Move } \textit{Ball}\left(c,r\right) \text{ on the geodesic until it touches} \\
&\qquad \text{the } \textit{farthest} \text{ density matrix } S; \\
&8.\ \ s = \max_i D_F\left(S_i\middle\|c\right) - r; \\
&9.\qquad\quad \textbf{if}\quad s \le \frac{3\delta}{4} \textbf{ then} \\
&10.\qquad\qquad\qquad \delta = \frac{3\delta}{4} \\
&11.\qquad \textbf{else} \\
&12.\qquad\qquad\qquad r = r + \frac{\delta}{4}; \\
&13.\qquad\qquad\qquad \delta = \frac{3\delta}{4}; \\
&14.\ \ \textbf{until } \delta \le \varepsilon.
\end{aligned}
\tag{F.4}
$$

Using the improved algorithm, the $\left(1+\varepsilon\right)$-approximation the superactivated quantum capacity $Q\left(\mathcal{N}_1 \otimes \mathcal{N}_2\right)$ of $\mathcal{N}_1 \otimes \mathcal{N}_2$ can be computed in a time $\mathcal{O}\left(\frac{dn}{\varepsilon}\right)$. The improved algorithm works with complexity $\mathcal{O}\left(\frac{dn}{\varepsilon}\right)$, instead of $\mathcal{O}\left(\frac{dn}{\varepsilon^2}\right)$, where $\varepsilon < 1$ to compute the quantum informational superball of the joint structure $\mathcal{N}_1 \otimes \mathcal{N}_2$.

# F.3 Performance Analysis

In order to highlight the efficiency of the above algorithms we compared the core-set algorithms for 30 center updates and the quality of the approximation with respect to quantum relative entropy was measured. The results of the simulation are depicted in Fig. F.6. Axis $x$ represents the number of center updates to find the center of the smallest quantum superball of $\mathcal{N}_1 \otimes \mathcal{N}_2$, while axis $y$ shows the quantum informational distance between the found center $\mathbf{c}$ and the optimal $\mathbf{c}^*$, using input parameters $n = 4000$ and $\varepsilon = 0.005$.

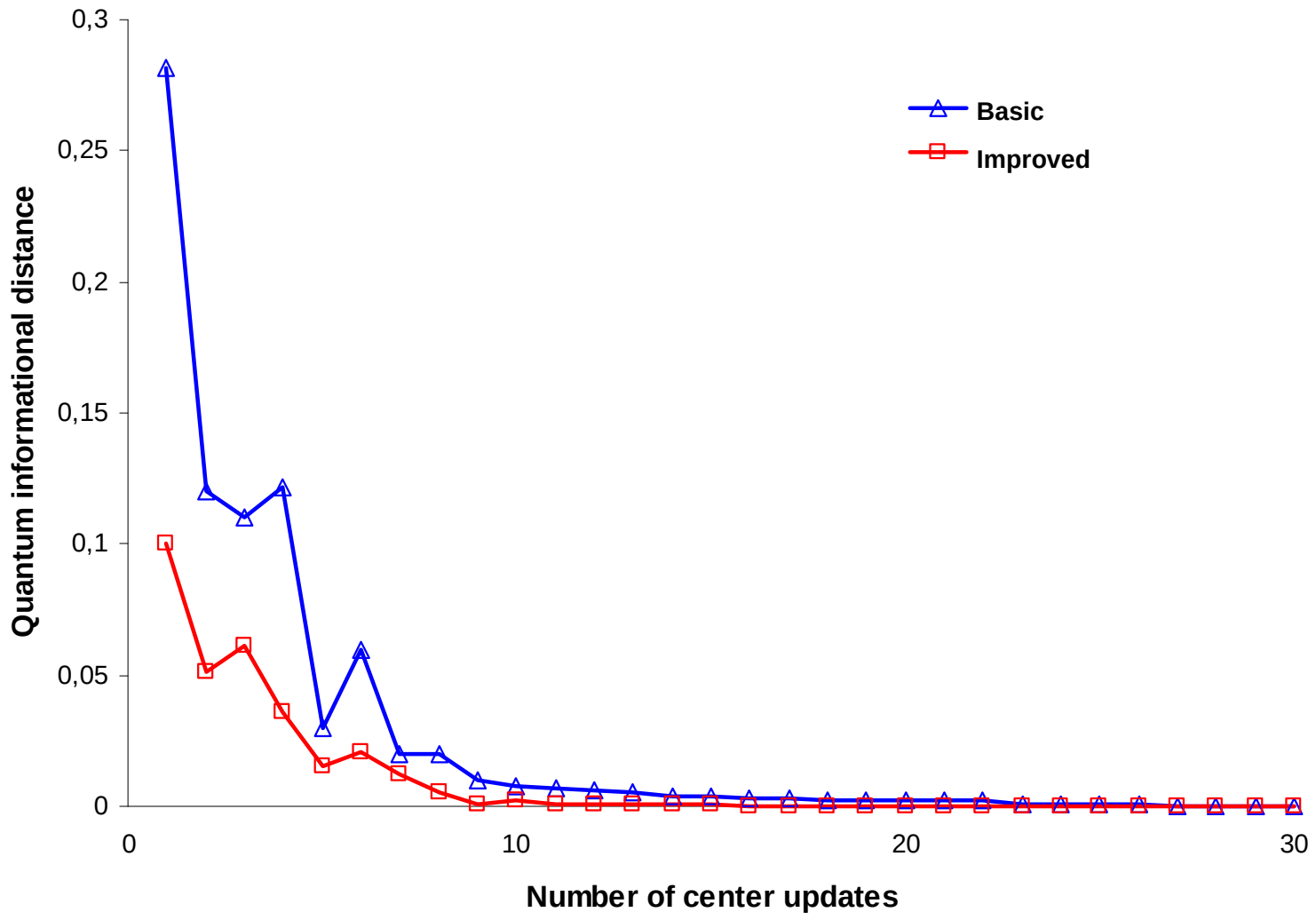

**Fig. F.6.** The convergence of the core-set algorithms.

From the results, it can be concluded that both algorithms find the approximate center $\mathbf{c}$ to the optimal center $\mathbf{c}^*$ very fast. The quantum relative entropy-based approximation algorithms have a very fast convergence of $\mathbf{c}$ towards $\mathbf{c}^*$.

# Appendix G

# Information Geometric Superactivation of Classical Zero-Error Capacity

## G.1 Superactivation of Zero-Error Capacities

Cubitt *et al.* [Cubitt09], [Cubitt09a] investigated an algebraic approach to study the superactivation property of the zero-error capacity, and as it has been found, there is also exists a stronger superactivation for the *asymptotic* zero-error capacity. In their work, both quantum channels can have zero zero-error capacity, while in the case of Duan's method [Duan09], one of the channels has to be equipped with a greater than zero zero-error capacity, otherwise the superactivation would not have worked. The usage of entanglement also implies the fact that superactivation is not possible in the case of classical communication channels. Next, we start to analyze the geometrical interpretation of superactivation of the classical zero-error capacity of quantum channels. The measure of the asymptotic classical zero-error capacity is based on the geometrical interpretation of classical HSW capacity. The proposed method can be used to verify both the results of Duan [Duan09] and, the stronger conditions of Cubitt *et al.* [Cubitt09], [Cubitt09a] on the asymptotic zero-error capacity. In the literature the methods of construction of Voronoi diagrams in the quantum space have been studied by Kato *et al.* [Kato06] and by Nielsen and Nock [Nielsen08b], however there were no further improvements made regarding the subject. The additivity problem of the classical capacity of quantum channels has been analyzed by an algorithmical framework by Gyongyosi and Imre [Gyongyosi10d-f, l-m], [Gyongyosi11p]. Our solution is the first approach to give an information geometric solution to the problem of superactivation of the zero-error capacity of quantum channels [Gyongyosi11a].

### G.1.1 Channel Setting for Superactivation of Classical Zero-Error Capacity

We show the encoder setting which requires EPR-states to encode the non-adjacent codewords and joint measurement to distinguish the codewords. We focus on the classical zero-error capacity $C_0\left(\mathcal{N}\right)$. Let assume the information source emits classical binary symbols $x_i \in \left\{0,1\right\}$ which are encoded into EPR pairs $X = \left[x_1, x_2, \ldots, x_n\right] \rightarrow \left[\left|\Psi_1\right\rangle \otimes \left|\Psi_2\right\rangle \otimes \left|\Psi_3\right\rangle \cdots \otimes \left|\Psi_n\right\rangle\right]$, where $n$ is the number of input EPR photon pairs in the quantum code, while

$$\left|\Psi_i\right\rangle \in \left\{\left|\beta_{00}\right\rangle, \left|\beta_{01}\right\rangle\right\} = \left\{\frac{1}{\sqrt{2}}\left(\left|00\right\rangle + \left|11\right\rangle\right), \frac{1}{\sqrt{2}}\left(\left|01\right\rangle + \left|10\right\rangle\right)\right\} \tag{G.1}$$

denotes the $i$-th entangled (EPR) input photon pair. The $n$ entangled states define an $\left|\Psi_{IN}\right\rangle$ input system which contains EPR states:

$$\left|\Psi_{IN}\right\rangle = \left[\left|\Psi_1\right\rangle \otimes \left|\Psi_2\right\rangle \otimes \left|\Psi_3\right\rangle \cdots \otimes \left|\Psi_n\right\rangle\right]. \tag{G.2}$$

The $i$-th sent codeword $\left|\Psi_{X_i}\right\rangle = \left[\left|\Psi_{i,1}\right\rangle \otimes \left|\Psi_{i,2}\right\rangle \otimes \left|\Psi_{i,3}\right\rangle \cdots \otimes \left|\Psi_{i,n}\right\rangle\right]$ encoded by the encoder will be decoded by $\mathcal{D}$, the decoder, using the POVM operators $\left\{\mathcal{M}_j\right\}$, where

$\sum_j \mathcal{M}_j = I$. For each $n$-length input codeword $X \in \{X_1, X_2, \ldots, X_K\}$, the output words $X_i' \in \{1, \ldots, m\}^n$ are generated by the measurement operators $\{\mathcal{M}_1, \ldots, \mathcal{M}_m\}$. The decoder associates each output word with integers *1* to $K$ representing input messages. In the encoding process, Alice chooses a message $X_i$ from the set of $K$ input messages, and with $\mathcal{E}$, her encoder, she prepares an $n$ length quantum blockcode $\left|\Psi_{X_i}\right\rangle = \left[\left|\Psi_1\right\rangle \otimes \left|\Psi_2\right\rangle \otimes \left|\Psi_3\right\rangle \cdots \otimes \left|\Psi_n\right\rangle\right]$. These entangled states are sent through the joint channel construction, in which each quantum channels have zero zero-error capacities individually. Bob uses his decoder, $\mathcal{D}$, to obtain an output number using the POVM operator, which will identify the input codeword [Gyongyosi11e-j], [Imre12].

The general view of the required channel setting for the superactivation of two quantum channels each with zero zero-error capacities is shown in Fig. G.1. The $i$-th entangled photon pair $\left|\Psi_i\right\rangle$ consists of entangled particles $\left|\Psi_i\right\rangle \in \left\{\left|\beta_{00}\right\rangle, \left|\beta_{01}\right\rangle\right\}$, where $\left|\beta_{00}\right\rangle$ and $\left|\beta_{01}\right\rangle$ are the Bell states (see Appendix B) and $\rho_i^{(1)} = \left|\Psi_i^{(1)}\right\rangle\left\langle\Psi_i^{(1)}\right|, \rho_i^{(2)} = \left|\Psi_i^{(2)}\right\rangle\left\langle\Psi_i^{(2)}\right|$ denote the first and the second qubits of the EPR state.

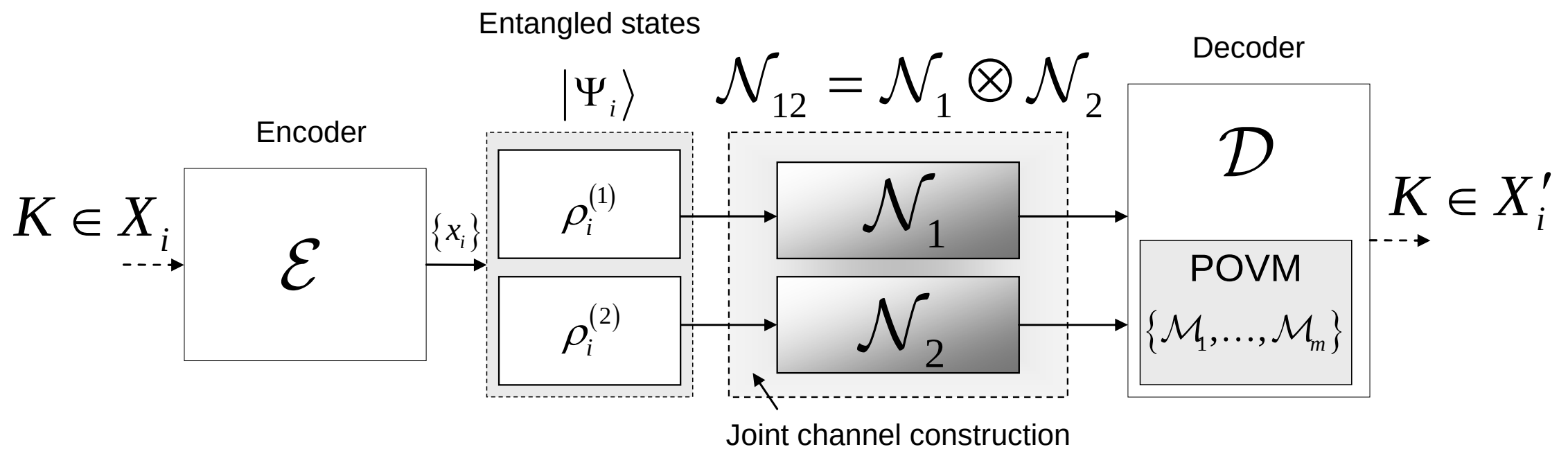

**Fig. G.1.** Information transmission with zero-error over superactivated quantum channels. Each quantum channels have zero zero-error capacities individually.

Using $n$ EPR states for the transmission the single-use and asymptotic superactivated zero error capacities for the joint structure $(\mathcal{N}_1 \otimes \mathcal{N}_2)$ can be expressed as follows:

$$C_0^{(1)}\left(\mathcal{N}\right)=\frac{1}{2}\log\left(K\left(\mathcal{N}_1\otimes\mathcal{N}_2\right)\right) \tag{G.3}$$

and

$$C_0\left(\mathcal{N}\right)=\lim_{n\to\infty}\frac{1}{2n}\max_n\log\left(K\left(\mathcal{N}_1\otimes\mathcal{N}_2\right)^{\otimes n}\right), \tag{G.4}$$

where $K\left(\left(\mathcal{N}_1\otimes\mathcal{N}_2\right)^{\otimes n}\right)$ is the maximum number of classical $n$-length messages that the superactivated joint channel construction $\left(\mathcal{N}_1\otimes\mathcal{N}_2\right)$ can transmit with zero error. Since we will encode the $n$-length input codewords with EPR input states, we use $2n$ in the denominator.

# G.2 Superactivation of Classical Zero-Error Capacity

In this section we derive the result on the decomposition of the superball radius for the analysis of the superactivation of classical zero-error capacity. The superball radius $r_{super\left(C_0\right)}^{(1)*}\left(\mathcal{N}_1\otimes\mathcal{N}_2\right)$ measures the superactivated zero-error classical capacity of the joint structure $\mathcal{N}_1\otimes\mathcal{N}_2$. Here we show that if the joint channel structure $\mathcal{N}_1\otimes\mathcal{N}_2$ is superactive then the decomposition of the superball radius $r_{super\left(C_0\right)}^{(1)*}\left(\mathcal{N}_1\otimes\mathcal{N}_2\right)$ cannot be made, otherwise the joint channel structure is not superactive.

Using the results derived in Chapter 5 and the results on the geometric interpretation of quantum channel capacities [Petz96,08], [Schumacher99,2000], [Cortese02,03], [Hayashi03,05] and Ruskai [Ruskai01] it also follows that if the classical zero-error capacity of the joint channel construction $\mathcal{N}_1\otimes\mathcal{N}_2$ is superactive (i.e., the channels can activate each other in the joint structure) then the superball radius $r_{super\left(C_0\right)}^{(1)*}\left(\mathcal{N}_1\otimes\mathcal{N}_2\right)$ cannot be decomposed in the following way:

$$
\begin{aligned}
& r_{super\left(C_0\right)}^{(1)*}\left(\mathcal{N}_1 \otimes \mathcal{N}_2\right)=C_0^{(1)}\left(\mathcal{N}_1 \otimes \mathcal{N}_2\right) \\
& =\min_{\sigma_{12}} \max_{\rho_{12}} D\left(\rho_{12}^{AB} \| \sigma_{12}^{AB}\right)=\min_{\sigma_1 \otimes \sigma_2} \max_{\rho_1 \otimes \rho_2} D\left(\rho_{12}^{AB} \| \sigma_{12}^{AB}\right) \\
& =\min_{\sigma_1 \otimes \sigma_2} \max_{\rho_1 \otimes \rho_2} Tr_{12}\left(\rho_{12}^{AB} \log\left(\rho_{12}^{AB}\right)-\rho_{12}^{AB} \log\left(\sigma_{12}^{AB}\right)\right) \\
& =\min_{\sigma_1} \min_{\sigma_2} \max_{\rho_1} \max_{\rho_2} Tr_{12}\left(\begin{array}{l}
\left(\rho_1^{AB} \otimes \rho_2^{AB}\right) \log\left(\rho_1^{AB} \otimes \rho_2^{AB}\right) \\
-\left(\rho_1^{AB} \otimes \rho_2^{AB}\right) \log\left(\sigma_1^{AB} \otimes \sigma_2^{AB}\right)
\end{array}\right) \\
& =\min_{\sigma_1} \min_{\sigma_2} \max_{\rho_1} \max_{\rho_2} Tr_{12}\left(\begin{array}{l}
\left(\rho_1^{AB} \otimes \rho_2^{AB}\right)\left(\log\left(\rho_1^{AB}\right) \otimes I_2\right) \\
+\left(\rho_1^{AB} \otimes \rho_2^{AB}\right)\left(I_1 \otimes \log\left(\rho_2^{AB}\right)\right)
\end{array}\right) \\
& \quad-\min_{\sigma_1} \min_{\sigma_2} \max_{\rho_1} \max_{\rho_2} Tr_{12}\left(\begin{array}{l}
\left(\rho_1^{AB} \otimes \rho_2^{AB}\right)\left(\log\left(\sigma_1^{AB}\right) \otimes I_2\right) \\
+\left(\rho_1^{AB} \otimes \rho_2^{AB}\right)\left(I_1 \otimes \log\left(\sigma_2^{AB}\right)\right)
\end{array}\right) \\
& =\min_{\sigma_1} \min_{\sigma_2} \max_{\rho_1} \max_{\rho_2} Tr_1\left(\rho_1^{AB} \log\left(\rho_1^{AB}\right)\right) Tr_2\left(\rho_2^{AB} I_2\right) \\
& \quad+\min_{\sigma_1} \min_{\sigma_2} \max_{\rho_1} \max_{\rho_2} Tr_1\left(\rho_1^{AB} I_1\right) Tr_2\left(\rho_2^{AB} \log\left(\rho_2^{AB}\right)\right) \\
& \quad-\min_{\sigma_1} \min_{\sigma_2} \max_{\rho_1} \max_{\rho_2} Tr_1\left(\rho_1^{AB} \log\left(\sigma_1^{AB}\right)\right) Tr_2\left(\rho_2^{AB} I_2\right) \\
& \quad-\min_{\sigma_1} \min_{\sigma_2} \max_{\rho_1} \max_{\rho_2} Tr_1\left(\rho_1^{AB} I_1\right) Tr_2\left(\rho_2^{AB} \log\left(\sigma_2^{AB}\right)\right) \\
& =\min_{\sigma_1} \min_{\sigma_2} \max_{\rho_1} \max_{\rho_2} Tr_1\left(\rho_1^{AB} \log\left(\rho_1^{AB}\right)\right)-Tr_1\left(\rho_1^{AB} \log\left(\sigma_1^{AB}\right)\right) \\
& \quad+\min_{\sigma_1} \min_{\sigma_2} \max_{\rho_1} \max_{\rho_2} Tr_2\left(\rho_2^{AB} \log\left(\rho_2^{AB}\right)\right)-Tr_2\left(\rho_2^{AB} \log\left(\sigma_2^{AB}\right)\right) \\
& =\min_{\sigma_1} \min_{\sigma_2} \max_{\rho_1} \max_{\rho_2}\left(D\left(\rho_1^{AB} \| \sigma_1^{AB}\right)+D\left(\rho_2^{AB} \| \sigma_2^{AB}\right)\right) \\
& =\min_{\sigma_1} \min_{\sigma_2} \max_{\rho_1} \max_{\rho_2} D\left(\rho_1^{AB} \| \sigma_1^{AB}\right)+\min_{\sigma_1} \min_{\sigma_2} \max_{\rho_1} \max_{\rho_2} D\left(\rho_2^{AB} \| \sigma_2^{AB}\right) \\
& =\min_{\sigma_1} \max_{\rho_1} D\left(\rho_1^{AB} \| \sigma_1^{AB}\right)+\min_{\sigma_2} \max_{\rho_2} D\left(\rho_2^{AB} \| \sigma_2^{AB}\right) \\
& =C_0^{(1)}\left(\mathcal{N}_1\right)+C_0^{(1)}\left(\mathcal{N}_2\right) \\
& =r_{super\left(C_0\right)}^{(1)*}\left(\mathcal{N}_1\right)+r_{super\left(C_0\right)}^{(1)*}\left(\mathcal{N}_1\right),
\end{aligned} \tag{G.5}
$$

where $I_1$ and $I_2$ are the $d$ dimensional identity matrices ($d$=2 for the qubit case), $\rho_{12}^{AB}$ is the optimal channel output state of the joint channel $\mathcal{N}_{12}=\mathcal{N}_1 \otimes \mathcal{N}_2$, and $\sigma_{12}^{AB}=\sum_i p_i \rho_{12}^{AB(i)}$ is the average output state of the joint structure $\mathcal{N}_{12}=\mathcal{N}_1 \otimes \mathcal{N}_2$ which can be obtained for the zero-error input codewords.

### G.2.1 Polynomial-Approximation of Classical Zero-Error Capacity

To construct the polynomial-approximation algorithm for the study of superactivation of classical zero-error capacity of quantum channels, in the first step we introduce a *core-set* method. We construct a $\left[\alpha,\beta\right]$ bicriteria approximation algorithm to get the set of median quantum states $M=\left\{\sigma_1,\sigma_2,\ldots\sigma_k\right\}$ of a $k$-median clustering of $\mathcal{S}_{IN}$, for which $error\left(\mathcal{S}_{IN},M\right)\leq\alpha opt_k\left(\mathcal{S}_{IN}\right)$ and $\left|M\right|=k\leq\beta k$. Using the results of [Chen07], [Badoiu03], [Ackermann08-09], [Zhang09] the bicriteria algorithm for the superactivation of the $C_0\left(\mathcal{N}_1\otimes\mathcal{N}_2\right)$ zero-error classical capacity of the joint structure $\mathcal{N}_1\otimes\mathcal{N}_2$ can be summarized as follows [Gyongyosi11a]:

$$\begin{array}{l}\textbf{Bicriteria algorithm to analysis of classical zero - error capacity}\\ \text{1. Choose an initial density matrices } \sigma_1 \text{ uniformly}\\ \quad\text{at random from } \mathcal{S}_{IN}\\ \text{2. Let } M \text{ be the set of chosen density matrices}\\ \quad\text{from } \mathcal{S}_{IN}\text{. State } \rho\in\mathcal{S}_{IN} \text{ is chosen with}\\ \quad\text{probability } \dfrac{D\left(\rho\|M\right)}{error\left(\mathcal{S}_{IN},M\right)} \text{ as the next density matrix of } M.\\ \text{3. Repeat step 2 until } M \text{ contains } k \text{ density matrices.}\end{array} \tag{G.6}$$

At the end of the bicriteria algorithm, we have a set of median quantum states $M=\left\{\sigma_1,\sigma_2,\ldots\sigma_k\right\}$, for which $error\left(\mathcal{S}_{IN},M\right)\leq\alpha opt_k\left(\mathcal{S}_{IN}\right)$ and $\left|M\right|=k\leq\beta k$.

After the application of the bicriteria algorithm [Chen07], we use the following core-set construction method to the density matrices [Gyongyosi11a] of the joint structure $\mathcal{N}_1\otimes\mathcal{N}_2$ [Badoiu03], [Ackermann08-09], [Zhang09]. The distance calculations between the density matrices to obtain $C_0\left(\mathcal{N}_1\otimes\mathcal{N}_2\right)$ of $\mathcal{N}_1\otimes\mathcal{N}_2$ are based on the quantum relative entropy function (see Chapter 5 and Appendix B).

**Core-set algorithm to analysis of classical zero - error capacity**

1. Partition $\mathcal{S}_{IN}$ into $\mathcal{S}_1, \mathcal{S}_2 \ldots, \mathcal{S}_k$ by assuming each density matrix $\rho \in \mathcal{S}_{IN}$ to their closest desnity matrix $\sigma_i \in M$.
2. Let $\rho \in \mathcal{S}_{IN}$ iff $\sigma_i = \arg\min_{\sigma \in M} D(\rho \| \sigma)$.
3. Let $R = \frac{1}{\alpha n} error(\mathcal{S}_{IN}, M)$.
4. Define quantum informational superball $\mathcal{B}(\sigma_i)$ with radius $r^*$ and center $\sigma_i$ as follows: $\mathcal{B}(\sigma_i) = D(x \| \sigma_i) \leq r$.
5. Define the partition of $\{\mathcal{S}_{ij}\}_{i,j}$ *of* $\mathcal{S}_{IN}$ by $\mathcal{S}_{ij} = \mathcal{S}_i \cap \mathcal{B}(\sigma_i)$ *for* $i = 1,2,\ldots,k$.
6. Let $\mathcal{S}_{ij} = \mathcal{S}_i \cap (\mathcal{B}_{2^j}(\sigma_i) \setminus \mathcal{B}_{2^{j-1}}(\sigma_i))$ *for* $i = 1,2,\ldots,k$ and $j = i = 1,2,\ldots\gamma$, where $\gamma = \lceil \log(\alpha n) \rceil$.
7. For $i,j$ let $\mathcal{S}_{ij}$ *be* a uniform set from $\mathcal{S}_{IN}$ of size $|\mathcal{S}_{ij}| = m$.
8. Let $w(\rho) = \frac{1}{m}|\mathcal{S}_{ij}|$ be the weight associated with density matrix $\rho \in \mathcal{S}_{ij}$.
9. Define the *weak coreset* $\mathcal{S}$ of input density matrices $\mathcal{S}_{IN}$ as follows: $\mathcal{S} = \bigcup_{i,j} \mathcal{S}_{ij}$ of size $|\mathcal{S}| = mk\gamma = m\beta k \lceil \log(\alpha n) \rceil$. (G.7)

Based on the conditions of the bicreteria algorithm and the proposed core-set method [Ackermann08-09], [Chen07], if $m = \Omega\left(\frac{\alpha^2}{\varepsilon^2} \log\left(\frac{\beta}{\delta} k |\mathcal{W}|^k \log(\alpha n)\right)\right)$, then the output set $\mathcal{S} = \bigcup_{i,j} \mathcal{S}_{ij}$ of the core-set algorithm is a $\mathcal{W}$-weak core-set of $\mathcal{S}_{IN}$ with probability $1-\delta$ [Gyongyosi11a]. Using these results, if the parameters $[\alpha, \beta]$ are found with the bicriteria algorithm (as shown previously in Section G.2.1), then an $[\alpha, \beta]$-approximate $k$-median clustering of $\mathcal{S}_{IN}$ in a $\mathcal{W}$-weak core-set of size $\mathcal{O}\left(\frac{1}{\varepsilon^2} k \log n \log\left(k |\mathcal{W}|^k \log n\right)\right)$ can be constructed in time

$$\mathcal{O}\left(dkn + \frac{1}{\varepsilon^2} k \log n \log\left(k |\mathcal{W}|^k \log n\right)\right). \tag{G.8}$$

Next, we design an algorithm for clustering density matrices of the quantum codewords, using the previously generated $\mathcal{W}$-weak core-set of set $\mathcal{S}_{IN}$ along with these results on the properties of weak core-sets [Gyongyosi11a]. The proposed algorithm is a core-set approach, i.e., it has approximation error $(1+\varepsilon)$, but the run time of the proposed method to obtain $C_0(\mathcal{N}_1 \otimes \mathcal{N}_2)$ of $\mathcal{N}_1 \otimes \mathcal{N}_2$ is more efficient since it uses the $\mathcal{W}$-weak core-set of set $\mathcal{S}_{IN}$ generated by the weak core-set algorithm, instead of the original input set $\mathcal{S}_{IN}$ of density matrices. Using the core-set and clustering algorithms from the tools of classical informational geometry along with the $\mathcal{W}$-weak core-set of the original input set $\mathcal{S}_{IN}$, the superactivation of classical zero-error capacity of quantum channels can be analyzed very efficiently. As will be shown in Section G.2.2, the $(1+\varepsilon)$-approximation can be obtained in a run time

$$\mathcal{O}\left(d^2 2^{\frac{k}{\varepsilon}} \log^{k+2} n + dkn\right). \tag{G.9}$$

In the next subsection a clustering algorithm to determine the superactivation of quantum channels is introduced.

## G.2.2 Determination of Median-Quantum States

In this section an algorithm for the determination of median-density matrices for the superactivation of $C_0(\mathcal{N}_1 \otimes \mathcal{N}_2)$ of the joint channel $\mathcal{N}_1 \otimes \mathcal{N}_2$ is shown. The median-density matrices for the superactivation analysis of the classical zero-error capacity are discovered as follows. The $\mathbf{CL}_{\text{superball}}$ clustering algorithm for a weak set of density matrices and $\mu$-similar [Ackermann08-09], [Badoiu03], [Chen06], [Zhang09], [Gupta06], [Sra08] quantum informational distances can be summarized as

**$\mathbf{CL}_{\text{superball}}$ : Clustering of quantum states for the quantum superball**

1. Let $\mathcal{S}_{IN}$ be the set of remaining input density matrices, with $w\left(\mathcal{S}_{IN}\right)=n$
2. Let $w$ be the weight function on input density matrices $\mathcal{S}_{IN}$
3. Let $m$ be the number of median-density matrices *yet to be* found
4. Let $C$ be the set of median density matrices *already* found
5. **if** $m=0$ then return $C$
6. **else**
7.     **if** $m \geq \left|\mathcal{S}_{IN}\right|$ then return $C \cup \mathcal{S}_{IN}$
8.     **else**
9.       Sample a multiset of density matrices $\mathcal{M}$ of size $\frac{96k^2}{\varepsilon^2\mu\delta}$ from $\mathcal{S}_{IN}$
10.       $T \leftarrow$ Let $c$ the weighted centroid of $\mathcal{M}' \subseteq \mathcal{M}$, with $\left|\mathcal{M}'\right|=\frac{3}{\varepsilon\mu\delta}$
11.       **for** all $c \in T$ do
12.       $\mathrm{C}^{(c)} \leftarrow \mathbf{CL}_{\text{superball}}\left(\mathcal{S}_{IN}, w, m-1, C \cup \left\{c\right\}\right)$
13.       **end for**
14.       Partition of the set of input density matrices $\mathcal{S}_{IN}$ into set $N$ and $\mathcal{S}_{IN} \setminus N$ such that:
15.       $\forall \rho \in N, \sigma \in \mathcal{S}_{IN} \setminus N : D\left(\rho \| C\right) \leq D\left(\sigma \| C\right)$ and
16.       $w\left(N\right)=w\left(\mathcal{S}_{IN} \setminus N\right)=\frac{n}{2}$
17.       Let $w^*$ the new weight function on $\mathcal{S}_{IN} \setminus N$
18.       Let $C^* \leftarrow \mathbf{CL}_{\text{superball}}\left(\mathcal{S}_{IN} \setminus N, q, m, C\right)$
19.       **return** $\mathrm{C}^{(c)}$ or $\mathrm{C}^*$ with *minimum error*
20.     **end if**
21. **end if**

(G.10)

In Fig. G.2, we illustrate the clustering of density matrices. In the clustering process, our algorithm computes the median-quantum states denoted by $\sigma_i$, using a fast weak core-set and clustering algorithms (shown in Sections G.2.1 and G.2.2). In the next step, we compute the convex hull of the median quantum states and, from the convex hull, the radius of the smallest quantum informational ball can be obtained. The quantum superball defined in Chapter 5 of the Ph.D Thesis, measures the superactivated classical zero-error capacity of $\mathcal{N}_1 \otimes \mathcal{N}_2$, (using the single channel view the radius is depicted by $\mathbf{r}^*$ in Fig. G.(b)). For simplicity we illustrate the algorithm in single channel view, however the su-peractivated classical zero-error capacity is analyzed by the proposed quantum superball ball, see the result of Chapters 5 and 7 and the proof of Thesis 3.1.

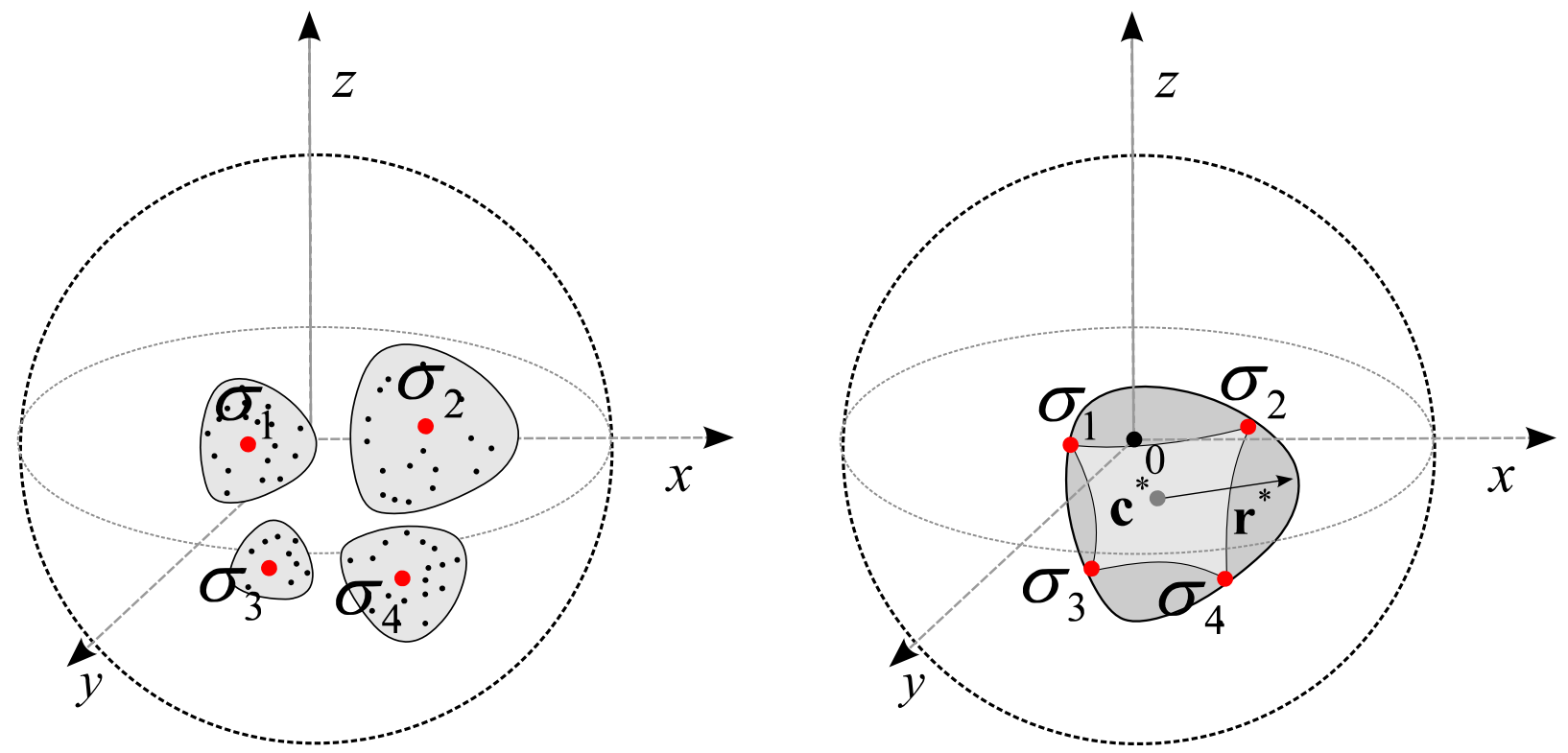

**Fig. G.2.** Clustering of quantum states. The smallest quantum informational ball contains the computed medians.

Using the modified weak core-set algorithm and the $\left(1+\varepsilon\right)$-approximation algorithm, the superactivation of $C_0\left(\mathcal{N}_1 \otimes \mathcal{N}_2\right)$ of the joint channel structure $\mathcal{N}_1 \otimes \mathcal{N}_2$ can be analyzed relative to $\mu$-similar quantum informational distances and $k$ median-quantum states with error $error\left(\mathcal{S}_{IN}, \mathcal{S}_{OUT}\right) \leq \left(1+7\varepsilon\right) opt_k\left(\mathcal{S}_{IN}\right)$.

# G.3 Performance Analysis

The algorithm presented in the this section, has lower complexity in comparison with other existing core-set and approximation algorithms, which can also be applied in quantum space. In Fig. G.3, we have compared the complexity of the algorithm presented by Nielsen and Nock in [Nielsen08b] for the tessellation of the quantum space and our advanced weak core-set approach, as a function of input size (number of input density matrices). Both algorithms result in $\left(1+\varepsilon\right)$-approximations, however the complexities of the proposed methods differ significantly.

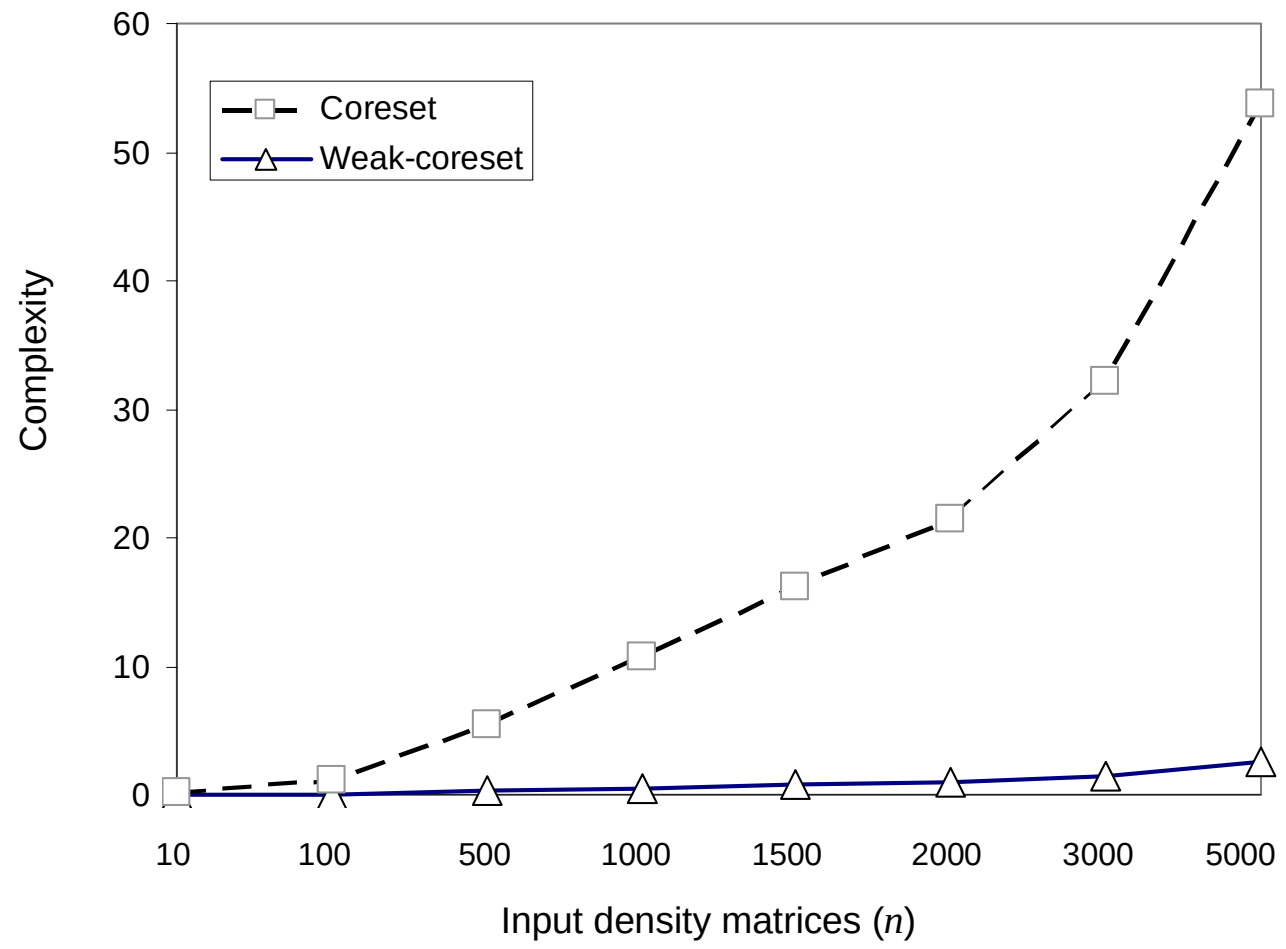

**Fig. G.3.** Complexity of core-set and weak core-set algorithms for the superactivation analysis of classical zero-error capacity.

As our results confirm, the complexity of the weak core-set method is significantly lower than for the core-set method, especially for a large number of input EPR quantum states $n$. The precision of the approximation is the same as the standard core-set approach, however the complexity of our weak core-set method is significantly lower, hence the quantum channels can be analyzed more efficiently.

# G.4 Superactivated Quantum Repeaters

The quantum repeater is based on the transmission of entangled quantum states between the repeater nodes. The entanglement creation uses the quantum communication channel; hence some noise is added to the transmitted states. In the next step, the created entanglement has to be purified. The purification is an error-correcting scheme, and it uses local quantum operations only – hence these operations can be realized in the separated base stations locally.

The working mechanism of the quantum repeater with the entanglement sharing and the swapping is illustrated in Fig. G.4.

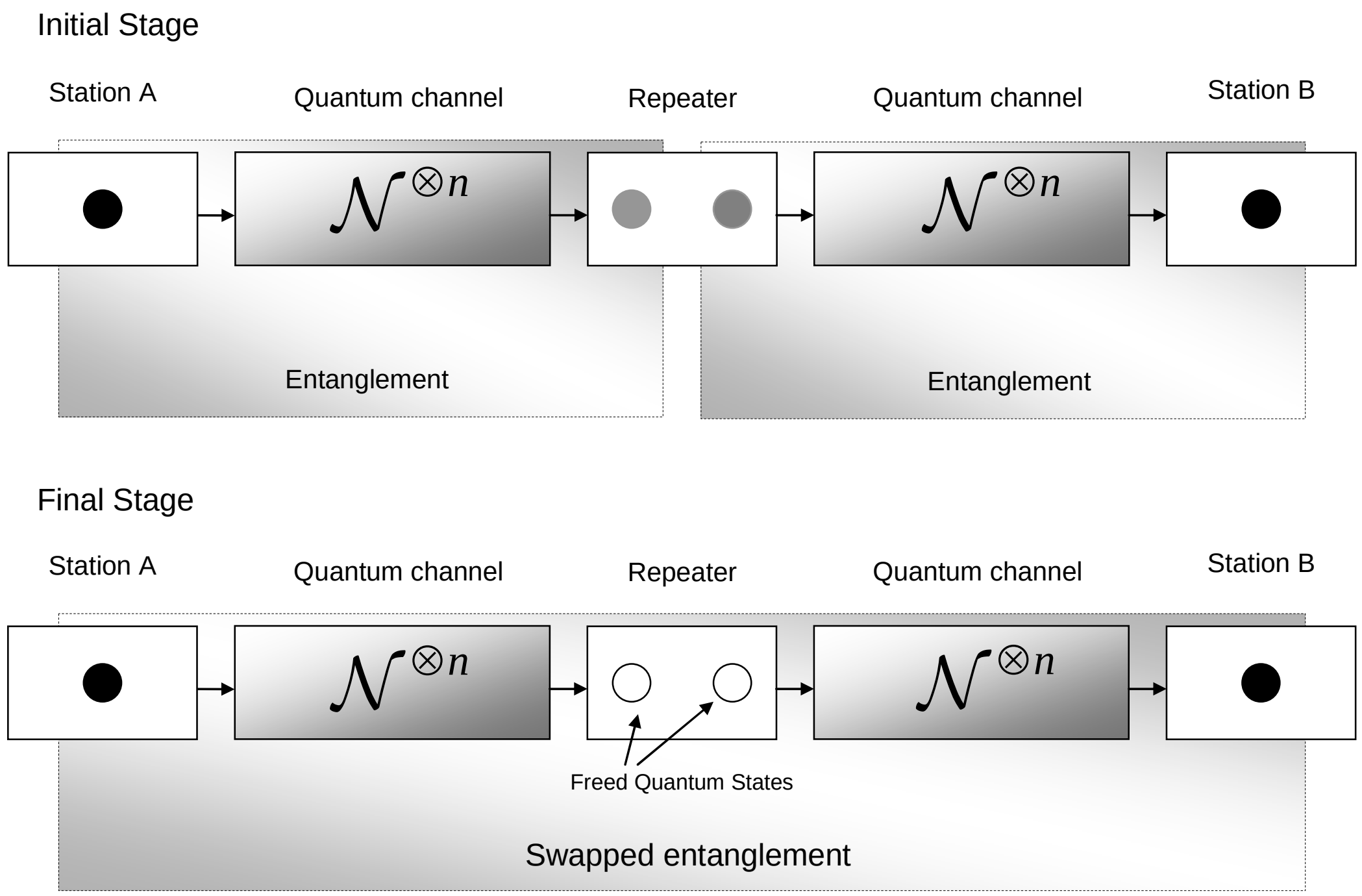

**Fig. G.4.** The standard model of the quantum repeater with noisy quantum channels. The structure of the quantum repeater consists of a chain of base stations and the information is transmitted via quantum teleportation.

The main task is the improvement of the fidelities of the shared entangled states. The most important part of quantum communication between quantum repeaters is the entanglement purification. This step purifies the noisy quantum states, however this process requires a lot of resources in the quantum nodes, and it cannot be implemented efficiently in current solutions. The creation of high-fidelity entanglement between the nodes requires a lot of entangled pairs, while the purification process is a computationally very complex. In order to recover fidelity of entanglement from noisy quantum states purification is needed. The efficiency of the quantum repeaters in future would be incremental only if the efficiency of the purification process could be increased.

The newly developed *superactivated quantum repeater* uses noisy quantum channels such as in the case of standard quantum repeaters. On the other hand, these quantum channels can be used for perfect information transmission, and the entangled quantum states can be sent through the channels with maximal fidelity. As a conclusion, no further purifications needed during the communication.

The basic model of the superactivated quantum repeater is depicted in Fig. G.5. [Gyongyosi11e-f].

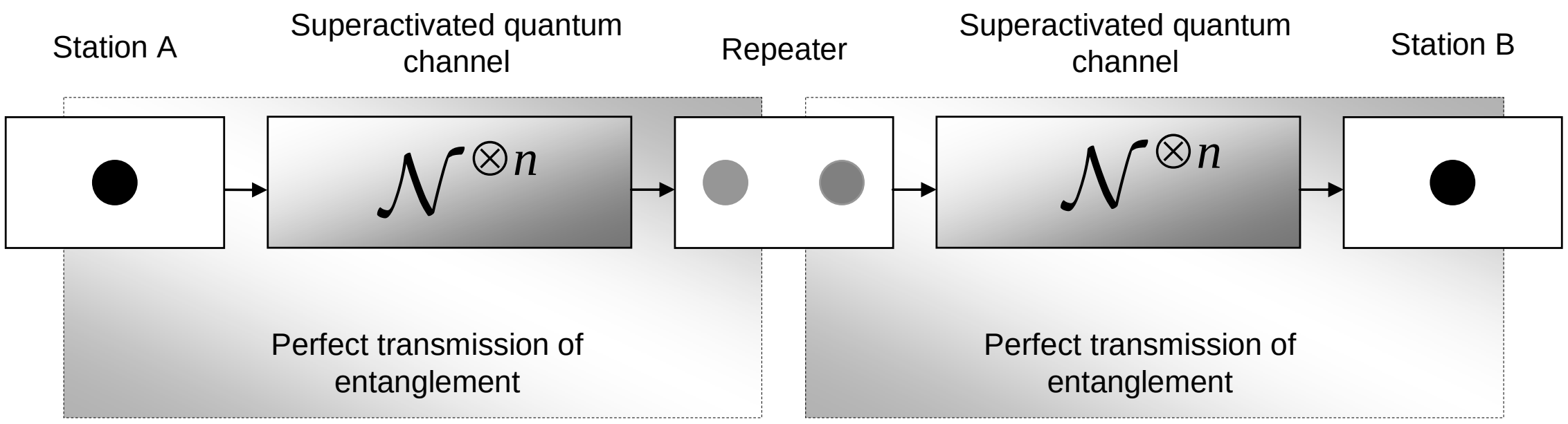

**Fig. G.5.** The newly developed quantum repeater with noisy quantum channels and perfect information transmission.

The proposed joint channel structure between the repeater nodes contains two different classes of Gaussian quantum channels with zero zero-error capacity individually. In the proposed model the quantum channels were Gaussian quantum channels, with zero zero-error capacities individually. The general setting of the zero-error information transmission over Gaussian quantum channel is similar to already shown Chapter 7. The input message is denoted by $X_i$. The encoder $\mathcal{E}$ and the decoder $\mathcal{D}$ are implemented by coherent optical devices [Wilde07], the quantum channel is a Gaussian quantum link [Gyongyosi11e], [Lupo11]. To realize the superactivation of the Gaussian quantum channel, we construct a joint channel denoted by $\mathcal{G}_{12} = \mathcal{G}_1 \otimes \mathcal{G}_2$, using two Gaussian quantum channels denoted by $\mathcal{G}_1$ and $\mathcal{G}_2$.

## G.4.1 Numerical Results

Using the results of the proposed algorithm in Chapters 7 and the theories of Chapter 5 we have discovered that the possibility of the superactivation of the zero-error capacity depends on the *length* of the input codewords $(N)$ encoded by the EPR photons, and on the *number* of input codewords $\left(M\right)$ that can be distinguished by the POVM (Positive Operator Valued Measure) operators (called the *non-adjacent* codewords) performed by the optical decoder.

We analyzed the classical zero-error capacity with using input blockcode length up to $N = 21$ EPR pairs, with $M\left(N\right) = 332$ non-adjacent input codewords, these results are shown in Fig. G.6(a). Axis $x$ represents the length of the input codewords encoded by EPR photon pairs, the $y$-axis illustrates the number of input state subspace length ($M$) with *non-adjacent input codewords N*. In Fig. G.6(b) the results for the quantum capacity are shown. In that case we used blockcode length up to $N = 45$ EPR pairs, with $M\left(N\right) = 2148$ non-adjacent input codewords [Gyongyosi11e].

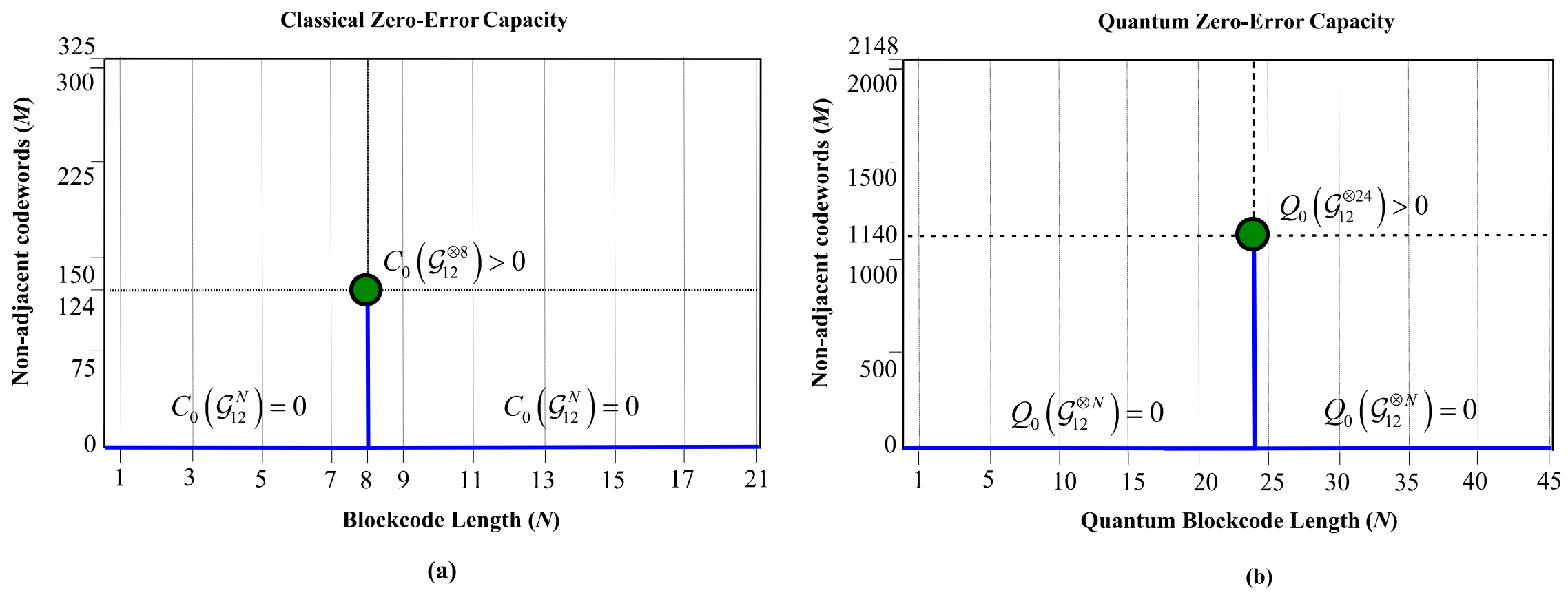

**Fig. G.6.** (a): Superactivated classical zero-error capacity. The non-adjacent input codewords as a function of the length of the input quantum blockcode. We have found only one possible constellation for the superactivation in the analyzed domain. (b) Results on the quantum zero-error capacity. The superactivation of quantum zero-error capacity requires different input settings.

The results demonstrate that it is possible to transmit classical and quantum information perfectly over very noisy Gaussian quantum channels. As we have found, within this large parameter domain, there is only one combination of input length of EPR photon pairs ($N$) and input state subspace length ($M$, the number of POVM elements of the joint measurement) exists for which the zero-error capacity of two Gaussian quantum channels can be superactivated. For larger values of $N$, and $M$, or different channel models, other solutions could be possible. Our goal is to discover these still unrevealed combinations in the near future.

## G.4.2 Application in Repeater Networks

The results of on the performance analysis of the superactivated quantum repeater are shown in Fig. G.7 [Gyongyosi11e]. The red line depicts the rate of entanglement generation between the nodes using superactivated quantum channels, with superactive quantum zero-error capacity.

In Fig. G.7(a) the *dashed* line depicts the entanglement generation rate between the repeater stations without entanglement purification using general noisy quantum channel, the *solid* line depicts the superactivated channel. In G.7(b), the lines represent one round of purification using standard quantum channels (*dashed* line) and superactivated channels (*solid* line). The $F$ represents the final fidelities in the repeater nodes.

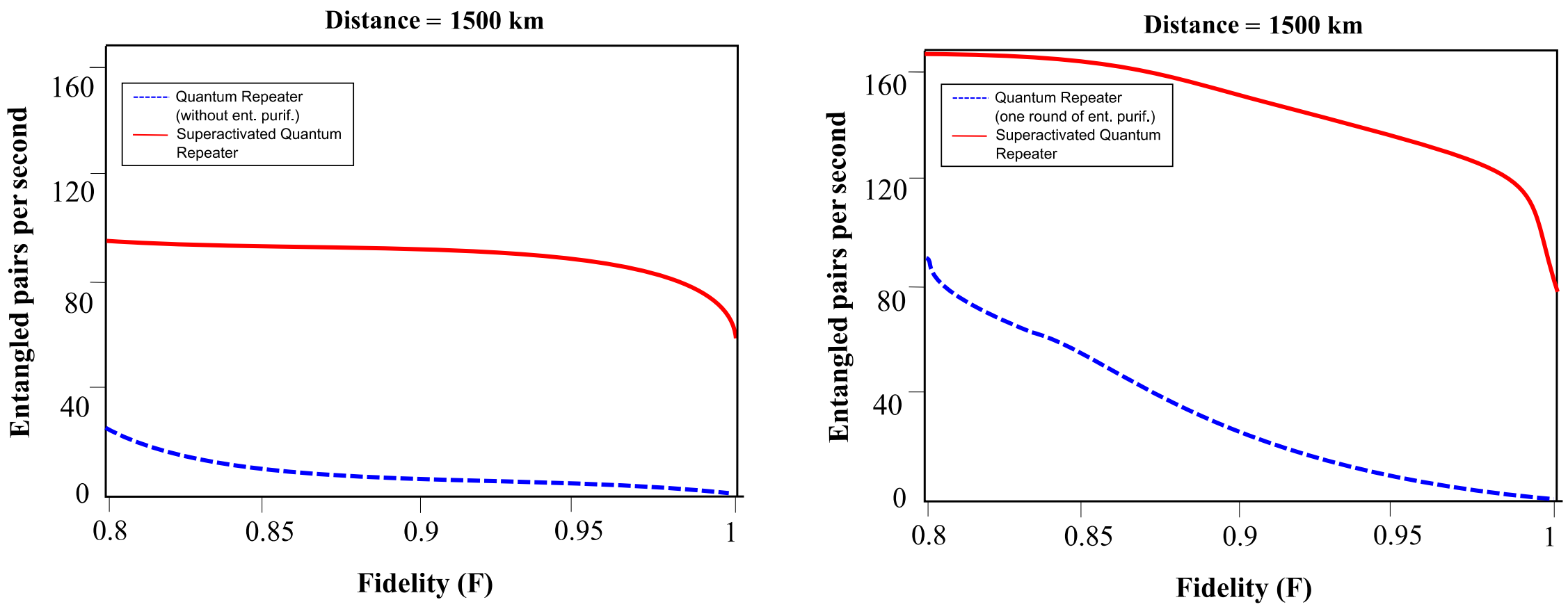

**Fig. G.7** (a): The rate of a standard quantum repeater (dashed line) and our newly designed quantum repeater (solid line) over a total distance of 1500 km, without purification using standard noisy quantum channels. (b): Comparison of superactivated zero-error quantum capacity with one-round of entanglement purification with standard quantum channels (dashed line) and superactivated quantum channel (solid line).

The superactivated channels preserved entanglement nearly perfectly and the rate of entanglement generation increased significantly. From the results follows that using standard noisy quantum channels, the entanglement generation rate without purification is very low. On the other hand, if the quantum states are not purified, but the noisy quantum channels between the repeater nodes are “superactivated”, then the rate of entanglement generation for the hybrid quantum repeater becomes about eight-times higher in average in comparison to the case if the quantum states are not purified and “standard” noisy quantum channels were used.

## G.5 Related Work on Quantum Repeaters

In this section we summarize the most important works regarding on the superactivation of quantum channel capacities.

### Further Superactive Channel Combinations

Smith *et al.* have also shown in 2011 that there exist a channel combination—using optical fiber quantum channels—for which superactivation of quantum capacity can be realized in practice [Smith11]. In 2010, a geometrical method for the discovery of further combinations of these superactive quantum channels was developed by Gyongyosi and Imre, [Gyongyosi11a-b].

Gyongyosi and Imre in 2011 showed that the phenomenon of superactivation can be used to develop more efficient quantum repeaters with the elimination of the inefficient entanglement purification process [Gyongyosi11c], [Gyongyosi11e-j].

### The Quantum Repeater

The biggest problem in the quantum communications of the future is long-distance quantum communication. Since quantum states cannot be copied, the amplification of quantum bits is a more complex question than in the case of classical communication. The success of global quantum communication very strongly depends on the development of a quantum repeater. The design of quantum repeater mainly based on optical elements [Imre12]. About long-distance quantum communication with optical elements, see the work of Duan *et al.* [Duan01]. The role of entanglement and the various encoding schemes can be applied in quantum repeaters can be found in the works of Ladd *et al.* [Ladd06] and Loock *et al.* [Loock08]. An analysis about the properties of quantum repeaters based on atomic ensembles and linear optics can be found in the article of Sangouard *et al.* [Sangouard09]. The role of the various photonics modules in the design of quantum repeater was studied by Stephens *et al.* [Stephens08].

The first entanglement purification approaches were developed by Dür *et al.* in 1999, [Dür99]. Later, many new purification algorithms were published, such as the greedy scheduling method, introduced by Ladd *et al.* in 2006 [Ladd06], or the method of entanglement pumping, which also was introduced by Dür *et al.* in 2007 [Dür07]. A more efficient purification algorithm was developed by Van Meter *et al.* [VanMeter08], called the banded purification scheme. Van Meter *et al.* have also studied the network integration of quantum repeaters, and the various algorithms for entanglement purification [VanMeter09]. On the entanglement purification and quantum error correction an article was published in 2007, by Dür and Briegel, for details see [Dür07]. For further information about the purification protocol to increase the fidelity of entanglement transmission, see [Fedrizzi09].

**Quantum Repeaters in Practice**

A practical approach of the quantum repeater is called the "hybrid quantum repeater", further information about the practical detail can be found in [VanMeter09], [Munro08], [Jiang08]. The *base stations* of the quantum repeaters could be connected by optical fibers, the entangled quantum states are sent through these fibers. The encoding schemes of quantum repeaters were studied by Jiang *et al.* in 2008, for details see [Jiang08]. Further information about the process of entanglement transmission between the repeater nodes can be found in [Devitt08], [Louis08], [Sangouard09]. About the connection of high-performance networks and the role of quantum repeaters, see the work of Munro *et al.* [Munro09] and Duan *et al.* [Duan01]. A very important work on the performance of practical quantum repeaters was published by Bernardes *et al.* in 2010, for details see [Bernardes10]. A practical implementation of optical-based quantum repeater was shown in the work of Munro *et al.* [Munro10]. A practical implementation of quantum repeater was presented by a group with researchers from the University of Vienna and from China [Yuan08]. In 2010, an other important analytical approach was presented with nearly maximally entangled EPR pairs over 1280 kilometers [Bernardes10].

# List of Notations

| | |
|---|---|
| $\otimes$ | Tensor product. |
| $*$ | Hadamard-product. |
| $\mathcal{H}$ | Hilbert space |
| $\mathcal{H}^d$ | $d$ dimensional Hilbert space |
| $\mathbb{C}^d$ | $d$ dimensional complex space |
| $\lvert\psi\rangle$,$\langle\psi\rvert$ | Dirac's ket and bra vectors, bra is dual to vector $\lvert\psi\rangle = \left(\lvert\psi\rangle\right)^{\dagger} = \left(\left(\lvert\psi\rangle\right)^{T}\right)^{*}$. |
| $\alpha^{*}$ | Complex conjugate of probability amplitude $\alpha$ . |
| $\lvert\psi\rangle^{\perp}$ | Vector orthogonal to vector $\lvert\psi\rangle$ . |
| $\alpha$ , $\beta$ | Probability amplitudes of generic state $\lvert\psi\rangle$ . |
| $\rho = \sum_i p_i \lvert\psi_i\rangle\langle\psi_i\rvert$ | Density matrix. |
| $Tr\left(\cdot\right)$ | Trace operation. |
| $i$ | The complex imaginary, $i^2 = -1$. |
| $\rho = \frac{1}{2}\begin{pmatrix} 1+z & x-iy \\ x+iy & 1-z \end{pmatrix}$ | Density matrix of two-level quantum system. |
| $P_m$ | Projector, if a quantum system is measured in an orthonormal basis $\lvert m\rangle$ , then we make a projective |
| | measurement with projector $P_m = \|m\rangle\langle m\|$. |
| $\mathcal{Z}$ | The von Neumann operator, $\mathcal{Z}=\sum_m \lambda_m P_m$, where $P_m$ is a projector to the eigenspace of $\mathcal{Z}$ with eigenvalue $\lambda_m$. For the projectors $\sum_m P_m = I$, and they are pairwise orthogonal. |
| $\mathcal{P}$ | Set of POVM operators $\{\mathcal{M}_i\}_{i=1}^{m+1}$, where $\mathcal{M}_i = \mathcal{Q}_i^\dagger \mathcal{Q}_i$, and $\mathcal{Q}_i$ is the measurement operator. For POVM operators $\mathcal{M}_i$ the completeness relation holds, $\sum_i \mathcal{M}_i = I$. |
| $H(\cdot)$ | Classical Shannon-entropy. |
| $H(A\|B)$ | Conditional Shannon entropy. |
| $\mathrm{S}(\rho) = -Tr(\rho\log\rho)$ | The von-Neumann entropy of $\rho$. |
| $D(\rho\\|\sigma)$ | Quantum relative entropy between quantum systems $\rho$ and $\sigma$, $Tr(\rho(\log\rho - \log\sigma))$. |
| $F = \langle\psi\|\rho\|\psi\rangle$ | Fidelity. |
| $U$ | Unitary transformation, $UU^{-1} = I$. |
| $U^\dagger$ | Operator dual (adjugate) to operator $U$. |
| $U^{-1}$ | Inverse of operator $U$. |
| $U^n$ | Operator $U$ for $n$-times. |
| $I$ | Identity operator, identity matrix. |
| $H$ | Hadamard transformation. |
| $X,\ Y,\ Z$ | Pauli $X$, $Y$ and $Z$ transformations. |
| $\|\beta_{00}\rangle, \|\beta_{01}\rangle, \|\beta_{10}\rangle, \|\beta_{11}\rangle$ | Bell states (EPR states). |
| $\sigma_x, \sigma_y, \sigma_z$ | Pauli operations $X$, $Y$ and $Z$. |
| $\{p_k, \rho_k\}$ | Ensemble. |
| $Tr_A(\rho_{AB})$ | Partial trace of the operator $\rho_{AB}$, tracing out system $A$ from the composite system $AB$. |
| $\{x_1,\ldots,x_n\} = \{x_i\}_{i=1}^{n}$ | The set containing $x_1,\ldots,x_n$. |
| $M_m$ | Measurement operator. |
| $N$ | Classical channel. |
| $\mathcal{N}$ | Quantum channel. |
| $\mathcal{N}_{EB}$ | Entanglement-breaking channel. |
| $\mathcal{N}_{H}$ | Horodecki channel. |
| $\mathcal{N}_{Had.}$ | Hadamard channel. |
| $\mathcal{A}_e$ | Erasure quantum channel. |
| $\mathcal{D}$ | Degrading quantum channel. |
| $\mathcal{N}_2 = \mathcal{D}\mathcal{N}_1$ | Degradable quantum channel, where $\mathcal{D}$ is the degrading quantum channel, and $\mathcal{N}_1$ has lower noise than $\mathcal{N}_2$. |
| $\mathcal{E}$ | Encoder operator. |
| $\mathcal{D}$ | Decoder operator. |
| $\mathcal{N}_1 \otimes \mathcal{N}_2$ | Joint structure of quantum channels $\mathcal{N}_1$ and $\mathcal{N}_2$. |
| $\mathcal{N}^{\otimes n}$ | Multiple uses ($n$) of quantum channel $\mathcal{N}$ (joint structure of $n$ parallel channels). |
| $C(\mathcal{N}_1) + C(\mathcal{N}_2)$ | Joint channel capacity of tensor product quantum channels $\mathcal{N}_1$ and $\mathcal{N}_2$. |
| $C(\mathcal{N}_1 \otimes \mathcal{N}_2)$ | Joint cannel capacity of entangled channel structure $\mathcal{N}_1 \otimes \mathcal{N}_2$. |
| $C_{PROD.}(\mathcal{N}_1 \otimes \mathcal{N}_2)$ | Joint channel capacity of quantum channels $\mathcal{N}_1$ and $\mathcal{N}_2$, for product state inputs. |
| $C_{ENT.}(\mathcal{N}_1 \otimes \mathcal{N}_2)$ | Joint channel capacity of quantum channels $\mathcal{N}_1$ and $\mathcal{N}_2$, for entangled input states. |
| $C(\mathcal{N})$ | Asymptotic classical capacity. |
| $C^{(1)}(\mathcal{N})$ | Single-use classical capacity. |
| $C_0^{(1)}(\mathcal{N})$, $C_0(\mathcal{N})$ | Classical zero-error capacity (single-use, asymptotic). |
| $Q(\mathcal{N})$ | Asymptotic quantum capacity. |
| $Q^{(1)}(\mathcal{N})$ | Single-use quantum capacity. |
| $Q_0^{(1)}(\mathcal{N})$, $Q_0(\mathcal{N})$ | Quantum zero-error capacity (single-use, asymptotic). |
| $P(\mathcal{N})$ | Asymptotic private classical capacity. |
| $P^{(1)}(\mathcal{N})$ | Single-use private classical capacity. |
| $C_E(\mathcal{N})$ | Entanglement assisted classical capacity. |
| $Q_{\mathcal{A}}(\mathcal{N})$ | Assisted quantum capacity, where channel $\mathcal{A}$ assists with $\mathcal{N}$. |
| $C_{ALL}(\mathcal{N})$ | Generalized capacity of quantum channels, involves the classical capacities and the quantum capacity. |
| $H(A,B)$ | Joint entropy. |
| $I(A:B)$ | Mutual information, the amount of information between two random variables $A$ and $B$. |
| $\mathrm{S}(A\|B)=\mathrm{S}(\rho_A\|\rho_B)$ | Quantum conditional entropy of quantum systems $\rho_A$ and $\rho_B$. |
| $\mathrm{S}(A)=\mathrm{S}(\rho_A)$ | The von Neumann entropy of density operator $\rho_A$. |
| $\mathrm{S}(AB)=\mathrm{S}(\rho_{AB})$ | Quantum joint entropy of quantum systems $\rho_A$ and $\rho_B$. |
| $I(A:B)=I(\rho_A:\rho_B)$ | Quantum mutual information, the amount of classical correlation between quantum systems $\rho_A$ and $\rho_B$. |
| $\chi$ | Holevo quantity. |
| $\chi_{\mathcal{N}}$ | Holevo quantity of quantum channel $\mathcal{N}$. |
| $\chi(\mathcal{N})=\max_{all\ p_i,\rho_i}\chi_{\mathcal{N}}$ | Maximized Holevo quantity of quantum channel $\mathcal{N}$. |
| $R$ | Channel rate of the communication channel. |
| $I_{coh}(\rho_A:\mathcal{N}(\rho_A))$ | Quantum coherent information between input and output quantum systems $\rho_A$ and $\mathcal{N}(\rho_A)$. |
| $\mathcal{L}$ | Eavesdropper's activity on the quantum channel. |
| $P$ | Purification state. |